\documentclass[12pt]{article}
\usepackage{graphics,cite,amssymb,epsfig,float,psfrag,axodraw}
\usepackage[usenames,dvips]{color}
\usepackage{rotating}

\def\red{\color{Red}}
\def\blue{\color{Blue}}

\begin{document}

\newcommand{\lsim}{\raisebox{-0.13cm}{~\shortstack{$<$ \\[-0.07cm] $\sim$}}~}
\newcommand{\gsim}{\raisebox{-0.13cm}{~\shortstack{$>$ \\[-0.07cm] $\sim$}}~}
\newcommand{\dx}{\mbox{\rm d}}
\newcommand{\ra}{\rightarrow}
\newcommand{\lra}{\longrightarrow}
\newcommand{\ee}{e^+e^-}
\newcommand{\gam}{\gamma \gamma}
\newcommand{\tb}{\tan \beta}
\newcommand{\s}{\smallskip}
\newcommand{\nn}{\noindent}
\newcommand{\non}{\nonumber}
\newcommand{\beq}{\begin{eqnarray}}
\newcommand{\eeq}{\end{eqnarray}}
\newcommand{\pslash}{\not\hspace*{-1.6mm}p}
\newcommand{\kslash}{\not\hspace*{-1.6mm}k}
\newcommand{\lslash}{\not\hspace*{-1.6mm}l}
\newcommand{\eslash}{\hspace*{-1.4mm}\not\hspace*{-1.6mm}E}
\newcommand{\bb}{\blue{\large $\bullet$}}
\newcommand{\rb}{\red{\large $\bullet$}}
\newcommand{\bH}{\blue{$H$}}
\newcommand{\ib}{{\it ibid.\ }}
\baselineskip=17pt
\thispagestyle{empty}

\hfill LPT--Orsay--05--17

\hfill March 2005

\vspace*{.5cm}

\begin{center}

{\sc\Large\bf The Anatomy of Electro--Weak Symmetry Breaking}

\vspace{0.5cm}

{\large\sc\bf Tome I: The Higgs boson in the Standard Model}

\vspace{0.7cm}

{\sc\large Abdelhak DJOUADI} 
\vspace*{7mm} 

Laboratoire de Physique Th\'eorique d'Orsay, UMR8627--CNRS,\\
Universit\'e Paris--Sud, B\^at. 210, F--91405 Orsay Cedex, France. 
\vspace*{2mm}

Laboratoire de Physique Math\'ematique et Th\'eorique, UMR5825--CNRS,\\
Universit\'e de Montpellier II, F--34095 Montpellier Cedex 5, France. 
\vspace*{2mm}

E--mail : {\tt Abdelhak.Djouadi@cern.ch}

\vspace*{2mm}

\end{center} 

\vspace*{5mm} 

\begin{abstract} 

\nn This review is devoted to the study of the mechanism of electroweak 
symmetry breaking and this first part focuses on the Higgs particle of the
Standard Model. The fundamental properties of the Higgs boson are reviewed
and its decay modes and production mechanisms at hadron colliders and at 
future lepton colliders are described in detail. 
\end{abstract}

\begin{figure}[!h]
\begin{center}
\vspace*{-1.8cm}
\hspace*{-2.2cm}
\mbox{
\epsfig{file=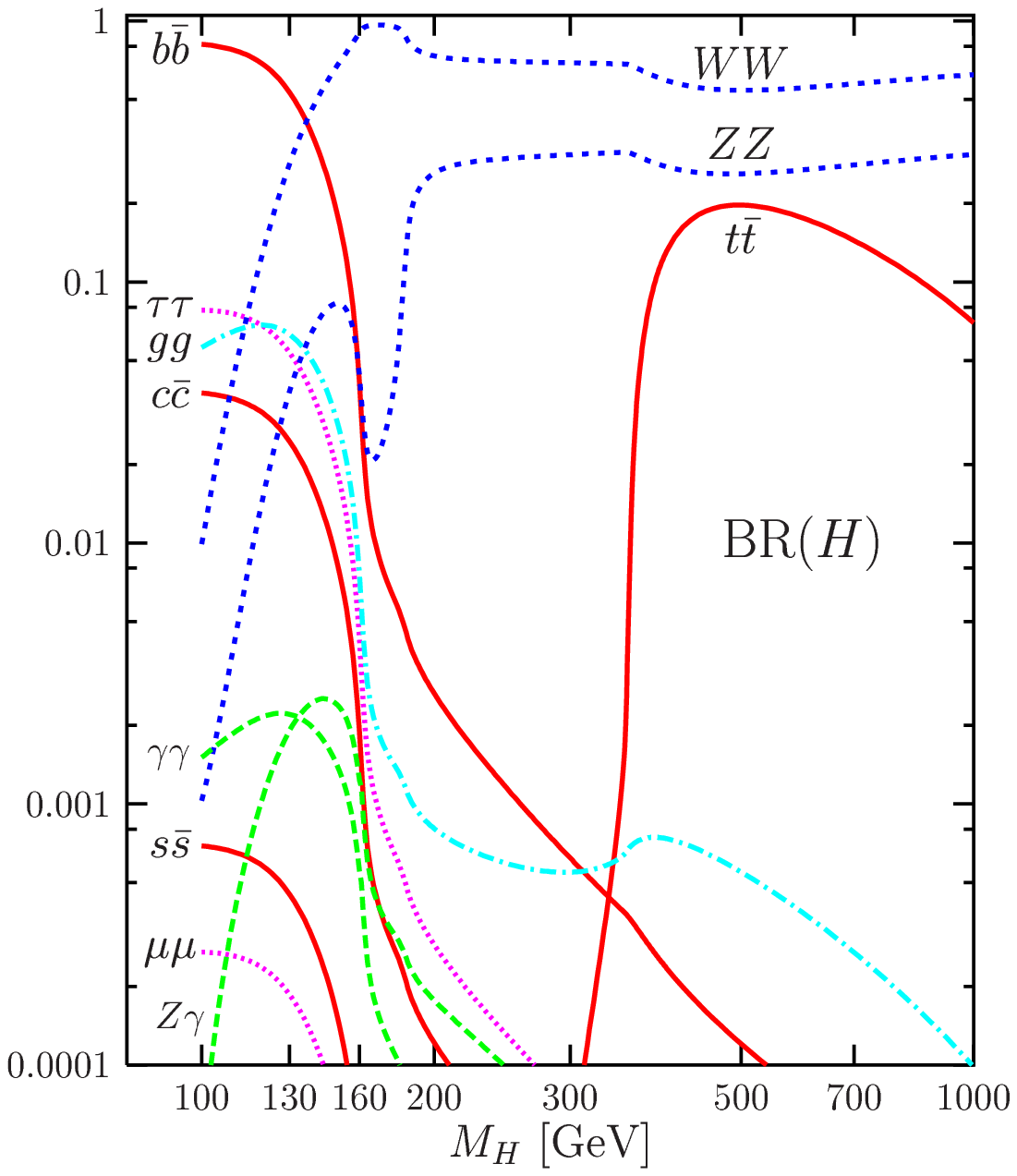,width=9.3cm}\hspace*{-4.3cm}
\epsfig{file=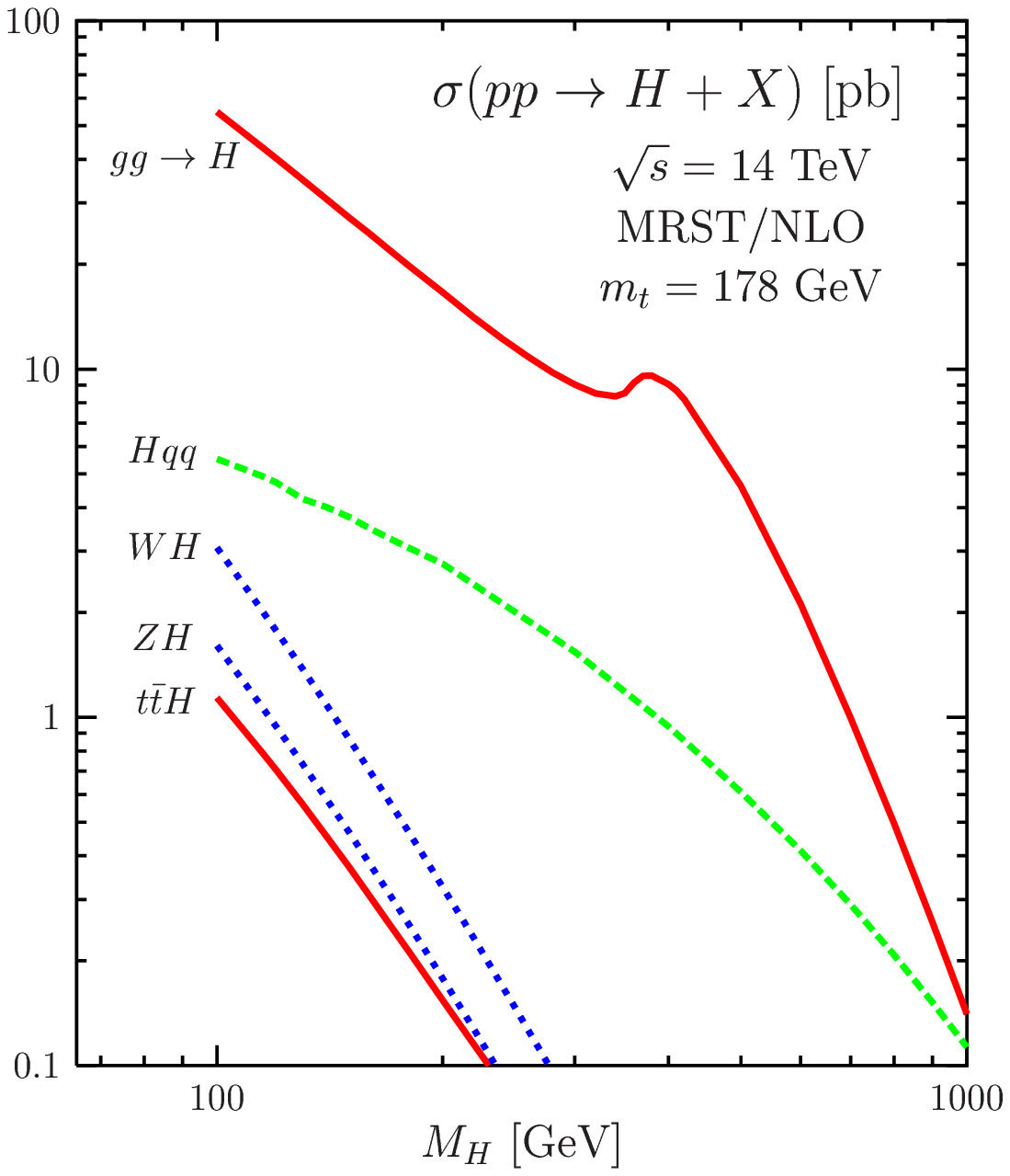,width=9.3cm}\hspace*{-4.3cm}
\epsfig{file=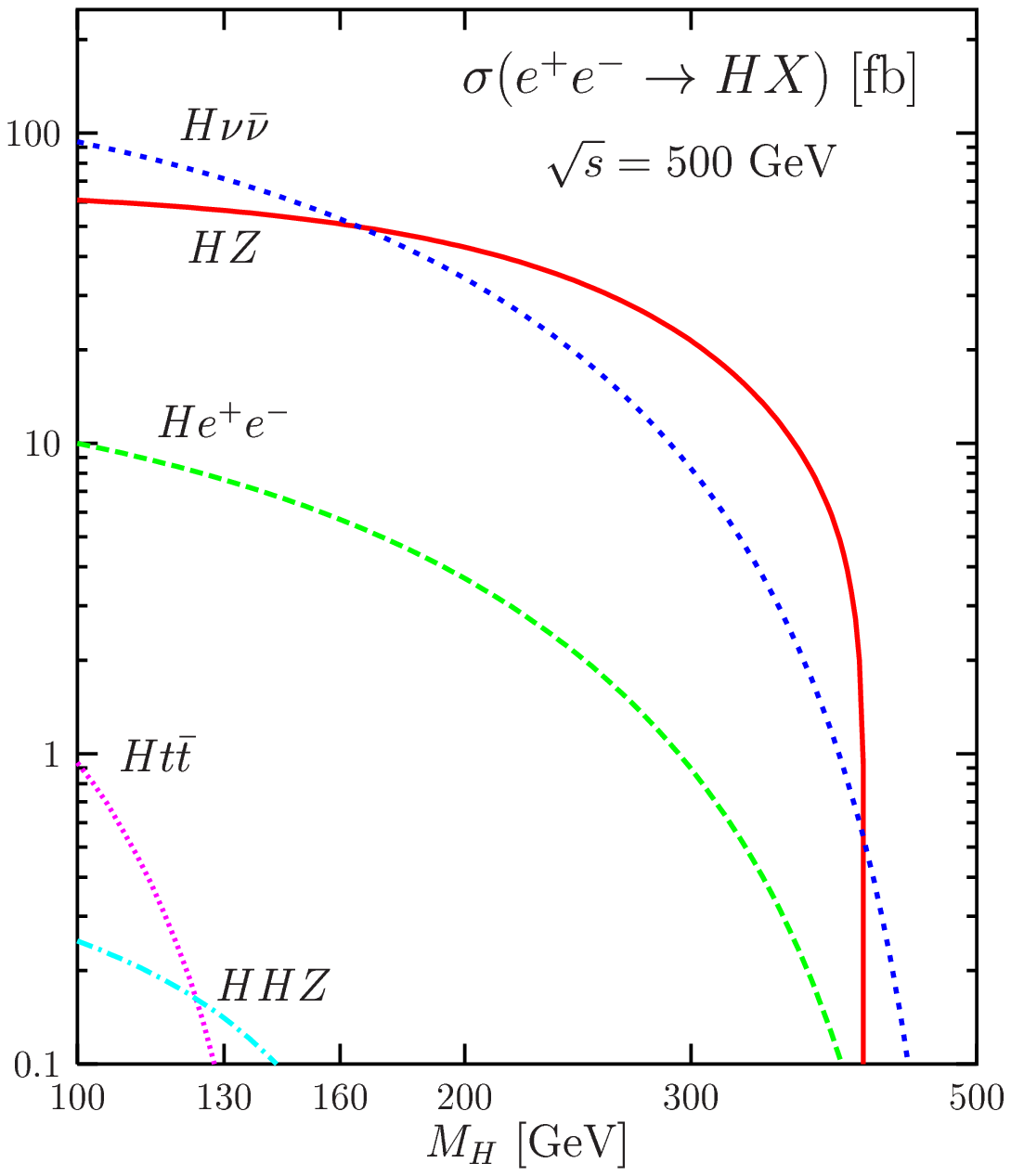,width=9.3cm}
} 
\end{center}
\vspace*{-6.8cm}
\begin{center}
\nn \small{\it 
The decay branching ratios of the Standard Model Higgs boson and its production 
cross sections in the main channels at the LHC and at a 500 GeV $\ee$ collider.}
\end{center}
\vspace*{-.6cm}
\end{figure}

\newpage

\setcounter{page}{2}

\baselineskip=16.7pt
\tableofcontents

\vspace*{.3cm}

\hspace*{.1cm} {\bf References} \hfill {\bf 295} 

\newpage
\baselineskip=17pt

\setcounter{section}{0} 
\renewcommand{\thesection}{}

\section{\hspace*{-.5cm} Pr\'eambule}

\subsubsection*{A short praise of the Standard Model}

The end of the last millennium witnessed the triumph of the Standard Model (SM)
of the electroweak and strong interactions of elementary particles
\cite{GSW,QCD}.  The electroweak theory, proposed by Glashow, Salam and
Weinberg \cite{GSW} to describe the electromagnetic \cite{QED} and weak
\cite{WEAK} interactions between quarks and leptons, is based on the gauge
symmetry group ${\rm SU(2)_L \times U(1)_Y}$ of weak left--handed isospin and
hypercharge. Combined with Quantum Chromo--Dynamics (QCD) \cite{QCD}, the
theory of the strong interactions between the colored quarks based on the 
symmetry group ${\rm SU(3)_C}$, the model provides a unified framework to
describe these three forces of Nature. The theory is perturbative at
sufficiently high energies \cite{QCD} and renormalizable \cite{RENORM}, and
thus describes these interactions at the quantum level.\s

A cornerstone of the SM is the mechanism of spontaneous electroweak symmetry
breaking (EWSB) proposed forty years ago by Higgs, Brout, Englert, Guralnik,
Hagen and Kibble \cite{Higgs} to generate the weak vector boson masses in a way
that is minimal and, as was shown later, respects the requirements of
renormalizability \cite{RENORM} and unitarity \cite{UNITARITY}.  An SU(2)
doublet of complex scalar fields is introduced and its neutral component
develops a non--zero vacuum expectation value. As a consequence, the
electroweak ${\rm SU(2)_L \times U(1)_Y}$ symmetry is spontaneously broken to
the electromagnetic ${\rm U(1)_Q}$ symmetry.  Three of the four degrees of
freedom of the doublet scalar field are absorbed by the $W^\pm$ and $Z$ weak
vector bosons to form their longitudinal polarizations and to acquire masses. 
The fermion masses are generated through a Yukawa interaction with the same
scalar field and its conjugate field.  The remaining degree of freedom
corresponds to a scalar particle, the Higgs boson. The discovery of this new
type of matter particle is unanimously considered as being of profound
importance.\s

The high--precision measurements of the last decade \cite{High-Precision,PDG}
carried out at LEP, SLC, Tevatron and elsewhere have provided a decisive test
of the Standard Model and firmly established that it provides the correct
effective description of the strong and electroweak interactions at present
energies. These tests, performed at the per mille level accuracy, have probed
the quantum corrections and the structure of the ${\rm SU(3)_C\times SU(2)_L
\times U(1)_Y}$ local symmetry.  The couplings of quarks and leptons to the
electroweak gauge bosons have been measured precisely and agree with those
predicted by the model. The trilinear couplings among electroweak vector bosons
have also been measured and agree with those dictated by the ${\rm
SU(2)_L\times U(1)_Y}$ gauge symmetry. The ${\rm SU(3)_C}$ gauge symmetric
description of the strong interactions has also been thoroughly tested at LEP
and elsewhere.  The only sector of the model which has not yet been probed in a
satisfactory way is the scalar sector. The missing and most important
ingredient of the model, the Higgs particle, has not been observed
\cite{PDG,LEP2-Higgs-exp} and only indirect constraints on its mass have
been inferred from the high--precision data \cite{High-Precision}.\s

\subsubsection*{Probing electroweak symmetry breaking: a brief survey of recent
developments} 

The SM of the electroweak interactions, including the EWSB mechanism for
generating particle masses, had been proposed in the mid--sixties; however, it
was only in the mid--seventies, most probably after the proof by 't Hooft and
Veltman that it is indeed a renormalizable theory \cite{RENORM} and the
discovery of the weak neutral current in the Gargamelle experiment
\cite{Gargamelle}, that all its facets began to be investigated thoroughly. 
After the discovery of the $W^\pm$ and $Z$ bosons at CERN \cite{WZ-Discovery},
probing the electroweak symmetry breaking  mechanism became a dominant theme of
elementary particle physics. The relic of this mechanism, the Higgs particle,
became the Holy Grail of high--energy collider physics and {\it l'objet de tous
nos d\'esirs}. Finding this particle and studying its fundamental properties
will be the major goal of the next generation of high--energy machines [and of
the upgraded Tevatron, if enough lumino\-sity is collected]: the CERN Large
Hadron Collider (LHC), which will start operation in a few years, and the next
high--energy and high--luminosity electron--positron linear collider, which
hopefully will allow very detailed studies of the EWSB mechanism in a decade.\s
 
In the seventies and eighties, an impressive amount of theoretical knowledge
was amassed on EWSB and on the expected properties of the Higgs boson(s), both
within the framework of the SM and  of its [supersymmetric and non 
supersymmetric]
extensions.  At the end of the eighties, the basic properties of the Higgs
particles had been discussed and their principal decay modes and main
production mechanisms at hadron and lepton colliders explored.  This
monumental endeavor was nicely and extensively reviewed in a celebrated book,
{\it The Higgs Hunter's Guide} \cite{HHG} by Gunion, Haber, Kane and
Dawson.  The constraints from the experimental data available at that time and
the prospects for discovering the Higgs particle(s) at the upcoming
high--energy experiments, the LEP, the SLC, the late SSC and the LHC, as well
as possible higher energy $\ee$ colliders, were analyzed and summarized. The
review indeed guided theoretical and phenomenological studies as well as
experimental searches performed over the last fifteen years.\s 

Meanwhile, several major developments took place. The LEP experiment, for which
the search for the Higgs boson was a central objective, was completed with mixed
results. On the one hand,  LEP played a key role in establishing the SM as the
effective theory of the strong and electroweak forces at presently accessible
energies. On the other hand, it unfortunately failed to find the Higgs particle
or any other new particle which could play a similar role.  Nevertheless, this
negative search led to a very strong limit on the mass of a SM--like Higgs
boson, $M_H \gsim 114.4$ GeV \cite{LEP2-Higgs-exp}.  This unambiguously ruled
out a broad low Higgs mass region, and in particular the range $M_H \lsim 5$
GeV, which was rather difficult to explore\footnote{This is mainly due to the
hadronic uncertainties which occur for such a small Higgs mass. Almost an
entire chapter of Ref.~\cite{HHG} was devoted to this mass range; see
pp.~32--56 and 94--129.}  before the advent of LEP1 and its very clean
experimental environment.  The mass range $M_H \lsim 100$ GeV would have been
extremely difficult to probe at very high--energy hadron colliders such as the
LHC.  At approximately the same period, the top quark was at last discovered at
the Tevatron \cite{Top-Discovery}. The determination of its mass entailed
that all the parameters of the Standard Model, except the Higgs boson mass, 
were then known\footnote{Another important outcome is due to the heaviness of 
the top quark \cite{Mt-Tevatron}: the search of the Higgs boson would have been
extremely more difficult at hadron colliders if the top quark mass were smaller
than $M_W$, a possibility for which many analyses were devoted in the past and
which is now ruled out. As a by--product of the large $m_t$ value, the cross
sections for some Higgs production channels at both hadron and $\ee$ machines
became rather large, thus increasing the chances for the discovery and/or 
study of the particle.}, implying that the profile of the Higgs boson will be
uniquely determined once its mass is fixed. \s 

Other major developments occurred in the planning and design of the
high--energy colliders. The project of the Superconducting Super Collider has
been unfortunately terminated and the energy and luminosity parameters of the
LHC became firmly established\footnote{The SSC was a project for a hadron
machine with a center of mass energy of $\sqrt{s}=40$ TeV and a yearly
integrated luminosity of 10 fb$^{-1}$ on which most of the emphasis for Higgs
searches at hadron colliders was put in Ref.~\cite{HHG}. Of course these
studies can be and actually have been adapted to the case of the LHC. Note
that in the late eighties, the c.m.  energy and the luminosity of the LHC were
expected to be $\sqrt{s}=17$ TeV and ${\cal L}=10^{33}$ cm$^{-2}$s$^{-1}$,
respectively, and the discovery range for the SM Higgs boson was considered to
be rather limited, $2M_W \lsim M_H \lsim 300$ GeV \cite{HHG}.}.  Furthermore,
the option of upgrading the Tevatron by raising the c.m.~energy and, more
importantly, the luminosity to a value which allows for Higgs searches in the
mass range $M_H \lsim 2M_Z$ was not yet considered. In addition, the path
toward future high--energy electron--positron colliders became more precise. 
The feasibility of the next generation machines, that is, $\ee$ linear
colliders operating in the energy range from $M_Z$ up to 1 TeV with very high
luminosities has been demonstrated [as in the case of the TESLA machine] and a
consensus on the technology of the future International Linear Collider (ILC)
has recently emerged. The designs for the next generation machines running at
energies in the multi--TeV range [such as the CLIC machine at CERN] also made
rapid developments.  Added to this, the option of turning future linear
colliders into high--energy and high--luminosity $\gamma \gamma$ colliders by
using Compton back--scattering of laser light off the high--energy electron
beams and the possibility of high--energy muon colliders have been seriously
discussed only in the last decade. \s

In parallel to these experimental and technological developments, a huge amount
of effort has been devoted to the detailed study of the decay and production
properties of the Higgs particle at these colliders. On the theoretical side,
advances in computer technology allowed one to perform almost automatically
very complicated calculations for loop diagrams and multi--particle processes
and enabled extremely precise predictions.  In particular, the
next--to--leading order radiative corrections to Higgs production in all the
important processes at hadron and $\ee$ colliders were calculated\footnote{This
started in the very late eighties and early nineties, when the one--loop QCD
corrections to associated Higgs production with $W/Z$ bosons and the $WW/ZZ$
and gluon--gluon fusion mechanisms at hadron colliders and the electroweak
corrections to the Higgs--strahlung production mechanism at $\ee$ colliders
have been derived, and continued until very recently when the QCD corrections
to associated Higgs production with heavy quarks at hadron colliders and the
electroweak corrections to all the remaining important Higgs production
processes at lepton colliders have been completed.}.  The radiative corrections
to the cross sections for some production processes, such as Higgs--strahlung
and gluon--gluon fusion at hadron colliders, have been calculated up to
next--to--next--to--leading order accuracy for the strong interaction part and
at next--to--leading order for the electroweak part, a development which
occurred only over the last few years. A vast literature on the higher order
effects in Higgs boson decays has also appeared in the last fifteen years and
some decay modes have been also investigated to next--to--next--to--leading
order accuracy and, in some cases, even beyond.  Moreover, thorough theoretical
studies of the various distributions in Higgs production and decays and new
techniques for the determination of the fundamental properties of the Higgs
particle [a vast subject which was only very briefly touched upon in
Ref.~\cite{HHG} for instance] have been recently carried out.\s

Finally, a plethora of analyses of the various Higgs signals and backgrounds,
many detailed parton--level analyses and Monte--Carlo simulations taking into
account the experimental environment [which is now more or less established, at
least for the Tevatron and the LHC and possibly for the first stage of the
$\ee$ linear collider, the ILC] have been performed to assess to what extent 
the Higgs particle can be observed and its properties studied in given 
processes at the various machines.

\subsubsection*{Objectives and limitations of the review}

On the experimental front, with the LEP experiment completed, we await the
accumulation of sufficient data from the upgraded Tevatron and the launch of
the LHC which will start operation in 2007. At this point, we believe that it
is useful to collect and summarize the large amount of work carried out over
the last fifteen years in preparation for the challenges ahead. This review is
an attempt to respond to this need. The review is structured in three parts. 
In this first part, we will concentrate on the Higgs boson of the Standard
Model, summarize the present experimental and theoretical information on the
Higgs sector, analyze the decay modes of the Higgs bosons including all the
relevant and important higher order effects, and discuss the production
properties of the Higgs boson and its detection strategies at the various
hadron and lepton machines presently under discussion. We will try to be 
as extensive and comprehensive as possible.\s 

However, because the subject is vast and the number of studies related to it is
huge\footnote{Simply by typing ``find title Higgs" in the search field of the
Spires database, one obtains more than 6.700 entries. Since this number does
not include all the articles dealing with the EWSB mechanism and not explicitly
mentioning the name of Prof. Higgs in the title, the total number of articles
written on the EWSB mechanism in the SM and its various extensions may, thus,
well exceed the level of 10.000.}, it is almost an impossible task to review
all its aspects. In addition, one needs to cover many different topics and each
of them could have [and, actually, often does have] its own review.  Therefore,
in many instances, one will have to face [sometimes Cornelian...] choices. The
ones made in this review  will be, of course, largely determined by the taste
of the author, his specialization and his own prejudice. I therefore apologize
in advance if some important aspects are overlooked and/or some injustice to
possibly relevant analyses is made.  Complementary material on the foundations
of the SM and the Higgs mechanism, which will only be briefly sketched here,
can be found in standard textbooks \cite{BOOKS} or in general reviews
\cite{SM-REVIEWS,Reviews-Higgs} and an account of the various calculations,
theoretical studies and phenomenological analyses mentioned above can be found
in many specialized reviews; see
Refs.~\cite{RCreviewEW,Reviews-AD,Review-Michael,RCreviewQCD,Review-CH} for
some examples. For the physics of the Higgs particle at the various colliders,
in particular for the discussion of the Higgs signals and their respective
backgrounds, as well as for the detection techniques, we  will simply summarize
the progress so far.  For this very important issue, we refer for additional
and more detailed informations to specialized reviews and, above all, to the
proceedings which describe the huge collective efforts at the various workshops
devoted to the subject.  Many of these studies and reviews will be referenced
in due time.  

\subsubsection*{Synopsis of the review}

The first part of this review (Tome I) on the electroweak symmetry breaking
mechanism is exclusively devoted to the SM Higgs particle. The discussion
of the Higgs sector of the Minimal Supersymmetric extension of the SM is 
given in an accompanying report \cite{Tome2}, while the EWSB mechanism in other
supersymmetric and non--supersymmetric extensions of the SM will be discussed
in a forthcoming report \cite{Tome3}. In our view, the SM incorporates an
elementary Higgs boson with a mass below 1 TeV and, thus, the very heavy or the
no--Higgs scenarios will not be discussed here and postponed to
Ref.~\cite{Tome3}.\s 

The first chapter is devoted to the description of the Higgs sector of the SM. 
After briefly recalling the basic ingredients of the model and its input
parameters, including an introduction to the electroweak symmetry breaking
mechanism and to the basic properties of the Higgs boson, we discuss the
high--precision tests of the SM and introduce the formalism which allows a
description of the radiative corrections which involve the contribution of the
only unknown parameter of the theory, the Higgs boson mass $M_H$ or,
alternatively, its self--coupling. This formalism will be needed when we
discuss the radiative corrections to Higgs decay and production modes. We then
summarize the indirect experimental constraints on $M_H$ from the
high--precision measurements and the constraints derived from direct Higgs
searches at past and present colliders. We close this chapter by discussing
some interesting constraints on the Higgs mass that can be derived from
theoretical considerations on the energy range in which the SM is valid before
perturbation theory breaks down and new phenomena should emerge.  The bounds on
$M_H$ from unitarity in scattering amplitudes, perturbativity of the Higgs
self--coupling, stability of the electroweak vacuum and fine--tuning in the
radiative corrections in the Higgs sector, are analyzed. \s

In the second chapter, we explore the decays of the SM Higgs particle. We
consider all decay modes which lead to potentially observable branching 
fractions: decays into quarks and leptons, decays into weak
massive vector bosons and loop induced decays into gluons and photons. We
discuss not only the dominant two--body decays, but also higher order decays,
which can be very important in some cases. We pay particular attention to the
radiative corrections and, especially, to the next--to--leading order QCD
corrections to the hadronic Higgs decays which turn out to be quite large. 
The higher order QCD corrections [beyond NLO] and the important electroweak
radiative corrections to all decay modes are briefly summarized. The expected
branching ratios of the Higgs particle, including the uncertainties which
affect them, are given.  Whenever possible, we compare the various decay
properties of the SM Higgs boson, with its distinctive spin and parity $J^{\rm
PC}=0^{++}$ quantum numbers, to those of hypothetical pseudoscalar Higgs bosons
with $J^{\rm PC}=0^{+-}$ which are predicted in many extensions of the SM Higgs
sector. This will highlight the unique prediction for the properties of the SM
Higgs particle [the more general case of anomalous Higgs couplings will be
discussed in the third part of this review]. \s

The third chapter is devoted to the production of the Higgs particle at hadron
machines. We consider both the $p\bar p$ Tevatron collider with a center of
mass energy of $\sqrt{s}=1.96$ TeV and the $pp$ Large Hadron Collider (LHC)
with a center of mass energy of $\sqrt{s}=14$ TeV. All the dominant production
processes, namely the associated production with $W/Z$ bosons, the weak vector
boson fusion processes, the gluon--gluon fusion mechanism and the associated
Higgs production with heavy top and bottom quarks, are discussed in detail.  In
particular, we analyze not only the total production cross sections, but also
the differential distributions and we pay  special attention to three 
important aspects: the QCD radiative corrections or the $K$--factors [and the
electroweak corrections when important] which are large in many cases, their
dependence on the renormalization and factorization scales which measures the
reliability of the theoretical predictions, and the choice of different sets of
parton distribution functions. We also discuss other production processes such
as Higgs pair production, production with a single top quark, production in
association with two gauge bosons or with one gauge boson and two quarks as
well as diffractive Higgs production. These channels are not considered as
Higgs discovery modes, but they might provide additional interesting
information.  We then summarize the main Higgs signals in the various detection
channels at the Tevatron and the LHC and the expectations for observing them 
experimentally.  At the end of this chapter, we briefly discuss the possible
ways of determining some of the properties of the Higgs particle at
the LHC: its mass and total decay width, its spin and parity quantum numbers
and its couplings to fermions and gauge bosons.  A brief summary of the
benefits that one can expect from raising the luminosity and energy
of hadron colliders is given.\s 

In the fourth chapter, we explore the production of the SM Higgs boson at
future lepton colliders. We mostly focus on future $\ee$ colliders in the
energy range $\sqrt{s}=350$--1000 GeV as planed for the ILC but we also discuss
the physics of EWSB at multi--TeV machines [such as CLIC] or by revisiting the
$Z$ boson pole [the GigaZ option], as well as at the $\gamma \gamma$ option of
the linear collider and at future muon colliders.  In the case of $\ee$
machines, we analyze in detail the main production mechanisms, the
Higgs--strahlung and the $WW$ boson fusion processes, as well as some
``subleading" but extremely important processes for determining the profile of
the Higgs boson such as associated production with top quark pairs and  Higgs
pair production.  Since $\ee$ colliders are known to be high--precision
machines, the theoretical predictions need to be rather accurate and we
summarize the work done on the radiative corrections to these processes [which
have been completed only recently] and to various distributions which allow to
test the fundamental nature of the Higgs particle.  The expectation for Higgs
production at the various possible center of mass energies and the potential of
these machines to probe the electroweak symmetry breaking mechanism in all its
facets and to check the SM predictions for the fundamental Higgs properties
such as the total width, the spin and  parity quantum numbers, the couplings to
the other SM particles [in particular, the important coupling to the top quark]
and the Higgs self--coupling [which allows the reconstruction of the scalar
potential which generates EWSB] are summarized. Higgs production at $\gamma
\gamma$ and at muon colliders are discussed in the two last sections, with some
emphasis on two points which are rather difficult to explore in $\ee$
collisions, namely, the determination of the Higgs spin--parity quantum numbers
and the total decay width.\s 

Since the primary goal of this review is to provide the necessary material to
discuss Higgs decays and production at present and future colliders, we present
the analytical expressions of the partial decay widths, the production cross
sections and some important distributions, including the higher order
corrections or effects, when they are simple enough to be displayed.  We
analyze in detail the main Higgs decay and production channels and also discuss
some channels which are not yet established but which can be useful and with
further effort might prove experimentally accessible. We also present summary
and updated plots as well as illustrative numerical examples [which can be used
as a normalization in future phenomenological and experimental studies] for the
total Higgs decay width and branching ratios, as well as for the cross sections
of the main production mechanisms at the Tevatron, the LHC and future $\ee$
colliders at various center of mass energies. In these updated analyses, we have
endeavored to include all currently available information. For collider Higgs
phenomenology, in particular for the discussion of the Higgs signals and
backgrounds, we  simply summarize, as previously mentioned, the main points and
refer to the literature for additional details and complementary discussions.  

\subsubsection*{Acknowledgments}

I would like first to thank the many collaborators with whom I shared the
pleasure to investigate various aspects of the theme discussed in this review. 
They are too numerous to be all listed here, but I would like at least to
mention Peter Zerwas with whom I started to work on the subject in an intensive
way.\s

I would also like to thank the many colleagues and friends who helped me during
the writing of this review and who made important remarks on the preliminary
versions of the manuscript and suggestions for improvements: Fawzi Boudjema,
Albert de Roeck, Klaus Desch, Michael Dittmar, Manuel Drees, Rohini Godbole,
Robert Harlander, Wolfgang Hollik, Karl Jakobs, Sasha Nikitenko, Giacomo
Polesello, Francois Richard, Pietro Slavich and Michael Spira.  Special thanks
go to Manuel Drees and Pietro Slavich for their very careful reading of large
parts of the manuscript and for their efficiency in hunting the many typos,
errors and awkwardnesses contained in the preliminary versions and for their
attempt at improving my poor English and fighting against my anarchic way of
``distributing commas". Additional help with the English by Martin Bucher and,
for the submission of this review to the archives, by Marco Picco are also
acknowledged. \s

I also thank the Djouadi {\it smala}, my sisters and brothers and their
children [at least one of them, Yanis, has already caught the virus of particle
physics and I hope that one day he will read this review], who bore my not
always joyful mood in the last two years. Their support was crucial for the
completion of this review. Finally, thanks to the team of {\it La Bonne
Franquette}, where in fact part of this work has been done, for their good
couscous and the nice atmosphere, as well as to Madjid Belkacem for sharing the
drinks with me.\s 

The writing of this review  started when I was at CERN as a scientific
associate, continued during the six months I spent at the LPTHE of Jussieu, 
and ended at the LPT d'Orsay. I thank all these institutions for their kind
hospitality. 

\newpage

\setcounter{section}{0} 
\renewcommand{\thesection}{\arabic{section}}

\setcounter{section}{0}
\renewcommand{\thesection}{\arabic{section}}

\setcounter{equation}{0}
\renewcommand{\theequation}{1.\arabic{equation}}

\section{The Higgs particle in the SM}
\subsection{The SM of the strong and electroweak interactions} 

In this section, we present a brief introduction to the Standard Model (SM)
of the strong and electroweak interactions and to the mechanism of electroweak 
symmetry breaking. This will allow us to set the stage and to fix the notation 
which will be used later on. For more detailed discussions, we refer the reader
to standard textbooks \cite{BOOKS} or reviews \cite{SM-REVIEWS}.

\subsubsection{The SM before electroweak symmetry breaking} 

As discussed in the preamble, the Glashow--Weinberg--Salam electroweak theory
\cite{GSW} which describes the electromagnetic and weak interactions between
quarks and leptons, is a Yang--Mills theory \cite{Yang-Mills} based on the
symmetry group ${\rm SU(2)_L \times U(1)_Y}$. Combined with the ${\rm SU(3)_C}$
based QCD gauge theory \cite{QCD} which describes the strong interactions
between quarks, it provides a unified framework to describe these three forces
of Nature: the Standard Model. The model, before introducing the electroweak
symmetry breaking mechanism to be discussed later, has two kinds of fields. \s

$\bullet$ There are first the matter fields, that is, the three generations of
left--handed and right--handed chiral quarks and leptons, $f_{L,R} =\frac{1}{2}
(1 \mp \gamma_5)f$. The left--handed fermions are in weak isodoublets, while
the right--handed fermions are in weak isosinglets\footnote{Throughout this 
review, we will assume that the neutrinos, which do not play any role here, are
massless and appear only with their left--handed components.}
\beq
I_f^{3L,3R}= \pm \frac{1}{2}, 0  \, : \, \begin{array}{l} 
L_1= \left( \begin{array}{c} \nu_e \\ e^- \end{array} \right)_L, \ e_{R_1}=
e^-_R  \, , \ Q_1= \left( \begin{array}{c} u \\ d \end{array} \right)_L,\ 
u_{R_1}=u_R \, ,\ d_{R_1}=d_R \\ 
L_2= \left( \begin{array}{c} \nu_\mu \\ \mu^- \end{array} \right)_L, \ 
e_{R_2} =\mu^-_R  \, , \ Q_2= \left( \begin{array}{c} c \\ s \end{array} 
\right)_L,\ u_{R_2}=c_R\ ,\ d_{R_2} = s_R \\ 
L_3= \left( \begin{array}{c} \nu_\tau \\ \tau^- \end{array} \right)_L, \ 
e_{R_3}=\tau^-_R  \, , \ Q_3= \left( \begin{array}{c} t \\ b \end{array} \right)_L,\ u_{R_3}=t_R\ ,\ d_{R_3}= b_R \\ 
\end{array}
\eeq
The fermion hypercharge, defined in terms of the third component of the weak 
isospin $I_f^3$ and the electric charge $Q_f$ in units of the proton charge 
$+e$, is given by ({\small $i$=1,2,3})
\beq
Y_f=2Q_f-2I_f^3 & \Rightarrow & Y_{L_i}=-1, \ Y_{e_{R_i}}=-2, \  
Y_{Q_i}=\frac{1}{3}, \ Y_{u_{R_i}}= \frac{4}{3}, \ Y_{d_{R_i}}= -\frac{2}{3} 
\eeq
Moreover, the quarks are triplets under the ${\rm SU(3)_C}$ group, while leptons
are color singlets. This leads to the relation 
\beq
\sum_f Y_f\! =\! \sum_f Q_f\!= \!0
\eeq
which ensures the cancellation of chiral anomalies \cite{Anomaly} within 
each generation, thus, preserving \cite{BIM-anomaly} the renormalizability 
of the electroweak theory \cite{RENORM}.  \s

$\bullet$ Then, there are the gauge fields corresponding to the spin--one bosons
that mediate the interactions. In the electroweak sector, we have the field 
$B_\mu$ which corresponds to the generator $Y$ of the U(1)$_{\rm Y}$ group and 
the three fields $W^{1,2,3}_\mu$ which correspond to the generators $T^{a}$  
[with {\small $a$=1,2,3}] of the SU(2)$_{\rm L}$ group; these generators are in 
fact equivalent  to half of the non--commuting $2 \times 2$ Pauli matrices
\beq
T^a= \frac{1}{2} \tau^a \, ; \quad 
\tau_1= \left( \begin{array}{cc} 0 & 1 \\ 1 & 0 \end{array} \right) \, , \ 
\tau_2= \left( \begin{array}{cc} 0 & -i \\ i & 0 \end{array} \right) \, , \ 
\tau_3= \left( \begin{array}{cc} 1 & 0 \\ 0 & -1 \end{array} \right)
\eeq
with the commutation relations between these generators given by
\beq
[T^a,T^b]=i\epsilon^{abc} T_c \ \ \ {\rm and} \ \ \  [Y, Y]=0 
\eeq
where $\epsilon^{abc}$ is the antisymmetric tensor. In the strong interaction 
sector, there is an octet of gluon fields $G_\mu^{1,\cdots,8}$ which correspond
to the eight generators of the ${\rm SU(3)_C}$ group [equivalent to half of the
eight $3\times 3$ anti--commuting Gell--Mann matrices] and which obey the 
relations
\beq
[T^a,T^b]=if^{abc} T_c \ \ \ {\rm with} \ \ \  {\rm Tr}[T^a T^b]=
\frac12 \delta_{ab} 
\eeq
where the tensor $f^{abc}$ is for the structure constants of the ${\rm 
SU(3)_C}$ group and where we have used the same notation as for the generators 
of SU(2) as little confusion should be possible. The field strengths are given 
by
\beq
G_{\mu \nu}^a &=& \partial_\mu G_\nu^a -\partial_\nu G_\mu^a +g_s \, 
f^{abc} G^b_\mu G^c_\nu \non \\
W_{\mu \nu}^a &=& \partial_\mu W_\nu^a -\partial_\nu W_\mu^a +g_2 \, 
\epsilon^{abc} W^b_\mu W^c_\nu \non \\ 
B_{\mu \nu} &=& \partial_\mu B_\nu -\partial_\nu B_\mu 
\eeq
where $g_s$, $g_2$ and $g_1$ are, respectively, the coupling constants of 
${\rm SU(3)_C}$,  ${\rm SU(2)_L}$ and  ${\rm U(1)_Y}$.  

Because of the non--abelian nature of the SU(2) and SU(3) groups, there are 
self--interactions between their gauge fields, $V_\mu \equiv W_\mu $ or $G_\mu$,
leading to 
\beq
{\rm triple\ gauge\ boson\ couplings} &:& i g_i \, {\rm Tr} (\partial_\nu V_\mu 
- \partial_\mu V_\nu)[V_\mu,V_\nu] \non \\
{\rm quartic\ gauge\ boson\ couplings} &:&  \frac12 g_i^2 \, {\rm Tr} 
[V_\mu,V_\nu]^2
\eeq 

The matter fields $\psi$ are minimally coupled to the gauge fields through the 
covariant derivative $D_\mu$ which, in the case of quarks, is defined as
\beq
D_{\mu} \psi = \left( \partial_\mu -ig_s T_a G^a_\mu -ig_2 T_a W^a_\mu -i 
g_1 \frac{Y_q}{2} B_\mu \right) \psi  
\label{CovariantDerivative}
\eeq
and which leads to unique couplings between the fermion and gauge fields 
$V_\mu$ of the form
\beq
{\rm fermion\ gauge\ boson\ couplings} &:& 
-g_i \overline \psi V_\mu \gamma^\mu \psi 
\eeq

The SM Lagrangian, without mass terms for fermions and gauge bosons is then 
given by 
\beq
\label{smlagrangian}
{\cal L}_{\rm SM}&=& -\frac{1}{4} G_{\mu \nu}^a G^{\mu \nu}_a 
-\frac{1}{4} W_{\mu \nu}^a W^{\mu \nu}_a -\frac{1}{4} 
B_{\mu \nu}B^{\mu \nu} \\ 
&& + \bar{L_i}\, i D_\mu \gamma^\mu \, L_i + \bar{e}_{Ri} \, i D_\mu 
\gamma^\mu \, e_{R_i} \ 
+ \bar{Q_i}\, i D_\mu \gamma^\mu \, Q_i + \bar{u}_{Ri} \, i D_\mu 
\gamma^\mu \, u_{R_i} \ + \bar{d}_{Ri} \, i D_\mu \gamma^\mu \, d_{R_i} \non 
\eeq
This Lagrangian is invariant under local ${\rm SU(3)_C \times SU(2)_L \times 
U(1)_Y}$ gauge transformations for fermion and gauge fields. In the case of 
the electroweak sector, for instance, one has  
\beq
L(x) \to L'(x)=e^{i\alpha_a(x) T^a + i \beta(x)Y } L(x) \ \ , \ \
R(x) \to R'(x)=e^{i \beta (x) Y} R(x) \non \\
\vec{W}_\mu (x) \to \vec{W_\mu}(x) -\frac{1}{g_2} \partial_\mu \vec{\alpha}(x)- 
\vec{\alpha}(x) \times \vec{W}_\mu(x) \ , \ B_\mu(x) \to B_\mu(x) -  \frac{1}
{g_1} \partial_\mu \beta (x) 
\eeq

Up to now, the gauge fields and the fermions fields have been kept massless. 
In the case of strong interactions, the gluons are indeed massless particles
while mass terms of the form  $-m_q\overline{\psi}\psi$ can be generated for
the colored quarks [and for the leptons] in an SU(3) gauge invariant way. In 
the case of the electroweak sector, the situation is more problematic:\s 

-- If we add mass terms, $\frac{1}{2} M_V^2 W_\mu W^\mu$, for the gauge bosons 
[since experimentally, they have been proved to be massive, the weak 
interaction being of short distance], this will violate local SU(2)$\times$U(1)
gauge invariance.
This statement can be visualized by taking the example of QED where the photon 
is massless because of the ${\rm U(1)_Q}$ local symmetry 
\beq
\frac{1}{2}M_A^2 A_\mu A^\mu \to \frac{1}{2}M_A^2 (A_\mu - \frac{1}{e} 
\partial_\mu \alpha) (A^\mu - \frac{1}{e} \partial^\mu \alpha) \neq
\frac{1}{2}M_A^2 A_\mu A^\mu 
\eeq  

-- In addition, if we include explicitly a mass term $-m_f \overline{\psi}_f 
\psi_f$ for each SM fermion $f$ in the Lagrangian, we would have for the 
electron for instance
\beq
- m_e \bar{e}e = -m_e \bar{e} \bigg( \frac{1}{2} (1-\gamma_5)+\frac{1}{2}(1+
\gamma_5) \bigg) e= -m_e(\bar{e}_R e_L+\bar{e}_Le_R) 
\eeq
which is manifestly non--invariant under the isospin symmetry transformations
discussed above, since $e_L$ is a member of an SU(2)$_{\rm L}$ doublet while 
$e_R$ is a member of a singlet. \s

Thus, the incorporation by brute force of mass terms for gauge bosons and for 
fermions leads to a manifest breakdown of the local ${\rm SU(2)_L\times U(1)_Y}
$ gauge invariance. Therefore, apparently, either we have to give up the fact 
that $M_Z\sim 90$ GeV and $m_e \sim 0.5$ MeV for instance, or give up the 
principle of exact or unbroken gauge symmetry. \s

The question, which has been asked already in the sixties, is therefore the
following: is there a  [possibly nice] way to generate the gauge boson and
the fermion masses without violating  SU(2)$\times$U(1) gauge invariance? 
The answer is yes: the Higgs--Brout--Englert--Guralnik--Hagen--Kibble mechanism
of spontaneous symmetry breaking \cite{Higgs} or the Higgs mechanism for short.
This mechanism will be briefly sketched in the following subsection and applied
to the SM case. 	

\subsubsection{The Higgs mechanism}

\subsubsection*{\underline{The Goldstone  theorem}} 

Let us start by taking a simple scalar real field $\phi$ with the usual 
Lagrangian 
\beq
{\cal L}= \frac{1}{2} \partial_\mu \phi \, \partial^\mu \phi - V(\phi) 
\ , \ V(\phi)= \frac{1}{2} \mu^2 \phi^2 + \frac{1}{4}\lambda \phi^4
\eeq
This Lagrangian is invariant under the reflexion symmetry $\phi \to -\phi$
since there are no cubic terms.  If the mass term $\mu^2$ is positive, the
potential $V(\phi)$ is also positive if the self--coupling $\lambda$ is
positive [which is needed to make the potential bounded from below], and the 
minimum of the potential is obtained for $\langle 0| \phi | 0 \rangle \equiv
\phi_0=0$ as shown in the left--hand side of Fig.~1.1. ${\cal L}$ is then 
simply the Lagrangian of a spin--zero particle of mass $\mu$. \s

\begin{figure}[htbp]
\begin{center}
\vspace*{-2.cm}
\hspace*{-2cm}
\epsfig{file=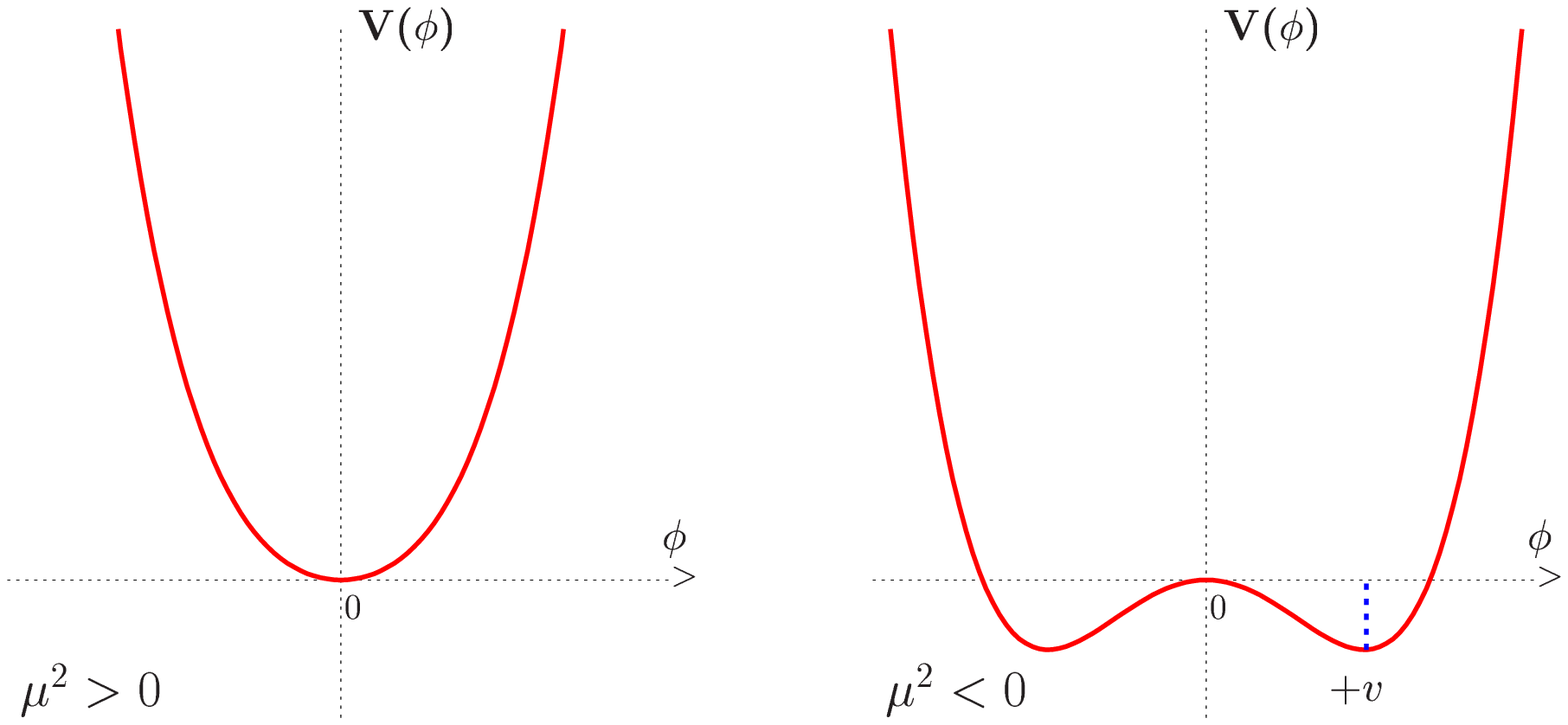,width=14.cm} 
\end{center}
\vspace*{-12.3cm}
\nn {\it Figure 1.1: The potential $V$ of the scalar field $\phi$ in 
the case $\mu^2>0$ (left) and $\mu^2 <0$ (right).}
\vspace*{-3mm}
\end{figure} 

\nn In turn, if $\mu^2 <0$, the potential $V(\phi)$ has a minimum when 
$\partial V/\partial \phi=\mu^2\phi +\lambda \phi^3=0$,  i.e. when  
\beq 
\langle 0| \phi^2 |0\rangle \equiv \phi_0^2=- \frac{\mu^2}{\lambda} \equiv v^2 
\eeq 
and not at $\phi_0^2=0$, as shown in the right--hand side of Fig.~1.1. The
quantity $\pm v \equiv \langle \, 0|\phi| 0 \,  \rangle$ is called the vacuum
expectation value (vev) of the scalar field $\phi$. In this case, ${\cal L}$ is
no more the Lagrangian of a particle with mass $\mu$ and to interpret correctly
the theory, we must expand around one of the minima $v$ by defining the field
$\sigma$ as $\phi= v + \sigma$. In terms of the new field, the Lagrangian
becomes 
\beq
{\cal L}= \frac{1}{2} \partial_\mu \sigma \, \partial^\mu \sigma - (-\mu^2) 
\, \sigma^2 - \sqrt{ - \mu^2 \lambda} \, \sigma^3 - \frac{\lambda}{4}  \, 
\sigma^4 + {\rm const.} 
\eeq
This is the theory of a scalar field of mass $m^2=-2\mu^2$, with $\sigma^3$ and 
$\sigma^4$ being the self--interactions. Since there are now cubic terms, the 
reflexion symmetry is broken: it is not anymore apparent in ${\cal L}$. 
This is the simplest example of a spontaneously broken symmetry.  \s

Let us make things slightly more complicated and consider four scalar fields 
$\phi_i$ with {\small $i=0,1,2,3$}, with a Lagrangian [the summation over the
index $i$ is understood]
\beq
{\cal L}= \frac{1}{2} \partial_\mu \phi_i \, \partial^\mu \phi_i - 
\frac{1}{2} \mu^2 \, (\phi_i \phi_i)- \frac{1}{4} \lambda (\phi_i \phi_i)^2 
\eeq
which is invariant under the rotation group in four dimensions O$(4)$, $\phi_i 
(x) = R_{ij}   \phi_j (x)$ for any orthogonal matrix $R$. \s  

Again, for $\mu^2<0$, the potential has a minimum at $\phi_i^2 = - \mu^2/
\lambda \equiv v^2$ where $v$ is the vev. As previously, we expand around one 
of the minima, $\phi_0= v+ \sigma$, and rewrite the fields $\phi_i=\pi_i$ 
with $i=1,2,3$ [in analogy with pion physics]. The Lagrangian in terms of the 
new fields $\sigma$ and $\pi_i$ becomes then
\beq
{\cal L} &=& \frac{1}{2} \partial_\mu \sigma \, \partial^\mu \sigma - 
\frac{1}{2} (- 2\mu^2) \sigma^2 - \lambda v \, \sigma^3- \frac{\lambda}{4} 
\, \sigma^4  \non \\ 
&+ &  \frac{1}{2} \partial_\mu \pi_i \, \partial^\mu \pi_i  
- \frac{\lambda}{4} (\pi_i \pi_i)^2 -  \lambda v \pi_i \pi_i 
\sigma -\frac{\lambda}{2} \pi_i \pi_i \sigma^2 
\eeq
As expected, we still have a massive $\sigma$ boson with $m^2=-2\mu^2$, but 
also, we have three massless pions since now, all the bilinear $\pi_i \pi_i$ 
terms  in the Lagrangian have vanished. Note that there is still an O(3) 
symmetry among the $\pi_i$ fields. \s

This brings us to state the Goldstone theorem \cite{Goldstone}: For every
spontaneously broken  continuous symmetry, the theory contains  massless scalar
(spin--0) particles called Goldstone bosons. The number of Goldstone bosons is
equal to the number  of broken generators. 
For an O$(N)$ continuous symmetry, there are $\frac{1} {2}N(N-1)$ generators;
the residual unbroken symmetry O$(N-1)$ has $\frac{1}{2} (N-1)(N-2)$ generators
and therefore, there are $N-1$ massless Goldstone bosons, i.e. 3 for the O$(4)$
group. \s 

Note that exactly the same exercise can be made for a complex doublet of scalar
fields
\beq
\phi = \left( \begin{array}{c} \phi^+ \\ \phi^0 \end{array} \right) = 
\frac{1}{\sqrt{2}} \left( \begin{array}{c} \phi_1 - i\phi_2 \\ \phi_3 - i 
\phi_4 \end{array} \right) 
\eeq
with the invariant product being $\phi^\dagger \phi= \frac{1}{2} (\phi_1^2+ 
\phi_2^2 +\phi_3^2+\phi_4^2) = \frac{1}{2} \phi_i \phi^i$. 

\subsubsection*{\underline{The Higgs mechanism in an abelian theory}} 

Let us now move to the case of a local symmetry and consider first the rather 
simple abelian U(1) case: a complex scalar field coupled to itself and to an 
electromagnetic field $A_\mu$ 
\beq
{\cal L} = - \frac{1}{4} F_{\mu \nu} F^{\mu \nu} + D_\mu \phi^* D^\mu \phi
- V (\phi) 
\eeq
with $D_\mu$ the covariant derivative $D_\mu= \partial_\mu - ie A_\mu$ and 
with the scalar potential 
\beq
V(\phi)=  \mu^2 \phi^* \phi + \lambda \ (\phi^* \phi)^2  
\eeq
The Lagrangian is invariant under the usual local U(1) transformation
\beq
\phi(x) \to e^{i \alpha(x)} \phi(x) \hspace*{1cm}
A_\mu (x) \to A_\mu (x) - \frac{1}{e} \partial_\mu \alpha(x) 
\eeq 
For $\mu^2>0$, ${\cal L}$ is simply the QED Lagrangian for a charged scalar 
particle of mass $\mu$ and with $\phi^4$ self--interactions. 
For $\mu^2<0$, the field $\phi(x)$ will acquire a vacuum expectation value and 
the minimum of the potential $V$ will be at
\beq
\langle \, \phi \, \rangle_0 \equiv \langle \, 0 | \, \phi \, | \, 0 \, \rangle 
= \left(- \frac{\mu^2}{2\lambda} \right)^{1/2} 
\equiv  \frac{v}{\sqrt{2}} 
\eeq
As before, we expand the Lagrangian around the vacuum state $\langle \phi 
\rangle$ 
\beq
\phi(x) = \frac{1}{\sqrt{2}} [ v + \phi_1(x) +i \phi_2 (x)] 
\eeq
The Lagrangian becomes then, up to some interaction terms that we omit for 
simplicity,  
\begin{eqnarray}
{\cal L} &=& - \frac{1}{4} F_{\mu \nu} F^{\mu \nu} + (\partial^\mu +ie A^\mu)
\phi^* (\partial_\mu -ie A_\mu) \phi -\mu^2 \phi^* \phi -\lambda 
(\phi^* \phi)^2 \hspace*{1cm} \non \\
&=& - \frac{1}{4} F_{\mu \nu} F^{\mu \nu} + \frac{1}{2} (\partial_\mu \phi_1)^2 
+ \frac{1}{2} (\partial_\mu \phi_2)^2 - v^2 \lambda \phi_1^2 
+ \frac{1}{2} e^2 v^2 A_\mu A^\mu - e v A_\mu \partial^\mu \phi_2 
\end{eqnarray}
Three remarks can then be made at this stage: \s

-- There is a photon mass term in the Lagrangian: $\frac{1}{2} M_A^2 A_\mu 
A^\mu$ with $M_A= e v = - e \mu^2/\lambda$.

-- We still have a scalar particle $\phi_1$ with a mass $M_{\phi_1}^2=- 
2\mu^2$.

-- Apparently, we have a massless particle $\phi_2$, a would--be Goldstone 
boson. \s

However, there is still a problem to be addressed. In the beginning, we had
four degrees of freedom in the theory, two for the complex scalar field $\phi$
and two for the massless electromagnetic field $A_\mu$, and now we have
apparently five degrees of freedom, one for $\phi_1$, one for $\phi_2$ and
three for the massive photon $A_\mu$. Therefore, there must be a field which
is not physical at the end and indeed, in ${\cal L}$ there is a bilinear term
$evA^\mu \partial_\mu \phi_2$ which has to be eliminated. To do so, we notice
that at first order, we have for the original field $\phi$
\beq
\phi= \frac{1}{\sqrt{2}} (v +\phi_1+i \phi_2)  \equiv \frac{1}{\sqrt{2}} 
[v +\eta (x) ] e^{i \zeta(x)/v} 
\eeq
By using the freedom of gauge transformations and by performing also the 
substitution 
\beq
A_\mu \to A_\mu - \frac{1}{ev} \partial_\mu \zeta(x)
\eeq
the $A_\mu \partial^\mu \zeta$ term, and in fact all $\zeta$ terms, disappear
from the Lagrangian. This choice of gauge, for which only the physical
particles are left in the Lagrangian, is called the unitary gauge.  Thus,   the
photon (with two degrees of freedom) has absorbed the would--be Goldstone boson
(with one degree of freedom) and became massive (i.e. with three degrees of
freedom): the longitudinal polarization is the Goldstone boson. The U(1) gauge
symmetry is no more apparent and we say that it is spontaneously broken.  This
is the Higgs mechanism \cite{Higgs} which allows to generate masses for
the gauge bosons.

\subsubsection*{\underline{The Higgs mechanism in the SM}}

In the slightly more complicated non--abelian case of the SM, we need to 
generate masses for the three gauge bosons $W^\pm$ and $Z$ but the  photon 
should remain massless and QED must stay an exact symmetry. Therefore, we need 
at least 3 degrees of freedom for the scalar fields. The simplest  choice is a 
complex SU(2) doublet of scalar fields $\phi$ 
\beq 
\Phi = \left( \begin{array}{c} \phi^+ \\ \phi^0 \end{array} \right) \, , 
\  \  Y_\phi=+1 
\eeq
To the SM Lagrangian discussed in the previous subsection, but where we ignore 
the strong interaction part
\beq
{\cal L}_{\rm SM}= -\frac{1}{4} W_{\mu \nu}^a W^{\mu \nu}_a -\frac{1}{4} 
B_{\mu \nu}B^{\mu \nu} + \overline{L}\, i D_\mu \gamma^\mu \, L + 
\overline{e}_R \, i D_\mu \gamma^\mu \, e_R \ \cdots  
\eeq
we need to add the invariant terms of the scalar field part 
\beq
{\cal L}_S = (D^\mu \Phi)^\dagger (D_\mu \Phi) - \mu^2 \Phi^\dagger \Phi
- \lambda ( \Phi^\dagger \Phi)^2 
\eeq
\nn For $\mu^2 <0$, the neutral component of the doublet field $\Phi$ will 
develop a vacuum expectation value [the vev should not be in the charged 
direction to preserve U$(1)_{\rm QED}$] 
\beq
\langle \, \Phi \, \rangle_0 \equiv  \langle \, 0 \, | \, \Phi \, | \, 0\, 
\rangle =\left( \begin{array}{c} 0\\ \frac{v}{\sqrt{2}} \end{array} \right) \ \ 
{\rm with} \ \ v= \left( - \frac{\mu^2}{\lambda} \right)^{1/2} 
\eeq
We can then make the same exercise as previously: \s

\nn --  write the field $\Phi$ in terms of four fields $\theta_{1,2,3}(x)$ and 
$H(x)$ at first order: 
\begin{eqnarray}
\Phi(x) = \left( \begin{array}{c} \theta_2 + i \theta_1 \\ 
\frac{1}{\sqrt{2}} ( v + H)  - i \theta_3 \end{array} \right) = 
e^{i \theta_a (x) \tau^a (x)/v} \,  \left( \begin{array}{c} 0 \\ 
\frac{1}{\sqrt{2}} (v + H(x) \, )  \end{array} \right) 
\end{eqnarray}
-- make a gauge transformation on this field to move to the unitary gauge:
\begin{eqnarray}
\Phi(x) \to  e^{- i \theta_a (x) \tau^a (x) } \,  \Phi(x) = \frac{1}{\sqrt{2}}
\left( \begin{array}{c} 0 \\  v + H (x)  \end{array} 
\right) 
\end{eqnarray}  
-- then fully expand the term $|D_\mu \Phi)|^2$ of the Lagrangian 
${\cal L}_S$: 
\beq
|D_\mu \Phi)|^2= \bigg| \bigg( \partial_\mu -i g_2 \frac{\tau_a}{2} W_\mu^a 
- i g_1 \frac{1}{2} B_\mu \bigg)\Phi  \bigg|^2 \hspace*{6cm} \non \\
= \frac{1}{2} \left| \left( \begin{array}{cc} \partial_\mu - \frac{i}{2}(
g_2 W_\mu^3 + g_1 B_\mu) & -\frac{ig_2}{2}(W_\mu^1 -iW^2_\mu) \\ -
\frac{ig_2}{2} (W_\mu^1 + iW^2_\mu) & \partial_\mu + \frac{i}{2} (g_2 W_\mu^3  
- g_1 B_\mu) \end{array} \right) \left( \begin{array}{c} 
0 \\ v+H  \end{array} \right) \right|^2 \non \\
= \frac{1}{2} (\partial_\mu H)^2 + \frac{1}{8}g_2^2(v+H)^2|W_\mu^1+iW_\mu^2|^2
+ \frac{1}{8}(v+H)^2 |g_2 W_\mu^3- g_1 B_\mu|^2 \non 
\eeq
-- define the new fields $W^\pm_\mu$ and $Z_\mu$ [$A_\mu$ is the field  
orthogonal to $Z_\mu$]:
\beq
W^\pm = \frac{1}{\sqrt{2}} (W^1_\mu \mp i W^2_\mu) \  , \ 
Z_\mu = \frac{g_2 W^3_\mu- g_1 B_\mu}{\sqrt{g_2^2+g_1^2}} \ , \ 
A_\mu = \frac{g_2 W^3_\mu+ g_1 B_\mu}{\sqrt{g_2^2+g_1^2}} \  
\label{WZA-fields}
\eeq
-- and pick up the terms which are bilinear in the fields $W^\pm, Z, A$:
\beq
M_W^2 W^+_\mu W^{-\mu} + \frac{1}{2} M_Z^2 Z_\mu Z^\mu + \frac{1}{2} M_A^2 
A_\mu A^\mu 
\eeq
The $W$ and $Z$ bosons have acquired masses, while the photon is still massless
\beq
M_W =\frac{1}{2}vg_2 \  , \ M_Z= \frac{1}{2} v  \sqrt{g^2_2+g_1^2} \ , 
 \ M_A=0 
\eeq
Thus, we have achieved (half of) our goal: by spontaneously breaking the 
symmetry ${\rm SU(2)_L \times}$ ${\rm U(1)_Y}$ ${\rm \to U(1)_{Q} }$, three 
Goldstone bosons  have been absorbed by the $W^\pm$ and $Z$ bosons to form their
longitudinal components and to get their masses. Since the ${\rm U(1)_{Q}}$ 
symmetry is still unbroken, the photon which is its generator, remains massless
as it should be.\bigskip

Up to now, we have discussed only the generation of gauge boson masses; but 
what about the fermion masses?  In fact, we can also generate the fermion
masses using the same scalar field $\Phi$, with hypercharge $Y$=1, and the
isodoublet  $\tilde{\Phi}= i\tau_2 \Phi^*$, which  has hypercharge $Y$=--1. For 
any fermion generation, we introduce the ${\rm SU(2)_L \times U(1)_Y}$ 
invariant Yukawa Lagrangian
\beq
{\cal L}_F= -\lambda_{e} \, \bar{L} \, \Phi \, e_{R}  
- \lambda_{d} \, \bar{Q} \, \Phi \, d_{R}
- \lambda_u \, \bar{Q} \, \tilde{\Phi} \, u_R    \ + \ { h.\, c.} 
\eeq
and repeat the same exercise as above. Taking for instance the case of the 
electron, one obtains
\beq
{\cal L}_F &=& -\frac{1}{\sqrt{2}} \lambda_e \, (\bar{\nu}_e, \bar{e}_L) \, 
\left( \begin{array}{c} 0 \\  v + H  \end{array} \right) \, e_R \ +\cdots 
\non\\ 
&=& -\frac{1}{\sqrt{2}} \, \lambda_e \, (v+H) \, \bar{e}_L e_R \ + \cdots 
\eeq
The constant term in front of $\bar{f}_L f_R$ (and h.c.) is identified
with the fermion mass
\beq
m_e= \frac{\lambda_e\, v}{\sqrt{2}} \ \ , \ 
m_u= \frac{\lambda_u\, v}{\sqrt{2} }\ \ , \ 
m_d= \frac{\lambda_d\, v}{\sqrt{2}} 
\eeq
 
Thus, with the same isodoublet $\Phi$ of scalar fields, we have generated
the masses of both the weak vector bosons $W^\pm, Z$ and the fermions, while 
preserving the SU(2)$\times$U(1) gauge symmetry, which is now spontaneously 
broken or hidden. The electromagnetic ${\rm U(1)_{Q} }$ symmetry, as well as 
the SU(3) color symmetry, stay unbroken. The Standard Model refers, in fact, 
to SU(3)$\times$SU(2)$\times$U(1) gauge invariance when combined with the 
electroweak symmetry breaking mechanism. Very often, the electroweak sector
of the theory is also referred to as the SM; in this review we will use this 
name for both options. 

\subsubsection{The SM Higgs particle and the Goldstone bosons}

\subsubsection*{\underline{The Higgs particle in the SM}}

Let us finally come to the  Higgs boson itself. The kinetic part of the Higgs
field,  $\frac{1}{2} (\partial_\mu H)^2$, comes from the term involving the
covariant derivative $|D_\mu \Phi|^2$, while the mass and self--interaction
parts,  come from the scalar potential $V(\Phi)=\mu^2 \Phi^\dagger \Phi+
\lambda (  \Phi^\dagger \Phi)^2$
\beq
V&=& \frac{\mu^2}{2} (0, v+H) \left( \begin{array}{c}0\\  v+H  \end{array} 
\right) +\frac{\lambda}{4} \Bigg| (0, v+H)\left( \begin{array}{c}0\\  v+H  
\end{array} \right) \Bigg|^2 
\eeq
Using the relation  $v^2=-\mu^2/\lambda$, one obtains 
\beq 
V &=& -\frac{1}{2} \lambda v^2\, (v+H)^2 + \frac{1}{4} \lambda (v+H)^4 \ \ 
\eeq
and finds that the Lagrangian containing the Higgs field $H$ is given by
\beq
{\cal L}_{H}&=& \frac{1}{2} (\partial_\mu H) (\partial^\mu H) -V \non \\
&=& \frac{1}{2} (\partial^\mu H)^2 - \lambda v^2 \, H^2 - \lambda v \, H^3 - 
\frac{\lambda}{4} \, H^4 
\eeq
From this Lagrangian, one can see that the Higgs boson mass simply reads
\beq
M_H^2=2 \lambda v^2 =- 2\mu^2
\eeq 
and the Feynman rules\footnote{\nn The Feynman rule for these vertices are 
obtained by  multiplying the term involving the interaction by a factor $-i$. 
One includes also a factor $n!$ where $n$ is the number of 
identical particles in the vertex.} for the Higgs self--interaction vertices 
are given by
\beq
g_{H^3}=(3!) i\lambda v = 3i \, \frac{M_H^2}{v} \ , \ g_{H^4}= (4!) i 
\frac{\lambda}{4} = 3i \frac{M_H^2}{v^2}
\eeq
As for the Higgs boson couplings to gauge bosons and fermions, they 
were almost derived previously, when the masses of these particles were 
calculated. Indeed, from the Lagrangian describing the gauge boson and fermion 
masses 
\beq
{\cal L}_{M_V} \sim  M_V^2 \left (1+ \frac{H}{v} \right)^2 \ , \ \
{\cal L}_{m_f} \sim - m_f \left(1+ \frac{H}{v} \right)   
\eeq
one obtains also the Higgs boson couplings to gauge bosons and fermions 
\beq
g_{Hff}= i \frac{m_f}{v} \ \ , \ 
g_{HVV}= -2 i \frac{M_V^2}{v} \ \ , \ 
g_{HHVV}= - 2i \frac{M_V^2}{v^2}
\eeq
This form of the Higgs couplings ensures the unitarity of the theory 
\cite{UNITARITY} as will be seen later.
The vacuum expectation value $v$ is fixed in terms of the $W$ 
boson mass $M_W$ or the Fermi constant $G_\mu$ determined from muon 
decay [see next section]
\beq
M_W=\frac{1}{2} g_2v = \left( \frac{\sqrt{2} g^2}{8 G_\mu} \right)^{1/2} 
\Rightarrow v= \frac{1}{(\sqrt{2} G_\mu)^{1/2} } \simeq 246~{\rm GeV}  
\label{MW-vs-v}
\eeq
We will see in the course of this review that it will be appropriate to use the
Fermi coupling constant $G_\mu$ to describe the couplings of the Higgs boson,
as some higher--order effects are effectively absorbed in this way. The Higgs 
couplings to fermions, massive gauge bosons as well as the self--couplings, are
given in Fig.~1.2 using both $v$ and $G_\mu$. This general form of the 
couplings will be useful when discussing the Higgs properties in extensions of 
the SM. \s

\begin{figure}[h]
\vspace*{-2mm}
\hspace*{0.5cm}
\SetWidth{1.}
\begin{picture}(300,100)(0,0)
\DashLine(0,50)(50,50){4}
\ArrowLine(50,50)(100,75)
\ArrowLine(50,50)(100,25)
\Text(50,50)[]{{\blue{\Large $\bullet$}}}
\Text(5,60)[]{${\blue{H}}$}
\Text(110,75)[]{$f$}
\Text(110,25)[]{$\bar{f}$}
\Text(260,50)[]{$g_{Hff} \ =\ m_f/ v\ =\ (\sqrt{2}G_\mu)^{1/2}\, m_f$}
\Text(400,50)[]{$ \times \ (i)$} 
\end{picture}
\vspace*{-.5cm}

\hspace*{0.5cm}
\begin{picture}(300,100)(0,0)
\DashLine(0,50)(50,50){4}
\Photon(50,50)(100,75){-3}{7}
\Photon(50,50)(100,25){3}{7}
\Text(50,50)[]{{\blue{\Large $\bullet$}}}
\Text(5,60)[]{${\blue{H}}$}
\Text(110,75)[]{$V_\mu$}
\Text(110,25)[]{$V_\nu$}
\Text(260,50)[]{$g_{HVV}\ =\ 2M_V^2/v\ =\ 2(\sqrt{2}G_\mu)^{1/2}\,M_V^2$}
\Text(400,50)[]{$\times \ (-i   g_{\mu \nu}) $}
\end{picture}
\vspace*{-.5cm} 

\hspace*{0.5cm}
\begin{picture}(300,100)(0,0)
\DashLine(0,25)(50,50){4}
\DashLine(0,75)(50,50){4}
\Photon(50,50)(100,75){-3}{7}
\Photon(50,50)(100,25){3}{7}
\Text(50,50)[]{{\blue{\Large $\bullet$}}}
\Text(5,60)[]{${\blue{H}}$}
\Text(5,40)[]{${\blue{H}}$}
\Text(110,75)[]{$V_\mu$}
\Text(110,25)[]{$V_\nu$}
\Text(260,50)[]{$g_{HHVV}\ =\ 2M_V^2/v^2\ =\ 2\sqrt{2}G_\mu \,M_V^2$}
\Text(400,50)[]{$\times \ (-i   g_{\mu \nu})$}
\end{picture}
\vspace*{-.5cm} 

\hspace*{0.5cm}
\begin{picture}(300,100)(0,0)
\DashLine(0,50)(50,50){4}
\DashLine(50,50)(100,75){4}
\DashLine(50,50)(100,25){4}
\Text(50,50)[]{{\blue{\Large $\bullet$}}}
\Text(5,60)[]{${\blue{H}}$}
\Text(110,75)[]{${\blue{H}}$}
\Text(110,25)[]{${\blue{H}}$}
\Text(260,50)[]{$g_{HHH}\ =\ 3M_H^2/v\ =\ 3(\sqrt{2}G_\mu)^{1/2}\,M_H^2$}
\Text(400,50)[]{$\times \ (i  ) $}
\end{picture}
\vspace*{-.5cm} 

\hspace*{0.5cm}
\begin{picture}(300,100)(0,0)
\DashLine(0,25)(50,50){4}
\DashLine(0,75)(50,50){4}
\DashLine(50,50)(100,75){4}
\DashLine(50,50)(100,25){4}
\Text(50,50)[]{{\blue{\Large $\bullet$}}}
\Text(5,60)[]{${\blue{H}}$}
\Text(5,40)[]{${\blue{H}}$}
\Text(110,75)[]{${\blue{H}}$}
\Text(110,25)[]{${\blue{H}}$}
\Text(260,50)[]{$g_{HHHH}=\ 3M_H^2/v^2\ =\ 3 \sqrt{2}G_\mu \,M_H^2$}
\Text(400,50)[]{$\times \ (i )$}
\end{picture}
\vspace*{-2mm}

\nn {\it Figure 1.2: The Higgs boson couplings to fermions and gauge bosons and
the Higgs self--couplings in the SM. The normalization factors of the Feynman 
rules are also displayed.}
\vspace*{-5mm}
\end{figure}

Note that the propagator of the Higgs boson is simply given, in momentum space,
by
\beq
\Delta_{HH}(q^2) = \frac{i}{q^2- M_H^2 + i\epsilon } 
\eeq

\subsubsection*{\underline{The Goldstone bosons}}

In the unitary gauge, the physical spectrum of the SM is clear:  besides the
fermions and the massless photon [and gluons], we have the massive $V=W^\pm$ 
and $Z$ bosons and the Goldstones do not appear. The propagators of the vector
bosons in this gauge are given by
\beq
\Delta_{VV}^{\mu \nu} (q)=  \frac{-i}{q^2 - M_V^2+ i \epsilon} \left[ 
g^{\mu \nu} - \frac {q^\mu q^\nu}{M_V^2}  \right]
\label{prop-unitary}
\eeq
The first term, $\propto g^{\mu \nu}$, corresponds to the propagation of the
transverse component of the $V$ boson [the propagator of the photon is simply 
$-i g^{\mu \nu}/q^2$], while the second term, $\propto q_\mu q_\nu$, corresponds
to the propagation of the longitudinal component which, as can be seen, does not
vanish $\propto 1/q^2$ at high energies. This terms lead to very complicated
cancellations in the invariant amplitudes involving the exchange of $V$ bosons
at high energies and, even worse, make the renormalization program very
difficult to carry out, as the latter usually makes use of four--momentum power
counting analyses of the loop diagrams. It is more convenient to work in
$R_\xi$ gauges where gauge fixing terms are added to the SM Lagrangian 
\cite{Gauge-Fix}
\beq
{\cal L}_{\rm GF}= \frac{-1}{2\xi}\left[2 (\partial^\mu W^+_\mu\! - \! i\xi M_W
w^+)(\partial^\mu W^-_\mu \!- \!i\xi M_W w^-)+(\partial^\mu Z_\mu \!- \!
i\xi M_Z w^0)^2 +(\partial^\mu A_\mu)^2 \right] 
\eeq
$w^0 \equiv G^0$ and $w^\pm \equiv G^\pm$ being the neutral and charged 
Goldstone bosons and where different choices of $\xi$ correspond to different
renormalizable gauges. In this case, the propagators of the massive gauge 
bosons are given by
\beq
\Delta_{VV}^{\mu \nu} (q)= \frac{-  i}{ q^2 - M_V^2+ i \epsilon}  \left[ 
g^{\mu \nu} +(\xi-1) \frac{q^\mu q^\nu}{q^2- \xi M_V^2} \right]
\label{Vpropagator}
\eeq
which in the unitary gauge, $\xi=\infty$, reduces to the expression 
eq.~(\ref{prop-unitary}). Usually, one uses the 't Hooft--Feynman gauge 
$\xi=1$, where the $q^\mu q^\nu$ term is absent, to simplify the calculations; 
another popular choice is the Landau gauge, $\xi =0$.  In renormalizable 
$R_{\xi}$ gauges, the propagators of the Goldstone bosons are given by
\beq
\Delta_{w^0 w^0}(q^2) &=& \frac{i}{q^2- \xi M_Z^2 + i\epsilon } \non \\
\Delta_{w^\pm w^\pm}(q^2) &=& \frac{i}{q^2- \xi M_W^2+ i\epsilon } 
\eeq
and as can be seen, in the unitary gauge $\xi=\infty$, the Goldstone bosons do
not propagate and decouple from the theory as they should, while in the 
Landau gauge they are massless and do not interact with the Higgs particle. In 
the 't Hooft--Feynman gauge,  the Goldstone bosons are part of the spectrum and
have ``masses" $\propto M_V$. Any dependence on $\xi$ should however be absent 
from physical matrix elements squared, as the theory must be gauge invariant.\s

Note that the couplings of the Goldstone bosons to fermions are, as in the 
case of the Higgs boson, proportional to the fermion masses 
\beq
g_{G^0 ff} &=& - 2 I_f^3 \, \frac{m_f}{v} \non \\ 
g_{G^- ud} &=& \frac{-i }{\sqrt 2 v} V_{ud} [m_d (1- \gamma_5) - m_u 
(1+\gamma_5) ] 
\eeq
where $V_{ud}$ is the CKM matrix element for quarks [see later] and which, in 
the case of leptons [where one has to set $m_d=m_\ell$ and $m_u=0$ in the 
equation above], is equal to unity. The couplings of the Goldstones to gauge 
bosons are simply those of scalar spin--zero particles. \s  

The longitudinal components of the $W$ and $Z$ bosons give rise to interesting
features which occur at high energies and that we shortly describe below. 
In the gauge boson rest frame, one can define the transverse and longitudinal
polarization four--vectors as
\beq
\epsilon_{T_1}^\mu=(0,1,0,0) \, , \quad  
\epsilon_{T_2}^\mu=(0,0,1,0) \, , \quad  
\epsilon_{L}^\mu=(0,0,0,1) 
\eeq
For a four--momentum $p^\mu=(E,0,0,|\vec{p}|)$, after a boost along the $z$ 
direction, the transverse polarizations remain the same while the longitudinal 
polarization becomes
\beq
\epsilon_{L}^\mu= \left( \frac{|\vec{p}|}{M_V}, 0, 0, \frac{E}{M_V} \right) 
\stackrel{\small E \gg M_V} \lra {p_\mu \over M_V}
\eeq 
Since this polarization is proportional to the gauge boson momentum, at very 
high energies, the longitudinal amplitudes will dominate in the scattering of 
gauge bosons.\s 

In fact, there is a theorem, called the Electroweak Equivalence Theorem
\cite{Equivalence-theorem,LQT,EffectiveWapprox}, which states that at very high
energies, the longitudinal massive vector bosons can be replaced by the
Goldstone bosons. In addition, in many processes such as vector boson
scattering, the vector bosons themselves can by replaced by their longitudinal
components. The amplitude for the scattering of $n$ gauge bosons in the initial
state to $n'$ gauge bosons in the final state is simply the amplitude for the
scattering of the corresponding Goldstone bosons 
\beq
A(V^1 \cdots V^n \to V^1 \cdots V^{n'}) &\sim &A(V_L^1 \cdots V_L^n \to 
V_L^1 \cdots V_L^{n'}) \non \\ 
& \sim & A(w^1 \cdots w^n \to w^1 \cdots w^{n'})  
\eeq
Thus, in this limit, one can simply replace in the SM scalar potential, the 
$W$ and $Z$ bosons by their corresponding Goldstone bosons $w^\pm, w_0$,
leading to 
\beq
V= \frac{M_H^2}{2v} (H^2+w_0^2+2w^+w^-)H +
\frac{M_H^2}{8v^2} (H^2+w_0^2+2w^+w^-)^2  
\label{Vequivalence}
\eeq
and use this potential to calculate the amplitudes for the processes involving
weak vector bosons. The calculations are then extremely simple, since one has 
to deal only with interactions among scalar particles.

\subsubsection{The SM interactions and parameters}

In this subsection, we summarize the interactions of the fermions and gauge 
bosons in the electroweak SM [for the strong interactions of quarks and gluons,
the discussion held in \S1.1.1 is sufficient for our purpose] and discuss the 
basic parameters of the SM and their experimental determination. \s

The equations for the field rotation which lead to the physical gauge bosons,  
eq.~(\ref{WZA-fields}), define the electroweak mixing angle $\sin 
\theta_W$
\beq
\sin \theta_W = \frac{g_1 }{\sqrt{g_1^2 + g_2^2}} = \frac{e}{g_2}
\eeq 
which can be written in terms of the $W$ and $Z$ boson masses as
\beq
\sin^2 \theta_W \equiv s_W^2 =1- c^2_W = 1- \frac{M_W^2}{M_Z^2}
\label{sw-definition}
\eeq 
Using the fermionic part of the SM Lagrangian, eq.~(\ref{smlagrangian}), 
written in terms of the new fields and writing explicitly the covariant 
derivative, one obtains 
\beq
{\cal L}_{\rm NC} &=& e J_\mu^{\rm A} A^\mu + \frac{g_2}{\cos\theta_W} 
J_\mu^Z Z^\mu \non \\
{\cal L}_{\rm CC} &=& \frac{g_2} {\sqrt{2}} (J_\mu^+ W^{+\mu}
+ J_\mu^- W^{-\mu}) 
\eeq
for the neutral and charged current parts, respectively. The currents $J_\mu$ 
are then given by 
\beq
J_\mu^A &=& Q_f \bar{f}\gamma_\mu f \non \\  
J_\mu^Z &=& \frac{1}{4}  \bar{f}
\gamma_\mu [ (2I_f^3 - 4 Q_f \sin^2 \theta_W) - \gamma_5 (2I^3_f) ]f \non \\ 
J_\mu^+ &=& \frac{1}{2} \bar{f}_u \gamma_\mu (1- \gamma_5) f_d
\eeq
where $f_u (f_d)$ is the up--type (down--type) fermion of isospin $+(-)\frac{1}
{2}$. \s

In terms of the electric charge $Q_f$ of the fermion $f$ and with $I_f^3 
= \pm \frac{1}{2}$ the left--handed weak isospin of the fermion and the weak 
mixing angle $s_W^2=1-c_W^2 \equiv \sin^2 \theta_W$, one can write the vector 
and axial vector couplings of the fermion $f$ to the $Z$ boson 
\beq
v_f = \frac{ \hat{v}_f} { 4 s_W c_W} = \frac{ 2I_f^3 -4 Q_f s_W^2}{ 4 s_W c_W}
\ \ , \ \ a_f = \frac{ \hat{a}_f} { 4 s_W c_W} = \frac{ 2I_f^3}{ 4 s_W c_W}
\label{Zffcouplings}
\eeq
where we also defined the reduced $Z f\bar f$ couplings $\hat{v}_f, \hat{a}_f$.
In the case of the $W$ boson, its vector and axial--vector couplings couplings 
to fermions are simply 
\beq
v_f = a_f = \frac{1}{2 \sqrt{2}s_W} = \frac{\hat a_f}{4 s_W} = 
\frac{\hat{v}_f}{4 s_W} 
\label{Wffcouplings}
\eeq
These results are only valid in the one--family approximation. While the
extension  to three families is straightforward for the neutral currents, 
there is a complication in the case of the charged currents: the current 
eigenstates for quarks $q'$ are not identical to the mass eigenstates $q$. If
we start by $u$--type quarks being mass eigenstates, in the down--type
quark sector, the two sets are connected by a unitary transformation \cite{CKM}
\beq
(d',s',b') = V (d,s,b)
\eeq
where $V$ is the $3\!\times\!3$ Cabibbo--Kobayashi--Maskawa matrix. The 
unitarity
of $V$ insures that the neutral currents are diagonal in both bases: this is 
the GIM mechanism \cite{GIM} which ensures a natural absence of flavor changing
neutral currents (FCNC) at the tree--level in the SM. For leptons, the mass and
current  eigenstates coincide since in the SM, the neutrinos are assumed to
be  massless, which is an excellent  approximation in most purposes. \s

Note that the relative strength of the charged and neutral currents, $J^\mu_Z 
J_{\mu Z}/J^{ \mu +} J_\mu^-$ can be measured by the parameter $\rho$ 
\cite{rho-definition} which, using previous formulae, is given by
\beq
\rho=  \frac{M_W^2}{c_W^2 M_Z^2}
\label{rho-MWMZ}
\eeq
and is equal to unity in the SM, eq.~(\ref{sw-definition}). This is a 
direct consequence of the choice of the representation of the Higgs field 
responsible of the breaking of the electroweak symmetry. In a model which makes
use of an arbitrary number of Higgs multiplets $\Phi_i$ with isospin $I_i$,  
third component $I_i^3$ and vacuum expectation values $v_i$, one obtains for
this parameter
\beq
\rho= \frac{\sum_i \left[ I_i (I_i+1) -(I_i^3)^2\right] v_i^2}
{2 \sum_i (I_i^3)^2 v_i^2}
\eeq
which is also unity for an arbitrary number of doublet [as well as singlet] 
fields. This is due to  the fact that in this case, the model has a
custodial SU(2) global symmetry. In the SM, this symmetry is broken at the
loop level when fermions of the same doublets have different masses and by the
hypercharge group. The radiative corrections to this parameter will be 
discussed in some detail in the next section.\s 

Finally, self--couplings among the gauge bosons are present in the SM as a 
consequence
of the non abelian nature of the ${\rm SU(2)_L\times U(1)_Y}$ symmetry. These
couplings are dictated by the structure of the symmetry group as discussed
in \S1.1.1 and, for instance, the triple self--couplings among the $W$ and the 
$V=\gamma,Z$ bosons are given by
\beq
{\cal L}_{WWV} = ig_{WWV} \left[ W^\dagger_{\mu \nu} W^{\mu} V^\nu - 
W^\dagger_{\mu} V_{\nu} W^{\mu \nu}  + W^\dagger_{\mu} W_{\nu} V^{\mu \nu}  
\right] \label{WWVcoupling}
\eeq
with $g_{WW\gamma}=e$ and $g_{WWZ}=e c_W/s_W$.\s

This concludes our description of the gauge interactions in the SM. We turn now
to the list of the model parameters that we will need in our subsequent 
discussions.   

\subsubsection*{\underline{The fine structure constant}}
 
The QED fine structure constant  defined in the classical Thomson limit 
$q^2 \sim 0$ of Compton scattering, is one of the best measured quantities
in Nature
\beq
\alpha (0) \equiv e^2/(4\pi) =1/137.03599976 \, (50)
\eeq
However, the physics which is studied at present colliders is at scales of 
the order of 100 GeV and the  running between $q^2 \sim 0$ and this 
scale must be taken into account. This running is defined as the difference 
between the [transverse components of the] vacuum polarization function of the 
photon at the two scales and, for $q^2=M_Z^2$ for instance, one has
\beq
\alpha(M_Z^2) = \frac{\alpha(0) }{1- \Delta \alpha}  \ \ , \ \ 
\Delta \alpha (M_Z^2) = \Pi_{\gamma \gamma}(0) - \Pi_{\gamma \gamma} (M_Z^2)
\eeq
Since QED is a vectorial theory, all heavy particles decouple from the photon
two--point function by virtue of the Appelquist--Carazzone theorem 
\cite{decoupling-theorem} and only the light particles, i.e. the SM light 
fermions, have to be taken into account in the running. [For instance, the top
quark contribution is $\Delta^{\rm top} \alpha \sim - 7 \cdot 10^{-5}$, while 
the small $W$ boson contribution is not gauge invariant by itself and has to be
combined with direct vertex and box corrections.] \s

The contribution of the $e, \mu$ and $\tau$ leptons to $\Delta \alpha$ simply 
reads \cite{alpha-leptons} 
\beq
\Delta \alpha^{\rm lept} (M_Z^2)=\sum_{\ell =e, \mu ,\tau} \frac{\alpha}{3\pi}
\left[ \log \frac{M_Z^2}{m^2_\ell} - \frac{5}{3} \right] + {\cal O} \left( 
\frac{ m_\ell^2 }{M_Z^2} \right) + {\cal O} (\alpha^2) + {\cal O} (\alpha^3) 
\simeq 0.0315
\eeq
For the contribution of light quarks, one has to evaluate $\Pi_{\gamma \gamma}$
at very low energies where perturbation theory fails for the strong
interaction. In fact, even if it were not the case, the light quark masses are
not known sufficiently precisely to be used as inputs. Fortunately, it is
possible to circumvent these complications and to derive the hadronic
contribution in an indirect way, taking all orders of the strong interaction
into account.  Indeed, one can use the optical theorem to relate the imaginary
part of the photon two--point function to the $\gamma f\bar{f}$ vertex 
amplitude and make use of the dispersion relation
\beq
\Delta \alpha^{\rm had}(M_Z^2) = - \frac{\alpha M_Z^2}{3 \pi} {\rm Re} \left( 
\int_{4 m_\pi^2}^\infty {\rm d}s' \frac{R_{\gamma \gamma} (s') }{s'(s'-M_Z^2 
)} \right) , \, R_{\gamma \gamma} (s) = \frac{ \sigma (\ee \to 
\gamma^* \to {\rm had.}) } {\sigma (\ee  \to \gamma^* \to \mu^+ \mu^-) }
\eeq
with the quantity $R_{\gamma \gamma} (s)$ measured in the problematic range 
using experimental data, and using perturbative QCD for the high energy range
\cite{alpha-hadrons0,alpha-hadrons}. 
Taking into account all available information from various experiments, one 
obtains for the hadronic contribution \cite{alpha-hadrons}
\beq
\Delta \alpha ^{\rm had} (M_Z^2) = 0.02761 \pm 0.00036
\eeq
This result is slightly  improved if one uses additional information
from $\tau$ decays $\tau^- \to \nu_\tau W^* \to \nu_\tau+$ hadrons, modulo
some reasonable theoretical assumptions. The latest world average value for 
the running electromagnetic coupling constant $\alpha$ at the scale $M_Z$ is 
therefore
\beq
\alpha^{-1} (M_Z^2) = 128.951 \pm 0.027
\label{Deltaalpha}
\eeq

\subsubsection*{\underline{The Fermi coupling constant}}

Another quantity in particle physics which is very precisely measured is the 
muon decay lifetime, which is directly related to the Fermi coupling constant 
in the effective four--point Fermi interaction but including QED corrections
\cite{GF-correction}
\beq
\frac{1}{\tau_\mu} &=& \frac{G_\mu^2 m_\mu^5}{192 \pi^3} 
\left( 1- \frac{8m_e^2}{m_\mu^2} \right) \left[1+ 1.810 \frac{\alpha}{\pi} 
+  (6.701 \pm 0.002) \left( \frac{\alpha} {\pi} \right)^2 \right]
\eeq
which leads to the precise value
\beq
G_\mu = (1.16637 \pm 0.00001) \cdot 10^{-5}~{\rm GeV}^{-2}
 \eeq    
In the SM, the decay occurs through gauge interactions mediated by $W$ boson 
exchange  and therefore, one obtains a relation between the $W,Z$  masses,
the QED constant $\alpha$ and $G_\mu$
\beq
\frac{G_\mu}{\sqrt{2}} = \frac{g_2}{2\sqrt{2}} \cdot \frac{1}{M_W^2}
\cdot \frac{g_2}{2\sqrt{2}} = \frac{ \pi \alpha}{2 M_W^2 s_W^2} 
= \frac{ \pi \alpha}{2 M_W^2 (1- M_W^2/M_Z^2)} 
\label{def:GF}
\eeq

\subsubsection*{\underline{The strong coupling constant}}

\nn The strong coupling constant has been precisely determined in various
experiments in $\ee$ collisions\footnote{Note that measurements of $\alpha_s$
have been performed at various energies, from $\sqrt s \sim 1.8$ GeV in 
$\tau$--lepton decays at LEP1 to $\sqrt{s} \sim 210$ GeV at LEP2, {\it en 
passant par} $\sqrt{s} \sim 20$ GeV at JADE, confirming in an unambiguous way 
the QCD prediction of asymptotic freedom. The non--abelian structure of QCD and
the three--gluon vertex has also been tested at LEP in four jet events.} and in
deep inelastic scattering; for a review, see Refs.~\cite{alphas-review,PDG}. The most
reliable results have been obtained at LEP where several methods can be used:
inclusive hadronic rates in $Z$ decays [$R_\ell, \sigma^0_{\rm had}$ and 
$\Gamma_Z$, see §1.2.1 later], inclusive rates in hadronic $\tau$ 
decays, event shapes and jet rates in multi--jet production. The world average 
value is given by \cite{PDG}
\beq
\alpha_s = 0.1172 \pm 0.002
\eeq
which corresponds to a QCD scale for 5 light flavors $\Lambda_{\rm QCD}^5=216
^{+25}_{-24}$ MeV.  Using this value of $\Lambda$, one can determine $\alpha_s$
at any energy scale $\mu$ up to three--loop order in QCD \cite{alphas-evol}
\beq
\alpha_s (\mu)= \frac{4 \pi}{\beta_0 \ell_\mu} \left[1 - \frac{2
\beta_1}{\beta_0^2} \frac{ \log \ell_\mu}{\ell_\mu} + \frac{4 \beta_1^2}
{\beta_0^4 \ell_\mu^2} \left( \left( \log \ell_\mu - \frac{1}{2} \right)^2 + 
\frac{ \beta_2 \beta_0}{8 \beta_1^2} - \frac{5}{4} \right)\right]
\eeq
with $\ell_\mu \equiv \log (\mu^2/\Lambda^2)$ and the $\beta_i$ coefficients 
given by
\beq
\beta_0 = 11 - \frac{2}{3} N_f \ , \ 
\beta_1 = 51 - \frac{19}{3} N_f \ , \ 
\beta_2 = 2857 - \frac{5033}{9} N_f + \frac{325}{27}N_f^2 
\eeq
with $N_f$ being the number of quarks with a mass smaller than the energy 
scale $\mu$. 

\subsubsection*{\underline{The fermion masses}} 

The top quark has been produced at the Tevatron in the reaction $p\bar{p}
\to q \bar{q}/ gg \to t\bar{t}$ and in the SM, it decays almost 100\% of the 
time into a $b$ quark and a $W$ boson, $t \to b W^+$. The top  quark mass is 
extracted mainly in the lepton plus jets and dilepton channels of the decaying 
$W$ bosons,  and combining CDF and D\O\ results, one obtains the average mass 
value \cite{Mt-Tevatron}
\beq
m_t= 178.0 \pm 4.3~{\rm GeV}
\eeq
The branching ratio of the decay $t \to Wb$ [compared to decays $t \to Wq]$ has
been measured to be BR$( t \to Wb)=0.94^{+0.31}_{-0.24}$ \cite{PDG}, allowing 
to extract the value of the $V_{tb}$ CKM matrix element, $|V_{tb}|=0.97^{+0.16}
_{-0.12}$. The top quark decay width in the SM is predicted to be 
\cite{Top-RC,Top-LHC,Top-width} 
\beq
\Gamma_t \simeq \Gamma (t \to b W^+) = \frac{G_\mu m_t^3} {8\sqrt{2} \pi} 
|V_{tb}|^2 \, \left( 1- \frac{M_W^2}{m_t^2} \right)^2 
\left( 1+ 2\frac{M_W^2}{m_t^2} \right) 
\left( 1- 2.72 \frac{\alpha_s}{\pi} \right) + {\cal O}(\alpha_s^2,\alpha) \ \
\eeq
and is of the order of $\Gamma_t \simeq 1.8$ GeV for $m_t \simeq 180$ GeV.\s

Besides the top quark mass, the masses of the bottom and charm quarks [and to 
a lesser extent the mass of the strange quark] are essential ingredients in 
Higgs physics. From many measurements, one obtains the following values
for the pole or physical masses $m_Q$ \cite{Narison}
\beq
m_b=4.88 \pm 0.07~{\rm GeV} \ , \ m_c=1.64 \pm 0.07~{\rm GeV} 
\eeq
However, the masses which are needed in this context are in general not the 
pole quark masses but the running quark masses at a high scale corresponding to
the Higgs boson mass. In the modified minimal subtraction or $\overline{\rm 
MS}$ scheme, the relation between  the pole masses  and the running 
masses  at the scale of the pole mass, ${\overline{m}}_{Q}(m_{Q})$, can be 
expressed as \cite{melnikov-ritbergen}
\beq \label{run-pole}
{\overline{m}}_{Q}(m_{Q})= m_{Q} \, \bigg[ 1- \frac{4}{3}
\frac{\alpha_{s}(m_Q)}{\pi} + (1.0414 N_f - 14.3323) \frac{\alpha_s^2(m_Q)}
{\pi^2} \bigg] \non \\
 +(-0.65269 N_f^2 +26.9239 N_f -198.7068)\frac{\alpha_s^3(m_Q)}{\pi^2} \bigg]
\eeq
where $\alpha_s$ is the $\overline{\rm MS}$ strong coupling constant evaluated 
at the scale of the pole mass $\mu=m_Q$.  The evolution of $\overline{m}_Q$ 
from $m_{Q}$ upward to a renormalization scale $\mu$ is 
\begin{eqnarray}
&& \hspace*{1.5cm}{\overline{m}}_{Q}\,(\mu )={\overline{m}}_{Q}\,(m_{Q})
\,\frac{c\,[\alpha_{s}\,(\mu)/\pi ]}{c\, [\alpha_{s}\,(m_{Q})/\pi ]}
\label{eq:msbarev} 
\eeq
with the function $c$, up to three--loop order, given by 
\cite{HqqQCD-2loop,runmass}
\beq
c(x)&=&\left(25x/6 \right)^{12/25} \,
[1+1.014x+1.389\,x^{2} + 1.091\, x^3]
\hspace{1.0cm} \mbox{for} \hspace{.2cm} m_{c}\,<\mu\,<m_{b} \non \\
c(x)&=&\left(23x/6 \right)^{12/23} \,
[1+1.175x+1.501\,x^{2} + 0.1725\, x^3]
\hspace{.8cm} \mbox{for} \hspace{.2cm} m_{b}\,<\mu \,< m_t \non \\
c(x)&=&\left(7x/2 \right)^{4/7} \,
[1+1.398x+1.793\,x^{2} - 0.6834\, x^3]
\hspace{1.35cm} \mbox{for} \hspace{.2cm} m_{t}\,<\mu \, 
\end{eqnarray}
For the charm quark mass for instance, the evolution is determined by the
equation for  $m_c < \mu < m_b$  up to the scale $\mu=m_b$, while for scales
above the bottom mass the evolution must be restarted at $\mu=m_b$.   Using
as starting points the values of the $t,b,c$ quark pole masses given previously
and for $\alpha_s(M_Z)=0.1172$ and $\mu=100$ GeV, the $\overline{\rm MS}$   
running $t,b,c$ quark masses are displayed in Table 1.1. As can be seen, the 
values of the running $b,c$ masses at  the scale $\mu \sim 100$ GeV are, 
respectively, $\sim 1.5$ and $\sim 2$ times smaller than the pole masses, 
while the top quark mass is only slightly different.\s

For the strange quark, this approach fails badly below scales of ${\cal O}$(1
GeV) because of the the too strong QCD coupling. Fortunately, $m_s$ will play 
only  a minor role in Higgs physics and whenever it appears, we will use the 
value $\overline{m}_s({\rm 1\, GeV})=0.2$ GeV.   

\begin{table}[hbt]
\renewcommand{\arraystretch}{1.5} \begin{center} \begin{tabular}{|c||c|c|c|}
\hline $Q$ & $m_Q$ & $\overline{m}_Q (m_Q)$ & $\overline{m}_Q 
(100~{\rm GeV})$ \\ \hline \hline 
$c$  & \ \ 1.64 GeV  \ \ & \ \ 1.23 GeV  \ \   & 0.63 GeV  \\ 
$b$ &  4.88 GeV  & 4.25 GeV    & 2.95 GeV  \\ 
$t$ &  178 GeV   & 170.3 GeV   & 178.3 GeV \\ \hline 
\end{tabular}\\[3mm] 
\end{center}
\label{tb:qmass} 
\nn{\it Table 1.1: The pole quark masses and the mass values in 
the $\overline{MS}$ scheme for the running masses at the scale $m_Q$ and 
at a scale $\mu=100$ GeV;  $\alpha_s(M_Z)=0.1172$.} 
\vspace*{-2mm}
\end{table}

The masses of the charged leptons are given by
\beq
m_\tau=1.777~{\rm GeV}\, ,\ m_\mu=0.1056~{\rm GeV}\, ,\ m_e=0.511~{\rm MeV} 
\eeq
with the electron being too light to play any role in Higgs physics. The 
approximation of massless neutrinos will also have no impact on our discussion.

\vspace*{-2mm}
\subsubsection*{\underline{The gauge boson masses and total widths}} 

Finally, an enormous number of $Z$ bosons has been produced at LEP1 and SLC
at c.m. energies close to the $Z$ resonance, $\sqrt{s} \simeq M_Z$, and of $W$ 
bosons at LEP2 and at the Tevatron. This allowed to make very precise 
measurements of the properties of these particles which provided 
stringent tests of the SM. This subject will be postponed to the next 
section. Here, we simply write the obtained masses and total decay
widths of the two particles \cite{High-Precision}
\beq
M_Z &=& 91.1875 \pm 0.0021 ~{\rm GeV} 
\label{Zmass-average} \\ 
\Gamma_Z &=& 2.4952 \ \pm 0.0023  ~{\rm GeV} 
\label{Zwidth-average}
\eeq
and, averaging the LEP2 \cite{MW-LEP2} and Tevatron \cite{MW-Tevatron} 
measurements, 
\beq
M_W &=& 80.425 \pm 0.034~{\rm GeV} 
\label{Wmass-average}
\\
\Gamma_W&=& 2.133 \ \pm 0.069~{\rm GeV} 
\label{Wwidth-average}
\eeq  
which completes the list of SM parameters that we will use throughout this 
review. 
 
\subsection{High--precision tests of the SM}

Except for the Higgs mass, all the parameters of the SM, the three gauge
coupling constants, the masses of the weak vector bosons and fermions as well
as the quark mixing angles, have been determined experimentally as seen in the
previous section. Using these parameters, one can in principle calculate any
physical observable and compare the result with experiment.  Because the
electroweak constants and the strong coupling constant at high energies are
small enough, the first order of the perturbative expansion, the tree--level or
Born term, is in general sufficient to give relatively good results for most of
these observables. However, to have a more accurate description, one has to
calculate the complicated higher--order terms of the perturbative series, the
so--called radiative corrections. The renormalizability of the theory insures
that these higher--order terms are finite once various formally divergent
counterterms are added by fixing a finite set of renormalization conditions. 
The theory allows, thus, the prediction of any measurable with a high degree of
accuracy.\s

Very precise experiments, which allow a sensitivity to these quantum
corrections, have been made in the last fifteen years. The $e^+ e^-$ colliders
LEP and SLC, which started operation in the late 80's, have collected an
enormous amount of electroweak precision data.  Measurements at the $Z$--pole
[where the production cross section is  extremely large, allowing to collect
more than ten million events at LEP1] of the $Z$ boson partial and total decay
widths, polarization and forward--backward asymmetries where made at the
amazing accuracy of one percent to one per mille \cite{High-Precision}. The $W$
boson properties have  been also determined at the $p \bar{p}$ collider
Tevatron with a c.m. energy  of $\sqrt{s}=1.8$ TeV \cite{MW-Tevatron} and at
LEP2 with a c.m.  energy up to $\sqrt{s}=209$ GeV \cite{MW-LEP2} with a
constant increase in accuracy. Many other high--precision measurements have
been performed at much lower energies.\s

At the same time, a large theoretical effort has been devoted to the
calculation of the radiative  corrections to the electroweak  observables, to
match the accuracies which have been or which could be reached experimentally
\cite{Z-Physics,Z-Precision,W-Physics,Z-Physics-H}.  The availability of both
highly accurate measurements and theoretical predictions, at the level of 0.1\%
precision and better, provides stringent tests of the SM. These high-precision
electroweak data are a unique tool in the search for indirect effects, through
possible small deviations of the experimental results from the theoretical
predictions of the minimal SM, and constitute an excellent probe of its still
untested scalar sector, as well as a probe of New Physics beyond the SM.\s

In this section, after summarizing the high--precision observables 
in the SM, we will describe the formalism needed to incorporate the radiative 
corrections and how the dominant part of the latter can be approximated. This
will allow to set the notation which will be used later and  the
framework which will be necessary to discuss the searches for the virtual 
effects of the Higgs bosons in electroweak observables, and to incorporate the 
important higher--order corrections in Higgs boson decay and production.

\subsubsection{Observables in Z boson decays}

A large variety of precision tests can be performed in $\ee$ experiments with
center--of--mass energies near the $Z$--resonance in  the process $\ee \to
f\bar{f}$ which is  mediated by the exchange of a photon and a $Z$ boson
\cite{Z-Physics2}. The differential cross section  is a binomial in $\cos
\theta$, where $\theta$ is the angle between the electron and the final fermion
$f$. At tree--level, for unpolarized initial beams and for massless final
state  fermions $f \neq e$, it is given by 
\beq
\frac{ {\rm d}\sigma}{ {\rm d} \cos \theta} = \frac{4 \pi \alpha^2}{3s} N_c 
 \left[ \frac{3}{8} (1+ \cos^2 \theta) \sigma_U + \frac{3}{4} \cos \theta 
 \sigma_F \right]
\label{eexsection}
 \eeq
where $\sigma_U$ and $\sigma_F$ are given by [$\Gamma_Z$ is the total decay 
width of the $Z$ boson] 
\beq
\sigma_U &=& Q_e^2 Q_f^2 + 2 Q_e Q_f v_e v_f \frac{s(s-M_Z^2)}{(s-M_Z^2)^2+ 
\Gamma^2_Z
M_Z^2} + (a_e^2+v_e^2)(v_f^2 + a_f^2) \frac{s^2}{(s-M_Z^2)^2+ \Gamma_Z^2M_Z^2}
\non \\
\sigma_F &=&  2 Q_e Q_f a_e a_f \frac{s(s-M_Z^2)}{(s-M_Z^2)^2+ \Gamma^2_Z
M_Z^2} + a_e v_e v_f a_f \frac{s^2}{(s-M_Z^2)^2+ \Gamma_Z^2M_Z^2}
\eeq
where the vector and axial vector couplings of the fermion $f$ to the $Z$ 
boson $v_f$ and $a_f$ [and the reduced couplings $\hat{v}_f$ and 
$\hat{a}_f$ to be used later on] have been given in eq.~(\ref{Zffcouplings}).\s

For center of mass energies near the $Z$ resonance, $\sqrt{s}\simeq M_Z$, the 
$Z$ boson exchange largely dominates. Integrating eq.~(\ref{eexsection}) over 
the entire range of the angle $\theta$, one obtains the total peak cross section
\beq
\sigma_0 (\ee \to Z \to f\bar{f} ) \equiv  \int_{-1}^{+1}
\frac{ {\rm d}\sigma}{ {\rm d} \cos \theta} = 
\frac{12 \pi}{ M_Z^2}  \times \frac{\Gamma_e \Gamma_f}{\Gamma_Z^2}
\eeq
with the partial $Z$ boson decay widths into massless fermion pairs given by  
\beq
\Gamma_f \equiv \Gamma (Z \to f\bar{f} ) = \frac{2\alpha}{3}  N_c M_Z  
(v_f^2 + a_f^2)
\eeq
Convenient measurable quantities which have been considered at LEP1 and SLC 
are, in this context,  the ratio of $Z$ boson partial widths
\beq
R_f = \frac{\Gamma (Z \to f\bar{f}) }{\Gamma (Z \to {\rm hadrons}) }
\eeq
If one integrates asymmetrically eq.~(\ref{eexsection}) and normalizes to the 
total cross section, one obtains the forward--backward asymmetry for the decay 
of a $Z$ boson into a fermion pair
\beq
A_{FB}^f \equiv  \left[  \int_{0}^{+1} \frac{ {\rm d}\sigma}{ {\rm d} \cos 
\theta} - \int_{-1}^{0} \frac{ {\rm d}\sigma}{ {\rm d} \cos \theta} \right] 
\times \sigma_0^{-1}  \ \stackrel{\small \sqrt s = M_Z} = \ \frac{3}{4} A_e A_f 
\label{eeafb}
\eeq
where the combinations $A_f$ are given, in terms of the vector and 
axial vector couplings of the fermion $f$ to the $Z$ boson, by
\beq
A_f = \frac{ 2 a_f v_f} {v_f^2 + a_f^2} \equiv \frac{ 2 \hat{a}_f \hat{v}_f} 
{\hat{v}_f^2 + \hat{a}_f^2}
\eeq
Note that if the initial $e^-$ beams are longitudinally polarized [as it was 
the case at the SLC], one can construct left--right asymmetries and left--right
forward--backward asymmetries; in addition one can measure the longitudinal
polarization in $\tau$ decays and also define a polarization asymmetry. In
terms of the combination of couplings $A_f$ defined above, these observables 
can be written as
\beq
A_{LR}^f = A_e \ \  , \ \  A_{LR, FB}^f = \frac{3}{4} A_f \ \ , \ \
A^\tau_{\rm pol} =A_\tau
\label{eealr}
\eeq
In particular $A_{LR}^f$ [which is the same for all $f \neq e$] and $A^\tau_{\rm
pol}$ are very sensitive to the precise value of $\sin^2 \theta_W$, being 
proportional to the factor $\hat{v}_e \equiv 1- 4 s_W^2 \sim 0$ for $s_W^2 
\sim 1/4$. \s

The tree--level expressions discussed above give results at the one percent
level and hold in most cases, except in the case of $b$--quark final states
where mass effects, ${\cal O}(4 m_b^2/M_Z^2) \sim 0.01$, have to be taken into
account, and in the production of $\ee$ final states where the complicated
$t$--channel gauge boson exchange contributions have to be included [this
process is particularly important since it allows to determine the absolute
luminosity  at $\ee$ colliders]. However, for a very precise description of the
$Z$ properties, one needs to include the one--loop radiative corrections and
possibly some important higher--order effects. These radiative corrections fall
into three categories [see Fig.~1.3]: \s

\begin{figure}[!h]
\begin{center}
\vspace*{-.3cm}
\hspace*{-10.cm}
\begin{picture}(300,100)(0,0)
\SetWidth{1.}
\SetScale{1.}
\ArrowLine(100,75)(130,50)
\ArrowLine(100,25)(130,50)
\Photon(130,50)(170,50){3}{5}
\Text(130,50)[]{{\blue{\large $\bullet$}}}
\Text(170,50)[]{{\blue{\large $\bullet$}}}
\ArrowLine(170,50)(210,25)
\ArrowLine(170,50)(210,75)
\Gluon(197,35)(195,65){3.}{4}
\Text(70,75)[]{{\red{\bf a)}}}
\Text(95,70)[]{$e^+$}
\Text(95,35)[]{$e^-$}
\Text(150,60)[]{$\gamma,Z$}
\Text(215,70)[]{$\bar{q}$}
\Text(215,35)[]{$q$}
\Text(205,50)[]{$g$}
\hspace*{5cm}
\ArrowLine(100,75)(130,50)
\ArrowLine(100,25)(130,50)
\Photon(130,50)(170,50){3}{5}
\Text(130,50)[]{{\blue{\large $\bullet$}}}
\Text(170,50)[]{{\blue{\large $\bullet$}}}
\ArrowLine(170,50)(210,25)
\ArrowLine(170,50)(210,75)
\Gluon(195,35)(215,60){3.2}{4}
\hspace*{5cm}
\ArrowLine(100,75)(130,50)
\ArrowLine(100,25)(130,50)
\Photon(130,50)(170,50){3}{5}
\Text(130,50)[]{{\blue{\Large $\bullet$}}}
\Text(170,50)[]{{\blue{\Large $\bullet$}}}
\ArrowLine(170,50)(210,25)
\ArrowLine(170,50)(210,75)
\Gluon(195,65)(215,40){3.2}{4}
\end{picture}
\end{center}
\vspace*{-1.2cm}
\begin{center}
\vspace*{-.3cm}
\hspace*{-10.cm}
\begin{picture}(300,100)(0,0)
\SetWidth{1.}
\SetScale{1.}
\ArrowLine(100,75)(130,50)
\ArrowLine(100,25)(130,50)
\Photon(130,50)(170,50){3}{5}
\ArrowLine(170,50)(210,25)
\ArrowLine(170,50)(210,75)
\Photon(195,35)(195,65){3.}{4.5}
\Text(130,50)[]{{\blue{\large $\bullet$}}}
\Text(170,50)[]{{\blue{\large $\bullet$}}}
\Text(70,75)[]{\red{\bf b)}}
\Text(95,70)[]{$e^+$}
\Text(95,35)[]{$e^-$}
\Text(150,60)[]{$\gamma,Z$}
\Text(215,70)[]{$\bar{f}$}
\Text(215,35)[]{$f$}
\Text(205,50)[]{$\gamma$}
\hspace*{5cm}
\ArrowLine(100,75)(130,50)
\ArrowLine(100,25)(130,50)
\Photon(130,50)(170,50){3}{5}
\ArrowLine(170,50)(210,25)
\ArrowLine(170,50)(210,75)
\Photon(195,35)(210,60){3.}{4.5}
\Text(130,50)[]{{\blue{\large $\bullet$}}}
\Text(170,50)[]{{\blue{\large $\bullet$}}}
\hspace*{5cm}
\ArrowLine(100,75)(130,50)
\ArrowLine(100,25)(130,50)
\Photon(130,50)(170,50){3}{5}
\ArrowLine(170,50)(210,25)
\ArrowLine(170,50)(210,75)
\Photon(115,63)(140,67){3}{4.5}
\Text(130,50)[]{{\blue{\large $\bullet$}}}
\Text(170,50)[]{{\blue{\large $\bullet$}}}
\end{picture}
\end{center}
\vspace*{-1.2cm}
\begin{center}
\vspace*{-.3cm}
\hspace*{-10.5cm}
\begin{picture}(300,100)(0,0)
\SetWidth{1.}
\SetScale{1.}
\ArrowLine(90,75)(120,50)
\ArrowLine(90,25)(120,50)
\Photon(120,50)(135,50){3}{3}
\ArrowArc(145,50)(10,0,180)
\ArrowArc(145,50)(10,180,360)
\Photon(155,50)(170,50){3}{3}
\ArrowLine(170,50)(210,25)
\ArrowLine(170,50)(210,75)
\Text(120,50)[]{{\blue{\large $\bullet$}}}
\Text(170,50)[]{{\blue{\large $\bullet$}}}
\Text(80,81)[]{\red{\bf c)}}
\Text(90,60)[]{$e^+$}
\Text(90,40)[]{$e^-$}
\Text(143,69)[]{$f$}
\Text(215,70)[]{$\bar{f}$}
\Text(215,37)[]{$f$}
\hspace*{5cm}
\ArrowLine(100,75)(130,50)
\ArrowLine(100,25)(130,50)
\Photon(130,50)(170,50){3}{5}
\ArrowLine(170,50)(210,25)
\ArrowLine(170,50)(210,75)
\Photon(195,35)(195,65){3}{4.5}
\Text(210,50)[]{$V$}
\Text(130,50)[]{{\blue{\large $\bullet$}}}
\Text(170,50)[]{{\blue{\large $\bullet$}}}
\hspace*{5cm}
\ArrowLine(100,75)(130,75)
\ArrowLine(100,25)(130,25)
\Photon(130,75)(170,75){3}{4.5}
\Photon(130,25)(170,25){3}{4.5}
\Text(130,75)[]{{\blue{\large $\bullet$}}}
\Text(170,75)[]{{\blue{\large $\bullet$}}}
\Text(130,25)[]{{\blue{\large $\bullet$}}}
\Text(170,25)[]{{\blue{\large $\bullet$}}}
\ArrowLine(130,75)(130,25)
\ArrowLine(170,75)(170,25)
\ArrowLine(170,25)(210,25)
\ArrowLine(170,75)(210,75)
\Text(150,65)[]{$V$}
\Text(150,35)[]{$V$}
\end{picture}
\end{center}
\vspace*{-.7cm}
\nn {\it Figure 1.3: Examples of Feynman diagrams for the radiative corrections
to the process $\ee \to f\bar{f}$: a) virtual and real QCD corrections for 
quark final states, b) virtual QED corrections and initial and final state 
photon radiation and c) genuine electroweak  corrections including 
self--energy, vertex and box corrections.} 
\end{figure}

{\bf a)} QCD corrections to final states quarks, where gluons are exchanged or
emitted in the final state. For massless quarks the correction factors  are
\beq
K^{\rm QCD}_{Z \to q\bar{q}} = 1+ \frac{\alpha_s}{\pi} + 1.41 
\left(\frac{\alpha_s}{\pi} \right)^2
\label{GammaQCD}
\eeq
for the partial decay widths $Z \to q\bar{q}$ or total cross section $\Gamma_q 
\propto \sigma( \ee \to q \bar{q})$, and 
\beq
K^{\rm QCD}_{A_{\rm FB}^{q}}= 1- \frac{\alpha_s}{\pi}
\eeq
for the forward--backward quark asymmetries. In fact these QCD factors are 
known to ${\cal O}(\alpha_s^3)$ for $\Gamma_q$ and  to ${\cal O}(\alpha_s^2)$ 
for $A_{FB}^q$; in the case of $b$--quarks one can include the mass effects at 
${\cal O} (\alpha_s)$ which are also known  [for a detailed
discussion of all these corrections, see Ref.~\cite{Z-Precision2} e.g.].
Note that $\Gamma_q$ allows for one of the cleanest and most precise 
determinations of $\alpha_s$ \cite{alphas-review}.\s

{\bf b)} Pure electromagnetic corrections. These consist of initial and final
state corrections where photons are exchanged in the $Zf\bar{f}$ vertices or
 emitted in the initial or final states. For final state corrections, it is
sufficient to include the small 
\beq
K^{\rm EM}_{Z \to f\bar{f}} = 1 + \frac{3}{4}Q^2_f \frac{\alpha}{\pi}
\ , \ 
K^{\rm EM}_{A_{FB}^{f}} = 1 - \frac{3}{4}Q^2_f \frac{\alpha}{\pi}
\eeq
correction factors, while for  initial state corrections, in particular the
photon radiation (ISR), one can use the  standard approach of structure
function where the corrections can be exponentiated. This is performed 
by convoluting the Born cross section eq.~(\ref{eexsection}) with
a radiator function $G(s')$ for the full accessible c.m. energies $s'=xs$ 
after photon radiation 
\beq
\sigma^{\rm ISR} (s)= \int_{x_0}^1 {\rm d} x \, G(xs) \sigma_{\rm Born} (xs)
\ , \ G(xs) = \beta (1-x)^{\beta-1} \delta_{V+S}(x) + \delta_H(x)
\label{QEDradiator}
\eeq
where $x_0$ is the minimum energy of the final state, $x_0= 4m_f^2/s$ for  $\ee
\to f\bar{f}$, and $G(x)$ is the radiator function, which is written in an
exponentiated form to resum the infrared sensitive and large corrections. In the
previous equation, $\beta= \alpha/\pi \times [\log s/m_e^2 -1]$ and
$\delta_{V+S}$, $\delta_H$ contain, respectively,  the virtual plus
soft--photon contributions, and  the  hard--photon contributions, which are
polynomials in  $\log(s/m_e^2)$. Their expressions, as well as many details on
ISR, FSR and their interference can be found in the reviews of 
Refs.~\cite{Z-Physics2,Z-Physics4}.  Note that all these corrections 
do not involve any other physics than well known QED.\s

{\bf c)} Electroweak corrections. They involve non--photonic ``direct" vertex
and box corrections which are in general rather small [except in a few cases to
be discussed later] as well as the ``oblique" $\gamma, W$ and $Z$ boson
self--energy corrections and the $\gamma$--$Z$ mixing, which give the bulk of
the contributions \cite{Z-Physics3,Z-Physics-H}. In particular, the top quark
[which was not yet discovered at the time LEP1 and SLC started] and the Higgs
boson will enter the electroweak observables through  their  contributions to
the $W$ and $Z$ boson self--energies. These electroweak corrections are
discussed in some detail in the next subsection.  

\subsubsection{The electroweak radiative corrections}

The electroweak radiative corrections can be cast into three main categories; 
Fig.~1.4:

\begin{itemize} 
\vspace*{-3mm}

\item[a)] The fermionic corrections to the gauge boson self--energies. They can
be divided themselves into the light fermion $f\neq t$ contributions and the 
contribution of the heavy top quark $f=t$. For the contributions of quarks, one
has to include the important corrections stemming from strong interactions.  
\vspace*{-2mm}

\item[b)] The contributions of the Higgs particle to the $W$ and $Z$ boson
self--energies both at the one--loop level and at the two--level when e.g. the
heavy top quark is involved. 
\vspace*{-2mm}

\item[c)] Vertex corrections to the $Z$ decays into fermions, in particular
into $b\bar{b}$ pairs, and vertex plus box contributions to muon decay [in 
which the bosonic contribution is not gauge invariant by itself and should be 
combined with the self--energy corrections]. There are also direct box 
corrections, but their contribution at the $Z$--peak is negligible. 
\end{itemize}

\begin{center}
\vspace*{-.3cm}
\hspace*{-10.cm}
\SetWidth{1.}
\SetScale{1.}
\begin{picture}(300,100)(0,0)
\Text(80,85)[]{\red{\bf a)}}
\Photon(85,50)(125,50){3.2}{5.5}
\Photon(175,50)(215,50){3.2}{5.5}
\ArrowArc(150,50)(25,0,180)
\ArrowArc(150,50)(25,180,360)
\Text(150,85)[]{$f$}
\Text(100,65)[]{$V$}
\Text(200,65)[]{$V$}
\Text(175,50)[]{{\blue{\large $\bullet$}}}
\Text(125,50)[]{{\blue{\large $\bullet$}}}
\hspace*{-2cm}
\Photon(285,50)(325,50){3.2}{5.5}
\Photon(375,50)(415,50){3.2}{5.5}
\Text(375,50)[]{{\blue{\large $\bullet$}}}
\Text(325,50)[]{{\blue{\large $\bullet$}}}
\ArrowArc(350,50)(25,0,180)
\ArrowArc(350,50)(25,180,360)
\Gluon(330,65)(370,65){3.5}{5}
\Text(350,54)[]{$g$}
\Text(350,85)[]{$q$}
\hspace*{-2cm}
\Photon(485,50)(525,50){3.2}{5.5}
\Photon(575,50)(615,50){3.2}{5.5}
\Text(575,50)[]{{\blue{\large $\bullet$}}}
\Text(525,50)[]{{\blue{\large $\bullet$}}}
\ArrowArc(550,50)(25,0,180)
\ArrowArc(550,50)(25,180,360)
\Gluon(526,60)(574,40){3.5}{6}
\Gluon(550,75)(550,25){3.5}{5}
\Text(550,85)[]{$q$}
\end{picture}
\vspace*{-1.2cm}
\end{center}
\begin{center}
\vspace*{-.3cm}
\hspace*{-10.cm}
\SetWidth{1.}
\SetScale{1.}
\begin{picture}(300,100)(0,0)
\Text(80,85)[]{\red{\bf b)}}
\Photon(85,50)(125,50){3.2}{5.5}
\Photon(175,50)(215,50){3.2}{5.5}
\DashCArc(150,50)(25,0,180){4}
\PhotonArc(150,50)(25,180,360){3.2}{9.5}
\Text(175,50)[]{{\blue{\large $\bullet$}}}
\Text(125,50)[]{{\blue{\large $\bullet$}}}
\Text(150,85)[]{$H$}
\Text(100,65)[]{$W/Z$}
\Text(200,65)[]{$W/Z$}
\hspace*{-2cm}
\Photon(285,50)(325,50){3.2}{5.5}
\Photon(375,50)(415,50){3.2}{5.5}
\Text(375,50)[]{{\blue{\large $\bullet$}}}
\Text(325,50)[]{{\blue{\large $\bullet$}}}
\DashCArc(350,50)(25,0,180){4}
\PhotonArc(350,50)(25,180,360){3.2}{9.5}
\DashLine(330,65)(370,65){4}
\Text(350,57)[]{$H$}
\hspace*{-2cm}
\Photon(485,50)(525,50){3.2}{5.5}
\Photon(575,50)(615,50){3.2}{5.5}
\Text(575,50)[]{{\blue{\large $\bullet$}}}
\Text(525,50)[]{{\blue{\large $\bullet$}}}
\ArrowArc(550,50)(25,0,180)
\ArrowArc(550,50)(25,180,360)
\DashLine(550,75)(550,25){4}
\Text(540,57)[]{$H$}
\Text(550,85)[]{$t$}
\vspace*{-2.5cm}
\end{picture}
\end{center}  
\vspace*{-1.2cm}
\begin{center}
\vspace*{-.9cm}
\hspace*{-7.cm}
\SetWidth{1.}
\SetScale{1.}
\begin{picture}(300,100)(0,0)
\Text(40,85)[]{\red{\bf c)}}
\Photon(85,50)(135,50){3.2}{6}
\Text(135,50)[]{{\blue{\large $\bullet$}}}
\ArrowLine(135,50)(185,85)
\ArrowLine(135,50)(185,15)
\Text(170,25)[]{{\blue{\large $\bullet$}}}
\Text(170,75)[]{{\blue{\large $\bullet$}}}
\Photon(170,25)(170,75){3.2}{6}
\Text(150,75)[]{$t$}
\Text(150,25)[]{$\bar{t}$}
\Text(190,80)[]{$b$}
\Text(190,20)[]{$\bar{b}$}
\Text(110,65)[]{$Z$}
\Text(190,50)[]{$W$}
\hspace*{-1cm}
\ArrowLine(280,50)(325,50)
\ArrowLine(325,50)(390,85)
\Text(325,50)[]{{\blue{\large $\bullet$}}}
\Text(365,30)[]{{\blue{\large $\bullet$}}}
\Photon(325,50)(365,30){3.2}{6}
\ArrowLine(365,30)(390,10)
\ArrowLine(365,30)(390,45)
\Photon(350,62)(380,42){3.2}{5}
\Text(350,62)[]{{\blue{\large $\bullet$}}}
\Text(380,42)[]{{\blue{\large $\bullet$}}}
\Text(310,62)[]{$\mu^-$}
\Text(400,82)[]{$\nu_\mu$}
\Text(400,10)[]{$e^-$}
\Text(400,40)[]{$\bar{\nu}_e$}
\Text(334,30)[]{$W$}
\Text(370,63)[]{$Z$}
\end{picture}
\vspace*{-5mm}
\end{center}
{\it Figure 1.4: Generic Feynman diagrams for the main electroweak radiative 
corrections: a) fermionic contributions to the two--point functions of the
$V=W/Z$ bosons, b) Higgs boson contributions to the two--point functions and  
c) vertex and box corrections.}\vspace*{2mm}

The contribution of the light fermions to the vector boson self--energies can
be essentially mapped into the running of the QED coupling constant which, as 
discussed in the previous section, is defined as the difference between the 
vacuum polarization function of the photon evaluated at low energies and at the 
scale $M_Z$, $\Delta \alpha (M_Z^2) = \Pi_{\gamma \gamma}(0) - \Pi_{\gamma 
\gamma} (M_Z^2) = 0.0590 \pm 0.00036$. Therefore, the only remaining fermionic 
contribution to the two--point functions is the one due to the top quark on 
which, besides the effects of the Higgs boson, we will mainly concentrate
by studying three important quantities, $\Delta \rho$, $ \Delta r$ and the
$Zb\bar{b}$ vertex.

\subsubsection*{\underline{The effective mixing angle and the $\rho$ 
parameter}}

\nn The effective weak mixing angle can be defined in the Born approximation in
terms of the  $W$ and $Z$ boson masses\footnote{When higher--order corrections
are included, different definitions of $s_W^2$ lead to different values. For
instance, $s_W^2$ as defined above is different from the effective leptonic
$s_W^2|^{\rm lept}_{\rm eff}$ defined in terms of $a_e$ and $v_e$.}, 
eq.~(\ref{sw-definition}). To include higher orders, one has to renormalize the
$V$ boson masses $M_V^2 \to M_V^2 - \Pi_{VV}(M_V^2)$ where $\Pi_{VV}$ is the 
real part of the transverse component of the  self--energy of $V$  at the 
scale $M_V$. One obtains an effective mixing angle \cite{Z-Physics3,Z-Physics-H}
\beq
\bar{s}_W^2 =   1- \frac{M_W^2}{M_Z^2} + c_W^2 \left( \frac{\Pi_{WW}(M_W^2)} 
{M_W^2} - \frac{\Pi_{ZZ}(M_Z^2)} {M_Z^2} \right) \sim 1- \frac{M_W^2}{M_Z^2} 
+ c_W^2 \Delta \rho
\label{swbar}
\eeq
This is in fact the correction to the $\rho$ parameter \cite{rho-definition}
which historically was used to measure the strength of the ratio of the
neutral current to the charged current at zero--momentum transfer in
deep--inelastic neutrino--nucleon scattering, eq.~(\ref{rho-MWMZ}). In the SM,
as already mentioned, because of a global or custodial  SU(2)$_{\rm R}$
symmetry of the Higgs Lagrangian [which survives the spontaneous breaking of
the EW symmetry], this parameter is equal to unity.  However, it receives
higher--order corrections usually parameterized by
\beq
\rho = \frac{ 1}{1 -\Delta \rho} \ \ , \ \ 
\Delta \rho = \frac{\Pi_{WW}(0)} {M_W^2}  - \frac{\Pi_{ZZ}(0)} {M_Z^2}
\label{rho-def}
\eeq
The main contribution to this parameter is due to the $(t,b)$ weak isodoublet. 
Indeed, the large mass splitting between the top and bottom quark masses
breaks the custodial ${\rm SU(2)_R}$ symmetry and generates a contribution  
which grows as the top mass squared\footnote{Because $m_t$ is large, the
contributions are approximately the same at the scale $q^2 \sim 0$ or $q^2
\sim  M_V^2$; in addition the light fermion contributions to $\Pi_{WW}$ and
$\Pi_{ZZ}$ almost cancel in the difference, $\sim \log M_W/M_Z$. This is the 
reason why one can approximate the correction to $s_W^2$ in eq.~(\ref{swbar}) 
by the one 
in eq.~(\ref{rho-def}).} \cite{rho-1loop}. Including the dominant higher--order 
QCD and electroweak corrections, one finds   
\beq 
\Delta \rho = 3 x_t \left[ 1+ (\Delta \rho)^{\rm QCD} + (\Delta \rho)^{\rm EW}
\right] \label{deltarho} \\
x_t= \frac{g_{Htt}^2}{(4 \pi)^2} = \frac{G_\mu m_t^2}{8 \sqrt{2} \pi^2} \sim 
0.3\%
\label{define:xt} 
\eeq
The higher--order QCD corrections are known at two--loop \cite{rho-2loopQCD}
and three--loop \cite{rho-3loopQCD} orders; with $\alpha_s$ defined at the 
scale $\mu = m_t$ with 6 flavors, they are given by
\beq
(\Delta \rho)^{\rm QCD} = - \frac{2}{3} \frac{\alpha_s}{\pi} \left( 
\frac{\pi^2}{3}+1 \right) - 14.59 \left( \frac{\alpha_s}{\pi} \right)^2 
\eeq
There are also two--loop electroweak corrections stemming from  fermion loops.
In particular, there is a correction where a Higgs or a Goldstone boson is
exchanged in loops containing top quarks and which grows as  $G_\mu^2 m_t^4$ and
$G_\mu^2 m_t^2 M_Z^2$. In the limit where the Higgs boson mass is much smaller
than $m_t$, the leading piece gives a tiny correction \cite{rho-2loopEW1} 
\beq
(\Delta \rho)^{\rm EW} \simeq (19 -2\pi^2)x_t \sim -x_t 
\eeq 
However, for the more realistic case of a finite Higgs mass, the correction 
can be much larger \cite{rho-2loopEW2}; in addition, the subleading ${\cal O} 
(G_\mu^2 m_t^2 M_Z^2)$ are also significant \cite{rho-2loopEW3,Paolo-approach}. 
Recently, the full fermionic contributions to $\Delta \rho$ and to $\sin^2
\theta_W$ have been derived at the two--loop level \cite{Awramicketal}. Other
higher--order corrections, such as the mixed QED--QCD contributions and the 
three--loop QCD corrections, are also available \cite{Mixed-QEDQCD}.   \s

At the one--loop level the Higgs boson will also contribute to the $\rho$ 
parameter \cite{rho-Veltman}
\beq
(\Delta \rho)^{\rm 1-Higgs}= -\frac{3G_\mu M_W^2}{8 \sqrt{2} \pi^2} 
\, f \bigg( \frac{M_H^2}{M_Z^2} \bigg) \ , \quad  
f(x) = x \left[ \frac{\ln c_W^2 - \ln x}{c_W^2-x} + \frac{\ln x}{ c_W^2
(1-x)} \right]
\eeq
This contribution vanishes in the limit $s_W^2 \to 0$ or  $M_W \to M_Z$, i.e. 
when the hypercharge group is switched off. For a very light Higgs boson
the correction also vanishes 
\beq
(\Delta \rho)^{\rm 1-Higgs} \to 0 \  \ \ {\rm for}~M_H \ll M_W
\eeq
while for a heavy Higgs boson, the contribution is approximately given by
\beq
(\Delta \rho)^{\rm 1-Higgs} \sim -  \frac{3G_\mu M_W^2}{8 \sqrt{2} 
\pi^2} \, \, \frac{s_W^2}{c_W^2}  \log \frac{M_H^2}{M_W^2}  \ \ \ 
{\rm for}~M_H \gg M_W
\eeq
This contribution has only a logarithmic dependence in the  Higgs boson mass.
This has to be contrasted with the general case, where the  contribution of two
particles with a large mass splitting grows with the mass of the heaviest
particle [as is the case of the  top/bottom  weak isodoublet] and thus, can be
very large.   This logarithmic dependence is due to what is called the
``Veltman screening theorem" \cite{rho-Veltman,Screening-others} which tells 
us that the quadratic corrections $\propto M_H^2$ appear only at the two--loop 
level, and are therefore screened or damped by an extra power of the  
electroweak coupling squared. \s

The two--loop Higgs corrections to the $\rho$ parameter stemming from the
exchange of the Higgs particles [and the Goldstone bosons] is known in the
large Higgs mass limit since quite some time \cite{rho-Higgs2loop}, but
recently the three--loop contribution has been also calculated \cite{Jochum}.
The sum of the two contributions reads for $M_H \gg M_W$
\beq
(\Delta \rho)^{\rm 2+3-Higgs} \sim  0.15  \left( \frac{G_\mu M_W^2}{2 \sqrt{2} 
\pi^2} \right)^2 \, \frac{s_W^2}{c_W^2} \, \frac{M_H^2}{M_W^2}
- 1.73 \left( \frac{G_\mu M_W^2}{2 \sqrt{2} \pi^2} \right)^4 \, \frac{s_W^2}
{c_W^2} \, \frac{M_H^4}{M_W^4}
\eeq
Both the two-- and three--loop corrections are extremely small for reasonable
values of $M_H$. However, for $M_H \sim 400$ GeV, the two corrections become of
the same size, ${\cal O} (10^{-5})$, but with opposite sign and cancel each
other. For $M_H \sim 1.2$ TeV, the three--loop correction is comparable with
the one--loop contribution and has the same sign. \s 

Nevertheless, for a relatively light Higgs boson and except when it comes to
very high--precision tests, one can neglect these Higgs boson corrections to
the $\rho$ parameter, and keep only the QCD and leading electroweak corrected
top quark contribution.  This $\Delta \rho$ correction will be the largest
contribution to the electroweak corrections after $\Delta \alpha (M_Z^2)$ since,
for $m_t \sim 180$ GeV, it is at the level of $\sim 1$\%.  

\subsubsection*{\underline{The $\mathbf{Zb\bar b}$ vertex}}

\nn In the context of precision tests, the $Z$ boson decays into bottom  quarks
has a special status. First of all, because of its large mass and relatively
large lifetime, the $b$ quark can be tagged and experimentally separated from
light quark and gluon jets allowing an independent measurement of the $Z \to
b\bar{b}$ partial decay width  and the forward backward asymmetry $A_{FB}^b$.
Since $m_b$ is sizable, mass effects of  ${\cal O} (4 m_b^2  /M_Z^2) \sim 1$\%
have to be incorporated in these observables at both the tree--level and in the
QCD corrections \cite{bquark-QCD}.  In addition, large radiative corrections 
involving the top quark and not contained in $\Delta \rho$ appear.  Indeed, the
latter can be exchanged  together with a $W$ boson in the $Zb \bar{b}$ vertex
and the longitudinal  components of the $W$ boson [or  the charged Goldstone
whose coupling is proportional to the fermion mass] leads to contributions that
are quadratic in the top quark mass.  These corrections can be accounted for
simply by shifting the reduced vector and axial--vector $Zb\bar{b}$ couplings
by the amount
\beq
\hat{a}_b \to 2I_b^3 (1+ \Delta_b) \ \ , \ \  \hat{v}_b \to  2I_b^3 
(1+  \Delta_b) -4 Q_b s_W^2 
\label{deltabv}
\eeq
where the rather involved expression of the vertex correction $\Delta_b$ is
given in Ref.~\cite{Zbbvertex}. In the limit of a heavy top quark, the 
correction can be cast into a rather simple form
\beq
\Delta_b = - \frac{ G_\mu m_t^2}{4 \sqrt{2} \pi^2}
- \frac{ G_\mu M_Z^2}{12 \sqrt{2} \pi^2}(1+c_W^2) \log \frac{m_t^2}{M_W^2} 
+ \cdots 
\eeq
This correction is large [note that the logarithmic piece is also 
important] being approximately of the same size as the $\Delta\rho$ correction. 
The $Zb\bar{b}$ vertex allows thus an independent probe of the top quark;
see for instance the discussions in Ref.~\cite{bquark-physics}. \s

The Higgs boson will contribute to the $Z b\bar b$ vertex in two ways. First,
at the one--loop level, it can be exchanged between the two bottom quarks, 
leading to a contribution that is proportional to \cite{Zbbvertex}
\beq
\Delta_b^{1-\rm Higgs} \propto   \frac{ G_\mu m_b^2}{4 \sqrt{2} \pi^2}
\eeq
Because the $b$--quark mass is very small compared to the $W$ boson mass,
$m_b^2/M_W^2 \sim 1/250$, this contribution is negligible in the SM. Another
contribution, similarly to what occurs in the $\Delta \rho$ case,  is simply 
due to the two--loop corrections of ${\cal O} (G_\mu^2 m_t^4)$ to $\Delta_b$, 
which in the limit of small Higgs boson masses is given by 
\cite{rho-2loopEW2,Zbbvertex2loop}
\beq
\Delta_b^{2-\rm EW} \propto  -2 x_t^2  \left( 9 - \frac{ \pi^2}{3} \right)
\eeq
which is again very small and can be safely neglected. Thus, only the Higgs
boson contributions to the two--point functions have to be taken into account.


\subsubsection{Observables in $W$ boson production and decay}

\subsubsection*{\underline{$W$ pair production in $\ee$ collisions}}

The pair production of $W$ bosons in $\ee$ collisions, $\ee \to W^+W^-$, is the
best suited process to test directly the gauge symmetry of the SM 
\cite{W-Physics}. 
Indeed, the process is mediated by $t$--channel neutrino exchange and by
$s$--channel photon and $Z$ boson exchanges, Fig.~1.5, which involve the triple
$\gamma WW$ and $ZWW$ couplings that are dictated by the ${\rm SU(2)_L \times
U(1)_Y}$ gauge symmetry, eq.~(\ref{WWVcoupling}). There is an additional
contribution from $s$--channel Higgs boson exchange but it is negligibly 
small, being proportional to the square of the electron mass\footnote{Note,
however, that at extremely high energies, this suppression is compensated  by
terms proportional to the c.m.  energies. In fact, the cross section with only
the two other channels included, violates unitarity at $\sqrt{s} \gg M_W^2$, 
and unitarity is restored only if the Higgs boson channel is included with the
couplings of the Higgs particle to electrons and $W$ bosons exactly as
predicted in the SM \cite{UNITARITY}.}.  

\begin{center}
\vspace*{-.5cm}
\hspace*{-9.5cm}
\begin{picture}(300,100)(0,0)
\SetWidth{1.}
\SetScale{1.1}
\ArrowLine(100,75)(130,50)
\ArrowLine(100,25)(130,50)
\Photon(130,50)(170,50){4}{5}
\Photon(170,50)(200,25){4}{5}
\Photon(170,50)(200,75){4}{5}
\Text(110,70)[]{$e^+$}
\Text(110,35)[]{$e^-$}
\Text(160,69)[]{$\gamma,Z$}
\Text(230,70)[]{$W^+$}
\Text(230,35)[]{$W^-$}
\ArrowLine(230,75)(265,75)
\Photon(265,75)(310,75){4}{5}
\ArrowLine(230,25)(265,25)
\Photon(265,25)(310,25){4}{5}
\ArrowLine(265,25)(265,75)
\Text(280,50)[]{$\nu_e$}
\hspace*{9cm}
\ArrowLine(100,75)(130,50)
\ArrowLine(100,25)(130,50)
\DashLine(130,50)(170,50){4}
\Photon(170,50)(200,25){4}{5}
\Photon(170,50)(200,75){4}{5}
\Text(160,65)[]{$H$}
\end{picture}
\vspace*{-1cm}
\end{center}
\centerline{\it Figure 1.5: Feynman diagrams for the pair production of $W$
bosons in $\ee$ collisions.} 
\vspace*{2mm}

The total production cross section for the process $\ee \to W^+ W^-$ is given 
by \cite{ee-WW-tree}
\beq
\sigma   & = & 
 \frac{\pi \alpha^2}{2s_W^4} \frac{\beta }{s}
 \left\{  \left[ 1+ \frac{2M_W^2}{s} + \frac{2M_W^4}{s^2} \right]
 \frac{1}{\beta } \log \frac{1+\beta }{1-\beta }
 - \frac{5}{4} \right. \nonumber \\
& +&  \frac{M_Z^2(1-2s_W^2)}{s-M_Z^2}
 \left[ 2  \left( \frac{M_W^4}{s^2} + \frac{2M_W^2}{s} \right) 
 \frac{1}{\beta } \log \frac{1+\beta }{1-\beta }
 - \frac{s}{12 M_W^2} - \frac{5}{3} - \frac{M_W^2}{s}  \right] 
\nonumber \\
&+ & \left. \frac{M_Z^4(8s_W^4-4s_W^2+1)\beta^2}{48(s-M_Z^2)^2}
 \left[ \frac{s^2}{M_W^4} + \frac{20s}{M_W^2} + 12 \right]  \right\}  
\end{eqnarray}
with $\beta= (1- 4M_W^2/s)^{1/2}$ being the velocity of the $W$ bosons. The
$W$ bosons will decay into almost massless fermion pairs with partial decay
widths given by
\beq
\Gamma (W \to f_i \bar{f}_j) = \frac{2}{3} N_c \, \alpha |V_{ij}|^2  \, M_W \,
(v_f^2 +a_f^2)
\eeq
where $V_{ij}$ are the Cabibbo--Kobayashi--Maskawa matrix elements and
$v_f$ and $a_f$ the vector and axial--vector couplings of the $W$ boson to
fermions given in eq.~(\ref{Wffcouplings}). The same type of radiative
corrections which affect the $Z \to f\bar f$ partial width appear also
in this case.\s

Of course, to have an accurate description of the process, one has to consider
many differential distributions, the possibility of off--shell $W$ bosons etc..,
and  higher--order effects including radiative  corrections, have to be
implemented. A large theoretical effort has been  devoted to this topic in the
last two decades; see e.g.~Ref.~\cite{W-Physics} for a review. Let us simply
note here that the contribution of the Higgs particle to the radiative 
corrections in this reaction \cite{ee-WW-RC,WWRC-rev} are too small to be 
measurable.\s

This process also allows to make a very precise measurement of the $W$ boson
mass. This can be performed not only via a scan in the threshold region where
the cross section rises steeply, $\sigma \sim   \beta$, but also in the
reconstruction of the $W$ bosons in mixed  lepton/jet final states for
instance. The $W$ boson width can also be measured by scanning around the
threshold.

\subsubsection*{\underline{$\mathbf{W}$ production in hadronic collisions}}

$W$ bosons can also be produced in hadronic collisions in the Drell--Yan
process, $q \bar{q}' \to W$, and detected  in their leptonic decay channel 
$W \to \ell \nu$ for instance \cite{MW-Tevatron-rev,EW-LHC}. The differential 
cross section for the subprocess $u \bar{d} \to W^+ \to \ell^+ \nu$ is  given by
\cite{Drell-Yan}
\beq
\frac{ {\rm d} \hat{\sigma} }{ {\rm d} {\hat \Omega}} =
\frac{\alpha^2 |V_{ud}|^2}{192 s_W^2} \, \frac{1}{\hat{s}} \, \frac{ \hat{u}^2 }
{ (\hat s-M_W^2)^2 + \Gamma_W^2 M_W^2} 
\eeq
where   $\hat{s}$ is the center--of--mass energy of the subprocess, $\hat{u}$
the square momentum difference between the up--type quark and the lepton 
and $\hat{\Omega}$ is the solid angle of the lepton $\ell$ in the parton
c.m. frame. The hadronic cross section can be obtained by convoluting the 
previous equation with the corresponding (anti)quark densities of the protons.  
Defining $\tau_0=M_W^2/s$  with $s$ being the total hadronic c.m.~energy 
squared, one would then have
\begin{eqnarray}
\sigma (pp \ra W) = \int_{\tau_0}^1 d\tau 
\sum_q \frac{{\rm d} {\cal L}^{q\bar q}}{{\rm d} \tau} \hat\sigma (\hat{s}
=\tau s)  
\end{eqnarray}
Here again, radiative corrections, in particular those due to the strong
interaction \cite{DYNLO,DYNNLO} to be discussed later, have to be implemented 
in order to describe accurately the process. \s

The $W$ boson mass can be determined \cite{MW-Tevatron-rev,EW-LHC} in the 
leptonic decay channels by 
measuring the transverse mass $m_T^W = \sqrt{ 2 p_T^\ell p_T^\nu (1- \cos 
\phi)}$ where  $\phi$ is the angle between the charged lepton and the  neutrino
in the transverse plane. While the lepton transverse momentum $p_T^\ell$ is
directly measured,  $p_T^\nu$ is obtained from the momentum of the system
recoiling against the $W$ in the transverse plane. The edge of the $m_T^W$
distribution is very sensitive to the $W$ boson mass.   By fitting the
experimental $m_T^W$ distribution with Monte--Carlo events generated with
different values of $M_W$, one can determine the $M_W$ value  which gives the
best result fit. The $W$ boson width can also be  measured  with a reasonable
accuracy since it enters the process.\s

However, besides the background problems, there are many uncertainties which
are involved in this measurement: the not very precise knowledge of the parton
distributions, the effect of the $W$ boson total width, the radiative decays
and the approximate knowledge of the $p_T$ spectrum and distribution of the $W$
boson. But most of these uncertainties can be strongly constrained by using the
much cleaner process $pp \to Z \to \ell^+ \ell^-$, with the $Z$ boson mass
accurately determined at LEP1/SLC. 

\subsubsection*{\underline{Muon decay and the radiative corrections to the
W boson mass}}

\vspace*{2mm}

As discussed in \S1.1.4, the $W$ boson mass is related to $\alpha, G_\mu$
and $M_Z$, eq.~(\ref{def:GF}).  Including the radiative corrections, one 
obtains the celebrated relation \cite{Deltar} 
\beq
M_W^2 \left( 1- \frac{M_W^2}{M_Z^2} \right) = \frac{ \pi \alpha}{\sqrt{2} G_\mu}
(1 + \Delta r)
\eeq
The $\Delta r$ correction can be decomposed into three main components  and 
can be written as \cite{Z-Physics3}
\beq
1+ \Delta r = \frac{1}{(1- \Delta \alpha)(1 + \frac{c_W^2}{s_W^2} \Delta \rho) 
- (\Delta r)_{\rm rem} }
\label{deltarsum}
\eeq
where the $\Delta \alpha$ and $\Delta \rho$ contributions have been discussed
previously and $( \Delta r)_{\rm rem}$ collects the remaining non--leading
contributions.  Among these are some non--quadratic but still sizable
corrections due to the  top quark, additional light fermions contributions, as
well as some vertex and  box corrections involved in muon decay 
\cite{Z-Physics3}
\beq
(\Delta r)_{\rm rem}^{\rm box+vertex} &=& \frac{\alpha}{4 \pi s_W^2} \left(6 +
\frac{7-4 s_W^2}{2s_W^2} \log c_W^2 \right) \non \\
(\Delta r)_{\rm rem}^{\rm light-fermions} &=& \frac{\alpha}{4 \pi s_W^2} 
\frac{N_f}{6} \left( 1- \frac{c_W^2}{s_W^2} \right) \log c_W^2 
\non \\
(\Delta r)_{\rm rem}^{\rm log-top} &=& \frac{G_\mu M_W^2}{4 \sqrt{2} \pi^2}
\left(\frac{c_W^2}{s_W^2} - \frac{1}{3} \right) \log \frac{m_t^2}{M_W^2} 
\eeq
Note that the factorization of the light and heavy fermion contribution and the
presence of the three terms  in the denominators of eq.~(\ref{deltarsum}) 
effectively sums many important higher--order terms \cite{Z-Physics3,Z-Physics-H}, 
such as those  of the form $(\Delta \rho)^2, (\Delta \rho \Delta \alpha), 
(\Delta \alpha \Delta r_{\rm rem})$ at the two--loop level and the light 
fermion contribution $(\Delta \alpha)^n$ to all orders. \s

At one--loop, the Higgs boson has a contribution to $(\Delta r)$ that is also 
only logarithmically dependent on $M_H$, as in the case of $\Delta \rho$. For a 
heavy Higgs, $M_H \gg M_W$, it reads \cite{Deltar,Deltar-Higgs1loop}
\beq
(\Delta r)^{\rm 1-Higgs}_{\rm rem} \simeq \frac{G_\mu M_W^2}{8 \sqrt{2} \pi^2}
\, \frac{11}{3} \left( {\rm log} \frac{M_H^2}{M_W^2} - 
\frac{5}{6} \right) 
\eeq
Again, the quadratic correction $\propto  M_H^2$ appears only at the two--loop 
level.\s

The complete two--loop bosonic corrections to $\Delta r$ have been calculated
recently \cite{EW-2loop} including the full $M_H$ dependence and were found to
be very small: a few times  $\times 10^{-5}$ for $M_H$ values in the range
between  100 GeV and 1 TeV. There are also two--loop electroweak corrections
stemming from  fermions; the main contribution is in fact contained in $\Delta
\rho$ but there is an extra piece contributing to $(\Delta r)_{\rm rem}$ which,
however, is small \cite{rho-2loopEW3,Paolo-approach}. Hence, the theoretical
knowledge of the $W$ mass is rather precise, being approximately the same as 
for the electroweak mixing angle.

\subsubsection*{\underline{The trilinear gauge boson couplings}}

As mentioned previously, the gauge structure of the ${\rm SU(2)_L\times
U(1)_Y}$ theory is best tested by the measurement of the triple gauge boson
vertices at LEP2 and eventually at the Tevatron [although possible, the test of
the quartic couplings is very limited in these experiments]. This can be
achieved by comparing the data with a general $WWV$ vertex with $V=\gamma,Z$,
which includes the possibility of anomalous gauge boson couplings that can be
induced, for instance, by radiative corrections in the SM or by New Physics
effects. We briefly discuss this aspect below, as it will be needed in another
part of this report.\s 

The most general Lorentz invariant $WWV$ vertex that is possible in processes
where the weak bosons couple to massless fermions, as is the case in the
reaction $\ee \to W^+ W^-$ for instance, can be written as\footnote{Additional
terms with higher derivatives may be present, but they are equivalent to a
dependence of the couplings on the vector boson momenta and thus, only lead to
a form factor behavior of these couplings.} \cite{LWWV-eff,TGC-LEP2}
\beq 
\label{GeneralWWV}
{\cal L}_{\rm eff}^{WWV} & = & i g_{WWV}\, \Bigl[ g_1^V V^\mu \left( W^{-}_{
\mu\nu}W^{+\nu}-W^{+}_{\mu\nu}W^{-\nu}\right) + \kappa_V\,  W^{+}_{\mu}W^{-}
_{\nu}V^{\mu\nu}  \\ \non
&+ & {\lambda_V \over M_W^2}\, V^{\mu\nu}W^{+\rho}_{\!\!\nu}W^-_{\rho\mu}
+ ig_5^V\varepsilon_{\mu\nu\rho\sigma}\left(
(\partial^\rho W^{- \mu})W^{+\nu} -
W^{- \mu}(\partial^{\rho}W^{+\nu}) \right) V^{\sigma} \\ \nonumber
&+ & ig_4^V W^{-}_{\mu}W^+_{\nu} (\partial^\mu V^\nu+\partial^\nu V^\mu)
-\frac{\tilde \kappa_V}{2} W^{-}_{\mu}W^+_{\nu}
\varepsilon^{\mu\nu\rho\sigma}
V_{\rho\sigma} -{\tilde \lambda_V\over {2 m_W^2}}\,
W^{-}_{\rho\mu}{W^{+\mu}}_{\nu}\varepsilon^{\nu\rho\alpha\beta}
V_{\alpha\beta}
\Bigr]
\eeq
with the overall couplings defined by $g_{WW\gamma}=e$ and $g_{WWZ}= e \cot
\theta_W$ and where the reduced field strengths $W_{\mu\nu}=\partial_\mu W_\nu 
- \partial_\nu W_\mu$ and $V_{\mu\nu}=\partial_\mu V_\nu - \partial_\nu V_\mu$
are used.
For on--shell photons, i.e. with $q^2=0$, the couplings $g_1^\gamma=1$ and 
$g_5^\gamma=0$ are fixed by ${\rm U(1)_{Q} }$ gauge invariance. In the 
Lagrangian
eq.~(\ref{GeneralWWV}), the couplings $g_1^V$, $\kappa_V$ and $\lambda_V$ 
separately conserve C and P symmetries, while $g_5^V$ violates them but 
conserves CP symmetry. The couplings $g_4^V$, $\tilde\kappa_V$ and $\tilde
\lambda_V$ parameterize a possible CP violation in the bosonic sector. \s  

Note that the C and P conserving terms in ${\cal L}_{\rm eff}^{WW\gamma}$ 
correspond to the lowest order terms in a multipole 
expansion of the $W$ boson--photon interactions, the charge $Q_W$, the magnetic
dipole moment $\mu_W$ and the electric quadrupole moment $q_W$ of the $W^+$ 
boson \cite{TGC-LEP2}
\beq 
\label{multipole-moments}
Q_W  =  e g_1^\gamma  \ , \ \ \mu_W  =  {e \over 2M_W} \left( g_1^\gamma + 
\kappa_\gamma + \lambda_\gamma \right) \ , \ \ q_W  =  -{e \over M_W^2} \left(
\kappa_\gamma-\lambda_\gamma \right) 
\eeq
In the SM, the Lagrangian eq.~(\ref{GeneralWWV}) reduces to the one given in 
eq.~(\ref{WWVcoupling}) and, thus, at the tree--level, the trilinear couplings 
are simply given by $g_1^Z = g_1^\gamma = \kappa_Z = \kappa_\gamma = 1$, while 
all the other couplings in eq.~(\ref{GeneralWWV}) are zero. It became common 
practice to introduce deviations of the former set of couplings from their 
tree--level SM  values
\beq
\label{kappas}
\Delta g_1^Z \equiv (g_1^Z - 1)  \ , \ \  \Delta \kappa_\gamma \equiv
(\kappa_\gamma-1) \ , \ \  \Delta \kappa_Z \equiv (\kappa_Z-1) 
\eeq
The rather precise measurement of these quantities is one of the big 
achievements of LEP. 

\newpage

\subsubsection{Approximating the radiative corrections}

The evaluation of the complete set of the previously discussed radiative
corrections is a very complicated task, in particular when initial and final
state photonic corrections or processes which need some special treatment, such
as Bhabha $\ee \to \ee$ scattering, are involved. This can be performed only
with the help of very sophisticated programs which fortunately exist
\cite{Z-Precision}. However, in most practical purposes, in particular when
effects of New Physics are analyzed, it is sufficient to probe some
quantities where the most important radiative correctives are expected to
occur. This is the case, for instance, of the $\Delta r$ and $\bar{s}_W^2$
observables. Here, we will shortly describe such approximations which will have
some application later. 

\subsubsection*{\underline{The improved Born approximation}}

One can express electroweak observables in the Born approximation in terms of
the QED constant $\alpha$, but to be accurate, one should use the running
$\alpha$ defined at the scale where  the considered process takes place, $M_Z$
or higher energies.  Since the running of $\alpha$ between the two latter scales
is rather small, one can make the substitution $\alpha(0) \to  \alpha(M_Z^2)=
\alpha(0)/(1- \Delta  \alpha)$ for scales larger than  $M_Z$. The $\Delta
\alpha$ corrections should in principle cancel the light  fermion contributions
in the two--point functions when the  radiative corrections to the observables
are calculated. This is in fact effectively done by using the Fermi decay
constant in the tree level expressions of the observables, $\alpha(0)  \to
\alpha(M_Z^2) = (\sqrt{2} G_\mu/\pi) M_W^2 s_W^2$, which implicitly includes
the $\Delta \alpha$ contribution. Since $\Delta \alpha$ is rather large being
at the level of 6\% [and which for $ 2\to 2 \, (3)$ processes that are
proportional to $\alpha^2 (\alpha^3)$, lead to contributions of the order of
$12\%\, (18\%)$], this gives  a more accurate description of the observable,
already at the tree--level.  This procedure is called the naive improved Born
approximation (naive IBA) \cite{IBA}.\s

The IBA is said to be naive because there are still residual  contributions 
from $\Delta \rho$ and $(\Delta r)_{\rm rem}$ and additional contributions to 
$\bar{s}_W^2$ which, despite of the fact that they are smaller than $\Delta 
\alpha$, should be taken into account. The dominant top quark contribution 
which is contained in the $\Delta \rho$ piece given in eq.~(\ref{deltarho}) 
can be  simply included by performing  the  shift 
\beq
\alpha \to  \frac{\sqrt{2} G_\mu}{\pi} M_W^2 \left( 1- \frac{M_W^2}{M_Z^2} 
\right) (1+ \Delta r )^{-1} \ , \ \Delta r \simeq  \Delta \alpha (M_Z^2) - 
3\Delta \rho 
\eeq

In the context of $Z$ physics, the IBA may be be sufficient in many purposes 
and can be implemented in the electroweak observables by simply performing 
the following substitutions:

\begin{itemize}
\vspace*{-2mm}

\item[$(i)$] replace the electromagnetic couplings of fermions by
\beq
Q_f e \to Q_f \sqrt{ 4\pi \alpha/ [1- \Delta \alpha (M_Z^2)]^{-1} }
\eeq

\item[$(ii)$] replace the Born couplings of the fermion to the $Z$ boson by
\beq
v_f \to (\sqrt{2}G_\mu M_Z^2 \rho )^{1/2}  \, \hat{v}_f \ \ , \ \ 
a_f \to (\sqrt{2}G_\mu M_Z^2 \rho )^{1/2}   \, \hat{a}_f
\eeq

\item[$(iii)$] replace everywhere $s_W^2$, in particular in the vector 
couplings $\hat{v}_f= 2I_3^f -4 Q_f s_W^2$, by the effective electroweak 
mixing angle for leptons   
\beq
s_W^2 \to \overline{s}_W^2 \equiv \sin^2\theta_{\rm eff}^{\rm lept}=  
\frac{1}{2} \left[ 1 - \sqrt{ 1- \frac{4 \pi \alpha(0)}
{\sqrt{2}M_Z^2 G_\mu} \frac{1}{\rho (1- \Delta \alpha)} } \right]
\eeq 
\item[$(iv)$] and for $b$ quark final states, perform in addition the 
substitution of eq.~(\ref{deltabv}) to include the large top quark mass 
corrections in the $Zb\bar{b}$ vertex.  
\end{itemize}

The remaining non--universal electroweak corrections are small and can be
safely neglected in most cases, but obviously not when probing the small 
Higgs boson effects. Of course, this IBA needs to be supplemented by the 
important QCD corrections to hadronic processes and QED corrections, in 
particular ISR corrections, whenever needed. \s

\subsubsection*{\underline{Model independent analyses and the STU and 
$\mathbf{\epsilon}$ approaches}}

In a more general context than the SM, it is often convenient to parametrize 
the radiative corrections to electroweak observables
in such a way that the contributions due to many kinds of New Physics beyond 
the SM are easily implemented and confronted with the experimental data. 
If one assumes that the symmetry group of New Physics is still the 
${\rm SU(3)_C \times SU(2)_L \times U(1)_Y}$ and, thus, there are no extra 
gauge bosons, and that it couples only weakly to light fermions so that one 
can neglect all the ``direct" vertex and box corrections, one needs to 
consider only the oblique corrections, that is, the ones affecting the $\gamma, 
Z,W$ two--point functions and the $Z\gamma$ mixing. If in addition the scale 
of the New Physics is much higher than $M_Z$, one can expand the complicated 
functions of the momentum transfer $Q^2$ around zero, and keep only the 
constant and the linear $Q^2/M_{\rm NP}^2$ terms of the series, which have
very simple expressions in general. The New Physics contributions can be then
expressed in terms of six functions: $\Pi_{\gamma\gamma}'(0), \Pi_{Z\gamma}'(0),\Pi_{ZZ}^{(')}$ and $\Pi_{WW}^{(')}$ ($\Pi_{\gamma \gamma}(0) = \Pi_{Z\gamma}
(0) =0$ because of the QED Ward Identity). Three of these functions will be 
absorbed in the renormalization of the three input parameters $\alpha, G_\mu$ 
and $M_Z$. This leaves three variables which one can choose as being 
ultraviolet finite and more or less related to physical observables.\s

A popular choice of the three independent variables is the STU linear 
combinations of self--energies introduced by Peskin and Takeuchi 
\cite{STU-approach}  
\begin{eqnarray} 
\alpha S &=& 4 s_W^2 c_W^2 \left[ \Pi_{ZZ}(0) - (c_W^2 - s_W^2)/(s_W c_W)
\cdot  \Pi'_{Z\gamma}(0) -\Pi'_{\gamma \gamma}(0) \right] \non \\
\alpha T &=& \Pi_{WW}(0)/M_W^2 - \Pi_{ZZ}(0)/M_Z^2 \non \\
\alpha U &=& 4 s_W^2  \left[ \Pi'_{WW}(0) - c_W^2 \Pi'_{ZZ}(0)
- 2 s_W c_W \Pi'_{Z\gamma}(0) - s_W^2 \Pi'_{\gamma \gamma}(0) \right] 
\end{eqnarray} 
These variables measure deviations with respect to the SM predictions and, for
instance, $S$ and $T$ are zero when the New Physics does not break the
custodial isospin symmetry. The variable $\alpha T$ is simply the shift of the
$\rho$ parameter due to the New Physics, $\alpha T = 1-\rho - \Delta
\rho|_{\rm SM}$.\s

Another parametrization of the radiative corrections, the $\epsilon$ approach 
of Altarelli and Barbieri \cite{epsilon-approach}, is closer to the IBA
approximation discussed previously and thus, more directly related to the 
precision electroweak observables. The three variables which parametrize the
oblique corrections are defined in such a way that they are zero in the 
approximation where only SM effects at the tree--level, as well as the pure QED 
and QCD corrections, are taken into account. Contrary to the STU approach, 
they are independent of the values of $m_t$ and $M_H$, the contributions of
which are included. In addition, a fourth variable is introduced to account 
for the quadratic top quark mass contribution to the $Zb\bar{b}$ vertex. \s 

In terms the two quantities $\Delta r_W$ and $\Delta k$ related to,
respectively, the gauge boson masses and $\sin^2\theta_{\rm eff}^{\rm lep}$ 
as measured from leptonic observables assuming universality, 
\beq
M_W^2/M_Z^2 \left(1- M_W^2/M_Z^2 \right)= s_0^2 c_0^2
(1- \Delta r_W) \ , \ v_e/a_e=1-4\sin^2\theta_{\rm eff}^{\rm lep}= 
1- 4(1+ \Delta k) s_0^2 \non
\eeq
with $ s_0^2 c_0^2= \pi \alpha (M_Z)/(\sqrt{2} G_\mu M_Z^2)$, the four 
variables $\epsilon_{1,2,3,4}$ 
are defined as
\beq
\epsilon_1= \Delta\rho \ , \ 
\epsilon_2= c_0^2 \Delta\rho +\frac{s_0^2}{c_0^2- s_0^2} \Delta r_W
-2 s_0^2 \Delta k \ ,  
\epsilon_3= c_0^2 \Delta\rho +(c_0^2- s_0^2) \Delta k \ ,  
\epsilon_4= \Delta_b
\eeq
The variables $\epsilon_2$ and $\epsilon_3$ are only logarithmic in $m_t$ and 
the $m_t^2$ terms appear only in $\epsilon_1$ and $\epsilon_b$;  the 
leading terms involving the Higgs boson mass are contained in $\epsilon_1$ and 
$\epsilon_3$. The relations between the $\epsilon$s and the electroweak 
observables, lead to very simple formulae.

\subsubsection*{\underline{Interpolation of the radiative corrections}}

Once the full set of radiative corrections has been made available,  one can
attempt to derive simple interpolation formulae to summarize the full result,
which is analytically complicated and numerically involved to handle. This is
possible since the only unknown parameter in the SM is the Higgs boson mass.
However, one has to include the experimental errors on some important SM input
parameters. For the values of $M_W$ and of the EW mixing angle
$\sin^2\theta_{\rm eff}^{\rm lep}$ as measured from lepton asymmetries which do
not involve final state QCD corrections, the interpolating formulae are indeed
rather simple. Using $G_\mu$ and $M_Z$ as inputs and taking into account the
possible variations of the measurements of $\Delta \alpha^{\rm had} (M_Z),
\alpha_s (M_Z)$ and $m_t$ from their central values, one obtains for a given
observable $X$ \cite{Paolo-approach}
\beq
X &=&  X^0 + a_1^X a_h  + a_2^X a_e + a_3^X a_t + a_4^X a_s + a_5^X a_h^2
\label{interpolationMW}
\eeq
with $X^0=(s_W^2)^0$ and $M_W^0$ being the [scheme dependent] theoretical 
results at the reference point $M_H=100$ GeV, $m_t=175$ GeV and the other 
parameters set at their experimentally measured central values, and the 
reduced quantities
\beq
a_h = \log\left(\frac{M_H}{100~{\rm GeV}}\right), \,
a_e = \frac{\Delta \alpha^{\rm had}}{0.028} - 1, \,
a_t= \left(\frac{m_t}{175~{\rm GeV}}\right)^2 - 1, \,
a_s = \frac{\alpha_s(M_Z)}{0.118} - 1 \hspace*{.7cm}
\eeq
In the on--shell scheme, the central values $X^0$ and the coefficients $a_i^X$ 
are displayed in Table 1.2 for the two observables $\sin^2\theta_{\rm eff}^{\rm
lep}$ and $M_W$. For a Higgs boson with a mass in the range 75 GeV $\lsim M_H 
\lsim 375$ GeV and for the SM input
parameters varying within their $1 \sigma$ allowed range, the previous formula
reproduces the complete result with errors $\Delta \sin^2\theta_{\rm eff}^{\rm 
lep} \lsim 1 \times 10^{-5}$ and $\Delta M_W \lsim 1$ MeV, which are well 
below the experimental accuracies on these observables as will be seen later. 

\begin{table}[hbt]
\renewcommand{\arraystretch}{1.5} 
\begin{center} 
\begin{tabular}{|c|c|c|c|c|c|c|c|} \hline 
Quantity & $X^0$ & $10^2a_1^X$ & $10 a_2^X$ & $10a_3^X$ & $10^2a_4^X$ & 
$10 a_5^X$ \\ \hline 
$\sin^2\theta_{\rm eff}^{\rm lep}$ & 0.231527 & 0.0519 & 0.986 & -- 0.277 
& 0.045 & 0 \\
$M_W$     & 80.3809  & --5.73   & --5.18  & 5.41  & --8.5   & --0.80 \\
\hline 
\end{tabular}\\[3mm] 
\end{center}
\centerline{\it Table 1.2: The central values and the deviation coefficients $a_i$
for $\sin^2\theta_{\rm eff}^{\rm lep}$ and $M_W$.} 
\vspace*{-5mm}
\end{table}

\subsubsection{The electroweak precision data}

Besides $\alpha(M_Z), G_\mu$ and $M_Z$ which are used as the basic input
parameters, there is an impressive list of electroweak observables which have
been measured with a very good accuracy and which can be predicted in the SM
with an equally good precision. These are:  

$\bullet$ Observables from the $Z$ lineshape at LEP1: the Z boson 
total width $\Gamma_Z$, the peak hadronic cross section $\sigma^0_{\rm had}$,
the partial decay widths of the $Z$ boson into leptons and $c,b$ quarks 
normalized to the hadronic $Z$ decay width, $R_{\ell, c,b}$, 
the forward--backward asymmetries $A_{FB}^f$ for leptons and heavy $c,b$ 
quarks, as well as the $\tau$  polarization asymmetry $A^\tau_{\rm pol}$;  
the asymmetries provide a determination of $\sin^2\theta_W$ as measured
from leptons and hadrons. \s

$\bullet$ The longitudinal polarization asymmetry $A_{LR}^f$ which has 
been measured at the SLC and which  gives the best individual measurement of
$\sin^2\theta_W$, as well as the left--right forward--backward asymmetries for
the heavy $b,c$ quarks,  $A^{b,c}_{LR,FB}$. \s

$\bullet$ The mass of the $W$ boson $M_W$ which is precisely  measured 
at LEP2 and at the Tevatron as well as the total decay width $\Gamma_W$,
eqs.~(\ref{Wmass-average}--\ref{Wwidth-average}). \s

$\bullet$  In addition there are high--precision measurements at low energies:
(i) the $\nu_\mu$-- and $\bar{\nu}_\mu$--nucleon deep--inelastic scattering
cross sections, the ratios of which measure the left-- and right--handed
couplings of fermions to the $Z$ boson which can be turned into a determination
of $s_W^2$, and (ii) the parity violation in the Cesium and Thallium atoms
which provide the weak charge $Q_W$ that quantifies the coupling of the nucleus
to the $Z$ boson and which can also be turned into a determination of the 
electroweak mixing angle via $s_W^2 =1-M_W^2/M_Z^2$.\s

One has in addition to include as inputs, the measurement of the top quark mass
at the Tevatron, the strong coupling constant at LEP and elsewhere, as well as
the value of $\Delta \alpha^{\rm had} (M_Z^2)$ as measured in $\ee$ collisions 
at low energies and in $\tau$--lepton decays at LEP1. The experimental values of
some of the electroweak observables mentioned above [as they were in summer
2004] are displayed in Table 1.3 together with the associated errors. Also shown
are the theoretical predictions of the SM [for the best fit of $M_H$ to be
discussed later] that have been obtained by including all known radiative 
corrections with the central values of $\Delta \alpha^{\rm had} (M_Z^2), m_t, 
\alpha_s$, {\it etc..} \s

As can be seen from Tab.~1.3, the theoretical predictions are in remarkable
agreement with the experimental data, the pulls being smaller than 2
standard deviations in all cases, except for $A_{FB}^b$ where the deviation is
at the 2.5 $\sigma$ level. A few remarks are in order here: \s

\begin{table}[tp]
\begin{center}
\psfig{figure=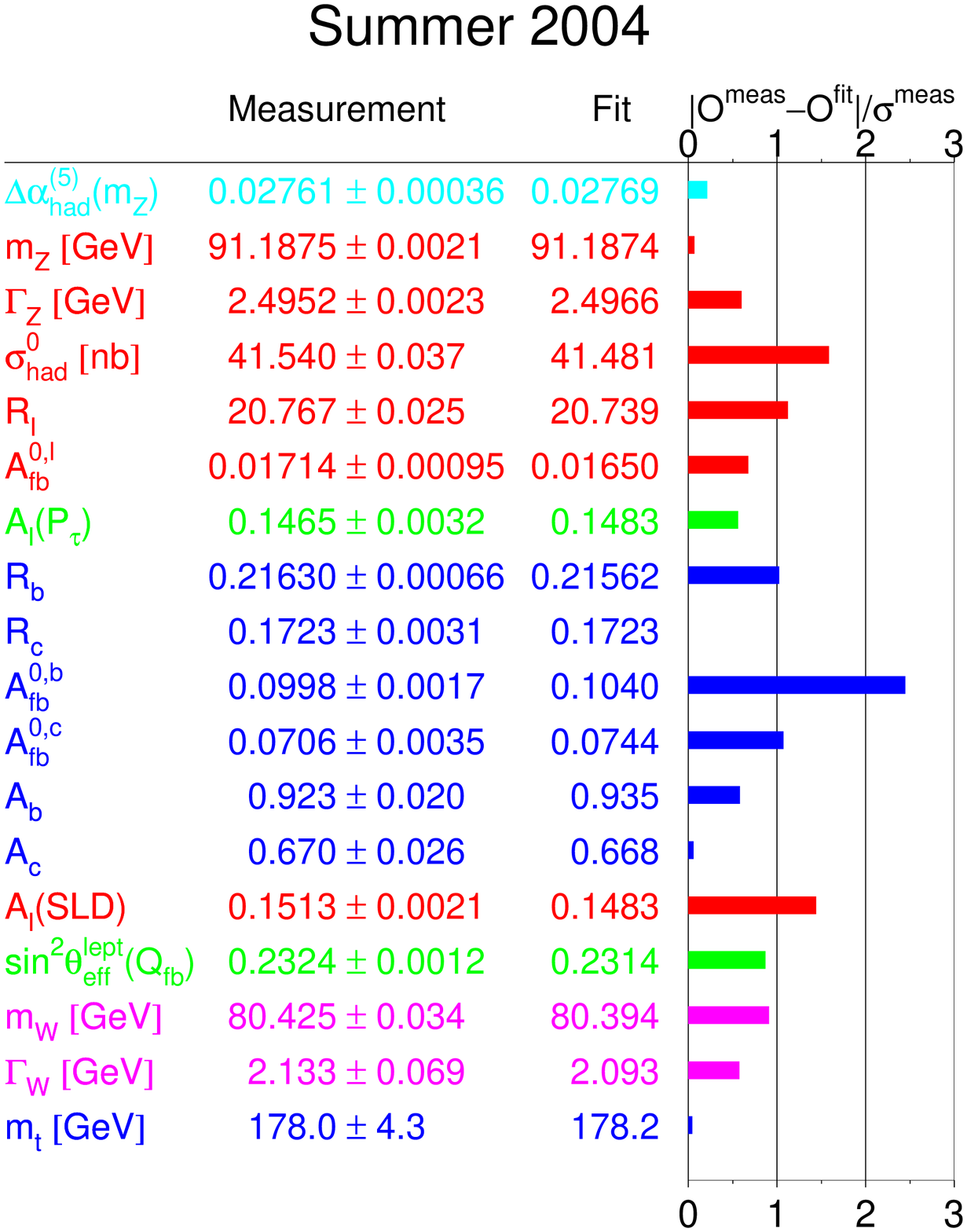,height=5.5in}
\end{center}
\vspace*{-27mm}
\nn {\it Table 1.3: Summary of electroweak precision measurements at LEP1, LEP2,
SLC and the Tevatron; from Ref.~\cite{High-Precision}. The SM fit results, 
which have been derived including all radiative corrections, and the standard
deviations are also shown. }
\vspace*{-3mm}
\end{table}

$i)$ From the $Z$ lineshape and partial width measurements, one obtains a 
determination of the number of light neutrino flavors contributing to the 
invisible $Z$ decay width   
\beq 
N_\nu=2.9841 \pm 0.0083
\eeq
which agrees with the SM expectation $N_\nu=3$ at the 2$\sigma$ level. \s
 
$ii)$ Using all these results, one derives the world average value for the 
effective weak mixing angle as measured from lepton asymmetries and partial 
widths and from hadronic asymmetries. The  status for the latter parameter, 
again as it was in summer 2004, is as follows \cite{High-Precision}
\beq
\sin^2\theta_{\rm eff}^{\rm lep} &=& 0.23150 \pm 0.00016 
\label{sineff_average}
\eeq

$iii)$ The values of $\sin^2\theta_{\rm eff}^{\rm lep}$ as measured from the
leptonic FB and $\tau$ polarization asymmetries at LEP1 and from the 
longitudinal
asymmetries at SLC are in a very good agreement. From these measurements, and 
from the measurement of the leptonic partial widths, the lepton universality of
the neutral weak current has been established with a high accuracy.\s

$iv)$ The forward--backward asymmetry $A_{FB}^b$ for $b$ quarks measured at 
LEP1 provides,
together with the longitudinal asymmetry $A_{LR}^f$ measured at the SLC, the
most precise individual measurement of $\sin^2\theta^{\rm lep}_{\rm eff}$, but
the result is 2.5 standard deviations away from the predicted value and the two
individual values differ by almost three standard deviations. This has led to
speculations about a signal of New Physics in the $Zb\bar b$ vertex.  However,
it turns out that this discrepancy cannot be easily explained without affecting
the $Z\to b\bar{b}$ partial width $R_b$ which is precisely measured and is
compatible with the SM expectation, and the hadronic asymmetries measured at
the SLC, although their errors are larger. It is likely that this anomaly
is a result of a large statistical fluctuation or some experimental problem.\s 

$v)$ While the value of weak charge as measured in the parity violation in Cs 
atoms, $Q_W (Cs)=-72.74 \pm 0.46$ \cite{Q-APV}, is in accord with the SM 
prediction $Q=-72.93$, the measurement of $\sin^2\theta_W$ from neutrino and 
antineutrino deep--inelastic scattering made by the NuTeV experiment gives 
$\sin^2\theta_W (\nu N) = 0.2277  \pm 0.0016$ \cite{NuTeV}, which is 3 standard 
deviations away from the predicted value in the SM, $\sin^2\theta_W=0.2227$. 
It becomes now apparent that the theoretical uncertainties in the higher--order 
analyses needed to extract the NuTeV value of $\sin^2\theta_W$ have been 
underestimated by the collaboration \cite{Paolo-NuTeV,NuTeV-new}. \s

In addition, the cross sections for the pair production of gauge bosons have
been rather accurately measured at LEP2 [and, to a lesser extent, at the
Tevatron]. In the case of the $\ee \to W^+ W^-$ process, the cross section
which depends on the triple self--coupling among the $W$ and the $V=\gamma,Z$
bosons, eq.~(\ref{WWVcoupling}), and on the $We\nu$--coupling given in
eq.~(\ref{Wffcouplings}), is shown in the left--hand side of Fig.~1.6 and it
agrees perfectly with the predicted value in the SM, with the $s$--channel
exchange of $\gamma,Z$ and the $t$--channel neutrino exchange contributions.\s 

This agreement can be turned into a strong constraint on the anomalous C,P 
conserving couplings of the effective Lagrangian of eq.~(\ref{GeneralWWV}) 
which are measured to be \cite{MW-LEP2}
\beq
\kappa_\gamma=0.943 \pm 0.055 \, , \  
\lambda_\gamma=-0.020 \pm 0.024 \, , \
g_1^Z = 0.998 \pm 0.025
\eeq
providing a stringent test of the ${\rm SU(2)_L\times U(1)_Y}$ gauge structure 
of the theory. The contours of the two parameter fits of these three C and P
conserving $W$ boson couplings are shown in the right--hand side of Fig.~1.6.\s

\begin{figure}[h!]
\vspace*{-.9cm}
\begin{center}
\begin{minipage}{8cm}
\vspace*{1cm}
\hspace*{-2mm}
\psfig{figure=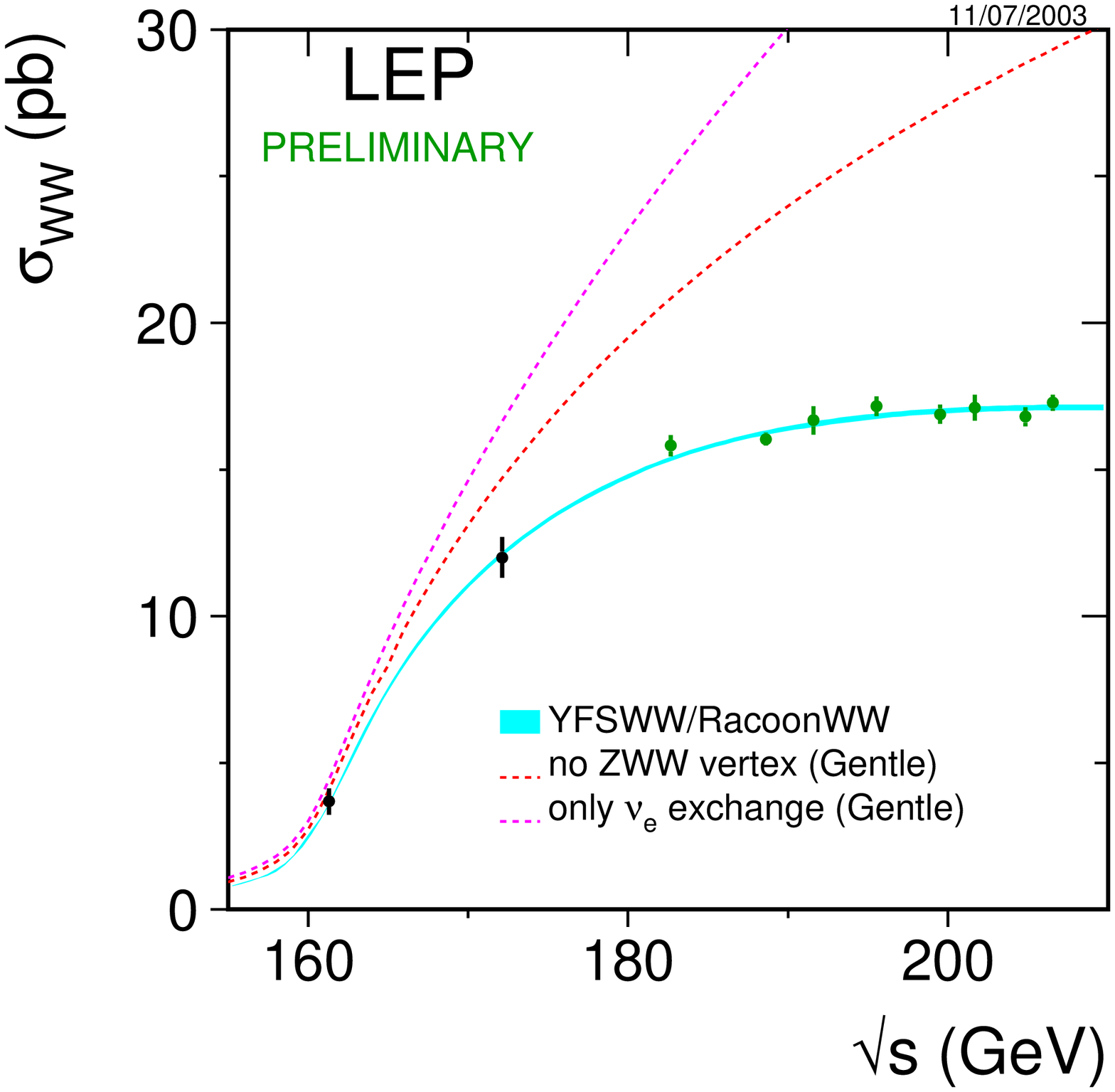,height=8.5cm,width=8.cm}
\end{minipage}
\begin{minipage}{8cm}
\psfig{figure=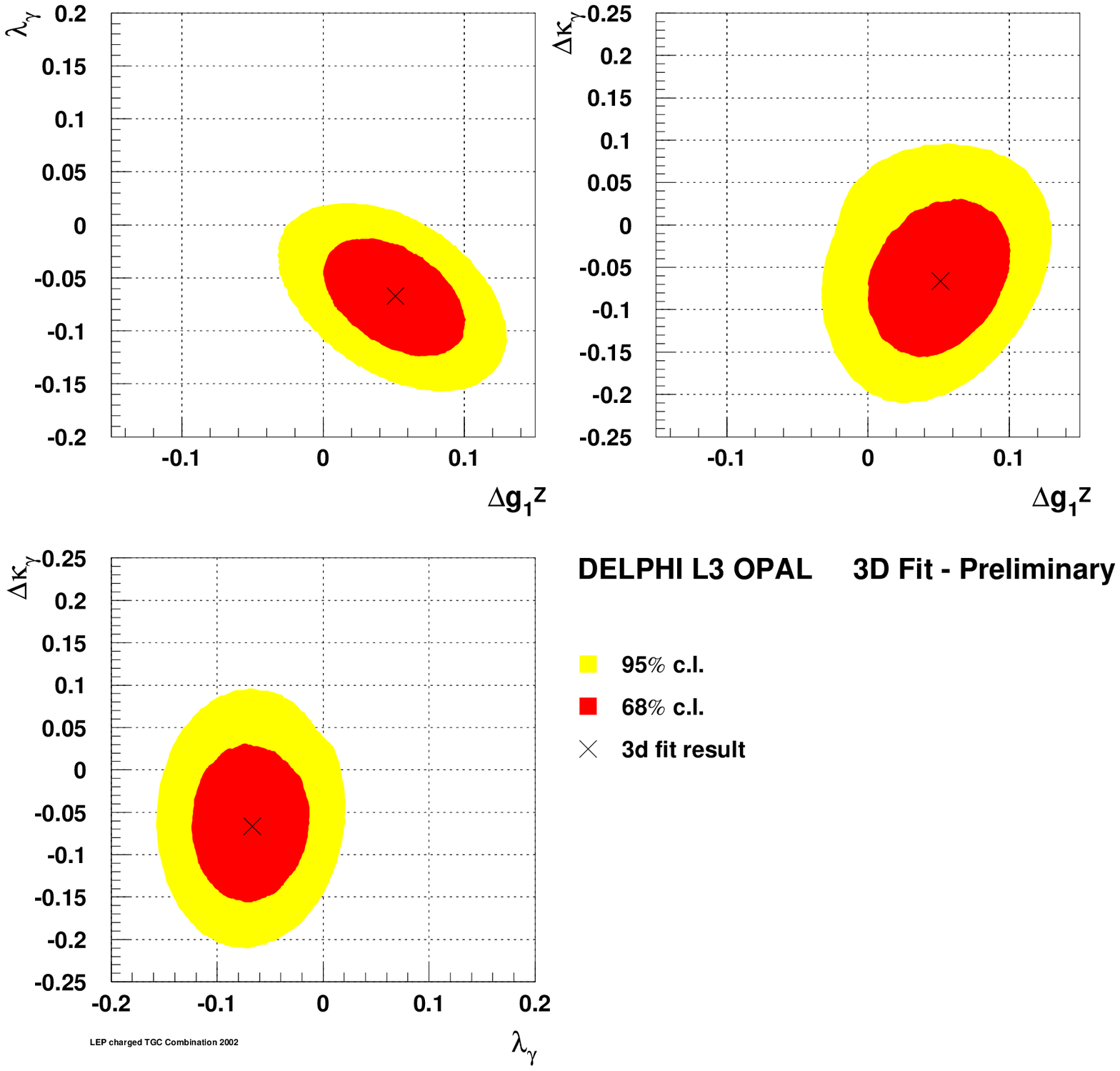,height=8.5cm,width=8.7cm}\hspace*{-5mm}
\end{minipage}
\end{center}
\vspace*{0mm}
\nn{\it Figure 1.6: Left: the measured value of the $\ee \to W^+ W^-$ cross
section at LEP2 and the prediction in the SM (full line) and when the
$s$--channel $Z$ boson or both $\gamma$ and $Z$ boson exchange diagrams are not
contributing. Right: the 68\% and 95\% confidence level contours of the three
two parameter fits to the $W$ boson C and P conserving trilinear couplings,
$g_1^Z$--$\lambda_\gamma$, $g_1^Z$--$\kappa_\gamma$ and 
$\lambda_\gamma$--$\kappa_\gamma$, as measured at LEP2 with a c.m. energy up to
$\sqrt{s}=209$ GeV and including systematical uncertainties; the fitted values
are indicated with a cross and the SM value for each fit is in the center of
the grid. From Ref.~\cite{MW-LEP2}.}
\label{fig:WWcross-section}
\vspace*{-1mm}
\end{figure}

In summary, the electroweak precision data have provided a decisive test of the
SM. These tests have been performed at the per mille level and have probed the
quantum corrections of the ${\rm SU(2)_L\times U(1)_Y}$ theory. The couplings
of quarks and leptons to the electroweak gauge bosons have been measured
precisely and found to be those predicted by the gauge symmetry.  The trilinear
couplings among electroweak gauge bosons have been also measured and found to
be those dictated by the gauge symmetry.  If, in addition, one recalls that the
${\rm SU(3)_C}$ gauge symmetry description of the strong interactions has been
thoroughly tested at LEP1 and elsewhere,  one concludes that the SM based on
the ${\rm SU(3)_C\times SU(2)_L\times U(1)_Y}$ gauge symmetry has been firmly
established as the theory of the strong and electroweak interactions at present
energies. The only missing ingredient of the model is the Higgs particle, which
has not yet been observed directly.  However, indirect constraints on this
particle can be obtained from the high precision data as we will discuss now.  

\newpage
\subsection{Experimental constraints on the Higgs boson mass}

Since the Higgs particle contributes to the radiative corrections to the 
high--precision electroweak observables discussed previously, there are 
constraints on its mass which, as discussed in \S1.1, is 
the only yet unknown free parameter in the SM. There are also constraints from
direct searches of the Higgs boson at colliders and in particular at LEP. 
These indirect and direct constraints on $M_H$ will be summarized in this 
section. 

\subsubsection{Constraints from high precision data}

The electroweak precision measurements allow rather stringent constraints on
the Higgs boson mass in the SM. Using for instance the LEP2 values of the $W$
boson mass and the effective weak mixing angle as measured in forward--backward
and polarization asymmetries, and the combined fit to the measurements giving
these parameters when the complete set of radiative corrections has been
included, one obtains the range where the Higgs boson mass should lie at the
$1\sigma$ level that is shown in Fig.~1.7.\s

\begin{figure}[htpb]
\begin{center}
\vspace*{-.1mm}
\mbox{\psfig{figure=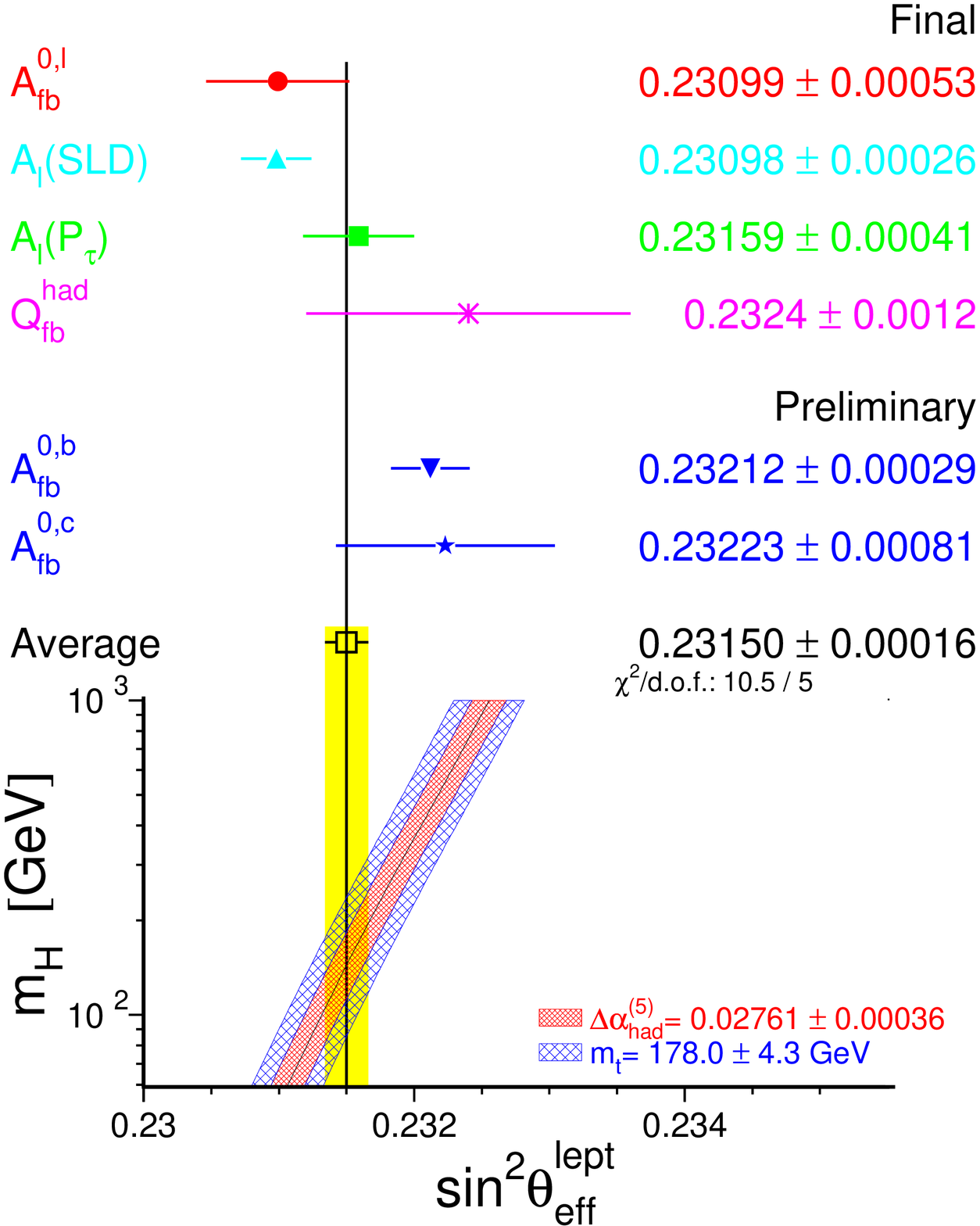,height=4.4in,width=2.7in}
\hspace*{2cm}
\psfig{figure=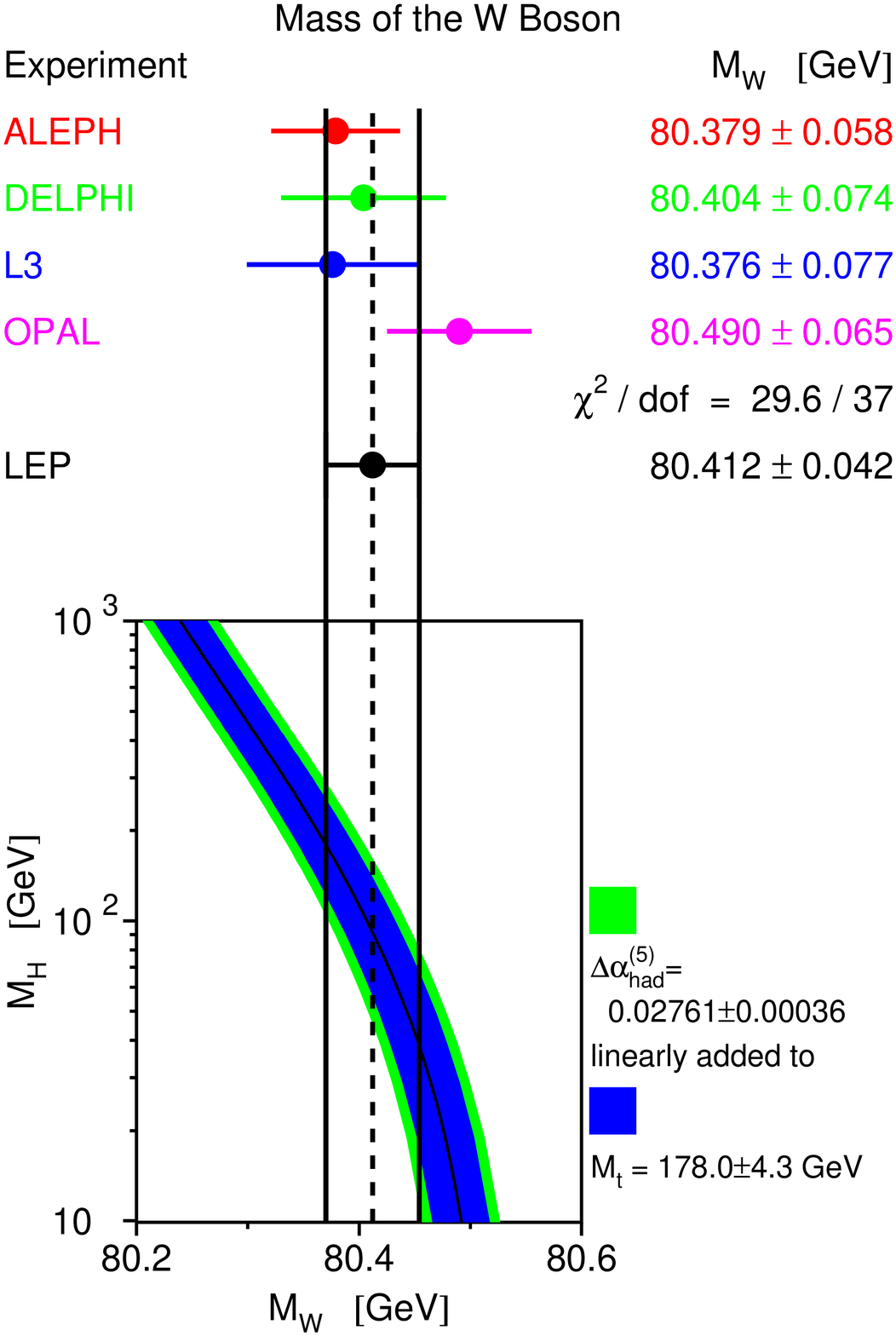,height=4.4in,width=2.7in} }
\end{center}
\vspace*{-.9mm}
\nn{\it Figure 1.7: The measurement [vertical band] and the theoretical 
prediction [the hatched bands] for $sin^2 \theta^{\rm lept}_{\rm 
eff}$ and $M_W$ as a function of the Higgs boson mass; 
from Ref.~\cite{High-Precision}.}
\label{fig:ewwg2002pulls}
\vspace*{-1mm}
\end{figure}

The vertical bands are due to the measurements and their errors,  while
the colored bands are for the theoretical prediction with the uncertainties
due to the SM input parameters, namely, $\Delta^{\rm had} \alpha (M_Z)
= 0.02761 \pm 0.00036, \alpha_s(M_Z)=0.118 \pm 0.002$ and $m_t=178.0 \pm 4.3$ 
GeV. The total width of the band is the linear sum of all these effects. 
As can be seen, the values of $\sin^2 \theta^{\rm lept}_{\rm eff}$ and
$M_W$ agree with the SM prediction only if the Higgs particle is rather 
light, a value of about $M_H \sim 100$ GeV being preferred by the 
experimental data.  \s

Taking into account all the precision electroweak data of Table 1.3 in a 
combined fit, one  can determine the constraint summarized in Fig.~1.8 which 
shows the  $\Delta  \chi^2$ of the fit to all measurements as a function of 
$M_H$, with the uncertainties on $\Delta^{\rm had}, \alpha (M_Z),\alpha_s(M_Z),
m_t$ as well as on $M_Z$ included \cite{High-Precision}. One then obtains  
the value of the SM Higgs boson mass
\beq
M_H=114^{+69}_{-45} ~{\rm GeV}
\eeq
leading to a 95\% Confidence Level (CL) upper limit in the SM
\beq
M_H <260~{\rm GeV}
\label{LEP_Mh_bound}
\eeq
\begin{figure}[htbp]
\begin{center}
\vspace*{-1.9cm}
\hspace*{-1cm}
\epsfig{file=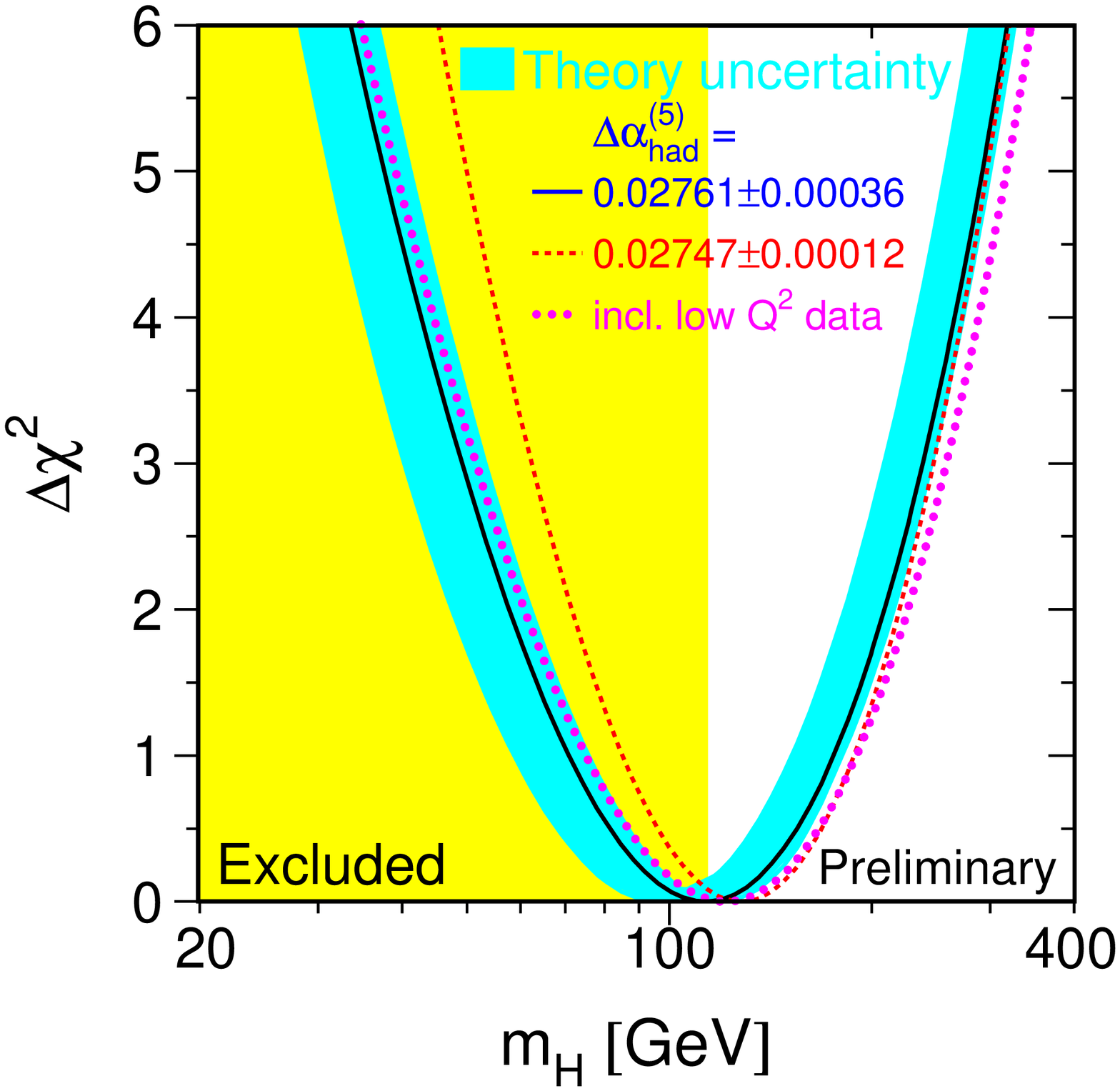,height=12.cm} 
\end{center}
\vspace{-1.cm}
{\it  Figure 1.8:
The $\Delta \chi^2$ of the fit to the electroweak  precision data as a function
of $M_H$. The solid line results when all data are included and  the blue/shaded
band is the estimated theoretical  error from unknown higher--order
corrections.  The effect of including the low $Q^2$ data and the use of a
different value for $\Delta  \alpha_{\rm had}$ are also shown; 
from Ref.~\cite{High-Precision}.}
\vspace*{-.1cm}
\end{figure} 

These values are relatively stable when the controversial NuTeV result is 
included in the fit, or when a slightly different value for $\Delta 
\alpha_{\rm had}^5$ is used. The area to the left to the vertical band 
which is very close to the minimum of the fit, shows the exclusion  limit 
$M_H >114.4$ GeV from direct searches at LEP2 to which we will turn our 
attention shortly. \s

It thus appears that the high--precision data, when confronted with the 
predictions of the SM after the radiative corrections have been
incorporated, lead to stringent constraints on the Higgs sector of the SM.   
 The data strongly disfavor a heavy Higgs boson with a mass $M_H
\gsim 700$ GeV for which perturbation theory breaks down anyway, as will be
seen in the next section. They clearly favor a light Higgs boson, $M_H \lsim
260$ GeV, with a central value that is very close to the present lower bound
from direct searches, $M_H \geq 114.4$ GeV. This is very encouraging for the
next generation of high--energy experiments. \s

However, there are two caveats to this statement, a theoretical and an
experimental one that we will discuss first. The most constraining observables,
besides the $W$ boson mass, are the LEP and SLC measurements of the leptonic
asymmetries, led by the longitudinal asymmetry $A_{LR}$, on the one hand, and of
the hadronic asymmetries, led by the forward--backward asymmetry for
$b$--quarks $A_{FB}^b$, on the other hand.  As can be seen from Fig.~1.7, while
the former set favors a light Higgs boson, as is also the case for the
measurement of $M_W$, the hadronic asymmetries favor a heavier Higgs particle. 
Because of the 3$\sigma$ difference of the value of $\sin^2\theta_W$ as measured
in the two sets of observables, it is only if one averages all the measurements
that one obtains the central value $M_H \simeq 114$ GeV.\s

Because of the 2.5 standard deviation of $A_{FB}^b$ from the theoretical
prediction and the smaller deviation of $A_{LR}$ but in the other direction,
the SM fit is in fact rather poor \cite{Poor-fit}: the weighted average leading
to the value $\sin^2\theta^{\rm lept}_{\rm eff}$ given in
eq.~(\ref{sineff_average}), corresponds only to a 6\% probability.  The fit can
be improved if one assumes New Physics effects which appear only in the
$Zb\bar{b}$ vertex.  However, as already mentioned, it is very difficult to
induce new effects in $A_{FB}^b$ without spoiling the agreement of $R_b$ and
$A_{LR,FB}^b$ with the data\footnote{Indeed, since $A_{FB}^b \propto A_e A_b$
and since $A_e \sim \hat{v}_e$ is small, one needs to alter significantly the
$Zb\bar{b}$ couplings to account for the discrepancy of the asymmetry with the
data: a 30\% change of the right--handed $Zb\bar{b}$ coupling, $g_{bR} \sim 
\hat{a}_b - \hat{v}_b$, is required, an effect that is too large not to disturb
the precise measurement of $R_b \sim  g_{bR}^2+ g_{bL}^2$ or $A^b_{LR,FB} \sim 
g_{bL}^2 - g_{bR}^2$. This 30\% change is anyway too large for a loop effect.}. 
On the other hand, if one assumes that the discrepancy in $A_{FB}^b$ is due to
some systematical errors which have been underestimated by the experiments and
remove this quantity from the global fit, one obtains a central value of $M_H$
which is lower than the mass bound obtained from the direct Higgs
boson searches at LEP2\footnote{In the past, when the top quark mass was
measured to be $m_t \simeq 175 \pm 5$ GeV, the situation was even worse since
the exclusion of $A_{FB}^b$ from the fit led to a rather low $M_H$ value, $M_H
\sim 45$ GeV, with only a 5\% probability that $M_H \geq 114$ GeV. This has led
to some justified speculations about the validity of the SM \cite{Poor-fit}.
The tension between the central value of the fit and the direct bound, has been
relaxed with the recent value of $m_t \simeq 178 \pm 4.3$ GeV, which increased
$M_H$ by several tens of GeV.}.\s

The bound on the Higgs mass, eq.~(\ref{LEP_Mh_bound}), is quite strong and 
there have been many speculations on how it can be relaxed or evaded. To do 
so, one has to introduce New Physics contributions which are of the same order 
as the one due to a heavy Higgs boson, and which conspire with the latter as 
to mimic the effect of a light SM Higgs particle. This has to be done without
spoiling the rest of the agreement of the SM with the high--precision data.\s

A way to look at these new contributions is to parametrize the Higgs sector by
an effective Lagrangian in which higher dimensional operators are added
\cite{HVV-Effective,Hff-Effective}.  These operators should respect the ${\rm 
SU(2)_L \times U(1)_Y}$ gauge symmetry, as well as some other constraints.  In 
this approach, one or a few higher dimensional operators which are damped by 
powers of the new scale $\Lambda$, produce corrections that counteract the one 
of a heavy Higgs boson, in such a way that the net result is compatible with 
the SM for $M_H \sim
100$ GeV. To produce such a conspiracy, the scale $\Lambda$ should range
between 2 to 10 TeV, depending on the nature of the operator or the combination
of operators which generate the effect \cite{High-operators}.\s 

However, this approach does not tell anything about the New Physics which is
behind the effective Lagrangian, and it is not actually clear whether it is 
possible to produce such a set of conspiring operators in a well motivated and 
consistent theoretical model. One therefore prefers to consider specific, and 
preferably well motivated, models. \s  

In general, because of decoupling, models which contain an elementary Higgs 
particle generate only small radiative corrections even if they involve a large
number of new particles. This is typically the case of supersymmetric 
extensions of the SM. In contrast, models where the Higgs boson is composite
or strongly interacting can generate large effects. However, in most cases
the new contributions add to the effect of a heavy Higgs boson, leading to a 
stronger disagreement with the precision data. This is, for instance, the case 
of early versions of Technicolor models which have been ruled out 
in the beginning of the nineties \cite{STU-approach}. \s

Nevertheless, there are still models of New Physics that are weakly interacting
and which induce corrections that are large enough, and with the adequate sign,
to accommodate a heavy Higgs boson. In Ref.\cite{Models-Heavy-H}, large classes
of models have been considered and their effects on the radiative corrections
have been analyzed. The conclusion of the study is that indeed, models with
a heavy Higgs boson exist, but they always need some conspiracy to
produce the required effect and more importantly, in most cases they predict
new degrees of freedom which should be sufficiently light to be observed at the
next generation of colliders\footnote{An example of such models are gauge
extensions of the SM [for instance based on the SO(10) group or on the
Superstrings--inspired $E_6$ symmetry] in which a heavy vector boson $Z'$ is
added. This particle will mix with the ${\rm SU(2)_L\times U(1)_Y}$ $Z$ boson
to produce the observed $Z$ particle; the mixing angle is inversely
proportional to the $Z'$ mass, $\theta_{\rm mix} \propto M_Z^2/M_{Z'}^2$.  It
has been shown in Ref.~\cite{Models-Heavy-H} that such a $Z'$ can indeed
generate any contribution to the $S$ and $T$ Peskin--Takeuchi parameters
discussed in \S1.2.4. However, to mimic the effect of a heavy Higgs boson, the
$Z'$ boson should have a rather low mass, $M_{Z'} \lsim 1.5$ TeV, making this
particle accessible at future colliders; see e.g.~\cite{Zprime-papers}.}.   

\subsubsection{Constraints from direct searches}

\subsubsection*{\underline{Searches at LEP1}}

The Higgs boson has been searched for at the LEP experiment, first at energies 
near the $Z$ boson resonance, $\sqrt{s}\simeq M_Z$. In this case, two 
channels allow to probe the Higgs boson \cite{Z-Physics5}. The dominant
production mode is the Bjorken process \cite{Bjorken-process}, where the $Z$ 
boson decays into a real Higgs boson and an off--shell $Z$ boson which goes
into two light fermions, $Z \to HZ^* \to H f\bar{f}$; the Feynman diagram is
shown in Fig.~1.9.

\begin{center}
\vspace*{-.5cm}
\hspace*{2.5cm}
\SetWidth{1.1}
\begin{picture}(300,90)(0,15)
\Photon(30,50)(100,50){3.2}{7}
\DashLine(100,50)(150,25){4}
\Photon(100,50)(150,75){3.2}{5.5}
\ArrowLine(150,75)(180,85)
\ArrowLine(150,75)(180,65)
\Text(100,50)[]{{\blue{\large $\bullet$}}}
\Text(150,75)[]{{\blue{\large $\bullet$}}}
\Text(185,85)[]{$f$}
\Text(185,65)[]{$\bar{f}$}
\Text(75,65)[]{$Z$}
\Text(160,20)[]{\blue{$H$}}
\Text(130,80)[]{$Z^*$}
\vspace*{-1.3cm}
\end{picture}
\mbox{\it Figure 1.9: The main production mechanism for Higgs bosons in $Z$ 
decays  at LEP1.} 
\end{center}

The partial decay width $\Gamma(Z \to H f\bar{f})$, when normalized to the $Z
\to f\bar f$ decay width where the fermion $f\neq t$ is considered as massless,
is given by \cite{Behrends-Kleiss}
\beq
{\rm BR}( Z \to H f\bar{f}) \equiv \frac{ \Gamma (Z \to H f\bar{f})}{\Gamma (Z
\to f \bar f)} = \frac{G_\mu M_Z^2  }{2 \sqrt{2} \pi^2} \int_{2a}^{1+ a^2} 
{\rm d}x \, \Gamma_0 (x)
\eeq
with the variable appearing in the integration bounds being $a=M_H/M_Z$
and $x$ is the reduced energy of the Higgs boson $x=2E_H/M_Z$.
The function in the integrand reads 
\beq
\Gamma_0( x) = \frac{\sqrt{x^2 -4a^2} } {( x- a^2)^2+ \gamma^2}
\left(1-  x + \frac{x^2}{12} + \frac{2a^2}{3} \right) 
\eeq
where $\gamma=\Gamma_Z/M_Z$ is the reduced total decay width of the $Z$ boson.
Neglecting the $Z$ width in $\Gamma_0$, the integration over the variable $x$ 
leads to a relatively simple analytical result \cite{HHG}
\beq
{\rm BR}( Z \to H f\bar{f}) &=& \frac{G_\mu M_Z^2  }{2 \sqrt{2} \pi^2}
\left[ \frac{ 3a(a^4-8a^2+20)}{\sqrt{4-a^2}} {\rm arcos} \left( \frac{1}{2}
a (3-a^2) \right) \right. \non \\
&& \left. -3(a^4-6a^2+4) \ln a - \frac{1}{2} (1-a^2)(2a^4-13a^2+47) \right]
\eeq 
This branching ratio follows that of the $Z$ decay into a given fermionic final 
state. For instance, ${\rm BR}( Z \to H  \mu^+ \mu^-)$ for muons and 
${\rm BR}( Z \to H  \nu \bar \nu)$ when summing over the three neutrino 
species are, respectively, $3\%$ and 18\% of the total Higgs sample.\s

The Higgs boson can also be produced in the decay $Z \to H \gamma$ 
\cite{Z-h-gamma1,Z-h-gamma2} which occurs through triangular loops built--up 
by heavy fermions and the $W$  boson; Fig.~1.10. The partial decay width, 
including only the dominant top quark and $W$ contributions, reads
\beq
\Gamma (Z \to H\gamma) = \frac{\alpha G_\mu^2 M_W^2}{48 \pi^4}  M_Z^3 
\left(1 - \frac{M_H^2}{M_Z^2} \right)^3 |  A_t + A_W|^2   
\eeq

\begin{center}
\vspace*{-.4cm}
\hspace*{1.cm}
\begin{picture}(300,100)(0,0)
\SetWidth{1.}
\SetScale{1.15}
\Photon(-20,50)(20,50){3.2}{5.5}
\Photon(20,50)(50,75){3}{5}
\Photon(20,50)(50,25){-3}{5}
\Photon(50,25)(50,75){3}{5.5}
\Photon(50,25)(85,25){3.2}{5}
\DashLine(50,75)(85,75){3}
\Text(20,57)[]{{\blue{\large $\bullet$}}}
\Text(56,85)[]{{\blue{\large $\bullet$}}}
\Text(56,30)[]{{\blue{\large $\bullet$}}}
\Text(-10,72)[]{$Z$}
\Text(45,56)[]{$W$}
\Text(85,76)[]{\blue{$H$}}
\Text(85,40)[]{$\gamma$}
\hspace*{1.5cm}
\Photon(100,50)(140,50){3.2}{5.5}
\Photon(170,25)(205,25){3.2}{5.5}
\DashLine(170,75)(205,75){3}
\Text(162,57)[]{{\blue{\large $\bullet$}}}
\Text(197,85)[]{{\blue{\large $\bullet$}}}
\Text(197,30)[]{{\blue{\large $\bullet$}}}
\ArrowLine(140,50)(170,25)
\ArrowLine(170,75)(140,50)
\ArrowLine(170,25)(170,75)
\Text(183,58)[]{$F$}
\Text(137,72)[]{$Z$}
\Text(220,76)[]{\blue{$H$}}
\Text(220,40)[]{$\gamma$}
\Text(80,60)[]{+}  
\end{picture}
\vspace*{-9mm}
\end{center}
\centerline{\it Figure 1.10: Feynman diagrams for the one--loop induced decay 
mode $Z \to H \gamma$ in the SM.}\s
 
The complete expressions of the form factors $A_t$ and $A_W$ will be given 
later, when the reverse decay $H \to Z \gamma$ will be discussed in detail. 
In the case of interest here, i.e. for $M_H \lsim M_W$, one can approximate the 
top quark form factors by its value in the vanishing $M_H$ limit, $A_t= N_c Q_t 
\hat{v}_t / (3 c_W) \sim 0.3$, but for the $W$ form factor, a good 
approximation in the Higgs boson mass range relevant at LEP1, is given by 
\cite{Z-h-gamma2}
\beq
A_W \simeq -4.6 + 0.3 M_H^2/M_W^2
\eeq
The two contributions interfere destructively, but the $W$ contribution is
largely dominating.  We show in Table 1.4, the number of Higgs particles
produced per $10^{7}$ $Z$ bosons, in both the loop induced process $Z \to
H\gamma$ and in the Bjorken process $Z \to H \mu^+ \mu^-$ [to obtain the rates
for any final state $f$ one has to multiply by a factor $\Gamma_Z / \Gamma_\mu
\sim 33$]. As can be seen, the number of produced $H$ bosons is much larger in
the Bjorken process for small Higgs masses but the loop decay process becomes
more important for masses around $M_H \sim 60$ GeV. However, in this case, only
a handful of events can be observed.\s

\begin{table}[hbt]
\renewcommand{\arraystretch}{1.6} 
\begin{center} \begin{tabular}{|c||c|c|c|c|c|c|c|} \hline 
$ M_H~({\rm GeV})$   & 10 & 20 & 30 &  40 & 50 & 60& 70 \\ \hline 
$ Z \to H\mu^+ \mu^-$ & 750 & 290 & 120 & 46 & 15.6 & 3.7 & 0.6 \\  \hline 
$ Z \to H \gamma$  & 20.4 & 18.4 & 15.3 & 11.6 & 7.8 & 4.4 & 1.8 \\ \hline 
\end{tabular}\\[4mm] 
\renewcommand{\arraystretch}{1.2}
{\it Table 1.4: The number of events for Higgs production at LEP1 per $10^7$ 
Z bosons.}
\vspace*{-6mm}
\end{center}
\end{table}

As will be discussed in great detail in the next chapter, the Higgs boson in
the mass range relevant at LEP1 [and also LEP2], decays dominantly into hadrons
[mostly $b\bar{b}$ final states for $M_H \gsim 10$ GeV], and less than $\sim
8\%$ of the time into $\tau$--lepton pairs. Thus, not to be swamped by the
large $\ee \to$ hadron background, the Higgs boson has been searched for at
LEP1 in the two topologies $Z \to (H \to {\rm hadrons}) (Z^* \to \nu \bar \nu)$
leading to a final state consisting of two acoplanar jets and missing energy
and $Z \to  (H \to {\rm hadrons}) (Z^* \to \ee,\mu^+\mu^-)$ with two energetic
leptons isolated from the hadronic system. The absence of any Higgs boson
signal by the four collaborations at LEP1 \cite{LEP1-Higgs}, allowed to set the
95\% Confidence Level limit of $M_H \gsim  65.2$ GeV on the SM Higgs boson mass
\cite{LEP1-Higgs-average}. \s

Before the advent of LEP1, the low Higgs mass range, $ M_H \lsim 5$ GeV, was
very difficult to explore. Indeed, the main probes were, for Higgs masses below
$20$ MeV, Nuclear Physics experiments which are very sensitive to the
theoretically uncertain Higgs--nucleon couplings and for larger masses, rare
meson [from pions to heavy $B$ mesons] decays which were plagued by various
theoretical and experimental uncertainties\footnote{For a very detailed
discussion of the SM Higgs boson searches in this low mass range, see Chapter
3.1 of {\it The Higgs Hunter's Guide} \cite{HHG}, pages 91-130.}. On the $Z$
resonance, this low mass range can be easily probed by considering the clean
final state $Z\to Z^*H \to \mu^+ \mu^- H$: since the invariant mass of the
system recoiling against the lepton pair is simply the Higgs boson mass, the
precise knowledge of the c.m.  energy and the accurate measurement of the
invariant mass and energy of the leptons allows an excellent resolution on
$M_H$. This process therefore definitely rules out any Higgs boson with a mass
below $\sim 60$ GeV, independently of its decay modes, provided that its
coupling to the $Z$ boson is as predicted in the SM.  

\vspace*{-2mm}
\subsubsection*{\underline{Searches at LEP2}}

The search for Higgs bosons has been extended at LEP2 with c.m.  energies up to
$\sqrt{s}=209$ GeV. In this energy regime, the dominant production process is
Higgs--strahlung  \cite{EGN,LQT,Petcov,Higgs-strahlung,Behrends-Kleiss} where
the $\ee$ pair goes into an off--shell $Z$ boson which then splits into a Higgs
particle and a real $Z$ boson, $\ee \to Z^* \to HZ$; see the digram of
Fig.~1.11. [The cross section for the $WW$ fusion process, to be discussed
later, is very small at these energies \cite{LEP2-Higgs-Th}.]

\begin{center}
\vspace*{-.6cm}
\hspace*{3cm}
\begin{picture}(300,90)(0,10)
\SetWidth{1.}
\ArrowLine(-10,25)(40,50)
\ArrowLine(-10,75)(40,50)
\Photon(40,50)(100,50){3.5}{7}
\DashLine(100,50)(150,25){4}
\Photon(100,50)(150,75){3.5}{5.5}
\Text(40,50)[]{{\blue{\large $\bullet$}}}
\Text(100,50)[]{{\blue{\large $\bullet$}}}
\Text(-10,20)[]{$e^-$}
\Text(-10,80)[]{$e^+$}
\Text(75,65)[]{$Z^*$}
\Text(160,20)[]{\blue{$H$}}
\Text(160,80)[]{$Z$}
\end{picture}
\vspace*{-6.mm}
\mbox{\it Figure 1.11: The production mechanism for SM Higgs bosons in $\ee$ 
collisions at LEP2.} 
\vspace*{5mm}
\end{center}

The production cross section for this Higgs--strahlung process [which will be 
discussed in more details later] is given by
\beq 
\sigma(\ee \ra ZH) = \frac{G_\mu^2 M_Z^4}{96 \pi s} [1+ (1-4s_W^2)^2] 
\lambda^{1/2} \frac{ \lambda+ 12M_Z^2/s}{(1-M_Z^2/s)^2} 
\eeq
It scales like $1/s$ and, therefore, is larger at low energies for light Higgs
bosons and is suppressed by  the usual two--particle phase space function
$\lambda^{1/2}=[(1-M_H^2/s-M_Z^2/s)^2-4M_H^2M_Z^2/s^2]^{1/2}$. At LEP2 and  
for the maximal c.m. energy that has been reached, $\sqrt{s}_{\rm max}
\sim 209$ GeV,  it is shown in Fig.~1.12 as a function of $M_H$. At $M_H \sim
115$ GeV, the cross section is of the order of 100 fb which, for the integrated
luminosity that has been collected, $\int {\cal L} \sim 0.1$ fb$^{-1}$, 
correspond to ten produced events. For a mass $M_H^{\rm max}  \sim \sqrt{s}-M_Z
\sim 117$ GeV, the $2 \to 2$ cross section vanishes, being suppressed by the 
phase--space factor $\lambda^{1/2}$. \s

\begin{figure}[htbp]
\begin{center}
\vspace*{-.8cm}
\hspace*{-2cm}
\epsfig{file=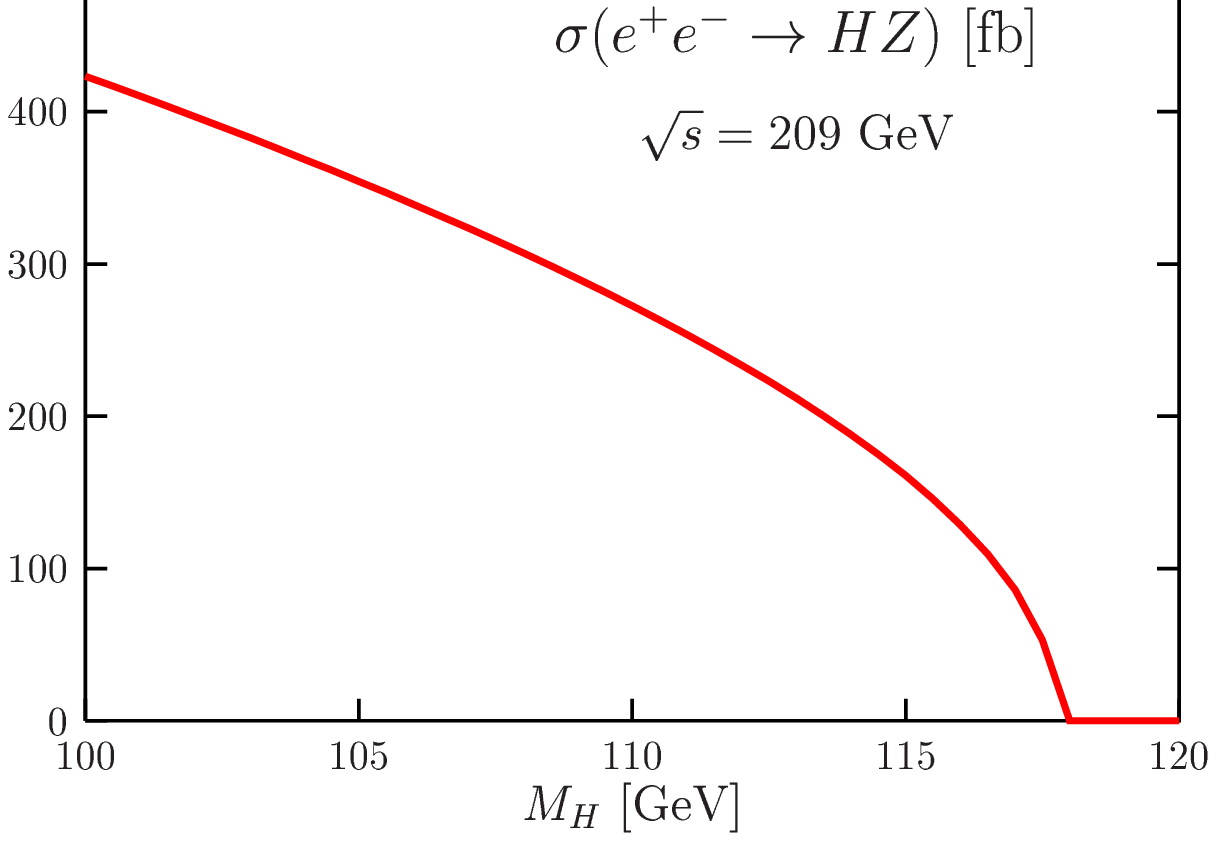,width=16.cm} 
\end{center}
\vspace*{-15.3cm}
{\it Figure 1.12: Production cross section for the SM Higgs boson at LEP2 [in
femtobarns] for a center of mass energy $\sqrt{s}=209$ GeV as a function of 
the Higgs boson mass.}
\vspace*{-3mm}
\end{figure} 

The searches by the LEP collaborations have been made in several topologies 
[recall that the Higgs boson decays mainly into $b\bar{b}$ final states and the 
branching ratio for the decays into $\tau$--lepton is a few percent]: 
$\ee \to  (H \to b\bar{b}) (Z^* \to \nu \bar \nu)$ and 
$\ee \to  (H \to b\bar{b}) (Z^* \to \ell^+ \ell^-)$ as at LEP1, as well as 
$\ee \to  (H \to \tau^+\tau^-) (Z^* \to b\bar{b})$ and
$\ee \to  (H \to b\bar{b}) (Z^* \to \tau^+ \tau^-)$.  
Combining the results of the four LEP collaborations, no significant excess 
above the expected SM background has been seen, and the exclusion  limit 
\cite{LEP2-Higgs-exp}
\beq 
M_H > 114.4~{\rm GeV} 
\eeq 
has been established at the 95\% CL from the non--observation of a signal, as 
shown in Fig.~1.13.  This upper limit, in the absence of additional 
events with respect to SM predictions, was expected to be $M_H>115.3$ GeV.  The
reason for the discrepancy is that there is a $1.7\sigma$ excess [compared to
the value  $2.9\sigma$ reported at the end of 2000] of events for a Higgs boson
mass in the vicinity of $M_H=116$ GeV \cite{LEP2-Higgs-exp}. But this excess 
is not significant enough, since  we need a $5\sigma$  signal to be sure that 
we have indeed discovered the Higgs boson.\s

Higgs bosons with SM couplings to the $Z$ boson have been searched for in
various decay modes, such as invisible decays \cite{LEP2-Higgs-ind} and flavor 
blind hadronic decays \cite{LEP2-Higgs-fbd} by considering the recoil of the 
$Z$ boson in the process $\ee \to  H (Z^* \to
\ell^+ \ell^-)$ for instance; Higgs boson masses close to the $M_H\sim 114$ GeV 
bound have been ruled out.  The bound $M_H \gsim 114.4$
GeV can be evaded only if the Higgs boson has non standard couplings to the $Z$
boson. Indeed a smaller value of the $g_{HZZ}$ coupling compared to the SM
prediction would suppress the  $\ee \to HZ$ cross section which is directly
proportional to $g_{HZZ}^2$. The 95\% CL bound on the Higgs boson mass as a
function of its coupling relative to the SM value, $\zeta=g_{HZZ}/g_{HZZ}^{\rm
SM}$ is shown in Fig.~1.14. For masses below $M_H \lsim 80$ GeV, Higgs bosons
with couplings to the $Z$ boson an order of magnitude smaller than in the SM
have thus also been ruled out \cite{LEP2-Higgs-exp}.

\begin{figure}[htbp]
\begin{center}
\vspace*{-.5cm}
\epsfig{figure=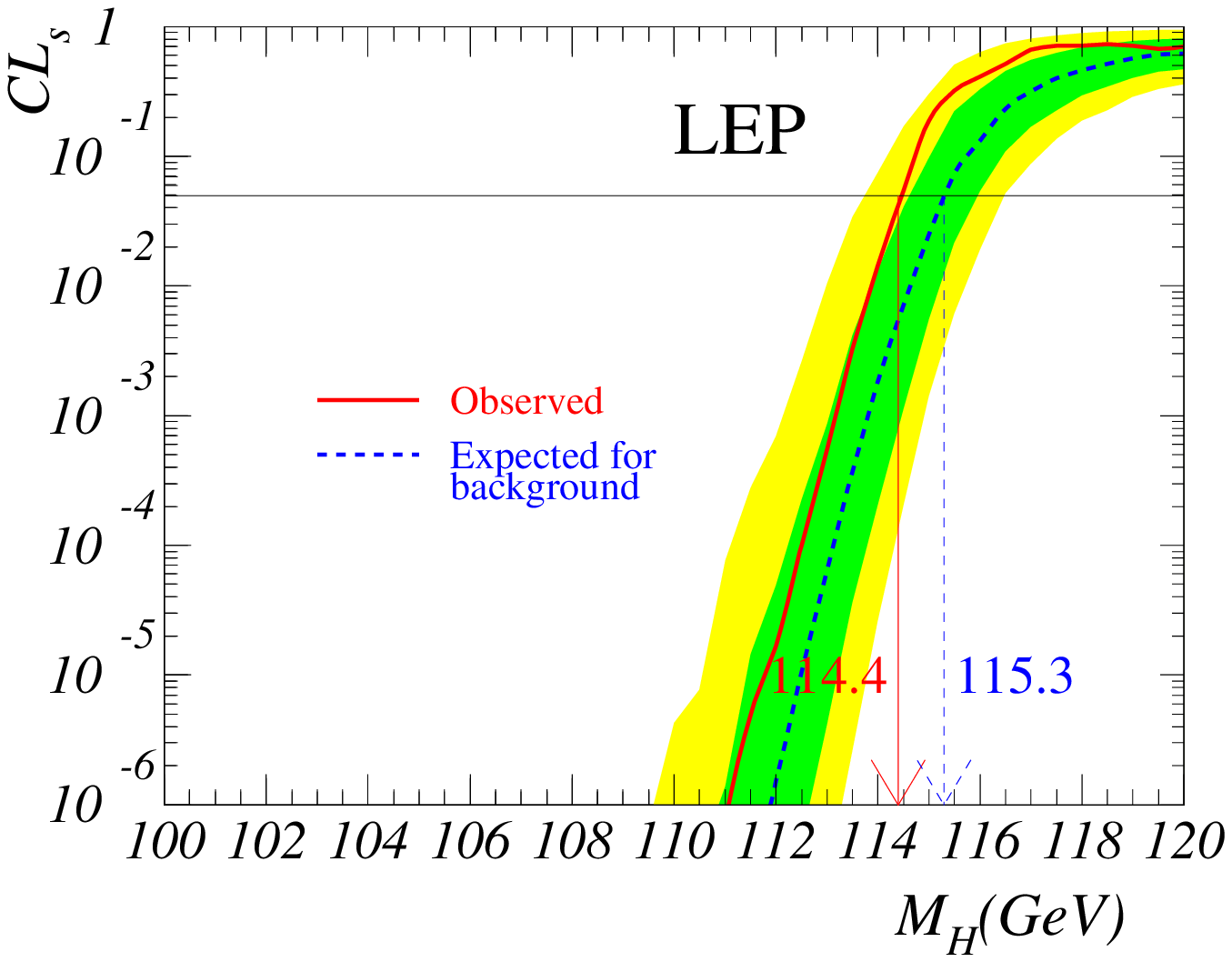,width=12cm}
\end{center}
\vspace*{-1cm}
\end{figure}
{\it Figure 1.13: Confidence Level $CL_s$ for the signal+background hypothesis
in Higgs production at LEP2. The solid/red line is for the observation, the 
dashed line is the median background expectation, the dark--grey/green and 
light--grey/yellow shaded bands around the median expected line correspond to 
the 68\% and 95\% simulated probability bands. The  intersection of the 
horizontal line at $CL_s = 0.05$ with the observed curve  defines the 95\% CL  
lower bound for $M_H$; from Ref.~\cite{LEP2-Higgs-exp}.}

\vspace*{.5cm}

\begin{figure}[htbp]
\begin{center}
\vspace*{-.5cm}
\epsfig{figure=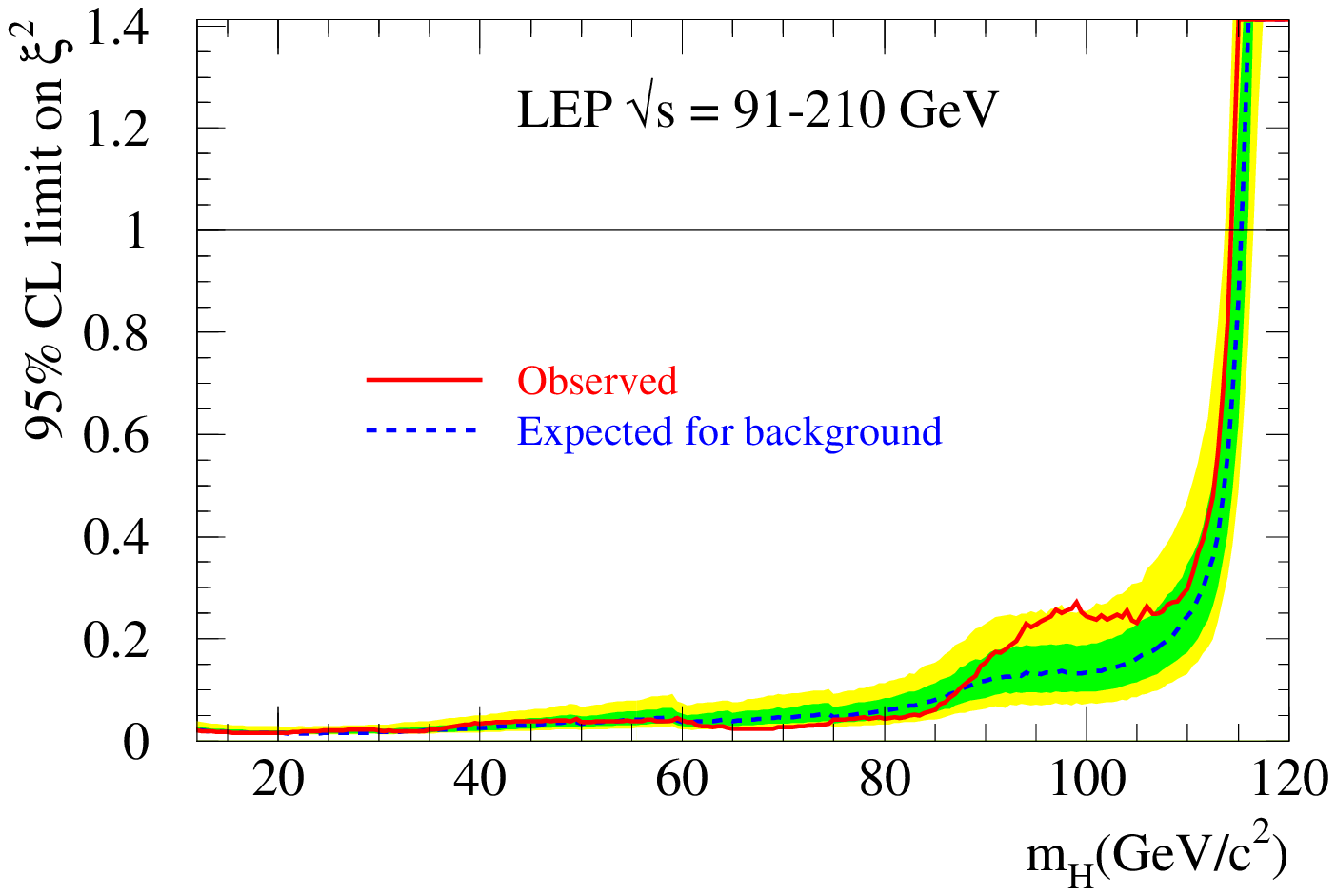,width=12cm}
\end{center}
\vspace*{-1cm}
\end{figure}
{\it Figure 1.14: The upper bound on the coupling $\zeta^2=(g_{HZZ}/g_{HZZ}^{
\rm SM})^2$ as a function of the Higgs mass. The solid line represents the 
observed limit while the dark (light) shaded band is for the 68\% (95\%) 
probability band; from Ref.~\cite{LEP2-Higgs-exp}.}

\subsection{Theoretical constraints on the Higgs boson mass}

In addition to the experimental constraints on the Higgs boson mass discussed
previously,  there are interesting theoretical constraints
which can be derived from assumptions  on the energy range in which the SM is
valid before perturbation theory breaks down and new phenomena should emerge.
These include constraints from unitarity in scattering amplitudes, 
perturbativity of the Higgs self--coupling, stability of the electroweak 
vacuum and fine--tuning. These constraints are summarized in this 
section. 

\subsubsection{Constraints from perturbativity and unitarity}

\subsubsection*{\underline{Perturbative unitarity}}
 
One of the main arguments to abandon the old Fermi theory for the weak
interaction was that it violates unitarity at energies close to the Fermi
scale.  Indeed, taking for  instance the reaction  $\nu_\mu e \to \nu_e \mu$
[which, in principle, proceeds through the $t$--channel exchange of a $W$ 
boson and has only the $J$=$1$ partial wave], the cross section at a high 
energy $\sqrt{s}$ behaves like $\sigma \sim  G_\mu^{-1/2} s$. However, 
unitarity requires that the cross section should be bounded by $s^{-1}$ and for
energies above $\sqrt{s} \sim G_\mu^{-1/2} \sim 300$ GeV, the cross section
will violate unitarity. This particular problem was cured in the intermediate
massive vector boson theory [i.e. including simply by hand the $W$ boson mass
and, hence, its longitudinal degree of freedom in the Lagrangian] but in other
processes, such as $\nu \bar \nu \to W^+ W^-$ through $t$--channel $e$ 
exchange, the amplitude had also a bad high energy behavior which called for the
introduction of the neutral $Z$ boson to be exchanged in the $s$--channel
to cancel it. In fact, if one demands that there is no such process
which  violates unitarity, one would end up with just the renormalizable
Lagrangian of the SM  discussed in \S1.1; see Ref.~\cite{UNITARITY}. \s

However, there is still a potential problem in the SM, but at much higher
energies than the Fermi scale. As discussed in \S1.1.3, the interactions of the
longitudinal components of the massive gauge bosons grow with their momenta. 
In processes involving the $W_L$ and $Z_L$ bosons, this would eventually lead
to cross sections which increase with the energy which would then violate
unitarity at some stage.  We will briefly  discuss this aspect in the
following, taking as an example the scattering process $W^+ W^- \to W^+ W^-$ at
high energies \cite{rho-Veltman,Equivalence-theorem,UNITARITYBounds}; for 
a detailed discussion,
see Ref.~\cite{WWreview} for instance.  Some contributing Feynman diagrams to 
this process are displayed in  Fig.~1.15; there are also additional diagrams 
involving the $s$-- and $t$--channel exchanges of $\gamma$ and $Z$ bosons.\s

The amplitude for the scattering of charged $W$ bosons, in the high--energy
limit $s \gg M_W^2$ and for heavy Higgs bosons, is given by
\beq
A(W^+ W^- \to W^+ W^-) \stackrel{\small s \gg M_W^2}\longrightarrow
\frac{1}{v^2} \left[ s+t - \frac{s^2}{s-M_H^2} - \frac{t^2}{t-M_H^2}\right]
\label{wlwlamp}
\eeq
where $s,t$ are the Mandelstam variables [the c.m. energy $s$ is the square
of the sum of the momenta of the initial or final states, while $t$ is the
square of the difference between the momenta of one initial and one final 
state]. In fact, this contribution is coming from longitudinal $W$ bosons 
which, at high energy, are equivalent to the would--be Goldstone bosons as 
discussed in \S1.1.3. One can then use the potential of eq.~(\ref{Vequivalence})
which gives the interactions of the Goldstone bosons and write in a very simple
way the three individual amplitudes for the scattering of longitudinal $W$ 
bosons 
\beq
A(w^+ w^- \to w^+ w^-) = - \left[ 2 \frac{M_H^2}{v^2} +  \left( \frac{M_H^2}{v} 
\right)^2 \frac{1}{s-M_H^2} + \left( \frac{M_H^2}{v} \right)^2\frac{1}{t-M_H^2}
\right]
\eeq
which after some manipulations, can be cast into the result of
eq.~(\ref{wlwlamp}) given previously.

\begin{center}
\vspace*{-.3cm}
\hspace*{-1.5cm}
\begin{picture}(300,100)(0,0)
\SetWidth{1}
\SetScale{1.1}
\Photon(0,25)(35,50){3}{5.}
\Photon(0,75)(35,50){3}{5}
\Photon(35,50)(70,75){3}{5}
\Photon(35,50)(70,25){3}{5}
\Text(37,53)[]{{\blue{\Large $\bullet$}}}
\Text(-15,20)[]{$W^-$}
\Text(-15,70)[]{$W^+$}
\Text(90,70)[]{$W^-$}
\Text(90,20)[]{$W^+$}
\hspace*{1cm}
\Photon(100,75)(130,50){3}{5}
\Photon(100,25)(130,50){3}{5}
\Text(140,53)[]{{\blue{\Large $\bullet$}}}
\Text(190,53)[]{{\blue{\Large $\bullet$}}}
\DashLine(130,50)(170,50){4}
\Photon(170,50)(200,25){3}{5}
\Photon(170,50)(200,75){3}{5}
\Text(160,65)[]{\blue{$H$}}
\Photon(230,75)(265,75){3}{5}
\Photon(265,75)(300,75){3}{5}
\Text(292,79)[]{{\blue{\Large $\bullet$}}}
\Text(292,25)[]{{\blue{\Large $\bullet$}}}
\Photon(230,25)(265,25){3}{5}
\Photon(265,25)(300,25){3}{5}
\DashLine(265,25)(265,70){4}
\Text(280,50)[]{\blue{$H$}}
\end{picture}
\vspace*{-6mm}
\end{center}
\centerline{\it Figure 1.15: Some Feynman diagrams for the scattering of $W$ 
bosons at high energy.} 
\vspace*{3mm}

These amplitudes will lead to cross sections $\sigma (W^+W^- \to W^+W^-) \simeq 
\sigma (w^+w^- \to w^+w^-)$ which could violate their unitarity bounds. To see 
this explicitly, we first decompose the scattering amplitude $A$ into partial 
waves $a_\ell$ of orbital angular momentum $\ell$
\beq 
A= 16 \pi \sum_{\ell=0}^{\infty} (2\ell+1) P_\ell (\cos \theta) \, a_\ell 
\eeq
where $P_\ell$ are the Legendre polynomials and $\theta$ the scattering angle.
Since for a $2\to 2$ process, the cross section is given by ${\rm d} \sigma 
/{\rm d} \Omega = |A|^2 /(64 \pi^2 s)$ with d$\Omega=2\pi$d$\cos\theta$, one 
obtains
\beq
\sigma &=& \frac{8\pi}{s} \sum_{\ell=0}^{\infty} 
\sum_{\ell'=0}^{\infty} (2\ell+1) (2\ell'+1) a_\ell a_{\ell'} \int_{-1}^1
{\rm d}\cos \theta  P_\ell (\cos \theta )  P_{\ell'} ( \cos \theta ) \non \\
&=& \frac{16 \pi}{s} \sum_{\ell=0}^{\infty}  (2\ell +1) |a_\ell|^2
\eeq
where the orthogonality property of the Legendre polynomials, $\int {\rm d} 
\cos \theta P_\ell P_{\ell'}=\delta_{\ell \ell'}$, has been used. The optical
theorem tells us also that the cross section is proportional to the imaginary
part of the amplitude in the forward direction, and one has the identity
\beq
\sigma = \frac{1}{s} \, {\rm Im}\, [\, A(\theta=0)\, ]
\, = \, \frac{16 \pi}{s} \sum_{\ell=0}^{\infty}  (2\ell +1) |a_\ell|^2 
\eeq
This leads to the unitary conditions \cite{WL-lattice}
\beq
|a_\ell|^2 = {\rm Im} (a_\ell) & \Rightarrow & [{\rm Re} (a_\ell)]^2 +
[{\rm Im}(a_\ell)]^2 = {\rm Im}(a_\ell) \non \\
&\Rightarrow & [{\rm Re} (a_\ell)]^2 + 
[{\rm Im}(a_\ell) -\frac{1}{2}]^2 = \frac{1}{4}
\eeq
This is nothing else than the equation of a circle of radius $\frac{1}{2}$
and center $(0, \frac{1}{2})$ in the plane $[{\rm Re} (a_\ell), {\rm Im}(a_\ell)
]$. The real part lies between $-\frac{1}{2}$ and $\frac{1}{2}$, 
\beq
|{\rm Re} (a_\ell)| < \frac{1}{2} 
\label{unitaritycondition} 
\eeq
If one takes the $J=0$ partial wave for the amplitude $A(w^+w^- \to w^+w^-)$
\beq
a_0  = \frac{1}{16 \pi s} \int_s^0 {\rm d}t |A| = 
 - \frac{M_H^2}{16 \pi v^2} \left[2 + \frac{M_H^2}{s-M_H^2} -  \frac{M_H^2}{s}
{\rm log} \left( 1 + \frac{s}{M_H^2} \right) \right]  
\eeq
and assumes the Higgs boson mass to be much smaller than $\sqrt{s}$, which 
leads to
\beq
a_0  \stackrel{\small s \gg M_H^2}\longrightarrow - \frac{M_H^2}{8\pi v^2} 
\eeq
From the requirement of the unitarity condition, 
eq.~(\ref{unitaritycondition}), one obtains the upper bound \cite{LQT}
\beq
M_H \lsim 870~{\rm GeV} 
\eeq
In fact the scattering channel $W_L^+W_L^-$ considered above can be coupled
with other channels: $Z_LZ_L$, $HH$  and $Z_L H $ [for a recent discussion, see
Ref.~\cite{Abdeslam} e.g.]. In addition to the four neutral particle initial
states, one can also consider the two charged channels $W_L^+H$ and $W_L^+Z_L$
which, because of charge conservation, are not coupled to the neutral ones. The
scattering amplitude is then given by a $6\times 6$ matrix which is diagonal by
block: a $4\times 4$ block for the neutral channels and a $2\times 2$ block for
the charged channels. At high energies, the matrix elements are dominated by
the quartic couplings and the full matrix in the basis
\beq
\left( W_L^+W_L^- \, ,  \ \frac{1}{\sqrt{2}} Z_LZ_L \, , \ 
\frac{1}{\sqrt{2}}HH \, , \ Z_L H \, , \ W_L^+H \, , \ W_L^+Z_L \right)
\eeq
with the factors $\frac{1}{\sqrt{2}}$ accounting for identical particle 
statistics, takes the form
\begin{eqnarray}
a_0 \propto \frac{M_{H}^2}{v^2} \left(
\begin{array}{cccccc}
1 & \frac{\sqrt{2}}{4} & \frac{\sqrt{2}}{4} & 0 & 0 & 0 \\
 \frac{\sqrt{2}}{4} & \frac{3}{4} & \frac{1}{4} & 0 & 0 & 0 \\
 \frac{\sqrt{2}}{4} & \frac{1}{4} & \frac{3}{4} & 0 & 0 & 0 \\
 0 & 0 & 0 & \frac{1}{2} & 0 & 0 \\
 0 & 0 & 0 & 0 & \frac{1}{2} & 0 \\
 0 & 0 & 0 & 0 & 0 & \frac{1}{2}\end{array}  \right) \label{msm}
\end{eqnarray}
The requirement that the largest eigenvalues of $a_0$, respects the unitarity 
constraint yields~\cite{Pert-HWcplg1}
\beq
M_{H} \lsim 710~{\rm GeV} 
\eeq

Thus, in the SM, if the Higgs boson mass exceeds values of ${\cal O}(700~{\rm
GeV})$, unitarity will be violated  unless new phenomena appear and restore it.
There is, however, a caveat to this conclusion. The analysis above has been
performed only at tree--level and since the Higgs boson  self--coupling
becomes strong for large masses, $\lambda = M_H^2/(2v^2)$, the  radiative
corrections can be very large and, eventually, render the theory non
perturbative; this tree--level result would be then lost. Thus, to apply the
previous  argument to set a bound on the Higgs boson mass, one has to assume
that the SM  remains perturbative and that higher--order corrections are not
large. The unitarity argument should therefore more properly be called, the
tree--level unitarity or perturbative unitarity argument. \s

In fact, one can use the unitarity argument in a different limit
\cite{Equivalence-theorem}: if one assumes that the Higgs  boson mass is much
larger than $\sqrt{s}$ [which in turn, is much  larger than $M_{W}$], the
unitarity constraint writes, if one takes into account only the $W_L^+ W_L^-
\to W_L^+ W_L^-$ channel, 
\beq
a_0  \stackrel{\small s \ll M_H^2}\longrightarrow - \frac{s}{32\pi v^2} 
\eeq
and with the condition $|{\rm Re}(a_0)| < {1 \over 2}$, one obtains
\beq
\sqrt{s}  \lsim 1.7~{\rm TeV}  
\eeq
Again, a more stringent bound  is obtained by considering all the coupled 
channels above
\beq
\sqrt{s} \lsim 1.2~{\rm TeV}
\eeq
This means that if the Higgs boson is too heavy [or, equivalently, not existing
at all], some New Physics beyond the SM should manifest itself at energies in 
the TeV range to restore unitary in the scattering  amplitudes of longitudinal 
gauge bosons.\s

Therefore, from the requirement that the tree--level contributions to the
partial waves of scattering processes involving gauge and Higgs bosons should 
not exceed the unitarity bound, one concludes that either: $(i)$ some New 
Physics, which plays a role similar to that of the Higgs particle should appear
in the TeV range to cancel this breakdown, or $(ii)$ the unitarity breakdown is
canceled by high--order terms which signal the failure of perturbation theory 
and the loss of the predictive power of the SM.

\subsubsection*{\underline{Perturbativity in processes involving the Higgs 
boson}}

In fact, it is known from a different context that for large values of the Higgs
boson mass, perturbation theory is jeopardized in the SM. This occurs for
instance in the decays of the Higgs boson into massive gauge bosons, which
will be discussed later in detail. Using the equivalence theorem and the
Lagrangian eq.~(\ref{Vequivalence}), one can write immediately the partial
decay width of the Higgs boson into two longitudinal $Z$ bosons [or $W$ bosons]
\beq
\Gamma(H \to ZZ) \sim  \Gamma(H \to w_0 w_0) = \left( \frac{1}{2M_H} 
\right) \ \left( \frac{2!\, M_H^2}{2v} \right)^2 \, \frac{1}{2}\, 
\left( \frac{1}{8\pi} \right) \to \frac{M_H^3}{32 \pi v^2} 
\eeq
where the first parenthesis is for the flux factor, the second for the 
amplitude squared, the factor $\frac{1}{2}$ is for the two identical final 
particles, and the last parenthesis is for the phase space factor. For the 
decay $H \to WW$, one simply needs to remove the statistical factor to
account for both $W^\pm$ states
\beq
\Gamma (H \to W^+W^-)  \simeq 2 \Gamma (H \to ZZ)
\eeq
The behavior, $\Gamma_H \propto M_H^3$, compared to  $\Gamma_H \propto M_H$ for
decays into fermions for instance, is due to the  longitudinal components that
grow with the energy [which is $M_H$ in this  context].\s

\begin{center}
\vspace*{-.6cm}
\hspace*{-3cm}
\begin{picture}(300,100)(0,0)
\SetWidth{1.1}
\SetScale{1.0}
\DashLine(-20,50)(20,50){4}
\Photon(20,50)(50,75){3.2}{5}
\Photon(20,50)(50,25){3.2}{5}
\Text(0,60)[]{\blue{$H$}}
\Text(59,75)[]{$V$}
\Text(59,25)[]{$V$}
\DashLine(100,50)(140,50){4}
\DashLine(140,50)(170,75){4}
\DashLine(140,50)(170,25){4}
\Photon(170,25)(170,75){3.2}{4}
\Photon(170,25)(200,25){3.2}{4}
\Photon(170,75)(200,75){3.2}{4}
\DashLine(230,50)(270,50){4}
\DashLine(270,50)(320,75){4}
\DashLine(270,50)(320,25){4}
\Photon(320,25)(320,75){3.2}{6}
\Photon(320,25)(350,25){3.2}{4}
\Photon(320,75)(350,75){3.2}{4}
\DashCArc(296,53)(27,190,290){4}
\Text(140,50)[]{\red{\large $\bullet$}}
\Text(270,50)[]{\red{\large $\bullet$}}
\Text(310,30)[]{\red{\large $\bullet$}}
\Text(80,50)[]{$+$}
\Text(210,50)[]{$+$}
\Text(370,50)[]{$+ \ \cdots $}
\end{picture}
\vspace*{-8mm}
\end{center}
\centerline{\it Figure 1.16: Generic diagrams for the one-- and two--loop 
corrections to Higgs boson decays.}
\vspace*{5mm}

Let us have a brief look at these decays when higher--order radiative 
corrections, involving the Higgs boson and therefore the quartic coupling
$\lambda$,  are taken into account. Including the one--loop and two--loop 
radiative corrections, with some generic Feynman diagrams shown in 
Fig.~1.16, the partial Higgs decay width into gauge bosons is given 
by \cite{Pert-HWcplg1,Pert-HWcplg2} 
\beq
\Gamma_{\rm tot} \simeq \Gamma_{\rm Born} \left[ 1 +  3 \hat \lambda +62 \hat 
\lambda^2 + {\cal O}(\hat \lambda^3) \right] 
\eeq
with $\hat \lambda= \lambda/(16 \pi^2)$.
If the Higgs boson mass is very large, $M_H \sim {\cal O}(10~{\rm TeV})$, the
one loop term becomes close to the Born term, $3\hat \lambda \sim 1$, and
the perturbative series is therefore not convergent. Even worse, already for a
Higgs boson mass in the TeV range, $M_H \sim {\cal O}(1~{\rm TeV})$, the 
two--loop contribution becomes as important as the one--loop contribution, $3
\hat \lambda \sim  62 \hat \lambda^2$. Hence, for perturbation theory
to hold,  $M_H$ should be smaller than about 1 TeV. \s

In addition, the partial decay widths become extremely large for a very heavy 
Higgs particle. Indeed, taking into account only $W$ and $Z$ decay modes, the
total width is
\beq
\Gamma(H \to WW+ZZ) \sim 500~{\rm GeV}~(M_H/1~{\rm TeV})^3
\eeq
and for a mass $M_H \sim 1.3~{\rm TeV}$, the total decay width becomes 
comparable to the mass: the Higgs boson is then ``obese" and cannot be 
considered as a ``true" resonance anymore. \s

The same exercise can be made in the case of the Higgs decays into
fermions. Including the one-- and two--loop corrections involving
the quartic interaction, one obtains \cite{rho-Veltman,Pert-Hfcplg2}
\beq
\Gamma_{\rm tot} \simeq \Gamma_{\rm Born} \left[ 1 +  2 \hat \lambda -32 \hat 
\lambda^2 + {\cal O}(\hat \lambda^3) \right] 
\eeq
Qualitatively, the situation is the same as for the decays into gauge bosons,
although the breakdown of perturbation theory is delayed because of the
smaller coefficients of the one-- and two--loop corrections. These features 
will be discussed in the chapter on Higgs decays. \s

The jeopardy of perturbation theory at large Higgs masses can also be seen in 
the scattering of longitudinal gauge bosons from which we have previously 
derived the upper bound on $M_H$ from perturbative unitarity. In the case of
the $W^+_L W^-_L \to W^+_L W^-_L$ scattering, the radiative corrections have 
been calculated at one and two loops in Refs.~\cite{RC-unitarity,WL-2loop} 
where it has been found that at high energy, the amplitude depends on the 
considered energy, contrary to what was occurring in the tree level case 
discussed 
previously. However, applying Renormalization Group methods, one can absorb the
logarithmic energy dependence by defining a running self--coupling $\lambda$ at
the energy scale $\sqrt{s}$ [see next subsection]. At two--loop order, one then
finds for the $W^+_L W^-_L \to W^+_L W^-_L$ scattering cross section at very
high energies \cite{WL-2loop}
\beq
\sigma (W^+_L W^-_L \to W^+_L W^-_L) \sim \frac{1}{s} \hat \lambda (s)
\left( 1- 48.64 \hat \lambda +333.21  \hat \lambda^2 \right)
\eeq
Here, the coefficients of the corrections are much larger than in Higgs decays
and in fact, the one--loop correction become of order unity already for $
\lambda (s)$ values close to 3. \s 

Using various criteria, such as the scheme and scale dependence of the
amplitudes, to estimate at which stage the breakdown of perturbation theory
occurs \cite{WL-2loop2} and a comparison with non--pertur\-ba\-ti\-ve
calculations on the lattice \cite{WL-lattice}, one arrives at the conclusion
that perturbation theory is lost for Higgs boson masses above $M_H \sim 700$
GeV. This result is remarkably close to what has been obtained by simply using
the [somewhat naive] perturbative unitarity argument.  

\subsubsection{Triviality and stability bounds}
 
As seen in previous discussions, because of quantum corrections, the couplings
as well as the masses which appear in the SM Lagrangian, depend on the
considered energy. This is also the case for the quartic Higgs coupling which 
will be monotonically increasing with the energy scale $|Q|$. This leads to
non--trivial constraints on this  coupling and, hence, on the Higgs boson mass,
that we summarize in this subsection. 

\subsubsection*{\underline{The triviality bound}}

Let us have a look at the one--loop radiative corrections to the Higgs 
boson quartic coupling, taking into account for the present moment only the 
contributions of  the Higgs boson itself. The Feynman diagrams for the 
tree--level and the one--loop corrections to the Higgs boson self--coupling are
depicted in Fig.~1.17. 

\begin{center}
\vspace*{-.5cm}
\hspace*{-3cm}
\begin{picture}(300,100)(0,0)
\SetWidth{1.}
\DashLine(-30,25)(40,75){4}
\DashLine(-30,75)(40,25){4}
\Text(5,50)[]{\red{\large $\bullet$}}
\DashLine(100,75)(140,60){4}
\DashLine(100,25)(140,40){4}
\DashLine(140,60)(180,75){4}
\DashLine(140,40)(180,25){4}
\DashCArc(140,50)(12,0,360){4}
\Text(140,62)[]{\red{\large $\bullet$}}
\Text(140,38)[]{\red{\large $\bullet$}}
\Text(-10,25)[]{\blue{$H$}}
\Text(-10,75)[]{\blue{$H$}}
\Text(20,25)[]{\blue{$H$}}
\Text(20,75)[]{\blue{$H$}}
\DashLine(200,75)(240,50){4}
\DashLine(200,25)(240,50){4}
\DashLine(280,50)(320,75){4}
\DashLine(280,50)(320,25){4}
\DashCArc(260,50)(17,0,360){4}
\Text(240,50)[]{\red{\large $\bullet$}}
\Text(280,50)[]{\red{\large $\bullet$}}
\DashLine(360,25)(410,75){4}
\DashLine(360,75)(410,25){4}
\DashCArc(385,30)(12,0,360){4}
\Text(376,40)[]{\red{\large $\bullet$}}
\Text(396,40)[]{\red{\large $\bullet$}}
\Text(80,50)[]{$+$}
\Text(180,50)[]{$+$}
\Text(340,50)[]{$+$}
\end{picture}
\vspace*{-7mm}
\end{center}
\centerline{\it Figure 1.17: Typical Feynman diagrams for the tree--level and 
one--loop Higgs self--coupling.} 
\bigskip

The variation of the quartic Higgs coupling with the energy scale $Q$ is 
described by the Renormalization Group Equation (RGE) \cite{RGE-Lambda}
\beq
\frac{ {\rm d}} { {\rm d}Q^2 }\, \lambda (Q^2) = \frac{3}{4\pi^2} \, \lambda^2 
(Q^2) \ + {\rm higher~orders}  
\eeq
The solution of this equation, choosing the natural reference energy point to 
be the electroweak symmetry breaking scale, $Q_0=v$, reads at one--loop
\beq
\lambda (Q^2) = \lambda (v^2) \left[ 1 - \frac{3}{4\pi^2} \, \lambda (v^2) \, 
{\rm log} \frac{Q^2}{v^2} \right]^{-1}  
\eeq
The quartic couplings varies logarithmically with the squared energy $Q^2$. 
If the energy is much smaller than the electroweak breaking scale, $Q^2 
\ll v^2$, the quartic coupling becomes extremely small and eventually vanishes,
$\lambda (Q^2) \sim  \lambda (v^2) /{\rm log}(\infty)  \to 0_+$. 
It is said that the theory is trivial, i.e. non interacting since the coupling 
is zero \cite{Triviality-term}. \s

In the opposite limit,  when the energy is much higher that weak scale,  
$Q^2 \gg v^2$, the quartic coupling grows and eventually becomes infinite,
$\lambda (Q^2) \sim \lambda (v^2)/ (1- 1)\gg 1$. The point, called Landau pole, 
where the coupling becomes infinite is at the energy 
\beq
\Lambda_C= v \, \exp \left ( \frac{4\pi^2}{3\lambda} \right) 
= v \, \exp \left ( \frac{4\pi^2 v^2}{M_H^2} \right)
\eeq
The general triviality argument \cite{TRIVIALITY,WL-lattice} states that 
the scalar sector of the SM is a $\phi^4$--theory, and for these theories to 
remain perturbative at all scales one needs to have a coupling $\lambda=0$ 
[which in the SM, means that the Higgs boson is massless], thus rendering the 
theory trivial, i.e. non--interacting. However,
one can view this argument in a different way:  one can use the RGE for the
quartic Higgs self--coupling to establish the energy domain in which the SM is
valid, i.e.  the energy  cut--off $\Lambda_C$ below which the self--coupling
$\lambda$ remains finite. In this case, and as can be seen from the previous 
equation, if $\Lambda_C$ is large, the Higgs mass should be small to avoid the 
Landau pole; for instance for the value $\Lambda_C \sim 10^{16}~{\rm GeV}$, 
one needs a rather light Higgs boson, $M_H \lsim 200~{\rm GeV}$. 
In turn, if the cut--off $\Lambda_C$ is small, the Higgs boson mass can
be  rather large and for $\Lambda_C \sim 10^{3}~{\rm GeV}$ for instance, the 
Higgs mass is allowed to be of the order of  $1~{\rm TeV}$.\s

In particular, if the cut--off is set at the Higgs boson mass itself,
$\Lambda_C = M_H$,  the requirement that the quartic coupling remains finite
implies that  $M_H \lsim  700~{\rm GeV}$. But again, there is a caveat in this
argument: when $\lambda$ is too large, one cannot use perturbation theory
anymore and this constraint is lost. However, from simulations of gauge
theories on the lattice, where the non--perturbative effects  are properly
taken into account, it turns out that one obtains the rigorous  bound $M_H <
640$ GeV \cite{LATTICE}, which is in a remarkable agreement with the bound
obtained by naively using perturbation  theory.  

\subsubsection*{\underline{The stability bound}}

In the preceding discussion, only the contribution of the Higgs boson itself
has been included in the running of the quartic coupling $\lambda$. This is
justified in the regime where $\lambda$ is rather large. However, to be
complete, one needs to also include the contributions from fermions and gauge 
bosons in the running. Since the Higgs boson couplings are proportional to the 
particle masses, only the contribution of top quarks and massive gauge bosons 
need  to be considered. Some generic Feynman diagrams for these additional 
contributions are depicted in Fig.~1.18. \s

The one--loop RGE for the quartic coupling, including  the fermion and gauge
boson contributions, becomes \cite{RGE-Lambda}
\beq
\frac{{\rm d} \lambda}{{\rm d log}Q^2} \simeq \frac{1}{16\pi^2} \left[
12 \lambda^2 + 6 \lambda \lambda_t^2  - 3\lambda_t^4  - \frac{3}{2}\lambda (3g_2^2+
g_1^2) + \frac{3}{16} \left(2 g_2^4+ (g_2^2+g_1^{2})^2 \right) \right]  
\eeq
where the top quark Yukawa coupling is given by $\lambda_t= \sqrt{2}m_{t}/v$.
The first effect of this extension is that for not too large $\lambda$ values,
the additional contributions will slightly alter the triviality bounds. In
particular, the scale at which the New Physics  should appear will depend on
the precise value of the top quark mass.\s

\begin{center}
\vspace*{-.7cm}
\hspace*{-3cm}
\begin{picture}(300,100)(0,0)
\SetWidth{1.}
\hspace*{2cm}
\DashLine(0,25)(40,25){4}
\DashLine(0,75)(40,75){4}
\Line(40,25)(80,25)
\Line(40,75)(80,75)
\Line(80,25)(80,75)
\Line(40,25)(40,75)
\DashLine(80,25)(120,25){4}
\DashLine(80,75)(120,75){4}
\Text(40,75)[]{\blue{\large $\bullet$}}
\Text(40,25)[]{\blue{\large $\bullet$}}
\Text(80,25)[]{\blue{\large $\bullet$}}
\Text(80,75)[]{\blue{\large $\bullet$}}
\Text(10,65)[]{\blue{$H$}}
\Text(10,35)[]{\blue{$H$}}
\Text(110,35)[]{\blue{$H$}}
\Text(110,65)[]{\blue{$H$}}
\Text(60,50)[]{$F$}
\hspace*{6cm}
\DashLine(0,25)(40,25){4}
\DashLine(0,75)(40,75){4}
\Photon(40,25)(80,25){3}{5}
\Photon(40,75)(80,75){3}{5}
\Photon(80,25)(80,75){3}{6}
\Photon(40,25)(40,75){3}{6}
\Text(40,75)[]{\blue{\large $\bullet$}}
\Text(40,25)[]{\blue{\large $\bullet$}}
\Text(80,25)[]{\blue{\large $\bullet$}}
\Text(80,75)[]{\blue{\large $\bullet$}}
\DashLine(80,25)(120,25){4}
\DashLine(80,75)(120,75){4}
\Text(60,50)[]{$V$}
\end{picture}
\vspace*{-.8cm}
\end{center}
\centerline{\it Figure 1.18: Diagrams for the one--loop contributions of
fermions and gauge bosons to $\lambda$.} \s 

However, it is for small values of the quartic couplings that the additional
contributions can have a large impact and give some new information. Indeed,
for $\lambda \ll \lambda_t, g_{1},g_{2}$, the RGE  can be approximated by
\beq
\frac{{\rm d} \lambda}{{\rm d log}Q^2} \simeq \frac{1}{16\pi^2} \left[
12 \lambda^2 - 12 \frac{m_{t}^4}{v^4} + \frac{3}{16} \left(2 g_2^4+ 
(g_2^2+g_1^{2})^2 \right) \right] 
\eeq
and its solution, taking again the weak scale as the reference point,
is 
\beq
\lambda(Q^2)=\lambda(v^2)+ \frac{1}{16 \pi^2} \left[-
12 \frac{m_{t}^4}{v^4} + \frac{3}{16} \left(2 g_2^4+ (g_2^2+g_1^{2})^2
\right) \right] {\rm log} \frac{Q^2}{v^2} 
\eeq
If the coupling $\lambda$ is too small, the top quark contribution can be
dominant and could  drive it to a negative value $\lambda(Q^2) <0$, leading to a
scalar potential $V(Q^2) < V(v)$. The vacuum is not stable anymore since it has
no minimum. The stability argument \cite{STABILITY,VACUUMbounds-PR,metas-Strumia} tells us that to have a scalar potential  which is bounded from below and,
therefore, to keep $\lambda (Q^2) >0$, the Higgs boson  mass should be larger
than the value
\beq 
M_H^2 >  \frac{v^2}{8 \pi^2} \left[-
12 \frac{m_{t}^4}{v^4} + \frac{3}{16} \left(2 g_2^4+ (g_2^2+g_1^{2})^2
\right) \right] {\rm log} \frac{Q^2}{v^2} 
\eeq
This puts a strong constraint on the Higgs boson mass, which  depends on the
value of the cut--off $\Lambda_C$. For relatively low and very high 
values for this cut--off, one obtains  
\beq
\Lambda_C \sim 10^{3}~{\rm GeV} &\Rightarrow& M_H \gsim 70~{\rm GeV} \non \\
\Lambda_C \sim 10^{16}~{\rm GeV} &\Rightarrow& M_H \gsim 130~{\rm GeV} 
\eeq
Note, however, that the stability bound on the New Physics scale can be relaxed
if the vacuum is metastable as discussed in Ref.~\cite{metastability}.  Indeed,
the SM effective potential can have a minimum which is deeper than the standard
electroweak minimum if the decay of the latter into the former, via thermal
fluctuations in the hot universe  or quantum fluctuations at zero temperature,
is suppressed. In this case, a lower bound on the Higgs mass follows from the
requirement that no transition between the two vacua occurs and we always
remain in the electroweak minimum. The obtained  bound on $M_H$ is in general
much weaker than in the case of absolute stability of the vacuum and even
disappears if the cut--off of the theory is at the TeV scale\footnote{Note that
the first argument, i.e. thermal fluctuations, relies on several cosmological
assumptions such as that the universe went through a phase of very high
temperature, which has been indirectly tested so far only for temperatures of
the order of a few MeV.  The second argument, quantum tunneling, where the only
cosmological input is the knowledge of the age of the universe which should be
larger than the lifetime of the instability of the vacuum,  gives less severe
bounds; see Ref.~\cite{metas-Strumia} for instance.}.

\subsubsection*{\underline{Higher order effects and combined triviality 
and stability bounds}}

Thus, the positivity and the finiteness of the self--coupling $\lambda$
impose, respectively,  a lower bound $M_H  \gsim 70$ GeV and an upper bound 
$M_H \lsim 1$ TeV, on the SM Higgs boson mass if the cut--off is set to 
${\cal O}(1~{\rm TeV})$. These bounds are only approximative and to have more 
precise ones, some refinements must, however,  be included 
\cite{VACUUMbounds-PR,VACUUMbounds,Riesselman}.\s

Since the $\beta$ functions of all SM couplings have been calculated up to
two loops, they can be included in the analysis. For the scalar sector 
for instance, one has at this order
\beq
\frac{{\rm d} \lambda}{{\rm d log}Q^2} \equiv 
\beta_\lambda = 24 \frac{\lambda^2}{(16 \pi^2)} - 312 \frac{\lambda^3}
{(16 \pi^2)^2} 
\eeq
While, at one--loop, the $\lambda (\mu)$ coupling monotonically increases with 
the scale $\mu$ until it becomes infinite when reaching the Landau pole at the 
scale $\Lambda_C$, at the two--loop--level, it approaches an ultraviolet
fixed--point corresponding to $\beta_\lambda=0$. From the previous equation at
two--loop, the resulting fixed--point value is $\lambda_{\rm FP} = 16 \pi^2 
\times 24 /  312 \simeq 12.1$  [however, top contributions cannot be neglected
and they modify the behavior of this fixed-point.] \s

To obtain the upper bound on $M_H$, we need to choose the cut--off value
for $\lambda$. Since $\lambda_{\rm FP}$ is large and perturbation theory is
lost even before reaching this value, one can choose a value smaller than
$\lambda_{\rm FP}$ as being this cut--off. An   estimate of the stability of
the bound can be made by varying the cut--off value for instance between
$\lambda_{\rm FP}/4$ and $\lambda_{\rm FP}/2$, which lead to two--loop
corrections which are about, respectively,  25\% and 50\%, of the one--loop
result. Therefore, one can consider the first value as leading to a well 
behaved perturbative series and the second value as being at the limit 
where perturbation theory is valid.\s

For the stability bound, one simply requires  that the coupling $\lambda$
remains positive at the cut--off scale, $\lambda(\Lambda_C)>0$. For an accurate
determination of the bound,  this requirement has to be made at the two--loop
level, including matching conditions, i.e. the precise relation between the
physical masses of the gauge bosons and the top quark and their corresponding 
couplings. The most important inputs are the Higgs and top quark masses
\beq
\lambda(\mu) = M_H^2/(2v^2) \times [1+ \delta_H (\mu) ] \ , \ 
\lambda_t (\mu) = \sqrt 2 m_t/v \times [1+ \delta_t (\mu)]
\eeq
Including the theoretical uncertainties by a variation of the cut--off
$\Lambda_C$ from $\lambda_{\rm FP}/2$ to  $\lambda_{\rm FP}/4$ using 
the  matching conditions for the top quark and Higgs boson masses, and the 
experimental errors mainly on $\alpha_s=0.118 \pm 0.002$ and $m_t =175\pm 6$
GeV, one obtains \cite{Riesselman} the modern version of the Roman plot shown
in Fig.~1.19  for the  stability [lower band] and triviality [upper band]
constraints, which give the  allowed range of $M_H$ as a function of the scale
of New Physics $\Lambda_C$ [between the bands]. The width of the bands 
corresponds
to the various experimental and theoretical errors.  As can be seen, if the New
Physics scale $\Lambda_C$ is at the TeV scale, the Higgs boson mass is allowed
to be in the range
\beq  
50~{\rm GeV}~ \lsim M_H \lsim ~800~{\rm GeV}  
\eeq 
while, requiring the SM to be valid up to the Grand Unification scale, 
$\Lambda_{\rm GUT} \sim 10^{16}$ GeV, the Higgs boson mass should lie in the 
range   
\beq  
130~{\rm GeV}~ \lsim M_H \lsim ~180~{\rm GeV}  
\eeq

\begin{figure}[htbp]
\begin{center}
\vspace*{-4.mm}
\epsfig{file=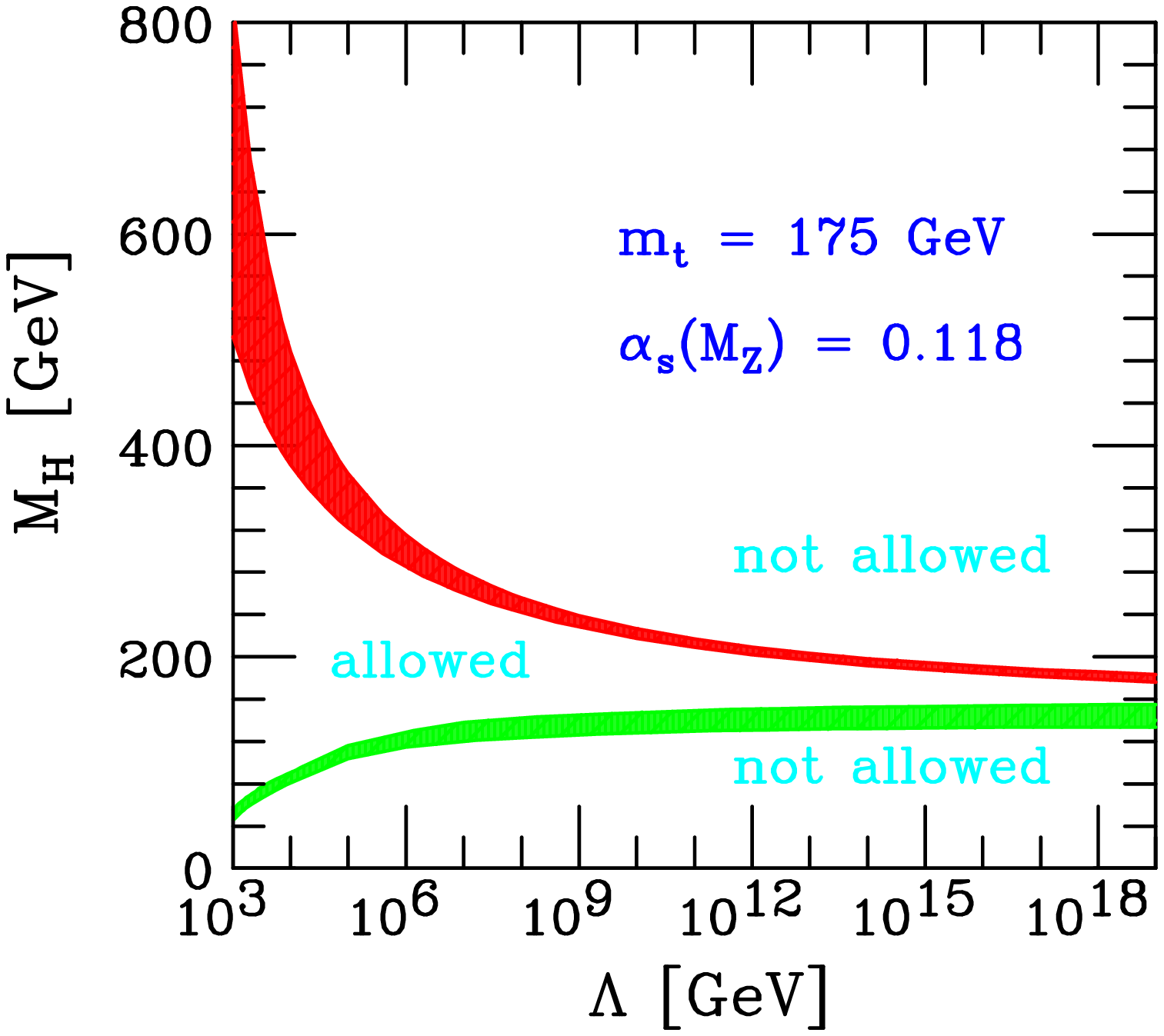,bbllx=18pt,
bblly=180pt,bburx=488pt,bbury=567pt,width=12.6cm}
\end{center}
\vspace*{2mm}
{\it Figure 1.19: The triviality (upper) bound and the vacuum stability (lower)
bound on the Higgs boson mass as a function of the New Physics or cut--off 
scale $\Lambda$ for a top quark mass $m_t=175 \pm 6$ GeV and $\alpha_s (M_Z)
=0.118 \pm 0.002$; the allowed region lies between the bands and the 
colored/shaded bands illustrate the impact of various uncertainties. From 
Ref.~\cite{Riesselman}.} 
\end{figure} 

\subsubsection{The fine--tuning constraint} 

Finally, a last theoretical constraint comes from the fine--tuning problem 
originating from the radiative corrections to the Higgs boson mass. The Feynman
diagrams contributing to the one--loop radiative corrections are depicted in 
Fig.~1.20 and involve Higgs boson, massive gauge boson and fermion loops. 

\begin{center}
\hspace*{-10.5cm}
\begin{picture}(300,100)(0,0)
\SetWidth{1.}
\DashLine(75,50)(125,50){4}
\DashLine(175,50)(225,50){4}
\ArrowArc(150,50)(25,0,180)
\ArrowArc(150,50)(25,180,360)
\Text(150,15)[]{$f$}
\Text(150,85)[]{$\bar{f}$}
\Text(125,50)[]{\blue{\large $\bullet$}}
\Text(175,50)[]{\blue{\large $\bullet$}}
\Text(100,60)[]{\blue{$H$}}
\Text(200,60)[]{\blue{$H$}}
\hspace*{7.5cm}
\DashLine(40,50)(160,50){4}
\DashCArc(100,75)(25,0,360){4}
\Text(100,50)[]{\blue{\large $\bullet$}}
\Text(220,50)[]{\blue{\large $\bullet$}}
\Text(270,50)[]{\blue{\large $\bullet$}}
\DashLine(180,50)(220,50){4}
\DashLine(270,50)(310,50){4}
\DashCArc(245,50)(25,0,360){4}
\Text(52,75)[]{$W,Z,H$}
\Text(247,85)[]{$W,Z,H$}
\end{picture}
\vspace*{-.5cm}
\end{center}
{\it Figure 1.20: Feynman diagrams for the one--loop corrections to the SM 
Higgs  boson mass.}
\vspace*{4mm}

Cutting off the loop integral momenta at a scale $\Lambda$, and keeping only
the dominant contribution in this scale, one obtains 
\beq
M_H^2 = (M_H^0)^2 + \frac{3 \Lambda^2}{8 \pi^2 v^2} \left[
M_H^2 + 2 M_W^2 + M_Z^2 - 4 m_t^2 \right]
\label{MHdivergences}
\eeq
where $M_H^0$ is the bare mass contained in the unrenormalized Lagrangian and
where we retained only the contribution of the top heavy quark for the fermion
loops. This is a completely new situation in the SM: we have a quadratic
divergence rather than the usual logarithmic ones. If the cut--off $\Lambda$ is
very large, for instance of the order of the Grand Unification scale $\sim
10^{16}$ GeV, one needs a very fine arrangement  of ${16}$ digits between
the bare Higgs mass and the radiative corrections to have a physical Higgs
boson mass in the range of the electroweak symmetry breaking scale, $M_H \sim
100$ GeV to 1 TeV, as is required for the consistency of the SM.  This is the
naturalness of fine--tuning problem\footnote{Note, however that the SM is a
renormalizable theory and this cancellation can occur in a mathematically
consistent way by choosing a similarly divergent counterterm.  Nevertheless,
one would like to give a physical meaning to this scale $\Lambda$ and view it
as the scale up to which the SM is valid.}. \s

However, following Veltman \cite{Veltman-conjecture}, one can note that by 
choosing the Higgs mass to be 
\beq
M_H^2 = 4 m_t^2 - 2M_W^2 - M_Z^2 \sim (320~{\rm GeV})^2
\eeq
the quadratic divergences can be canceled and this would be even a prediction 
for the Higgs boson mass. But the condition above was given only at the  
one--loop level and at higher orders, the general form of the correction to
the Higgs  mass squared reads \cite{Two-loop-FT1,Two-loop-FT2}
\beq
\Lambda^2 \sum_{n=0}^{\infty} c_n (\lambda_i) \log ^n (\Lambda/Q)
\eeq
where $(16\pi^2) c_0= (3/2  v^2)(M_H^2 + 2 M_W^2 + M_Z^2 - 4 m_t^2)^2$ and
the remaining coefficients $c_n$ can be calculated recursively from the
requirement that $M_H^2$ should not depend on the renormalization scale $Q$.
For instance, for the two--loop coefficient, one finds \cite{Two-loop-FT1}
\beq
(16 \pi^2)^2 c_1 &=& \lambda (114 \lambda -54 g_2^2- 18 g_1^2 +72 \lambda_t)^2 
+ \lambda_t^2 (27 g_2^2 + 17 g_1^2 + 96 g_s^2 -90 \lambda_t^2) \non \\
&& - \frac{15}{2} g_2^4 + \frac{25}{2} g_1^4 + \frac{9}{2} g_1^2 g_2^2
\eeq
The higher--order coefficients have more powers of $1/(16\pi^2)$ and should
therefore be more and more suppressed.  The Veltman condition requires that the
fine cancellation occurs to all perturbative orders, i.e. for any value 
of $n$.  Given the fact that the various $c_n$ terms of the 
perturbative series are independent, there is obviously no solution for $M_H$.\s

A priori, one can then conclude that the Veltman condition is not useful and
cannot solve the fine--tuning problem. However, as it has been discussed in
Refs.~\cite{Kolda+Murayama,CasasFT}, this is only true if the scale of New
Physics is extremely large. For scales not much larger that the electroweak
scale, one does not need very large cancellations. For instance, at the
one--loop level, the fine--tuning problem appears only if $\Lambda \gsim 4\pi v
\sim 2$ TeV. If the Veltman solution is by chance satisfied, then the scale
$\Lambda$ can be pushed at the two loop level to a much higher value,
$\Lambda^2 \log \Lambda \gsim (16 \pi^2)^2 v^2$, that is, for $\Lambda \sim 15$
TeV. If again the Veltman conjecture is satisfied, then the three--loop
quadratic divergences start to be problematic only at a scale $\Lambda \gsim 50$
TeV.  One can thus have almost no, or only a small amount of fine--tuning, up
to rather high scales. \s

For such a scale, one simply needs to manage such that $\sum_{n=0}^{1} c_n 
(\lambda_i) \log ^n (\Lambda/M_H)=0$ at two--loop. It appears that first, such 
a solution exists and second, that the predicted $M_H$ value  becomes cut--off
dependent. As mentioned previously, this prediction assumes exact cancellation
and this is not required for rather low scales $\Lambda$. Following again
Ref.~\cite{Kolda+Murayama}, a more adequate condition would be
\beq
\sum_{n=0}^{1} c_n (\lambda_i) \log ^n (\Lambda/M_H) \lsim v^2/\Lambda^2
\eeq
and if it is satisfied, the fine--tuning might be acceptable. But, as is well
known, there is a problem with the definition of the amount of fine--tuning, 
that is largely a subjective matter. Following again Ref.~\cite{Kolda+Murayama},
one can define it as the sensitivity  of the electroweak scale to
the cut--off $\Lambda$, $\Delta M_W^2 (\Lambda) /M_W^2$. This leads then
to the measure
\beq
\Delta_{\rm FT}= \left| \frac{\Delta M_W^2}{M_W^2} \right| = 
\left| \frac{ \Delta M_H^2} {M_H^2} \right| = \frac{2 \Lambda^2}{M_H^2}
\left| \sum_n c_n \log^n (\Lambda/M_H) \right| 
\eeq
For a given value of $\Delta_{\rm FT}$, the weak scale is fine--tuned to one
part in $\Delta_{\rm FT}$: the larger than unity is the value of $\Delta_{\rm
FT}$, the more fine--tuning we have and there is no fine--tuning if
$\Delta_{\rm FT} \leq 1$.  One can see from the previous equation that the
fine--tuning is large not only when $\Lambda$ increases but also when the Higgs
boson is light. \s

The Higgs boson mass is shown in Fig.~1.21 as a function of the maximal value
of the cut--off scale $\Lambda$. Also shown, are the regions not allowed by the
triviality and stability bounds on $M_H$, as well as the (``electroweak") area
ruled out by high--precision measurements\footnote{More details on how these
constraints have been obtained can be found in Ref.~\cite{Kolda+Murayama}.}. 
The regions of fine--tuning less than 10 and 100 are given, respectively, by
the light and dark hatched regions. The white region corresponds to the one
where all constraints are fulfilled and where the Veltman condition is
approximately satisfied. \s

For low values of the scale, $\Lambda \lsim 1$ TeV,  there is no fine--tuning 
problem for any reasonable Higgs boson mass value. But as $\Lambda$ increases,  
the range of Higgs masses where the fine--tuning is smaller than 10\% or 1\%
becomes narrow. For instance, with $\Lambda \sim 3$ TeV, the Higgs boson mass 
must be above $\sim 150$ GeV while with $\Lambda \sim 10$ TeV, only a narrow
range around $M_H \sim 200$ GeV for $ \Delta_{\rm FT}=10$, sometimes called the
Veltman throat, is allowed. For even higher scales, only the line with 
$M_H \sim 200$ GeV, where the Veltman condition is approximately satisfied, 
survives. 

\begin{figure}[htbp]
\begin{center}
\epsfxsize=4.2truein
\vspace*{-3mm}
\hspace*{0in}
\epsffile{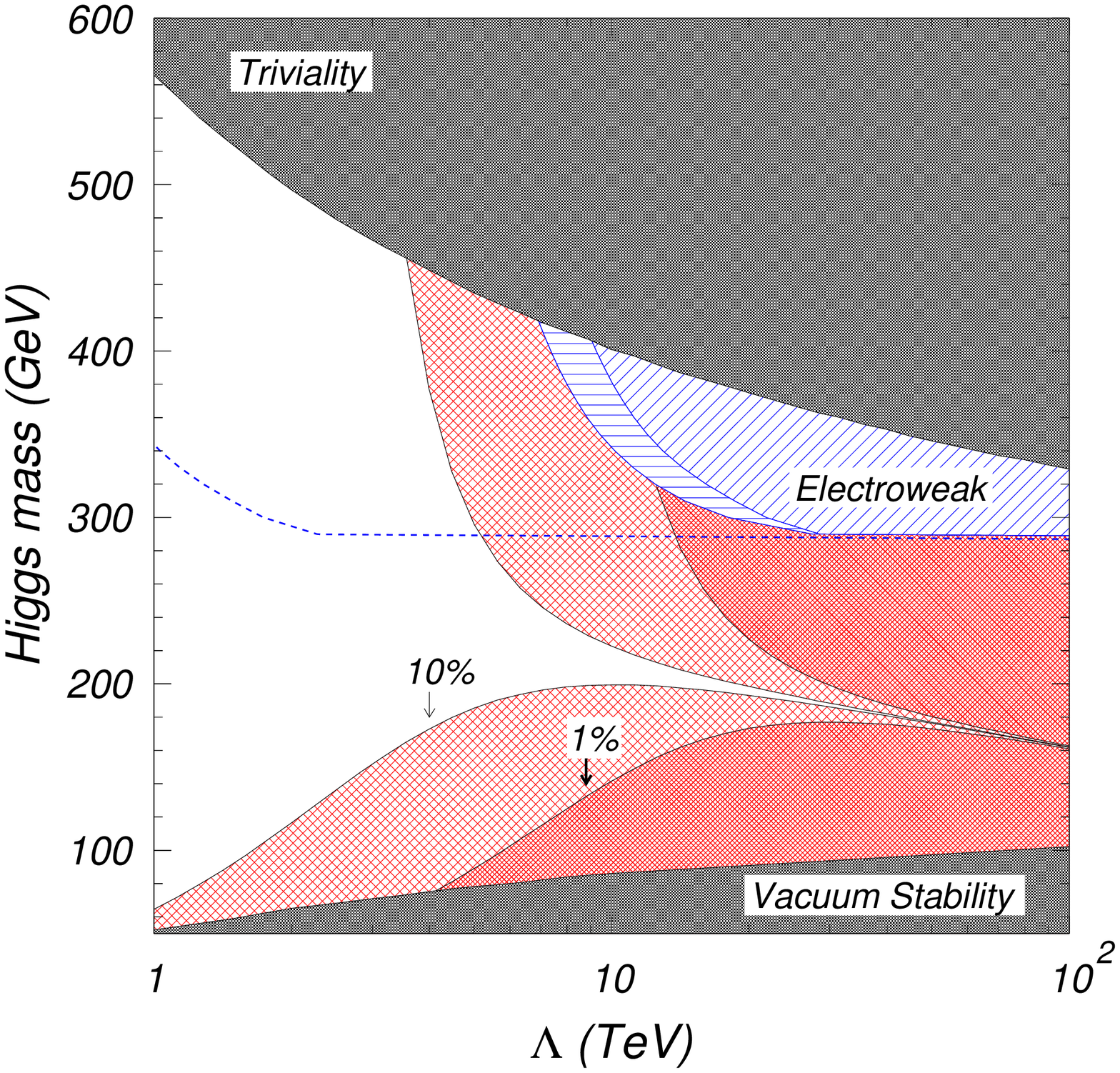}
\end{center}
\vspace*{-5mm}
{\it Figure 1.21: The contours for the fine--tuning parameter $\Delta_{\rm FT}$
in the  plane $(M_H, \Lambda)$. The dark (light) hatched region marked ``1\%" 
(``10\%") represents fine--tunings of greater than 1 part in 100 (10). The 
constraints from triviality, stability and electroweak precision data are also 
shown. The empty region is consistent with all constraints and has  
$\Delta_{\rm FT}$ less than 10\%. From Ref.~\cite{Kolda+Murayama}.}
\label{all-murayama}
\vspace*{-3mm}
\end{figure}

Thus, one can obtain a very useful information by considering the fine--tuning
problem in the SM at scales of a few tens of TeV. In the vicinity of these 
scales, a Higgs boson with a mass $M_H \sim 200$ GeV can still allow for an 
acceptable amount of fine--tuning. 
 
\newpage

\section{Decays of the SM Higgs boson}
\setcounter{equation}{0}
\renewcommand{\theequation}{2.\arabic{equation}}

In the Standard Model, once the Higgs mass is fixed, the profile of the
Higgs particle is uniquely determined. The Higgs couplings to gauge bosons 
and fermions are directly proportional to the masses of the particles and
the Higgs boson will have the tendency to decay into the heaviest ones 
allowed by phase space. Since the pole masses of the gauge bosons and 
fermions are known [the electron and light quark masses are too small to be
relevant]
\beq 
M_Z=91.187~{\rm GeV} \ , \ M_W=80.425~{\rm GeV} \ , \ m_\tau=1.777~{\rm GeV} 
\ , \  m_\mu= 0.106~{\rm GeV} \, , \non \\ 
m_t=178 \pm 4.3~{\rm GeV} \ , \
m_b=4.88 \pm 0.07~{\rm GeV} \ , \
m_c=1.64 \pm 0.07~{\rm GeV} \hspace*{1cm}
\label{allmasses}
\eeq
all the partial widths for the Higgs decays into these particles can be 
predicted.\s 

The decay widths into massive gauge bosons $V=W,Z$ are directly proportional 
to the $HVV$ couplings, which in the SM are given in terms of the fields by 
\begin{eqnarray}
{\cal L}(HVV)&=&\left(\sqrt2G_\mu\right)^{1/2}M_V^2HV^\mu V_\mu 
\label{HVVcp}
\end{eqnarray}
These are S--wave couplings and even under parity and charge conjugation,
corresponding to the $J^{\rm PC}=0^{++}$ assignment of the Higgs  spin and
parity quantum numbers. The decay widths into fermions are proportional to
the $H f\bar{f}$ couplings which are of the scalar type
\beq
g_{H\bar{f}f} \propto \frac{m_f}{v} =  (\sqrt{2}G_\mu)^{1/2} m_f 
\label{Hffcp}
\eeq
In this section, we will discuss all the decay modes of the SM Higgs boson: 
decays into quarks and leptons, into real or virtual gauge bosons and loop 
induced decays into photons [including $Z\gamma$ final states] and gluons, and
summarize the important QCD and electroweak radiative corrections to 
these processes. \s 

The $J^{\rm PC}=0^{++}$ quantum numbers of the SM Higgs particle lead also to
unique predictions for the angular and energy distributions of the partial
decay widths.  Whenever possible, we will confront these properties with those
of an  hypothetical CP--odd Higgs particle\footnote{The decays of the Higgs
bosons \cite{Hdecay-Effective,Eff-decays,Eff-Romao} in the general case of anomalous Higgs couplings 
\cite{HVV-Effective,Hff-Effective} will be discussed in another part of this 
review.}, 
that we will denote by $A$, and  which is predicted in many extensions of the 
SM. In this case, the Higgs coupling to vector gauge bosons is a P--wave 
coupling corresponding to the $J^{\rm PC}=0^{+-}$ assignment and, if CP 
symmetry is conserved, does not occur at the tree--level and is only induced by 
higher loop effects.  With $\eta$ being a dimensionless factor, the effective 
point--like coupling can be written as 
\begin{eqnarray}
{\cal L}(AVV)= {1 \over 4} \eta \left(\sqrt2 G_\mu \right)^{1/2}M_V^2 A
V^{\mu\nu} \widetilde V_{\mu\nu} \ , \ \ \ \ \widetilde V^{\mu\nu}= 
\epsilon^{\mu\nu\rho
\sigma} V_{\rho\sigma}
\label{AVVcp}
\end{eqnarray}
In the presence of fermions, the couplings of the $A$ boson are of the 
pseudoscalar type
\beq
g_{A\bar{f}f} \propto \frac{m_f}{v} \gamma_5 =  (\sqrt{2}G_\mu)^{1/2} m_f 
\gamma_5
\label{Affcp}
\eeq

\subsection{Decays to quarks and leptons} 

\subsubsection{The Born approximation}

In the Born approximation, the partial width of the Higgs boson decay into
fermion pairs, Fig.~2.1, is given by \cite{HffBorn,EGN}
\beq
\Gamma_{\rm Born} (H \ra f \bar{f})= \frac{G_\mu N_c}{4 \sqrt{2} \pi} \,  
M_H \, 
m_f^2 \, 
\beta_f^3
\eeq
with $\beta=(1- 4m_f^2/M_H^2)^{1/2}$ being the velocity of the fermions in the 
final state and $N_c$ the color factor $N_c=3\, (1)$ for quarks (leptons). In
the lepton case, only decays into $\tau^+ \tau^-$ pairs and, to a much lesser 
extent, decays into muon pairs are relevant.

\vspace*{-8mm}
\begin{center}
\hspace*{6cm}
\begin{picture}(300,100)(0,0)
\SetWidth{1.}
\SetScale{1.2}
\DashLine(0,50)(40,50){4}
\ArrowLine(40,50)(70,75)
\ArrowLine(40,50)(70,25)
\Text(49,60)[]{{\blue{\large $\bullet$}}}
\Text(27,70)[]{\blue{$H$}}
\Text(90,80)[]{$f$}
\Text(90,40)[]{$\bar{f}$}
\end{picture}
\end{center}
\vspace*{-1.4cm}
\centerline{\it Figure 2.1: The Feynman diagram for the Higgs boson decays into 
fermions.}
\vspace*{1mm}

The partial decay widths exhibit a strong suppression near threshold, $\Gamma
(H \ra f \bar{f}) \sim \beta_f^3 \to 0$ for $M_H \simeq 2m_f$. This is typical 
for the decay of a Higgs  particle with a scalar coupling eq.~(\ref{Hffcp}).
If the Higgs boson were a pseudoscalar $A$ boson  with couplings given in 
eq.~(\ref{Affcp}), the partial decay width would have been suppressed only by a 
factor $\beta_f$ \cite{MSSMcplgs} 
\beq
\Gamma_{\rm Born} (A \ra f \bar{f})= \frac{G_\mu N_c}{4 \sqrt{2} \pi} \, M_H \, 
m_f^2 \, \beta_f
\eeq
More generally,  and to anticipate the discussions that we will have on the 
Higgs CP--properties, for a $\Phi$ boson with mixed CP--even and CP--odd 
couplings $g_{\Phi \bar{f}f} \propto a+ ib\gamma_5$, the differential
rate for the fermionic decay $\Phi (p_+) \to f (p,s) \bar f (\bar p, \bar s)$
where $s$ and $\bar s$ denote the polarization vectors of the fermions and 
the four--momenta are such that $p_\pm=p\pm \bar p$, is given by
[see Ref.~\cite{Hff-spin-cor} for instance]
\beq
\frac{ {\rm d} \Gamma}{ {\rm d} \Omega } (s,\bar s) &=& \frac{\beta_f}{64 
\pi^2 M_\Phi} \bigg[ ( |a|^2 + |b|^2) \bigg( {1 \over 2} M_\Phi^2 - m_f^2 +
m_f^2 s \! \cdot \! \bar s \bigg) \non \\ 
&& \hspace*{1.4cm} + (|a|^2 - |b|^2) \bigg( p_+ \! \cdot \! s \, p_+
\! \cdot \! \bar s- {1 \over 2} M_\Phi^2 s \! \cdot \! \bar s+ m_f^2 
s \! \cdot \! \bar  s - m_f^2 \bigg) \non \\
&& \hspace*{1.4cm}  - {\rm Re}(ab^*) \epsilon_{\mu \nu \rho \sigma} p_+^\mu 
p_-^\nu s^\rho \bar s^\sigma -2 {\rm Im}(ab^*) m_f p_+ \! \cdot \! (s+\bar s) 
\bigg]
\label{dGammaHff}
\eeq 
The terms proportional to ${\rm Re}(ab^*)$ and ${\rm Im}(ab^*)$ represent the
CP--violating part of the couplings. Averaging over the polarizations of the
two fermions, these two terms disappear and we are left with  the two 
contributions $\propto \frac{1}{2} |a|^2 (M_\Phi^2 -2m_f^2 - 2m_f^2)$ 
and $\propto \frac{1}{2} |b|^2 (M_\Phi^2 -2m_f^2+ 2m_f^2)$ which reproduce
the $\beta^3_f$ and $\beta_f$ threshold behaviors of the pure CP--even ($b=0)$ 
and CP--odd ($a=0)$ states noted above.

\subsubsection{Decays into light quarks and QCD corrections}

In the case of the hadronic decays of the Higgs boson, the QCD corrections turn
out to be quite  large and, therefore, must be included. At the one--loop level,
the Feynman diagrams for the corrections are shown in Fig.~2.2: one has to
include gluon--exchange [which multiplies the Born term] and the emission of a
gluon in the final state [which has to be squared and added to the former]. In
the limit where $M_H$ is much larger than the quark masses, $M_H \gg 2m_f$, one
obtains for the next--to--leading order (NLO) decay width [the quark mass is
kept only in the Yukawa coupling and in the leading logarithmic term]
\cite{HqqQCD-1loop,Drees+Hikasa}
\beq 
\Gamma_{\rm NLO} (H \ra q \bar{q}) \simeq \frac{3G_\mu}{4 \sqrt{2} \pi} \, M_H 
\, m_q^2 \, \left[ 1+ \frac{4}{3} \frac{\alpha_s}{\pi} \left( \frac{9}{4} + 
\frac{3}{2} \log  \frac{m_q^2}{M_H^2} \right) \right]
\eeq
As can be seen, there is a large logarithmic ${\rm log}(m_q/M_H)$ contribution
which, for very light quarks, might render the partial decay width very small 
and even drive it to negative values [a definitely not physical situation].
However, these large logarithms  can be absorbed in the redefinition of the
quark masses: by using the running quark masses in the $\overline{\rm MS}$
scheme at the scale of the Higgs mass, as discussed in \S1.1.4, these logarithms
are summed to all orders in the strong interaction coupling constant 
\cite{HqqQCD-1loop}. \s

\begin{center}
\hspace*{-5cm}
\begin{picture}(300,100)(0,0)
\SetWidth{1.}
\SetScale{1.2}
\DashLine(0,50)(40,50){4}
\Text(49,60)[]{{\blue{\large $\bullet$}}}
\ArrowLine(40,50)(90,75)
\ArrowLine(40,50)(90,25)
\GlueArc(65,68)(10,0,230){3}{5}
\Text(20,70)[]{\blue{$H$}}
\Text(112,85)[]{$q$}
\Text(112,40)[]{$\bar{q}$}
\Text(85,70)[]{$g$}
\hspace*{.8cm}
\DashLine(100,50)(140,50){4}
\Text(169,60)[]{{\blue{\large $\bullet$}}}
\ArrowLine(140,50)(170,75)
\ArrowLine(140,50)(170,25)
\Gluon(170,25)(170,75){3.2}{6.5}
\ArrowLine(170,25)(200,25)
\ArrowLine(170,75)(200,75)
\Text(216,60)[]{$g$}
\Text(324,60)[]{{\blue{\large $\bullet$}}}
\DashLine(230,50)(270,50){4}
\ArrowLine(270,50)(310,75)
\ArrowLine(270,50)(310,25)
\Gluon(300,69)(310,43){3}{3.5}
\Text(374,80)[]{$g$}
\Text(100,60)[]{$+$}
\Text(250,60)[]{$+$}
\Text(400,60)[]{$+ \  \cdots $}
\end{picture}
\vspace*{-11mm}
\end{center}
\centerline{\it Figure 2.2: Generic diagrams for the one--loop QCD corrections 
to Higgs decays into quarks.}
\vspace*{3mm}

Including the ${\cal O}(\alpha_{s}^{2})$ \cite{HqqQCD-2loop} and  ${\cal O}
(\alpha_{s}^{3} )$ \cite{HqqQCD-3loop} QCD radiative corrections, the partial 
Higgs  decay widths into light quarks can be then written as
\beq
\Gamma (H \to q\bar{q}) &=& 
\frac{3 G_\mu}{4\sqrt{2}\pi} \,  M_H \, \overline{m}_q^2 (M_H) \,  
\bigg[ 1+ \Delta_{qq} +  \Delta^2_H \bigg]
\eeq
with the running quark mass $\overline{m}_{q} (M_H^2)$ and the strong coupling 
constant $\bar \alpha_s \equiv \alpha_s(M_H^2)$  both defined at the scale 
$M_{H}$. In the ${\overline{\rm MS}}$ renormalization scheme, with $N_f$ the 
number of light quark flavors, one has for the QCD correction factor 
$\Delta_{qq}$ 
\beq
\Delta_{qq}= 5.67 \frac{\bar \alpha_s}{\pi} + (35.94 - 
1.36 N_f) \frac{\bar \alpha_s^2}{\pi^2} + (164.14 - 25.77 N_f + 
0.26 N_f^2) \frac{\bar \alpha_s^3}{\pi^3} \ \ 
\label{Delta-qq}
\eeq
Since the values of the running $b$ and $c$ quark masses at the scale $\mu \sim
M_H =100$ GeV are typically, respectively, a factor $\sim 1.5$ and a factor of
$\sim 2$ smaller than the pole masses, the partial decay widths are suppressed
by large factors compared to the case where the pole masses are used. This is
shown in Fig~2.3 where $\Gamma(H \to b\bar{b})$ and $\Gamma(H\to c\bar{c})$ are
displayed as functions of the Higgs mass $M_H$ in the Born approximation, using
only the running quark masses and with the full set of QCD corrections
implemented. Note that the latter increase the partial widths by approximately 
20\%. \s

The additional correction at ${\cal O}(\alpha_s^2)$ involves logarithms of the 
masses of the light quarks and the heavy top quark and is given by 
\cite{HqqQCD-2mass}
\beq
\Delta_H^2 &=&  \frac{\bar \alpha_s^2 }{\pi^2} \, \left( 1.57 - \frac{2}{3} 
\log \frac{M_H^2}{m_t^2} + \frac{1}{9} \log^2 \frac{\overline{m}_q^2}
{M_H^2} \right) 
\label{Delta-Sqq}
\eeq
Because of chiral symmetry, all this discussion holds true if the
Higgs particle were a pseudoscalar boson; the only exception is that the
correction $\Delta_H^2$ would be different, since it involves the quark masses 
which break the symmetry.


\begin{figure}[htbp]
\begin{center}
\vspace*{-2.8cm}
\hspace*{-3cm}
\epsfig{file=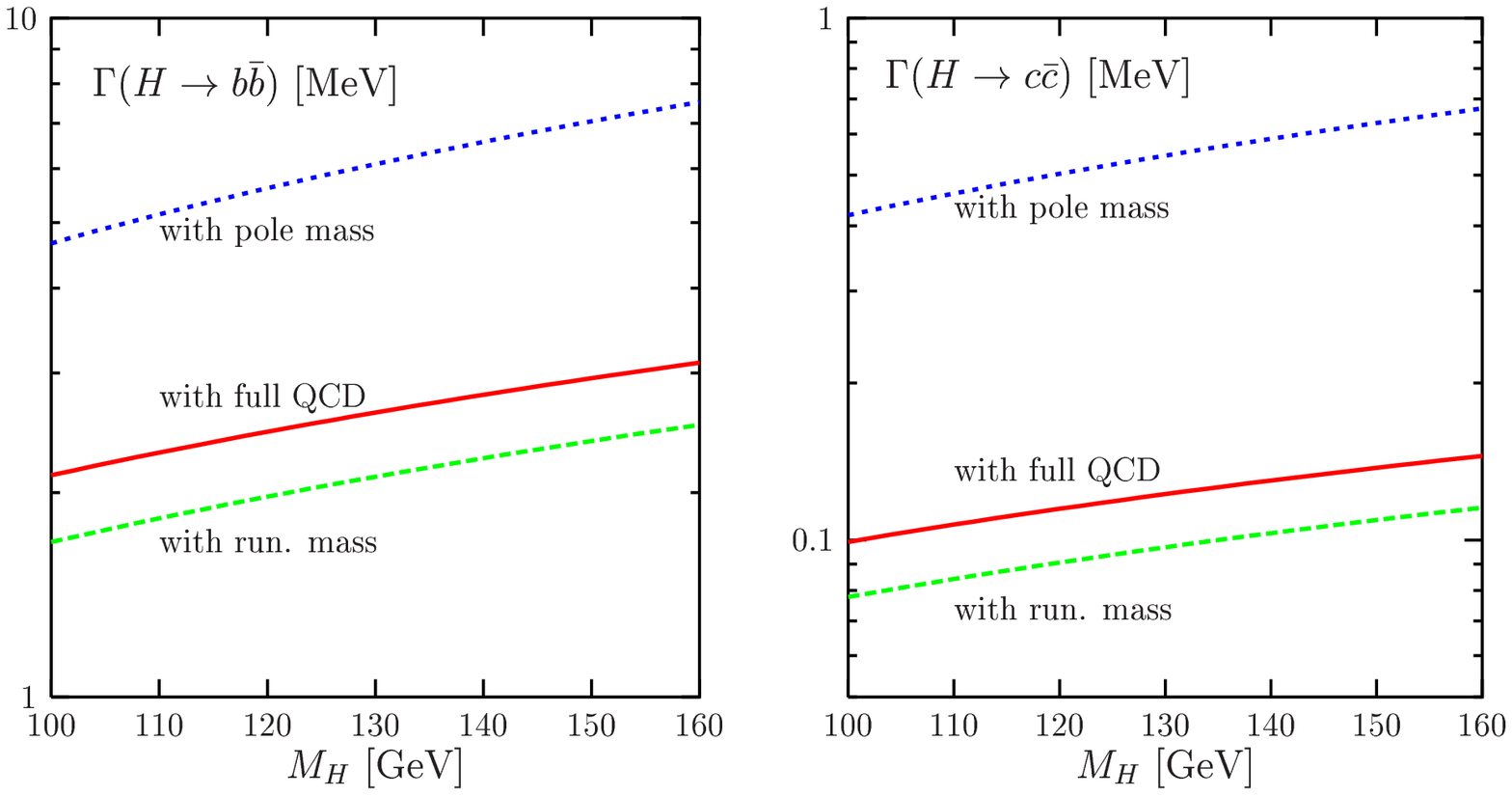,width=18.cm} 
\end{center}
\vspace*{-14.6cm}
\nn {\it Figure 2.3: The partial widths for the decays $H \to b\bar{b}$
(left) and $H \to c\bar{c}$ (right) as a function of $M_H$. They are shown 
in the Born approximation (dotted lines), including only the running quark
masses (dashed lines) and with the full set of QCD corrections (solid lines).
The input pole masses are $m_b=4.88$ GeV and $m_c=1.64$ GeV and the running 
strong coupling constant is taken at the scale of the Higgs mass and is 
normalized to $\alpha_s(M_Z)=0.1172$.}
\end{figure} 
\vspace*{-6mm}

\subsubsection{The case of the top quark}

For Higgs bosons decaying into top quarks, the QCD corrections do not lead to
large logarithms since $m_t$ is comparable to $M_H$. However, 
these corrections can be sizable, in particular near the threshold $M_H 
\sim 2m_t$. At next--to--leading--order, they are given by 
\begin{eqnarray}
\Gamma (H \ra t\bar{t}\,)= \frac{3 G_\mu}{4 \sqrt{2} \pi} \,
\, M_H \,  m_t^2 \, \beta_t^3 \, \left[ 1 +\frac{4}{3} \frac{\alpha_s}{\pi}
\Delta_H^t  (\beta_t) \right]
\end{eqnarray}
Using the Spence function defined by Li$_2(x)=  -\int_0^x dy y^{-1} \log(1-y)$,
the QCD correction factor in the massive case reads 
\cite{HqqQCD-1loop,Drees+Hikasa,AD+PG}
\begin{eqnarray}
\Delta_H^t(\beta) = \frac{1}{\beta}A(\beta) + \frac{1}{16\beta^3}(3+34\beta^2-
13 \beta^4)\log \frac{1+\beta}{1-\beta} +\frac{3}{8\beta^2}(7 \beta^2-1) 
\label{eq:dqcdmass}
\end{eqnarray}
with 
\begin{eqnarray}
A(\beta) &= & (1+\beta^2) \left[ 4 {\rm Li}_2 \left( \frac{1-\beta}{1+\beta}
\right) +2 {\rm Li}_2 \left( -\frac{1-\beta}{1+\beta} \right) -3 \log
\frac{1+\beta}{1-\beta} \log \frac{2}{1+\beta} \right. \non \\
& & \left. -2 \log \frac{1+\beta}{1-\beta} \log \beta \right] -
3 \beta \log \frac{4}{1-\beta^2} -4 \beta \log \beta 
\label{Abeta-tt}
\end{eqnarray}
Part of the full massive two--loop corrections, i.e. corrections  of ${\cal O}
(N_f \alpha_s^2)$ which are expected to provide the largest contribution, have 
been computed some time ago \cite{HqqQCD-massive1} and the full two--loop 
corrections have been derived slightly after \cite{HqqQCD-massive2}. \s

The left--hand side of Fig.~2.4 shows the partial $H \to t\bar{t}$ decay width
in the Born approximation, with the running top quark mass and including the
full set of one--loop QCD corrections. As can be seen, and contrary to the 
$b\bar b$ and $c\bar c$ cases, the corrections are rather moderate in
this case.\s

\begin{figure}[htbp]
\begin{center}
\vspace*{-2.7cm}
\hspace*{-3cm}
\epsfig{file=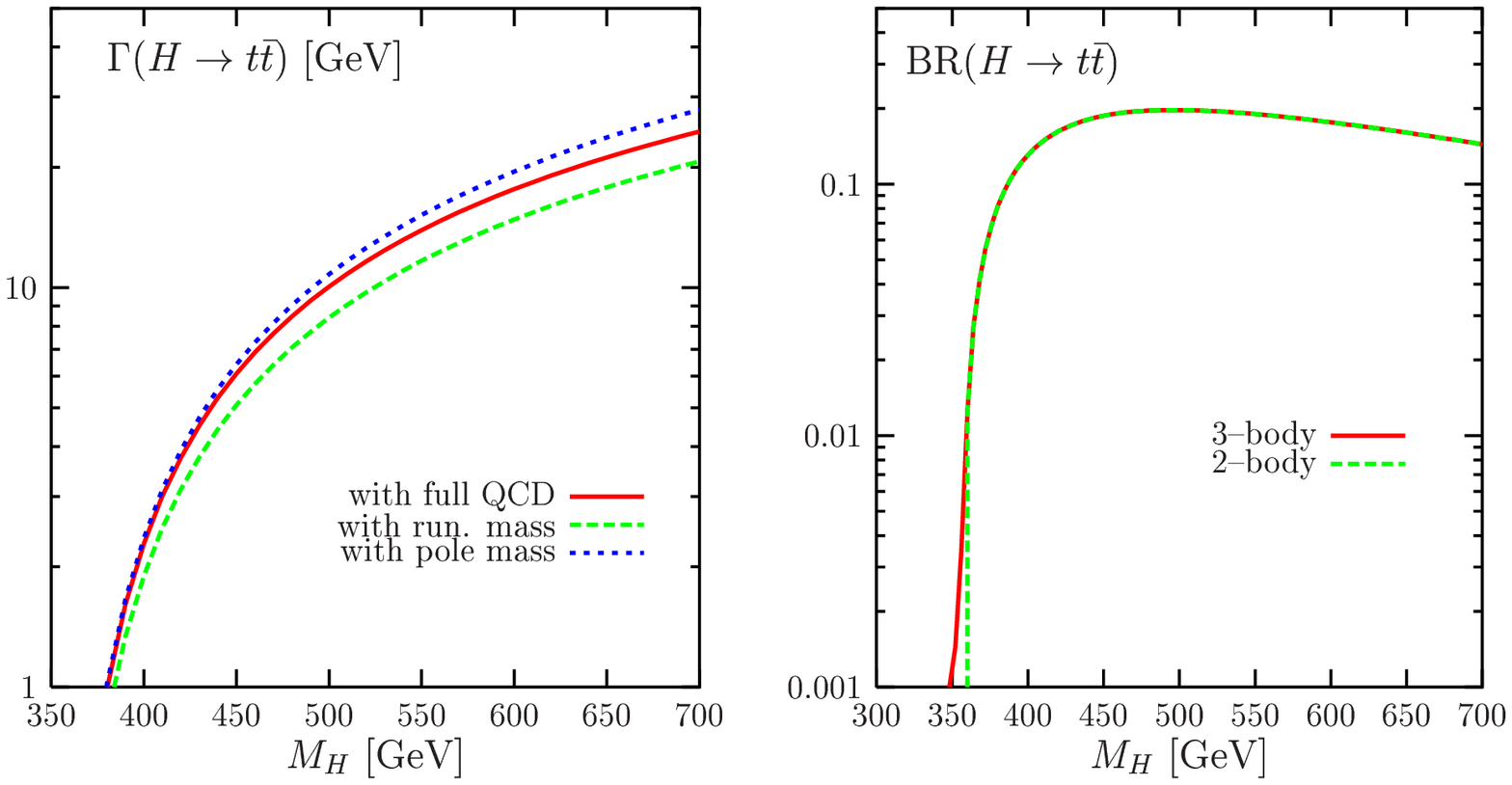,width=18.cm} 
\end{center}
\vspace*{-14.6cm}
\nn {\it Figure 2.4: The partial width for the decay $H \to t\bar{t}$ as a 
function of $M_H$. In the left figure, it is shown  in the Born approximation
(dotted line), with the running top mass (dashed lines) and with the full set 
of QCD corrections (solid lines). In the right figure the partial width is 
shown with (solid line) and without (dashed line) the inclusion of the 
three--body decay. The inputs are $m_t=178$ GeV and $\alpha_s(M_Z)=0.1172$.}
\end{figure}
\vspace*{-1mm}

Another special feature in the case of top quarks is that the three--body 
decays $H\to t\bar  t^* \to t\bar b W^-$ into on--shell and off--shell top 
states are possible \cite{Three-Body2,Three-Body,Three-Body1}, see Fig.~2.5. 
These three--body  decays reach the  percent level slightly 
below the $2m_t$ threshold, when compared to the two--body decay as shown 
in the right--hand side of Fig.~2.4. A smooth transition from below to above 
threshold occurs when the top quark width is included. \s

\begin{center}
\hspace*{-8.5cm}
\begin{picture}(300,100)(0,0)
\SetWidth{1.}
\SetScale{1.2}
\DashLine(100,50)(140,50){4}
\ArrowLine(140,50)(170,75)
\ArrowLine(140,50)(170,25)
\ArrowLine(170,25)(200,10)
\Photon(170,25)(200,45){3}{5.5}
\Text(125,90)[]{\red{${\bf a)}$}}
\Text(170,60)[]{\bb}
\Text(140,70)[]{\blue{$H$}}
\Text(195,75)[]{$t$}
\Text(195,48)[]{$\bar{t}$}
\Text(245,15)[]{$b$}
\Text(248,50)[]{$W$}
\hspace*{6.5cm}
\DashLine(100,50)(140,50){4}
\Photon(140,50)(170,75){3}{5.5}
\Photon(140,50)(170,25){3}{5.5}
\ArrowLine(170,25)(200,15)
\ArrowLine(170,25)(200,45)
\Text(170,60)[]{\bb}
\Text(125,90)[]{\red{${\bf b)}$}}
\Text(140,70)[]{\blue{$H$}}
\Text(205,75)[]{$W$}
\Text(205,45)[]{$W$}
\Text(245,25)[]{$t$}
\Text(245,50)[]{$\bar{b}$}
\end{picture}
\vspace*{-6mm}
\end{center}
\centerline{\it Figure 2.5: Diagrams for the three--body decays of the Higgs 
boson into $tbW$ final states.}
\vspace*{3mm}

Taking into account only the diagram of Fig.~2.5a where the top quark is
off--shell and which provides the dominant contribution [the virtuality of the 
$W$ boson in the other diagram is too large, thus strongly suppressing the
contribution], the differential partial width or Dalitz density for this 
decay can be written as 
\begin{equation}
\frac{{\rm d} \Gamma}{{\rm d}x_1 {\rm d} x_2} (H\to t\bar{t}^*\to t 
\bar{b}W^-) = \frac{3G_\mu^2} {32\pi^3} \, M_H^3 \, m_t^2 \, \frac{\Gamma_H^t}
{y_1^2 + \gamma_t \kappa_t}
\label{eq:httdalitz}
\end{equation}
with the reduced energies $x_{1,2}=2E_{t,b}/M_H$, the scaling variables
$y_{1,2} = 1-x_{1,2}$, $\kappa_i = M_i^2/M_H^2$ and the reduced decay width
of the virtual top quark $\gamma_t=\Gamma_t^2/M_H^2$. The squared amplitude 
is given by \cite{Three-Body}
\begin{eqnarray}
\Gamma_H^t & = & y_1^2(1-y_1-y_2+\kappa_W-5\kappa_t) + 2\kappa_W(y_1y_2-\kappa_W
-2\kappa_ty_1+4\kappa_t\kappa_W) \nonumber \\
& & -\kappa_ty_1y_2+\kappa_t(1-4\kappa_t)(2y_1+\kappa_W+\kappa_t) 
\end{eqnarray} 
The differential decay width has to be integrated over the allowed range of the
$x_1, x_2$ variables. The boundary condition is 
\begin{equation}
\left| \frac{2(1-x_1-x_2+\kappa_t+\kappa_b-\kappa_W) + x_1x_2}
{\sqrt{x_1^2-4\kappa_t} \sqrt{x_2^2-4\kappa_b}} \right| \leq 1 
\label{eq:dalitzbound}
\end{equation}
The additional diagram leading to the same final state,  with the Higgs boson
decaying into two $W$ bosons with one of them being off--shell and
decays into $t\bar{b}$ final states, $H \to WW^* \to t\bar{b}W$, 
gives very small contributions and can be safely neglected. \s

\subsubsection{Distinction between scalar and pseudoscalar Higgs bosons}

The distinction between a scalar and a pseudoscalar Higgs particle can be made 
by investigating the angular
correlations in the decays into heavy fermions 
\cite{CPHff2,Bargeretal,CPHff1,CP-ee-HZChang,CP-H-Pois,CP-GunionGrz}.
In the processes $H/A \to t\bar{t} \to (W^+b)(W^-\bar{b}$), denoting the spin
vector of the $t$ and $\bar{t}$ states in their respective rest frames by $s$
and $\bar{s}$, and orienting the $z$ axis along the $t$ flight direction, the
spin dependence is different in the two cases; from eq.~(\ref{dGammaHff}) one
obtains \cite{Bargeretal}
\beq
\Gamma (H/A \to t\bar{t}) \propto 1 -s_z \bar{s}_z \pm s_\perp \bar{s}_\perp
\eeq
Denoting by $\theta^*_\pm$ the polar angle between the $W^\pm$ bosons and the  
$t$--quark in the $W^\pm$ rest frames and by $\phi^*$ the relative azimuthal 
angle between the decay planes of the two $W$ bosons, Fig.~2.6, and using the 
abbreviations $c_{\theta_+^*}=\cos\theta_+^*$ {\it etc}, the angular  
distributions of the $W^\pm$ bosons in the decays of scalar and  pseudoscalar 
Higgs particles are given by \cite{CPHff1,Kuhn-Wagner}
\beq
\frac{ {\rm d}\Gamma( H/A \to W^+ W^- b\bar{b} )}{ \Gamma_{H/A}
{\rm d}c_{\theta_+^*} {\rm d} c_{\theta_-^*} {\rm d}\phi^*}= \frac{1}{8\pi}
\left[ 1 + \left(\frac{m_t^2-2M_W^2}{m_t^2+2M_W^2}\right)^2 \left( 
c_{\theta_+^*} c_{\theta_-^*} \mp s_{\theta_+^*} s_{\theta_-^*} 
c_{ \phi^*} \right) \right]
\label{Httangular}
\eeq
\vspace*{-6mm}
\begin{figure}[htbp]
\vspace*{-9mm}
\begin{center}
\begin{picture}(550,200)(0,0)
\SetWidth{1.1}
\Line(230,100)(310,100)
\LongArrow(310,100)(317,100)
\Line(230,100)(150,100)
\LongArrow(150,100)(143,100)
\DashLine(320,100)(410,100){5}
\DashLine(140,100)(50,100){5}
\Line(230,170)(230,100)
\Line(230,50)(230,30)
\Line(230,100)(230,50)
\Line(230,170)(410,170)
\Line(230,30)(410,30)
\Line(410,170)(410,30)
\Line(205,160)(230,100)
\DashLine(230,100)(255,40){5}
\Line(25,160)(75,40)
\Line(25,160)(205,160)
\Line(75,40)(230,40)
\DashLine(230,40)(255,40){5}
\SetWidth{0.8}
\LongArrowArc(320,100)(23,0,51.34)
\LongArrowArcn(140,100)(23,180,141.34)
\LongArrowArc(245,100)(45,110,130)
\SetWidth{1.1}
\Text(230,100)[]{\Large\color{blue} $\bullet$}
\Text(222,88)[]{\large\color{blue} $H$}
\Text(270,88)[]{\large\color{red} $t$}
\Text(190,88)[]{\large\color{red} $\bar t$}
\Text(327,130)[c]{\color{red} $W^+$}
\Text(310,70)[c]{\large\color{red} $b$}
\Text(129,131)[c]{\color{red} $W^-$}
\Text(155,70)[c]{\large\color{red} $\bar{b}$}
\Text(353,112)[c]{\color{black} $\theta^*_+$}
\Text(106,112)[c]{\color{black} $\theta^*_-$}
\Text(240,140)[c]{\large\color{black} $\phi^*$}
\Photon(140,100)(90,140){4}{5.5}
\LongArrow(140,100)(190,60)
\Photon(320,100)(360,150){4}{5.5}
\LongArrow(320,100)(280,50)
\end{picture}
\end{center}
\vspace*{-1.2cm}
{\it Figure 2.6: The definition of the polar angles ${\theta^*_\pm}$  and the
azimuthal angle $\phi^*$ for the sequential decay $H \rightarrow t \bar{t}
\rightarrow (b W^+) (\bar{b}W^-)$. The polar angles are defined in the 
$t, \bar{t}$ rest frames, with respect to the $t$ flight direction. The angle
$\phi^*$ stays the same after boost along the $t\bar{t}$ directions.}
\end{figure}
\vspace*{-4mm}

[The QCD corrections to the angular distributions can be found in 
Ref.~\cite{Hff-spinQCD} for instance]. 
If the Higgs boson mass is precisely known, the Higgs rest frame can be 
reconstructed. Because the boost of the Higgs boson to quarks is not too large
and the mass ratio between daughter--to--parent parent particles in the decay
is significant, the kinematical reconstruction of the full event should not be
very difficult. \s

If the integral over the polar angles is performed, one obtains a simple 
asymmetry in the azimuthal angle which projects out the parity of the Higgs 
boson \cite{Bargeretal,CPHff1}
\beq
\frac{1}{\Gamma_{H/A}} \frac{ {\rm d}\Gamma( H/A \to W^+ W^- b\bar{b}) }{ {\rm 
d}\phi^*}= \frac{1}{2\pi} \left[ 1 \mp \frac{\pi^2}{16} \left(\frac{m_t^2-2
M_W^2}{m_t^2+2M_W^2} \right)^2 c_{\phi^*}  \right]
\eeq
allowing to determine the azimuthal angle up to a two--fold ambiguity. The
distribution of the decays $H/A \to t\bar{t} \to b\bar{b}W^+W^-$ as a 
function of the azimuthal angle is shown in Fig.~2.7. One sees that the 
separation between the scalar and pseudoscalar cases can clearly be made.\s

\begin{figure}[!h]
\begin{center}
\vspace*{-2.4cm}
\hspace*{-3cm}
\epsfig{file=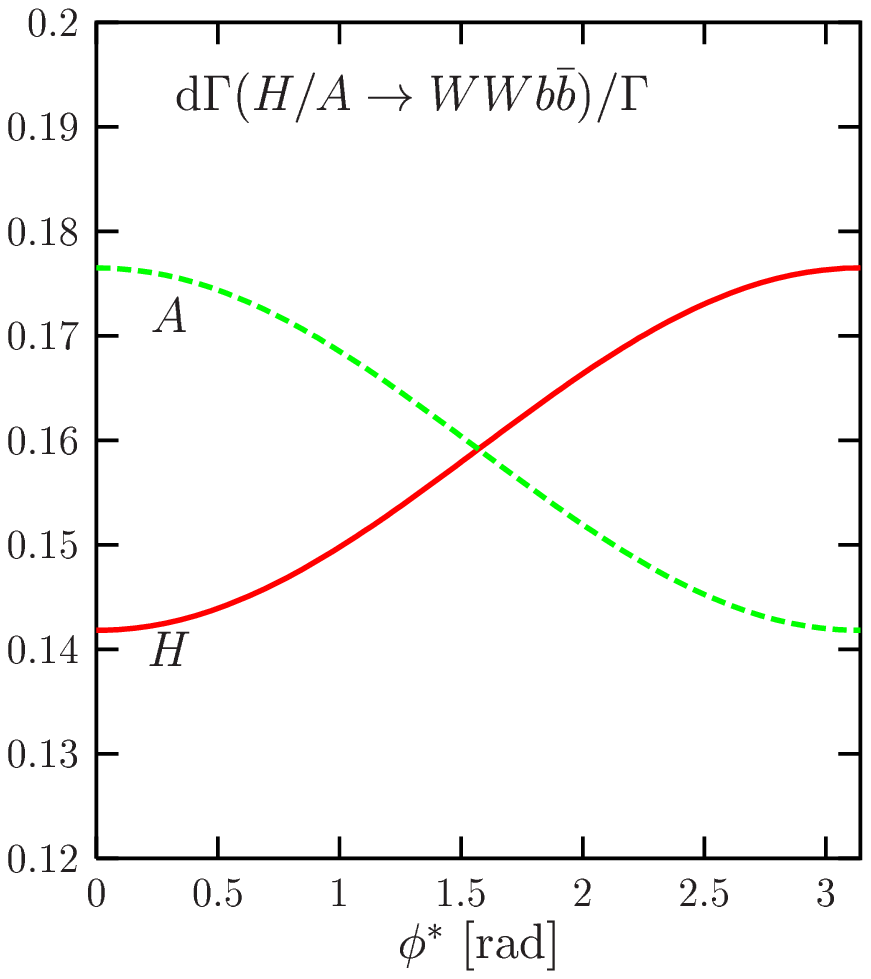,width=15.cm} 
\end{center}
\vspace*{-12.3cm}
\centerline{\it Figure 2.7: Distribution of the decays  $H/A \to t\bar{t} \to 
b\bar{b}W^+ W^-$ in the azimuthal angle $\phi^*$.}
\end{figure}
\vspace*{-.1cm}

One can perform the same study when integrating over the $b$--quark directions
and consider the $W$ bosons decaying into leptons $W^\pm \to \ell^\pm \nu_e$. 
The angular distribution is still given by eq.~(\ref{Httangular}) but with
$\theta_\pm^*$ denoting this time the polar angles between the charged leptons
and the top quarks in the rest frame of the latter, and with the mass factor
$(m_t^2-2M_W^2)^2/(m_t^2+2M_W^2)^2$ omitted. \s

CP quantum number studies of the Higgs particles can also be performed for
smaller Higgs masses, in the decays into light fermions. In the case of
$b\bar{b}$ final state decays [which are dominant for relatively light Higgs
bosons], it is unfortunately very difficult, because of depolarization effects,
to extract the spin information of the bottom quark. A much cleaner channel is
provided by the Higgs decays into $\tau^+ \tau^-$ pairs
\cite{CPHff1,CPHff4,CPHff3}, although the rates are suppressed by an order of
magnitude compared to the $b\bar{b}$ case. A possible channel would be the
decays $H/A \to \tau^+ \tau^- \to \pi^+ \bar{\nu} \pi^- \nu$. \s

Defining again the polar angles $\theta_\pm^*$ as those giving the $\pi^\pm$
and $\tau^-$ directions and the azimuthal angle $\phi^*$ as the angle between
the decays planes of $\tau^\pm$, the angular distribution will be as in the
case of $H/A \to t\bar{t} \to WW b\bar{b}$ with $W^\pm \to \ell^\pm \nu_e$ 
\cite{CPHff1}
\beq
\frac{1}{\Gamma_{H/A}} \frac{ {\rm d}\Gamma( H/A \to \pi^+\bar{\nu}_\tau \pi^- 
\nu_\tau )}{ {\rm d} c_{\theta_+^*} {\rm d}c_{\theta_-^*} {\rm d}\phi^*}= 
\frac{1}{8\pi}\left[ 1 +  c_{\theta_+^*} c_{\theta_-^*} \mp s_{\theta_+^*} 
s_{\theta_-^*} c_{\phi^*} \right]
\label{Htauangular}
\eeq
leading, once the polar angles are integrated out, to an asymmetry in the 
azimuthal angle 
\beq
\frac{{\rm d}\Gamma_{H/A} }{\Gamma_{H/A} {\rm d} \phi^*} = \frac{1}{2\pi} 
\left[ 1 \mp \frac{\pi^2}{16} \ c_{\phi^*} \right]
\eeq
The asymmetry is shown in Fig.~2.8 and the distinction between the scalar and
pseudoscalar cases is even easier than in the case of top quarks in Fig.~2.7, 
since the suppression factor $(m_t^2-2M_W^2)^2/(m_t^2+2M_W^2)^2$ is absent. \s

\begin{figure}[!h]
\begin{center}
\vspace*{-2.4cm}
\hspace*{-3cm}
\epsfig{file=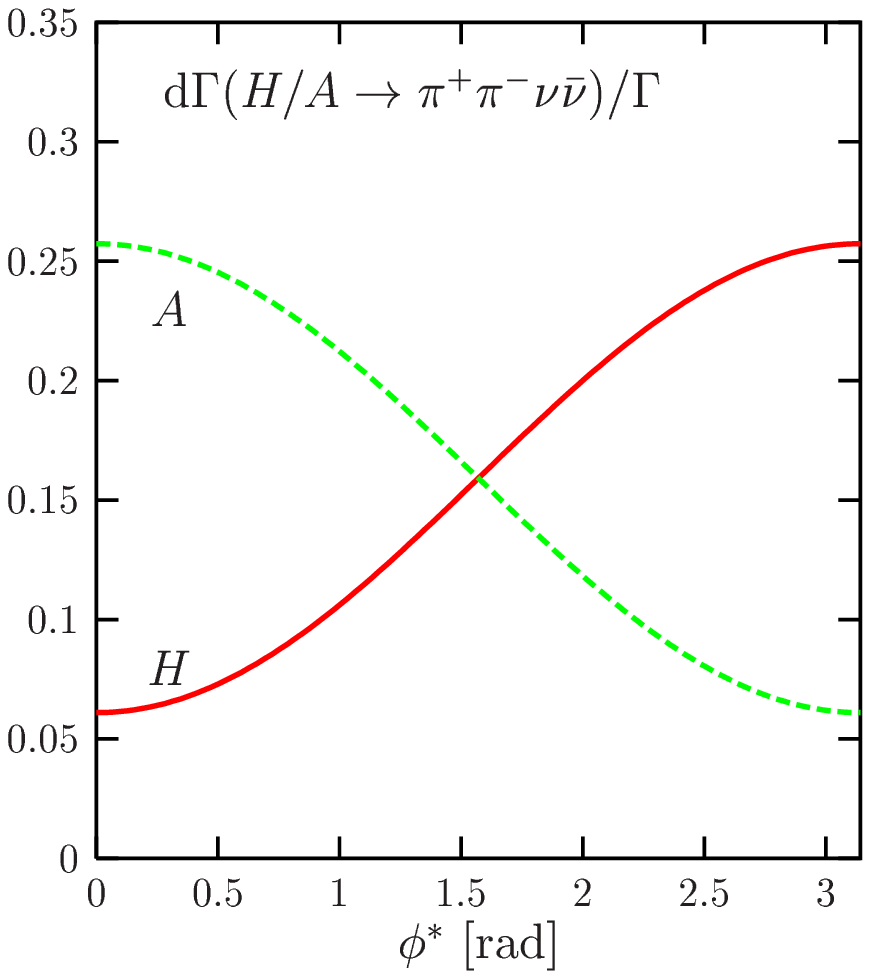,width=15.cm} 
\end{center}
\vspace*{-12.3cm}
\nn {\it Figure 2.8: Distribution of the decays  $H/A \to \tau^+ \tau^- \to 
\pi^+\bar{\nu}_\tau \pi^- \nu_\tau$ in the azimuthal angle $\phi^*$.}
\end{figure}

An observable which is sensitive to the Higgs parity is the angle $\delta$ 
between the pions in the rest frame of the Higgs boson 
\cite{CPHff1,Kuhn-Wagner,CPHff4,CPHff3}
\beq
16 \vec{\pi}^+ \cdot \vec{\pi}^- =  M_H^2 \left[ (1 + \beta_\tau \beta_\pi 
c_{\theta^*_-} )^2- 16 \frac{m_\pi^2}{M_H^2} \right]^{\frac{1}{2}}
\left[ (1 - \beta_\tau \beta_\pi c_{\theta^*_+})^2- 16 \frac{m_\pi^2}
{M_H^2}\right]^{\frac{1}{2}} \cos \delta
\eeq
where $\beta_\tau = (1- 4m_\tau^2/M_H^2)^{1/2}$ and $\beta_\pi = (m_\tau^2
-m_\pi^2)/(m_\tau^2+m_\pi^2)$ are the rest frame boosts of, respectively, 
the Higgs to the $\tau$--lepton and the $\tau$--lepton to the pions. The 
azimuthal angle $\phi^*$ can be then written in terms of the angles 
$\theta_\pm^*$ and $\delta$ and, integrating over the polar angles, one obtains 
for the distributions a rather complicated function of $\delta$. However, 
for $\delta=\pi$, the distributions  are rather simple and very different for 
$0^{++}$ and $0^{+-}$ states. For a scalar Higgs boson decay, it reaches its 
maximum for $\delta=\pi$ 
\beq
\frac{1}{\Gamma_H} \frac{ {\rm d}\Gamma( H) }{ {\rm d} \cos
\delta} \simeq \frac{2}{15} \frac{5+\beta_\tau^2}{1-\beta_\tau^2} 
\eeq
while for a pseudoscalar Higgs boson, it peaks at a small value of $\pi-\delta$
for $m_\pi \sim 0$ 
\beq
\frac{1}{\Gamma_A} \frac{ {\rm d}\Gamma(A)}{{\rm d} \cos \delta} \simeq (1+\cos 
\delta) \frac{1}{20} \frac{5+10\beta_\tau^2+\beta_\tau^4}{(1-\beta_\tau^2)^2} 
\eeq
The analysis for Higgs decays into multi--pion final states, such as  $H/A \to
\tau^+ \tau^- \to \rho^+ \bar{\nu}_\tau \rho^- \nu_\tau $ $\to \pi^+  \pi^0
\bar{\nu}_\tau \pi^- \pi^0 \nu_\tau$  follows the same line if the hadron 
system is treated as a single particle; see Refs.~\cite{CPHff1,CPHff3} for 
more details.

\subsection{Decays into electroweak gauge bosons} 

\subsubsection{Two body decays}

Above the $WW$ and $ZZ$ kinematical thresholds, the Higgs boson will decay 
mainly into pairs of massive gauge bosons; Fig.~2.9a. The decay widths are 
directly proportional to the $HVV$ couplings given in eq.~(\ref{HVVcp}) which, 
as discussed in the beginning of this chapter, correspond to the $J^{\rm PC}=
0^{++}$ assignment of the SM Higgs boson spin and parity quantum numbers. These 
are S--wave couplings, $\sim \vec\epsilon_1\cdot\vec\epsilon_2$ in the 
laboratory frame, and linear in $\sin \theta$, with $\theta$ being the angle 
between the Higgs and one of the vector bosons.\s

\begin{center}
\hspace*{-2.5cm}
\begin{picture}(300,100)(0,0)
\SetWidth{1.}
\SetScale{1.2}
\DashLine(-20,50)(20,50){4}
\Photon(20,50)(50,75){3.}{5}
\Photon(20,50)(50,25){3.}{5}
\Text(-25,90)[]{\red{${\bf a)}$}}
\Text(25,60)[]{\bb}
\Text(0,70)[]{\blue{$H$}}
\Text(65,75)[]{$V$}
\Text(65,39)[]{$V$}
\hspace*{-1cm}
\DashLine(100,50)(140,50){4}
\Photon(140,50)(170,75){3}{5}
\Photon(140,50)(170,25){3}{5}
\ArrowLine(170,25)(200,15)
\ArrowLine(170,25)(200,45)
\Text(173,60)[]{\bb}
\Text(125,90)[]{\red{${\bf b)}$}}
\Text(140,70)[]{\blue{$H$}}
\Text(210,78)[]{$V$}
\Text(245,25)[]{$f$}
\Text(245,50)[]{$\bar{f}$}
\DashLine(230,50)(270,50){4}
\Photon(270,50)(300,75){3}{4}
\Photon(270,50)(300,25){3}{4}
\ArrowLine(300,25)(330,15)
\ArrowLine(300,25)(330,35)
\ArrowLine(300,75)(330,85)
\ArrowLine(300,75)(330,55)
\Text(325,60)[]{\bb}
\Text(275,90)[]{\red{${\bf c)}$}}
\Text(299,70)[]{\blue{$H$}}
\Text(410,40)[]{$f_3$}
\Text(410,20)[]{$\bar{f}_4$}
\Text(410,100)[]{$f_1$}
\Text(410,80)[]{$\bar{f}_2$}
\end{picture}
\vspace*{-6mm}
\end{center}
\centerline{\it Figure 2.9: Diagrams for the Higgs boson decays into real and/or
virtual gauge bosons.}
\vspace*{3mm}

The partial width for a Higgs boson decaying into two real gauge bosons, 
$H \to VV$ with $V=W$ or $Z$, are given by \cite{LQT,HffBorn}
\beq
\Gamma (H \ra VV) = \frac{G_\mu M_{H}^3}{16 \sqrt{2} \pi} \, \delta_V \,
\sqrt{1-4x} \, (1-4x +12x^2) \ , \ \ x= \frac{M_V^2}{M_{H}^2} 
\label{HVV-2body}
\eeq
with $\delta_{W}=2$ and $\delta_Z =1$. For large enough Higgs boson masses, 
when the phase space factors can be ignored, the decay width into $WW$ bosons is
two times larger than the decay width into $ZZ$ bosons and the branching 
ratios for the decays would be, respectively, 2/3 and 1/3 if no other decay
channel is kinematically open. \s

For large Higgs masses, the vector bosons are longitudinally polarized
\cite{Bargeretal}
\begin{eqnarray}
\frac{\Gamma_L}{\Gamma_L+\Gamma_T} =  \frac{1-4x+4x^2}{1-4x+12x^2} 
\stackrel{M_H \gg M_V} \longrightarrow 1
\end{eqnarray}
while the ${\small L,T}$ polarization states are democratically populated near 
the threshold, at $x=1/4$. Since the longitudinal wave functions are linear  in
the energy, the width grows as the third power of the Higgs mass,  $\Gamma (H
\to VV) \propto M_H^3$. As discussed in \S1.4.1, a heavy Higgs boson would be 
obese since its total decay width becomes comparable to its mass
\beq
\Gamma (H \to WW+ZZ) \sim 0.5~{\rm TeV} \, [ M_H/1~{\rm TeV}]^3 
\eeq
and behaves hardly as a resonance.

\subsubsection{Three body decays} 

Below the $WW/ZZ$ kinematical thresholds, the Higgs boson decay modes into
gauge bosons, with one of them being off--shell, Fig.~2.9b, are also important.
For instance, from $M_{H}  \gsim 130$ GeV, the Higgs  boson decay into  $WW$
pairs with one off--shell $W$ boson, starts to dominate over the  $H \ra
b\bar{b}$ mode.  This is due to the fact that in these three--body decays,
although suppressed by an additional power of the  electroweak coupling squared
compared to the dominant $H \to b\bar{b}$ case and by the virtuality of the
intermediate vector boson state,  there is a compensation since the Higgs
couplings to $W$ bosons are much larger than the Higgs Yukawa coupling to $b$
quarks.\s

The partial width for the decay $H \to VV^* \to V f\bar{f}$, the charges of the
vector bosons $V$ summed over and assuming massless fermions, is given by 
\cite{HVV-3body}
\begin{eqnarray} 
\Gamma (H \ra VV^*) = \frac{3 G_\mu^2 M_V^4}{16 \pi^3} M_H \delta_V' R_T(x) 
\label{HVV-3body}
\end{eqnarray} 
with $\delta'_W=1$, $\delta_Z' = \frac{7}{12} - \frac{10}{9}
\sin^2\theta_W+ \frac{40}{9}\sin^4\theta_W$ and 
\begin{eqnarray}
R_T(x) & = & \frac{3(1-8x+20x^2)}{(4x-1)^{1/2}} \arccos \left( \frac{3x-1}
{2x^{3/2}} \right) -\frac{1-x}{2x} (2-13x+47x^2) \non \\ && 
- \frac{3}{2}(1-6x+4x^2) \log x 
\label{HVVrt}
\end{eqnarray}

The invariant mass ($M_*$) spectrum of the off--shell vector boson peaks close 
to the kinematical maximum corresponding to zero--momentum of the on--shell and 
off--shell final state bosons 
\begin{eqnarray}
\frac{{\rm d} \Gamma (H \to VV^*)}{{\rm d}M_*^2} = \frac{3G_\mu^2 M_V^4} {16 \pi^3M_H}
\delta_V' \ \frac{\beta_V (M_H^4\beta_V^2 + 12M_V^2 M_*^2)}{(M_*^2-M_V^2)^2+
M_V^2 \Gamma_V^2} 
\label{dGHVV*}
\end{eqnarray}
with $\beta_V^2=[1-(M_V+M_*)^2/M_H^2][1-(M_V-M_*)^2/M_H^2]$.
Since both $V$ and $V^*$ preferentially have small momenta, the
transverse and longitudinal polarization states are populated with almost equal
probabilities. Neglecting the widths of the vector bosons, $\Gamma_V$, one finds
after summing over all $M_*$ values 
\begin{eqnarray}
{\Gamma_L \over \Gamma_L+\Gamma_T}= {R_L(M_V^2/M_H^2) \over R_T(M_V^2/M_H^2)}
\end{eqnarray}
where $R_T$ is given in eq.~(\ref{HVVrt}) and $R_L$ reads \cite{Bargeretal} 
\begin{eqnarray}
R_L(x)&=&\frac{3-16x+20x^2} {(4x-1)^{1/2}} \arccos \left( \frac{3x-1}{2x^{3/2}}
\right) -\frac{1-x}{2x} (2-13x+15x^2) \non \\ &&
- \frac{1}{2}(3-10x+4x^2) \log x
\end{eqnarray}
[Note that for heavy Higgs bosons, the three--body modes $H \to W^+W^- Z$ 
and $H \to t\bar{t}Z$ have been considered \cite{Three-Body2,Three-Body3}; 
they lead to marginal branching ratios.]
 
\subsubsection{Four body decays}

In fact, even Higgs decays into two off--shell  gauge bosons, Fig.~2.9c, can be
relevant \cite{HVV-4body,HVV-4bodyAll}; see also Ref.~\cite{Eff-Romao}. The 
branching ratios for the latter reach the percent level for
Higgs masses above about 100 (110) GeV for both $W\, (Z)$ boson pairs 
off--shell. For higher masses, it is sufficient to allow for one off--shell 
gauge boson only. The decay width can be cast into the compact form 
\cite{HVV-4body}
\begin{eqnarray} 
\Gamma(H \ra V^*V^*) = \frac{1}{\pi^2}\int_0^{M_{H}^2} \hspace*{-0.4cm}  
\frac{{\rm d} q_1^2 M_V \Gamma_V}{(q_1^2 - M_V^2)^2 + M_V^2 \Gamma_V^2} 
\int_0^{(M_{H}-q_1)^2} \hspace*{-1cm} \frac{{\rm d}q_2^2 M_V \Gamma_V} {(q_2^2  
- M_V^2)^2 + M_V^2 \Gamma_V^2} \, \Gamma_0 
\label{HV*V*}
\end{eqnarray} 
with $q_1^2, q_2^2$ being the squared invariant masses of the virtual gauge 
bosons, $M_V$ and $\Gamma_V$ their masses and total decay widths, and in terms 
of $\lambda(x,y;z)=(1-x/z-y/z)^2-4xy/z^2$ with $\delta_V =2(1)$ for $V=W(Z)$,
the  matrix element  squared $\Gamma_0$ is 
\begin{eqnarray} 
\Gamma_0 = \frac{G_\mu M_{H}^3}{16\sqrt{2}\pi} \delta_V 
\sqrt{\lambda(q_1^2,q_2^2;M_{H}^2)} \left[ \lambda(q_1^2,q_2^2;M_{H}^2) +
\frac{12 q_1^2q_2^2}{ M_{H}^4} \right]  
\end{eqnarray}
Taking into account the total decay width of the vector bosons in the
denominators of eq.~(\ref{HV*V*}), this expression for the four--body decay 
mode can be in fact used to reproduce the partial widths of the two--body 
and three--body decay modes, once the thresholds are crossed. Fig.~2.10 
shows the branching ratios for the decays $H \to WW$ and $H \to ZZ$ in the 
three cases of two--body,  three--body and four--body modes. \s

\begin{figure}[htbp]
\begin{center}
\vspace*{-2.7cm}
\hspace*{-3cm}
\epsfig{file=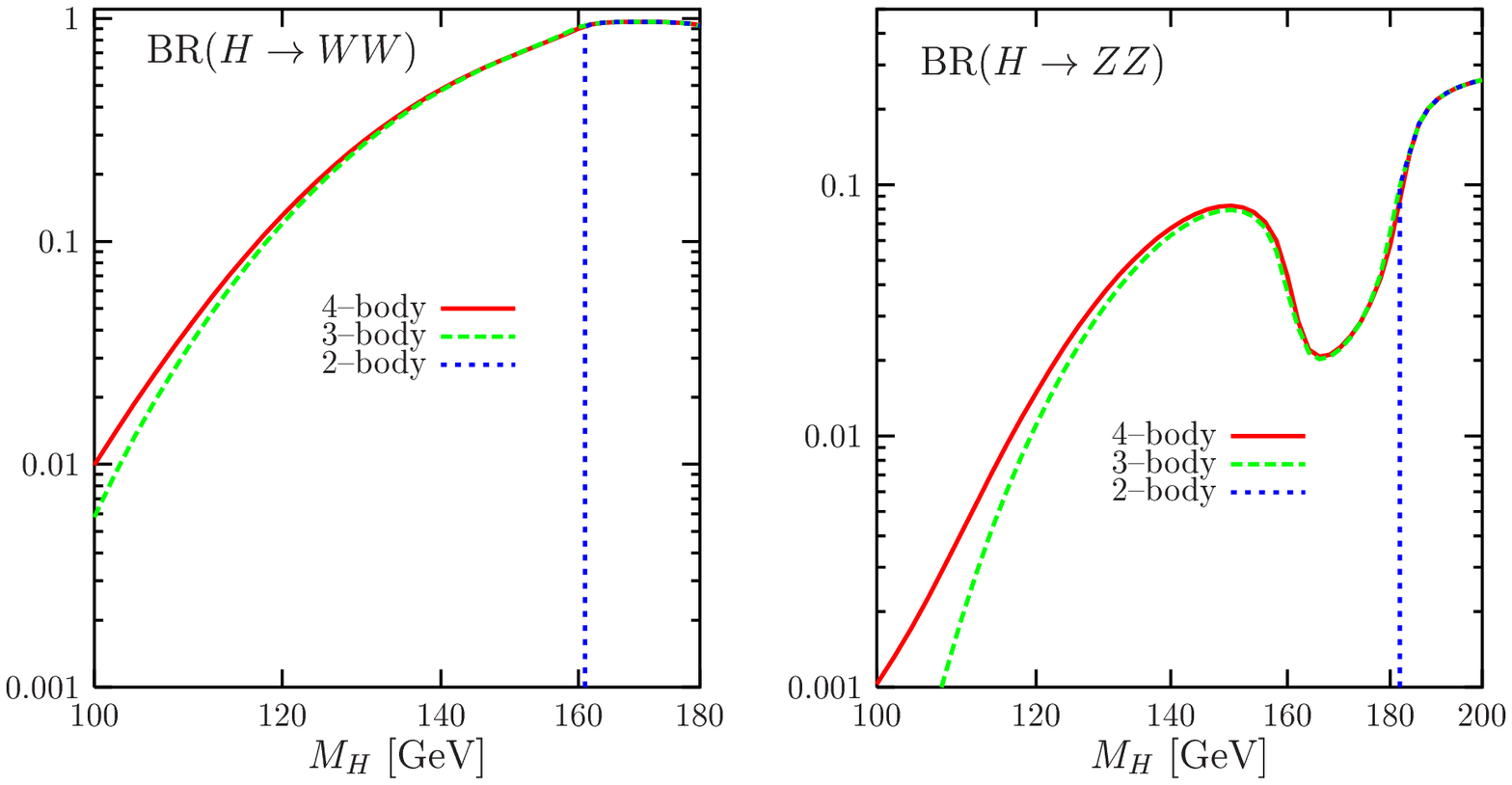,width=18.cm} 
\end{center}
\vspace*{-14.6cm}
\nn {\it Figure 2.10: The branching ratios for the decays $H \to W^+W^-$  (left)
and $ZZ$ (right) as a  function of $M_H$ at the two-- (dotted), 
three-- (dashed) and four--body  (solid) levels. }
\end{figure}

\subsubsection{CP properties and comparison with the CP--odd case}

Let us now confront the angular distributions of the final state fermions in 
the decay processes $H/A \rightarrow  V \ V^* \rightarrow  (f_1 \bar{f}_2) (f_3 
\bar{f}_4)$, which are different for a CP--even and a CP--odd Higgs particle
\cite{CPHVVold,Bargeretal,CP-Gunion-GP,CPHVVchoi}. Denoting the polar and 
azimuthal angles of
the fermions $f_1,f_3$ in the rest frames of the vector bosons by $(\theta_1,0
)$ and $(\theta_3,\phi_3)$ [see Fig.~2.11 for the conventions and definitions], 
the angular  distribution is given by \cite{Bargeretal}
\begin{eqnarray}
\frac{ {\rm d}\Gamma (H \rightarrow  VV)} { {\rm d} c_{\theta_1} {\rm
d}c_{\theta_3}
{\rm d} \phi_3} &\sim & \ \ s^2_{\theta_1} s^2_{\theta_3} +\frac{1}{2 \gamma_1
\gamma_3(1+\beta_1 \beta_3)} s_{2 \theta_1} s_{2\theta_3} c_{\phi_3}
 \\
& &+{1\over2\gamma_1^2\gamma_3^2(1+\beta_1\beta_3)^2}
\left[\left(1+c_{\theta_1}^2\right)\left(1+c_{\theta_3}^2\right)
+s_{\theta_1}^2s_{\theta_3}^2c_{2\phi_3}\right]
\nonumber \\
& &-
\frac{4 A_{f_1} A_{f_3} }{\gamma_1\gamma_3(1+\beta_1 \beta_3)}
\left[ s_{\theta_1} s_{\theta_3} c_{\phi_3}
+\frac{1}{\gamma_1\gamma_3(1+\beta_1 \beta_3)}
c_{\theta_1}c_{\theta_3}\right] \non 
\label{HVV*distr}
\end{eqnarray}
\vspace*{-1.3cm}
\begin{figure}[htbp]
\begin{center}
\begin{picture}(550,200)(0,0)
\SetWidth{1.1}
\Photon(230,100)(310,100){2}{8}
\LongArrow(310,100)(317,100)
\Photon(230,100)(150,100){2}{8}
\LongArrow(150,100)(143,100)
\DashLine(320,100)(410,100){5}
\DashLine(140,100)(50,100){5}
\Line(230,170)(230,100)
\Line(230,50)(230,30)
\Line(230,100)(230,50)
\Line(230,170)(410,170)
\Line(230,30)(410,30)
\Line(410,170)(410,30)
\Line(205,160)(230,100)
\DashLine(230,100)(255,40){5}
\Line(25,160)(75,40)
\Line(25,160)(205,160)
\Line(75,40)(230,40)
\DashLine(230,40)(255,40){5}
\SetWidth{0.8}
\LongArrowArc(320,100)(23,0,51.34)
\LongArrowArcn(140,100)(23,180,141.34)
\LongArrowArc(245,100)(45,110,130)
\SetWidth{1.1}
\Text(230,100)[]{\Large\color{blue} $\bullet$}
\Text(222,88)[]{\large\color{blue} $H$}
\Text(270,88)[]{\large\color{red} V}
\Text(190,88)[]{\large\color{red} V}
\Text(330,130)[c]{\large\color{red} $f_1$}
\Text(310,70)[c]{\large\color{red} $\bar{f}_2$}
\Text(125,130)[c]{\large\color{red} $f_3$}
\Text(155,70)[c]{\large\color{red} $\bar{f}_4$}
\Text(353,112)[c]{\large\color{black} $\theta_1$}
\Text(109,112)[c]{\large\color{black} $\theta_3$}
\Text(240,140)[c]{\large\color{black} $\phi_3$}
\LongArrow(140,100)(90,140)
\LongArrow(140,100)(190,60)
\LongArrow(320,100)(360,150)
\LongArrow(320,100)(280,50)
\end{picture}
\end{center}
\vspace*{-1.2cm}
\label{fig:angles}
{\it Figure 2.11: The definition of the polar angles ${\theta_{1,3}}$  and the
azimuthal angle $\phi_3$ for the sequential decay $H \rightarrow V V
\rightarrow (f_1\bar{f}_2) (f_3\bar{f}_4)$ in the rest frame of the Higgs
particle.}
\end{figure}

\nn where the combination of $Vf\bar f$ couplings is $A_f=2\hat{v}_f \hat{a}_f/
(\hat{v}_f^2+\hat{a}_f^2)$; for $V=W$, the weak charges are as usual $\hat{v}_f
=\hat{a}_f=\sqrt{2}$ while for $V=Z$, $\hat{v}_f=2I_f^{3}-4Q_f \sin^2\theta_W$
and $\hat{a}_f=2I_f^{3}$. $\beta_i, \gamma_i= (1-\beta_i^2)^{-1/2}$ are the 
velocities and $\gamma$
factors of the [on/off--shell] vector bosons and $s_\theta \equiv \sin \theta$,
{\it etc}.  The dependence on the azimuthal angle between the decay planes
disappears for large Higgs masses, $\sim 1/\gamma$, a consequence of the 
asymptotic longitudinal $V$ polarization. After integrating out the polar
angles, we are left with \cite{Bargeretal} 
\begin{eqnarray}
\frac{{\rm d}\Gamma (H \rightarrow  VV)}{  {\rm d} \phi_3 }\sim 1+a_1 
c_{\phi_3} + a_2 c_{2\phi_3} \hspace*{3cm} \non \\
a_1 = -\frac{9 \pi^2}{32}\, \frac{\gamma_1 \gamma_3 (1+\beta_1 \beta_3)}
{\gamma_1^2\gamma_3^2 (1+\beta_1 \beta_3)^2+2}\,
A_{f_1} A_{f_3} \ , \ a_2 =  \frac{1}{2}\,\frac{1}
{\gamma_1^2 \gamma_3^2 (1+\beta_1 \beta_3)^2+2} 
\label{HVV*azimuth}
\end{eqnarray}
where the coefficient $a_1$ measures the P--odd amplitude. \s

These are unique predictions for the SM Higgs boson with $J^{\rm PC}=0^{++}$ 
quantum  numbers.  One can again confront these predictions with what is 
expected in the case of a $J^{\rm PC}= 0^{+-}$ CP--odd Higgs 
boson\footnote{The more general case where both CP--even and CP--odd couplings
are present can be found in Ref.~\cite{CP-full}.}.  
The $AVV$ coupling has been defined in eq.~(\ref{AVVcp}), and  reduces to 
$(\vec{\epsilon}_1 \times \vec{\epsilon}_2)\cdot(\vec{p}_1 -\vec{p}_2)$ in the 
laboratory frame.  The CP--odd angular distributions in the decays $ A  
\rightarrow V V \rightarrow  (f_1 \bar{f}_2) \, (f_3 \bar{f}_4)$ are  given 
by \cite{Bargeretal}
\begin{eqnarray}
\frac{ {\rm d}\Gamma (A \rightarrow  VV)} { {\rm d} c_{\theta_1} {\rm
d}c_{\theta_3} {\rm d} \phi_3} & \sim & 1+ c^2_{\theta_1} c^2_{\theta_3} 
-\frac{1}{2} s^2_{\theta_1} s^2_{\theta_3} -\frac{1}{2} s^2_{\theta_1} 
s^2_{\theta_3} c_{2\phi_3} -2 A_{f_1} A_{f_3} c_{\theta_1} c_{\theta_3}
\end{eqnarray}
and simply reduces, after integrating over the polar angles, to 
\begin{eqnarray}
\frac{{\rm d}\Gamma (A \rightarrow  VV)}{  {\rm d} \phi_3 }\sim 1 -
\frac{1}{4} c_{2\phi_3}
\end{eqnarray}
The normalization follows from the total and differential decay widths. Since
the $A$ boson does not decay into longitudinal gauge bosons,
the partial width for the two--body decay is
\begin{eqnarray}
\Gamma (A \rightarrow  VV) = \frac{G_\mu M_H^3}{16 \pi^3 M_A} \, \delta_V  \, 
\eta^2 \, ( 8 x^2 )  \, \sqrt{1-4x} \, 
\end{eqnarray}
while for the three--body decay, one has
\begin{eqnarray}
\Gamma (A \rightarrow  VV^*) = \frac{3 G_\mu^2 M_V^6}{8 \pi^3 M_A} \delta_V'
\eta^2 R_A \left( \frac{M_V^2}{M_A^2} \right)
\end{eqnarray}
with 
\begin{eqnarray}
R_A(x)&=&(1-7x)(4x-1)^{1/2} \arccos \left( \frac{3x-1}{2x^{3/2}}
\right) -\frac{1-x}{6}(17-64x-x^2) \non \\ && + \frac{1}{2}(1-9x+6x^2) \log x
\end{eqnarray}
The invariant mass spectrum of the off--shell vector bosons reads
\begin{eqnarray}
\frac{{\rm d} \Gamma (A \rightarrow  VV^*)}{{\rm d}M_*^2} = \frac{3 G_\mu^2
M_V^6}{8 \pi^3 M_A} \delta_V' \eta^2 \frac{M_*^2 \beta_V^3}{(M_*^2-M_V^2)^2+ 
M_V^2 \Gamma_V^2}
\end{eqnarray}

The fraction of the decay of the Higgs bosons into longitudinal vector bosons
[which is zero in the CP--odd Higgs case] and  the distributions with
respect  to the invariant mass of the off--shell gauge boson in the decays $H/A
\to Z^*Z$ for $M_{H/A}=150$ GeV are shown in Fig.~2.12. The mass and momentum 
distributions of the decay width are determined by the P--wave decay 
characteristics and 
the transverse polarization of the gauge bosons. The dependence on the
azimuthal angle is shown in Fig.~2.13 for the decays $H/A \to ZZ \to 4\mu$ and 
$H/A \to WW \to 4f$ with $M_{H/A}=300$ GeV. Again, the difference between the 
CP--even and CP--odd cases is noticeable. In the case of $H \to ZZ$ decays, 
the variation with the azimuthal angle is small since the factor in front 
of $\cos\phi_3$ is tiny, $a_1 \propto v_e^2 \ll 1$ [while $v_f=\sqrt{2}$ for 
$W$ bosons]; the coefficient of  $\cos2\phi_3$ drops like $1/\gamma^4$ in the 
scalar case.

\begin{figure}[htbp]
\begin{center}
\vspace*{-2.5cm}
\hspace*{-3cm}
\epsfig{file=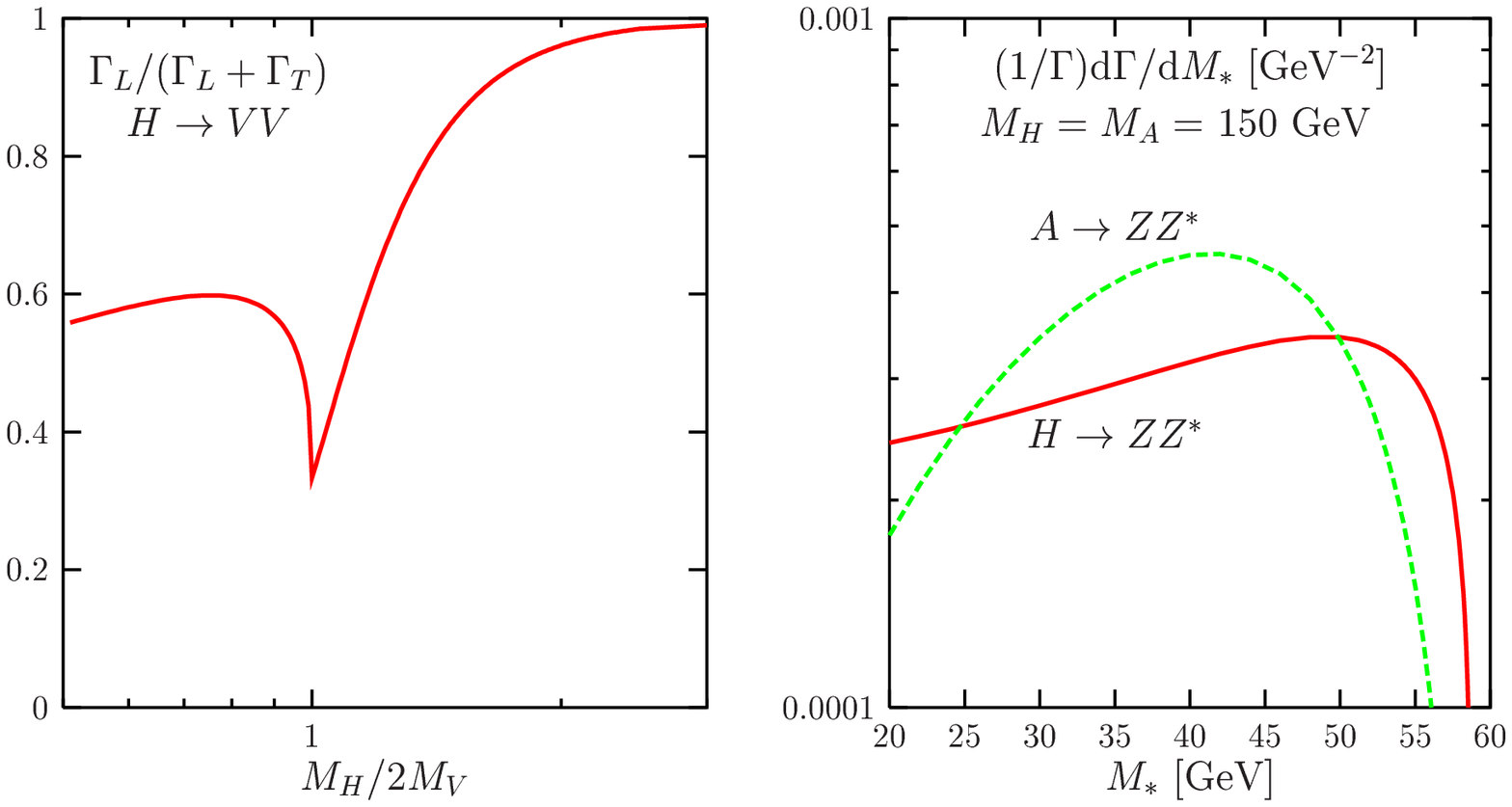,width=18.cm} 
\end{center}
\vspace*{-14.6cm}
\nn {\it Figure 2.12: The decay width of the Higgs boson into longitudinal gauge
bosons as a function of the ratio $M_H/2M_V$ (left) and the distribution with
respect to the invariant mass of the off--shell gauge boson in the decays
$H/A \to ZZ^*$ for $M_H=M_A=150$ GeV (right).}
\vspace*{-.6cm}
\end{figure}

\begin{figure}[htbp]
\begin{center}
\vspace*{-2.9cm}
\hspace*{-3cm}
\epsfig{file=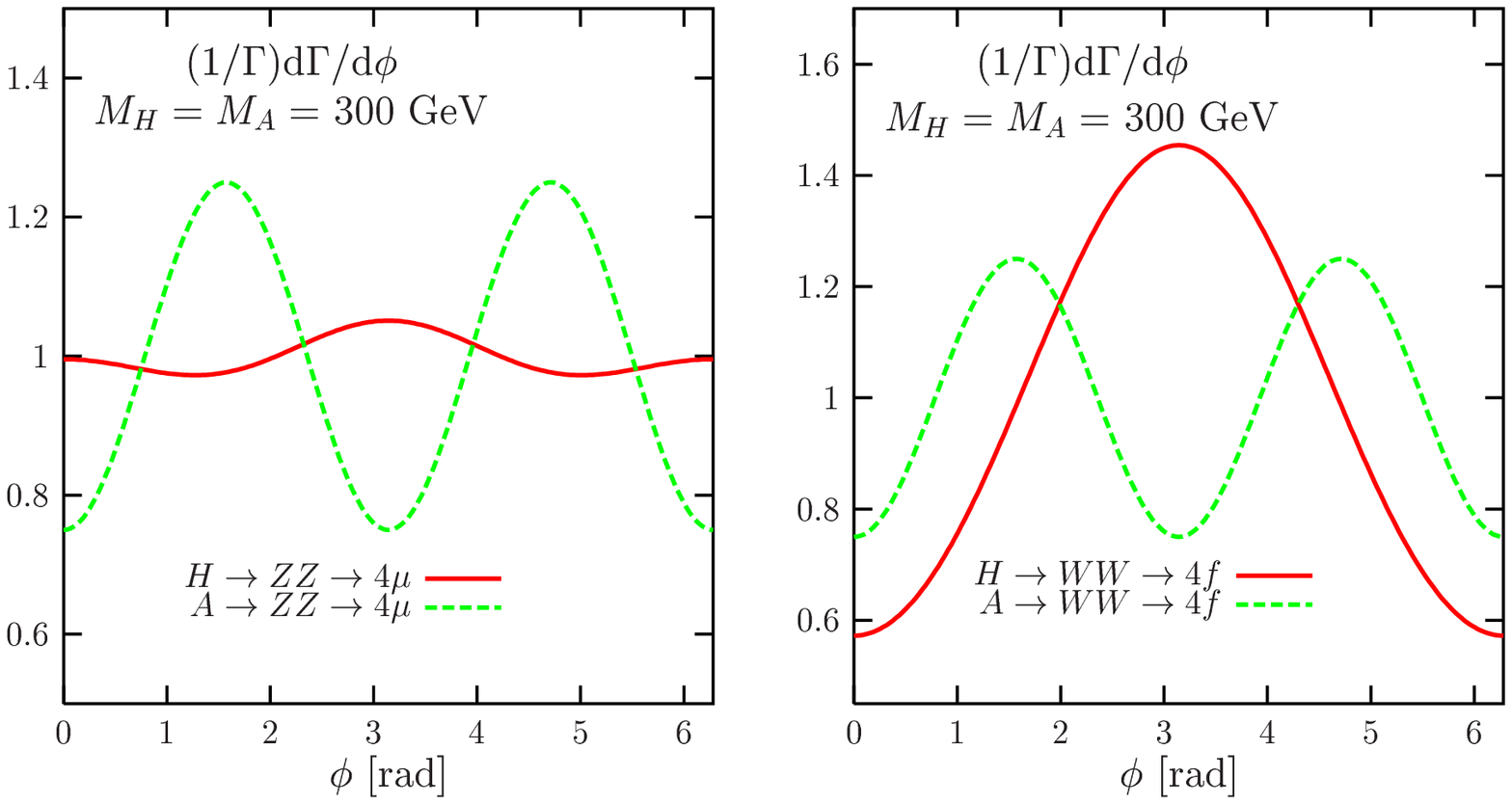,width=18.cm} 
\end{center}
\vspace*{-14.6cm}
\nn {\it Figure 2.13: The azimuthal dependence in the decays $H/A \to ZZ \to 
4\mu^\pm$ (left) and $H/A \to WW \to 4f$ for CP--even and CP--odd Higgs 
bosons with masses $M_H=M_A=300$ GeV.}
\vspace*{-.6cm}
\end{figure}

\subsection{Loop induced decays into $\gamma \gamma, \gamma Z$ and $gg$}

Since gluons and photons are massless particles, they do not couple to the
Higgs boson directly. Nevertheless, the $Hgg$ and $H\gamma\gamma$ vertices, as
well as the $HZ\gamma$ coupling, can be generated at the quantum level with
loops involving  massive [and colored or charged] particles which couple to the
Higgs boson. The $H \gamma \gamma$ and $HZ\gamma$ couplings are mediated by $W$
boson and charged fermions loops, while the  $Hgg$ coupling is mediated only by
quark loops; Fig.~2.14. For fermions, only the heavy top quark and, to a lesser
extent, the bottom quark contribute substantially for Higgs boson masses $M_H
\gsim 100$ GeV.

\begin{center}
\hspace*{-1cm}
\begin{picture}(300,100)(0,0)
\SetWidth{1.}
\SetScale{1.2}
\DashLine(-20,50)(20,50){4}
\Photon(20,50)(50,75){3}{4.5}
\Photon(20,50)(50,25){-3}{4.5}
\Photon(50,25)(50,75){3}{5.5}
\Photon(50,25)(89,25){-3}{5}
\Photon(50,75)(89,75){3}{5}
\Text(-50,90)[]{\red{a)}}
\Text(20,60)[]{\bb}
\Text(0,70)[]{\blue{$H$}}
\Text(46,60)[]{$W$}
\Text(122,90)[]{$\gamma(Z)$}
\Text(112,30)[]{$\gamma$}
\hspace*{1cm}
\DashLine(110,50)(140,50){4}
\Photon(170,25)(210,25){-3}{5}
\Photon(170,75)(210,75){3}{5}
\ArrowLine(140,50)(170,25)
\ArrowLine(140,50)(170,75)
\ArrowLine(170,75)(170,25)
\Text(170,60)[]{\bb}
\Text(190,60)[]{$F$}
\Text(146,70)[]{\blue{$H$}}
\Text(270,90)[]{$\gamma(Z)$}
\Text(260,30)[]{$\gamma$}
\Text(110,60)[]{+}  
\end{picture}
\begin{picture}(300,80)(0,0)
\SetWidth{1.}
\SetScale{1.2}
\hspace*{-8cm}
\DashLine(250,50)(290,50){4}
\Text(350,60)[]{\bb}
\Gluon(320,25)(359,25){-3}{5}
\Gluon(320,75)(359,75){3}{5}
\ArrowLine(290,50)(320,25)
\ArrowLine(290,50)(320,75)
\ArrowLine(320,25)(320,75)
\Text(330,70)[]{\blue{$H$}}
\Text(370,60)[]{$Q$}
\Text(440,90)[]{$g$}
\Text(440,30)[]{$g$}
\Text(240,75)[]{\red{b)}}
\end{picture}
\vspace*{-9mm}
\end{center}
\centerline{\it Figure 2.14: Loop induced Higgs boson decays into a) two 
photons $(Z\gamma$) and b) two gluons.}
\vspace*{5mm}

For masses much larger than the Higgs boson  mass, these virtual particles do
not decouple since their couplings to the Higgs boson grow with the masses,
thus compensating the loop mass suppression. These decays are thus extremely
interesting since their strength is sensitive to scales far beyond the Higgs
boson mass and can be used as a possible probe for new charged and/or colored
particles whose masses are generated by the Higgs mechanism and which are too
heavy to be produced  directly. \s

Unfortunately, because of the suppression by the additional electroweak or
strong coupling constants, these loop decays are  important only for Higgs
masses below $\sim 130$ GeV when the total Higgs decay  width is rather small. 
However, these partial widths will be very important when we will discuss the
Higgs production at hadron and photon colliders, where the cross sections will
be directly proportional to, respectively, the gluonic and photonic partial 
decay widths. Since the entire Higgs boson mass range can be probed in these 
production processes, we will also discuss the amplitudes for heavy Higgs 
bosons.\s

In this section, we first analyze the decays widths both at leading order (LO) 
and then including the next--to--leading order (NLO) QCD corrections. The 
discussion of the LO electroweak corrections and the higher--order QCD 
corrections will be postponed to the next section. 
 
\subsubsection{Decays into two photons}

\subsubsection*{\underline{The partial width at leading order}}

The decay of the SM Higgs boson into two photons is mediated by $W$ boson and 
heavy  charged fermion loops. The partial decay width  can be cast into the 
form \cite{EGN,HppBorn,HppBorn0,HppAnnecy}
\begin{eqnarray}
\Gamma\, (H\ra \gamma\gamma) = \frac{G_{\mu}\, \alpha^{2}\,M_{H}^{3}}
{128\,\sqrt{2}\,\pi^{3}} \left| \sum_{f} N_{c} Q_f^2 A_{1/2}^H(\tau_f) +
A^H_1(\tau_W) \right|^2
\label{eq:hgaga}
\end{eqnarray}
with the form factors for spin--$\frac{1}{2}$ and spin--1 particles given by
\begin{eqnarray}
A_{1/2}^H(\tau) & = & 2 [\tau +(\tau -1)f(\tau)]\, \tau^{-2}  \nonumber \\   
A_1^H(\tau) & = & - [2\tau^2 +3\tau+3(2\tau -1)f(\tau)]\, \tau^{-2}
\label{eq:Af+Aw}
\end{eqnarray}
and the function $f(\tau)$ defined as
\begin{eqnarray}
f(\tau)=\left\{
\begin{array}{ll}  \displaystyle
\arcsin^2\sqrt{\tau} & \tau\leq 1 \\
\displaystyle -\frac{1}{4}\left[ \log\frac{1+\sqrt{1-\tau^{-1}}}
{1-\sqrt{1-\tau^{-1}}}-i\pi \right]^2 \hspace{0.5cm} & \tau>1
\end{array} \right.
\label{eq:ftau}
\end{eqnarray}
The parameters $\tau_i= M_H^2/4M_i^2$ with $i=f,W$ are defined by the
corresponding masses of the heavy loop particles. The electromagnetic 
constant in the coupling should be taken at the scale $q^2=0$ since the 
final state photons are real. \s

Since the $Hf\bar f$ coupling is proportional to $m_f$, the contribution
of light fermions is negligible so that in the SM with three families, only the
top quark and the $W$ boson effectively contribute to the $\gamma \gamma$
width. If the Higgs boson mass is smaller than the $WW$ and $f \bar f$ pair
thresholds, the amplitudes are real and above the thresholds they are complex;
Fig.~2.15. Below thresholds, the $W$ amplitude is always dominant, falling
from $A_1^H =-7$ for very small Higgs masses to $A_1^H=-5 - 3 \pi^2/4$ at the 
$WW$ threshold; for large Higgs masses the $W$ amplitude approaches $A_1^H \to
-2$. The fermionic contributions increase from $A^H_{1/2} = 4/3$ for small 
$\tau_f$ values to $A^H_{1/2} \sim 2$ at the $2m_f$ threshold; far above the 
fermion threshold, the amplitude vanishes linearly in $\tau_f$ modulo 
logarithmic coefficients, 
\beq 
M_H^2 \gg 4 m_f^2  &:& A_{1/2}^H (\tau_f) \to   - [\log(4\tau_f)-i\pi]^2  
/(2 \tau_f)  \non \\
M_H^2 \ll 4 m_f^2  &:& A_{1/2}^H (\tau_f) \to 4/3
\eeq

In Fig.~2.16, we display the partial decay width $\Gamma( H\to \gamma \gamma)$.
The width varies rapidly from a few KeV  for $M_H \sim 100$ GeV to $\sim 100$ 
KeV for $M_H \sim 300$ GeV as a consequence of the  growth  $\propto M_H^3$.
The contribution of the $W$ boson  loop interferes destructively with the quark
loop and for Higgs masses of about 650 GeV, the two contributions nearly cancel
each other. The contribution of the $b$--loop is negligible, while the 
$t$ quark contribution with $m_t \to \infty$ is a good approximation for
Higgs masses below the $2m_t$ threshold.  

\begin{figure}[!h]
\begin{center}
\vspace*{-2.7cm}
\hspace*{-3cm}
\epsfig{file=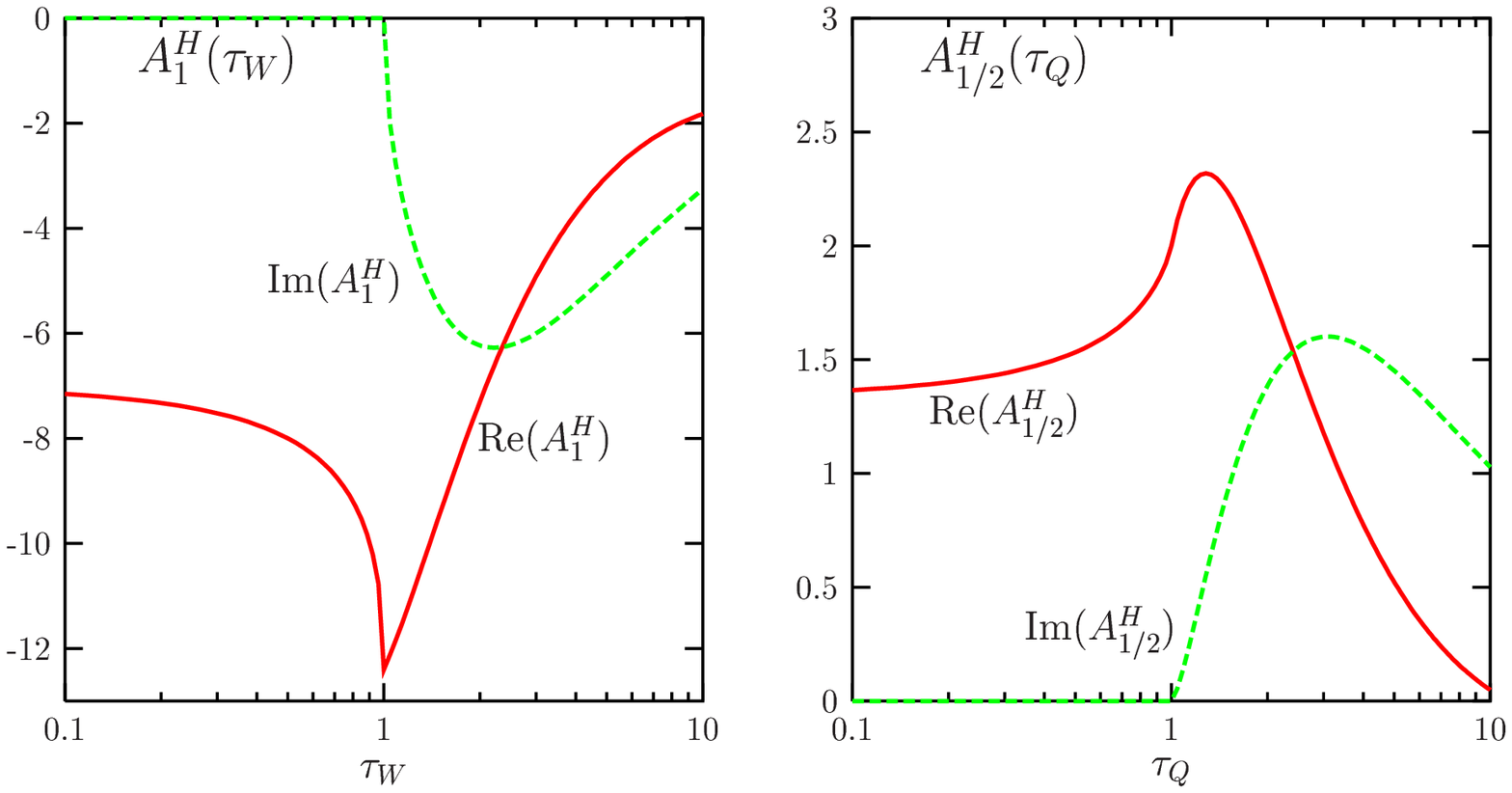,width=16.5cm} 
\end{center}
\vspace*{-13.5cm}
\nn {\it Figure 2.15: Real and imaginary parts of the $W$ boson (left) and
heavy fermion (right) amplitudes in the decay $H \to \gamma \gamma$ as a 
function of the mass ratios $\tau_i=M_H^2/4M_i^2$.}
\end{figure}
\vspace*{-.5cm}
\begin{figure}[!h]
\begin{center}
\vspace*{-2.5cm}
\hspace*{-3cm}
\epsfig{file=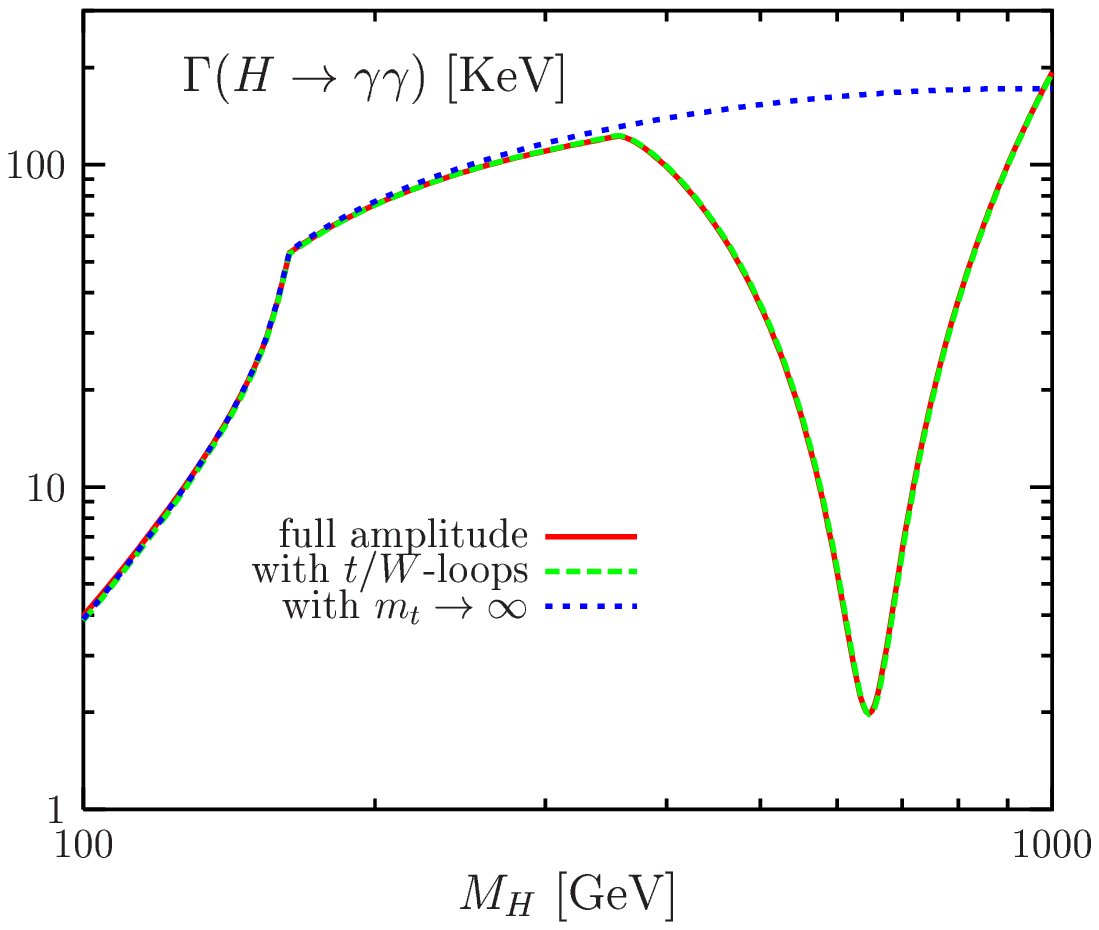,width=17.cm} 
\end{center}
\vspace*{-14.2cm}
\nn {\it Figure 2.16: The partial width  for the decay $H \to \gamma \gamma$ 
as a function of $M_H$ with the $W$ and all third generation 
fermion contributions (solid) and with $W$ and only the top quark 
contribution (dashed) and with the $W$ and $t$ quark contributions for
$m_t \to \infty$ (dotted lines).} 
\vspace*{-.7cm}
\end{figure}

\subsubsection*{\underline{The NLO QCD corrections}}

The QCD corrections to the quark amplitude in the decay $H \to \gamma \gamma$
consist only of two--loop virtual corrections and the corresponding 
counterterms; some generic diagrams are shown in Fig.~2.17. There are no real
corrections since the decay $H \to \gamma \gamma +g$ does not occur due to color
conservation. The calculation can be done in the on--shell scheme, in which  
the quark mass $m_Q$ is defined as the pole of the propagator and the quark 
wave function is renormalized with a renormalization constant $Z_2^{1/2}$ such 
that the residue at the pole is equal to unity.  The photon--quark vertex is 
renormalized at zero--momentum transfer and the standard QED Ward identity 
renders the corresponding renormalization factor equal to the one of the wave
function.  Since in the SM  the fermion masses are generated by the interaction
with the Higgs field, the renormalization factor $Z_{HQQ}$ associated with the
Higgs--quark vertex  is fixed unambiguously by the renormalization factors
$Z_m$ for the mass and $Z_2$ for the wave function.  From the bare Lagrangian
[the subscript $0$ stands for bare quantities]
\begin{eqnarray}
{\cal L}_0 & = & -m_0 \bar Q_0 Q_0 \frac{H}{v} = -m_Q \bar Q Q \frac{H}{v}
                 + Z_{HQQ} m_Q \bar Q Q \frac{H}{v}
\end{eqnarray}
one finds $Z_{HQQ} = 1-Z_2 Z_m$ \cite{HqqQCD-1loop,Drees+Hikasa}. Thus, in 
contrast to the photon--fermion vertex, the scalar $HQQ$ vertex is renormalized
at zero momentum transfer by a finite amount $\gamma_m$ after subtracting 
$Z_{HQQ}$ due to the lack of a corresponding Ward identity.\s 

\vspace*{5mm}
\begin{center}
\begin{picture}(300,80)(0,0)
\SetWidth{1.}
\SetScale{1.2}
\hspace*{-13cm}
\DashLine(250,50)(290,50){4}
\Photon(320,25)(350,25){3}{4}
\Photon(320,75)(350,75){3}{4}
\Line(290,50)(320,25)
\Line(290,50)(320,75)
\Line(320,25)(320,75)
\Gluon(303,40)(303,60){-3}{3.2}
\Text(350,60)[]{\bb}
\Text(330,70)[]{\blue{$H$}}
\Text(430,90)[]{$\gamma$}
\Text(430,30)[]{$\gamma$}
\Text(377,57)[]{$g$}
\hspace*{5cm}
\DashLine(250,50)(290,50){4}
\Text(350,60)[]{\bb}
\Photon(320,25)(350,25){3}{4}
\Photon(320,75)(350,75){3}{4}
\Line(290,50)(320,25)
\Line(290,50)(320,75)
\Line(320,25)(320,75)
\Gluon(303,60)(320,40){-3}{4}
\hspace*{5cm}
\DashLine(250,50)(290,50){4}
\Text(350,60)[]{\bb}
\Photon(320,25)(350,25){3}{4}
\Photon(320,75)(350,75){3}{4}
\Line(290,50)(320,25)
\Line(290,50)(320,75)
\Line(320,25)(320,75)
\GlueArc(320,50)(10,90,270){3}{4}
\end{picture}
\vspace*{-11mm}
\end{center}
\centerline{\it Figure 2.17: QCD corrections to the quark amplitude for the $H 
\to \gamma \gamma$ decay.}
\vspace*{2mm}

The two--loop amplitudes for the $H \to \gamma \gamma$ decay  have
been calculated in Refs.~\cite{Hpp1loop,SDGZ,Hpp1loopmassive}. In the 
general massive case, the five--dimensional Feynman parameter integrals have 
been reduced analytically down to one--dimensional integrals over 
polylogarithms which were evaluated numerically \cite{SDGZ}. Very recently 
\cite{Hpp1loopmassive}, these integrals have been derived analytically.
The QCD corrections of the quark contribution to the
two--photon Higgs decay amplitude can be parameterized as
\beq
A_{1/2}^H (\tau_Q)=A_{1/2}^H (\tau_Q)|_{\rm LO} \left[ 1+ \frac{\alpha_s}{\pi} 
\, C_H (\tau_Q) \right] 
\eeq
In principle, the scale in $\alpha_s$ is arbitrary to this order although, in
practice, it should be chosen to be, typically, of order $M_H$. However, the
renormalization scale should be defined at $\mu_Q={1 \over 2} M_H$ for two
reasons: $(i)$ the $Q \overline{Q}$ decay threshold is defined at the correct
position $2m_Q(m_Q)=2m_Q$ and $(ii)$ it turns out a posteriori that all
relevant large logarithms are effectively absorbed into the running mass for
the entire range of the variable $\tau$.  Note that near the threshold
\cite{AppCoulomb}, within a margin of a few GeV, the perturbative analysis is
in principle not valid since the formation of a P--wave $0^{++}$ resonance,
interrupted by the rapid quark decay modifies the amplitude in this range. 
Since $Q \overline{Q}$ pairs cannot form $0^{++}$ states at the threshold,
Im$C_H$ vanishes there and  Re$C_H$ develops a maximum very close to this
threshold.  \s

The real and imaginary parts of the correction factor $C_H$ are shown in 
Fig.~2.18 as a function of $\tau_Q$ with the scale set to $\mu_Q={1\over 2}
M_H$ (left) and $\mu_Q=m_Q$ (right). In the limit $m_Q \rightarrow \infty$, 
the correction factor can be evaluated analytically and one finds 
\cite{Hpp1loop}
\begin{eqnarray}
M_H^2/4m_Q^2 \to 0\,: \hspace{0.5cm} 1+
C_H \frac{\alpha_s}{\pi} \to  1 - \frac{\alpha_s}{\pi}
\end{eqnarray}
In the opposite limit $m_Q (\mu_Q^2) \rightarrow 0$ the leading and subleading
logarithms of the correction factor can also be evaluated analytically
\begin{equation}
m_Q(\mu_Q^2) \to 0\,: \hspace{0.5cm} \left\{
\begin{array}{l} {\rm Re} C_H \to -\frac{1}{18} [\log^2 (4\tau)-\pi^2] - 
\frac{2}{3} \log (4\tau) + 2\log\frac{\mu_Q^2}{m_Q^2} \\
{\rm Im} C_H \to \frac{\pi}{3} \left[ \frac{1}{3} \log(4\tau) + 2 \right]
\end{array} \right. 
\end{equation}

\begin{figure}[hbtp]
\vspace*{-1.2cm}
\hspace*{-1.7cm}
\centerline{
\epsfig{file=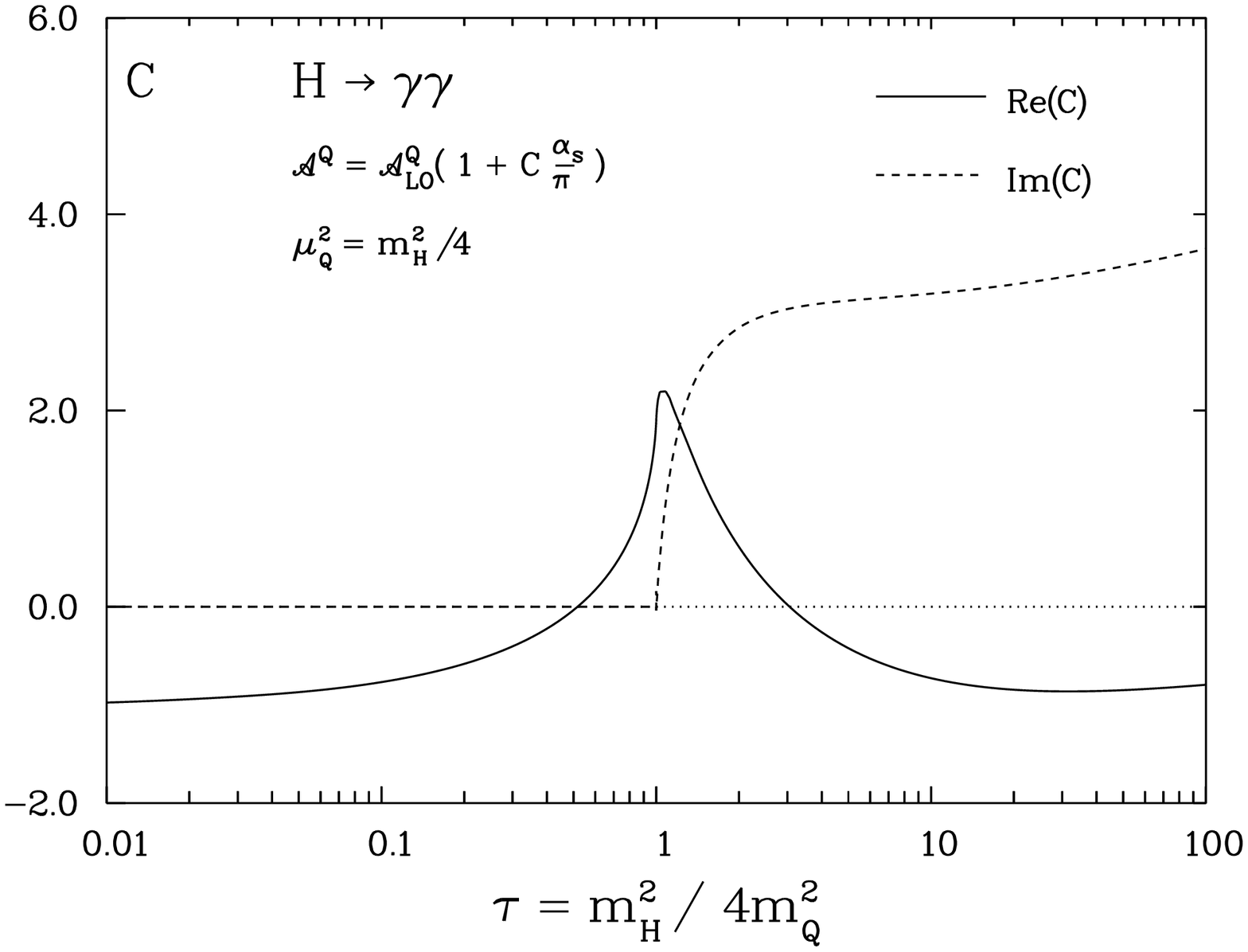,width=8.4cm,angle=-90}\hspace*{.9cm}
\epsfig{file=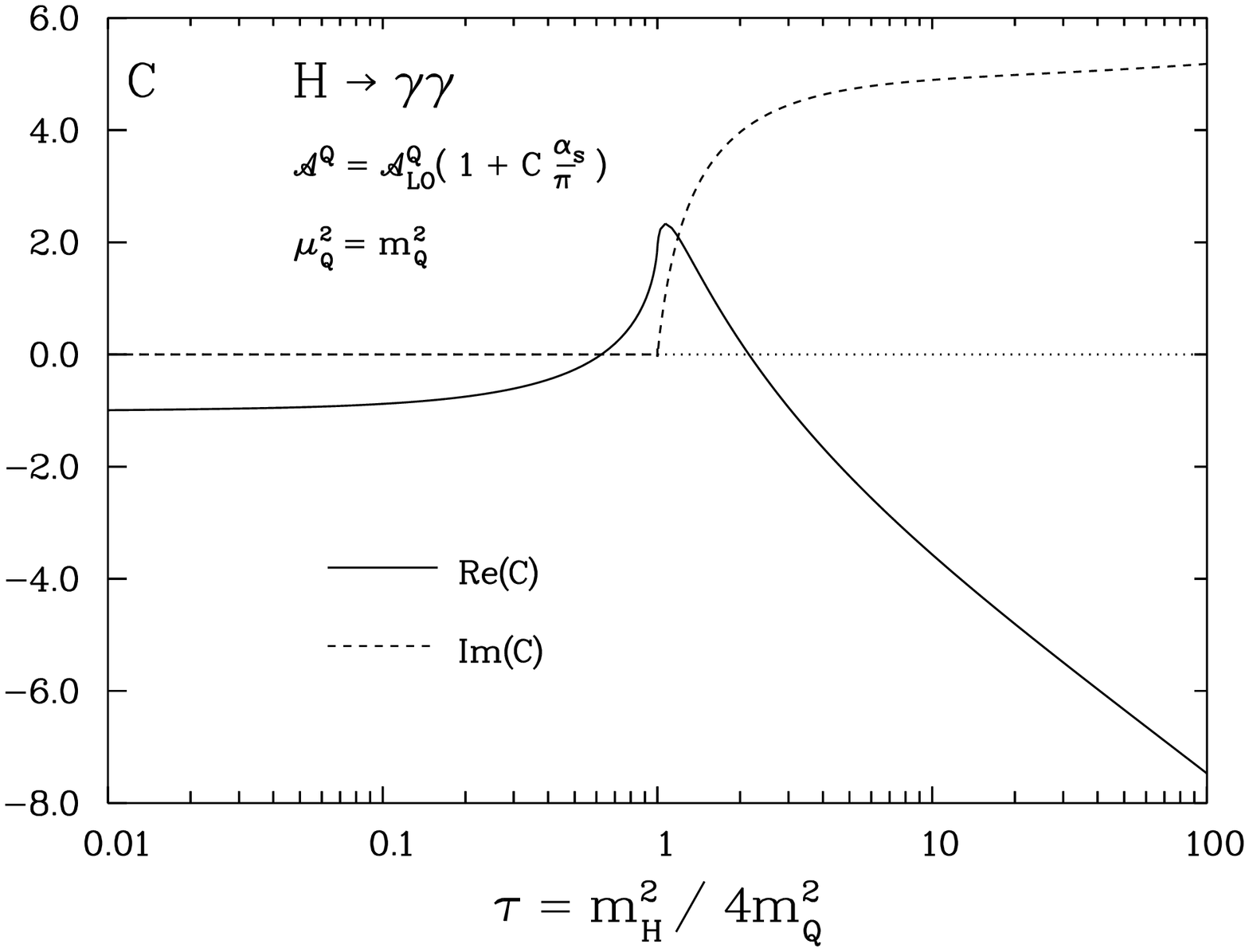,width=8.4cm,angle=-90}
}\\[-1.cm]
\nn {\it Figure 2.18: The QCD correction factor to the real and imaginary parts
of the quark amplitude $A_{1/2}^H$ in the $H \to \gamma \gamma$ decay as a 
function of  $\tau_Q=M_H^2/4m_Q^2$. The scale at which the correction is 
evaluated is $\mu_Q=\frac{1}{2}M_H$ (left) and $\mu_Q=m_Q$ (right).}
\vspace*{-.1cm}
\end{figure}

The QCD correction factor to the partial decay width relative to the lowest
order result, $\Gamma=\Gamma_{\rm LO} (1+\delta)$ is shown in Fig.~2.19 as a
function of the Higgs boson mass.  The correction is very large slighly above
the $t\bar t$ threshold and in the area $M_H \sim 650$ GeV where the
destructive $W$-- and $t$--loop interference makes the decay amplitude nearly
vanish.

\begin{figure}[hbtp]
\vspace*{-1.cm}
\hspace*{1cm}
\begin{turn}{-90}%
\epsfxsize=10cm \epsfbox{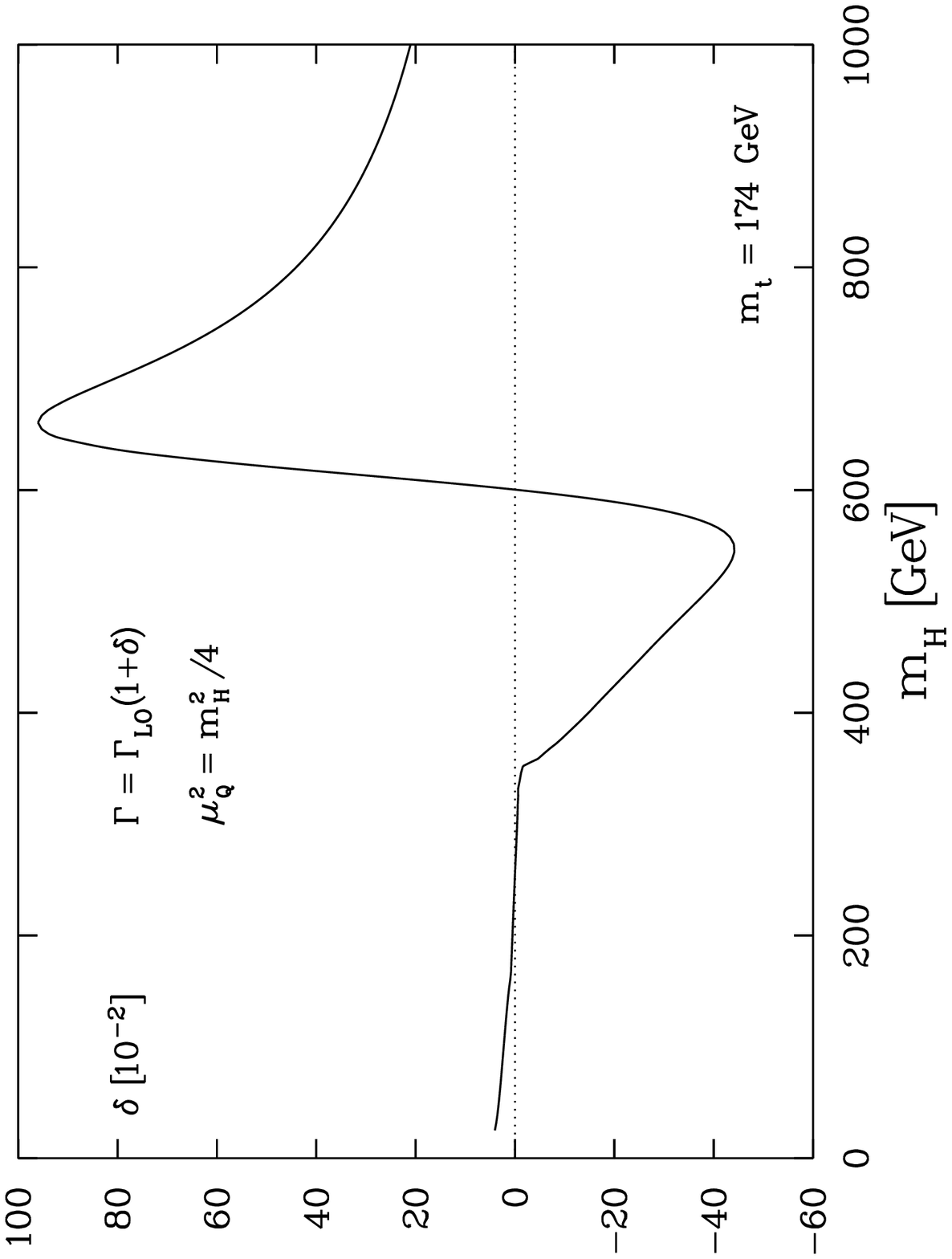}
\end{turn}\\[-1.2cm]
\nn {\it Figure 2.19: The QCD correction factor for the partial width 
$\Gamma(H \to \gamma \gamma)$  as a function of $M_H$. The pole
quarks masses are $m_t=174$ GeV and $m_b=5$ GeV and the QCD couplings is
normalized at $\alpha_s(M_Z)=0.118$. The renormalization scale is 
set to $\mu_Q= \frac{1}{2}M_H$.}
\vspace*{-.3cm}
 \end{figure}

\subsubsection{Decays into a photon and a $Z$ boson}

Similarly to the $\gamma \gamma$ case, the $H\ra Z\gamma$ coupling is built up 
by the heavy top quark and $W$ boson loops. The partial decay width is given by
\cite{Z-h-gamma1,Z-h-gamma2}
\begin{eqnarray}
\Gamma (H\ra Z\gamma ) = \frac{G^2_{\mu}M_W^2\, \alpha\,M_{H}^{3}} 
{64\,\pi^{4}} \left( 1-\frac{M_Z^2}{M_H^2} \right)^3 \left|
\sum_{f} N_{f} \frac{Q_f \hat{v}_f}{c_W} A_{1/2}^H(\tau_f,\lambda_f) + 
A^H_1(\tau_W,\lambda_W) \right|^2 
\label{eq:hzga}
\end{eqnarray}
with now $\tau_i= 4M_i^2/M_H^2$, $\lambda_i = 4M_i^2 /M_Z^2$ and the form 
factors
\begin{eqnarray}
A_{1/2}^H (\tau,\lambda) & = & \left[I_1(\tau,\lambda) - I_2(\tau,\lambda)
\right]  \\
A_1^H (\tau,\lambda) & = & c_W \left\{ 4\left(3-\frac{s_W^2}{c_W^2} \right)
I_2(\tau,\lambda) + \left[ \left(1+\frac{2}{\tau}\right) \frac{s_W^2}{c_W^2}
- \left(5+\frac{2}{\tau} \right) \right] I_1(\tau,\lambda) \right\} \non 
\label{eq:hzgaform}
\end{eqnarray}
with $\hat{v}_f=2I_f^3-4 Q_f s_W^2$ as usual. The functions $I_1$ and $I_2$ 
are given by
\begin{eqnarray}
I_1(\tau,\lambda) & = & \frac{\tau\lambda}{2(\tau-\lambda)}
+ \frac{\tau^2\lambda^2}{2(\tau-\lambda)^2} \left[ f(\tau^{-1})-f(\lambda^{-1}) 
\right] + \frac{\tau^2\lambda}{(\tau-\lambda)^2} \left[ g(\tau ^{-1}) - 
g(\lambda^{-1}) \right] \non \\
I_2(\tau,\lambda) & = & - \frac{\tau\lambda}{2(\tau-\lambda)}\left[ f(\tau
^{-1})- f(\lambda^{-1}) \right]
\end{eqnarray}
where the function $f(\tau)$ is defined in eq.~(\ref{eq:ftau}) while the 
function $g(\tau)$ can be expressed as
\begin{equation}
g(\tau) = \left\{ \begin{array}{ll}
\displaystyle \sqrt{\tau^{-1}-1} \arcsin \sqrt{\tau} & \tau \ge 1 \\
\displaystyle \frac{\sqrt{1-\tau^{-1}}}{2} \left[ \log \frac{1+\sqrt{1-\tau
^{-1}}}{1-\sqrt{1-\tau^{-1}}} - i\pi \right] & \tau  < 1
\end{array} \right.
\label{eq:gtau}
\end{equation}
Due to charge conjugation invariance, only the vectorial $Z$ coupling 
contributes to the fermion loop so that in the limit $M_H \gg M_Z$, the
$HZ\gamma$ amplitude reduces to the $H\gamma\gamma$ amplitude modulo
the different $Z$ and $\gamma$ couplings to fermions and $W$ bosons. \s

The partial width for this decay is shown in Fig.~2.20 as a function of
$M_H$. As mentioned in \S1.3.2 where the reverse decay $Z \to H\gamma$ was
discussed, the $W$ loop contribution is by far dominating. Below the $WW$
threshold, where this decay might have a visible branching ratio, it can be
approximated by $A_1^H  \simeq -4.6 + 0.3 M_H^2/M_W^2$. The top quark 
contribution 
interferes destructively with the $W$ loop but is very small; for low Higgs 
boson masses it can be approximated by $A_{1/2}^H= N_c Q_t \hat{v}_t /
(3 c_W) \sim 0.3$. The partial decay width, varies from a few KeV for $M_H \sim
120$ GeV to $\sim 100$ KeV for $M_H \sim 2M_W$. \s

\begin{figure}[!h]
\begin{center}
\vspace*{-2.7cm}
\hspace*{-3cm}
\epsfig{file=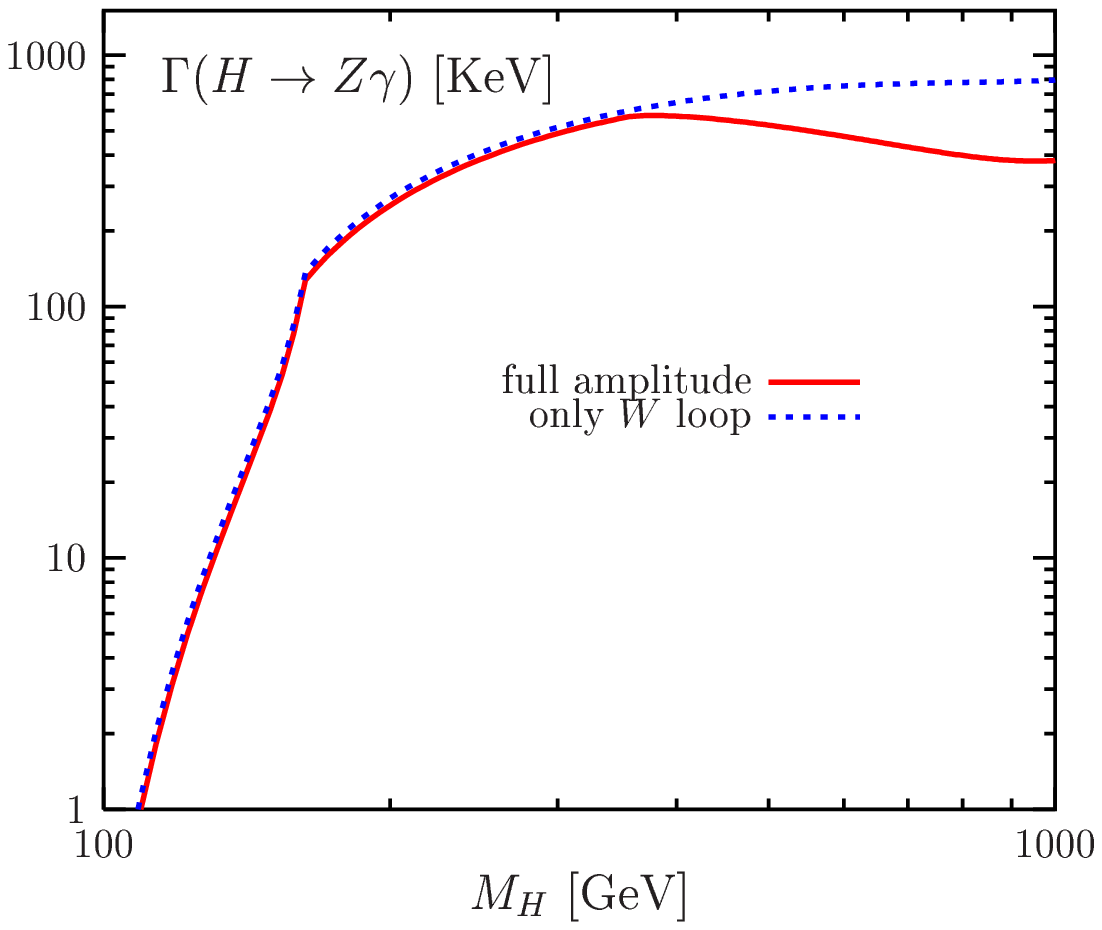,width=17.cm} 
\end{center}
\vspace*{-14.2cm}
\nn {\it Figure 2.20: The partial width  for the decay $H \to Z \gamma$ 
as a function of $M_H$ with the full $W$ boson and top quark contributions 
(solid line) and with the $W$ and top quark contribution but with $m_t \to 
\infty$ (dotted line).} 
\vspace*{-.3cm}
\end{figure}

The QCD corrections to the quark loop, calculated in Ref.\cite{HZpQCD}, are 
rather small in the interesting mass range, $M_H \lsim 2M_W$. In the heavy top 
quark  limit, which can be used here,  the correction factor for the top 
quark amplitude is exactly as in the  $H \to \gamma \gamma$ case
\beq
A_{1/2}^H(\tau_t,\lambda_t)  \to  A_{1/2}^H(\tau_t,\lambda_t) \times \left[
1- \frac{\alpha_s}{\pi} \right] \hspace*{1cm} \mbox{for $M_H^2\ll 4m_t^2$}
\label{eq:hgagaqcd}
\eeq

\subsubsection{Decays into gluons}

\subsubsection*{\underline{The partial width at leading order}}

The decay of the Higgs boson into two gluons is mediated by loops involving
heavy quarks, with the main contribution coming from top quarks and a small 
contribution from bottom quarks. At the one--loop (leading) order, the partial
decay width reads \cite{HggBorn,pp-ggH-LO}
\begin{eqnarray}
\Gamma\, (H\ra gg) = \frac{G_{\mu}\, \alpha_{s}^{2}\,M_{H}^{3}}
{36\,\sqrt{2}\,\pi^{3}}\left| \frac{3}{4} \sum_{Q} A_{1/2}^H(\tau_Q) \right|^2
\label{eq:hgglo}
\end{eqnarray}
The parameter $\tau_Q=M_H^2/4m_Q^2$ is defined by the pole mass $m_Q$ of the
heavy quark. The form factor $A_{1/2}^H(\tau_Q)$, similarly to the $H \to
\gamma \gamma$ case, is given in eq.~(\ref{eq:Af+Aw}) and is again normalized
such that for $m_Q \gg M_H$, it  reaches ${4\over 3}$,  while it approaches
zero in the chiral limit $m_Q \ra 0$. When crossing the quark threshold,
$M_H=2m_Q$, the amplitude develops an imaginary part.\s

The gluonic decay  width is shown as a function of the Higgs mass in Fig.~2.21
in the exact  case where top and bottom quark loops, with $m_t=178$ GeV and
$m_b=5$ GeV, are  included (solid line), when only the top quark contribution 
is included (dashed line) and when the top quark mass is sent to infinity 
(dotted line). As can be seen, keeping only the top quark contribution is a 
good approximation, better than 10\% even for $M_H \sim 100$ GeV, and below the 
$M_H=2m_t$ threshold, the heavy top--quark approximation is quite reliable.\s

\begin{figure}[!h]
\begin{center}
\vspace*{-2.7cm}
\hspace*{-3cm}
\epsfig{file=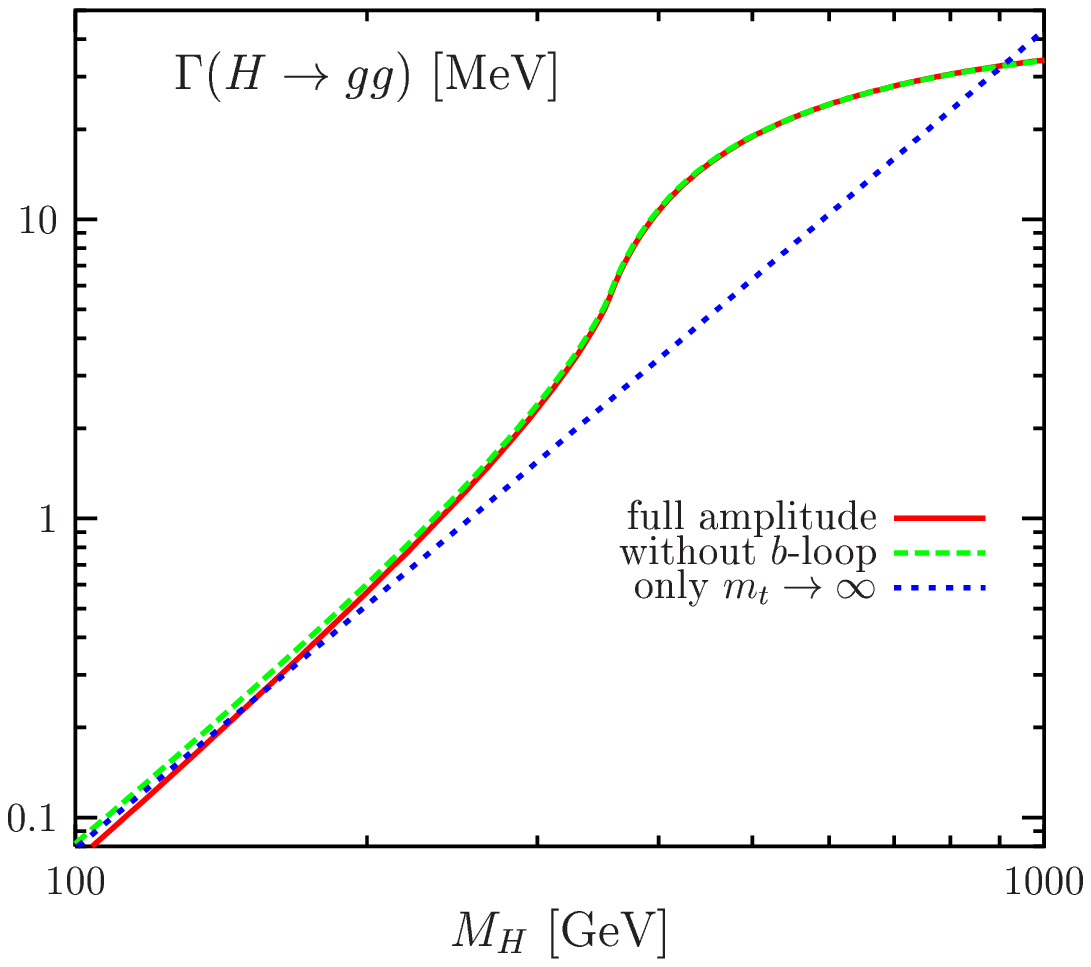,width=17.cm} 
\end{center}
\vspace*{-13.8cm}
\nn {\it Figure 2.21: The partial width  for the decay $H \to gg$ 
as a function of the Higgs boson mass with the top and bottom quark
contributions included (solid line), with only the top quark contribution 
included (dashed line) and in the limit of infinite top quark mass
(dotted line).} 
\vspace*{-.5cm}
\end{figure}

\subsubsection*{\underline{The QCD corrections at NLO}}

To incorporate the QCD corrections to the gluonic Higgs boson decay width, one
needs to consider virtual corrections where the gluons are attached to the
quark lines, as in the case of the $H \to \gamma \gamma$ decay at NLO, but also
corrections involving the triple and  quartic gluon vertices; Fig.~2.22a. These
corrections are finite in the ultraviolet [since the complementary virtual
corrections involved in the $H \to \gamma \gamma$ amplitude are also finite]
once the proper counterterms associated with the renormalization of the QCD
coupling [$Z_g-1= (Z_1- 1) -\frac{3}{2}(Z_3-1)]$ have been added;  $\alpha_s$
can be defined in the $\overline{\rm MS}$ scheme with five active quark flavors
and the heavy top quark decoupled. However, there are left--over infrared and
collinear singularities which are canceled only if the real corrections with
three gluon and a gluon plus a quark--antiquark pair final states $H \to gg+ g$
and $g+ q\bar{q}$ are added, Fig.~2.22b. The $q\bar{q}$ final states will be
assumed to be massless and, as a consequence of chiral symmetry, there is no
interference of the amplitude for $H \to g+ q\bar{q}$ and the one $H \to
q\bar{q}^* \to q\bar{q} g$ in which the Higgs boson couples directly to quarks
[this interference will be discussed in more detail later].

\begin{center}
\vspace*{1mm}
\begin{picture}(300,100)(0,0)
\SetWidth{1.}
\SetScale{1.2}
\hspace*{-13.5cm}
\Text(300,10)[]{\red{\bf b)}}
\Text(300,90)[]{\red{\bf a)}}
\DashLine(250,50)(290,50){4}
\Text(350,60)[]{\bb}
\Gluon(320,25)(360,25){3}{5}
\Gluon(320,75)(360,75){3}{5}
\Line(290,50)(320,25)
\Line(290,50)(320,75)
\Line(320,25)(320,75)
\Gluon(320,50)(340,75){-3}{3.2}
\Text(330,70)[]{\blue{$H$}}
\Text(440,90)[]{$g$}
\Text(440,30)[]{$g$}
\Text(400,60)[]{$g$}
\hspace*{5cm}
\DashLine(250,50)(290,50){4}
\Text(350,60)[]{\bb}
\Gluon(320,25)(360,25){3}{5}
\Gluon(320,75)(360,75){3}{5}
\Line(290,50)(320,25)
\Line(290,50)(320,75)
\Line(320,25)(320,75)
\Gluon(303,40)(340,75){-3}{6}
\hspace*{5cm}
\Text(350,60)[]{\bb}
\DashLine(250,50)(290,50){4}
\Gluon(320,25)(360,25){3}{5}
\Gluon(320,75)(360,75){3}{5}
\Gluon(340,25)(340,75){3}{6}
\Line(290,50)(320,25)
\Line(290,50)(320,75)
\Line(320,25)(320,75)
\end{picture}
\begin{picture}(300,80)(0,0)
\SetWidth{1.}
\SetScale{1.2}
\hspace*{-13.5cm}
\DashLine(250,50)(290,50){4}
\Text(350,60)[]{\bb}
\Gluon(320,25)(350,25){3}{4}
\Gluon(320,75)(350,75){3}{4}
\Line(290,50)(320,25)
\Line(290,50)(320,75)
\Line(320,25)(320,75)
\Gluon(320,50)(350,50){-3}{4}
\Text(330,70)[]{\blue{$H$}}
\Text(430,90)[]{$g$}
\Text(430,30)[]{$g$}
\Text(430,57)[]{$g$}
\hspace*{5cm}
\Text(350,60)[]{\bb}
\DashLine(250,50)(290,50){4}
\Gluon(320,75)(360,75){3}{6}
\Gluon(320,25)(340,25){3}{3}
\Gluon(340,25)(360,40){3}{3}
\Gluon(340,25)(360,10){3}{3}
\Line(290,50)(320,25)
\Line(290,50)(320,75)
\Line(320,25)(320,75)
\hspace*{5cm}
\DashLine(250,50)(290,50){4}
\Text(350,60)[]{\bb}
\Line(290,50)(320,25)
\Line(290,50)(320,75)
\Line(320,25)(320,75)
\Gluon(320,75)(360,75){3}{6}
\Gluon(320,25)(340,25){3}{3}
\ArrowLine(340,25)(360,40)
\ArrowLine(340,25)(360,10)
\Text(440,50)[]{$q$}
\Text(440,15)[]{$\bar q$}
\end{picture}
\end{center}
\vspace*{-7mm}
\nn {\it Figure 2.22: Typical Feynman diagrams for the QCD corrections to 
the process $H\to gg$ at NLO: a) virtual corrections not present in the decay 
$H \to \gamma \gamma$ and b) real corrections.}\vspace*{3mm}

The calculation of the NLO QCD correction in the full massive case has been 
performed in Ref.~\cite{SDGZ} where the rather complicated analytical 
expressions can be found. The total correction can be cast into the form 
\begin{equation}
\Gamma(H\rightarrow gg(g),~gq\bar q) = \Gamma_{\rm LO}(H\rightarrow gg)
\left[ 1 + E_H(\tau_Q) \frac{\alpha_s}{\pi} \right]
\end{equation}
and one obtains for the correction factor
\begin{equation}
E_H(\tau_Q) = \frac{95}{4} - \frac{7}{6} N_f
+ \frac{33-2N_f}{6}\ \log \frac{\mu^2}{M_H^2} + \Delta E_H (\tau_Q)
\label{Hgg-EH}
\end{equation}
where $\mu$ is the renormalization point and defines the scale of $\alpha_s$. 
The first three terms survive in the limit of large loop masses while $\Delta
E_H$ vanishes in this limit \cite{Inami+Kubota,HggQCD,AggQCD,HggExp}. \s
 
The QCD radiative corrections turn out to be quite important, nearly doubling 
the gluonic partial decay width; Fig.~2.23.  In the mass range $M_H \lsim
2M_W$, assuming $N_f=5$ light quarks and a scale $\mu=M_H$, the leading order 
term is corrected by a factor 
\beq
K= 1+ \frac{215}{12} \, \frac{\alpha_s^{N_f=5} (M_H)}{\pi}
\eeq
leading to an increase of the partial width by $\sim 70\%$. Near the $t\bar{t}$ 
threshold, when the $Hgg$ form factor develops an imaginary part, the 
correction is also at the level of 70\%. It decreases slowly with the Higgs 
mass to reach 40\% at $M_H \sim 1$ TeV. Also shown in Fig.~2.23 are the QCD 
corrections in the heavy top quark limit, but where the LO amplitude
includes the full $m_t$ dependence. As can be seen, this procedure approximates 
quite well the full result in the mass range $M_H \lsim 300$ GeV, the 
difference being less than a ten percent.\s

\begin{figure}[!h]
\begin{center}
\vspace*{-2.5cm}
\hspace*{-3cm}
\epsfig{file=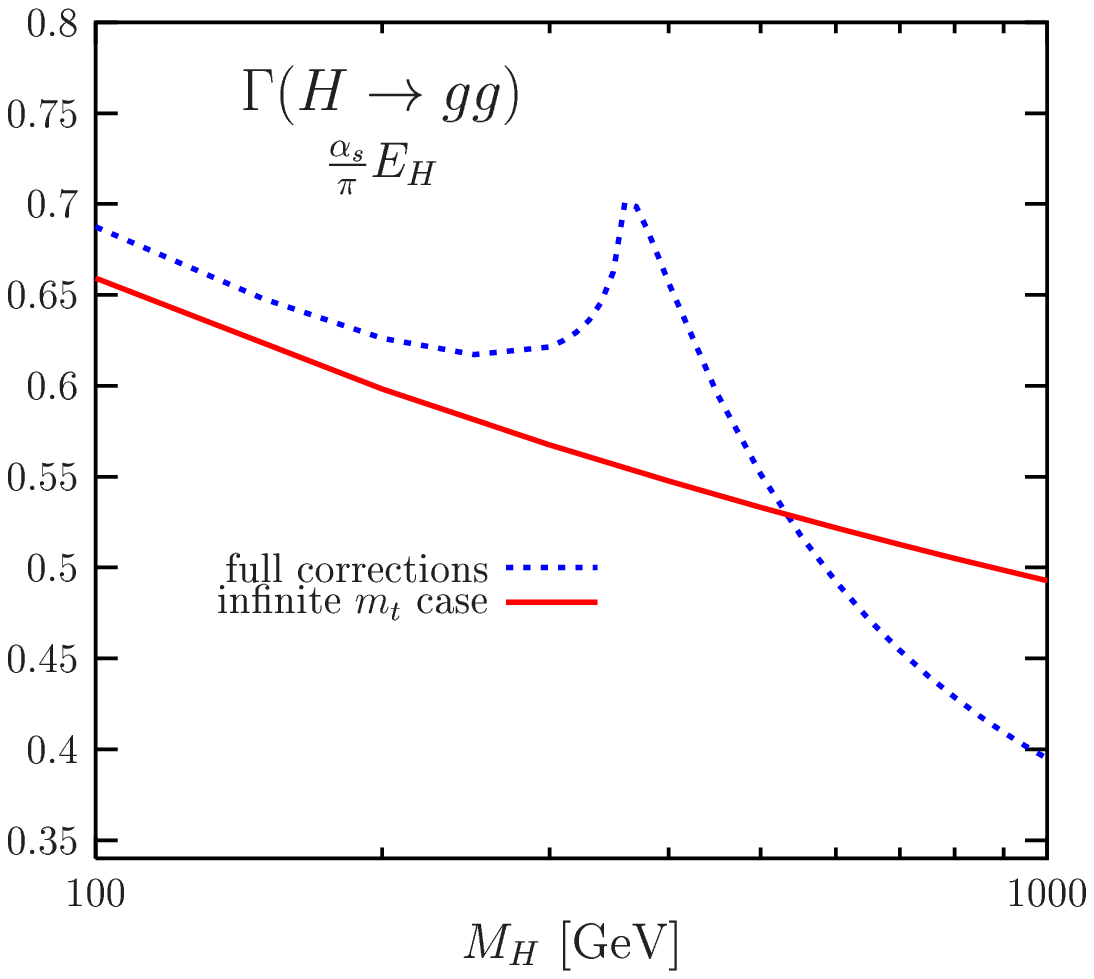,width=17.cm} 
\end{center}
\vspace*{-13.8cm}
\nn {\it Figure 2.23: The QCD correction factor for the partial width 
$\Gamma(H \to gg)$  as a function of the Higgs boson mass in the full massive
case with $m_t=178$ GeV (dotted line) and in the heavy top quark limit
(solid line). The strong coupling constant is $\alpha_s (M_Z) =0.118$.}
\vspace*{-.3cm}
\end{figure}

Since $b$ quarks, and eventually $c$ quarks, can in principle be tagged 
experimentally, it is physically meaningful to include gluon splitting $g^*\,
\ra\, b{\overline{b}} \; (c{\overline{c}})$ in $H \ra\,gg^*\ra\, gb{
\overline{b}} \, (c{\overline{c}})$ decays to the inclusive
decay probabilities $\Gamma(H \ra b\bar{b}+ \dots)$ {\it etc.} \cite{DSZ,SDGZ}.
The contribution of the $b,c$ quark final states in $H \to g+q\bar{q}$ reads
\beq
-\frac{7}{3}+ \frac{1}{3} \left[\log \frac{M_{H}^{2}}{ m_{b}^{2}} 
+\log \frac{M_{H}^{2}}{m_{c}^{2}} \right]
\label{ggsub}
\eeq
Separating this contribution generates large logarithms, which can be
effectively absorbed by redefining the number of active flavors in the gluonic
decay mode, i.e.~by evaluating $\alpha_s$ with $N_f=3$ when both the charm and
bottom quark contributions are subtracted. The contributions of the 
subtracted flavors have then to be added to the corresponding heavy quark decay 
modes discussed in \S2.1 [some details will be given in the next section]. 

\subsection{The electroweak corrections and QCD improvements}

In this section, we discuss the electroweak radiative corrections and the 
higher--order QCD corrections to the Higgs decay modes. Some of these 
corrections have been reviewed in Refs.~\cite{RCreviewEW} and 
\cite{Review-Michael,RCreviewQCD} for, respectively, the electroweak and 
higher--order  QCD parts.\s

The electroweak radiative corrections to the decays of Higgs bosons into
fermions and gauge bosons can be classified in three categories:

\begin{itemize}

\item[$(i)$] The fermionic corrections, which can be separated into the loop
contributions of the light fermions and those due to the heavy top quark. Most
of the former corrections are involved in the running of $\alpha$ and can be 
readily taken into account by using the improved Born approximation discussed 
in \S1.2.4. For the top quark correction, a universal part is due to 
the renormalization of the Higgs wave function and vev and appears for all 
fermion species and for gauge bosons. These corrections are in general the 
dominant electroweak corrections for a SM Higgs boson with a mass $M_H \lsim 
2m_t$.

\item[$(ii)$] Corrections due to the Higgs boson itself that are proportional
to the Higgs self--coupling $\lambda$. These corrections are important only
when $M_H \gg M_W$, when the coupling $\lambda$ becomes sizable. We have
seen in \S1.4.1 that for $M_H \sim {\cal O}(1~{\rm TeV})$, they can be so 
large that perturbation theory breaks down. 

\item[$(iii)$] The electromagnetic and the remaining weak corrections which 
do not depend on $\lambda$ and which are not quadratic in the top--quark 
mass. These corrections are process dependent and, in general, they lead to 
  small contributions, except in very special cases such as the $H \to 
t\bar{t}$ decay where the heavy top quark limit cannot be applied. 
\end{itemize}

Collecting all these electroweak  contributions, the correction factor for a
given Higgs decay channel $ H \to XX$ [also including the decay $H \to
Z\gamma$], can be then written as
\beq
K_{H\to XX}^{\rm EW}= 1+ \delta^t_{HXX} + \delta^\lambda_{HXX}+ \delta^e_{HXX}  
+\delta^w_{HXX} 
\label{ewcorfac}
\eeq

The present knowledge of the electroweak radiative corrections to the SM Higgs
decays is as follows.  The complete one--loop calculations of the $H \to f\bar
f$ and $H\to VV$ decays have been carried out in the massive cases in
Refs.~\cite{RCcha,RChff} and \cite{RCcha,RChvv}, respectively. The knowledge of
the partial widths for these decays has been improved by considering
higher--order corrections either in $\alpha_s$ or in the dominant electroweak
coupling $G_\mu m_t^2$.   The two--loop ${\cal O}(\alpha_s G_\mu m_t^2)$
heavy--top corrections to the light--fermion and bottom Yukawa couplings have
been calculated in Refs.~\cite{RCalb,RCadj} and \cite{RChbb}, respectively, and
those to the $HVV$ couplings in Ref.~\cite{RCsea}. The three--loop ${\cal
O}(\alpha_s^2 G_\mu m_t^2)$ corrections may be found in Ref.~\cite{RCmat} for
the the $H \to \ell^+ \ell^-$ and $H\to VV$ decays and in Ref.~\cite{RCkos} for
the decay $H\to q\bar q$, including the $b\bar b$ case.  The two--loop ${\cal
O}(G_\mu^2 m_t^4)$ pure electroweak corrections for the $H\to f \bar{f}$ and
$H\to VV$ decays have been derived in Ref.~\cite{RCdgk}.  The radiative
corrections due to the Higgs self--couplings have been calculated at one and
two loops in Refs.~\cite{Pert-HWcplg1,Pert-HWcplg2} for decays into massive
gauge bosons and in Refs.~\cite{Pert-HWcplg1,Pert-Hfcplg2} for decays into
fermions.\s

As for the loop induced Higgs boson vertices, the leading two--loop electroweak
corrections, which are of ${\cal O}(G_\mu m_t^2)$ relative to the one--loop
results,  have been calculated in  Refs.~\cite{RCdjo} for the  $Hgg$
coupling and in Refs.~\cite{RCdgk,RChppnew} for the $H\gamma \gamma$ and
$HZ\gamma$ couplings.  Recently, the two--loop electroweak corrections induced
by light fermion loops have been calculated for the $H\to \gamma \gamma$ and
$H\to gg$ decays \cite{RCita,PepeHgg}. Furthermore, still in the heavy top 
quark limit, the NNLO QCD corrections to the decays $H \to \gamma \gamma$ 
\cite{RCste} and $H \to gg$ \cite{RChgg} have been evaluated. Other
corrections \cite{HqqQED-QCD,RCstegg,Hpp-MH2} are also available and will
be discussed. \s

The dominant heavy top--quark corrections, including the two--loop order in 
$G_\mu m_t^2$ and in $\alpha_s$, as well as the NNLO QCD corrections to the 
loop induced decays, can be derived using a low energy theorem in which the top
quark has been integrated out by sending its mass to infinity. The results can 
nevertheless be extrapolated to Higgs boson masses up to the $M_H \sim 2m_t$
threshold in principle. In the following, we first discuss this low energy 
theorem and its applications for SM Higgs boson decays. 

\subsubsection{The low energy theorem} 

In the case of the top quark loop contributions to the interactions of a light
Higgs boson with $M_H \ll 2m_t$, a rather simple and efficient way of deriving
the corrections is to construct an effective Lagrangian where the top quark is
integrated out.  This can be done by considering the limit of a massless Higgs
boson or, equivalently, of a very heavy top quark and using a low energy theorem
proposed in Refs.~\cite{EGN,HppBorn,LET} and extended to higher orders in
Refs.~\cite{SDGZ,LET2}. The low--energy theorem relates the amplitudes of two 
processes which
differ only by the emission of a Higgs boson with vanishing momentum. Indeed, if
one recalls the discussion in \S1.1.3, the coupling of a Higgs boson  to a
fermion with a mass $m_i$ is generated by simply performing the substitution 
\beq
m_i^0  \to m_i^0 (1+H^0/v^0)
\eeq
in the bare Lagrangian [the index $0$ stands for bare quantities], where the 
Higgs boson is a constant field.  This implies the following relation between 
two matrix elements with and without the attachment of a Higgs field with 
zero--momentum $p_H$
\beq
\lim_{p_H\to 0} {\cal M}(X \to Y+H) = \frac{1}{v_0} m_i^0 \frac{\partial}
{\partial m_i^0} {\cal M}(X \to Y)
\eeq

However, in higher orders, there is a subtlety in the use of this relation: when
renormalizing the $H f \bar{f}$ interaction, the counterterm for the
Higgs--fermion Yukawa coupling is not the $Hf\bar{f}$ vertex with a subtraction
at zero momentum transfer, $\Gamma_{Hf \bar{f}}(q^2=0)$ [which is implicitly
used in the low--energy theorem] but, rather, is determined by the counterterms
for the fermion mass $Z_m$ and wave--function $Z_2$ as discussed previously.
This has to be corrected for and, in fact, this can be done by replacing the
differentiation with respect to the bare mass with a differentiation with
respect to the renormalized mass, which gives rise to a finite contribution
which is simply the anomalous mass dimension of the fermion 
\beq
m_0 \frac{\partial}{\partial m_0} = \frac{ m}{ 1+ \gamma_m} \frac{\partial}
{\partial m}
\eeq
which relates the bare mass $m_0$ and the renormalized mass $m$, 
d$ \log m_0 = (1 + \gamma_m) {\rm d}\log m$. \s

It is well known that this low energy theorem can be exploited to derive the 
$H \gamma \gamma$ coupling in lowest order \cite{HppBorn,LET}, but the theorem 
is also valid if radiative QCD corrections are included \cite{SDGZ,LET2}. The 
contribution of a heavy quark to the vacuum polarization of the photon at 
zero--momentum transfer is given in dimensional regularization, with 
$n=4-\epsilon$ being the number of space dimensions, by
\begin{eqnarray}
\Pi = - Q_Q^2\frac{\alpha}{\pi} \Gamma(\epsilon)
\left(\frac{4\pi\mu^2}{m_Q^2}\right)^\epsilon \left[ 1+\frac{\alpha_s}{2\pi}
\Gamma(1+\epsilon) \left(\frac{4\pi\mu^2}{m_Q^2}\right)^\epsilon
+ {\cal O} (\epsilon) \right]
\end{eqnarray}
so that $m_Q (\partial \Pi/ \partial m_Q) = 2 Q_Q^2\frac{\alpha}{\pi} \left( 1+
\frac{\alpha_s}{\pi} \right)$. From the anomalous quark mass dimension to
lowest  order, $\gamma_m = 2\alpha_s/\pi$, one immediately obtains the 
correction $C_H$ of the $H \gamma \gamma$ coupling in agreement with what
has been discussed in the previous section, \S2.3.1, 
\begin{eqnarray}
M_H^2/4m_Q^2 \to 0\,: \hspace{0.5cm} 1+
C_H\frac{\alpha_s}{\pi} \to \frac{1+\alpha_s/\pi} {1+2 \alpha_s/\pi}
\ = 1 - \frac{\alpha_s}{\pi}
\end{eqnarray}

The same result can also be derived by exploiting well--known results on the 
anomaly in the trace of the energy--momentum tensor \cite{LET3}
\begin{equation}
\Theta_{\mu\mu} = (1+\gamma_m) m_0 \overline{Q}_0 Q_0 +
\frac{1}{4}\frac{\beta_\alpha}{\alpha} F_{\mu\nu} F_{\mu\nu}
\end{equation}
with $\beta_\alpha$ denoting the mixed QED/QCD $\beta$ function defined by 
$\partial \alpha(\mu^2)/\partial \log \mu$=$\beta_\alpha$. Since the matrix
element $\langle \gamma \gamma | \Theta_{\mu \mu} |0 \rangle$ vanishes at 
zero--momentum transfer, the coupling of the two--photon state to the Higgs 
source $(m_0/v) \overline{Q}_0 Q_0$ is simply given by the effective Lagrangian
\beq
{\cal L} (H \gamma \gamma) = \frac{H}{v} F^{\mu \nu} F_{\mu \nu} 
\frac{1}{4} \frac{\beta_\alpha^Q }{ \alpha} \frac{1}{1+\gamma_m}
\eeq
including only the heavy quark contribution to the QED/QCD $\beta$ function.  
With $\beta_\alpha^Q = 2 Q_Q^2 \alpha^2/\pi (1 + \alpha_s/\pi)$ and
$\gamma_m =2\alpha_s/\pi$, one recovers again the previous result for the QCD 
correction to the $H\gamma \gamma$ coupling. 

\subsubsection{EW corrections to decays into fermions and
massive gauge bosons} 

\subsubsection*{\underline{Heavy top quark corrections}}

If one only wishes to extract  the leading correction to the Higgs couplings
due to a heavy top quark, one may work in the framework of a Yukawa Lagrangian
where it couples only to the Higgs boson and to the longitudinal components of
the gauge bosons, a situation which corresponds to the gaugeless limit of the
SM; of course, the interactions due to light quarks and gluons have to be kept
when considering the QCD corrections. \s

The bare Lagrangian describing the interactions of the  Higgs boson with 
fermions and vector bosons
\begin{equation}
{\cal L}=\frac{H_0}{v_0}\left(-\sum_f m_{f0}\bar f_0f_0
+2M_{W0}^2W_{0\mu}^\dag W_0^\mu+M_{Z0}^2Z^\dag_{0\mu}Z_0^\mu\right)
\end{equation}
contains the overall factor $H_0/v_0$, which undergoes a finite renormalization.
Working in the on--shell scheme, where $G_\mu$ and the physical $W$ boson mass 
are used as inputs, and performing the renormalization of all the
fields and couplings that are involved, one obtains a universal electroweak 
correction which appears in the Higgs boson couplings to all particles
\beq
\frac{H_0}{v_0} &\to & (\sqrt{2}G_\mu)^{1/2} H \left( 1- \frac{\Delta M_W^2}
{M_W^2} \right)^{-1/2}  [1+ {\rm Re} \Pi'_{HH}(M_H^2)]^{-1/2} \non \\
& \to & (\sqrt{2}G_\mu)^{1/2} H \left( 1+ \delta_u \right)
\label{univ}
\eeq
In the heavy top quark limit, one sets the momentum transfer to zero in the 
boson propagators, since $m_t \gg M_W$ and $M_H$, and extracts the leading 
components which grow as $m_t^2$. Including  the QCD corrections up
to ${\cal O}(\alpha_s^2)$  and electroweak corrections to ${\cal O} (G_\mu
m_t^2)$ to this terms, one obtains results similar to what has been obtained
for $\Delta \rho$ at this order, eq.~(\ref{deltarho}), with the Higgs boson mass
set to zero in the corrections $(\Delta \rho)^{\rm EW}$. Using the
abbreviations $x_t= G_\mu m_t^2/(8\sqrt{2} \pi^2)$ and $a_s=\alpha_s^{N_f=6}
(m_t)/\pi$, the end result for the contribution $\delta_u$ will be then 
\cite{RCalb,RCadj}
\beq
\delta_u  = x_t \left[ \frac{7}{2} + 3 \left( \frac{149}{8} -\pi^2 \right)x_t
-  \left( 3 + \frac{\pi^2}{3} \right) a_s - 56.7 a_s^2 \right]
\eeq

For the Higgs boson couplings to leptons, this is in fact the only heavy top 
quark correction which is involved, unless one moves to higher orders in the
electroweak coupling.   For the couplings to light quarks $q\neq b,t$ the 
same correction $\delta_u$  appears, except from the small ${\cal O}(x_t a_s^2)$
term which is different. However, in the case of the bottom quarks as well as
for the massive gauge bosons, there are extra contributions due to the 
exchange of the top quark in the vertices. As previously mentioned, to derive
these additional terms, one can use again the  low--energy theorem with the
additional information provided by the knowledge of the particle 
self--energies. In the case of $b$ quarks, one obtains the
non--universal correction from the Lagrangian 
\beq
{\cal L} (Hb\bar{b})= - m_b \, \bar{b} b \, \frac{H^0}{v_0} \left(1+ 
\delta_{Hbb}^{\rm non-univ} \right) = - m_b \bar{b}b \frac{H^0}{v^0} 
\left (1 -\frac{m_{t0} \partial \Sigma_{bb} }{\partial m_{t0}} \right)
\eeq
where $\Sigma_b$ is the two--point function of the $b$ quark, which receives
contributions from the top quark when exchanged together with a $W$ boson in
the propagator loop. The non universal corrections in this case is obtained to 
be \cite{RChbb}
\beq
\delta_{Hbb}^{\rm non-univ}= -3 x_t \left( 1- \frac{1}{3} a_s -11.2 a_s^2  
\right) 
\eeq
Combined to the universal corrections, $\delta_u \sim  \frac{7}{2} x_t$, this
leads to a large cancellation which gives a rather small total correction, 
$\delta_{b}=\frac{1}{2} x_t$, at the one--loop  level. \s

In the case of the massive gauge bosons, besides the correction $\delta_u$, one
has also to include a non--universal vertex correction, which is different for 
$W$ and $Z$ bosons at higher orders. Again, using the knowledge on the $W$ and 
$Z$ boson two--point functions and setting their momentum transfer to zero, 
the non--universal correction is obtained from the differentiation with 
respect to the top mass of the bare $M_V^2 V_\mu V^\mu$ interaction
\beq 
\delta_{HVV}= (1+ \delta_u) \left(1 - \frac{m_{t}^2 \partial} {\partial 
m^2_{t}} \right) \frac{\Pi_{VV}(0)}{M_V^2} 
\eeq 
One then obtains for the total heavy top quark correction at the same 
order as for the correction $\delta_u$ \cite{RCsea}
\beq
\delta_w&=& x_t \left[ -\frac{5}{2} +  \left( \frac{39}{8} - 3\pi^2 \right)x_t
+  \left(9 - \frac{\pi^2}{3} \right) a_s + 27.0 a_s^2 \right] \non \\
\delta_z&=& 
x_t \left[ -\frac{5}{2} +  \left( \frac{177}{8} +3 \pi^2 \right)x_t
+  \left( 15 - \frac{\pi^2}{3} \right) a_s + 17.1 a_s^2 \right]
\label{WWHvertex}
\eeq
Adding up all the previous results, one finds for the heavy top correction 
factor  $\delta_{HXX}^t$ in eq.~(\ref{ewcorfac}), for the fermionic and bosonic
decay widths of the Higgs boson [in which the $HXX$ coupling appears squared]
\beq
\delta^t_{HXX} &=& (1+ \delta_x)^2 -1 
\eeq

\subsubsection*{\underline{The remaining electroweak corrections}}

In the case of light fermions, the electromagnetic corrections are simply
given by \cite{RChff}
\beq
\delta_{Hff}^e =  \frac{3}{2} \frac{\alpha}{\pi} Q_f^2 \left(\frac{3}{2} -
\log \frac{M_H^2}{m_f^2} \right)
\eeq
For quark final states, the large logarithms $\log M_H^2/m_q^2$ can be absorbed 
in the running quark masses analogously to the QCD corrections. In this case, 
the electromagnetic correction, supplemented by the NLO QCD correction,  
reads \cite{HqqQED-QCD} 
\beq
\delta_{Hqq}^e =  4.2 \, Q_q^2 \, \frac{\overline{\alpha}(M_H)}{\pi} \, \left[
1+5.2 \frac{\alpha_s}{\pi} \right]
\label{HqqQED}
\eeq
The remaining weak corrections can be approximated by [the reduced vector and 
axial couplings $\hat{v}_f$ and $\hat{a}_f$ have been defined previously] 
\cite{RCreviewEW} 
\beq
\delta_{Hff}^w=   \frac{G_\mu M_Z^2}{8\sqrt{2} \pi^2} \left[
c_W^2 \left( - 5 + \frac{3}{s_W^2} \log c_W^2 \right) - \frac{6 \hat{v}_f^2 
- \hat{a}_f^2}{2} \right]
\eeq
In the case of Higgs decays into massive gauge bosons, the electromagnetic
corrections for $H \to ZZ$ are absent, while the vertex corrections and the
photon real emission in the decay  $H \to W^+ W^-$ do not form a gauge
invariant and meaningful set,  and must be combined with the photonic
contributions to the self--energies \cite{RChvv}. The remaining electroweak 
corrections [except for the ones involving the self--coupling $\lambda$] are in 
general small.\s

For Higgs boson decay into top quarks, since $m_t$ cannot be set to zero or
infinity anymore, the situation is more complicated. The electromagnetic
corrections with virtual photon exchange and  real photon emission [the
running of $\alpha$ is again taken care of by using the IBA with $G_\mu$ as
input] are the same as the QCD corrections discussed in \S2.1.3 if the strong
coupling $\alpha_s$ is replaced by the proper electromagnetic factor
\beq
\delta_{Htt}^e = \frac{3}{4} Q_t^2 \frac{\alpha}{\pi} \Delta_H^t (\beta_t)
\eeq
Because of the Coulomb singularity,  these corrections are large near
threshold, $M_H \sim 2m_t$, but are small  far above threshold leading to 
a correction less than 1\%. \s

For the electroweak corrections, which are interesting since they  involve the
Higgs contributions [and if $M_H \sim 2m_t$, mixing between the Higgs boson and
the spin zero $t\bar{t}$ bound state would occur], the expression is rather
complicated since $m_t \neq 0$ \cite{RChff}. However a simple interpolating
formula can be obtained, which approximates the full result to the level of
1\% even in the threshold region. In terms of $h_t = M_H^2 /4 m_t^2$ and
$\ell_t=\log M_H/m_t$, one has \cite{RCreviewEW}
\beq
\delta_{Htt}^w = \frac{G_\mu m_t^2}{2\sqrt{2} \pi^2}  
\left( 1+ \frac{5}{2 h_t} \right) \ell_t  \left( \ell_t -2  \right) 
+ 1.059 h_t + 3.477 + \frac{0.272}{ h_t} - \frac{1.296} {h_t^2} - 
\frac{0.182}{ h_t^3} \ \ \
\eeq
Numerically, this correction is extremely small near the threshold and 
increases monotonically to reach the level of $\sim 15$\% for $M_H \sim 1$ 
TeV.

\subsubsection*{\underline{Higgs self--coupling corrections}}

Finally, one has to include the corrections due to the triple and quartic Higgs
boson couplings. In the regime where the Higgs boson mass is large, one
obtains at two--loop order in the on--shell scheme 
\cite{Pert-HWcplg2,Pert-Hfcplg2}
\beq
\delta_{Hff}^\lambda &=& (13 - 2 \sqrt{3} \pi) \left( \frac{\lambda} {16 \pi^2}
\right) -32.66 \left( \frac{\lambda}{16 \pi^2} \right)^2 \non \\
\delta_{HVV}^\lambda &=&  \left( 19 -6 \sqrt{3} \pi - \frac{5 \pi^2}{3} \right) 
 \left( \frac{\lambda}{ 16 \pi^2} \right) +62.0 \left( 
\frac{\lambda}{16 \pi^2}\right)^2 
\eeq
Numerically, the result as a function of the Higgs boson mass is
\beq
\delta_{Hff}^\lambda &=& 0.11 \left( M_H/ {\rm 1~TeV} \right)^2
- 0.09 \left(M_H/{\rm 1~TeV} \right)^2 \non \\
\delta_{HVV}^\lambda &=& 0.15 \left( M_H/{ \rm 1~TeV} \right)^2
+ 0.17 \left( M_H/{\rm 1~TeV} \right)^2 
\eeq
As discussed in \S1.4.1, if the Higgs boson mass is very large, $M_H \sim
{\cal O}(10~{\rm TeV})$, the one loop terms of these expansions become close to
the Born terms and the perturbative series does not converge. In fact, already
for a Higgs boson mass close to $M_H \sim 1~{\rm TeV}$, the  two--loop 
contributions become as important as the one--loop contributions. Hence, for 
perturbation theory to hold,  $M_H$ should be smaller than about one TeV. In 
this mass regime, however, the total correction $\delta^\lambda_{HVV}$ is 
moderate, being at the level of $\delta_{HVV}^\lambda \sim 20$\% for $M_H \sim 
1$ TeV. In the case of fermionic decays, the total correction is even smaller, 
$\delta_{Hff}^\lambda \simeq 2$\% for $M_H \simeq 1$ TeV, because of the 
accidental cancellation of the one--loop and two--loop contributions. 

\subsubsection{NNLO QCD and EW corrections to the loop induced decays} 

\subsubsection*{\underline{The NNLO QCD corrections}}

One can use the low energy theorem discussed in \S2.4.1 to derive the 
higher--order QCD corrections to the $H\gamma \gamma$ and $Hgg$ couplings 
in the heavy top quark limit.  In the case of the $H\gamma \gamma$ operator, 
the QED/QCD $\beta$ function and the anomalous mass dimension $\gamma_m$ 
are known to four loops.  The contribution of the top quark to the $H \gamma
\gamma$ coupling at ${\cal O}(\alpha_s^2)$, with $N_f=6$ flavors and a
renormalization scale $\mu$, is found to be \cite{RCste}
\begin{equation}
{\cal L}_{\rm eff} (H\gamma\gamma) = \frac{Q_t^2\alpha}{2\pi}~\left( \sqrt{2}
G_F \right)^{1/2} \left[ 1-\frac{\alpha_s}{\pi} - \left( \frac{31}{4}
+ \frac{7}{4} \log \frac{\mu^2}{m_t^2} \right) \left( \frac{\alpha_s}{\pi} 
\right)^2 \right] F_{\mu\nu} F_{\mu\nu}~H
\label{eq:LHgam}
\end{equation}

In the case of the $Hgg$ operator in the heavy top quark limit
\beq
{\cal L}_{\rm eff} (Hgg) = \frac{H}{v} \,  G_{\mu \nu}^a 
G^{a \mu \mu} \, C_g
\eeq 
the QCD correction can be again expressed in terms of  the heavy quark 
contribution $\beta_Q(\alpha_s)$ to the QCD $\beta$ function and to the 
anomalous quark mass dimension $\gamma_m$ as 
\beq
{\cal L}_{\rm eff} (Hgg) = - \frac{\alpha_s}{4}\, \frac{H}{v} \,  G_{\mu \nu}^a 
G^{a \mu \nu} \, \frac{\beta_Q(\alpha_s)}{\alpha_s^2} \frac{1}{ 1+
\gamma_m(\alpha_s)}
\eeq
which is valid at two loops [at three loops, some subtelties appear and are
discussed in Ref.~\cite{Review-Michael} for instance]. At ${\cal O}(\alpha_s^2
)$, the anomalous quark mass dimension is given by 
\cite{anommass}
\begin{equation}
\gamma_m(\alpha_s) = 2\frac{\alpha_s}{\pi} + \left( \frac{101}{12} -
\frac{5}{18} (N_f+1) \right) \left(\frac{\alpha_s}{\pi}\right)^2
\label{eq:anommass}
\end{equation}
while the QCD $\beta$ function at NNLO in the $\overline{\rm MS}$ scheme is 
given by 
\cite{alphas-evol}
\beq
\beta_Q (\alpha_s) =  \frac{\alpha_s^2}{3\pi}\left[ 1+
\frac{19}{4} \frac{\alpha_s}{\pi} + \frac{7387 - 325 N_f}{288}
\left(\frac{\alpha_s}{\pi} \right)^2 \right] 
\end{eqnarray}
From these expressions and taking care of the fact that the $\overline{\rm MS}$
strong coupling $\alpha_s$ of the effective theory should include only the $N_f
= 5$ light flavors [see again Ref.~\cite{Review-Michael} for details], one 
arrives, using a consistent $\alpha_s$ expansion at the final result for the 
coefficient function $C_g$ at NNLO with a scale taken to be $\mu=m_t$, to 
\cite{RCkos}
\beq
C_g = - \frac{\alpha_s}{12\pi} \left[ 1+ \frac{11}{4} \frac{\alpha_s}{\pi} +
\frac{2777-201 N_f}{288} \left(\frac{\alpha_s}{\pi} \right)^2 + \cdots \right]
\label{Cg:two-loop}
\eeq
However, contrary to the two--photon case,  ${\cal L}_{\rm eff}(Hgg)$ does not
describe the $Hgg$ interaction in total: it accounts  only for the interactions
mediated by the heavy quarks directly but it does  not include the
interactions of the light fields. It must be added to the  light--quark and
gluon part of the basic QCD Lagrangian, i.e. the effective coupling has to be
inserted into the blobs of the effective diagrams shown in Fig.~2.24 for the
interaction of the Higgs boson with gluons and massless quarks. \s

\begin{figure}[hbt]
\begin{center}
\setlength{\unitlength}{1pt}
\SetWidth{1.1}
\begin{picture}(500,100)(-15,0)
\DashLine(0,50)(50,50){5}
\Gluon(50,50)(75,65){3}{4}
\Gluon(75,65)(100,80){3}{4}
\Gluon(50,50)(75,35){-3}{4}
\Gluon(75,35)(100,20){-3}{4}
\Gluon(75,65)(75,35){3}{4}
\GCirc(50,50){10}{0.5}
\put(-15,46){\bH}
\put(105,18){$g$}
\put(105,78){$g$}

\DashLine(140,50)(190,50){5}
\GlueArc(215,50)(25,0,180){3}{10}
\GlueArc(215,50)(25,180,360){3}{10}
\Gluon(240,50)(290,80){3}{7}
\Gluon(240,50)(290,20){3}{7}
\GCirc(190,50){10}{0.5}
\put(125,46){\bH}
\put(295,18){$g$}
\put(295,78){$g$}

\DashLine(330,50)(380,50){5}
\Gluon(380,50)(405,65){3}{5}
\Gluon(405,65)(430,80){3}{3}
\Gluon(405,65)(430,50){-3}{3}
\Gluon(380,50)(430,20){-3}{7}
\GCirc(380,50){10}{0.5}
\put(315,46){\bH}
\put(435,18){$g$}
\put(435,78){$g$}
\put(435,48){$g$}
\end{picture}
\end{center}
\vspace*{-5mm}
\nn {\it Figure 2.24: Effective diagrams contributing to the $Hgg$ interaction
in the limit where the top quark is heavy and has been integrated out. The 
blob represents the effective $Hgg$ coupling.}
\vspace*{-1mm}
\end{figure}

For instance, for the Higgs decay into gluons at NLO, one  adds to the 
contribution to the effective $Hgg$ coupling squared $(1+ \frac{11}{4} 
\frac{\alpha_s}{\pi})^2$,  the  gluon and light quarks contributions from
the pure gluonic virtual corrections and the real correction from $H \to ggg$
and $H \to g q\bar{q}$ with $N_f$ light quarks, $\left( \frac{73}{4} - 
\frac{7 N_f}{6} \right)$, leading to the  total contribution for $\mu^2=M_H^2$
\beq
\frac{11}{2}\frac{\alpha_s}{\pi} \ + \  \left( \frac{73}{4}-
\frac{7N_f}{6} \right) \frac{\alpha_s}{\pi}=
\left( \frac{95}{4} - \frac{7 N_f}{6} \right) \frac{\alpha_s}{\pi}
\eeq
which was given in eq.~(\ref{Hgg-EH}) for the gluonic Higgs partial width at 
NLO. \s

At NNLO, the calculation has also been done for the interaction of the
Higgs boson with the light fields and this will be discussed later when we will
address the question of Higgs production in the $gg \to H$ fusion mechanism. 
Here, we simply give the final result for the correction factor for the 
partial $H \to gg$ decay width at NNLO, for a number of light flavors $N_f=5$ 
and with a scale $\mu=M_H$, which reads \cite{RChgg}
\beq
K_{H \to gg}^{\rm QCD}= 1+ \frac{215}{12} \frac{\alpha_s(M_H)}{\pi} + 
\frac{\alpha_s^2 (M_H)}{\pi^2} \left(156.8 -5.7\log \frac{m_t^2}{M_H^2} \right)
\eeq
The three--loop correction amounts to $\sim 20$\% of the [one--loop] Born term
and $\sim 30$\% of the two--loop term, therefore showing a good convergence
behavior of the perturbative series.

\subsubsection*{\underline{Electroweak and self--coupling corrections}}

We now turn to the dominant electroweak corrections to the Higgs boson loop 
induced decays, those which are proportional to $G_\mu m_t^2$.  Again,
one can use a variant of the low energy theorem discussed previously to 
calculate the two--loop ${\cal O}(G_\mu m_t^2)$ correction to the $Hgg$
coupling \cite{HppBorn,LET}. The obtained effective $Hgg$ coupling at this 
order is given by
\beq
{\cal L}(Hgg)= (\sqrt{2}G_\mu )^{1/2} \, \frac{\alpha_s}{12\pi} \,
\, HG_{\mu \nu} G^{\mu \nu} \, (1+\delta_1+ \delta_2 +\delta_3)
\eeq
Here, $\delta_1$ is the contribution of the top quark to the QCD $\beta$ 
function at ${\cal O} (\alpha_s G_\mu m_t^2)$, which can be evaluated by
considering the two--loop diagrams where Higgs and Goldstone bosons are 
exchanged in the heavy quark loop
\beq
\frac{\beta (\alpha_s)}{g_s} = \frac{\alpha_s}{6\pi} (1+ \delta_1) \Rightarrow
\delta_1 = - 12 \frac{G_\mu m_t^2}{8\sqrt{2} \pi^2} 
\eeq
The term $\delta_2$ is simply the contribution of the anomalous quark mass 
dimension 
\beq
\delta_2 = (Z_2^Q -1) - \frac{\delta m_Q}{m_Q}  + \Gamma_{H Q \bar{Q}} (q^2=0)
= 6 \frac{G_\mu m_t^2}{8\sqrt{2} \pi^2} 
\eeq
Finally, $\delta_3$ represents  the renormalization of the Higgs wave
function and vev
\beq
\delta_3= - \frac{1}{2} \left[ \frac{ \Pi_{WW}(0)}{M_W^2} +  \frac{
\partial \Pi_{HH} (M_H^2=0)}{\partial M_H^2} \right]
= 7\frac{G_\mu m_t^2}{8\sqrt{2} \pi^2} 
\eeq
Due to the large cancellation between the three components, $\delta_1 =-12, 
\delta_2= + 6$ and $\delta_3 =7$ in units of $ \frac{1}{2}x_t=G_\mu m_t^2 
/(16  \sqrt{2} \pi^2)$, the total correction factor at this order is rather 
small. The ${\cal O}(G_\mu m_t^2)$ correction to the NLO QCD term has also 
been also calculated \cite{RCstegg} and the total correction factor for the 
gluonic decay  width is then
\beq
\delta_{Hgg}^{t} = x_t \left( 1+ 30.3 \frac{\alpha_s}{\pi} \right) 
\eeq

For $m_t \sim 180$ GeV, the total factor is very small being at the level of
0.5\%. Recently, these top quark corrections to the $H\to gg$ decays have been
calculated exactly in the mass range $M_H \lsim 2M_W$ \cite{PepeHgg}. The
numerical result turned out to be quite different from the one obtained in the
infinite top mass limit, even for a low mass Higgs boson. However, the
correction factor is still rather small. \s

The electromagnetic corrections to the $Hgg$ amplitude can be straightforwardly
adapted from those of the NLO QCD corrections to the $H\gamma \gamma$ coupling. 
Indeed, the only contributions which are involved are those in which a photon
is exchanged in the internal quark lines. One then obtains, after the 
appropriate change of the QCD and electric charge factors 
\beq
\delta_{Hgg}^{e} = - \frac{3}{4} \, Q_t^2 \, \frac{\alpha}{\pi} = -
\frac{1}{3} \frac{\alpha}{\pi}
\eeq
a correction which is extremely small, being at the per mille level.\s 

In the case of the $H\gamma \gamma$ coupling, while the correction to the 
fermionic loop can be carried out along the same lines as in the case of
the $Hgg$ coupling, for the $W$ boson loop several subtleties arise. First,
the application of the low--energy theorem is restricted to the mass range $M_H
\lsim 160$ GeV in this case. A second complication is due to the fact that 
when considering the leading $m_t$ correction, owing to QED--like Ward 
identities, there is no ${\cal O}(G_\mu^2 m_t^4)$ correction [as one notices 
from the $Hgg$ case] and the largest correction scales only quadratically 
with the top mass. In the calculation of this ${\cal O}(G_\mu^2 m_t^2)$ 
correction, one cannot simply use the gaugeless limit of the SM 
since the contributions involving virtual $W$ bosons cannot be neglected.
In fact, after integrating out the heavy fermion contribution, one has
two dimension four operators which produce ${\cal O}(G_\mu^2 m_t^2)$
corrections to the $H\gamma\gamma$ amplitude
\beq
{\cal L}(H\gamma \gamma) = (\sqrt{2}G_\mu)^{1/2} H \left(
c_1 F_{\mu \nu} F^{\mu \nu} + c_2 M_{W}^2 W^\dagger_\mu W^\mu  + \cdots \right)
\eeq
with the dots standing for the contribution of higher--order operators. While
the coefficient $c_2$ has been previously derived, one
needs to perform an explicit two--loop calculation to derive the coefficient
$c_1$.  This can be done again by considering only diagrams involving, along
with top quarks, virtual Goldstone bosons minimally coupled to photons. Once
the relevant contribution to the photon self--energy has been calculated,
one can use the low--energy theorem to relate it to the $H \gamma \gamma$
amplitude in the kinematical regime where $M_H \lsim 2M_W \lsim 2m_t$.
The calculation has been performed in Ref.~\cite{RChppnew} and the obtained
correction factor can be attributed to the $W$ amplitude and written as
\beq
A_1^H (\tau_W) \to A_1^H (\tau_W) \left( 1- 2.9 x_t \right)  
\eeq
The total correction decreases the $H\to \gamma \gamma$ decay width by 
approximately 2.5\% and, thus, fully cancels the positive ${\cal O} 
(\alpha_s/\pi)$ QCD correction in the heavy top limit.\s

In the loop induced decays, there are also corrections due to the light 
fermions, $f\neq t$. At the one--loop level, these contributions are suppressed
by their couplings to the Higgs boson and are thus negligible. However, at the
two--loop level, one can avoid this suppression by coupling them to the $W$ and
$Z$ bosons which are then directly attached to the Higgs boson. These 
corrections have been calculated only recently \cite{RCita}. \s

In the case of the $H\to gg$ decay, the light quark contributions generate a
correction to the partial decay width that is positive and increases from $\sim
4.5\%$ at $M_H \sim 115$ GeV to $\sim 9 \%$ at $M_H \sim 2M_W$ [the correction
varies from 4.5\% to 7.5\% in this mass range, if the the heavy top
contribution is included]. Above this value, the correction decreases sharply
and stays below $-2\%$ for $M_H \lsim 2m_t$. In the case of the $H \to
\gamma \gamma$ decay, below the $2M_W$ threshold, the light fermion
contribution leads to a correction of the same size as the QCD correction, i.e.
$\sim 1\%$, but with opposite sign. Above the $2M_W$ threshold, the corrections 
are larger and lead to a suppression of the decay width by a few percent. \s

Finally, one has to include the Higgs self--coupling corrections which appear 
only in the bosonic contribution to the $H\gamma\gamma$ amplitude at two loops. 
The calculation can be done using the equivalence theorem 
where the $W$ boson is replaced by its corresponding Goldstone boson which
can be taken as massless [but only at the end of the calculation, since it 
serves as an infrared cut--off in intermediate steps]. In this limit
one obtains \cite{Hpp-MH2} 
\beq
\delta_{H \gamma \gamma}^{\lambda} =  - 12.1 \frac{\lambda^2}{16 \pi^2} 
\eeq
The correction is small for $M_H \lsim 500$ GeV, but is significant for 
values $M_H \sim 650$ GeV where the amplitude almost vanishes because
of the $t$ and $W$ negative interference. For $M_H \sim 1$ 
TeV, the correction becomes large and decreases the partial width by 
approximately $-30\%$. 

\subsubsection{Summary of the corrections to hadronic Higgs decays} 

Let us finally reconsider the QCD corrections to the hadronic Higgs boson
decays in the light of all the corrections that have been discussed previously.
As already mentioned, at higher orders, the Higgs decays into gluons and light
quarks are mixed and already at the next--to--leading order, the two decays $H
\to  gg^* \to g \bar{q}q$ and $H \to \bar{q}^* q \to g\bar{q}q$ lead the same
final states.  The two decays cannot therefore be considered separately at
higher orders.  The present knowledge of the higher--order QCD corrections [and
the leading electroweak corrections] to the full decay $H \to {\rm hadrons}$
has been discussed in detail in Ref.~\cite{RCreviewQCD}. In this subsection, we 
will simply give the full result for the hadronic Higgs decay width that one 
obtains for $M_H \lsim 2M_W$ by including all the corrections which are known 
up to ${\cal O}(\alpha_s^3), {\cal O}(\alpha \alpha_s), {\cal O}(G_\mu m_t^2 
\alpha_s^2)$ and ${\cal O}(\lambda^2)$.\s

Writing the interaction Lagrangian of the Higgs boson with quarks and gluons as
\beq
{\cal L}_{\rm had} = \sqrt{2} G_\mu H \bigg[ m_q \bar{q} q C_{q} 
+ G_{\mu \nu}^a G^{\mu \nu}_a  C_g  \bigg] 
\eeq
the decay width of the Higgs boson, summing the gluonic and light--quark decays
and working in the approximation of an infinitely heavy top quark, can be 
written as \cite{RCreviewQCD}
\beq
\Gamma (H \to {\rm hadrons}) = \sum_q  A_{q\bar{q}} K^{\rm EW}_{H \to 
qq} \left[ \left( 1+ \Delta_{qq} \right) \left( C_{q} \right)^2 +\Delta_{qg} 
C_g C_{q} + \delta_{q}^{\rm me}  \right]  \non \\
+ A_{gg}  K^{\rm EW}_{H \to gg} \left[ \Delta_{gg} (C_g)^2 + \delta_g^{\rm me} 
\right] 
\hspace*{3.5cm}
\label{decay-hadrons}
\eeq
where the tree--level $q\bar q$ and $gg$ squared amplitudes are given by
\beq
A_{qq} =  \frac{3G_\mu M_H}{4\sqrt{2} \pi} \overline{m}_q^2 (M_H^2) \ \ , \ \ 
A_{gg} =  \frac{4G_\mu M_H^3}{\sqrt{2} \pi} 
\eeq
and the coefficients of the operators appearing in the Lagrangian, by
\beq
&& \hspace*{-1cm}
C_{g} = - \frac{1}{12} \frac{\alpha_s}{\pi}  \left[ 1  
+  \left( \frac{11}{4} - \frac{1}{6} \ell_t \right) \frac{\alpha_s}{\pi} 
+ \bigg( 9.35 -0.7 N_f +(0.33 N_f-0.52 \ell_t) + 0.028 \ell_t^2 \bigg) \left( 
\frac{\alpha_s}{\pi} \right)^2 \right] 
\non \\ && \hspace*{-1cm}
C_{q} = 1+ \left( \frac{5}{18} -\frac{1}{3} \ell_t \right) \left( \frac{ 
\alpha_s}{\pi} \right)^2 + \bigg( 1.35+0.25N_f -2.9 \ell_t +  (0.056 N_f - 0.8) 
\ell_t^2 \bigg)  \left( \frac{\alpha_s}{\pi} \right)^3  
\eeq
with $\alpha_s \equiv \alpha_s^5 (M_H^2)$ defined at the scale $M_H$ with 
$N_f=5$ light quarks and $\ell_t=\log(M_H^2/m_t^2)$. The various terms 
appearing in equation eq.~(\ref{decay-hadrons}), are as follows: \s

$\bullet$ $\Delta_{qq}$ is the pure QCD corrections to the decays into quarks
eq.~(\ref{Delta-qq}) up to ${\cal O}(\alpha_s^3)$, supplemented by the 
contributions of order $\alpha$ and the mixed QCD/QED contribution at 
${\cal O}(\alpha \alpha_s)$ eq.~(\ref{HqqQED})
\beq
\Delta_{qq} &=&  \frac{\alpha_s}{\pi} \bigg[ \frac{17}{3} + (35.94- 1.36 
N_f) \frac{\alpha_s}{\pi} + (164.14-25.77 N_f + 0.26 N_f^2) \left( 
\frac{\alpha_s}{\pi} \right)^2 \bigg] \non \\
&& + \frac{\alpha (M_H)}{\pi}Q_q^2 \bigg[ 4.25 + 11.71
\frac{\alpha_s} {\pi} \bigg]  
\eeq

$\bullet$ $\Delta_{gg}$ is the QCD correction to the gluonic decay mode
due to the light quark and gluon fields
\beq
\Delta_{gg} =  1 +  (18.25-1.17 N_f) \frac{\alpha_s}{\pi} +
(243 -39.4 N_f +0.9N_f^2) \left(\frac{\alpha_s}{\pi} \right)^2 
\eeq

$\bullet$ $\Delta_{gq}$ is the mixed contribution in quark and gluon Higgs 
decays
\beq
\Delta_{qg} = - \frac{\alpha_s}{\pi} \left[ 30.67 + (524.85 -20.65 N_f) 
\frac{\alpha_s}{\pi} \right]
\eeq 

If one considers final states involving quarks only, one has to subtract  
from the previous equation the gluonic contribution as discussed previously;
at ${\cal O}(\alpha_s)$, one has for instance
\beq
\Delta_{gg}' =  \frac{\alpha_s}{\pi} \left[13.56 -\frac{4}{3}\log^2 
(m_q^2/M_H^2)+ {\cal O} \left( \frac{\alpha_s^2}{\pi^2} \right) \right]
\eeq

$\bullet$ $K^{\rm EW}_{H\to qq}$ and $K^{\rm EW}_{H\to gg}$ are the sum of the 
electroweak corrections  for the quark and gluonic decays discussed previously
[but without the electromagnetic corrections for the former decay since they
are included in $\Delta_{qq}$]. Note that in this case, $\alpha_s$ is defined
at the scale $m_t$.  \s

$\bullet$ Finally, $\delta_q^{\rm me}$ and $\delta_g^{\rm me}$ are the 
remaining contributions that contain the light quark masses and non--leading  
terms in $m_t$ in fermionic and gluonic Higgs decays. Since higher--order terms 
${\cal O}(M_H^4/m_t^4)$ and ${\cal O}(\bar{m}_b^4/M_H^4)$ are very 
small for $M_Z \lsim M_H \lsim 2M_W$, one can simply retain the first terms
in the $M_H^2/m_t^2$ and $\bar{m}_q^2/M_H^2$ expansions 
\beq
\delta_q^{\rm me} &=& \left( \frac{\alpha_s}{\pi} \right)^2 \frac{M_H^2}
{m_t^2} \left[0.241 -0.07 \log \frac{M_H^2}{m_t^2} \right] 
- 6 \frac{\bar{m}_q^2}{M_H^2} \left[ 1+ 6.67 \frac{\alpha_s}{\pi} \right]
\non \\
\delta_g^{\rm me} &=& 0.1167 \frac{M_H^2} {m_t^2} \left( \frac{\alpha_s}{\pi} 
\right)^2 \left[ 1+ \left (17.85-2 \log \frac{M_H^2}{m_t^2} \right) 
\frac{\alpha_s}{\pi} \right]
\eeq
This completes the discussion of the main QCD and electroweak radiative 
corrections to the hadronic decays of an intermediate mass Higgs boson.

\subsection{The total decay width and the Higgs branching ratios}

The decay branching ratios and the total width of the SM Higgs boson  are shown
in Figs.~2.25 and 2.26, respectively, as a function of the Higgs mass.  They
have been obtained using the {\sc Fortran} code {\tt HDECAY} \cite{HDECAY} with
the fermion and gauge boson mass inputs of eq.~(\ref{allmasses}) and with the
strong coupling constant normalized to $\alpha_s(M_Z)=0.1172$. Included are all
decay channels that are kinematically allowed and that have branching ratios
larger than $10^{-4}$, {\it y compris} the loop mediated, the three body
$\bar{t}t^*$ and $VV^*$ decay modes and the double off--shell decays of the 
Higgs boson into massive gauge bosons which then decay into four massless 
fermions. 
In addition,  all relevant two--loop QCD corrections to the decays into quark
pairs and  to the quark loop mediated decays into gluons [and photons] are
incorporated; the smaller leading electroweak
radiative corrections are also included. To be as complete as possible, we also
present in Table  2.1 the numerical values of the branching ratios and total
decay width for selected values of $M_H$, as it might be useful to have a
normalization as close as possible to the state of the art, to be used in other
theoretical or experimental studies.\s

To discuss the Higgs decays, it is useful to consider three distinct mass 
ranges: 
\begin{itemize}
\vspace*{-2mm}

\item[$\bullet$] the ``low mass" range 110 GeV $ \lsim M_H \lsim 130$ GeV,
\vspace*{-2mm}

\item[$\bullet$] the ``intermediate  mass" range 130 GeV $\lsim M_H \lsim 180$ 
GeV, \vspace*{-2mm}

\item[$\bullet$] the ``high mass" range 180 GeV $\lsim M_H \lsim 1$ TeV.
\vspace*{-2mm}
\end{itemize}

\nn The main features of the branching ratios and total width can
be summarized as follows.\s

In the ``low mass" range, 100 GeV $\lsim M_{H}\lsim 130$ GeV,  the main decay
mode of the Higgs boson  is by far $ H \ra b\bar{b}$ with a branching ratio  of
$\sim \, $75--50\% for $M_H = 115$--130 GeV, followed by the decays into 
$\tau^+\tau^-$ and $c\bar{c}$ pairs with branching ratios of the order of 
$\sim$ 7--5\% and $\sim$ 3--2\%, respectively. Also of significance is the 
$H\to gg$ decay with a branching fraction of $\sim \, $7\% for $M_{H} \sim 120$
GeV. The $\gamma \gamma$ and $Z \gamma$ decays are rare, with branching ratios 
at the level of a few per mille, while the decays into pairs of muons and 
strange quarks [where $\bar{m}_s (1~{\rm GeV})=0.2$ GeV is used as input] are 
at the level of a few times $10^{-4}$. The $H \to WW^*$ decays, which are 
below the 1\% level for $M_H \sim 100$ GeV, dramatically increase with $M_H$ 
to reach $\sim 30\%$ at $M_H \sim 130$ GeV; for this mass value, $H\to 
ZZ^*$ occurs at the percent level.\s 

In the ``intermediate mass"  range, the Higgs boson decays mainly  into $WW$
and $ZZ$ pairs, with one virtual gauge boson below the $2M_V$ kinematical 
thresholds. The only other decay mode which survives is the $b\bar{b}$ decay 
which has a branching ratio that drops from 50\% at $M_H \sim 130$ GeV to the 
level of a few  percent for $M_H \sim 2M_W$. The $WW$ decay starts to dominate
at $M_H \sim 130$ GeV and becomes gradually overwhelming, in particular for 
$2M_W \lsim M_H \lsim 2M_Z$ where the $W$ boson is real [and thus the decay 
$H \to WW$ occurs at the two--body level] while the $Z$ boson is still virtual,
strongly suppressing the $H \to ZZ^*$ mode and leading to a $WW$ branching ratio
of almost 100\%.  \s

\begin{figure}[!h]
\begin{center}
\vspace*{-2.cm}
\hspace*{-3cm}
\epsfig{file=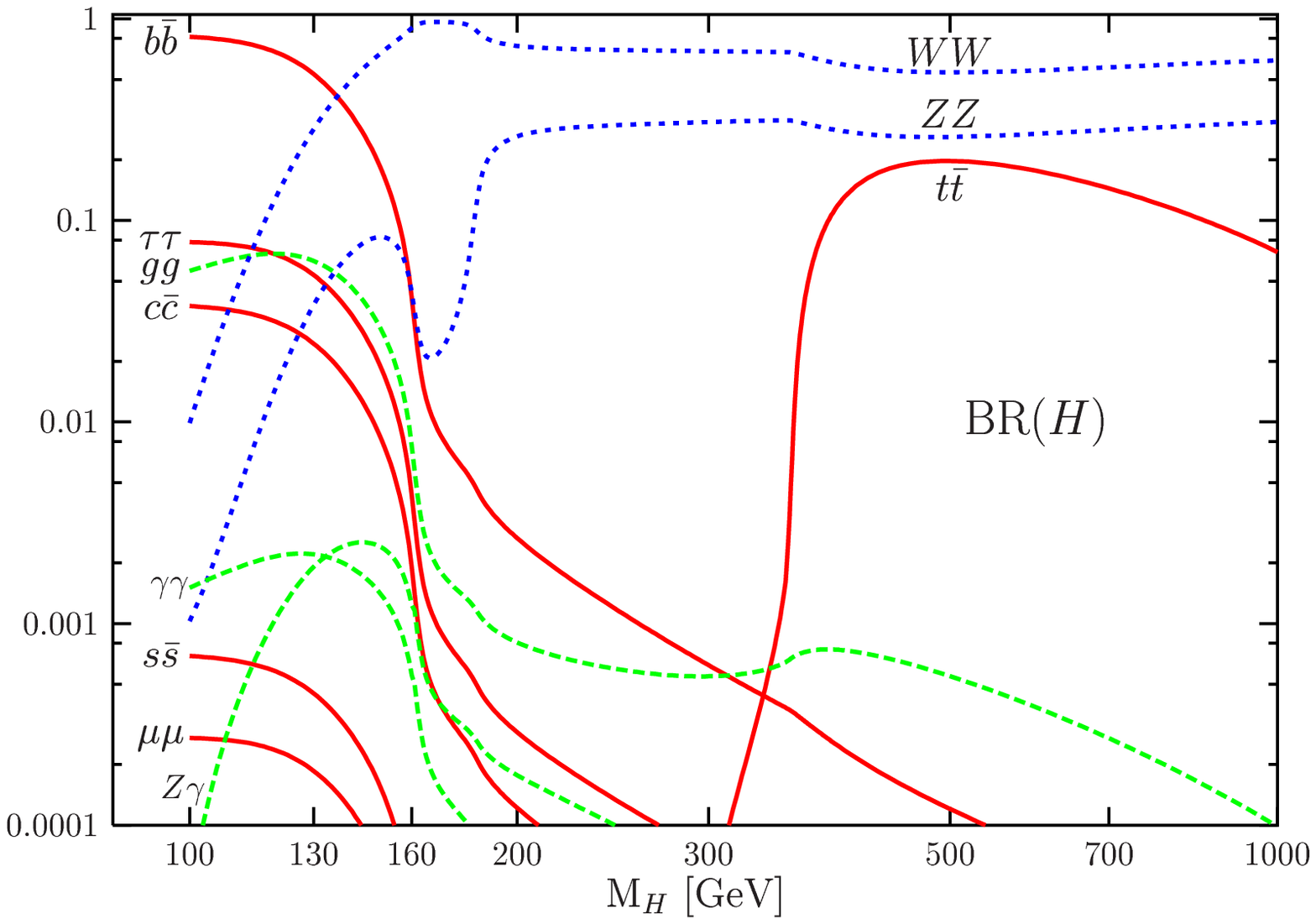,width=17.cm} 
\end{center}
\vspace*{-12.5cm}
\centerline{\it Figure 2.25: The SM Higgs boson decay branching ratios as a 
function of $M_H$.}
\end{figure}

\begin{figure}[!h]
\begin{center}
\vspace*{-2.cm}
\hspace*{-3cm}
\epsfig{file=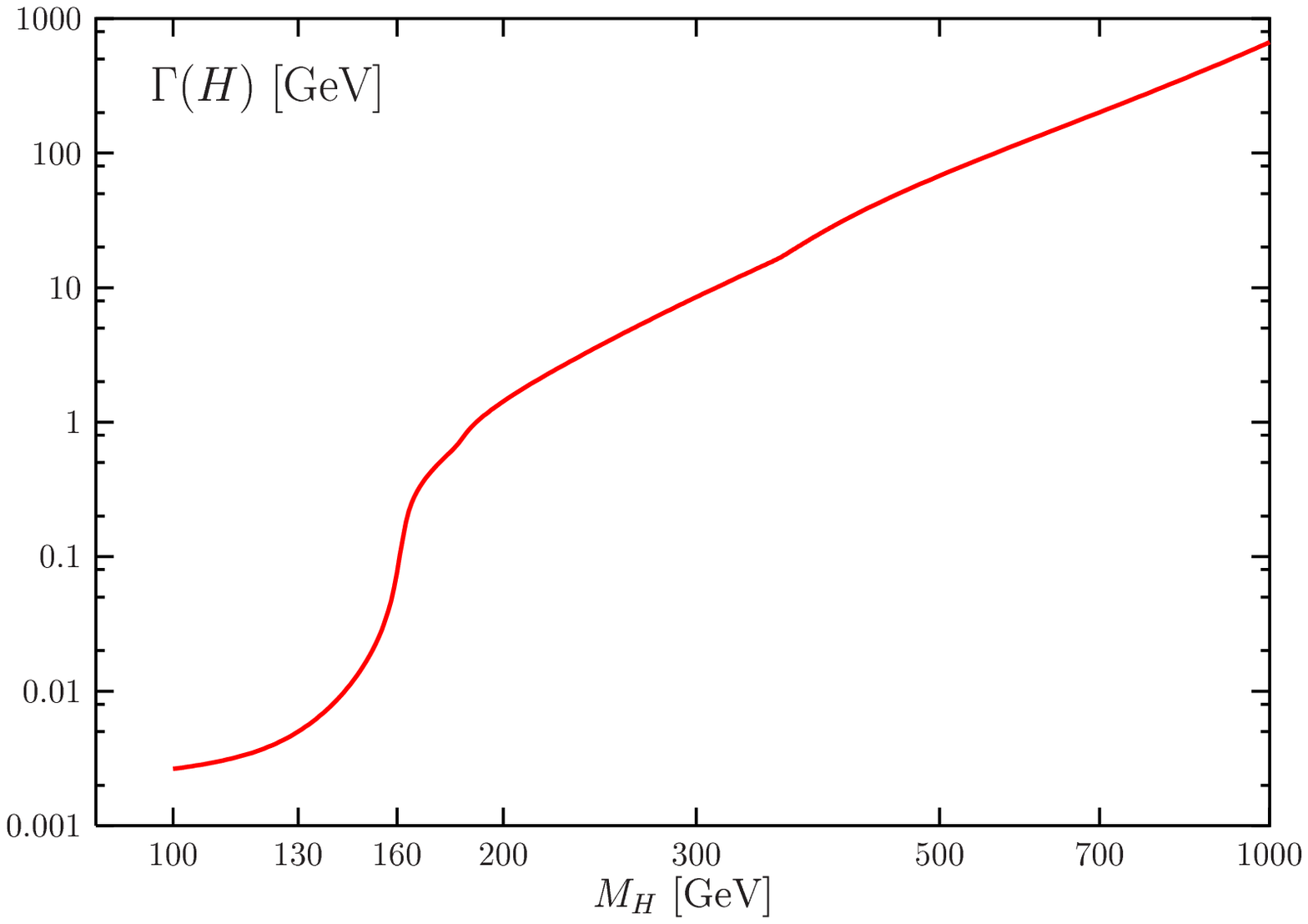,width=17.cm} 
\end{center}
\vspace*{-12.5cm}
\centerline{\it Figure 2.26: The SM Higgs boson total decay width as a function
of $M_H$.} 
\end{figure}

\begin{table}[htbp]
\begin{center}
\renewcommand{\arraystretch}{1.25}
\begin{tabular}{|c||c|c|c|c|c|c|}\hline
$M_H$ (GeV) & \ BR$(b\bar b)$ \ & \ BR($\tau \tau)$ \ &   \ BR$(\mu \mu)$ \ & 
BR($s \bar s)$ & \  BR$(c \bar c)$ \ & BR$(t \bar t)$ \\ \hline \hline
115 & 0.736 & $7.21 \cdot\! 10^{-2}$ & $2.51 \cdot\! 10^{-4}$ & $6.23 \cdot\! 
10^{-4}$ & $3.39 \cdot\! 10^{-2}$ & -- \\
120 & 0.683 & $6.78 \cdot\! 10^{-2}$ & $2.35 \cdot\! 10^{-4}$ & $5.79 \cdot\! 
10^{-4}$ & $3.15 \cdot\! 10^{-2}$ & -- \\
130 & 0.533 & $5.36 \cdot\! 10^{-2}$ & $1.86 \cdot\! 10^{-4}$ & $4.51 \cdot\! 
10^{-4}$ & $2.45 \cdot\! 10^{-2}$ & -- \\
140 & 0.349 & $3.56 \cdot\! 10^{-2}$ & $1.23 \cdot\! 10^{-4}$ & $2.95 \cdot\! 
10^{-4}$ & $1.60 \cdot\! 10^{-2}$ & -- \\
150 & 0.179 & $1.85 \cdot\! 10^{-2}$ & -- & $1.51 \cdot\! 10^{-4}$ & $8.23 
\cdot\! 10^{-3}$ & -- \\
160 & $4.11 \cdot \! 10^{-2}$ & $4.30 \cdot\! 10^{-3}$ & -- & -- & $1.89 
\cdot\! 10^{-3}$ & -- \\
170 & $8.64 \cdot\! 10^{-3}$ & $9.13 \cdot\! 10^{-4}$ & -- & -- & $3.97 
\cdot\! 10^{-4}$ & -- \\
180 & $5.53 \cdot\! 10^{-3}$ & $5.90 \cdot\! 10^{-4}$ & -- & -- & $2.54 
\cdot\! 10^{-4}$ & -- \\
200 & $2.65 \cdot\! 10^{-3}$ & $2.89 \cdot\! 10^{-4}$ & -- & -- & $1.22 
\cdot\! 10^{-4}$ & -- \\
300 & $6.21 \cdot\! 10^{-4}$ & -- & -- & -- & -- & -- \\
400 & $2.35 \cdot\! 10^{-4}$ & -- & -- & -- & -- & $0.131$ \\
500 & $1.20 \cdot\! 10^{-4}$ & -- & -- & -- & -- & $0.197$ \\
600 & -- & -- & -- & -- & -- & $0.176$ \\
700 & -- & -- & -- & -- & -- & $0.144$ \\
1000 & -- & -- & -- & -- & -- & $0.070$ \\ \hline
\end{tabular}
\end{center}
\vspace*{0mm}
\end{table}

\begin{table}[htbp]
\begin{center}
\renewcommand{\arraystretch}{1.25}
\begin{tabular}{|c||c|c|c|c|c||c|}\hline
$M_H$ (GeV) & BR$(gg)$ \ & BR $(\gamma \gamma)$ & BR($Z \gamma)$ & BR$(WW)$ & 
$BR(ZZ)$  & $\Gamma_H$ (GeV) \\ \hline \hline
115 & $6.74 \cdot\! 10^{-2}$ & $2.04 \cdot\! 10^{-3}$ & $6.75 \cdot\! 10^{-4}$ 
& $7.48 \cdot\! 10^{-2}$ & $8.04 \cdot\! 10^{-3}$ & $3.27 \cdot \! 10^{-3}$\\
120 & $6.84 \cdot\! 10^{-2}$ & $2.16 \cdot\! 10^{-3}$ & $1.06 \cdot\! 10^{-3}$ 
& 0.130& $1.49 \cdot\! 10^{-2}$ & $3.65 \cdot \! 10^{-3}$\\
130 & $6.30 \cdot\! 10^{-2}$ & $2.21 \cdot\! 10^{-3}$ & $1.91 \cdot\! 10^{-3}$ 
& $0.283$ & $3.80 \cdot\! 10^{-2}$ & $5.00 \cdot \! 10^{-3}$\\
140 & $4.82 \cdot\! 10^{-2}$ & $1.93 \cdot\! 10^{-3}$ & $2.47 \cdot\! 10^{-3}$ 
& 0.480 & $6.71 \cdot\! 10^{-2}$ & $8.11 \cdot \! 10^{-3}$\\
150 & $2.87 \cdot\! 10^{-2}$ & $1.39 \cdot\! 10^{-3}$ & $2.39 \cdot\! 10^{-3}$ 
& 0.679 & $8.27 \cdot\! 10^{-2}$ & $1.67 \cdot \! 10^{-2}$\\
160 &  $7.57 \cdot\! 10^{-3}$ & $5.54 \cdot\! 10^{-4}$ & $1.23 \cdot\! 10^{-3}$ 
& 0.900 & $4.36 \cdot\! 10^{-1}$ & $0.77 \cdot \! 10^{-1}$\\
170 &  $1.82 \cdot\! 10^{-3}$ & $1.50 \cdot\! 10^{-4}$ & $3.97 \cdot\! 10^{-4}$ 
& 0.965 & $2.25 \cdot\! 10^{-2}$ & $0.383$ \\
180 & $1.32 \cdot\! 10^{-3}$ & $1.02 \cdot\! 10^{-4}$ & $2.98 \cdot\! 10^{-4}$ 
& 0.934 & $5.75 \cdot\! 10^{-1}$ & $0.628$ \\
200 & $8.06 \cdot\! 10^{-4}$ & -- & $1.77 \cdot\! 10^{-4}$ 
& 0.735 & $0.261$ & $1.425$ \\
300 & $5.47 \cdot\! 10^{-4}$ & -- & --& 0.691 & $0.307$ & $8.50$ \\
400 & $7.37 \cdot\! 10^{-4}$ & -- & --& 0.592 & $0.276$ & $28.65$ \\
500 & $5.48 \cdot\! 10^{-4}$ & -- & --& 0.542 & $0.260$ & $67.81$ \\
600 & $3.84 \cdot\! 10^{-4}$ & -- & --& 0.554 & $0.269$ & $123.3$ \\
700 & $2.70 \cdot\! 10^{-4}$ & -- & --& 0.575 & $0.281$ & $201.3$ \\
1000 & -- & -- & --& 0.622 & $0.308$ & $667.2$ \\
\hline
\end{tabular}
\end{center}
\vspace*{0mm}
\centerline{\it Table 2.1: The Higgs decay branching ratios and total widths in the SM.}
\end{table}

\newpage

In the ``high mass" range, $M_H \gsim 2M_Z$, the Higgs boson decays exclusively
into the massive gauge boson channels with a branching ratio of $\sim 2/3$ for
$WW$ and $\sim 1/3$ for $ZZ$ final states, slightly above the $ZZ$ threshold. 
The opening of the $t\bar{t}$ channel for $M_H \gsim 350$ GeV does not alter
significantly this pattern, in particular for high Higgs masses: the $H \to
t\bar{t}$ branching ratio is at the level of 20\% slightly above the $2m_t$
threshold and starts decreasing for $M_H \sim 500$ GeV to reach a level
below 10\% at $M_H \sim 800$ GeV.  The reason is that while the $H \to
t\bar{t}$ partial decay width grows as $M_H$, the partial decay width
into (longitudinal) gauge bosons increases as $M_H^3$. \s

Finally, for the total decay width, the Higgs boson is very narrow in the low
mass range, $\Gamma_{H} <10$ MeV, but the width becomes rapidly wider for
masses larger than 130 GeV, reaching $\sim 1$ GeV slightly above the $ZZ$
threshold. For larger Higgs masses,  $M_{H} \gsim 500$ GeV, the Higgs boson
becomes obese: its decay width is comparable to its mass because of the
longitudinal gauge boson contributions in the decays $H \to WW,ZZ$. For $M_H
\sim 1$ TeV, one has a total decay width of $\Gamma_H \sim 700$ GeV, resulting 
in a very broad resonant structure.  However, as previously discussed, for this
large Higgs mass value, perturbation theory is jeopardized anyway.\s

A final word must be devoted to the uncertainties on these Higgs decay
branching ratios. As discussed at length in this section, the strong coupling
constant $\alpha_s$ and the quark masses play a prominent role in Higgs
physics. However, these parameters are affected by relatively large experimental
errors which then translate into sizable uncertainties in the Higgs boson
decay branching ratios and in the total decay width\footnote{Thus, contrary to
what is sometimes claimed in the literature, these as not ``theoretical errors"
but mostly a reflection of the poor knowledge of the quark masses and QCD
coupling constant.}. Following Ref.~\cite{DSZ}, and using the updated values of
the quark masses given in eq.~(\ref{allmasses}) and of $\alpha_s (M_Z)=0.1172
\pm 0.002$, we show in Fig.~2.27 the effect of varying the input parameters
[but only one at a time] by one standard deviation from their central values.\s

In the low to intermediate mass range where the Higgs decays into light quarks
and gluons are significant, these errors are rather large. In particular, the
branching ratios for the charm and gluonic decays have uncertainties at the
level of 20\% and 10\%, respectively. The main reason for these errors is
the $\sim 2\%$ uncertainty in $\alpha_s$, which translates into a $4\%$ ($6\%$)
error in $\Gamma(H \to gg) \propto \alpha_s^2 \, (\alpha_s^3)$ at the one--
(two)--loop level, and in a very strong variation of the charm quark mass, 
$m_c(\mu)
\sim [\alpha_s(\mu)]^{12/13}$, at the high scales. The error on $m_t$ does
not affect substantially the $H \to gg$ branching ratio since, as already
noticed, the heavy top  quark limit is a good approximation for these Higgs
mass values.  The uncertainty on the dominant $H \to b\bar{b}$ branching ratio
is small since the experimental error on the $b$--quark mass is relatively
smaller and its running is less important than in the case of charm quarks; in
addition for low Higgs masses, $\Gamma(H \to b\bar{b})$ controls the total
width and most of the uncertainty cancels in the branching ratio. The error on
the $H\to \tau^+ \tau^-$ branching ratio is simply due to that of $\Gamma(H \to
b\bar{b})$ in the total Higgs decay width.\s 

[Note that, in the high mass range above the $t\bar{t}$ threshold, the errors
on the top quark mass and the strong coupling constant do not affect
significantly the branching fraction of the $H\to t\bar{t}$ decay, the error
being at the percent level for $M_H \gsim 500$ GeV, and {\it a fortiori} the
branching ratios for $H \to WW,ZZ$ which dominate in this Higgs mass range.]\s

Thus, although the expected hierarchy of the Higgs decay modes is still visible
from Fig.~2.27, a more precise measurement of $\alpha_s$ and the quark masses 
will be necessary to check completely the predictions of the SM for the Higgs
decay branching ratios which, as will be discussed in the next sections, can be 
measured at the level of a few percent. In turn, if we are confident enough 
that the observed Higgs is the SM Higgs particle, one can turn the experimental
measurement of the branching ratios into a determination of the light quark
masses and $\alpha_s$ at the scale of the Higgs mass, in much the same 
way as the running $b$--quark mass has been determined in $Z$ decays at
LEP1 \cite{Narison}.  

\begin{figure}[!h]
\begin{center}
\vspace*{-3.cm}
\hspace*{-2cm}
\epsfig{file=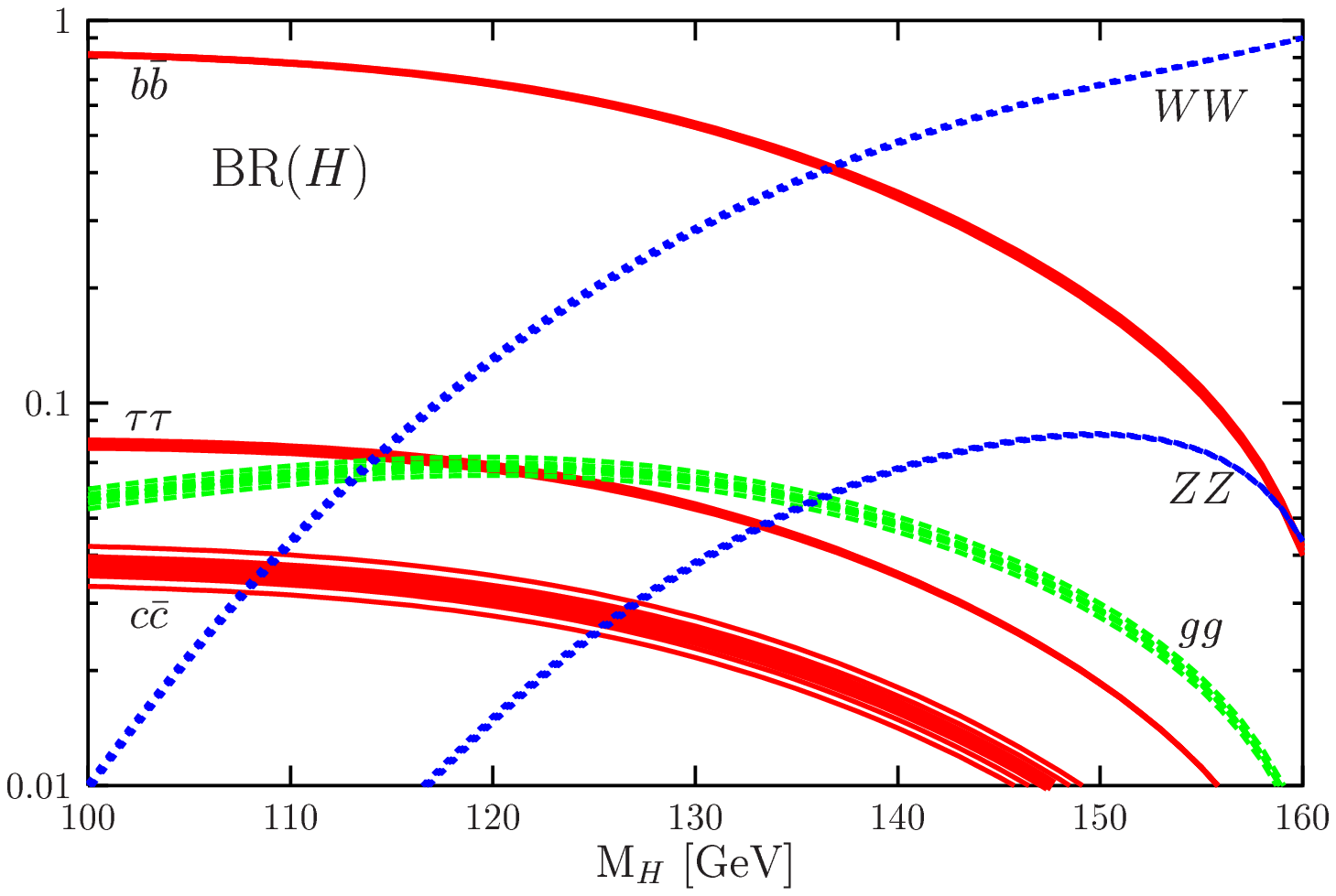,width=19.cm} 
\end{center}
\vspace*{-15.5cm}
\nn {\it Figure 2.27: The SM Higgs boson decay branching ratios in the low and
intermediate Higgs mass range including the uncertainties from the quark masses
$m_t=178 \pm 4.3$ GeV, $m_b=4.88 \pm 0.07$ GeV and $m_c=1.64 \pm 0.07$ GeV as 
well as from $\alpha_s(M_Z)=0.1172 \pm 0.002$.}
\end{figure}

\newpage

\section{Higgs production at hadron colliders}
\setcounter{equation}{0}
\renewcommand{\theequation}{3.\arabic{equation}}

\vspace*{-2mm}
\subsection{Higgs bosons at hadron machines}

\subsubsection{Generalities about hadron colliders}

The $p\bar{p}$ collider\footnote{For simplicity, we will use sometimes the
notation $pp$ for both $pp$ and $p\bar{p}$ collisions in this review.} Tevatron
at Fermilab is the highest energy accelerator available today. In the previous
Run I, the collider was operating at an energy of $\sqrt{s}=1.8$ TeV in the
$p\bar p$ center of mass, and the CDF and D\O\ experiments each have collected
data corresponding to about $\int {\cal L}{\rm d}t\sim 100$ pb$^{-1}$ of
integrated luminosity\footnote{Also for simplicity, we will denote by ${\cal
L}$ both the instantaneous and integrated luminosities.}. The upgrade with the
Main Injector allows the machine to possibly deliver an order of magnitude more
instantaneous luminosity. In Run II, it is expected that 5 fb$^{-1}$ of data
will be collected, with the possibility of increasing the sample to 10
fb$^{-1}$ if the machine runs efficiently until the end of the decade
\cite{Garbincius-Moriond};  see Ref.~\cite{Lumi-TeV} for the luminosity
delivered by the machine.  In Run II, the energy of the machine has been raised
from $\sqrt s = 1.8$ TeV to $\sqrt s = 1.96$ TeV which, typically, increased
the cross sections for some physics processes by about 30\%. The CDF and D\O\
detectors have also been upgraded, allowing them to make more sensitive
searches than previously \cite{pp-HV-expT,Higgs-TeV}.\s

The CERN Large Hadron Collider (LHC) under construction is a $pp$ collider
designed to run at an energy $\sqrt{s}=14$ TeV in the $pp$ center of mass and a
luminosity of ${\cal L}=10^{34}\, {\rm cm}^{-2}\, {\rm s}^{-1}$ (high
luminosity regime). The first collisions are expected in June 2007 but only
with an instantaneous luminosity of ${\cal L}=10^{33\, }{\rm cm}^{-2} \, {\rm
s}^{-1}$ (low luminosity regime); see \cite{LHC-status,Vienna}. At the end of
the decade, the accumulated integrated luminosity is expected to be ${\cal L}=
30$ fb$^{-1}$, to be increased to 100 fb$^{-1}$ per year when the machine runs
at the design luminosity. The hope is to collect at least 300 fb$^{-1}$ of data
per experiment during the entire LHC operation \cite{LHC-status}. There are
plans, the so-called SLHC, to operate the LHC at still the same energy
$\sqrt{s} \sim 14$ TeV, i.e. retaining the present magnets and dipoles, but at
the luminosity of ${\cal L}=10^{35}\, {\rm cm}^{-2} \, {\rm s}^{-1}$ leading to
1 ab$^{-1}$ integrated luminosity per year \cite{SLHC,SLHC+VLHC,Snowmass2001}. 
With new magnets with field strengths of approximately 16 Tesla (which do not
currently exist), the energy of the collider could be raised to $\sqrt{s}=28$
TeV.  Designs for a very large hadron collider (VLHC), with a c.m. of mass
energy of the order of 40 TeV to 200 TeV [a revival of the ancient Eloisatron
idea, see Ref.~\cite{LaThuile} for instance], are currently studied
\cite{VLHC0,VLHC}.  The SLHC and VLHC options will only be briefly discussed in
this report.\s

The two general purpose experiments under construction, ATLAS
\cite{ATLAS-TP,ATLAS-TDR} and CMS \cite{CMS-TDR,CMS-TDR-True}, have been
optimized to cover a large spectrum of possible signatures in the LHC
environment \cite{CMS+ATLAS-Vienna}.  However, the Higgs search, together with
Supersymmetry, has been the major guide to define the detector requirements and
performances for the experiments, and most of the simulation studies have been
performed for these two physics cases.\s

The total cross section at hadron colliders is extremely large. It is about 100
mb at the LHC, resulting in an interaction rate of $\approx$ 10$^{9}$ Hz at the
design luminosity. In this hostile environment, the detection of processes with
signal to total hadronic cross section ratios of about 10$^{-10}$, as is the
case for the production a SM Higgs boson in most channels, will be a difficult
experimental challenge
\cite{HiggsLHC,HiggsLHC0,ATLAS-review,CMS-review,ATLAS+CMS,LHC-Vienna,Houches19
99,Houches 2001,Houches2003}.  The huge QCD--jet backgrounds prevents from
detecting the produced Higgs boson [and any particle in general] in fully
hadronic modes.  Recalling that when ignoring the light quark and gluon modes,
the Higgs decays mostly into $b\bar b, \tau \tau, WW, ZZ$ and $\gamma \gamma,
Z\gamma$ final states in the mass range below $M_H \lsim 160$ GeV and into
$WW,ZZ$ and $t\bar t$ final states above  this mass value, the following
general requirements have to be met in order to extract a signal in the entire
Higgs mass range: \s

-- In the decay $H\to WW, ZZ$, at least one of the $W/Z$ bosons has to be
observed in its leptonic decays which have small branching ratios, BR$(W
\to \ell \nu) \simeq 20\%$ with $\ell =\mu,e$ and BR$(Z \to \ell^+ \ell^-)
\simeq 6\%$; in the latter case the invisible neutrino decays, BR$(Z \to \nu
\nu) \simeq 18\%$, can also be sometimes used to increase the statistics. A
very good detection of isolated high transverse momentum muons and electrons
and an accurate calorimetry with hermetic coverage to measure the transverse
energy of the missing neutrinos is thus required. \s

-- A very high resolution on the photons is necessary to isolate the narrow 
$\gamma \gamma$ signal peak in the decay $H\to \gamma \gamma$ from the large
continuum $\gamma \gamma$ background.  Since the Higgs boson width is small, a
few MeV for $M_H\simeq$120--140 GeV, the measured mass peak is entirely 
dominated by the experimental resolution.  Furthermore, the very large number 
of high transverse momentum $\pi^0$ decaying into two photons should be 
rejected efficiently. \s

-- In the dominant Higgs decay mode in the low mass range, $H\to b \bar b$,
excellent micro--vertex detectors are needed to identify the $b$--quark jets
with a high efficiency and a high purity. $\tau$--lepton identification is also
important to detect the decays $H\to \tau^+ \tau^-$ and the invariant mass of
the final state should be reconstructed with a good resolution. \s

Together with good granularity and hermeticity coverage for jet resolution and 
missing transverse energy, these requirements are apparently met by the CDF 
and D\O\ detectors at Tevatron \cite{Higgs-TeV} and are expected to be met 
by the ATLAS and CMS detectors at LHC.\s

The most unambiguous signal for a Higgs boson [and for any new particle] is a
peak in the invariant mass distribution of its decay products. The narrow mass
peak can be discovered without any Monte--Carlo simulation for the backgrounds,
since the latter can be precisely measured from the side bands. In addition,
the discovery can be made even if the signal is rather low and the background
large, since the significance is $\propto S/\sqrt{S+B}$. This however is not
true when it comes to study some properties of the Higgs boson, such as its
couplings and its spin--parity quantum numbers. In this case, Monte--Carlo
simulations are needed to determine the cross sections and the various
characteristics distributions of the signal and backgrounds. The most precise 
theoretical predictions are therefore required.  

\subsubsection{Higgs production at hadron machines}
 
In the Standard Model, the main production mechanisms for Higgs particles at
hadron colliders make use of the fact that the Higgs boson couples
preferentially to the heavy  particles, that is the massive $W$ and $Z$ vector
bosons, the top quark and, to a lesser extent, the bottom quark. The four main
production processes, the Feynman diagrams of which are displayed in Fig.~3.1,
are thus: the associated production with $W/Z$ bosons
\cite{pp-HV-LO,pp-EHLQ},  the weak vector boson fusion processes 
\cite{Petcov,VVH-Cahn,VVH-DW,VVH-Altarelli,VVH-Kilian}, 
the gluon--gluon fusion mechanism 
\cite{pp-ggH-LO} and the associated Higgs production with heavy top 
\cite{pp-Htt-LO,pp-Htt-LO1}
or bottom  \cite{pp-Hbb-LO1,pp-Hbb-LO} quarks:
\beq
{\rm associated~production~with}~W/Z: & & q\bar{q} \lra V + H \\
{\rm ~vector~boson~fusion}: & & qq \lra V^*V^* \lra   qq+ H \\
{\rm gluon-gluon~fusion}: & & gg  \lra H \hspace*{2cm} \\
{\rm associated~production~with~heavy~quarks}: & & gg,q\bar{q}\lra Q\bar{Q}+H
\eeq

\begin{center}
\SetWidth{1.1}
\vspace*{2mm}
\begin{picture}(300,100)(0,0)
\ArrowLine(0,25)(40,50)
\ArrowLine(0,75)(40,50)
\Photon(40,50)(90,50){4}{5.5}
\DashLine(90,50)(130,25){4}
\Photon(90,50)(130,75){3.5}{5.5}
\Text(-10,20)[]{$q$}
\Text(-10,80)[]{$\bar{q}$}
\Text(70,65)[]{$V^*$}
\Text(90,50)[]{\blue{\Large\bf $\bullet$}}
\Text(139,20)[]{\bH}
\Text(139,80)[]{$V$}
\ArrowLine(200,25)(240,25)
\ArrowLine(200,75)(240,75)
\ArrowLine(240,25)(290,0)
\ArrowLine(240,75)(290,100)
\Photon(240,25)(280,50){3.5}{5}
\Photon(240,75)(280,50){-3.5}{5}
\DashLine(280,50)(320,50){4}
\Text(280,50)[]{\blue{\Large\bf $\bullet$}}
\Text(190,20)[]{$q$}
\Text(190,80)[]{$q$}
\Text(275,72)[]{$V*$}
\Text(275,27)[]{$V^*$}
\Text(300,60)[]{\bH}
\Text(300,10)[]{$q$}
\Text(300,90)[]{$q$}
\end{picture}
\vspace*{-3.mm}
\end{center}
\begin{center}
\SetWidth{1.1}
\begin{picture}(300,100)(0,0)
\Gluon(0,25)(40,25){4}{5.5}
\Gluon(0,75)(40,75){4}{5.5}
\ArrowLine(40,75)(90,50)
\ArrowLine(40,25)(90,50)
\Line(40,75)(40,25)
\DashLine(90,50)(130,50){4}
\Text(90,50)[]{\blue{\Large\bf $\bullet$}}
\Text(-10,25)[]{$g$}
\Text(-10,75)[]{$g$}
\Text(110,60)[]{\bH}
\Text(60,50)[]{$Q$}
\Gluon(200,25)(240,25){4}{5.5}
\Gluon(200,75)(240,75){4}{5.5}
\ArrowLine(240,25)(290,25)
\ArrowLine(240,75)(290,75)
\Line(240,75)(240,25)
\DashLine(240,50)(280,50){4}
\Text(240,50)[]{\blue{\Large\bf $\bullet$}}
\Text(190,25)[]{$g$}
\Text(190,75)[]{$g$}
\Text(289,50)[]{\bH}
\Text(306,75)[]{$Q$}
\Text(306,25)[]{$\bar{Q}$}
\end{picture}
\vspace*{-7mm}
\end{center}
\centerline{\it Figure 3.1: The dominant SM Higgs boson production mechanisms 
in hadronic collisions.} 
\vspace*{4mm}

There are also several mechanisms for the pair production of the
Higgs particles
\beq
{\rm Higgs~pair~~production}: &  pp \lra HH + X
\eeq
and the relevant sub--processes are the $gg \to HH$ mechanism, which proceeds
through heavy top and bottom quark loops \cite{pp-ggHH-LO,pp-ggHH-LO1}, the 
associated double production with massive gauge bosons \cite{pp-HHV,pp-DKMZ}, 
$q\bar{q} \to HHV$, and the vector boson fusion mechanisms $qq \to  V^* V^* 
\to HHqq$ \cite{pp-VVHH,pp-VVH-Abas}; see also Ref.~\cite{pp-DKMZ}. However, 
because of the suppression by the additional electroweak couplings, they have 
much smaller production cross sections than the single Higgs production 
mechanisms listed above.\s

Also suppressed are processes where the Higgs is produced in association with 
one \cite{pp-Hgg-PT,pp-Hgg-PT2}, two \cite{pp-ggHqq-Limit,pp-ggHqq} or three 
\cite{pp-ggHqqq} hard jets in gluon--gluon
fusion, the associated Higgs production with gauge boson pairs
\cite{pp-HVV,DWP}, the production with a vector  boson and two jets
\cite{pp-HVqq,pp-HVqq-Rain,DWP}. Other production processes  exist which have
even smaller production cross sections
\cite{pp-Hgamma,pp-qqHH,pp-qqttHH0,pp-qqttHH,pp-t-H,pp-t-H2,Three-Body2}. 
Finally, Higgs bosons can also be
produced in diffractive processes 
\cite{BL-diffr,diffr1,diffr2,Valery-myths,diffr-Houches}. For the interesting 
exclusive central
diffractive processes \cite{diffr2,Valery-myths,diffr-Houches}, the
mechanism is mediated by color singlet exchanges leading to the diffraction of
the incoming hadrons and a centrally produced Higgs boson. A
mixture of perturbative and non perturbative aspects of QCD is needed to
evaluate the cross sections, leading to uncertainties in the 
predictions.\s

In this chapter, we discuss all these processes in detail, analyzing not only 
the total production cross sections but also the differential  distributions 
and, in particular, the Higgs boson transverse momentum and rapidity 
distributions. In addition, we pay a special attention to three very important 
points: the
QCD radiative corrections or the $K$--factors, the residual cross sections
dependence on the renormalization and factorization  scales, and the choice of
different sets of parton distributions functions (PDFs) with which one has to
convolute the partonic cross sections to obtain the total hadronic cross
sections.

\vspace*{-2mm}
\subsubsection{The higher--order corrections and the $K$--factors} 

It is well known that for processes involving strongly interacting particles,
as is the case for the ones that we will consider here, the lowest order (LO)
cross  sections are affected by large uncertainties arising from higher--order
(HO) corrections. If at least the next--to--leading order (NLO) QCD corrections
to these processes are included, the total cross sections can be defined
properly and in a reliable way in most cases: the renormalization scale
$\mu_R$ at which one defines the strong coupling constant and the factorization
scale $\mu_F$ at which one performs the matching between the perturbative
calculation of the matrix elements and the non perturbative part which resides
in the parton distribution functions, are fixed and the generally
non--negligible radiative corrections are taken into account. \s

The impact of higher--order QCD corrections is usually quantified by calculating
the $K$--factor, which is defined as the ratio of the cross section for the
process [or its distribution] at HO with the value of $\alpha_s$ and the PDFs
evaluated also at HO, over the cross section [or distribution] at LO with
$\alpha_s$ [for those processes which are QCD processes at LO] and the PDFs
consistently also evaluated at LO\footnote{Note that if the $K$--factor is
defined as the ratio of NLO to LO cross sections both evaluated with $\alpha_s$
and PDFs at NLO, it would be in many cases larger since the value of the strong
coupling constant, which appears in both the matrix element squared of the hard
process and in the parton distribution functions, is smaller at NLO,
$\alpha_s^{\rm NLO}(M_Z) \sim 0.12$, than at LO, $\alpha_s^{\rm LO} (M_Z)\sim
0.13$, thereby decreasing the LO cross section.} 
\beq 
K= \frac{\sigma_{\rm HO} (pp \to H+X) }{\sigma_{\rm LO}( pp \to H+X) } 
\eeq 
All the dominant Higgs production processes which are addressed here will be
discussed at least at NLO \cite{Michael-Web}. At this order, the QCD 
corrections are known since
more than a decade for the associated production with $W/Z$ bosons
\cite{HVNLO,HVNLOrest,HVNLO-DS}, the vector boson fusion processes
\cite{pp-Hqq-NLO1,pp-Hqq-NLO2,MC-WWNLO,HVNLO-DS} and the gluon--gluon mechanism
\cite{HggQCD,ggH-Dawson,ggH-GSZ,SDGZ}, while the NLO corrections to the
associated production with heavy quarks have been calculated only recently
\cite{Htt-NLO-DESY,Htt-NLO-US,Htt-NLO-Tev,Hbb-NLO1,Hbb-NLO2}.
To improve further the theoretical predictions
for the cross sections, one can also resum  the soft and collinear gluon
radiation parts which in general lead to large logarithms and include the 
dominant electroweak radiative corrections which however, are much smaller than
the QCD corrections, in particular when the improved Born approximation 
of \S1.2.4 is used.\s 

The QCD corrections to the transverse momentum and rapidity distributions are
also available in the case of vector boson fusion \cite{pp-Hqq-NLO2,MC-WWNLO} 
and gluon--gluon fusion 
\cite{pp-ggH-PT0,pp-ggH-PT1,Pt-eta-distrib,pp-ggH-Ital,pp-ggH-eta1,pp-ggH-eta2,pp-ggH-distrib}.  
In the latter
case, the resummation of the large logarithms for the $P_T$ distribution has
been performed at next--to--next--to--leading--logarithm (NNLL) accuracy. The 
QCD corrections to the various distributions in the associated Higgs production
with $t\bar t$ are discussed in \cite{Htt-NLO-DESY}.\s

In two cases, the associated $HV$ production \cite{pp-HV-NNLO}  and the $gg \to
H$ fusion mechanism in the approximation where the top quark is very heavy
\cite{ggH-NNLO1,ggH-NNLO2,ggH-NNLO3,ggH-NNLO-resum}, the calculation of
the production cross sections at NNLO has been performed recently and will be
discussed.  However, these calculations are not sufficient to obtain a full
NNLO prediction: the cross sections must be folded with the NNLO evolved PDFs,
which are also necessary. The latter require the calculation of the
Altarelli--Parisi splitting functions \cite{apsplit} up to three loops
and until very recently the latter were not completely known at this order.
Nevertheless, a large number of moments of these functions were available
\cite{van-neerven} which, when combined  with additional information on the
behavior at small $x$, allowed to obtain an approximation of the splitting
functions at the required order. The NNLO MRST \cite{MRSTNNLO} parton
distributions followed this approach and have been therefore adopted for NNLO
calculations\footnote{The calculation of the $N_f$ part of the non--singlet
structure function in DIS, from which one can extract the corresponding
splitting function, is available since some time and has been compared to the
approximate result of Ref.~\cite{van-neerven} and full agreement has been
obtained, giving a great confidence that the approximate NNLO PDFs are rather
accurate. Recently, the full calculation of the NNLO splitting function has
been completed \cite{NNLO-AP} and they alter the NNLO MRST PDFs only by a small
amount \cite{Thorne-rev}.}.

\vspace*{-2mm}
\subsubsection{The scale dependence}

The evaluation of the residual theoretical uncertainties in the production
cross  sections or distributions, due to the not yet calculated higher--order
corrections, is generally based on the exploration of the cross section
dependence on the renormalization scale $\mu_R$ and on the factorization scale
$\mu_F$. Starting from a median scale $\mu_0$ which, with an educated guess, is 
considered as the ``natural scale" of the process and is expected to
absorb the large logarithmic corrections, the by now standard convention is to 
vary the two scales, either  collectively or independently 
[i.e. keeping one scale fixed at the reference value], within
\beq 
\mu_0/a \, \leq \, \mu_F, \mu_R \, \leq  \, a\mu_0 
\eeq 
The value of the constant $a$ is in general chosen to be 2 or 3, the
latter case being more conservative and will be adopted in most cases. 
In some situations in which widely different scales are involved in the 
processes, it is more prudent to use larger values for $a$, as will be seen in 
the case of Higgs production in bottom quark fusion for instance. \s

Note that the scale dependence at leading order can be studied by defining a 
kind of $K$--factor for the LO cross section, $K_{\rm LO}$, by  evaluating the 
latter at given factorization and renormalization scales $\mu_F$ and $\mu_R$, 
and normalizing to the LO cross sections evaluated at the median  scale $\mu_0$
\beq
K_{\rm LO}=\sigma_{\rm LO} (\mu_F, \mu_R)/\sigma_{\rm LO}(\mu_F= \mu_R= \mu_0)
\eeq

By varying the scales $\mu_R$ and $\mu_F$, one then obtains an uncertainty 
band: the narrower the band is, the smaller the higher--order corrections 
are expected to be. Note that the scale uncertainty should be in principle
reduced when higher--order corrections are  included, that is, the scale 
variation should be smaller at NNLO, than at NLO, than at LO. However, this is 
not the case all the time, and a counter--example  will be discussed 
later. \s

One should nevertheless caution that the variation of the cross section with
respect to the scale choice is unphysical: it is just a reflexion of the
truncation of the perturbative series; if the cross sections are  known to all
orders, they will not exhibit this dependence. The scale variation is thus, by 
no means a rigorous way to estimate the theoretical uncertainty. At best,  it
might only give an indication of the ``full" uncertainty. This
can be seen in many cases, where for instance the NLO and LO uncertainty
bands for some production cross sections do not overlap at all, as will be  
shown later. 

\vspace*{-2mm}
\subsubsection{The parton distribution functions}

Parton distribution functions (PDFs), which describe the momentum distribution
of a parton in the proton,  play a central role at hadron colliders.  A precise
knowledge of the PDFs over a wide range of the proton momentum fraction $x$
carried by the parton and the squared center of mass energy $Q^2$ at which the
process takes place, is mandatory to precisely predict the production cross
sections of the various signal and background processes. However, they
are plagued by uncertainties, which arise either from the starting
distributions obtained from a global fit  to the available data from
deep--inelastic scattering, Drell--Yan and hadronic data, or from the DGLAP
evolution \cite{apsplit,DGLAP} to the higher $Q^2$ relevant to the scattering
processes.  Together with the effects of unknown perturbative higher--order
corrections, these uncertainties dominate the theoretical error on the
predictions of the cross sections.\s

The CTEQ \cite{CTEQ6} and MRST \cite{MRST2001E} collaborations, as well as
Alekhin \cite{ALEKHIN} and others \cite{Others-PDFs}, recently introduced new
schemes, which provide the possibility of estimating the  intrinsic and spread
uncertainties on the prediction of physical observables at hadron colliders.
The CTEQ and MRST schemes are based on the Hessian matrix method which enables a
characterization of a parton parametrization in the neighborhood of the global
$\chi^2$ minimum fit and gives an access to the uncertainty estimation through
a set of PDFs that describes this neighborhood. The corresponding PDFs are
constructed as follows: (i)  a global fit of the data is performed using the
free parameters $N_{\rm PDF}=20$ for CTEQ and $N_{\rm PDF}=15$ for MRST; this
provides the nominal PDF (reference set) denoted by $S_0$ and corresponding to
CTEQ6M and MRST2001C, respectively; (ii) the global $\chi^2$ of the fit  is
increased by $\Delta \chi^2\!=\!100$ for CTEQ and $\Delta \chi^2\!=\!50$ for
MRST, to obtain the error matrix; (iii) the error matrix is diagonalized to
obtain $N_{\rm PDF}$ eigenvectors corresponding to $N_{\rm PDF}$ independent
directions in the parameter space; (iv) for each eigenvector, up and down
excursions are performed in the tolerance gap, leading to $2N_{\rm PDF}$ sets
of new parameters, corresponding to 40 new sets of PDFs for CTEQ and 30 sets
for MRST.  They are denoted by $S_i$, with $i=1, 2N_{\rm PDF}$. \s 

To build the Alekhin PDFs \cite{ALEKHIN}, only light--target  deep--inelastic
scattering data are used. This PDF set involves 14 parameters, which are fitted
simultaneously with $\alpha_s$ and the structure functions, leading to $2N_{\rm
PDF}=30$  sets of PDFs for the uncertainty estimation. Note that the three PDF
sets use different values for  $\alpha_s$: at NLO, the central sets CTEQ6M, 
MRST2001C and A02 use, respectively, $\alpha_s^{\rm NLO}(M_Z)= 0.118$, $0.119$  and 0.117.\s 

The three sets of PDFs discussed above can be used to calculate the uncertainty
on a cross section $\sigma$ in the following way \cite{Samir}: one first
evaluates the cross section with the nominal PDF $S_0$ to obtain  the central
value $\sigma_0$. One then calculates the cross section with  the $S_i$ PDFs,
giving $2N_{\rm PDF}$ values $\sigma_i$, and defines, for each $\sigma_i$
value, the deviations  $\sigma_i^\pm =\mid \sigma_i -\sigma_0\mid$ when
$\sigma_i \ ^{>}_{<}  \sigma_0$. The uncertainties are summed quadratically to
calculate {\bf $\Delta  \sigma^\pm  = \sqrt{ \sum_i \sigma_i^{\pm 2} }$}.  The
cross section, including the error, is then given by $\sigma_0|^{+\Delta
\sigma^+}_{- \Delta \sigma^-}$.  This procedure will be applied to estimate the
uncertainties in the cross sections for SM Higgs production in the four main
mechanisms.  The spread in the cross section prediction will depend on the
considered partons and their $x$ regime that we will briefly summarize below. \s

The differences between the PDFs originate from three main sources: (i) the
choice of the data used in the global fit, (ii) the theoretical assumptions
made for the fit and (iii) the choice of the tolerance used to define the error
in the PDFs. Thus, for example, the MRST and CTEQ differences arise from points
(ii) and (iii) only, with point (iii) dominating in most cases. The differences
between the two approaches \cite{CTEQ6,MRST2001E} are explained in detail in 
Ref.~\cite{MRST2001E}, and for instance the CTEQ6 high--$x$ gluon is larger 
than the MRST2001 one. The differences with the Alekhin analysis, which does 
not use the Tevatron data, are larger.\s 

To be more qualitative, we present in Fig.~3.2, the MRST and Alekhin densities
for the gluon and for the up and down quarks and antiquarks, normalized to the
CTEQ6 ones, for a wide range of $x$ values and for a fixed c.m.  energy
$Q^2=(100\ {\rm GeV})^2$. One notices the following main features: $(i)$ the
MRST gluon PDF is smaller than the CTEQ one, except for values $x\sim 0.1$; in
contrast, the Alekhin gluon PDF is larger than the  CTEQ one for all $x$
values, except for $x \sim 0.01$ and for very high $x$. $(ii)$ The MRST
(anti)quark PDFs are practically equal in magnitude and are smaller than the
CTEQ ones for low $x$, while they are in general slightly larger for higher
$x$, except for values near unity; in the Alekhin case, all (anti)quark PDFs
are larger than the CTEQ ones, except for the $\bar{u}$ density above $x \sim
0.05$. For values, $x \gsim 10^{-4}$, the differences between the Alekhin and
the CTEQ6 PDFs are more pronounced than the differences between the MRST and
the CTEQ ones.\s

\begin{figure}[h]
\begin{center}
\vspace*{-2.5cm}
\hspace*{-1cm}
\psfig{figure=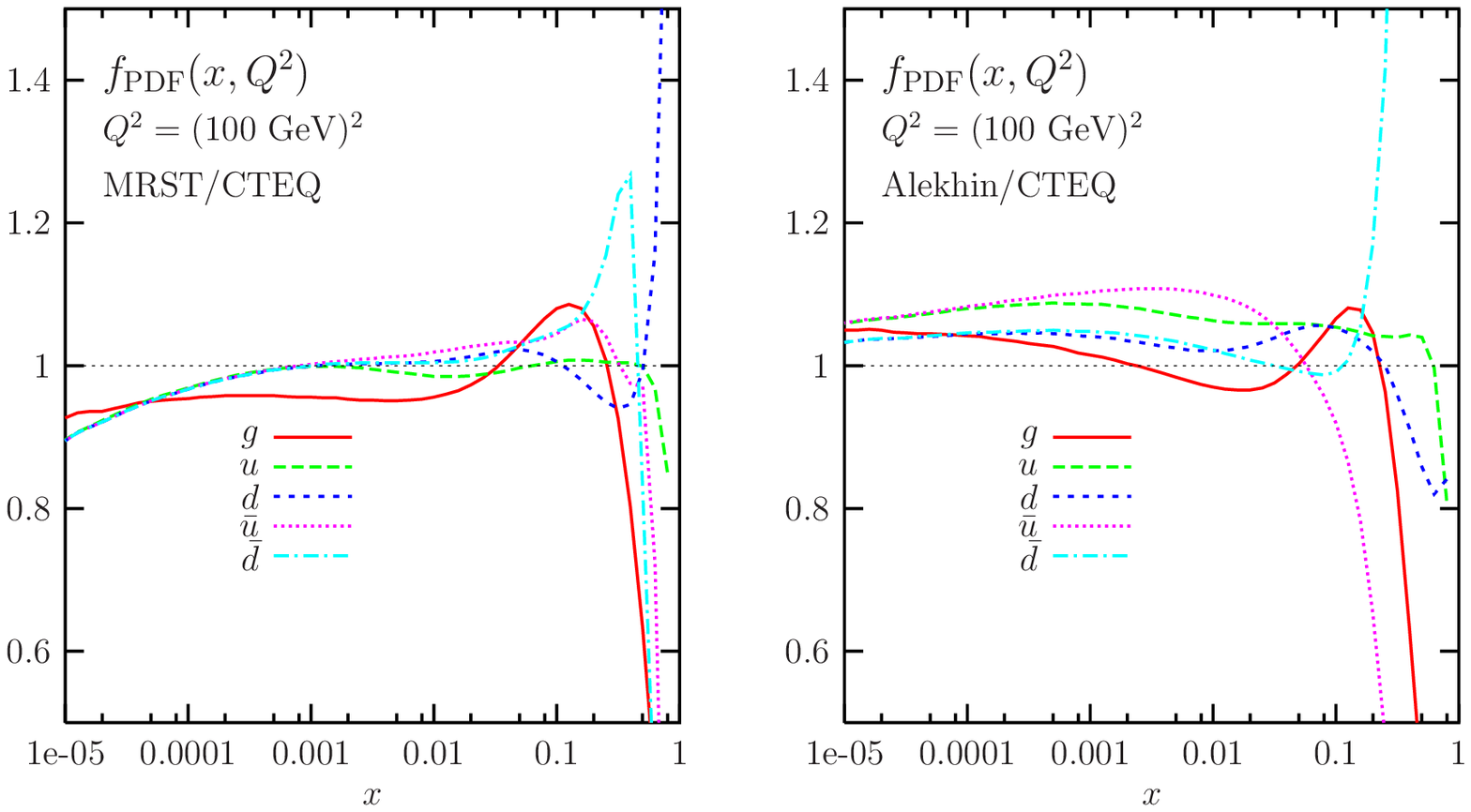,width=18cm}
\vspace*{-15.9cm}
\end{center}
{\it Figure 3.2: MRST and Alekhin densities for the gluon, up quark/down
quark and antiquarks, normalized to the CTEQ6 ones, as a function of $x$ and
for $Q^2=(100\ {\rm GeV})^2$; from Ref.~\cite{Samir}.}
\vspace*{-6mm}
\end{figure}

As for the CTEQ and MRST parameterizations,  three different behaviors of the
uncertainty bands according to three $x$ ranges can be distinguished:
decreasing uncertainties at low $x$, constant or slightly oscillating ones at
intermediate $x$, and increasing  ones at high $x$. The magnitude of these
uncertainties depends on the considered parton  and on the c.m. energy $Q^2$.
In the case of quarks, the three behaviors are observed: the low-$x$ behavior
extends up to $x \sim$ few $10^{-3}$, and the high--$x$ one starts in the
neighborhood of $x=0.7$. At high $Q^2$, the uncertainties at high and low--$x$
values exceed a few tens of a percent and in the intermediate regime, they are
less than a few  percent. In the gluon case and at high $Q^2$, the low--$x$ and
the intermediate--$x$ bands are not well separated as in the case of quarks;
the uncertainty band reaches also the few percent level. The high--$x$  regime
starts in the neighborhood of $x \sim 0.3$, i.e earlier than in the case of
quarks.  

\newpage
\subsection{The associated production with W/Z bosons}
\subsubsection{The differential and total cross sections at LO}

It is useful to consider the cross section for the associated production of the
Higgs particle with massive gauge bosons, which then decay into two massless
fermions, in a completely differential form so that various distributions can
be presented and cuts can be imposed on the final decay products. For the Higgs
boson, since it is a scalar particle, the incorporation of its decays into a
given final  state, $H\to X$, is simply done by multiplying the matrix element
squared by the branching ratio BR$(H \to X)$ and generating the final state $X$
isotropically  in the rest frame of the $H$ boson. \s

The general  form of the matrix element squared for the process
\beq
q_1 (p_1) \bar{q}_2 (p_2) \to V^*(k=p_1+p_2) \to V(k_1=p_3+ p_4)
H(k_2) \to f_3 (p_3) \bar{f}_4 (p_4) H(k_2)
\eeq 
where the momenta of the particles are explicitly written, with $\hat{s}=k^2 = 
(p_1+ p_2)^2$ being the c.m. energy of the partonic subprocess, can be 
expressed as
\beq
|{\cal M}|^2 &=& 2 \sqrt{2} N_{c}^f G_\mu^3 M_V^8 
\frac{1} {(k^2-M_V ^2)^2+\Gamma_V^2 M_V^2} 
\frac{1} {(k_1^2-M_V^2)^2+\Gamma_V^2 M_V^2} \bigg[  \\
&&+ \bigg( (\hat v_{q_1} + \hat a_{q_1} )^2 (\hat v_{f_3}+ \hat 
a_{f_3} )^2 + (\hat v_{q_1} - \hat a_{q_1} )^2 (\hat v_{f_3}- \hat a_{f_3} )^2 
\bigg) (p_1 \cdot p_4) (p_2 \cdot p_3) \non \\
&&+  \bigg( (\hat v_{q_1} + \hat a_{q_1} )^2 (\hat v_{f_3} - \hat 
a_{f_3} )^2 + (\hat v_{q_1} - \hat a_{q_1} )^2 (\hat v_{f_3} + \hat a_{f_3} )^2
 \bigg)  (p_1 \cdot p_3) ( p_2 \cdot p_4) \bigg] \non
\eeq
where the reduced fermion couplings to gauge bosons are as usual: $\hat a_f=
2I_f^3, \hat v_f=2I_f^3 -4 Q_fs_W^2$ for $V=Z$ and $\hat v_f=\hat a_f=\sqrt{2}$
for $V=W$. Averaging over the quark spins  and colors, dividing by the flux
factor, and integrating over the three--particle phase--space, one obtains the
total cross section of the subprocess. In the case where the decay products
of the final vector boson are ignored, one would have a simple $2 \to 2$
subprocess, with an integrated cross section at lowest order given by
\cite{pp-HV-LO,pp-EHLQ}
\beq
\hat{\sigma}_{\rm LO}(q\bar{q} \ra V H)= \frac{G_\mu^2 M_V^4}{288 \pi \hat{s}}
(\hat v_q^2 + \hat a_q^2) \lambda^{1/2} (M_V^2, M_H^2; \hat{s}) \frac{
\lambda(M_V^2, M_H^2; \hat{s})+12 M_V^2/\hat{s}}{(1-M_V^2/\hat{s})^2}
\label{sigma-HV-SM}
\eeq
with $\lambda$ being the usual two--body phase space function $\lambda(x,y;z)$
=$(1-x/z- y/z) ^2-4xy/z^2$.\s

Note that the Higgs and the vector bosons have opposite transverse momenta and
the differential partonic distribution with respect to the $p_T$ is given by
\beq
\frac{ {\rm d}\hat{\sigma}_{\rm LO}}{ {\rm d}p_T^2} = 
\frac{G_\mu^2 M_V^4}{24 \pi}  \frac{v_q^2 + a_q^2} {(\hat s-M_Z^2)^2} 
\frac{2 M_Z^2+p_T^2} {2(M_Z^2+ M_H^2)-\hat{s} \sqrt{\lambda -4p_T^2/ \hat{s}}} 
\eeq
The partonic cross section can be recovered by integrating $p_T$ in the range 
$0 \leq p_T \leq \frac{\sqrt{\hat s \lambda}}{2}$.\s

In fact, this process can be viewed simply as the Drell--Yan production of a  
virtual vector boson with $k^2 \neq M_V^2$, which then splits into a real 
vector boson and a  Higgs particle. The energy distribution of the full 
subprocess can be written at leading order as
\beq
\hat{\sigma} (q\bar{q} \to H V) =  \hat \sigma
(q\bar{q} \to V^*) \times \frac{ {\rm d} \Gamma }{ {\rm d}k^2 }(V^* \to H V)
\label{DY-factorization}
\eeq
where, in terms of $0\leq k^2\leq Q^2=\hat{s}$ and the two-body phase-space
function $\lambda$, one has
\beq 
\frac{ {\rm d} \Gamma }{ {\rm d}k^2 } (V^* \to H V) = \frac{ G_\mu M_V^4}{
2\sqrt{2} \pi^2}  \frac{\lambda^{1/2} (M_V^2, M_H^2; k^2)}{(k^2-M_V^2)^2}
\left(1 + \frac{\lambda(M_V^2, M_H^2;k^2)}{12M_V^2/k^2} \right) \ .
\label{HV-dGamma}
\eeq
The total production cross section is then obtained  by convoluting with the
parton densities and summing over the contributing partons
\beq
\sigma_{\rm LO} ( pp \to VH)  = \int_{\tau_0}^1  {\rm d}\tau \,
\sum_{q,\bar{q}} \, \frac{ {\rm d} {\cal L}^{q \bar{q}} }{ {\rm d} \tau}
\, {\hat \sigma}_{\rm LO} (\hat{s}= \tau s) 
\eeq
where $\tau_0= (M_V+M_H)^2/s$, $s$ being the total hadronic c.m. energy and the
parton luminosity is defined in terms of the parton densities $q_i(x_i,
\mu_F^2)$ defined at a factorization scale $\mu_F$, by  
\beq
\sum_{q,\bar{q}} \frac{ {\rm d} {\cal L}^{q \bar{q}} }{ {\rm d} \tau } =
\sum_{q_1,\bar{q}_2}   \int_{\tau}^1 \frac{{\rm d} x}{x} \, \left[ q_1 (x, 
\mu_F^2) \, \bar{q}_2 (\tau/x, \mu_F^2)  \right] 
\eeq

\begin{figure}[!h]
\begin{center}
\vspace*{-2.9cm}
\hspace*{-3cm}
\epsfig{file=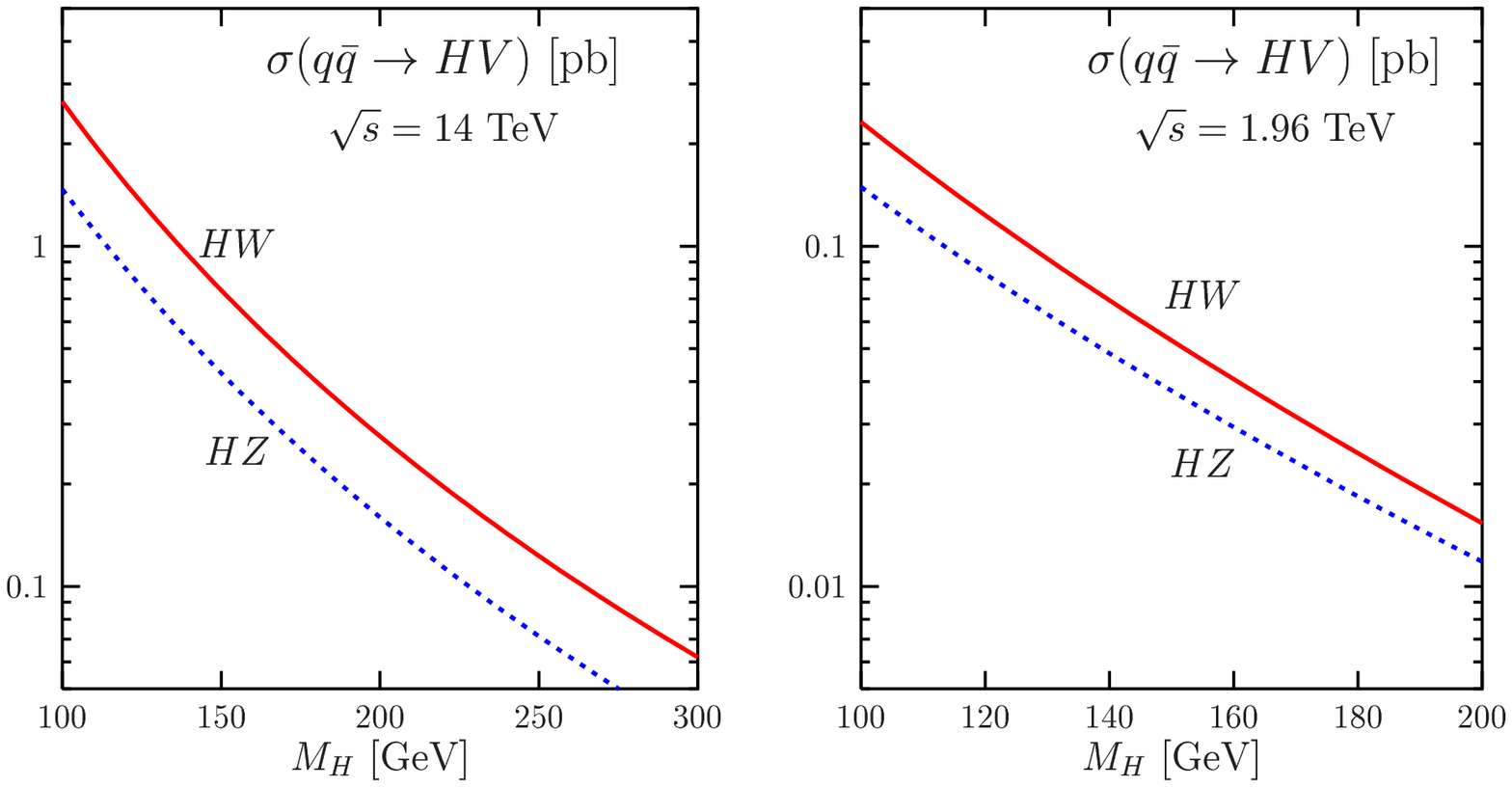,width=17.cm} 
\end{center}
\vspace*{-13.7cm}
{\it Figure 3.3: Total production cross sections of Higgs bosons in the 
strahlung $q\bar q\to H+W/Z$ processes at leading order at the LHC (left) 
and at the Tevatron (right). For $q\bar q \to HW$, the final states with both 
$W^+$ and $W^-$ have been added. The MRST set of PDFs has been used.} 
\vspace*{-.2cm}
\end{figure}

The total production cross sections are shown as a function of the Higgs boson
mass for the Tevatron and the LHC in both the $HW^\pm$ and $HZ$ channels in
Fig.~3.3; the MRST parton densities are used. The cross sections for $W^\pm$
final states are approximately two times larger than the ones for the $HZ$
final state at both colliders. If, in addition, one requires the gauge bosons
to decay into charged leptons $\ell= \mu+e$, the charged channel is much more
interesting since BR($W^\pm \to \ell^\pm \nu) \sim 20$\% while BR($Z \to \ell^+
\ell^-) \simeq  6$\%. The various detection channels at the LHC 
\cite{pp-Galison,pp-HW-laa0,pp-HW-laa1,pp-HW-bb-LHC,pp-HW-lWW-LHC+Tev}
and at the Tevatron \cite{pp-HW-bb-TeV,Gunion-Han,pp-HW-Mrenna,pp-HW-laaTeV} 
and \cite{pp-HW-lWW-LHC+Tev} will be discussed in \S3.7.

\vspace*{-2mm} 
\subsubsection{The QCD radiative corrections}

\subsubsection*{\underline{The NLO corrections}}

The factorization of the $pp\to HV$ cross section eq.~(\ref{DY-factorization}) 
holds in principle at any order of perturbation theory in the strong 
interaction and one can thus write
\beq
\frac{ {\rm d} \hat{\sigma} }{ {\rm d}k^2 } (pp \to H V+X) =  \sigma
(pp \to V^*+X ) \times \frac{ {\rm d} \Gamma }{ {\rm d}k^2 } (V^* \to H V) \ ,
\eeq
where d$\Gamma/$d$k^2$ is given by eq.~(\ref{HV-dGamma}). Therefore, the QCD 
corrections to the Higgs--strahlung process, derived at NLO in 
Refs.~\cite{HVNLO,HVNLOrest,HVNLO-DS}, are  simply the corrections to the  
Drell--Yan process \cite{DYNLO,DYNNLO}, as pointed out in 
Ref.~\cite{pp-HW-laa0,DYequiv}.\\[3mm]

\begin{figure}[!h]
\begin{center}
\setlength{\unitlength}{.8pt}
\SetWidth{1.1}
\begin{picture}(180,100)(-50,0)
\ArrowLine(0,100)(50,50)  
\ArrowLine(50,50)(0,0)
\Gluon(25,75)(25,25){-3}{5}
\Photon(50,50)(100,50){-3}{5}
\put(85,46){$V^*$}
\put(-10,109){$q$}
\put(-10,4){$\bar{q}$}
\put(10,56){$g$}
\end{picture}
\begin{picture}(180,100)(-30,0)
\ArrowLine(0,100)(50,50)
\ArrowLine(50,50)(0,0)
\Photon(50,50)(100,50){-3}{5}
\GlueArc(27.5,72.5)(12.5,-45,135){3}{4}
\put(85,46){$V^*$}
\put(-10,109){$q$}
\put(-10,4){$\bar{q}$}
\put(40,119){$g$}
\end{picture}
\begin{picture}(180,100)(-10,0)
\ArrowLine(0,100)(50,50)
\ArrowLine(50,50)(0,0)
\Photon(50,50)(100,50){-3}{5}
\Gluon(27.5,72.5)(50,100){-3}{4}
\put(85,46){$V^*$}
\put(-10,109){$q$}
\put(-10,4){$\bar{q}$}
\put(50,90){$g$}
\vspace*{-3mm}
\end{picture}
\end{center}
\vspace*{-3mm}
\centerline{\it Figure 3.4: NLO QCD corrections to the vector 
boson--quark--antiquark vertex.} 
\vspace*{-3mm}
\end{figure}

At NLO, the QCD corrections to the Drell--Yan process consist of virtual
corrections with gluon exchange in the $q \bar{q}$ vertex and quark self-energy
corrections, which have to be multiplied by the tree-level term, and the
emission of an additional gluon, the sum of which has to be squared and added
to the corrected tree--level term; see Fig.~3.4. \s

Including these contributions, and taking into account the virtuality of the
vector boson, the LO cross section is modified in the following way 
\begin{eqnarray}
\sigma_{\rm NLO} & = & \sigma_{\rm LO} + \Delta\sigma_{q\bar q} +
\Delta\sigma_{qg} 
\eeq
\vspace*{-2mm}
with
\vspace*{-2mm}
\beq
\Delta\sigma_{q\bar q} & = & \frac{\alpha_s(\mu_R)}{\pi} \int_{\tau_0}^1 d\tau
\sum_q \frac{d{\cal L}^{q\bar q}}{d\tau} \int_{\tau_0/\tau}^1 dz~\hat
\sigma_{\rm LO}(\tau z s)~\omega_{q\bar q}(z) \nonumber \\
\Delta\sigma_{qg} & = & \frac{\alpha_s(\mu_R)}{\pi} \int_{\tau_0}^1 d\tau
\sum_{q,\bar q} \frac{d{\cal L}^{qg}}{d\tau} \int_{\tau_0/\tau}^1 dz~\hat
\sigma_{\rm LO}(\tau z s)~\omega_{qg}(z) 
\end{eqnarray}
with the coefficient functions \cite{DYNLO}
\begin{eqnarray}
\omega_{q\bar q}(z) & = & -P_{qq}(z) \log \frac{\mu_F^2}{\tau s}
+ \frac{4}{3}\left[ \left(\frac{\pi^2}{3} -4\right)\delta(1-z) +
2(1+z^2) \left(\frac{\log(1-z)}{1-z}\right)_+ 
\right] \nonumber \\
\omega_{qg}(z) & = & -\frac{1}{2} P_{qg}(z) \log \left(
\frac{\mu_F^2}{(1-z)^2 \tau s} \right) + \frac{1}{8}\bigg[ 1+6z-7z^2 \bigg] 
\label{AP-functions}
\end{eqnarray}
where $\mu_R$ denotes  the renormalization scale and $P_{qq}, P_{qg}$  are the
well--known Altarelli--Parisi splitting functions which are given by
\cite{apsplit,pp-APabs}
\begin{eqnarray}
P_{qq}(z)  &= & \frac{4}{3} \left[ \frac{1+z^2}{(1-z)_+}+\frac{3}{2}\delta(1-z)
\right]  \non \\ 
P_{qg}(z) & = & \frac{1}{2} \bigg[ z^2 + (1-z)^2 \bigg] 
\label{AP-function}
\end{eqnarray}
The index $+$ denotes the usual distribution ${\cal F}_+(z)={\cal F}(z)-
\delta(1-z)\int_0^1 dz' {\cal F}(z')$. Note that the cross section depends 
explicitly on $\log(\mu_F^2/Q^2)$; the scale choice $\mu_F^2 =
Q^2$ therefore avoids
the occurrence of these potentially large logarithms. The  renormalization
scale dependence enters in the argument of $\alpha_s$ and is rather weak. In
most of our discussion, we will set the two scales at the invariant mass
of the $HV$ system $\mu_F= \mu_R= M_{HV}$.
For this choice, the NLO corrections increase the LO cross section by
approximately 30\%.

\subsubsection*{\underline{The NNLO corrections}}

The NNLO corrections, i.e. the contributions at ${\cal O}(\alpha_s^2)$, to the
Drell--Yan process $pp \to V^*$  consist of the following set of corrections
besides the one--loop squared terms [see also Fig.~3.5a--c]: $a)$ two-loop
corrections to $q\bar{q}\to V^*$, which have to be multiplied by the Born term;
$b)$ one--loop corrections to the processes $qg \to qV^*$ and $q\bar{q} \to
gV^*$, which have to be multiplied by the tree--level $g q$ and $q\bar{q}$ terms
initiated by the diagrams shown in Fig.~3.4; $c)$ tree--level contributions from
$q\bar{q}, qq, qg, gg \to V^*+$ 2 partons in all possible ways, with the sums
of these diagrams for a given initial and final state to be squared and  added.
\begin{center}
\setlength{\unitlength}{1pt}
\SetWidth{1.1}
\begin{picture}(450,100)(-30,0)
\ArrowLine(0,100)(50,50)
\ArrowLine(50,50)(0,0)
\Photon(50,50)(100,50){-3}{5}
\GlueArc(27.5,72.5)(12.5,-45,135){3}{4}
\Gluon(21,20)(21,80){4}{4}
\put(80,35){$V^*$}
\put(-10,10){$q$}
\put(-10,90){$\bar{q}$}
\put(50,-5){\red{${\bf a)}$}}
\hspace*{1mm}
\Gluon(170,80)(170,20){4}{5}
\Line(140,80)(210,80)
\Line(140,20)(210,20)
\ArrowLine(210,20)(210,80)
\Gluon(210,20)(260,20){4}{4}
\Photon(210,80)(260,80){-3}{5}
\put(130,80){$q$}
\put(130,20){$\bar{q}$}
\put(170,-5){\red{${\bf b)}$}}
\hspace*{1mm}
\ArrowLine(290,80)(340,50)
\ArrowLine(290,20)(340,50)
\Photon(340,50)(380,50){-3}{5}
\Gluon(322,60)(360,80){3.5}{4.5}
\Gluon(322,40)(350,20){-3.5}{4.5}
\put(290,70){$q$}
\put(290,30){$\bar{q}$}
\put(320,-5){\red{${\bf c)}$}}
\end{picture} 
\end{center}
\vspace*{0mm}
\centerline{\it Figure 3.5: Diagrams for the NNLO QCD corrections to the 
process $q\bar{q} \to W^*$.}
\vspace*{1mm}

These corrections have been calculated a decade ago in Ref.~\cite{DYNNLO} and
recently updated \cite{ggH-NNLO1}.  However, these calculations are not
sufficient to obtain a full NNLO prediction: in the case of $ pp \to HZ$
production, because the final state is electrically neutral, two additional
sets of corrections need to be  considered at ${\cal O}(\alpha_s^2)$
\cite{pp-HV-NNLO}. \s

Indeed, contrary to charged $W$ bosons, the neutral $Z$ bosons can be produced via an effective $Z$--gluon--gluon coupling induced by quark loops. This can
occur at  the two--loop level in a box+triangle diagram in  $q\bar{q} \to Z^*$
[to be multiplied by the Born term], or at the one--loop level where
vertex diagrams appear for the $q\bar{q} \to gZ^*$ and $qg \to qZ^*$ processes
[to be multiplied by the respective ${\cal O}(\alpha_s)$ tree--level terms]. 
Because gluons have only vector couplings to quarks and the effective $Zgg$
coupling must be a color singlet, only the axial--vector part $a_q=2I_Q^3$ of 
the $Zq\bar{q}$ coupling will contribute as a consequence of Furry's theorem
\cite{Furry-theorem}. 
Since $a_q$ differs only by a sign for isospin up-- and down--type quarks, 
their contribution vanishes in the case of quarks that are degenerate in mass. 
Thus, in the SM, only the top and bottom quarks will contribute to these
topologies. These corrections have been evaluated in Refs.~\cite{Scott,Zgg} 
and have been shown to be extremely small and can be safely neglected. \s 

Another set of diagrams that contribute at  ${\cal O}(\alpha_s^2)$ to $ZH$ and
not to $WH$ production [again because of charge conservation] is the $gg$
initiated mechanism $gg \to HZ$ \cite{ggZH,BK}. It is mediated by quark loops 
[see Fig.~3.6] which enter in two ways. There is first a triangular
diagram with $gg \to Z^* \to HZ$, in which only the top and bottom quark
contributions are present, since  because of C--invariance, the  $Z$ boson 
couples only axially  to the internal quarks and the contribution of a mass 
degenerate
quark weak--isodoublet vanishes. There are also box diagrams where both the $H$
and $Z$ bosons are emitted from the internal quark lines and where only the
contribution involving heavy quarks which couple strongly to the Higgs boson  
[the top quark and, to a lesser extent, the bottom quark] are important.
It turns out that the two contributing triangle and box amplitudes interfere
destructively. 

\begin{center}
\setlength{\unitlength}{1pt}
\SetWidth{1.1}
\vspace*{-2mm}
\begin{picture}(450,100)(-10,0)
\Gluon(20,20)(60,20){4}{4}
\Gluon(20,80)(60,80){4}{4}
\ArrowLine(60,20)(60,80)
\ArrowLine(60,80)(100,50)
\ArrowLine(100,50)(60,20)
\Photon(100,50)(150,50){4}{5}
\Photon(150,50)(190,80){3.5}{4.5}
\DashLine(150,50)(190,20){5}
\put(120,60){$Z^*$}
\put(68,46){$Q$}
\put(10,18){$g$}
\put(10,78){$g$}
\put(200,20){$H$}
\put(200,80){$Z$}
\hspace*{9cm}
\Gluon(0,20)(40,20){4}{4}
\Gluon(0,80)(40,80){4}{4}
\ArrowLine(40,20)(90,20)
\ArrowLine(40,80)(90,80)
\ArrowLine(40,80)(40,20)
\ArrowLine(90,80)(90,20)
\Photon(90,80)(140,80){3.5}{4.5}
\DashLine(90,20)(140,20){5}
\put(145,20){$H$}
\put(58,46){$Q$}
\put(-10,18){$g$}
\put(-10,78){$g$}
\put(145,80){$Z$}
\end{picture} 
\vspace*{-2mm}
\nn {\it Figure 3.6: Diagrams for the $gg\to HZ$ process, which contributes to 
${\cal O}(\alpha_s^2)$.}
\end{center}

At the LHC, the contribution of this  gluon--gluon fusion mechanism to the $pp
\to HZ$ total production cross section can be substantial. This is due to the
fact that the suppression of the cross section  by a power $(\alpha_s/\pi)^2$
is partly compensated by the increased gluon luminosity  at high energies. In
addition, the tree--level cross section for $q\bar{q} \to HZ$  drops for
increasing c.m. energy and/or $M_H$ values, since it is mediated by
$s$--channel gauge boson exchange. Note that the cross section for this process 
is negligible at the Tevatron because of the low gluon luminosity and the 
reduced phase space. 

\subsubsection*{\underline{Numerical results}}

The $K$--factors, defined as the ratios of the cross sections at higher order
with $\alpha_s$ and the PDFs evaluated also at higher order, relative to the LO
order cross sections with  $\alpha_s$ and the PDFs consistently evaluated also
at LO,  are shown at NLO and NNLO in Figs.~3.7 in solid black lines for 
the LHC (left--hand side) and the Tevatron (right--hand side) as a function of 
the Higgs  mass for the process $pp \to HW$.  The scales have been fixed to
$\mu_F=\mu_R=M_{HV}$, where $M_{HV}$ is the invariant mass of the $HV$ system,  
and the MRST sets of PDFs for each perturbative order are used in a consistent 
manner. \s

The NLO $K$--factor is practically constant at the LHC, increasing only from
$K_{\rm NLO}=1.27$ for $M_H=110$ GeV to $K_{\rm NLO}=1.29$ for $M_H=300$ GeV.
The NNLO contributions increase the $K$--factor by a mere 1\% for the low $M_H$
value and by 3.5\% for the high value. At the Tevatron, the NLO $K$--factor is 
somewhat higher than at the LHC, enhancing the cross section by  $K_{\rm
NLO}=1.35$ for $M_H=110$ GeV and $K_{\rm NLO}=1.3$ for $M_H=300$ GeV with a
monotonic decrease.  The NNLO corrections increase the $K$--factor uniformly by
about 10\%. Thus, these NNLO corrections are more important at the Tevatron
than at the LHC. \s

Because of the slightly different phase space and scale, the $K$--factor for $pp
\to ZH$ is not identical to the $K$--factor for $pp \to WH$. However, since
$(M_Z^2-M_W^2)/\hat{s}$ is small and the dependence  of d$\Gamma$ in
eq.~(\ref{DY-factorization}) on $k^2$ is not very strong in the range that we
are considering, the $K$--factors for the two processes are very  similar when
the contribution of the $gg \to HZ$ component to be discussed later is not
included. \s

The bands around the $K$--factors in Fig.~3.7 represent the variation of the
cross sections when they are evaluated at renormalization and factorization
scale values that are independently varied from $\frac{1}{3} M_{HV} \leq \mu_F
\, (\mu_R) \leq 3 M_{HV}$, while the other is fixed to $\mu_R \, (\mu_F) =
M_{HV}$; the normalization is provided by the production cross section
evaluated at scales $\mu_F=\mu_R=M_{HV}$. A $K$--factor for the LO cross
section, $K_{\rm LO}$, has also been defined by evaluating the latter at given
factorization and renormalization scales and normalizing to the LO cross
sections evaluated at the central scale, which, in our case, is given by
$\mu_F=\mu_R=M_{HV}$. As can be seen, except from the accidental cancellation
of the scale dependence of the LO cross section at the LHC for $M_H \sim 260$
GeV, the decrease of the
scale variation is strong when going from LO to NLO and then to NNLO. For
$M_H=120$ GeV, the uncertainty from the scale choice at the LHC drops from 10\%
at LO, to 5\% at NLO, and to 2\% at NNLO. At the Tevatron and for the same
Higgs boson mass, the scale uncertainty drops from 20\% at LO, to 7\% at NLO,
and to 3\% at NNLO. \s

If this variation of the cross section with the two scales is taken as an
indication of the uncertainties due to the not yet calculated higher--order
corrections, one concludes that once the NNLO contributions are included in the
prediction, the cross section for the $pp \to HV$ process is known at the
rather accurate level of a few percent.  

\begin{figure}
\begin{center}
{ \unitlength 1cm
\begin{picture}(15.5,6.0)
\put(-2.2,-5.7){\includegraphics{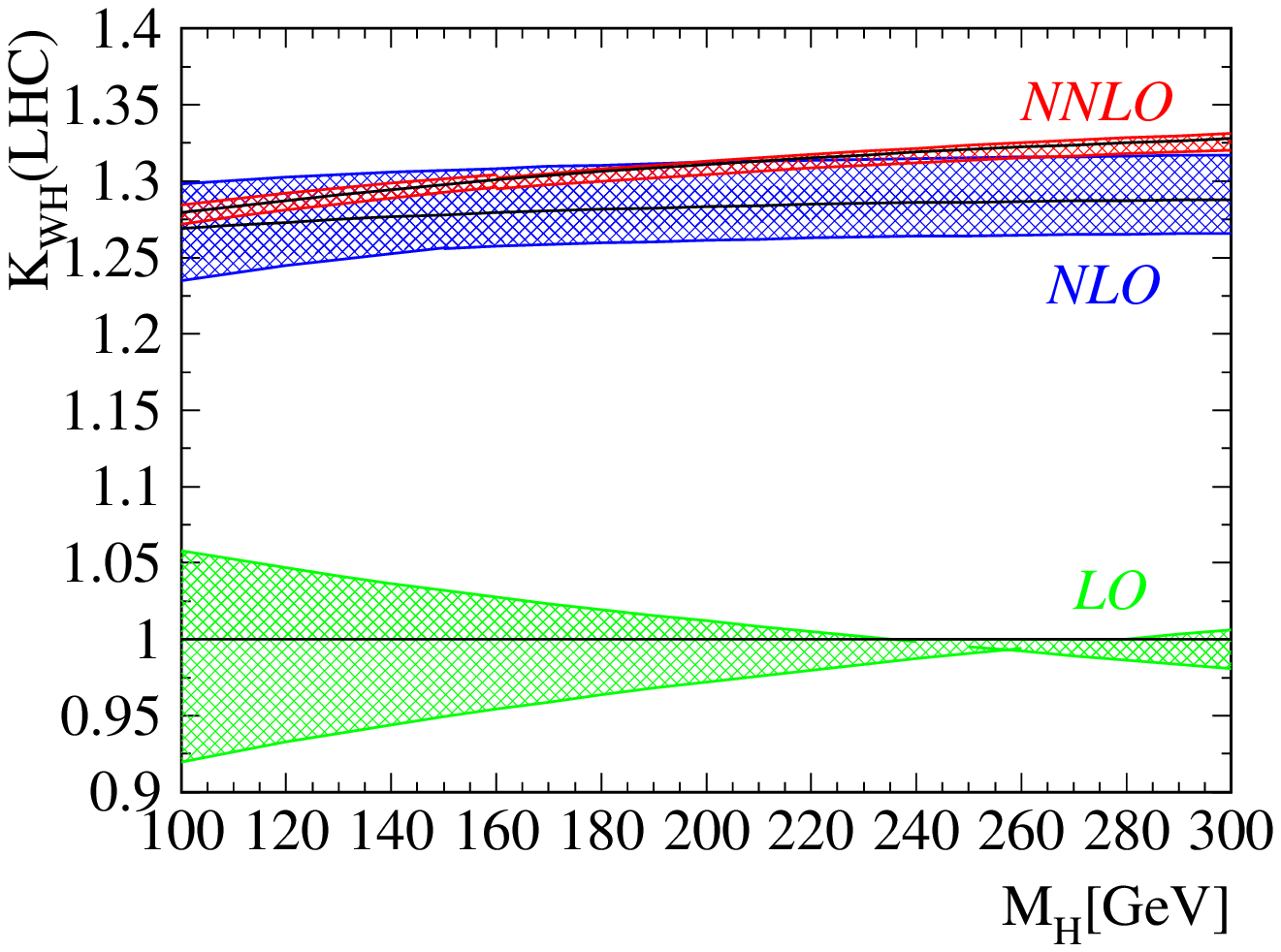}}
\put( 6.0,-5.7){\includegraphics{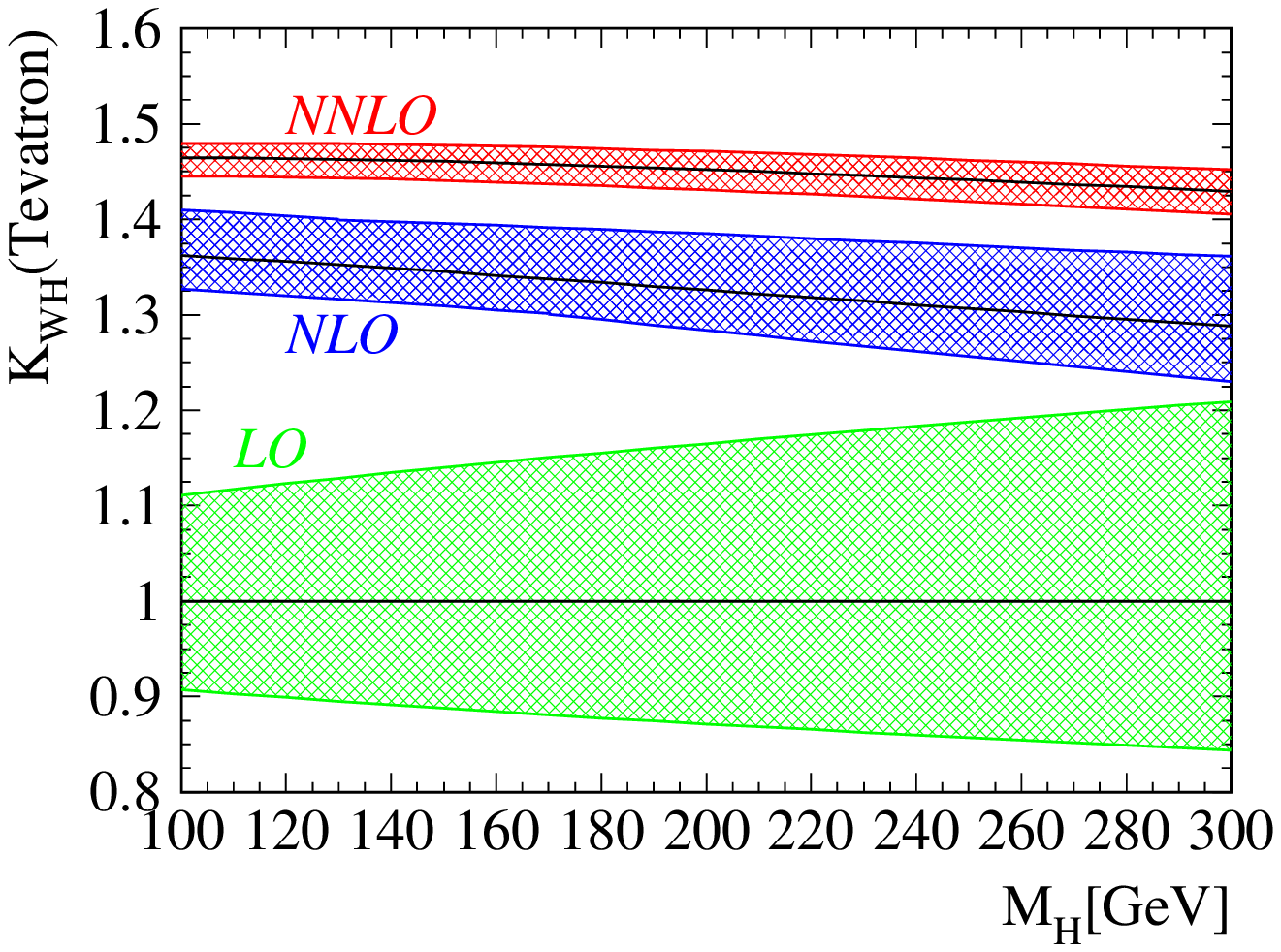}}
\end{picture} }
\vspace*{-2mm}

\end{center}
{\it Figure 3.7: The $K$--factors for $pp \to HW$ at the LHC (left) and the
Tevatron (right) as a function of $M_H$ at LO, NLO and NNLO (solid black 
lines). The bands represent the spread of the cross section when the 
renormalization and factorization scales are varied in the range $\frac{1}{3}
M_{HV} \leq \mu_R\, (\mu_F) \leq 3M_{HV}$, the other scale being fixed at 
$\mu_F (\mu_R)= M_{HV}$; from Ref.~\cite{pp-HV-NNLO}.}
\vspace*{-4mm}
\end{figure}

\subsubsection{The electroweak radiative corrections}

The associated $W/Z+H$ process is the only Higgs production mechanism for which
the complete calculation of the ${\cal O}(\alpha)$ electroweak corrections has
been performed \cite{pp-HV-EW}. There are a few hundred Feynman diagrams
contributing at the one--loop level, and some generic ones are shown in
Fig.~3.8. The radiative corrections can be cast into three categories.

\begin{center}
\vspace*{-.2cm}
\hspace*{-12.5cm}
\SetWidth{1.1}
\begin{picture}(300,80)(0,0)
\ArrowLine(100,25)(140,50)
\ArrowLine(100,75)(140,50)
\Photon(140,50)(165,50){3.2}{4.5}
\ArrowLine(165,50)(200,75)
\ArrowLine(165,50)(200,25)
\Line(200,75)(200,25)
\DashLine(200,75)(240,75){4}
\Photon(200,25)(240,25){3.2}{4.5}
\Text(202,75)[]{\bb}
\Text(100,60)[]{$q$}
\Text(100,40)[]{$\bar q$}
\Text(150,63)[]{$\gamma,Z,W$}
\Text(190,50)[]{$f$}
\Text(250,30)[]{$V$}
\Text(250,70)[]{\bH}
\ArrowLine(270,25)(310,25)
\ArrowLine(270,75)(310,75)
\ArrowLine(310,25)(310,75)
\Photon(310,25)(355,25){3.2}{5.5}
\Photon(310,75)(355,75){3.2}{5.5}
\Photon(355,75)(355,25){3.2}{5.5}
\DashLine(355,75)(390,75){4}
\Text(357,75)[]{\bb}
\Photon(355,25)(390,25){3.2}{5}
\ArrowLine(420,25)(460,50)
\ArrowLine(420,75)(460,50)
\Photon(460,50)(500,50){3.2}{5.5}
\Photon(500,50)(540,25){3.2}{5.5}
\DashLine(500,50)(540,75){4}
\Text(502,50)[]{\bb}
\Photon(440,65)(470,75){3.2}{5.}
\Text(477,70)[]{$\gamma$}
\Text(330,-3)[]{\it Figure 3.8: Generic diagrams for the ${\cal O}(\alpha)$
corrections to the $pp \to HV$ production process.}
\vspace*{0.mm}
\end{picture}
\end{center}
\vspace*{-2mm}

There are first QED corrections in which photons are exchanged in the initial
quark--antiquark states and, in order to obtain infrared finite corrections,
real--photon bremsstrahlung has to be added. Having done this, ${\cal
O}(\alpha)$ corrections due to collinear photon emission and involving
logarithms of the initial state quark masses are still present. These mass
singularities are absorbed into the PDFs in exactly the same way as in QCD by
$\overline{\mbox{MS}}$ factorization.  This, however, also requires the
inclusion of the corresponding ${\cal O}(\alpha)$ corrections into the DGLAP
evolution of these distributions and into their fit to experimental data, which
has not been performed yet.  Nevertheless, an approximate inclusion of
these corrections to the DGLAP evolution shows \cite{Kripfganz} that the
impact of these corrections on the quark distributions is well below 1\%, at
least in the $x$ range that is relevant at the Tevatron and the LHC. This is
also supported by a recent analysis of the MRST collaboration \cite{Stirling}
which took into account these effects into the DGLAP equations. \s

The bulk of the electroweak corrections can be in principle incorporated by
using the improved Born approximation discussed in \S1.2.4. Using the Fermi
coupling constant $G_\mu$ rather than $\alpha(0)$ as input in the tree--level
cross section, $\pi \alpha \to  \sqrt{2} G_\mu M_W^2 ( 1- M_W^2/M_Z^2)$, takes
into account the contribution $\Delta r \simeq \Delta \alpha (M_Z^2) - 3\Delta
\rho$. In this case, the large universal corrections originating from the light
fermion contributions to the running of $\alpha$ [$2 \times \Delta \alpha (M_Z)
\sim 12\%$, since the cross section is proportional to $\alpha^2$] and those
which are quadratic in the top quark [$2 \times 3\Delta \rho  \sim 6\%$] are
automatically included. One has also to include the contributions that are
quadratic in the top mass and which are contained in  the $HVV$ vertex as it
was discussed in \S2.4.2, i.e.  $\delta_{HVV} \sim -5 x_t$ with $x_t= G_\mu^2
m_t^2/( 8 \sqrt{2} \pi^2)$ at this order. \s

Finally, one has to include the bosonic one--loop corrections which involve
many self--energy, vertex and box correction diagrams and which have to be
calculated by brute force using standard techniques. The calculation of
Ref.~\cite{pp-HV-EW} has been performed in the on--shell renormalization
scheme. It turns out that the non--universal bosonic contributions are rather
large and negative and, in fact, dominate over the fermionic corrections and
even over the photonic initial state corrections.  

\vspace*{3.cm}
\begin{figure}[!h]
\begin{center}
{ \unitlength 1cm
\begin{picture}(15.5,6.5)
\put(-2.2,-2){\includegraphics{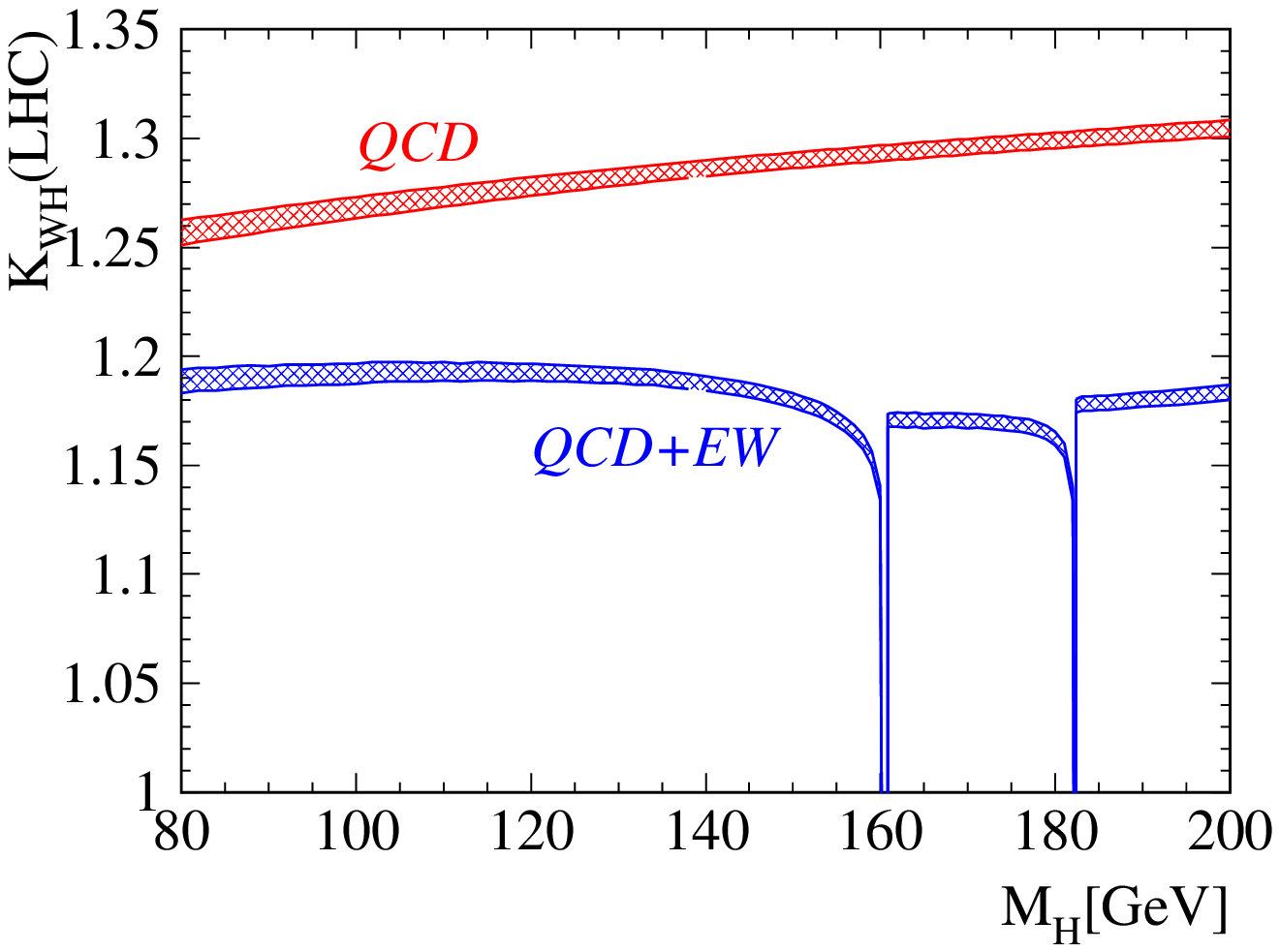}}
\put( 6.0,-2){\includegraphics{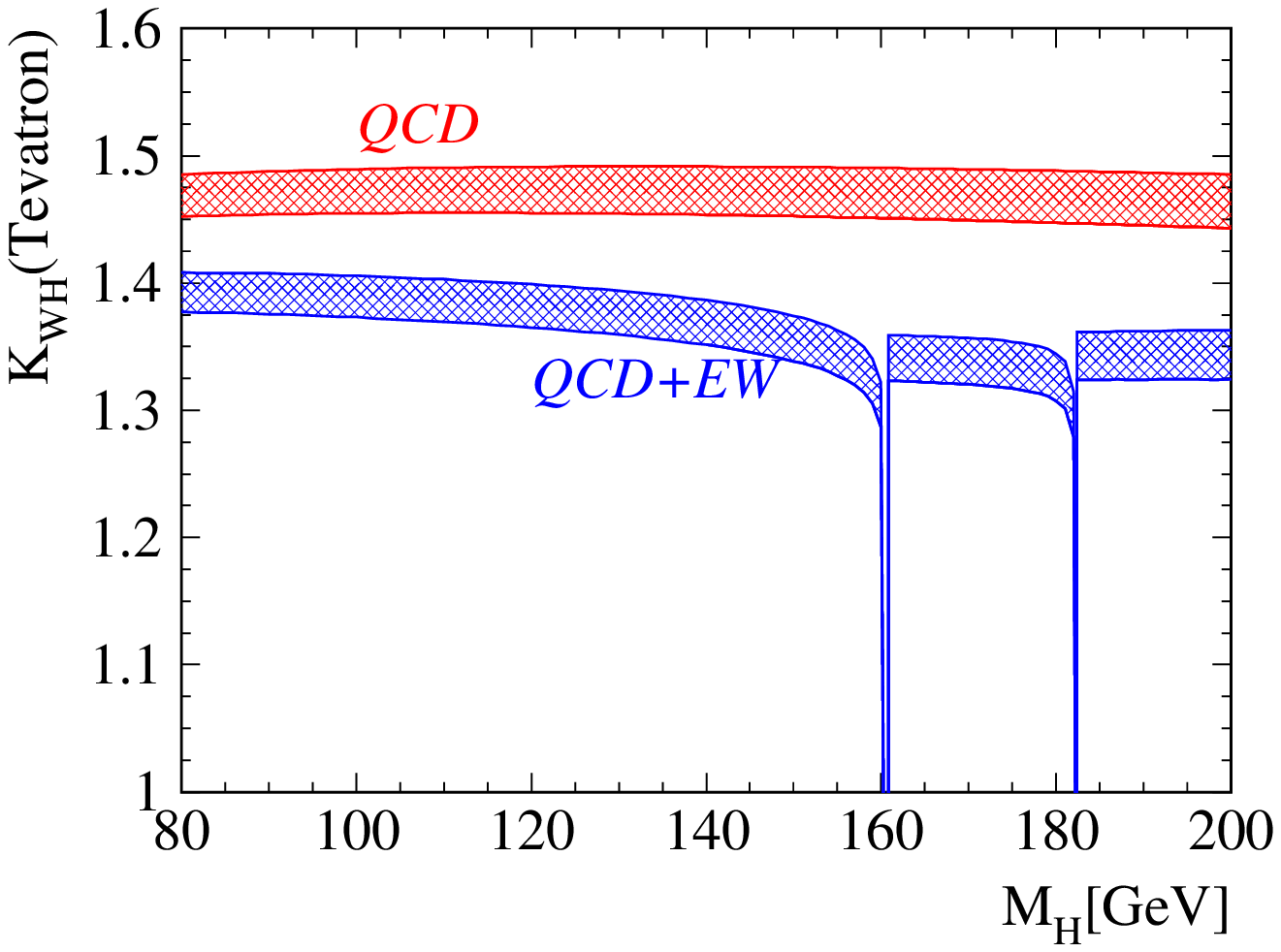}}
\end{picture} }
\end{center}
\vspace*{-3.7cm}
{\it Figure 3.9: $K$--factors for $WH$ and $ZH$ production at the LHC (left 
figure) and the Tevatron (right figure) after including  the NNLO QCD and
the electroweak ${\cal O}(\alpha)$ corrections \cite{pp-HV-EW+QCD}. }
\vspace*{-2mm}
\end{figure}   

The fermionic contributions being positive and the bosonic ones negative, there
is a partial cancellation of the two contributions and, since the bosonic
corrections are more important, the net effect is that the total electroweak
corrections decrease the $q\bar{q} \to HV$ production cross section at both the
Tevatron and the LHC by approximately 5 to 10\% for Higgs masses in the range
100--200 GeV where the production rates are large enough.  This is shown in
Fig.~3.9 where we display the $K$--factors for $pp \to HW$ at the Tevatron and
LHC as functions of $M_H$, when only NLO+NNLO QCD corrections are included
(upper bands) and when the electroweak corrections are also taken into account
(lower bands). The thickness of the bands is due to the scale variation as
discussed previously. The unphysical singularities in the electroweak
corrections at the $M_H=2M_W$ and $2M_Z$ thresholds can be removed by including
the finite width of the particles. Note that at the LHC, the
electroweak correction is almost the same for $pp \to HW$ and $pp\to HZ$, the
difference being less than 2\%.  

\subsubsection{The total cross section and the PDF uncertainties}

In Fig.~3.10, we present the total production cross sections for the processes
$q\bar{q} \to HW$ and $HZ$ at the Tevatron and the LHC as a function of
$M_H$, when both the NNLO QCD and the electroweak corrections are added. In the
case of the $HZ$ process, the contribution of the $gg \to ZH$ subprocess to the
total cross section is not included, but it is displayed separately in the  LHC
case. For Higgs masses in the range 100 GeV $\lsim M_H \lsim 250$ GeV where
$\sigma(q\bar{q} \to HZ)$ is significant, $\sigma(gg \to HZ)$ is at the level
of 0.1 to 0.01 pb and represents about 10\% of the total cross section for low
$M_H$ values. The $gg \to HZ$ cross section is thus much larger than the 
contribution of the NNLO correction and, therefore, generates a scale 
uncertainty that is larger than in the $HW$ production case.

\begin{figure}[!h]
\begin{center}
{ \unitlength 1cm
\begin{picture}(15.5,7)
\put(-4.4,-8.5){\includegraphics{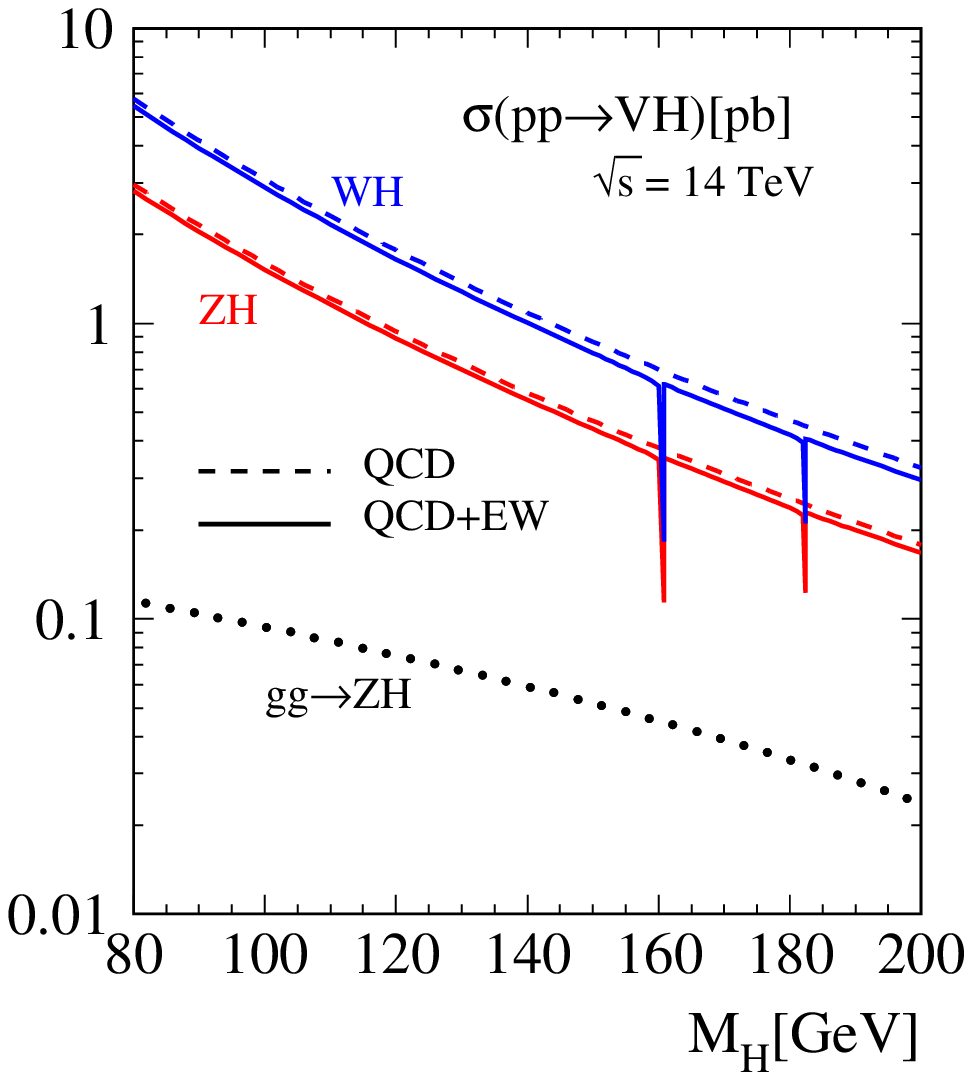}}
\put( 3.8,-8.5){\includegraphics{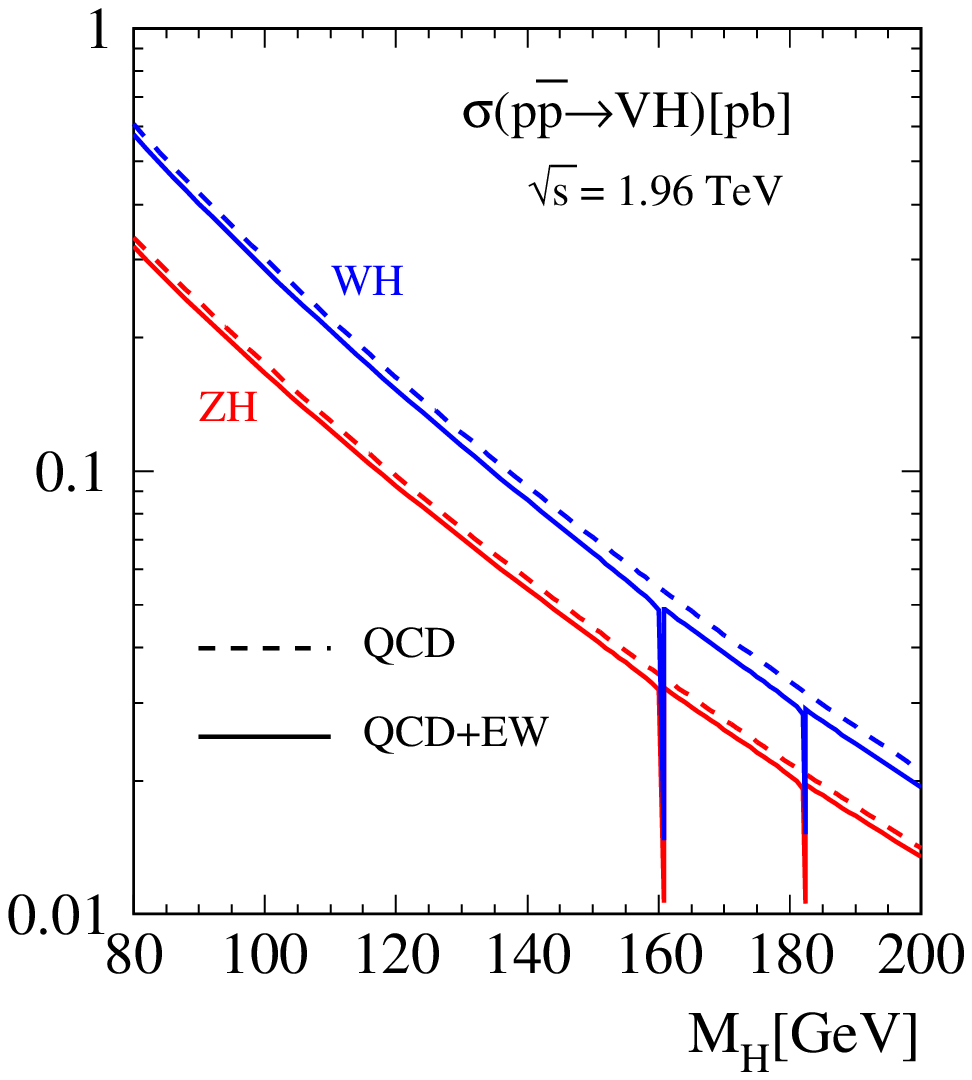}}
\end{picture} }
\end{center}
\vspace*{10mm}
{\it Figure 3.10: The total production cross sections for $pp \to HW$ and 
$HZ$ at the  LHC (left) and the Tevatron (right) as a function of $M_H$ when 
the NNLO QCD and the electroweak corrections are included. The MRST  parton 
densities have been used. The contribution of the $gg \to HZ$ process is shown 
separately in the case of the LHC; from Ref.~\cite{pp-HV-EW+QCD}.}
\vspace*{-3mm}
\end{figure}

Finally, let us discuss the PDF uncertainties in the $pp \to HV$ cross
sections, following the lines introduced in \S3.1.5. In Fig.~3.11, we show as a
function of $M_H$ and for the LHC and the Tevatron, the central values and the
uncertainty band limits of the NLO QCD $q\bar{q} \to HW$ cross section for the
CTEQ, MRST and Alekhin parameterizations. In the inserts to these figures, we
show the spread uncertainties in the predictions for the cross sections, when
they are normalized to the prediction of the reference CTEQ6M set. \s

At the LHC, the uncertainty band is almost constant and for CTEQ, is of the
order of 4\% over the Higgs mass range between 100 and 200 GeV.  At the
Tevatron, the uncertainty band increases with the Higgs mass and exceeds 6\% at
$M_{H}\sim 200$ GeV. The uncertainty in the MRST parameterization is twice
smaller. To produce a vector plus a Higgs boson in this mass range,  the
incoming quarks originate from the intermediate--$x$ regime at the LHC, at
Tevatron energies, however, some of the participating quarks originate from the
high--$x$ regime, which explains the increasing  behavior of the uncertainty
bands observed in this case.  The different magnitude of the cross sections,
$\sim 12$\% ($\sim 8$\%) larger in the Alekhin case than for CTEQ  at the LHC
(Tevatron), is due to the larger quark and antiquark densities of the former
parameterization. For this particular PDF set, the difference in the shifts of
the central values in the LHC and Tevatron cases is due to the different
initial states, $pp$ [where $\bar q$  comes from the sea] versus $p\bar{p}$
[where it is valence+sea $\bar q$]; see Fig.~3.2. \s

\begin{figure}[hbtp]
\begin{center}
\vspace*{-2.5cm}
\hspace*{-1cm}
\psfig{figure=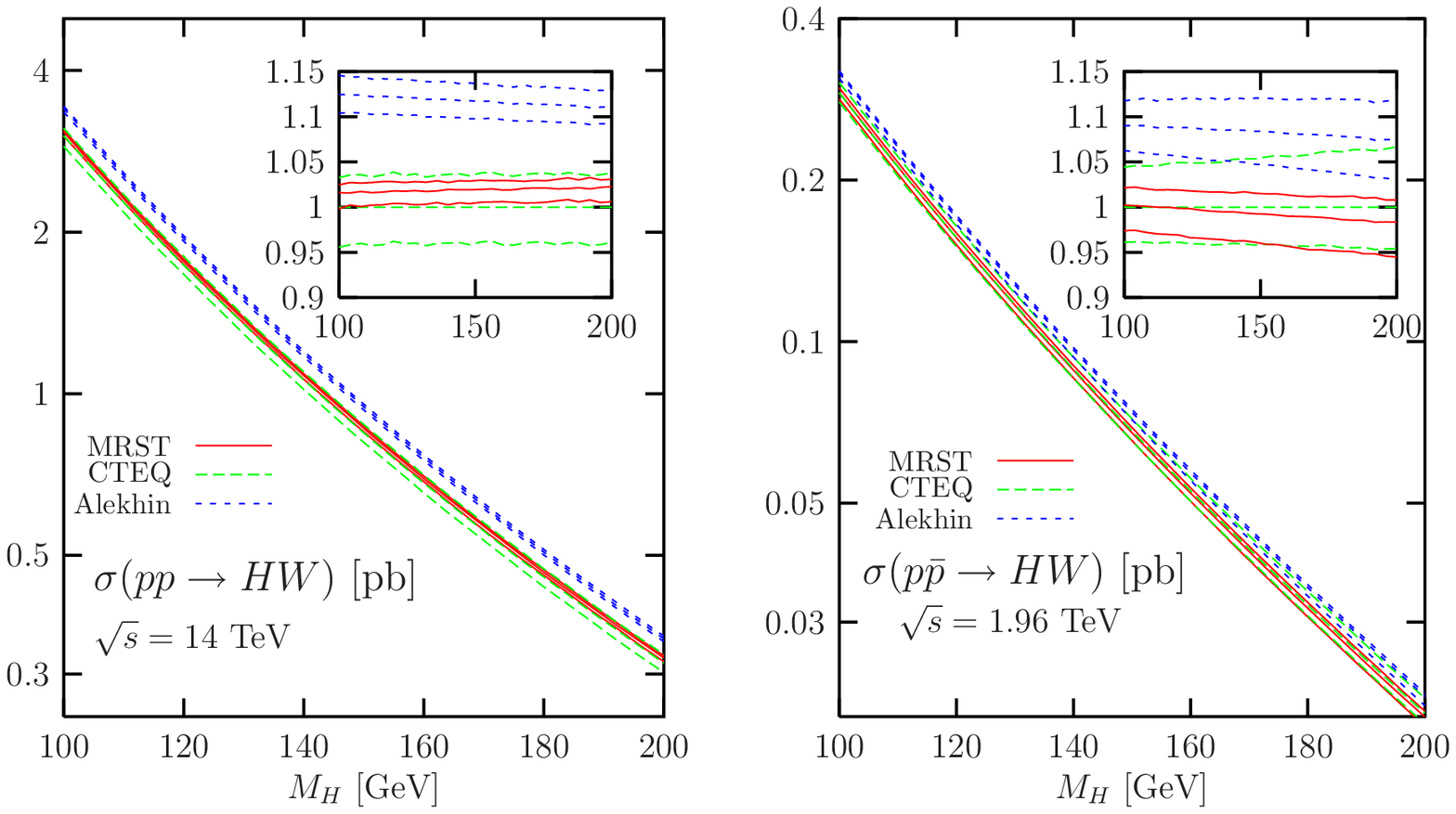,width=18cm}
\vspace*{-15.5cm}
\end{center}
{\it Figure 3.11: The CTEQ, MRST and Alekhin PDF uncertainty bands for the NLO
cross sections for the production of the Higgs boson at the LHC (left) and at
the Tevatron (right) in the $q\bar{q} \to HW$  process.  The inserts 
show the spread in the predictions; from Ref.~\cite{Samir}.}
\vspace*{-1mm} 
\end{figure}

Note that an additional systematic error of about 5\% arises from the $pp$
luminosity. If one uses the Drell--Yan processes to measure directly the $q$ and
$\bar q$ luminosities at hadron colliders, the errors on the cross sections for
associated $HV$ production when normalized to this rate would lead to a total 
systematical uncertainty of less than 1\% \cite{Dittmar}. In this case, the
dominant part the of the $K$--factor will also drop out in the ratio.  

\subsection{The vector boson fusion processes}

\subsubsection{The differential and total cross sections at LO}

The matrix element squared for the massive vector boson fusion process
\cite{VVH-Cahn,VVH-DW,VVH-Altarelli,VVH-Kilian}, in terms of the momenta 
of the involved particles
\beq
q_1 (p_1) \, q_2 (p_2) \to V^*(q_1=p_3-p_1)  \, V^*(q_2=p_4-p_2)  \, q_3(p_3) 
 \, q_4(p_4) \to q_3 (p_3)  \, q_4 (p_4)  \, H(k)
\eeq
with $V=W,Z$, is given by
\beq
|{\cal M}|^2  &=& 4  \sqrt{2} N_{c}^f G_\mu^3 M_V^8 
\frac{C_+ (p_1 \cdot p_2) (p_3 \cdot p_4)+C_- (p_1 \cdot p_4) (p_2 \cdot p_3)} 
{(q_1^2-M_V^2)^2 (q_2^2-M_V^2)^2}
\label{Hqq:amp2}
\eeq
where, in terms of the usual vector and axial-vector couplings of the gauge
bosons to fermions $\hat a_f=2I_f^3, \hat v_f=2I_f^3 -4 Q_f s_W^2$ for $V=Z$ 
and $\hat v_f= \hat a_f=\sqrt{2}$ for $V=W$, $C_\pm$ read
\beq 
C_\pm= (\hat v_{q_1}^2 + \hat a_{q_1}^2) ( \hat v_{q_3}^2+ \hat a_{q_3}^2) 
\pm 4 \hat v_{q_1} \hat a_{q_1} \hat v_{q_3} \hat a_{q_3} 
\eeq
giving rise to the differential distribution
\beq
{\rm d}\hat{\sigma}_{\rm LO} = \frac{1}{4}\frac{1}{9} \frac{1}{2\hat{s}}
\times |{\cal M}|^2 \times \frac{1}{(2\pi)^5} \, \frac{ {\rm d}^3k }
{ 2 {\rm d}E_H} \, \frac{ {\rm d}^3p_3 }{ 2 {\rm d}E_3}
\, \frac{ {\rm d}^3p_4 }{ 2 {\rm d}E_4} \, \delta^4 (p_1+p_2 -p_3-p_4-k) 
\eeq
The integration over the variables $p_3$ and $p_4$ are conveniently
performed in the rest frame of the two quarks $\vec{p}_3+\vec{p}_4=0$, 
and one finds \cite{VVH-Altarelli,VVH-Kilian}
\beq
\frac{ {\rm d}\hat{\sigma}_{\rm LO} } {{\rm d}E_H {\rm d}\cos \theta } = 
\frac{G_\mu^3 M_V^8}{9 \sqrt2\,\pi^3 \hat s}\, \frac{p_H}{32s_1 s_2 r}\bigg[ 
C_+ {\cal H}_+ + C_- {\cal H}_- \bigg]
\label{Hqq:Edistr}
\eeq
with
\beq
{\cal H}_+ &=& (h_1+1)(h_2+1) \left[
\frac{2}{h_1^2-1} + \frac{2}{h_2^2-1} - \frac{6s_\chi^2}{r}
        + \left(\frac{3t_1t_2}{r}-c_\chi\right)
          \frac{\ell}{\sqrt{r}}\right]
    \nonumber\\
    && \qquad{}
        - \left[ \frac{2t_1}{h_2-1} + \frac{2t_2}{h_1-1}
                + \left(t_1+t_2+s_\chi^2\right)
                \frac{\ell}{\sqrt{r}}\right] \non \\
{\cal H}_- &=& 	2(1 - c_\chi) \left[
        \frac{2}{h_1^2-1} + \frac{2}{h_2^2-1} - \frac{6s_\chi^2}{r}
        + \left(\frac{3t_1t_2}{r}-c_\chi\right)
          \frac{\ell}{\sqrt{r}}\right]		
\label{pp:VVH+H-}
\eeq		
In these equations, $p_H=\sqrt{E_H^2-M_H^2}$ is the Higgs boson momentum,
$\theta$ is the scattering angle, while $\epsilon_\nu=\sqrt{\hat s} -E_H$ and
$s_\nu=\epsilon_\nu^2-p_H^2$ are the energy and the  invariant mass of the
final state quark pair. The other abbreviations are
\beq
\label{pp:VVHabrev}
   s_{1,2} = \sqrt{\hat s}(\epsilon_\nu\pm p_H\cos\theta)\ , \ 
   h_{1,2} = 1 + 2M_V^2/s_{1,2} \ , \
  t_{1,2} = h_{1,2} + c_\chi h_{2,1} \hspace*{2cm} \\[1mm]
   c_\chi = 1 - \frac{2 \hat s s_\nu}{s_1s_2} = 1 - s_\chi^2 , \ 
  r = h_1^2 + h_2^2 + 2c_\chi h_1h_2 - s_\chi^2\ , \ \non 
  \ell = {\displaystyle \log\frac{h_1h_2 + c_\chi + \sqrt{r}}
                        {h_1h_2 + c_\chi - \sqrt{r}}}
\eeq	
To derive the partonic total cross section, $\hat{\sigma}_{\rm LO}(qq \to qq 
H)$, the differential cross section must be integrated over the region
\beq
-1< \cos \theta <1 \quad \mbox{and}\quad
M_H < E_H < \frac{\sqrt{\hat s}}{2}\left(1+\frac{M_H^2}{\hat s}\right)
\label{pp:VVHbound}
\eeq 
Summing over the contributing partons, including both the $WW$ and $ZZ$ fusion
channels and folding with the parton luminosities, one obtains the  total
hadronic cross section $\sigma(pp \to V^* V^* \to qq H)$ at LO.
The cross sections, using the CTEQ set of parton densities, are shown in
Fig.~3.12 as a function of $M_H$ for $p\bar{p}$ at the Tevatron and for $pp$ at
the LHC. In the latter case, the separate $WW$ and $ZZ$ contributions, as
well as their total sum, are displayed; the interference between the $WW$
and $ZZ$ contributions is less than 1\% and can be neglected. \s

While they are rather large at the LHC, in particular for Higgs bosons in the
mass range 100 GeV $\lsim M_H \lsim 200$ GeV where they reach the level of a few
picobarns, the total cross sections are very small at the Tevatron and they
barely reach the level of 0.1 pb even for $M_H=100$ GeV. This is due to the
fact that the main contribution originates from longitudinal gauge bosons
[which as, discussed previously, have interactions which grow with energy], and
the partonic cross sections  rise logarithmically with the c.m. energy of the
subprocess, $\hat{\sigma} \propto \log \hat s/M_V^2$, giving much larger rates 
at high energies. In our subsequent discussion, we will therefore consider this
process only in the case of the LHC. \s

Note also that the main contribution to the cross section is due to the $WW$
fusion channel, $\sigma (WW\to H) \sim 3 \sigma(ZZ \to H)$ at the LHC,  a
consequence of the fact that the $W$ boson couplings to fermions are larger
than those of the $Z$ boson.  

\begin{figure}[htbp]
\begin{center}
\vspace*{-2.5cm}
\hspace*{-3cm}
\epsfig{file=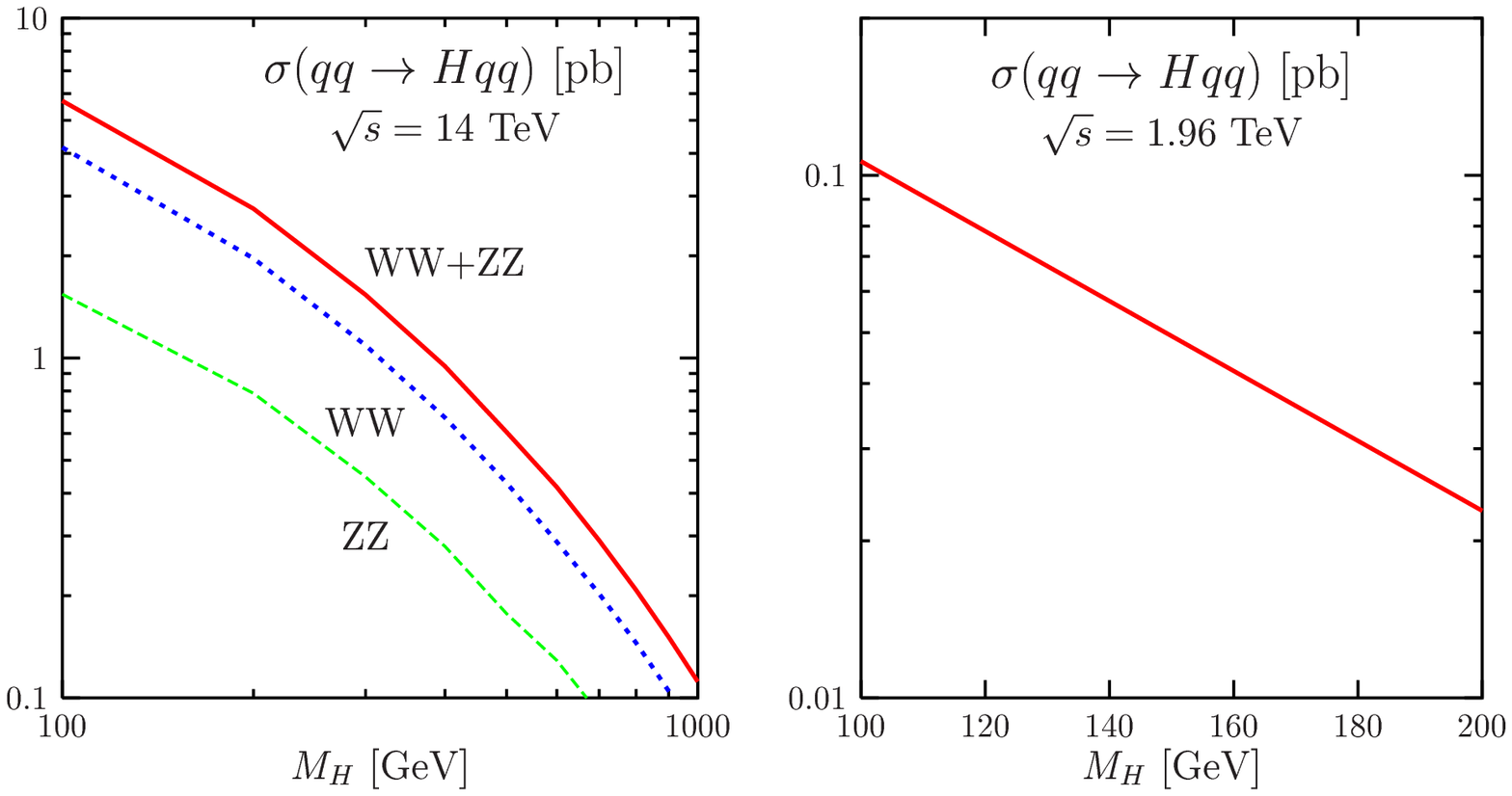,width=16.cm} 
\end{center}
\vspace*{-13cm}
{\it Figure 3.12: Individual and total cross sections in the vector fusion $q 
q\to V^*V^* \to Hqq$ processes at leading order at the LHC (left) and 
total cross section at the Tevatron (right).} \label{ppHqq:sigtot}
\end{figure}
\vspace*{-3mm}

\subsubsection{The cross section at NLO}

The QCD corrections to the vector boson fusion process, $q q\ra qqV^* V^* \ra 
qqH$ consist of the virtual quark self energy and vertex corrections and the 
additional gluon emission from the initial and final states, $qq \to Hqq+g$; 
the gluon initiated subprocess $gq \to Hqq+q$ has also to be taken into 
account. Some generic Feynman diagrams are shown in Fig.~3.13.

\vspace*{1cm}
\begin{figure}[!h]
\begin{center}
\setlength{\unitlength}{.8pt}
\SetWidth{1.1}
\begin{picture}(180,100)(-50,0)
\ArrowLine(0,100)(50,50)
\ArrowLine(50,50)(0,0)
\Gluon(25,75)(25,25){-3}{5}
\Photon(50,50)(100,50){-3}{5}
\put(85,46){$V^*$}
\put(-10,109){$q$}
\put(-10,4){$\bar{q}$}
\put(10,56){$g$}
\end{picture}
\begin{picture}(180,100)(-30,0)
\ArrowLine(0,100)(50,50)
\ArrowLine(50,50)(0,0)
\Photon(50,50)(100,50){-3}{5}
\GlueArc(27.5,72.5)(12.5,-45,135){3}{4}
\put(85,46){$V^*$}
\put(-10,109){$q$}
\put(-10,4){$\bar{q}$}
\put(40,119){$g$}
\end{picture}
\begin{picture}(180,100)(-10,0)
\ArrowLine(0,100)(50,50)
\ArrowLine(50,50)(0,0)
\Photon(50,50)(100,50){-3}{5}
\Gluon(27.5,72.5)(50,100){-3.5}{4.5}
\put(85,46){$V^*$}
\put(-10,109){$q$}
\put(-10,4){$\bar{q}$}
\put(50,90){$g$}
\vspace*{-3mm}
\end{picture}
\end{center}
\vspace*{-3mm}
\centerline{\it Figure 3.13: Feynman diagrams for NLO QCD corrections to the 
$V^*qq$ vertex. } 
\vspace*{-3mm}
\end{figure}

Since at the lowest order the incoming/outgoing quarks are in color singlets, 
at NLO no
gluons will be exchanged between the first and the second incoming (outgoing)
quark line [this will be no longer true at ${\cal O}(\alpha_s^2)$] and, hence,
the QCD corrections only consist of the well--known corrections to the
structure functions $F_i(x,M^2)$. The NLO corrections can therefore be more
conveniently calculated in the structure function approach. In this case,  
the differential LO partonic cross section  can be cast into the 
form \cite{pp-Hqq-NLO1,HVNLO-DS,Review-Michael} 
\begin{eqnarray}
{\rm d}\sigma_{\rm LO} &=& \frac{1}{4} \frac{\sqrt{2}G_\mu^3M_V^8
q_1^2 q_2^2} {[q_1^2-M_V^2]^2 [q_2^2-M_V^2]^2} \left\{
F_1(x_1,\mu_F^2) F_1(x_2,\mu_F^2) \left[ 2+\frac{(q_1 q_2)^2}{q_1^2q_2^2} 
\right] \right. \nonumber \\
&+& \frac{F_1(x_1,\mu_F^2)F_2(x_2,\mu_F^2)}{P_2q_2} \left[\frac{(P_2q_2)^2}
{q_2^2}- m_P^2+\frac{1}{q_1^2}\left(P_2q_1-\frac{P_2q_2}{q_2^2}q_1q_2\right)^2 
\right] \nonumber \\
&+&\! \frac{F_2(x_1,\mu_F^2)F_1(x_2,\mu_F^2)}{P_1q_1}\left[\frac{(P_1q_1)^2}
{q_1^2}- m_P^2+\frac{1}{q_2^2}\left(P_1q_2-\frac{P_1q_1}{q_1^2}q_1q_2\right)^2 
\right] \nonumber \\
&+& \frac{F_2(x_1,\mu_F^2)F_2(x_2,\mu_F^2)}{(P_1q_1)(P_2q_2)}\left[P_1P_2\!-\! 
\frac{(P_1q_1)(P_2q_1)}{q_1^2}\!-\!\frac{(P_2q_2)(P_1q_2)}{q_2^2} 
\!+\! \frac{(P_1q_1)(P_2q_2)(q_1q_2)}{q_1^2q_2^2}\right]^2 \hspace*{-1cm}
\nonumber \\ 
&+&\left. \frac{F_3(x_1,\mu_F^2)F_3(x_2,\mu_F^2)}{2(P_1q_1)(P_2q_2)}\left[
(P_1P_2)(q_1q_2) - (P_1q_2)(P_2q_1) \right] \right\} {\rm d} x_1 {\rm d}x_2
\frac{{\rm dPS}_3}{\hat s}
\label{eq:vvhlo}
\end{eqnarray}
where dPS$_3$ denotes the three--particle phase space, $m_P$ the proton mass, 
$P_{1,2}$ the proton momenta and $q_{1,2}$ the momenta of the virtual vector 
bosons $V^*$. The functions $F_i(x,\mu_F^2)$, with {\small $i=1,2,3$}, are the 
usual structure functions from deep--inelastic scattering processes at the 
factorization scale $\mu_F$ and read 
\begin{eqnarray}
F_1(x,\mu_F^2) & = & \sum_q (\hat v_q^2+\hat a_q^2) [q(x,\mu_F^2) + \bar 
q(x,\mu_F^2)] \nonumber \\
F_2(x,\mu_F^2) & = & 2x \sum_q (\hat v_q^2+\hat a_q^2) [q(x,\mu_F^2) + 
\bar q(x,\mu_F^2)] \nonumber \\
F_3(x,\mu_F^2) & = & 4 \sum_q \hat v_q \hat a_q [-q(x,\mu_F^2) + 
\bar q(x,\mu_F^2)]
\label{eq:stfu}
\end{eqnarray}
The QCD corrections only consist of the well--known corrections to the
structure functions $F_i(x,M^2)$ and the final result for the corrected cross
section at ${\cal O}(\alpha_s)$ can be simply obtained from the replacements
\cite{pp-Hqq-NLO1,HVNLO-DS,Review-Michael} 
\begin{eqnarray}
F_i(x,\mu_F^2)  \to  F_i(x,\mu_F^2) + \Delta F_i(x,\mu_F^2,Q^2) 
\end{eqnarray}
%
\vspace*{-7mm}
\begin{eqnarray}
\Delta F_1(x,\mu_F^2,Q^2) & = & \frac{\alpha_s(\mu_R)}{\pi}\sum_q (\hat v_q^2+
\hat a_q^2)
\int_x^1 \frac{dy}{y} \left\{ \frac{2}{3} [q(y,\mu_F^2) + \bar q(y,\mu_F^2)]
\right. \nonumber \\
& &  \hspace*{-2cm}
\left[ -\frac{3}{4} P_{qq}(z) \log \frac{\mu_F^2z}{Q^2} + (1+z^2) {\cal D}_1(z)
- \frac{3}{2} {\cal D}_0(z) 
+ 3 - \left(
\frac{9}{2} + \frac{\pi^2}{3} \right) \delta(1-z) \right]
\nonumber \\
& & \hspace*{-2cm} \left. + \frac{1}{4} g(y,\mu_F^2) \left[ -2 P_{qg}(z) \log \frac{\mu_F^2z}
{Q^2(1-z)}
 + 4z(1-z) - 1 \right] \right\} \\
\Delta F_2(x,\mu_F^2,Q^2) & = & 2x\frac{\alpha_s(\mu_R)}{\pi}\sum_q (\hat v_q^2
+\hat a_q^2)
\int_x^1 \frac{dy}{y} \left\{ \frac{2}{3} [q(y,\mu_F^2) + \bar q(y,\mu_F^2)]
\right. \nonumber \\ & &  \hspace*{-2cm}
\left[ -\frac{3}{4} P_{qq}(z) \log \frac{\mu_F^2z}{Q^2} + (1+z^2) {\cal D}_1(z)
- \frac{3}{2} {\cal D}_0(z) 
+ 3 + 2z - \left(
\frac{9}{2} + \frac{\pi^2}{3} \right) \delta(1-z) \right]
\nonumber \\
& &  \hspace*{-2cm} \left. + \frac{1}{4} g(y,\mu_F^2) \left[ -2P_{qg}(z) \log \frac{\mu_F^2z}
{Q^2(1-z)}
+ 8z(1-z) - 1 \right] \right\} \\
\Delta F_3(x,\mu_F^2,Q^2) & = & \frac{\alpha_s(\mu_R)}{\pi} \sum_q 4 \hat v_q 
\hat a_q \int_x^1 \frac{dy}{y} \left\{ \frac{2}{3} [-q(y,\mu_F^2) + 
\bar q(y,\mu_F^2)]\right.  \\ 
& &  \hspace*{-2cm}
\left[ -\frac{3}{4} P_{qq}(z) \log \frac{\mu_F^2 z}{Q^2} + (1+z^2) {\cal D}_1(z)
- \frac{3}{2} {\cal D}_0(z) 
\left. + 2 + z - \left(
\frac{9}{2} + \frac{\pi^2}{3} \right) \delta(1-z) \right] \right\} \non
\end{eqnarray}
where $z=x/y$ and the Altarelli--Parisi splitting functions $P_{qq}, P_{qg}$ 
are as given in eq.~(\ref{AP-function});   the notation ${\cal D}_i(z)
= \left[ \log^i(1-z)/(1-z)\right]_+$ with ${\small i=0,1}$ has been introduced
before. 
$\mu_R$ is the renormalization scale at which $\alpha_s$ is evaluated and the 
physical scale $Q$ is given by $Q^2 = -q_i^2$ for $x=x_i$ with $i=1,2$.
These expressions have to be inserted in the LO differential cross section
eq.~(\ref{eq:vvhlo}) and the full result expanded up to NLO. The typical
renormalization and factorization scales are fixed by the corresponding
vector--boson momentum transfer at each leg, $\mu_R^2=\mu_F^2=-q_i^2$ for  
$x=x_i$. \s

The correcting $K$--factor, again defined as $K =\sigma_{\rm NLO}/\sigma_{\rm
LO}$ with $\alpha_s$ and the PDFs consistently taken at the respective order, 
where the renormalization and factorization scales are set to $\mu_R=\mu_F
=Q$, is practically constant at the LHC in the entire Higgs mass range 
100 $\lsim M_H \lsim 1$ TeV, and increases the LO cross section by about 5
to 10\%. More details on the $K$--factor and the scale dependence at LO and 
NLO will be given later, after the discussion of the specific kinematics  of 
the  vector boson fusion process to which we turn now. 

\subsubsection{Kinematics of the process}

Because weak vector boson fusion is a three--body production process and is
mediated by $t$--channel gauge boson exchange, its kinematics is rather
complicated. However, its characteristics distributions play an extremely
important role once it comes to discriminate the signal from the many large QCD
backgrounds. In particular, forward jet--tagging
\cite{jet-tagging,pp-jettag-Dicus,jet-tag-veto} and central jet--vetoing
\cite{jet-vetoing,jet-tag-veto} are essential ingredients. We therefore
summarize the main features of the various distributions; for more details, see
the reviews of Refs.~\cite{Zepp-review,Rainwater-thesis}.\s

To study the kinematics of the $pp \to Hqq$ process, it is more convenient to 
write the differential partonic cross section, eq.~(\ref{Hqq:Edistr}), in terms
of the transverse momentum and rapidity of the Higgs boson. The latter, in 
terms of $p_H, E_H$ and $\cos \theta$, are given by
\beq
E_H = \sqrt{M_H^2+ p_T^2} \ {\rm ch}(y) \ \ , \ \ p_H \cos \theta = \sqrt{
M_H^2+ p_T^2} \ {\rm sh}(y)
\eeq
The total partonic cross section is obtained by integrating the double 
differential distribution [which is given in eq.~(\ref{Hqq:Edistr}) 
and where the above changes have been performed]
\beq
\hat{\sigma}_{\rm LO} (qq \to Hqq) = \int_{y_-}^{y_+} {\rm d}y \int_0^{p_T^{
\rm
max}} {\rm d}p_T \, (2\pi p_T) \, \frac{ {\rm d}^2 \hat{\sigma}_{\rm LO} }
{ {\rm d}y {\rm d}p_T }
\eeq
the integration bounds on the rapidity and the transverse momentum being 
\beq
y_\pm = \pm \log \frac{ \sqrt{\hat s} }{M_H} \ \ , \ \ p_T^{\rm max}= 
\left[ \left( \frac{\hat{s}+M_H^2} {2\sqrt{\hat{s}} \, {\rm ch}(y) }\right)^2
-M_H^2 \right]^{1/2}
\eeq 

Similarly to the emission of a Weizs\"acker--Williams photon from an energetic  
electron or positron beam, the intermediate vector bosons in the fusion
process tend to carry only a small fraction of the initial parton energies. At
the same time, they must have an energy of ${\cal O}(\frac{1}{2}M_H)$ to 
produce the Higgs boson. Thus, the two quarks in the final state have very large
energies, of order 1 TeV at the LHC. In contrast, they have small transverse 
momenta, $p_{T} \sim M_V$, which are set by the vector boson propagators in 
the amplitude squared eq.~(\ref{Hqq:amp2}), $1/(q_{1,2}^2 -M_V^2) \lsim 1/
(p_{T3,4}^2 +M_V^2)$, and which suppress the cross section for $p_T$ values 
larger than $M_V$. The relatively small transverse momenta and high energies of
the final state quarks correspond to rather small scattering angles 
$\theta_{3,4}$. In terms of  the pseudo--rapidity
\beq
 \eta = \frac{1}{2} \log \frac{1+ \cos \theta}{1-\cos \theta} 
\eeq
one obtains typically, $1 \lsim \eta \lsim 5$.  This is exemplified in
Fig.~3.14 where the transverse momenta and rapidity distributions of the two
scattered quarks are shown at the LHC for a Higgs boson mass $M_H=120$ GeV. One
can see that the rapidity distributions tend to be central, in particular in
the case of one of the jets. One also sees that the average transverse
momentum of one of the quarks  is substantially smaller,  a factor of two less,
than for the other quark and that small values, $p_T \sim 35$ GeV, are 
possible.\s

\begin{figure}[!h]
\begin{center}
\leavevmode
\psfig{figure=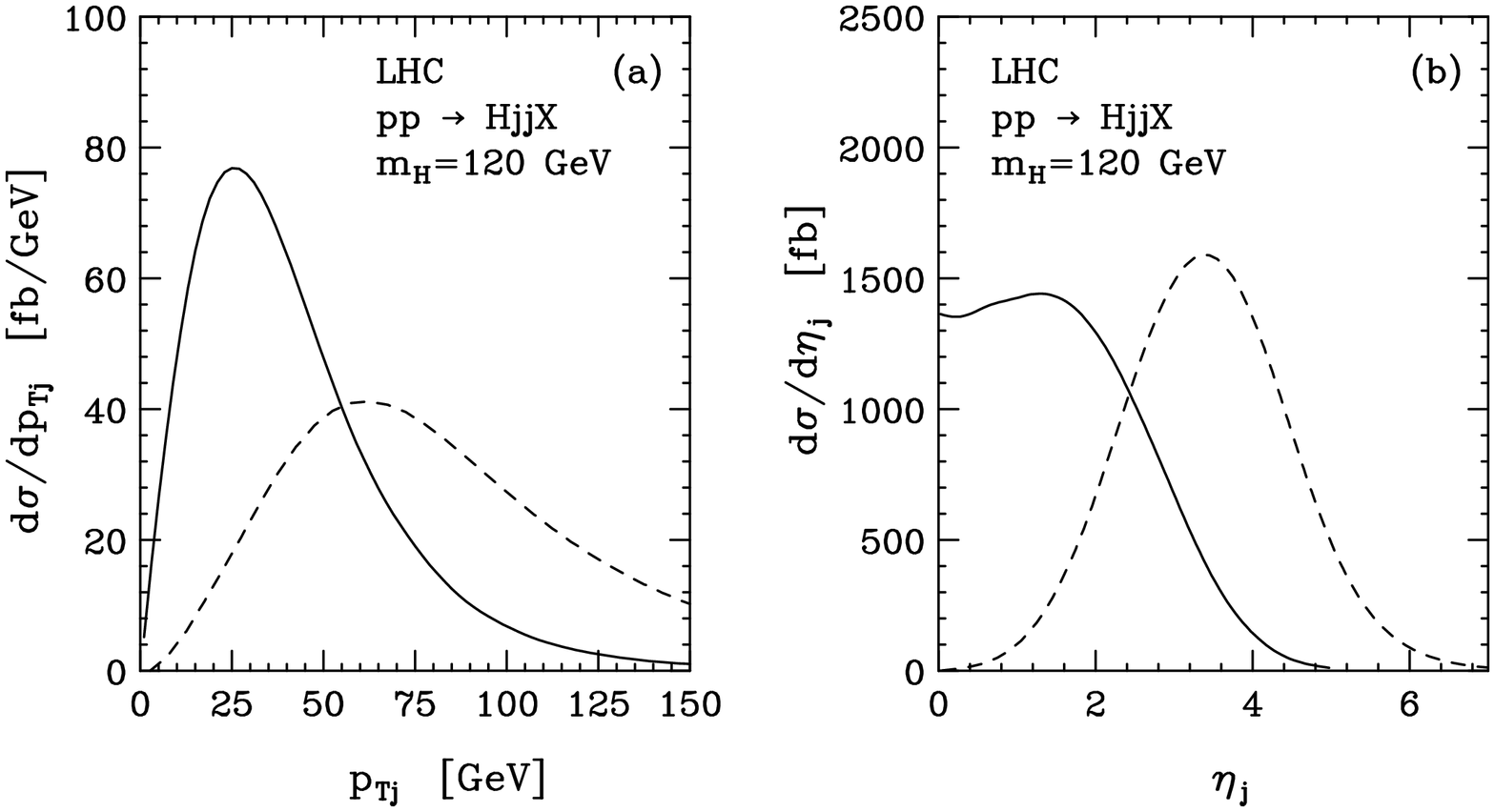,height=15.5cm,angle=90}
\vspace*{-5mm}
\end{center}
{\it Figure 3.14: The transverse momentum (left) and pseudo--rapidity (right)
distributions of the two scattered jets in the fusion process $qq \to Hqq$ at
the LHC with $M_H=120$ GeV. Shown are the $p_T$ distributions for the lowest
(solid) and highest (dashed) jets and the $|\eta|$ distribution for the most 
central (solid) and most forward (dashed) jets; from Ref.~\cite{Zepp-review}}
\vspace*{-3mm}
\end{figure}

Therefore, requiring that the two scattered jets have a large invariant 
mass, a sizable $p_T$ and  rapidity distributions which are central, 
will substantially reduce the backgrounds
\beq
{\rm Cut~1}: \ \ \ m_{q_3, q_4} \gsim 1~{\rm TeV} \ , \ p_{T_{q_3, q_4}} 
\gsim  20~{\rm GeV} \ , \ |\eta_{q_3,q_4}| \lsim 5  
\eeq

Because of the scalar nature of the Higgs boson, its decay $H \to X_1 X_2$ is
isotropic and can be treated separately from the production process. One can 
then discuss the kinematics of Higgs production in the vector boson fusion 
channel,
independently of the detection channel. Nevertheless, the Higgs decay products
should be observable, i.e. they must have a substantial $p_T$ and they must be
well separated from the jets. The decay products tend to be very central as is
exemplified in Fig.~3.15 in the case of the $H \to \gamma \gamma$ decay 
\cite{Zepp-gamma}, where
the normalized pseudo--rapidity of the most forward photon is shown for
$M_H=120$ GeV. In contrast, the photons in the irreducible QCD background $pp
\to jj\gamma \gamma$ are more forward.  Thus, a second cut will reduce the
background without affecting too much the signal
\beq
{\rm Cut~2}: \ \ p_{T_{X1,X2}} \gsim 20~{\rm GeV} \  , \  |\eta_{X_1,X_2}| 
\lsim 2.5 \ ,  \   \Delta R_{qX} \gsim 0.7 
\eeq
where $\Delta R_{qX} = \sqrt{(\eta_X - \eta_q)^2+ (\phi_X -\phi_q)^2 }$ is the 
separation between one of the jets and one of the Higgs decay products in the 
rapidity--azimuthal angle. 

\begin{figure}[hbtp]
\begin{center}
\leavevmode
\mbox{
\psfig{figure=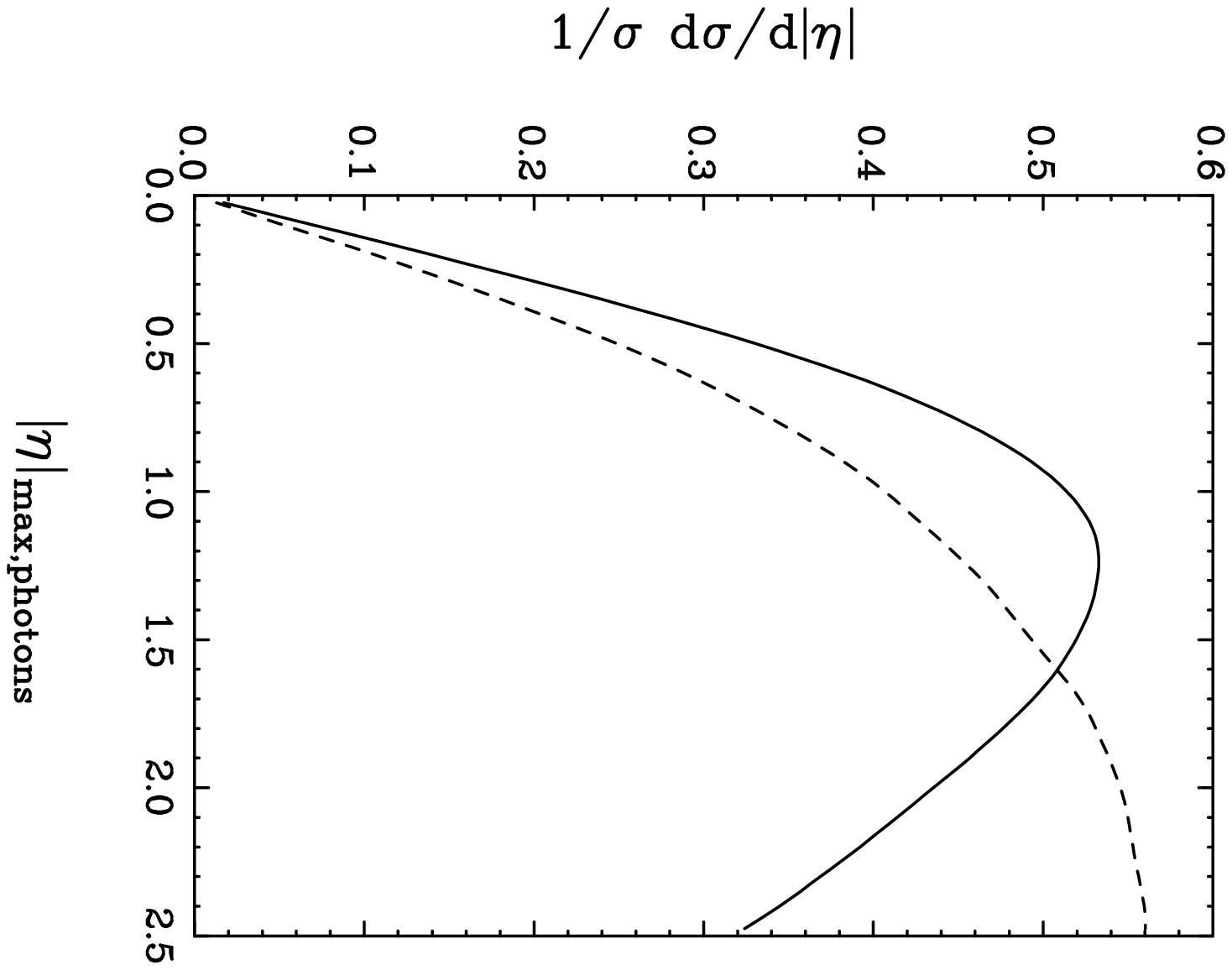,height=5.4cm,width=6.7cm,angle=90}
\psfig{figure=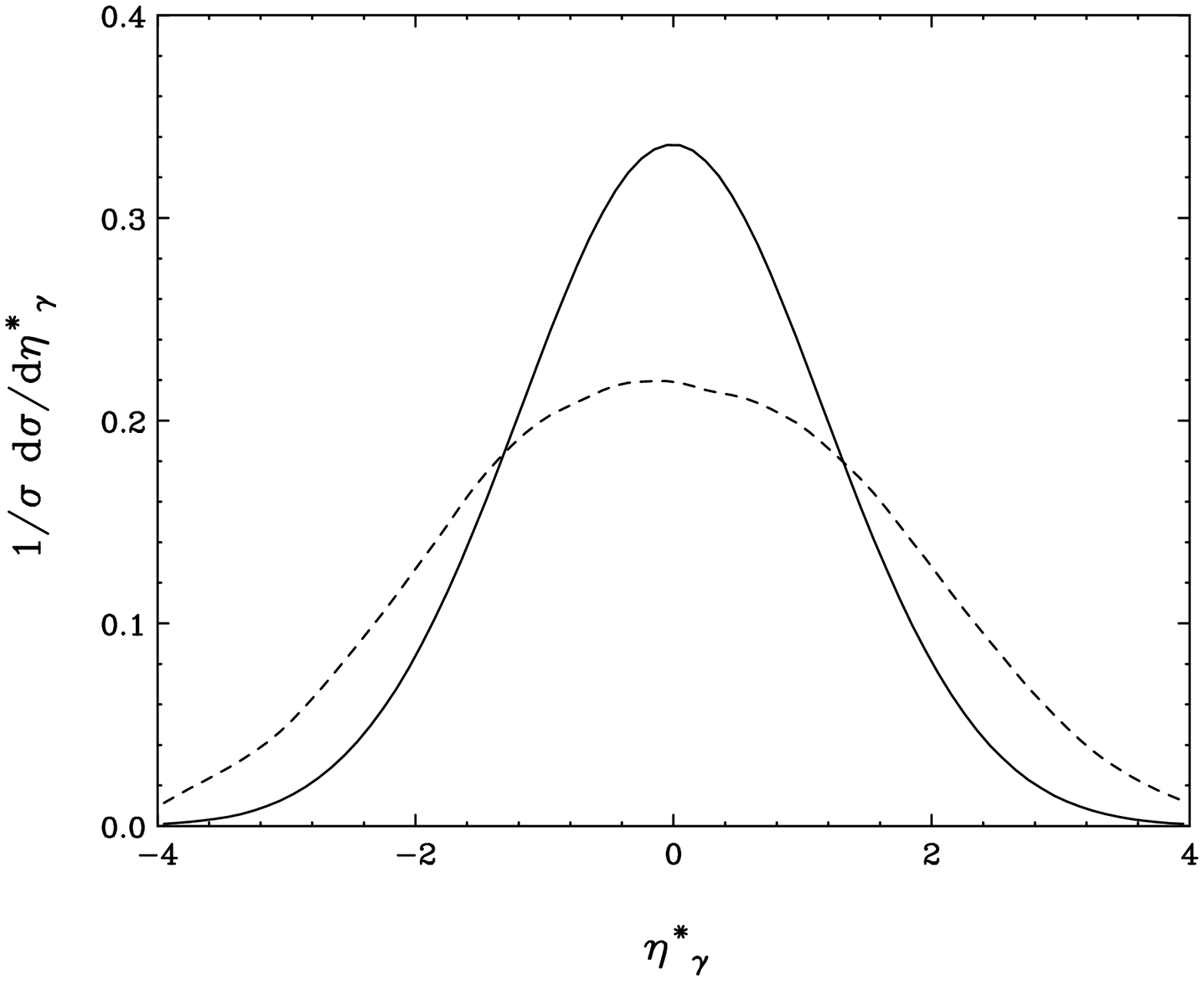,height=5.2cm,width=6.7cm,angle=90}
\psfig{figure=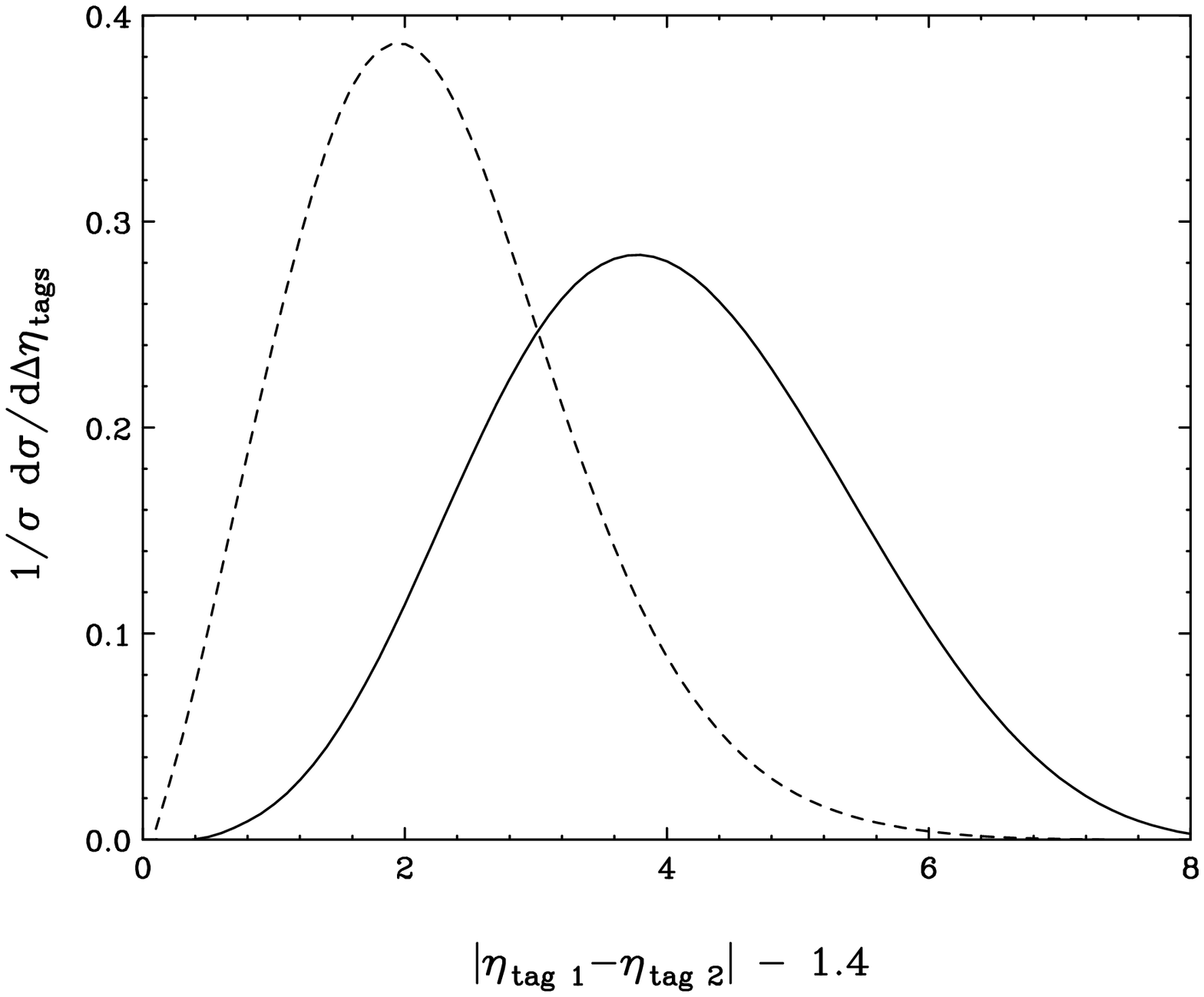,height=5.2cm,width=6.7cm,angle=90} }
\vspace*{-2mm}
\end{center}
{\it Figure 3.15: The normalized pseudo--rapidity distributions of the most 
forward photon (left), of both photons with respect to the center of the tagging
jets (center), and of the two jet rapidity gap  (right) in $jj \gamma \gamma$ 
events at the LHC; the solid lines are for the $H\to \gamma \gamma$ signal with 
$M_H=120$ GeV and the dotted lines are for the QCD background;
from Ref.~\cite{Rainwater-thesis}.}
\vspace*{-3mm}
\end{figure}

In fact, Higgs production takes place in the central region and its decay 
products will also tend to be central. This is again in contrast to the
QCD background which gives a higher rapidity for the $X$ final states.  To 
visualize more clearly this feature, one can define a shifted rapidity
$\eta_X^*$ which is the rapidity of $X$ with respect to the center of
the two jets, $\eta_X^* = \eta_X - \frac{1}{2} (\eta_{q_3} +\eta_{q_4})$. 
As shown in the central plot of Fig.~3.15, where the example of $H \to 
\gamma \gamma$ with $M_H=120$ GeV is again used, this pseudo--rapidity is
more central in the signal than in the QCD background. One can thus make 
the additional requirement that the decay products $X_{1,2}$ fall between 
the two tagged jets in rapidity, with a minimum separation in $ \eta$. 
Typically one can demand that
\beq
{\rm Cut~3}: \ \  \  \eta_{q, \, \rm min} +0.7 \lsim 
\eta_{X_{1,2}} \lsim \eta_{q, \, \rm max} -0.7 \ , \ \eta_{q_3} \cdot 
\eta_{q_4} <0  
\eeq
where it is also required that the two jets are produced in opposite 
hemispheres and, thus, the product of their pseudo--rapidities is negative. \s

In addition, the two forward tagging jets tend to be very well separated 
in pseudo--rapidity. This is shown in the right--hand side of Fig.~3.15
in the case of the $jj \gamma \gamma$ events for both the $H \to \gamma
\gamma$ signal with again $M_H=120$ GeV  and the QCD background. Requiring a 
rapidity gap between the two forward jets, the QCD backgrounds  are 
significantly suppressed  
\beq
{\rm Cut~4}: \ \  \  \Delta \eta_{qq} = | \eta_{q_3} - \eta_{q_4}| \gsim 4.4
\eeq

The cuts 1--4 form the basic ingredients to isolate the vector boson fusion
signal at the LHC from the various QCD backgrounds. For Higgs masses in the 
range 100--200 GeV, approximately 30\% of the Higgs signal events from 
the initial sample are left over after these cuts have been imposed; for 
a detailed discussion see Ref.~\cite{Rainwater-thesis}.  Additional and more 
specialized cuts can be applied for specific Higgs decays, in particular 
for the $H \to \tau^+ \tau^-$ \cite{Zepp-tau}, $H \to W^+ W^- \to \ell 
\ell \nu \nu$ \cite{Zepp-WW,Zepp-WW-old}, and even $H \to \mu^+ \mu^-$
\cite{VVH-mu} or $b \bar b$ \cite{VVH-bb} final states.\s

Finally, another important discriminant between the Higgs signal and the
backgrounds is the amount of hadronic activity in the central region. 
Indeed, and as mentioned when studying the QCD corrections, the vector boson
fusion process proceeds without color exchange between the scattered quarks, 
and gluons will be preferentially emitted at rather small angles in the 
forward and backward directions and not in the central region. This is opposite
to the QCD background which proceeds via color exchange of the 
incident partons and where the gluons are very often in the central region. 
Therefore vetoing any jet activity in the central region will substantially 
reduce the backgrounds. \s

The forward jet--tagging and the central jet vetoing techniques have been
discussed in numerous papers and have been shown to
efficiently allow to isolate a Higgs production signal in the vector
boson fusion process [there are, however, still some experimental issues
such as the central jet veto efficiencies and to a lesser extent, the forward 
jet reconstruction, which need further detailed studies]. Combined with the 
possibility of having large production rates at the LHC for a Higgs boson in 
the 100 to 200 GeV mass range, this process offers therefore a very promising 
channel not only for the production of the SM Higgs boson but also 
for the study of its properties. 

\subsubsection{Dependence on the scale and on the PDFs at NLO}

Since rather stringent cuts have to be applied to the vector boson fusion
process in order to suppress the various backgrounds, one may wonder if the NLO
corrections and their residual scale dependence are the same as in the case of
the inclusive cross section, i.e. without applying the cuts.  This question has
been addressed recently \cite{pp-Hqq-NLO2,MC-WWNLO} by implementing the full
one--loop QCD corrections to the $qq \to Hqq$ process into a parton--level
Monte--Carlo program \cite{pp-MCFM}. With cuts similar to those discussed in
the previous subsection [see the original reference for the details], the
output for the production cross section is shown in Fig.~3.16 for a Higgs boson
in the mass range between 100 and 200 GeV.\s

In the left--hand side of the figure, the cross section is displayed at LO
(dotted line) and at NLO for two methods of tagging the forward jets: one
chooses the tagging jets as being either the two highest $P_T$ jets ($P_T$
method, solid line) or the two highest energy jets ($E$ method, dashed line).
One first notices that with the cuts of Ref.~\cite{pp-Hqq-NLO2}, the acceptance
is less than $\sim 25\%$ of the initial cross section, {\it c.f.} Fig.~3.12. 
The corrections are modest and, in the chosen Higgs mass range, they are of the
order of 3\% to 5\% in the $P_T$ method and 6\% to 9\% in the $E$ method, the
largest variation being for low Higgs masses.\s 
   
To illustrate the impact of the choice of the factorization and renormalization
scales on the $qq \to Hqq$ production cross section at the LHC, we show in the
right--hand side of Fig.~3.16 the LO and NLO $K$--factors as functions of the
Higgs mass when the central value of the scales $\mu_F=\mu_R = Q_V$ is divided
or multiplied by a factor of two, $\mu_F=\mu_R=\frac{1}{2} Q_V$ and $2Q_V$
[note that the variation with the renormalization scale $\mu_R$ is small since
$\alpha_s$ enters only at NLO and the contribution of this order to the total
production cross section is tiny]. Again, the $K$--factor at leading order is
defined as $K_{\rm LO} = \sigma_{\rm LO} (\mu_F, \mu_R)/\sigma_{\rm LO}(\mu_F=
\mu_R= Q_V)$. As can be seen, the uncertainty on the total cross section that
is generated by the scale variation is relatively large at LO, the spread being
of the order of $\Delta\sigma/\sigma \simeq \pm 3\%$ for low Higgs masses and 
reaching
the level of $5\%$ at high Higgs masses. At NLO, the cross section varies only
slightly, with a spread smaller than $\sim 2\%$ for the displayed Higgs mass
range. This implies that the vector boson fusion cross section at NLO is well
under control and that the higher--order QCD corrections are presumably very
small\footnote{The electroweak corrections to this process have not been
calculated yet. However, if one  uses the IBA discussed in \S1.2.4, the bulk of 
these corrections is incorporated and the remaining piece should be rather 
small. See the discussion in the next chapter, when this process will be 
considered in $\ee$ collisions.}.

\begin{figure}[h] 
\centerline{ 
\epsfig{figure=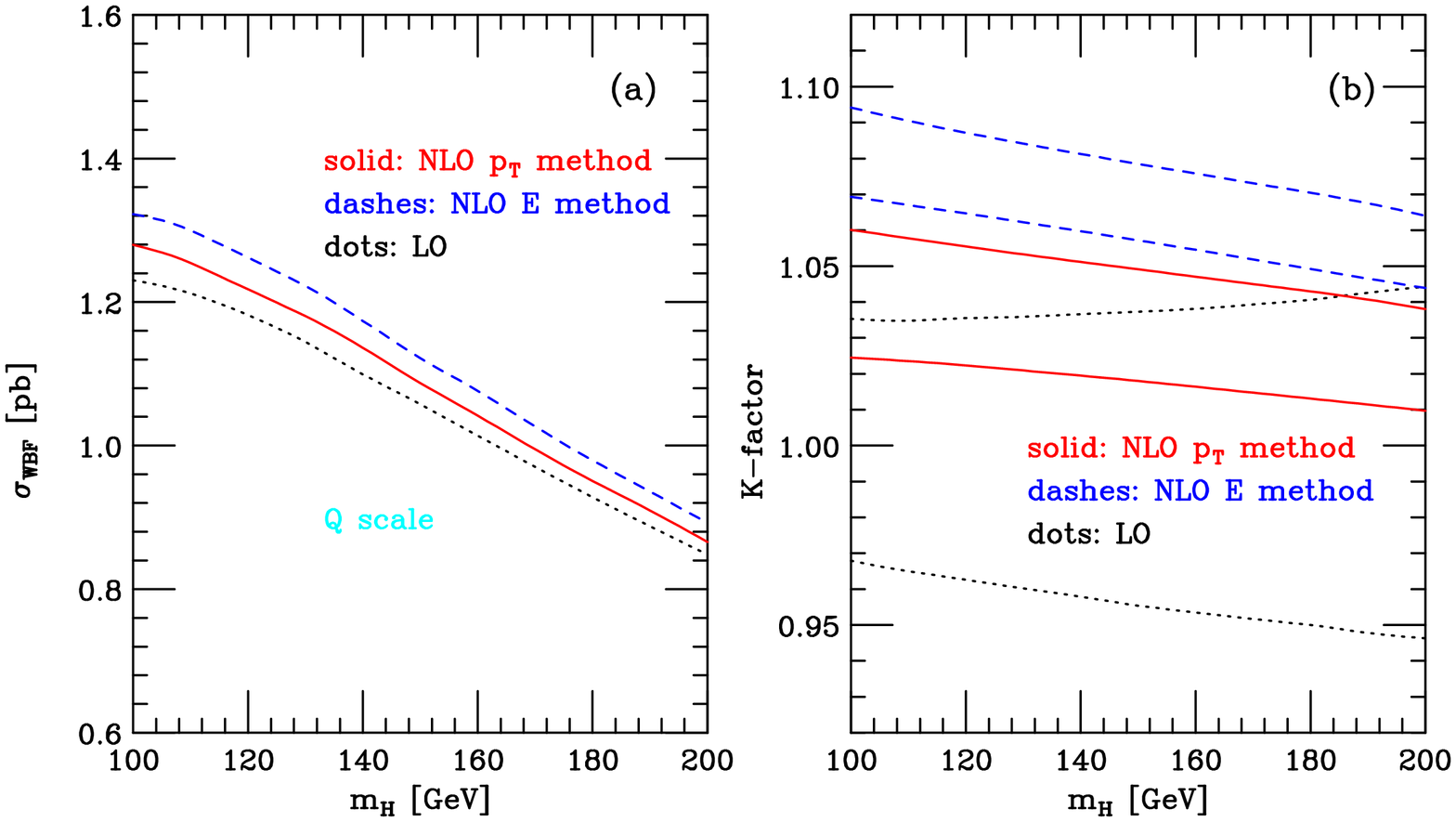,width=0.5\textwidth,angle=90,clip=} 
} 
\nn {\it Figure 3.16: Left: the $pp\to Hqq$ cross section at the LHC after cuts
as a function of $M_H$ at LO (dotted line) and NLO with the tagging jets 
defined in the $P_T$ (solid line) and $E$ (dashed line) methods. Right: 
The scale variation of the LO and NLO cross sections for Higgs production in 
the $qq\to qqH$ fusion process as a function of $M_H$ at the LHC  
\cite{pp-Hqq-NLO2}.}
\vspace*{-.2cm}
\end{figure} 

Note that the NLO QCD corrections for the $p_T$ and $\eta$ distributions in $pp
\to Hqq$ have also been calculated in this reference. In general, they are 
of the same size as the corrections to the total cross section, $\sim 10\%$, 
but they can reach larger values depending on the phase--space regions; see 
Ref.~\cite{pp-Hqq-NLO2} for details. \s

Turning to the PDF uncertainties in the prediction for the $qq \to Hqq$
cross section at NLO, we will follow again the procedure outlined in \S3.1.5. 
The  central values and the uncertainty band limits of the NLO cross sections 
are shown for the CTEQ, MRST and Alekhin parameterizations in Fig.~3.17 as a 
function of $M_H$ at LHC energies. We also show in the insert to this 
figure, the spread uncertainties in the predictions when the cross sections 
are normalized to the values obtained using the reference CTEQ6M set.\s

In the entire Higgs mass range from 100 GeV  to 1 TeV, the incoming
quarks involved in this process originate from the intermediate--$x$ regime and
the uncertainty band is almost constant, ranging between 3\% and 4\% in the 
CTEQ parameterization; as usual, the uncertainty is twice smaller in the MRST 
case. When using the Alekhin set of PDFs, the behavior is different, because 
the quark PDF behavior is different, as discussed in the case of the $q\bar{q}
\to HV$  production channel. The decrease in the central value with higher 
Higgs masses [which is absent in the $q\bar{q} \to HV$ case, since we 
stopped the $M_H$ variation at 200 GeV] is due to the fact that we reach here 
the high--$x$ regime, where the Alekhin $\bar{u}$ PDF drops steeply; see 
Fig~3.2.  Thus, as in the case of the $q\bar q \to HV$ process, the PDF 
uncertainties are below the 5\% level if the Alekhin parametrisation is 
ignored and, therefore, rather small. In view of the small QCD corrections and
scale dependence, weak boson fusion can thus also be considered as a rather 
clean Higgs production process. \s

\begin{figure}[h!]
\begin{center}
\vspace*{-2.6cm}
\hspace*{-2cm}
\psfig{figure=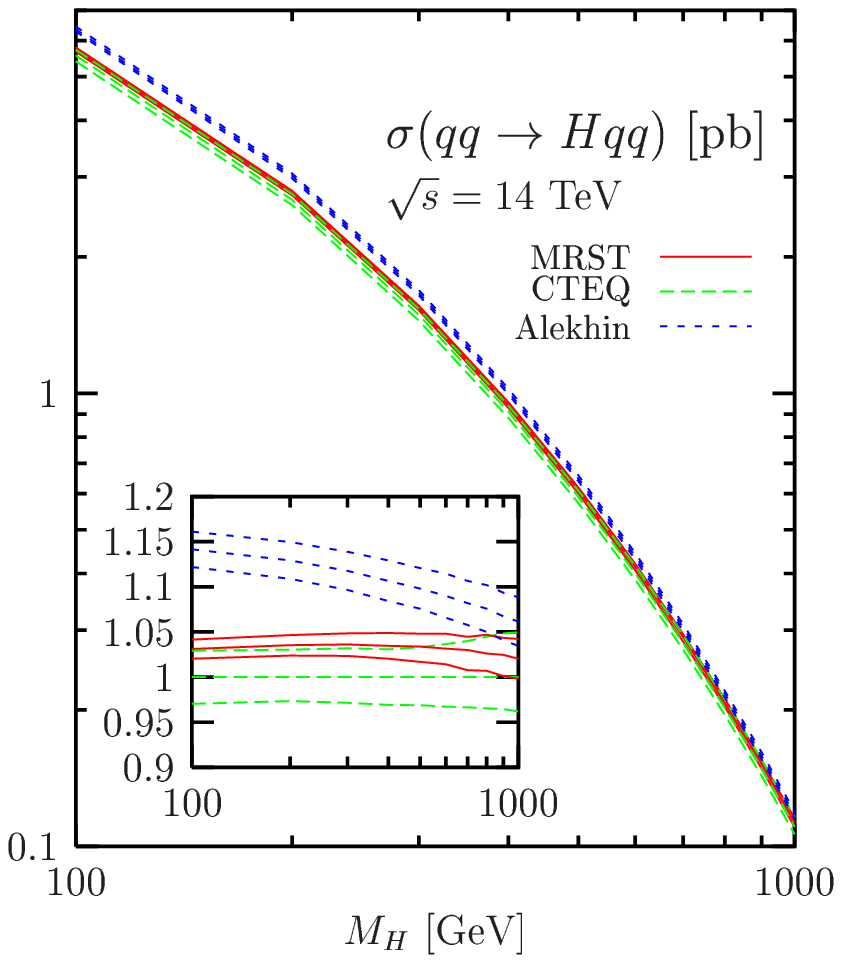,width=18cm}
\end{center}
\vspace*{-14.8cm}
{\it Figure 3.17: The CTEQ, MRST and Alekhin PDF uncertainty bands for the NLO
cross section of the vector boson fusion process $pp\to Hqq$ at the LHC. In the
insert is shown the spread uncertainty, when the cross sections are normalized
to the default CTEQ PDF set; from Ref.~\cite{Samir}.}
\vspace*{-1mm} 
\end{figure}

\subsubsection{The effective longitudinal vector boson approximation}

Before closing this section, let us reconsider the total $pp\to Hqq$ production
cross section in the light of the previous discussion. Following 
Ref.~\cite{VVH-Altarelli} and recalling that the transverse 
momenta of the scattered quarks are small, one may write the parton 
four--momenta as 
\beq
p_{3/4} = \left( x_{3/4}E + p_{T3/4}^2/(2x_{3/4} E),\vec{p}_{T3/4},
 \pm x_{3/4} E \right) 
\, , \, 
p_{1/2}= (E, \vec{0}, \pm E ) 
\eeq
with $E$ being half of the parton c.m. energy,  and neglect terms of the order 
of $p_{T3,4}^2/E^2 \ll 1$ in the amplitude squared. On then immediately  obtains
for the invariants of eq.~(\ref{Hqq:amp2})
\beq 
(p_1 \cdot p_2) (p_3 \cdot p_4) \simeq (p_1 \cdot p_4) (p_2 \cdot p_3) \simeq 
4 E^4 x_3 x_4
\eeq
leading to an amplitude squared for the process that is simply given by
\beq
|{\cal M}|^2  &=&   \sqrt{2} N_{c}^f G_\mu^3 M_V^8 
\frac{(C_+ +C_-) (x_3 x_4)^3 \hat{s}^2} 
{(p_{T3}^2+ x_3M_V^2)^2 (p_{T4}^2+ x_4 M_V^2)^2}
\eeq
The three--body phase space also simplifies to
\beq
{\rm dPS}_3 \simeq \frac{1}{8 (2\pi)^5} \, \frac{ {\rm d}x_3} {x_3} \frac{ 
{\rm d}x_4}{ x_4} {\rm d}^2 \vec{p}_{T3} {\rm d}^2 \vec{p}_{T4} \, 
\frac{2} {\hat s} \, \delta \left( (1-x_3)(1-x_4) - \frac{M_H^2}{\hat s} \right)
\eeq
The integrations on the transverse momenta can therefore be easily done,
leading to 
\beq
\int \frac{{\rm d}^2 \vec{p}_{Ti}} {(p_{Ti}^2+ x_i M_V^2)^2}
\simeq \pi \int_0^\infty \frac{{\rm d} p^2} {(p^2+ x_i M_V^2)^2}
= \frac{\pi} {x_{i}M_V^2} 
\eeq
and, with the help of the delta function,  the integrations on $x_{3,4}$ are 
straightforward. One finally obtains for the total partonic cross section
\beq
\hat{\sigma}_{\rm LO} (qq \to qq H) \simeq \frac{G_\mu^3 M_V^4 N_c}{128 
\sqrt{2} \pi^3} (C_+ + C_-) \, \left[ \left( 1+ \frac{M_H^2}{\hat s} \right) 
\log \frac{ \hat s}{M_H^2} -2 + 2\frac{M_H^2}{ \hat s} \right]
\label{Hqq-EWA}
\eeq
This is nothing else than the cross section for Higgs boson production in the
effective longitudinal vector boson approximation
\cite{Equivalence-theorem}, where one calculates the
cross section for the subprocess where the Higgs boson is produced in the
fusion of $V_L V_L$ [which according to the equivalence theorem can be replaced
by their corresponding Goldstone bosons] and then folds the result with the
$V_L$ spectra \cite{WWA-Higgs,VVH-Cahn,pp-VVH-Abas}. Since we will use this 
approximation in the course of our discussion, we briefly summarize its salient
features.\s

Just as in the Weizs\"acker--Williams approximation in the processes $\ee \to 
e^\pm X$, where the final state $X$ particle is produced at small angles 
through the exchange of a photon, and where the bulk of the production rate 
is described by the cross section $\hat{\sigma}$ for the subprocess $\gamma 
e^\pm \to X$ folded by the probability of the the initial $\ee$ state to 
radiate a photon \cite{WWA-photon}
\beq 
\sigma( \ee \to e^\pm X) = \int {\rm d}z P_{\gamma/e^\pm} (z) \hat{\sigma}
(\hat{s} =zs) \ , \  P_{\gamma/e^\pm} (z) =\frac{\alpha}{2\pi} \frac{1+ 
(1-z)^2}{z} \log \frac{s}{m_e^2}
\label{WW-photon}
\eeq
where $\sqrt{s}$ is the total c.m. energy and $m_e$ the electron mass, the
process $qq \to  qqV^* V^* \to qqH$ at very high energies can be viewed as
originating from the subprocess $VV \to H$ with the real vector bosons being
radiated from the initial quarks.  The only difference with the
Weizs\"acker--Williams approximation is that the $W/Z$ bosons are massive and
thus have a longitudinal degree of polarization.  The distribution functions
for the transverse and longitudinal polarizations in this case are given by
\beq
P_{V_{\pm}/q} (z) &= &\frac{\alpha}{4\pi} \frac{1}{z} \left[ (v_q \mp a_q)^2 
+ (v_q \pm a_q)^2 (1-z)^2 \right] \log \frac{\hat s}{M_V^2} \non \\
P_{V_L/q} (z) &= & \frac{\alpha}{\pi} \frac{1-z}{z} \, (v_q^2+a_q^2) 
\label{WW-spectra}
\eeq
One recovers the photon case in eq.~(\ref{WW-photon}) by appropriately
replacing the quark weak charge by the electron electric charge, $v_q \to 1,
a_q \to 0$. The $VV$ luminosity in the process $VV \to X$
\beq
\left. \frac{ {\rm d}{\cal L} } {{\rm d} \tau} \right|_{VV/qq} = \int_\tau^1
P_{V/q} (z) P_{V/q} (\tau/z)  \frac{ {\rm d} z} { z} 
\eeq
with $\tau=M_X^2/\hat{s}$ where $\hat{s}$ is the $qq$ c.m. energy, is then 
given by 
\beq
\left. \frac{ {\rm d}{\cal L} } {{\rm d} \tau} \right|_{V_TV_T/qq} &=& 
\frac{\alpha} {8 \pi^3} (v_q^2+a_q^2)^2 \frac{1}{\tau} \log \frac{\hat s}
{M_V^2} \bigg[ (2+\tau)^2  \log(1/\tau) -2(1-\tau) (3+\tau) \bigg] \non \\
 \left. \frac{ {\rm d}{\cal L} } {{\rm d} \tau} \right|_{V_LV_L/qq} &=& 
\frac{\alpha} {4 \pi^3} (v_q^2+a_q^2)^2 \frac{1}{\tau} \bigg[ (1+\tau) 
\log(1/\tau) -2(1-\tau)\bigg] 
\label{WW-effective}
\eeq 
In principle, at high energies, the luminosity for transverse gauge bosons is 
much larger than for longitudinal ones because of the $\log^2(M_V^2/\hat
s)$ term. However, for large masses, the Higgs boson is produced in the 
subprocess $VV \to H$ mainly through the longitudinal components which give
rates $\propto M_H^3$. The effective cross section in this case is simply
given by
\beq
\sigma_{\rm eff} = \frac{16 \pi^2}{M_H^3} \, \Gamma (H\to V_L V_L)  
 \left. \frac{ {\rm d}{\cal L} } {{\rm d} \tau} \right|_{V_LV_L/qq}
\eeq
which, when the expression of the luminosity is inserted reproduces the 
result of eq.~(\ref{Hqq-EWA}).\s

In the case of the partonic process [at the hadronic level, a difference is
generated by the parton densities], the contribution of the $WW$ fusion channel
is one order of magnitude larger than the one of the $ZZ$ channel because of
the larger charged current couplings.  However, in practice, the effective
longitudinal approximation approaches the exact result only by a factor 2 to 5,
depending on the considered c.m. energy and the Higgs mass. For light Higgs
bosons, it can be improved by including the transverse vector boson components,
see Ref.~\cite{WWA-trans}. This approximation should therefore be used only as 
an indication of the order of magnitude of the cross sections.  

\subsection{The gluon--gluon fusion mechanism}
\subsubsection{The production cross section at LO}

Higgs production in the gluon--gluon fusion mechanism is mediated by triangular
loops of heavy quarks. In the SM, only the top quark and, to a lesser extent,
the bottom quark will contribute to the amplitude. The decreasing $Hgg$ form
factor with rising loop mass is counterbalanced by the linear growth of the
Higgs coupling with the quark mass. In this section we discuss the analytical
features of the process. The relevant phenomenological aspects at the LHC 
\cite{pp-EHLQ,pp-Galison,pp-Wudka,pp-HWW-Theory,pp-HZZ-llnnTheory,pp-ggH-tau-old,pp-ggH-mu,pp-ggH-mu-others} and the Tevatron \cite{pp-WW-TeV,pp-WW-TeV-E,pp-tau-TeV} will 
be presented in \S3.7.\s

To lowest order, the partonic cross section can be expressed by the gluonic 
width of the Higgs boson discussed in \S2.3.3,
\begin{eqnarray}
\hat\sigma_{\rm LO} (gg \to H) & = & {\sigma_0^H}\, {M_H^2} \, \delta
(\hat s -M_H^2)  \ = \  \frac{\pi^2}{8 M_H}\,  \Gamma_{\rm LO} (H \to gg) 
\, \delta (\hat s -M_H^2)
\end{eqnarray}
where $\hat{s}$ is the $gg$ invariant energy squared. Substituting in this LO 
approximation the Breit--Wigner form of the Higgs boson width, in place of the 
zero--width $\delta$ distribution 
\begin{eqnarray}
\delta(\hat s - M_H^2) \to \frac{1}{\pi}~\frac{\hat s \Gamma_H/M_H}
{(\hat s - M_H^2)^2 + (\hat s \Gamma_H/M_H)^2}
\end{eqnarray}
recalling the lowest--order two--gluon decay width of the Higgs boson, one 
finds for the cross section \cite{pp-ggH-LO}
\begin{equation}
\sigma_0^H = \frac{G_{\mu}\alpha_{s}^{2}(\mu_R^2)}{288 \sqrt{2}\pi} \
\left| \, \frac{3}{4} \sum_{q} A_{1/2}^H (\tau_{Q}) \, \right|^{2}
\end{equation}
The form factor $A_{1/2}^H (\tau_Q)$ with $\tau_Q = M_H^2/4m_Q^2$ is given in 
eq.~(\ref{eq:Af+Aw}) and is normalized such that for $m_Q \gg M_H$, it reaches 
$\frac{4}{3}$  while it approaches zero in the chiral limit $m_Q \ra 0$.\s

The proton--proton cross section at LO in the narrow--width 
approximation reads
\begin{equation}
\sigma_{\rm LO}(pp\to H) = \sigma_0^H \tau_H \frac{d{\cal L}^{gg}}{d\tau_H}
\ \ {\rm with} \ 
\frac{d{\cal L}^{gg}}{d\tau} = \int_\tau^1 \frac{dx}{x}~g(x,\mu_F^2)
g(\tau /x,\mu_F^2)
\end{equation}
where the Drell--Yan variable is defined as usual by $\tau_H = M^2_H/s$ with 
$s$ being the invariant collider energy squared. The expression of the 
luminosity $\tau_H d {\cal L}^{gg}/d \tau_H$ is only mildly divergent for
$\tau_H \rightarrow 0$.\s 

The total hadronic cross sections at LO are shown in Fig.~3.18 as a function of
the  Higgs boson mass for the LHC and the Tevatron energies. We have
chosen  $m_t=178$ GeV, $m_b=4.88$ GeV and $\alpha_s (M_Z)=0.13$ as inputs and
used the CTEQ parametrization for the parton densities. For the Tevatron, the
cross section is monotonically decreasing with the Higgs boson mass, starting
slightly below  1 pb for $M_H \sim 100$ GeV and reaching $\sigma \sim 0.01$ 
pb for $M_H \sim 300$ GeV. At the LHC, the cross section is two orders of
magnitude larger, being at the level of $\sim 30$ pb for $M_H \sim 100$ GeV 
and is still sizable, $\sigma \sim 1$ pb, for $M_H \sim 700$ GeV. There is a 
kink at $M_H \sim 350$ GeV, i.e. near the $t\bar{t}$ threshold where the $Hgg$ 
amplitude develops an imaginary part. \s

\begin{figure}[h!]
\begin{center}
\vspace*{-.5cm}
\hspace*{-2cm}
\psfig{figure=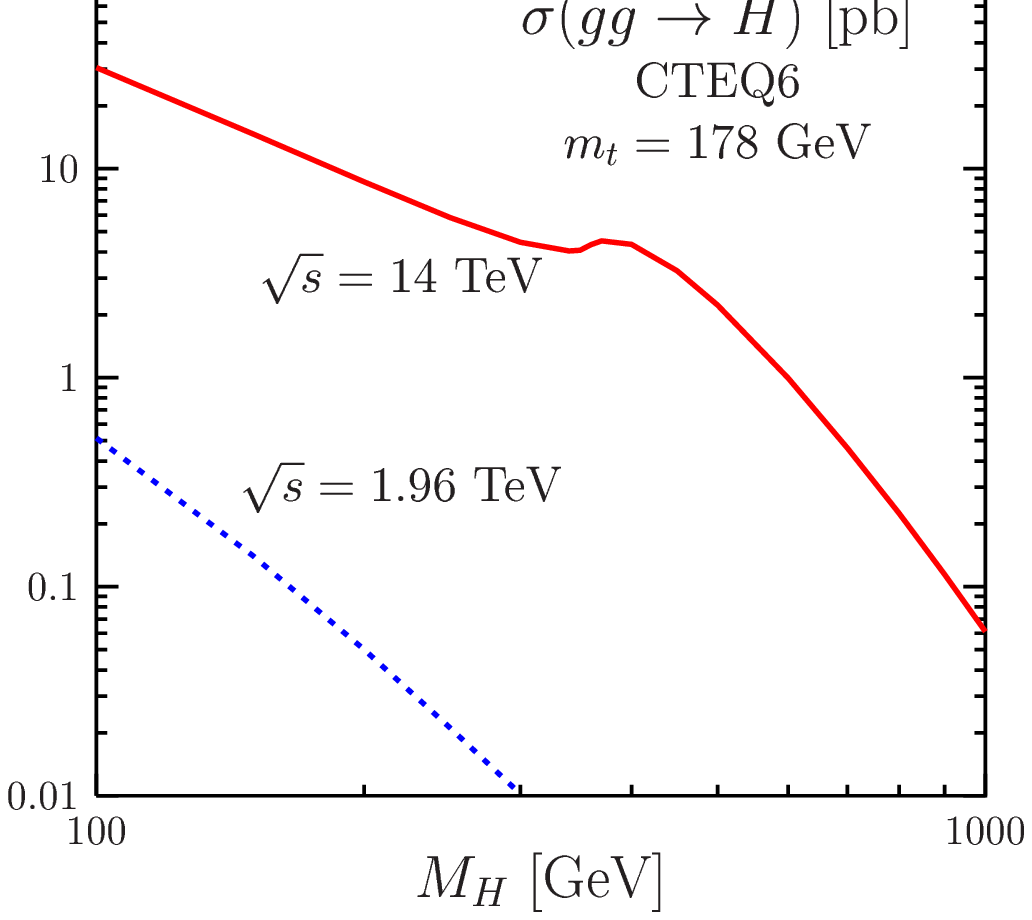,width=18.cm}
\end{center}
\vspace*{-16.3cm}
{\it Figure 3.18: The hadronic production cross section for the $gg$ fusion
process at LO as a function of $M_H$ at the LHC and the Tevatron. The inputs
are  $m_t=178$ GeV, $m_b=4.88$ GeV, the CTEQ set of PDFs has been used
and the scales are fixed to $\mu_R=\mu_F=M_H$.} 
\vspace*{-.2cm}
\end{figure}

As discussed in \S2.3.3, the cross section in the case where the internal quark
is assumed to have an infinite mass, $m_q \to \infty$, i.e. when the form
factor $\frac{3}{4}A_{1/2}^H$ is equal to unity, is a rather good approximation
for Higgs masses below the $t\bar{t}$ threshold, and it reproduces the
exact result at the level of 10\%. For low Higgs masses, the difference
is in fact due to the contribution of the bottom quark loop: although the
$b$--quark mass is small, the form factor $A_{1/2}^H(\tau_b)$   exhibits a
dependence on $m_b^2/M_H^2 \times \log^2(m_b^2/M_H^2)$ which is not that small.
Together with the $\pi^2$ terms and the imaginary part, the
$b$--quark loop  generates a non--negligible contribution which interferes
destructively with the contribution of the top--quark loop. Above the
$t\bar{t}$ threshold, $M_H \gsim 350$ GeV, the approximation of an  infinite
loop quark mass fails since it cannot reproduce the imaginary part of the form
factor.

\subsubsection{The cross section at NLO}

To incorporate the QCD corrections to $\sigma (pp \ra H + X)$, one has to
consider the processes
\begin{equation}
gg \ra H (g)  \hspace{5mm} \mbox{and} \hspace{5mm}
gq \ra H q,   \hspace{5mm} q \overline{q} \ra H g
\end{equation}
Characteristic diagrams of the QCD radiative corrections are shown in 
Fig.~3.19. They involve the virtual corrections to the $gg \to H$ subprocess,
which modify the LO fusion cross section by a coefficient linear in $\alpha_s$,
and the radiation of gluons in the final state. In addition, Higgs bosons can
be produced in gluon--quark collisions and quark--antiquark annihilation which
contribute to the cross section  at the same order of $\alpha_s$. 

\begin{figure}[!h]
\begin{center}
\vspace*{-2mm}
\hspace*{5mm}
\setlength{\unitlength}{1pt}
\SetWidth{1.}
\begin{picture}(450,100)(-10,0)
\Gluon(0,20)(30,20){4}{4}
\Gluon(0,80)(30,80){4}{4}
\Gluon(30,20)(60,20){4}{4}
\Gluon(30,80)(60,80){4}{4}
\Gluon(30,20)(30,80){4}{6}
\ArrowLine(60,20)(60,80)
\ArrowLine(60,80)(90,50)
\ArrowLine(90,50)(60,20)
\DashLine(90,50)(120,50){5}
\Text(90,50)[]{\blue{\Large\bf $\bullet$}}
\put(100,55){\bH}
\put(65,46){$Q$}
\put(-10,18){$g$}
\put(-10,78){$g$}
\put(15,48){$g$}
\Gluon(160,80)(200,80){4}{4}
\Gluon(160,20)(200,20){4}{4}
\Gluon(215,30)(215,70){4}{4}
\ArrowLine(200,20)(200,80)
\Line(200,80)(240,50)
\Line(240,50)(200,20)
\DashLine(240,50)(270,50){5}
\Text(240,50)[]{\blue{\Large\bf $\bullet$}}
\Gluon(310,20)(360,20){4}{4}
\Gluon(310,80)(360,80){4}{4}
\Gluon(330,80)(360,50){4}{5}
\Line(360,20)(360,80)
\ArrowLine(360,80)(390,50)
\ArrowLine(390,50)(360,20)
\DashLine(390,50)(420,50){5}
\Text(390,50)[]{\blue{\Large\bf $\bullet$}}
\end{picture} 
\vspace*{-3mm}
\hspace*{5mm}
\begin{picture}(450,100)(-10,0)
\Gluon(0,20)(40,20){4}{4}
\Gluon(0,80)(40,80){4}{4}
\ArrowLine(40,20)(80,20)
\ArrowLine(40,80)(80,80)
\ArrowLine(40,80)(40,20)
\ArrowLine(80,80)(80,20)
\Gluon(80,80)(120,80){4}{4}
\DashLine(80,20)(120,20){5}
\Text(80,20)[]{\blue{\Large\bf $\bullet$}}
\put(100,25){\bH}
\put(58,46){$Q$}
\put(-10,18){$g$}
\put(-10,78){$g$}
\put(100,65){$g$}
\Gluon(180,80)(210,70){4}{4}
\ArrowLine(160,80)(180,80)
\Gluon(160,20)(210,20){4}{4}
\ArrowLine(180,80)(210,99)
\ArrowLine(210,20)(210,70)
\ArrowLine(210,70)(240,50)
\ArrowLine(240,50)(210,20)
\DashLine(240,50)(270,50){5}
\Text(240,50)[]{\blue{\Large\bf $\bullet$}}
\put(150,80){$q$}
\put(150,20){$g$}
\ArrowLine(300,70)(330,50)
\ArrowLine(300,30)(330,50)
\Gluon(330,50)(360,50){4}{4}
\ArrowLine(390,20)(390,80)
\ArrowLine(360,50)(390,80)
\ArrowLine(360,50)(390,20)
\DashLine(390,20)(420,20){5}
\Text(390,20)[]{\blue{\Large\bf $\bullet$}}
\Gluon(390,80)(420,80){4}{4}
\put(290,70){$q$}
\put(290,30){$\bar{q}$}
\end{picture} 
\vspace*{-1mm}
\nn {\it Figure 3.19: Typical diagrams for the virtual and real QCD corrections
to  $gg\to H$.}
\vspace*{-3mm}
\end{center}
\end{figure}

The cross sections for the subprocesses $ij \rightarrow H + X$,
$i,j=g,q,\overline{q}$, can be written as
\begin{equation}
\hat\sigma_{ij} = \sigma_0 \left\{
\delta_{ig}\delta_{jg}\left[ 1+C^H (\tau_Q)\frac{\alpha_s}{\pi} \right]
\delta(1-\hat{\tau}) + D_{ij}^H (\hat{\tau},\tau_Q) \frac{\alpha_s}{\pi}
\Theta (1- \hat{\tau}) \right\}
\label{eq:sigij}
\end{equation}
where the new scaling variable $\hat{\tau}$, supplementing $\tau_H=M_H^2/s$ 
and $\tau_Q=M_H^2/4m_Q^2$ introduced earlier, is defined at the parton level as 
$\hat{\tau}=M^2_H/\hat{s}$; $\Theta$ is the step function.  \s

The coefficients $C^H(\tau_Q)$ and $D_{ij}^H (\hat{\tau},\tau_Q)$ have been
determined in Refs.~\cite{SDGZ,ggH-GSZ} for arbitrary Higgs boson and quark
masses and the lengthy analytical expressions have been given there [see also
\S2.3.3 for some details on the calculation and on the renormalization scheme].
If all the corrections eq.~(\ref{eq:sigij}) are added up, ultraviolet and
infrared divergences cancel.  However collinear singularities are left over and
are absorbed into the renormalization of the parton densities
\cite{DYNLO,pp-APabs} where the $\overline{\rm MS}$ factorization scheme can be
adopted. \s

The final result  for the hadronic cross section at NLO can be cast into the 
form
\begin{equation}
\sigma(pp\rightarrow H +X)=\sigma_0^H
         \left[
            1+C^H \frac{\alpha_s}{\pi}
         \right] \tau_H
         \frac{d{\cal L}^{gg}}{d\tau_H}
         +\triangle \sigma_{gg}^H
         +\triangle \sigma_{gq}^H
         +\triangle \sigma_{q \overline{q}}^H
\end{equation}
The coefficient $C^H$ denotes the  contributions from the virtual two--loop
quark corrections regularized by the infrared singular part of the cross
section for real gluon emission. It splits into the infrared term $\pi^2$, a 
term depending on the renormalization scale $\mu_R$ of the coupling constant, 
and a piece $c^H$ which depends on the mass ratio $\tau_Q$.
\begin{eqnarray}
C^H = \pi^2 + c^H + \frac{33-2 N_f}{6} \log \frac{\mu_R^2}{M^2_H } 
\eeq
\vspace*{-2mm}
with
\vspace*{-2mm}
\beq
c^H =  {\rm Re}\, \sum_Q A_{1/2}^H (\tau_Q) \, c^H_Q (\tau_Q) / 
\sum_Q A_{1/2}^H (\tau_Q)
\end{eqnarray}

The (non--singular) contributions from gluon radiation in $gg$ scattering, from
$gq$ scattering and $q \overline{q}$ annihilation, depend on the
renormalization scale $\mu_R$ and the factorization scale $\mu_F$ of the 
parton densities
\begin{eqnarray}
\triangle \sigma_{gg}^H & = &
       \int_{\tau_H}^1 d\tau \frac{d{\cal L}^{gg}}{d\tau}
       \frac{\alpha_s (\mu_R) }{\pi} \sigma_0^H
       \left\{-z P_{gg}(z)\log\frac{\mu_F^2}{\tau s}
       +d_{gg}^H (z,\tau_Q) \right.  \nonumber \\
& & \hspace{4cm} \left.
                  +12 \left[\left(\frac{\log(1-z)}{1-z}\right)_{+}
                  -z\left[2-z(1-z)\right]\log(1-z) \right]
       \right\} \nonumber \\
   \triangle\sigma_{gq}^H & = &
       \int_{\tau_H}^1 d\tau \sum_{q,\overline{q}}
       \frac{d{\cal L}^{gq}}{d\tau}
       \frac{\alpha_s (\mu_R) }{\pi} \sigma_0^H
       \left\{ \left[ -\frac{1}{2}\log\frac{\mu_F^2}{\tau s}+\log(1-z)
       \right] z P_{gq}(z) +d_{gq}^H (z,\tau_Q) \right\} \nonumber\\
   \triangle\sigma_{q\overline{q}}^H & = &
       \int_{\tau_H}^1 d\tau \sum_q
           \frac{d{\cal L}^{q\overline{q}}}{d\tau}
       \frac{\alpha_s (\mu_R) }{\pi} \sigma_0^H
       d_{q\overline{q}}^H (z,\tau_Q)
\label{ggNLOreal}
\end{eqnarray}
with $z=\tau_H/\tau$ and the standard Altarelli--Parisi splitting functions
given by
\beq
P_{gg}(z) & = & 6 \left[ \left( \frac{1}{1-z} \right)_+ + \frac{1}{z} -2 +
z (1-z) \right] + \frac{33-2N_f}{6} \, \delta(1-z) \non \\
P_{gq}(z) & = & \frac{4}{3} \frac{1+ (1-z)^2}{z} 
\eeq
where $F_+$ denotes the usual $+$ distribution such that  $F(\hat{\tau})_+ = 
F(\hat{\tau}) - \delta (1 - \hat{\tau}) \int_0^1 {\rm d}
\hat{\tau}'F(\hat{\tau}')$.\s

The coefficients $d_{gg}^H, d_{gq}^H$ and $d_{q \bar{q}}^H$, as well as $c^H$, 
have been evaluated for arbitrary quark masses \cite{ggH-GSZ,AggQCD,SDGZ}.
In the limit where the Higgs mass is very large compared with the quark mass,
$\tau_Q = M_H^2/4m_Q \gg 1$, as is the case of the bottom quark contribution, 
a compact analytic result can be derived, which is valid to leading and  
subleading logarithmic accuracy  
\begin{eqnarray}
c^H(\tau_Q) & \to & \frac{5}{36} \left[ \log^2 (4\tau_Q)-\pi^2\right] 
-\frac{4}{3} \log (4\tau_Q ) \non \\ \non \\
d^H_{gg}(\hat{\tau},\tau_Q) & \to & -\frac{2}{5} \log(4\tau_Q)
\bigg[ 7-7\hat{\tau} +5\hat{\tau}^2 \bigg]
- 6 \log (1-\hat{\tau}) \bigg[ 1-\hat{\tau} +
\hat{\tau}^2  \bigg] \non \\
& & +2\frac{\log \hat{\tau}}{1-\hat{\tau}}
\bigg[ 3-6\hat{\tau} -2\hat{\tau}^2
+5\hat{\tau}^3 - 6\hat{\tau}^4 \bigg] \non \\ \non \\
d^H_{gq}(\hat{\tau},\tau_Q) & \to & \frac{2}{3} \left[ \hat{\tau}^2 -
\left( 1+(1-\hat{\tau})^2 \right) \left( \frac{7}{15} \log (4\tau_Q) +
\log\left( \frac{1-\hat{\tau}}{\hat{\tau}} \right)
\right) \right] \non \\ 
d^H_{q\bar q}(\hat{\tau},\tau_Q) & \to & 0
\label{dij:light}
\end{eqnarray}
In the limit of large quark masses, $\tau_Q = M_H^2/4m_Q^2 \ll 1$, as is the 
case for the top quark when the Higgs mass is small, one also obtains very 
simple expressions for the coefficients 
\begin{eqnarray}
c^H(\tau_Q)  \rightarrow \frac{11}{2} \, , \
d_{gg}^H  \rightarrow  -\frac{11}{2}(1-z)^3 \, , \
d_{gq}^H \rightarrow  -1+2z-\frac{1}{3}z^2 \, , \ 
d_{q\overline{q}}^H \rightarrow  \frac{32}{27}(1-z)^3
\label{dij:heavy}
\end{eqnarray}
In this heavy quark case, the corrections of ${\cal O} (M_H^2/m_Q^2)$ in a
systematic Taylor expansion have been shown to be very small \cite{HggExp}. In 
fact, the leading term provides an excellent approximation up to the quark 
threshold $M_H \sim 2 m_Q$.\s 

The results for the $K$--factors, defined as the ratios $K_{\rm tot}=
\sigma_{\rm NLO}/\sigma_{\rm LO}$, with the cross section $\sigma_{\rm NLO}$ 
normalized to the LO cross section $\sigma_{\rm LO}$, evaluated consistently  
for parton densities and an $\alpha_s$ value at LO, are displayed in Fig.~3.20 
as a function of $M_H$ for the LHC (left) and the Tevatron (right). Again the 
CTEQ6 parametrization for the structure functions defined in the $\overline{\rm
MS}$ scheme is used and the top and bottom quark pole masses are fixed to 
$m_t= 178$ GeV and $m_b=4.88$ GeV. Both the renormalization and the 
factorization scales have been set to the Higgs mass $\mu_R=\mu_F=M_H$.\s

The $K$--factors have been decomposed into their various components: $K_{\rm
virt}$ accounts for the virtual corrections after regularization
[corresponding to the coefficient $C^H$], while $K_{ij}$ with $i,j=g,q, \bar
q$ stand for the real corrections in the three channels given in 
eq.~(\ref{ggNLOreal}). One sees that  $K_{\rm virt}$ and $K_{gg}$  are rather
large, being both of the order of 50\%, while $K_{q\bar{q}}$ and $K_{gq}$
are tiny, the latter being negative. The total $K$--factor is large, increasing 
the total production cross section by about 60\% and 90\% for the low and high 
range of the Higgs mass at the LHC and by a factor 2.2 to 2.8 for 
$M_H=100$--300 GeV at the Tevatron.  \s

\begin{figure}[h!]
\begin{center}
\vspace*{-1.2cm}
\hspace*{-2cm}
\psfig{figure=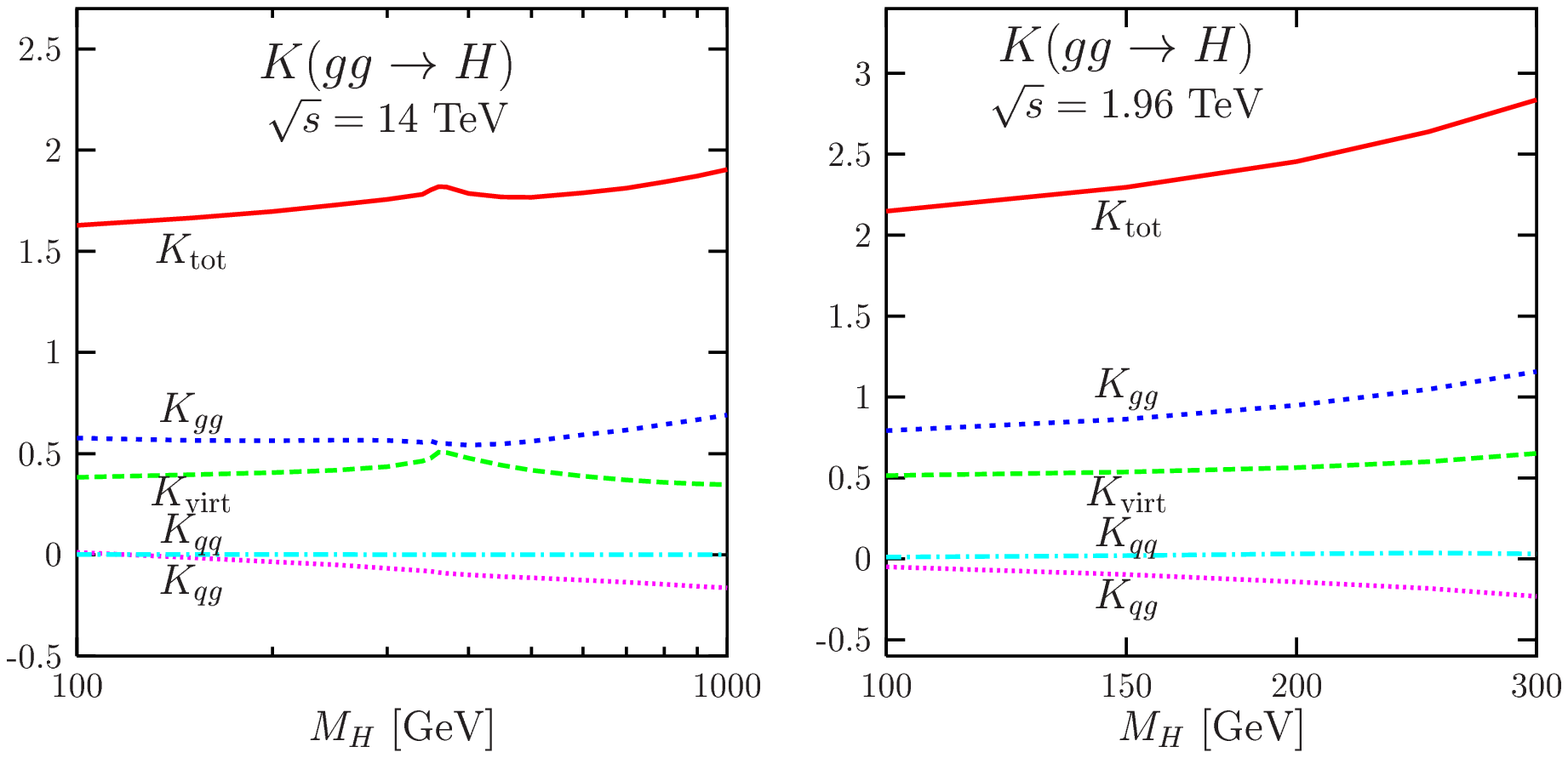,width=18cm}
\end{center}
\vspace*{-16.5cm}
{\it Figure 3.20: The total $K$ factor and its various components, $K_{\rm
virt}$, $K_{gg}$  and $K_{q\bar{q}}$,  for Higgs production in the $gg$
fusion process as a function of $M_H$ at the LHC (left) and the Tevatron
(right). The CTEQ6 parton densities have been adopted and the renormalization
and factorization scales are fixed to $\mu_R\!=\!\mu_F\!=\!M_H$; $m_t\!=\!178$ 
GeV and $m_b\!=\!4.88$ GeV.}
\vspace*{-3mm} 
\end{figure}

Apart for the small kink in the $M_H \sim 2m_t$ threshold region, $K_{\rm 
tot}$ is only mildly depending on the Higgs mass. In fact, if one compares the
exact numerical results for the cross section at NLO with the approximation
of a very heavy top quark, it turns out that multiplying the LO cross
section, which includes the full $m_t$ and $m_b$ dependence, with the $
K$--factor taken in the asymptotic limit $m_t \to \infty$ and where the 
$b$--quark contribution has been neglected, provides a good approximation
\beq
\sigma_{\rm NLO} \simeq K_{\rm tot}|_{ m_t \to \infty} \times \sigma_{\rm LO}
(\tau_t, \tau_b)
\eeq
The difference between this approximation and the exact result is less 
than 10\% even for Higgs boson masses beyond the $M_H =2m_t$ threshold and
up to $ M_H \sim 700$ GeV \cite{ggH-Laenen}.\s

Finally, note that the two--loop electroweak corrections to the $gg \to H$
production cross section are the same as the ones discussed previously in
\S2.4.3 for the decay $H \to gg$. While the top quark correction is rather
small, being less than one percent \cite{RCdjo}, the light fermion electroweak
contributions \cite{RCita,PepeHgg} are much larger in the $M_H \lsim 2M_W$ range
where they reach the level of 5--9\%; for $M_H \gsim 2M_W$ these corrections
become again very small. \s

\vspace*{-6mm} 
\subsubsection*{\underline{Dependence on the PDFs}}

The  central values and the uncertainty band limits of the NLO cross sections 
are shown for the CTEQ, MRST and Alekhin parameterizations in Fig.~3.21 for
the $gg \to H$ process. As usual, in the inserts to these figures, we show
the spread uncertainties in the predictions for the cross sections, when
normalized to the prediction of the reference CTEQ6M set.\s

At the LHC, the uncertainty band for the CTEQ set of PDFs  decreases from the
level of about 5\% at $M_{H} \sim 100$ GeV, down to the 3\% level at $M _H
\sim$ 300 GeV.  This is because Higgs bosons with relatively small masses  are
mainly  produced by  asymmetric  low--$x$--high--$x$ gluons with a low effective
c.m. energy. To produce heavier Higgs bosons, a symmetric process in which the
participation of intermediate--$x$ gluons with high density is needed,
resulting in a smaller  uncertainty band. At higher masses, $M_H \gsim 300$
GeV, the participation of  high--$x$ gluons becomes more important, and the
uncertainty band increases to reach the 10\% level at Higgs masses of about 1
TeV. At the Tevatron, because of  the smaller c.m.  energy, the high--$x$ gluon
regime is already reached for low Higgs masses and the uncertainties increase
from 5\% to 15\% for $M_H$ varying between 100 GeV and 200 GeV. As discussed
previously and shown in Fig.~3.2, the MRST  gluon PDF is smaller than the CTEQ 
one for low $x$ and larger for relatively high $x$ ($\sim 0.1$): this explains 
the increasing cross section obtained with MRST compared to the one obtained 
with CTEQ, for increasing Higgs masses  at the LHC.  At the  Tevatron the gluons
are already in the high--$x$ regime. \s

\begin{figure}[hbtp]
\begin{center}
\vspace*{-2.2cm}
\hspace*{-1cm}
\psfig{figure=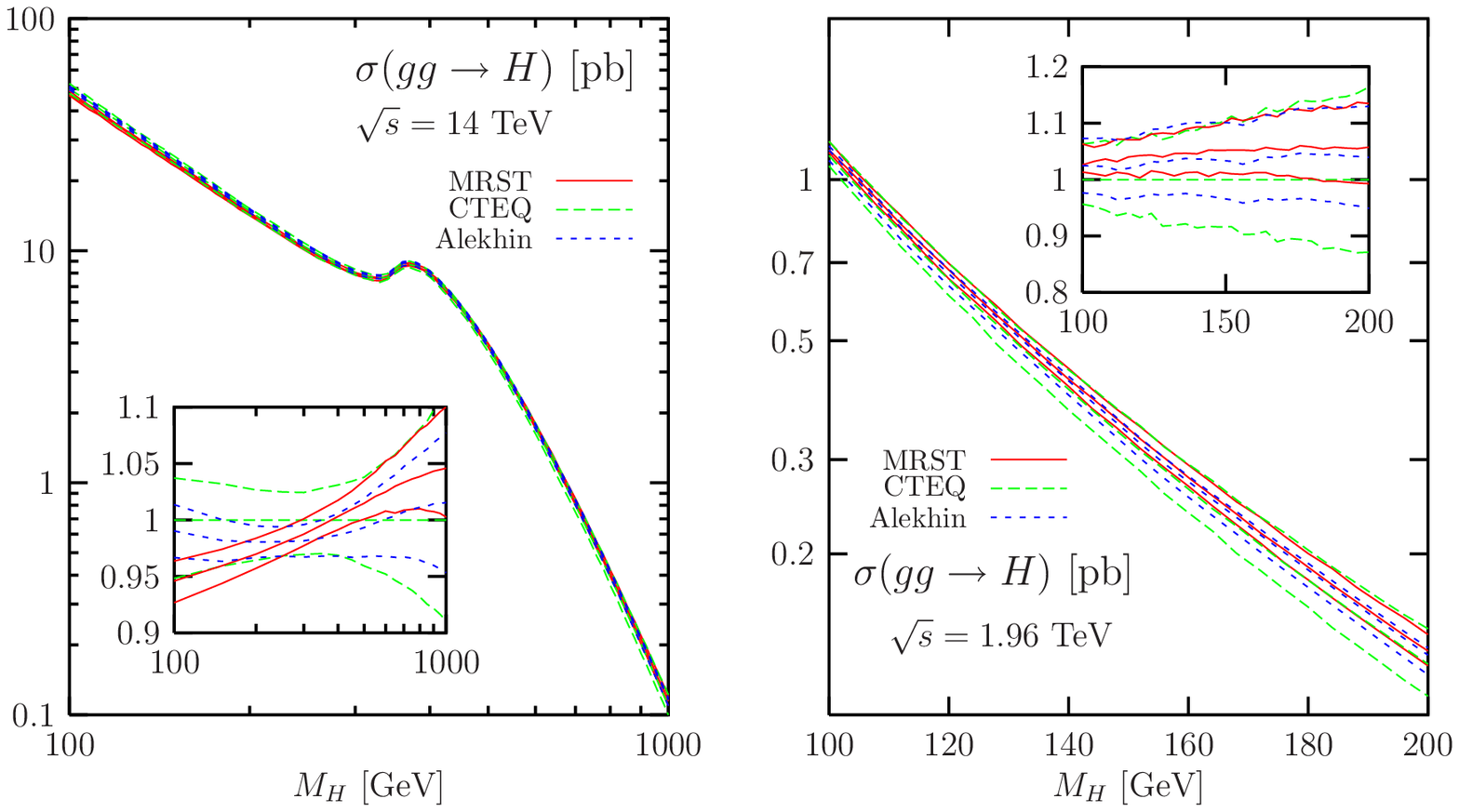,width=17cm}
\vspace*{-14.3cm}
\end{center}
{\it Figure 3.21: The CTEQ, MRST and Alekhin PDF uncertainty bands for the NLO
$gg \to H$ cross sections at the LHC (left) and Tevatron (right). The inserts 
show the spread in the predictions, when the NLO cross sections are
normalized to the CTEQ6 reference set \cite{Samir}.}
\vspace*{-5mm} 
\end{figure}

The variation of the cross section with the renormalization and factorization
scales will be discussed later after inclusion of the NNLO corrections to which
we turn now. 

\vspace*{-3mm} 
\subsubsection{The cross section beyond NLO in the heavy top quark limit}

\subsubsection*{\underline{The calculation at NNLO}}

Recently, the very complicated three--loop NNLO QCD corrections to the $gg \to
H$ fusion process have been calculated  by three different groups
\cite{ggH-NNLO1,ggH-NNLO2,ggH-NNLO3} in the limit of a very heavy top quark. In
this limit, the Feynman diagrams contributing to the process factorize into two
pieces: a massive component where the heavy quark has been integrated out and
which represents an effective coupling  constant which multiplies the $Hgg$
vertex, and a massless component involving only gluons and light quarks, which
describes the short distance effects  and where the finite momenta of the
particles have to be taken into account.  The calculation effectively  reduces
then to a two--loop calculation with massless particles. \s

However, many Feynman diagrams, some of which are displayed in Fig.~3.22, have 
to be evaluated at this order and they can be cast into three categories 
[which lead to more than one thousand square and interference terms] besides
the one--loop squared contribution:  
$a)$ two loop virtual corrections for the process $gg \to H$ which have 
to be multiplied by the effective Born amplitude; 
$b)$ one loop single real emission diagrams for the $gg \to Hg,\ gq \to 
Hq$ and $q\bar q \to Hg$ processes, which have to be multiplied by the Born 
amplitude for the same processes; 
$c)$ tree--level double real emission diagrams for the processes $gg 
\to Hgg, \ gg \to Hq\bar q, \ gq \to Hgq, \ qq \to Hqq$ and $q\bar q \to H 
q\bar q$, which have to be squared.

\begin{figure}[!h]
\vspace*{-.5cm}
\begin{center}
\setlength{\unitlength}{1pt}
\SetWidth{1.1}
\begin{picture}(450,100)(-10,0)
\Gluon(0,20)(30,20){3}{4}
\Gluon(0,80)(30,80){-3}{4}
\Gluon(30,20)(30,80){3}{8}
\Gluon(30,80)(90,50){3}{8}
\Gluon(30,20)(90,50){-3}{8}
\Gluon(50,32)(50,68){3}{4}
\DashLine(90,50)(130,50){5}
\put(84,44){\red{\Huge $\bullet$}}
\put(100,55){$H$}
\put(-10,18){$g$}
\put(-10,78){$g$}
\put(15,48){$g$}
\hspace*{5.5cm}
\Gluon(0,20)(40,20){3}{4}
\Gluon(0,80)(40,80){3}{4}
\Gluon(40,20)(80,20){3}{4}
\Gluon(40,80)(80,80){3}{4}
\Gluon(40,80)(40,20){3}{7}
\Gluon(80,80)(80,20){3}{7}
\Gluon(80,80)(120,80){3}{4}
\DashLine(80,20)(120,20){5}
\put(76,14){\red{\Huge $\bullet$}}
\put(125,20){$H$}
\put(-10,18){$g$}
\put(-10,78){$g$}
\put(125,80){$g$}
\hspace*{6cm}
\Gluon(0,20)(50,20){3}{5}
\Gluon(0,80)(50,80){3}{5}
\Gluon(50,80)(90,80){3}{5}
\Gluon(50,50)(90,50){3}{5}
\Gluon(50,20)(50,80){3}{7}
\DashLine(50,20)(90,20){5}
\put(46,14){\red{\Huge $\bullet$}}
\put(95,20){$H$}
\put(95,50){$g$}
\put(-10,18){$g$}
\put(-10,78){$g$}
\put(95,80){$g$}
\end{picture} 
\end{center}
\vspace*{-1.4cm}
\begin{center}
\setlength{\unitlength}{1pt}
\SetWidth{1.1}
\begin{picture}(450,100)(-10,0)
\ArrowLine(10,20)(40,20)
\ArrowLine(10,80)(40,80)
\Line(40,20)(40,80)
\Gluon(40,80)(90,50){3}{8}
\Gluon(40,20)(90,50){-3}{8}
\Gluon(60,32)(60,68){3}{4}
\DashLine(90,50)(130,50){5}
\put(84,44){\red{\Huge $\bullet$}}
\put(100,55){$H$}
\put(-10,18){$q$}
\put(-10,78){$\bar q$}
\put(15,48){$g$}
\hspace*{5.5cm}
\Gluon(0,20)(40,20){3}{4}
\ArrowLine(0,80)(40,80)
\Gluon(40,20)(80,20){3}{4}
\ArrowLine(40,80)(80,80)
\Gluon(40,80)(40,20){3}{7}
\Gluon(80,80)(80,20){3}{7}
\ArrowLine(80,80)(120,80)
\DashLine(80,20)(120,20){5}
\put(76,14){\red{\Huge $\bullet$}}
\put(125,20){$H$}
\put(-10,18){$g$}
\put(-10,78){$q$}
\put(125,80){$q$}
\hspace*{6cm}
\Gluon(0,20)(50,20){3}{5}
\ArrowLine(0,80)(50,80)
\ArrowLine(50,80)(90,80)
\Gluon(50,50)(90,50){3}{5}
\Gluon(50,20)(50,80){3}{7}
\DashLine(50,20)(90,20){5}
\put(46,14){\red{\Huge $\bullet$}}
\put(95,20){$H$}
\put(95,50){$g$}
\put(-10,18){$g$}
\put(-10,78){$q$}
\put(95,80){$q$}
\end{picture} 
\end{center}
\vspace*{-6mm}
\nn {\it Figure 3.22: Typical diagrams for the QCD corrections to $gg\to H$ at 
NNLO in the heavy quark limit. {\red{\Huge $\bullet$}} denotes the effective
$Hgg$ vertex where the quark has been integrated out.}
\vspace*{-3mm}
\end{figure}

This {\it tour de force} has been made possible thanks to two simplifying
features: the possibility of using the low energy theorem discussed in
\S2.4.1, which allows to calculate the corrections to the effective $Hgg$
vertex,  and the development of new techniques \cite{Baikov+Smirnov} to 
evaluate massless
three--point functions at the two--loop level in complete analogy to massless 
three--loop propagator diagrams which are standard and can be done fully
automatically. \s

As already discussed in \S2.4.3, the NNLO QCD corrected $Hgg$ effective operator
in the heavy quark limit, ${\cal L}_{\rm eff} (Hgg)$, can  be obtained 
\cite{RChgg,ggH-Laenen,Review-Michael} 
by means of the low--energy theorem, eq.~(\ref{Cg:two-loop}). This operator does
not describe the $Hgg$ interaction in total: it accounts  only for the
interactions mediated by the heavy quarks directly, but it does  not include
the interactions of the light fields. It must be added to the  light--quark and
gluon part of the basic QCD Lagrangian, i.e. the effective  coupling has to be
inserted into the blobs of the effective two--loop diagrams shown in Fig.~3.22.
The NNLO corrections to inclusive Higgs production in $gg \to H$
can be cast then into the three categories which have been already
encountered when we discussed the NLO case. In terms of the variable $\hat\tau$
defined as $\hat{\tau}=M_H^2/\hat{s}$, one has $\delta$ function terms $\propto
\delta (1- \hat \tau)$, large logarithms of the form $\log^n (1-\hat
\tau)/(1-\hat \tau)$, and hard  scattering terms that have at most a
logarithmic singularity  in the limit  $\hat \tau \to 1$
\beq
\hat{\sigma}_{ij}^{(2)} = a^{(2)} \delta(1-\hat \tau) + \sum_{k=0}^3 b^{(2)}_k 
{\cal D}_k(\hat \tau) + \sum_{l=0}^\infty \sum_{k=0}^3 c_{lk}^{(2)} (1-\hat
\tau)^l  \ell^k
\eeq

where $\ell_k= \log^k(1-\hat \tau)$ and ${\cal D}_k(\hat \tau)$, with now $i=1,
2,3$, are the usual $+$ distributions defined earlier. The virtual corrections
\cite{ggH-virtual}, which are of course UV finite when all contributions are
added up, and in particular the coefficient function $C_g$ of the $Hgg$
effective operator contribute only to the coefficient $a^{(2)}$ in front of the
delta function \cite{ggH-virtual,ggH-virtual-E}.  The soft  corrections to
the $gg \to H$ cross section, i.e. when the momenta of the  final state gluons
or quarks tend to zero, contribute to both  the $a^{(2)}$  and $b^{(2)}$ terms;
they have been evaluated in Refs.~\cite{ggH-SSL,ggH-SVC} and, when added to the
virtual corrections, the infrared divergences cancel out after mass
factorization. The combination of the virtual+soft with the collinear terms
$\propto \ell^3$ gives the ``soft+subleading" \cite{ggH-SSL} or
``soft+virtual+collinear corrections" \cite{ggH-SVC} approximations which
include also the contributions  to the coefficient $c^{(2)}_{03}$ which has
been evaluated  in Ref.~\cite{ggH-Laenen} using resummation techniques. \s

The remaining pieces which have to be evaluated at NNLO \cite{ggH-NNLO1} are
then the coefficients $c_{lk}^{(2)}$ with $k=0, \cdots 3$ and $l \geq 0$ which
receive contributions from all sub--processes. One can perform this calculation
by making a systematic expansion of the partonic cross section around the soft
limit $\hat \tau \sim 1$, leading to a series in $(1-\hat \tau)^n$ whose
coefficients  depend on $\ell^n \equiv \log^n(1-\hat \tau)$ with $n=0,1,2,3$ at
NNLO.  However, because the bulk of the cross section is at the threshold $\hat
\tau \to 1$, the series converges very rapidly and it is sufficient to keep
only the contributions of the terms up to order $(1-\hat \tau)^1$. The
convergence can be improved \cite{ggH-HSVC} by pulling  out a factor $\hat
\tau$ before expanding in $(1-\hat \tau)$. In practice, the expansion to order
$(1-\hat \tau)^1$ reproduces the exact result, with all terms up to order
$(1-\hat \tau)^{16}$ or equivalently with the exact calculation as performed in
Refs.~\cite{ggH-NNLO2,ggH-NNLO3}, with an accuracy of order 1\%.\s

This approach leads to a rather simple analytical result. Summing the
soft and hard contributions, one obtains the following partonic cross sections
up to NNLO [we display the LO and NLO contributions for completeness] in the 
various production channels, normalized to $\sigma_0^H=G_\mu \alpha_s^2/(288
\sqrt{2} \pi)$  introduced before and using $\ell_H=\log(M_H^2/m_t^2)$
\cite{ggH-NNLO1}
\beq
\hat{\sigma}^{(2)}_{gg} &=& \delta(1-\hat \tau)+ \frac{\alpha_s}{\pi} \bigg[ 
15.37\, \delta(1-\hat \tau)+ 6 - 24 \ell -9(1+4 \ell) (1-\hat \tau) + 
12 {\cal D}_1 (\hat \tau) \bigg] \non \\
&+& \left( \frac{\alpha_s}{\pi} \right)^2 \bigg[ 87.76 \, \delta(1-\hat \tau) + 
5.71 \ell_H -531.134 + 39.92 \ell +185.5 \ell^2 +144 \ell^3 
\non \\
&& \hspace*{1.5cm} + (632.06 + 632.87 \ell - 559.58 \ell^2 +216 \ell^3)(1-
\hat \tau) \non \\
&& \hspace*{1.5cm} +222.91 {\cal D}_0(\hat \tau) -31.71 {\cal D}_1(\hat \tau) 
-23 {\cal D}_2 (\hat \tau) +72 {\cal D}_3 (\hat \tau) \bigg] \non \\ 
\hat{\sigma}^{(2)}_{qg} &=& \frac{2}{3} \frac{\alpha_s}{\pi} \bigg[ 1 + 2
\ell - (1-\hat \tau) \bigg] \non \\ 
&+& \left(\frac{\alpha_s}{\pi} \right)^2 \bigg[ 29.93+ 6.47 \ell +2.63 \ell^2 
+6.79 \ell^3 (-40.19 + 50.33 \ell - 16.5 \ell^2)(1-\hat \tau ) \bigg] \non \\
\hat{\sigma}^{(2)}_{qq} &=& \left(\frac{\alpha_s}{\pi} \right)^2 \bigg[ 
-0.70 - 1.78 \ell +1.78 \ell^2 \bigg]
\eeq
where the scale dependence has been explicitly suppressed by setting  the
factorization and renormalization scales to $\mu_R=\mu_F=M_H$ [the dependence 
can be reconstructed by requiring the total cross section to be scale 
invariant] and the number of light quarks has been set to $N_f=5$. The 
component $\hat{\sigma}^{(2)}_ {qq}$ denotes the flavor singlet and 
non--singlet contributions in both the channels $qq$ and $q\bar{q} \to H+X$, 
the contributions of which are equal at order $(1-\hat \tau)$
\beq
\hat{\sigma}^{(2)}_{qq,S}=
\hat{\sigma}^{(2)}_{qq,NS}=
\hat{\sigma}^{(2)}_{q\bar q,S }=
\hat{\sigma}^{(2)}_{q\bar q,NS}
\eeq

\subsubsection*{\underline{The $K$--factors and the scale dependence up to NNLO}}

The cross sections $\sigma( pp \to H+X)$ at the three orders LO,  NLO and NNLO,
are  shown in Fig.~3.23 at the LHC and the Tevatron as a function of the Higgs
mass, using the MRST parton distributions which include the
approximated NNLO PDFs. The factorization and renormalization scales are set to
$\mu_R =\mu_F=\frac{1}{2}M_H$ (upper curves) and  $\mu_R=\mu_F=2M_H$ (lower
curves).  To improve the heavy quark approximation, the LO cross section
contains the full top mass dependence where $m_t=175$ GeV has been used.
Considering first the relative magnitude of the cross sections at the different
orders of perturbation theory, one can see that  while from LO to NLO, the
cross section increases at the LHC by 70\% for moderate Higgs boson masses, the
increase from NLO to NNLO of about 30\%, is more modest. This explicitly shows 
the nice convergence behavior of the perturbative series. The $K$--factors are
larger at the Tevatron, since they increase the cross section by a factor of 
about three at NNLO, the bulk of which is provided by the NLO correction.  \s

\begin{figure}[htbp]
\begin{center}
    \begin{tabular}{cc}
      \epsfxsize=19em
      \epsffile[110 265 465 560]{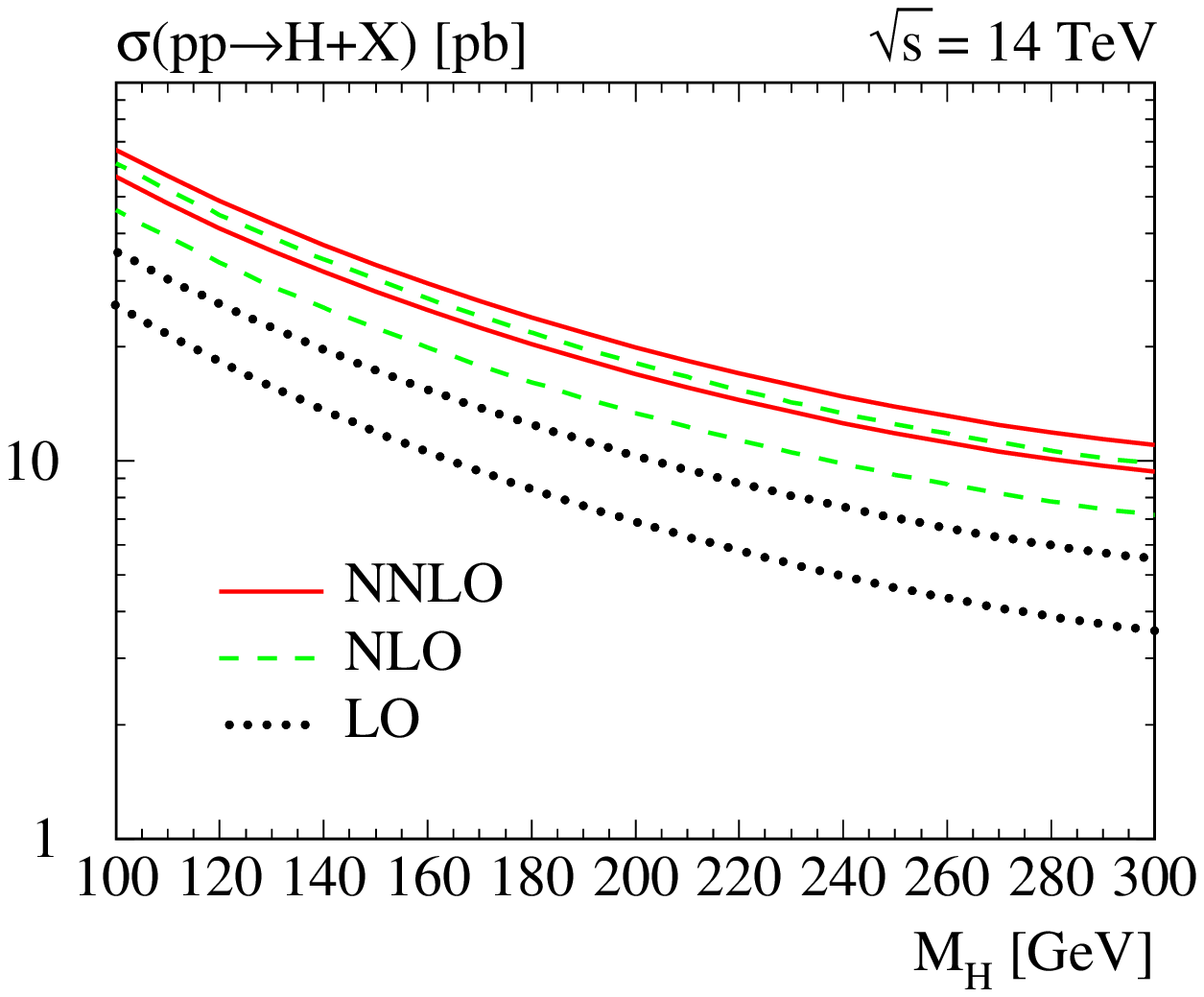}
      &
      \epsfxsize=19em
      \epsffile[110 265 465 560]{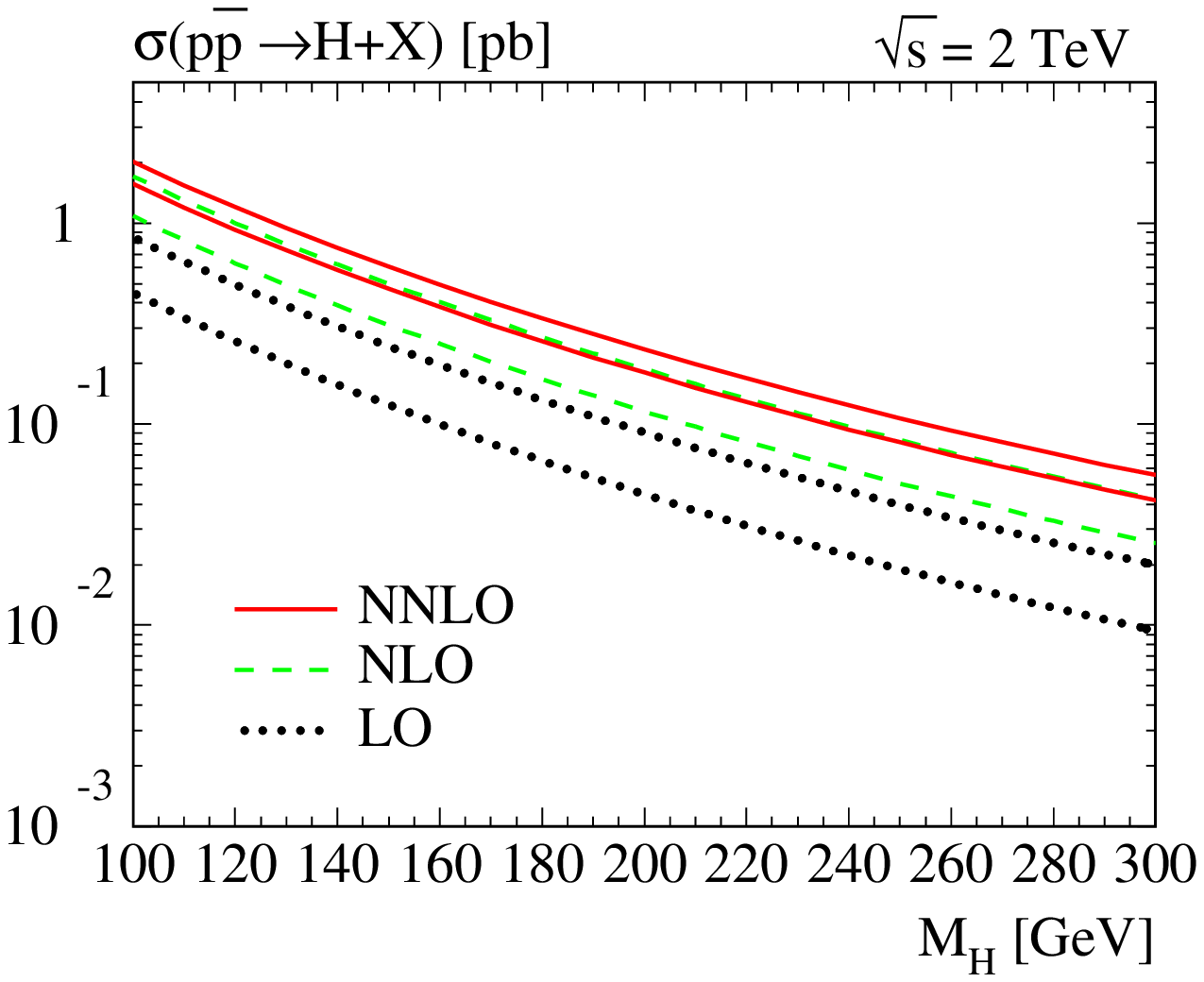}\\[-1mm]
    \end{tabular}
\end{center}
{\it Figure 3.23:  The cross sections for Higgs production in the $gg \to H+X$
fusion mechanism  at the LHC (left) and Tevatron (right) at LO (dotted), NLO
(dashed) and NNLO (solid) for two factorization and renormalization scales:
$\mu_R=\mu_F=\frac{1}{2}M_H$ (upper curves) and $\mu_R=\mu_F=2M_H$ (lower 
curves). The MRST  PDFs are used; from Ref.~\cite{ggH-Robert}.}
\vspace*{-1.mm}
\end{figure}

When considering the effect of the variation of the renormalization and 
factorization scales on the cross section, by multiplying and dividing by a
factor of two the median scale $\mu_F=\mu_R=M_H$, one first sees that 
globally, the scale dependence is reduced when going from LO, to NLO and then 
to NNLO. The residual scale dependence at NNLO is 25\% at the LHC and 15\% at 
the Tevatron, a factor two and a factor of four smaller than the dependence 
on the scale choice, at respectively, NLO and LO. \s

It has been noticed in  Refs.~\cite{ggH-NNLO-resum,ggH-Robert} that at the LHC
the dependence on the renormalization and factorization scales have different
signs: the cross section increases (decreases) with increasing $\mu_F \,
(\mu_R)$ values when the other scale is fixed, to $\mu_R\, (\mu_F)=M_H$ for
instance [at the Tevatron the dependence on $\mu_R$ and $\mu_F$ go the same
direction]; the decrease with $\mu_R$ is much  stronger. It is thus more
appropriate to choose smaller values for the scale than the standard choice
$\mu_R=\mu_R=M_H$.  This is shown in Fig.~3.24 where the scales are varied
within a factor $\frac{1}{4}$ and 4 with respect to the default scale
$\mu_F=\mu_R=M_H=115$ GeV, first collectively and then by varying $\mu_F \,
(\mu_R)$ while the other scale is fixed at the default value.\s

With the choice $\mu_R=\mu_F= \frac{1}{2}M_H$ e.g., the NLO correction
increases  while the NNLO correction decreases, with a total cross section
which increases compared to the choice $\mu_R=\mu_F=M_H$. Therefore, since the
difference between the NLO and NNLO contributions is small, the convergence of
the perturbative series is improved for $\mu_R=\mu_F= \frac{1}{2}M_H$. This
choice is supported by the fact that these fixed order results are in a better
agreement with recent estimates of the cross section with a resummation of the
dominant corrections which are due the contribution near the threshold $\hat
\tau \to 1$ to which we turn now. 

\begin{figure*}[h!]
  \begin{center}
    \leavevmode
    \begin{tabular}{ccc}
      \epsfxsize=11em
      \epsffile[184 210 411 610]{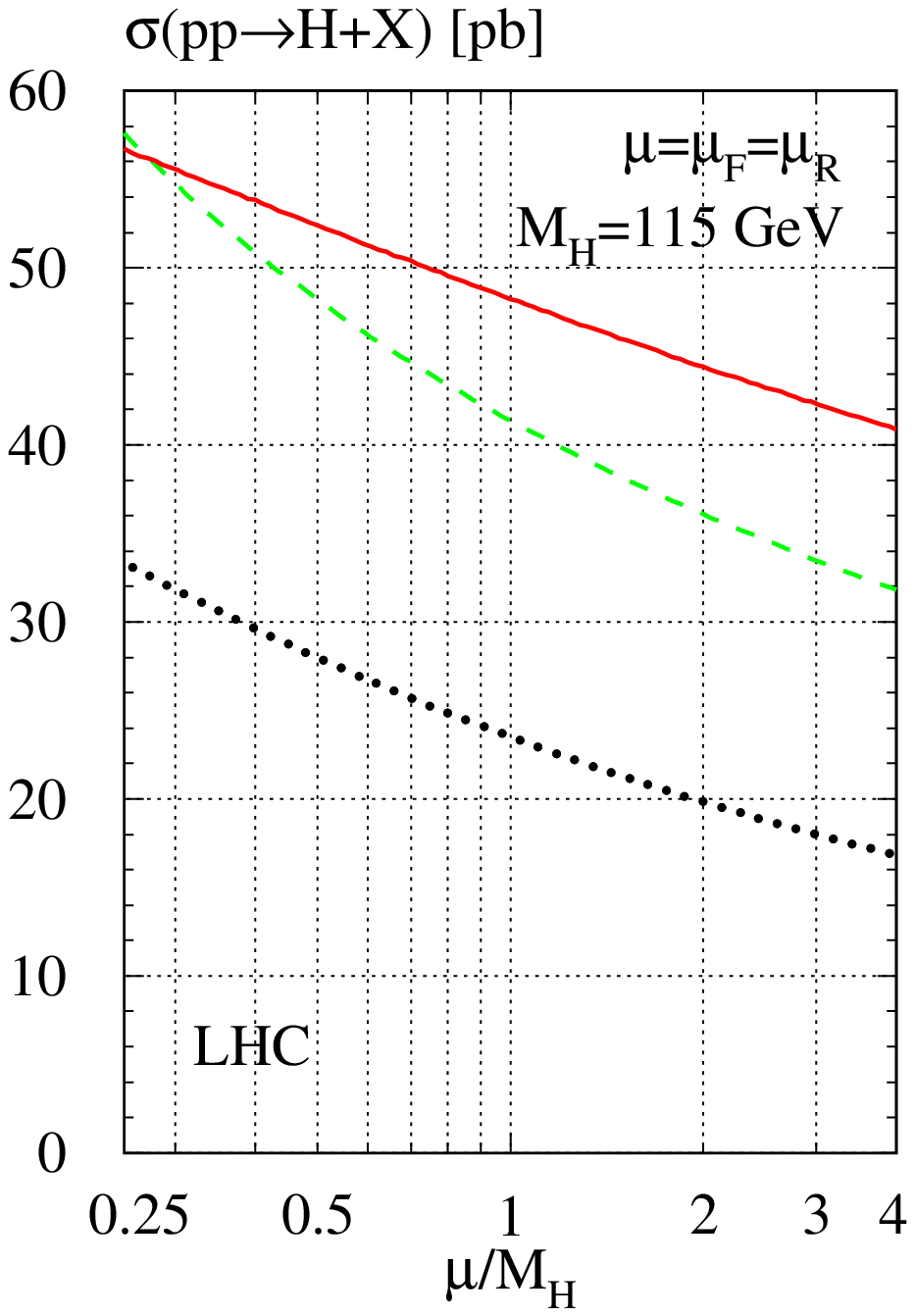} &
      \epsfxsize=11em
      \epsffile[184 210 411 610]{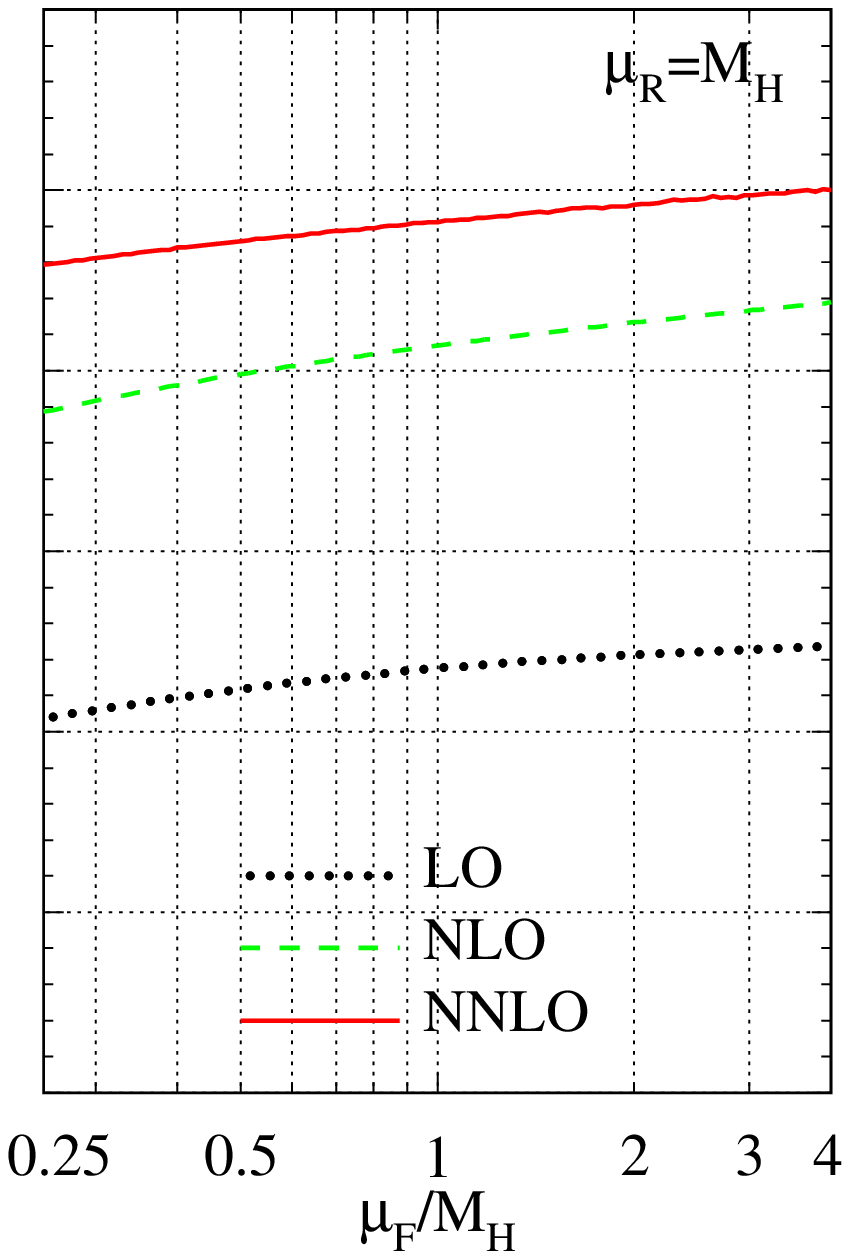} &
      \epsfxsize=11em
      \epsffile[184 210 411 610]{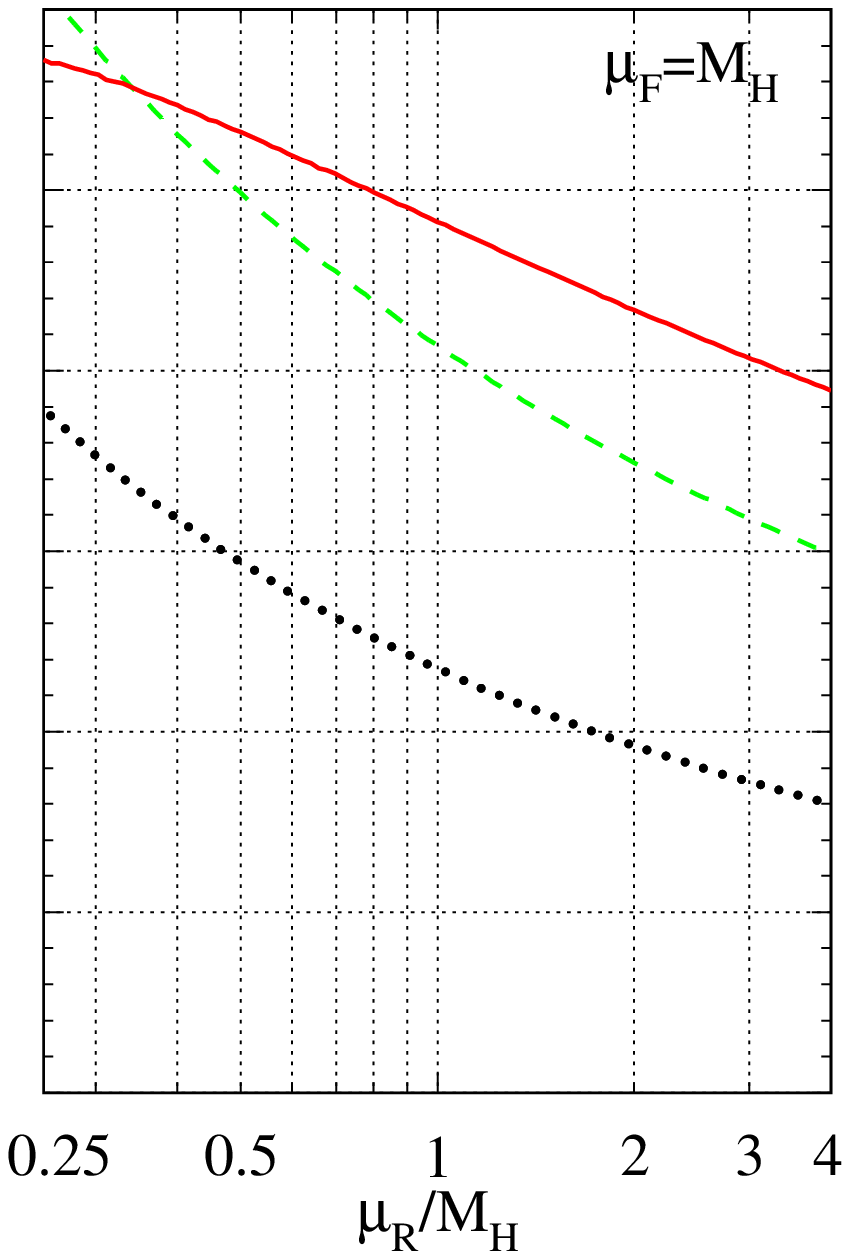}
    \end{tabular}
  \end{center}
\vspace*{-.5cm}
{\it Figure 3.24: The scale dependence of $\sigma(gg \to H)$ at LHC for 
$M_H=115$ GeV: variation of $\mu\equiv\mu_R=\mu_F$ (left), $\mu_F$ with 
$\mu_R=M_H$ (center) and $\mu_R$ with $\mu_F=M_H$ (right); from 
\cite{ggH-Robert}.}
\vspace*{-.5cm}
\end{figure*}

\subsubsection*{\underline{The soft--gluon resummation up to NNLL}}

As mentioned when we discussed the necessary ingredients to perform the $gg \to
H$ calculation at NNLO, the corrections to the cross section,
eq.~(\ref{eq:sigij}), fall into three categories: virtual and soft corrections
which generate the $\delta (1- \hat \tau)$ terms and the ${\cal D}_k$
distributions, collinear logarithmic contributions that are controlled by the
regular part of the Altarelli--Parisi splitting kernels and the hard scattering
terms.  The soft gluon corrections contribute to the most singular terms above
and they involve only the $gg$ initial state which, as already seen at NLO, is
the channel where the most important part of the correction originates from.\s

The soft gluon contributions in the $gg \to H$ process can be resummed up to
the next--to--next--to--leading logarithm (NNLL) order in the heavy top quark
limit \cite{ggH-NNLO-resum}, that is, all large logarithmic terms $\alpha_s^n
\log^m (1 - \hat \tau)$ in the $+$ distributions with $1 \leq m \leq 2n$ in the
limit $\hat \tau \to 1$ can be exponentiated. The resummation relies on the
basic factorization theorem for partonic cross sections into soft, collinear
and hard parts near the phase--space boundary \cite{Collins}, and can be
performed in the Mellin or N--moment space \cite{Sterman+Catani+Luca} for
instance. The formalism and the calculation's technique have been presented in
detail in Refs.~\cite{ggH-NNLO-resum,ggH-Laenen}. \s

The resummation of the logarithms in the soft gluon contributions is formally
justified only near the thresholds $\hat \tau \to 1$. However, it can be used
away from the threshold and the expectation is that the soft+virtual
corrections, eventually supplemented by the collinear parton radiation (SVC), is
a good approximation of the exact result for the cross section. Indeed, owing
to the suppression of the gluon densities at large $x$, the partonic c.m. energy
$\sqrt{\hat{s}}$ is much smaller than the c.m. energy of the hadron collider,
$s=x_1x_2\hat{s}$, and the dominant value of $\hat \tau$ which appears in the
hard scattering terms of the partonic cross section can be close to unity also
when $\sqrt s$ is not close to $M_H$ \cite{ggH-SVC,ggH-HSVC}. This has been 
verified both at NLO and NNLO:  SVC approximates the exact result quite well, 
in particular at LHC energies.  \s

The results for the resummed cross sections, in terms of the $K$--factors,
are shown in Fig.~3.25 for the LHC as a function of $M_H$, for the LL,
NLL and NNLL approximations (right) and are compared with the fixed order 
results at LO, NLO and NNLO (left). The bands result from a scale variation 
$\frac12 M_H \leq \mu_{F,R} \leq 2M_H$. One can note that the scale dependence  
after resummation is smaller than at fixed order and that, at NNLO, the 
resummation increases the central value of the cross section by $\sim 5\%$ 
in the low Higgs mass range. 

\begin{figure}[htb]
\begin{center}
\begin{tabular}{c}
\epsfxsize=10truecm
\hskip -0.5cm\epsffile{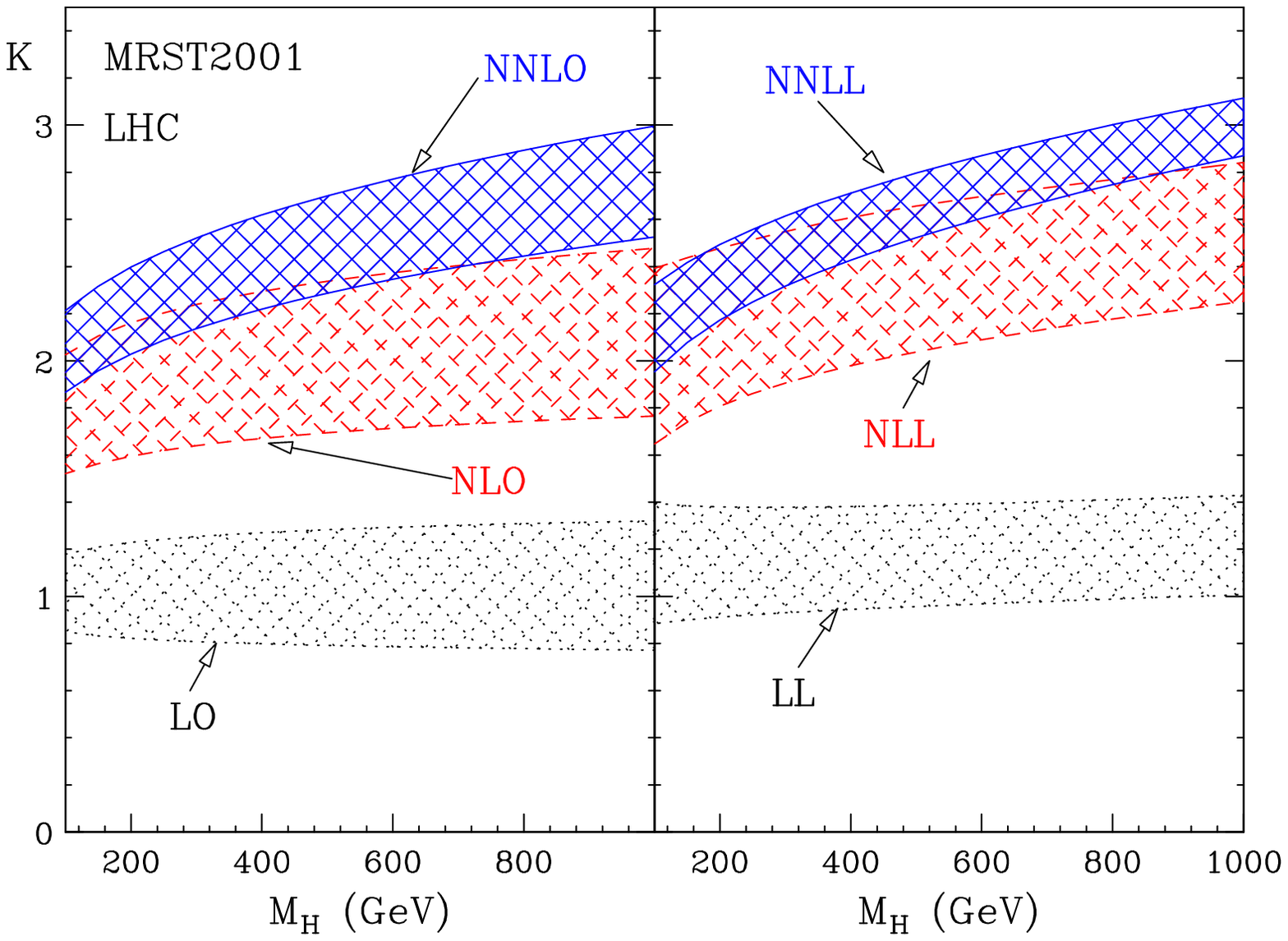}
\end{tabular}
\end{center}
\vspace*{-.4cm}
{\it Figure 3.25: Fixed order (left) and resummed (right) $K$--factors 
for $gg \to H+X$ at the LHC as a function of $M_H$. The MRST2001 parton 
distributions have been used; from Ref.\,\cite{ggH-NNLO-resum}.}
\vspace*{-.2cm}
\end{figure}

\subsubsection{The distributions and Higgs + n jet production}

\vspace*{-2mm}
\subsubsection*{\underline{The transverse momentum and rapidity distributions}}

At leading order, the Higgs boson produced in the fusion process $gg \to H$ has
no transverse momentum. The $p_T$ of the Higgs boson is generated at higher
orders, when additional partons are radiated and balance the Higgs $p_T$
\cite{pp-Hgg-PT,pp-Hgg-PT2,pp-ggH-PT0,pp-ggH-PT1,Pt-eta-distrib,pp-ggH-Ital}. 
The leading order for the Higgs boson transverse momentum and rapidity is
therefore part of the NLO for the production cross section, when
the processes  responsible for them, $gg \to Hg, gq \to Hq$ and $q\bar{q} \to
Hg$, take place. The $p_T$ and $y_H$ distributions have been calculated in the
full massive case at LO \cite{pp-Hgg-PT,pp-Hgg-PT2} and it was shown that the
heavy top quark limit is a reasonably good approximation,  provided of course 
that $M_H \lsim 2m_t$, but more importantly in this case, that $p_T \lsim m_t$,
which is typically the case  as will be seen shortly. We therefore restrict
ourselves to the heavy quark limit and summarize the salient features of these
distributions. \s

Defining the momenta of the initial particles involved in the process $ij \to H
k$,  with $i,j,k=g, q, \bar{q}$, as $p_{i,j} = x_{i,j} \, p_{1,2}$ with
$p_{1,2}$ the incoming hadron momenta, and as $p_k$ the momentum of the final
parton, the differential partonic cross section in terms of the Higgs 
transverse momentum $p_T$ and rapidity  $y_H$ can be written in the heavy quark
limit as
\begin{eqnarray}
\frac{{\rm d}^2 \hat\sigma (ij \to kH)}{{\rm d} p_T^2 {\rm d}y_H }=  
\frac{G_\mu \alpha_s^3} {576 \sqrt{2} \pi^2}~{\cal H}_{ij \to kH} ( p_T, y_H) 
\non \hspace*{3cm} \\
{\cal H}_{gg\to Hg} = 3 \frac{\hat s^4+\hat t^4+\hat u^4+M_H^8}{\hat{s}^2
\hat t\hat u} \, , \ {\cal H}_{gq\to Hq} = -\frac{4}{3} \frac{\hat s^2+
\hat u^2}{ \hat s \hat t} \, ,  \ {\cal H}_{q\bar q\to Hg} = \frac{32}{9}
\frac{\hat t^2+\hat u^2} {\hat{s}^2}
\label{sigma:ij-kH}
\end{eqnarray}
with the Mandelstam variables $\hat{s}, \hat{t}, \hat{u}$ given in terms of
$y_H$ and the transverse mass squared $m_T^2=M_H^2+p_T^2$ as $\hat s =
(p_i+p_j)^2 = (p_k+p_H)^2 = s x_i x_j$ and $\hat t/ \hat u  = (p_{i/j}-p_k)^2
=(p_{j/i}- p_H)^2=M_H^2 -\sqrt{s} x_j m_T e^{\pm y_H}$, with $s$ being the
total hadronic c.m.~energy.  The expressions are singular for $\hat{t}, \hat{u}
\ra 0$ and, in particular, ${\cal H}_{gg \to Hg}$ is singular in both $\hat{t}$
and $\hat{u}$. The singularities can be regularized by moving to
$n=4-2\epsilon$ space--time dimensions.\s

To include the NLO corrections to the differential distribution, and similarly
to part of the NNLO corrections for the total cross section, one has to
calculate: $(i)$ the virtual corrections to Higgs production 
with a parton, which has to be multiplied by the Born term of the same process,
and $(ii)$ the real corrections due to the production of the Higgs boson with 
two partons, the sum of  which has to be squared. In addition, one has to add 
the corrections to the Altarelli--Parisi splitting functions from the parton 
densities  at NLO.\s
 
These corrections have been calculated by several  groups \cite{pp-ggH-PT1,Pt-eta-distrib,pp-ggH-Ital,pp-ggH-eta1,pp-ggH-eta2,pp-ggH-distrib}, using
different methods and different schemes. In all cases, the heavy top quark
limit has been used. We summarize below the main results at NLO, 
concentrating on the case of the LHC where the transverse momentum and rapidity
distributions of the Higgs boson are very important ingredients. Unless
otherwise stated, the Higgs mass is set to $M_H=120$ GeV and the heavy top
limit is assumed; the renormalization and factorization scales are set equal and
fixed to the transverse Higgs mass, $\mu_R=\mu_F = m_T= \sqrt{M_H^2+ p_T^2}$. \s

The left--hand side of Fig.~3.26 shows the $p_T$ distribution of the Higgs
boson at NLO for several fixed rapidity values. One first notices that the
differential distribution decreases with increasing rapidity and with
increasing  $p_T$ and that at small values of the latter, $p_T \to 0$, it
diverges to $-\infty$ [while at LO it diverges to $+ \infty$]. In the low
$p_T$ regime, $p_T\lsim 30$ GeV, the spectrum is unstable due to occurrence of 
large logarithms; the perturbative treatment  is therefore not reliable and
resummation techniques, to be discussed later, are required. Note that at small
and moderate $p_T$, the cross section is dominated by the gluonic $gg \to H+X$
contribution, while for $p_T$ values beyond 200 GeV  the contribution of the
$gq \to HX$ process becomes comparable; the (anti)quark initiated processes
give very small contributions. \s

The NLO corrections increase the $p_T$ distribution except for small $p_T$. 
While the increase is very strong for $p_T$ values below 30 GeV [recall that
the  distribution at LO was diverging in the opposite direction], it becomes
moderate for $p_T$ values in the range of applicability of perturbation theory.
The $K$--factor, defined as $K = {\rm d} \sigma_{\rm NLO}/ {\rm d} \sigma_{\rm
LO}$, rises slowly from $K \sim 1.6$ at $p_T=30$ GeV to $K \sim 1.8$ for
$p_T=200$ GeV when the total rate becomes too small.

\begin{figure}[h!]
\begin{center}
\mbox{
\epsfxsize=7.9cm \epsfbox{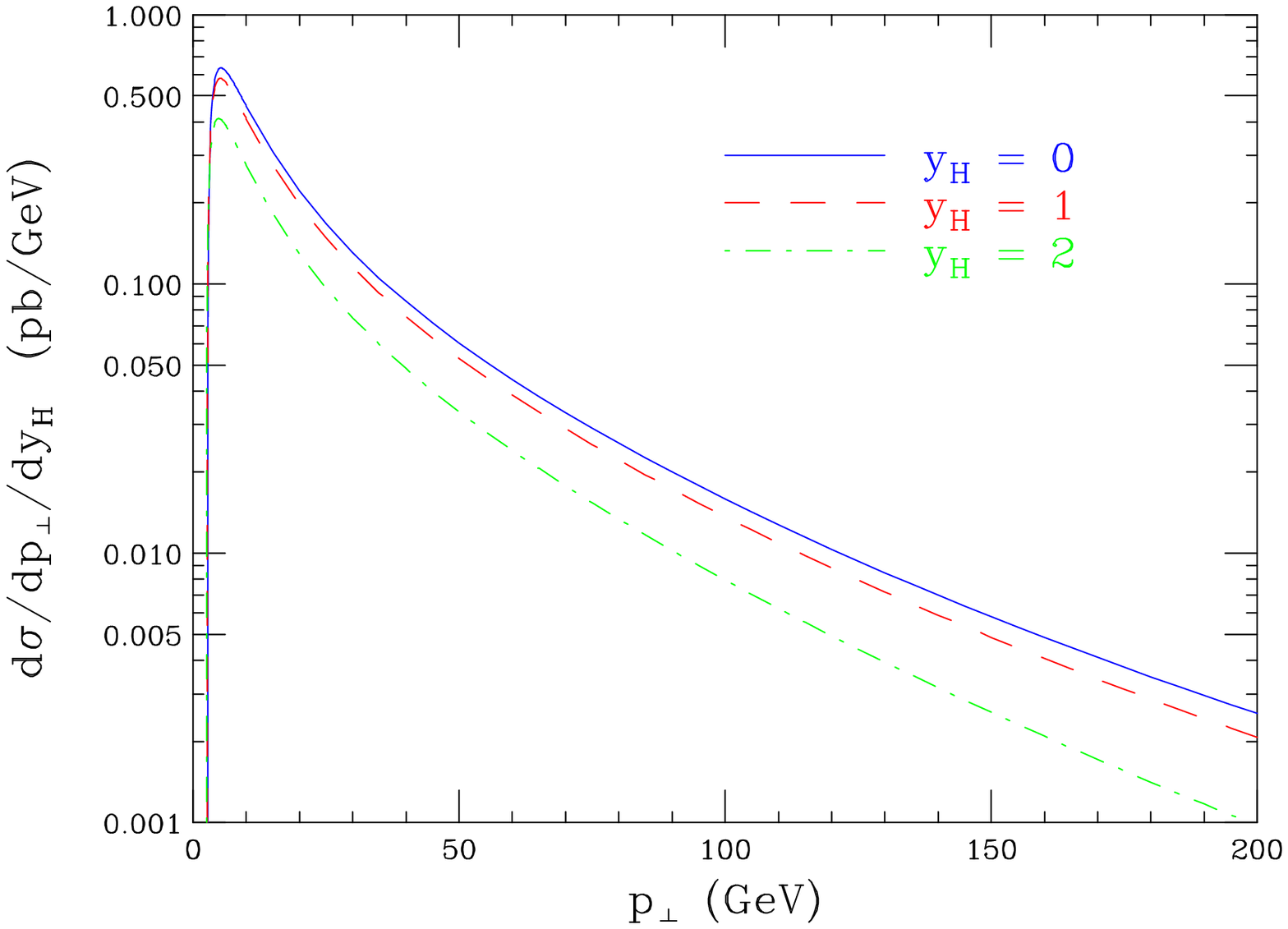}\hspace*{3mm}
\epsfxsize=7.9cm \epsfbox{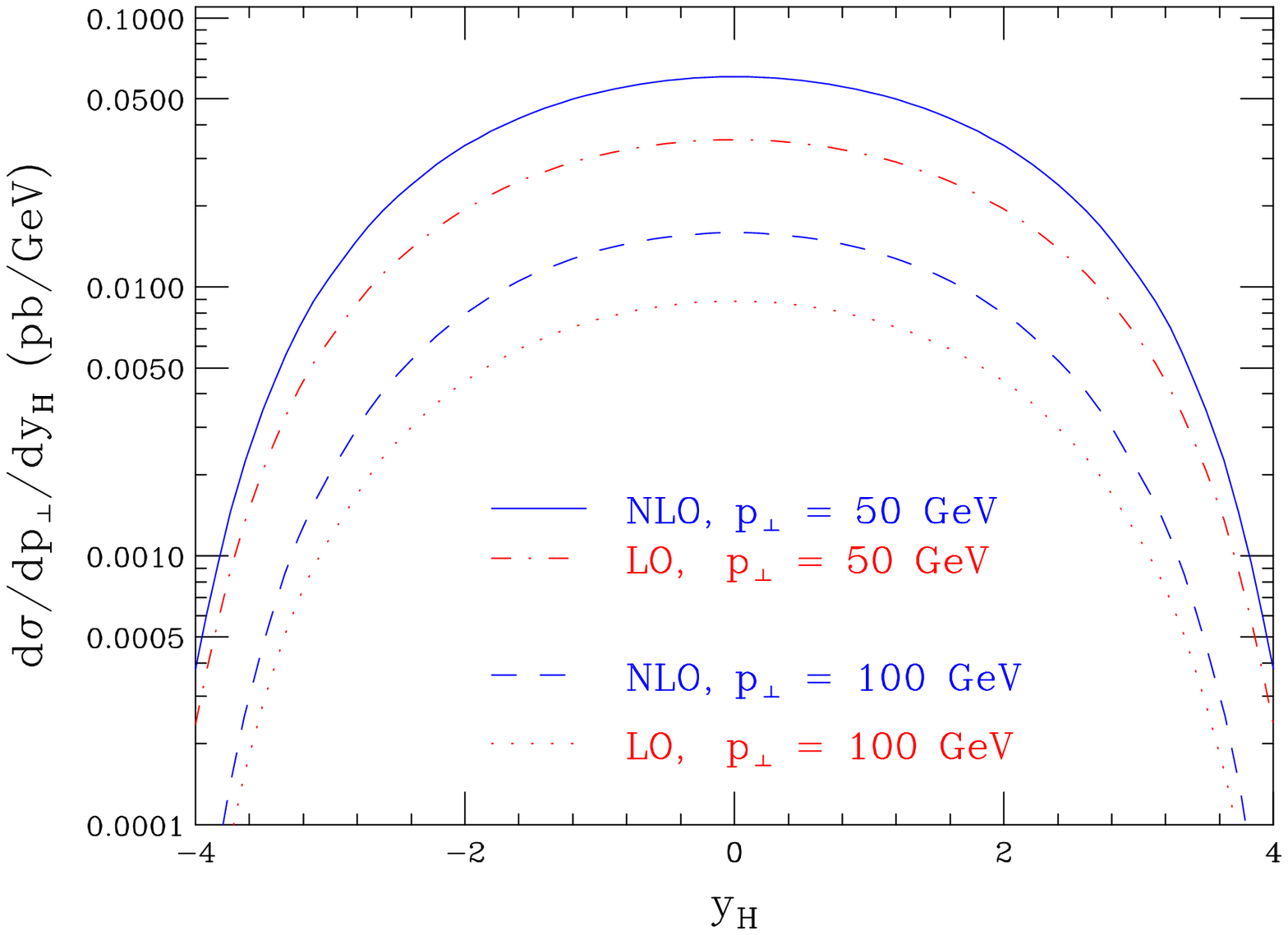}} \\[-5mm]
\end{center}
\vspace*{-2mm}
{\it Figure 3.26: The Higgs transverse momentum dependence at NLO for three 
values of the rapidity $y_H=0,1,2$ (left) and the rapidity dependence for two 
different transverse momenta $p_T=50$ and 100 GeV at both LO and NLO (right). 
The CTEQ5 set of PDFs has been used while $M_H=120$ GeV and the scales are 
set to $\mu_R=\mu_F= m_T$; from Ref.~\cite{Pt-eta-distrib}.}
\vspace*{-2mm}
\end{figure}

The right--hand side of Fig.~3.26 shows the rapidity dependence of the cross
section for fixed values of the transverse momentum, $p_T=50$ and 100 GeV, at
both LO and NLO. As usual, the differential cross section is smaller for higher
$p_T$ values. It is maximal at $y_H=0$ and falls off steeply  for large
rapidity values due to the restriction of the available phase, reaching zero
for  $|y_H| \gsim 4$. The NLO corrections increase the distribution: the
$K$--factor for reasonable $p_T$ values is at the level $\sim 1.6$ and is
almost independent on the value of the rapidity, except at the boundary of the
phase space where it drops slightly.\

Thus, the NLO corrections acquire a size of about 60\% to 80\% over the entire
perturbative range of $p_T$ and $y_H$ values.  The variation with the
renormalization and factorization scale has also  been discussed and found to
follow the same trend as in the production cross section at NLO: a variation
from the central value $\mu_F=\mu_R=m_T$ by a factor of two generates an
uncertainty of about 20\%.  There is also an uncertainty originating from the
choice of the PDFs, which is similar to what has been observed for the total
cross section and which is thus smaller than the scale uncertainty.\s

Let us make a final comment on the low $p_T$ case. As already mentioned,  the
distribution diverges to $+ \infty$ at LO and to $- \infty$ at NLO for $p_T \to
0$. This is because in the region  $p_T \ll M_H$, where the cross section is in
fact the largest, the expansion parameter is not $\alpha_s/\pi$ but rather,
$(\alpha_s/\pi) \log^2 (M_H^2/p_T^2)$ which is close to unity and invalidates
perturbation theory. However, the large logarithms, as singular as $(1/p_T^2)
\alpha_s^n \log^m (M_H^2/p_T^2)$,  with 1 $<$$m$$<2n-1$, can be systematically
resummed to all orders \cite{resum-PT} as in the case of the total cross 
section, resulting in a well behaved spectrum for $p_T \to 0$; see for 
instance Refs.~\cite{Pt-eta-distrib,pp-ggH-Ital} for details.\s

In the case of the $gg\to H$ process, the resummation has been performed at the 
NNLL level of accuracy. This resummation for the low $p_T$ region, and the fixed
order calculation at NLO for the high $p_T$ region, have been consistently
matched at intermediate $p_T$ values to provide a smooth transition. The result
for the $p_T$ distribution is shown in Fig.~3.27 at the LHC for a Higgs mass of
$M_H=125$ GeV. In the left figure, the NLO and NLO+NNLL approximations are
displayed and, as can be seen, the divergent behavior of this distribution is
removed by the resummation, the effects of which are relevant up to values $p_T
\sim 100$ GeV. The scale variation is shown in the right figure in the NLO+NNLL
case: the spread  is at the level of 10\%  near the peak and increases to 20\%
for lower $p_T$ values, $p_T \sim 100$ GeV.

\begin{figure}[h]
\vspace*{-2mm}
\begin{center}
\begin{tabular}{c}
\epsfxsize=11truecm
\epsffile{./sm3/nnll-PT.epsi}
\end{tabular}
\end{center}
\vspace*{-3mm}
{\it Figure 3.27: The $p_T$ distribution in $gg \to H$ at the LHC for $M_H=125$
GeV: at NLO and NLO+NNLL for scales $\mu_R=\mu_F= M_H$ (left) and at NLO+NNLL
when the scales are collectively varied by a factor of two; from 
Ref.~\cite{pp-ggH-Ital}.}
\vspace*{-5mm}
\end{figure}

\subsubsection*{\underline{Higgs boson plus $n$ jet production}}

It has been suggested that Higgs production with one high $p_T$ jet might have
a much smaller background at the LHC, in particular in the decay channel $H \to
\gamma \gamma$ \cite{ggHp-dubinin}, than the $gg \to H$ channel alone. At LO, 
the process is just the $gg \to Hg, qg \to Hq$ and  $q \bar q \to Hg$ processes
that we have discussed previously and for which we have displayed the partonic
differential cross sections in the heavy top limit, eq.~(\ref{sigma:ij-kH}). In
this limit, the NLO corrections to $pp \to H+j$ are those which appear in the
${\cal O}(\alpha_s^2)$ real corrections to the NNLO $gg \to H$ cross section. \s

The Higgs plus 2 jet production process is  generated by $q\bar q$ scattering 
mediated by triangles involving top quarks, $gq$ scattering mediated by boxes 
and triangles and $gg$ fusion mediated by triangles up to pentagon diagrams, 
and is known exactly at LO \cite{pp-ggHqq}. This mechanism has been discussed 
in connection with the vector boson fusion process, since it leads to the same 
final states, the gluons and the light quarks being indistinguishable, and may 
act as a background in the study of the former process. Characteristics which 
discriminate between the processes have been worked out and summarized in 
Fig.~3.28. The main points have been already discussed in \S3.3.3: with
basic (inclusive) cuts $p_{Tj} \gsim 20$ GeV, $|\eta_j| <5$ and $R_{jj} >0.6$, 
gluon fusion dominates, while the specific additional cuts $m_{jj} >0.6$ TeV, 
$|\eta_{j1}- \eta_{j2}| >4.2$ and $\eta_{j1} \cdot \eta_{j2} <0$, select the 
vector boson fusion (WBF) process \cite{Vittorio}.  

\begin{figure}[h]
\hbox to\hsize{
\includegraphics[width=.48\hsize]{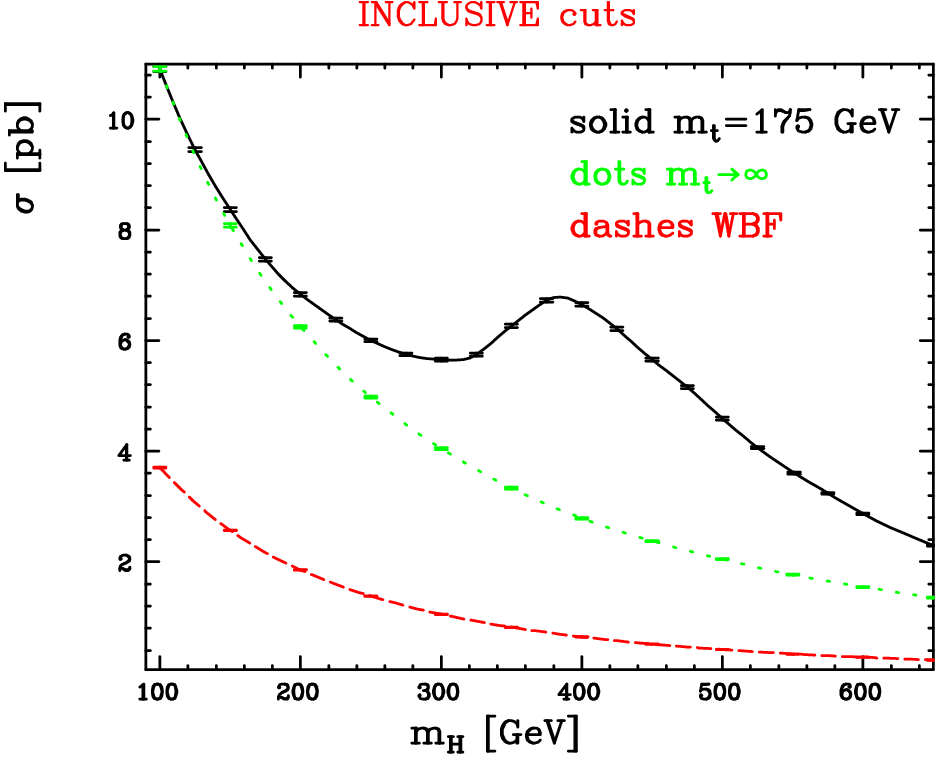}
\includegraphics[width=.48\hsize]{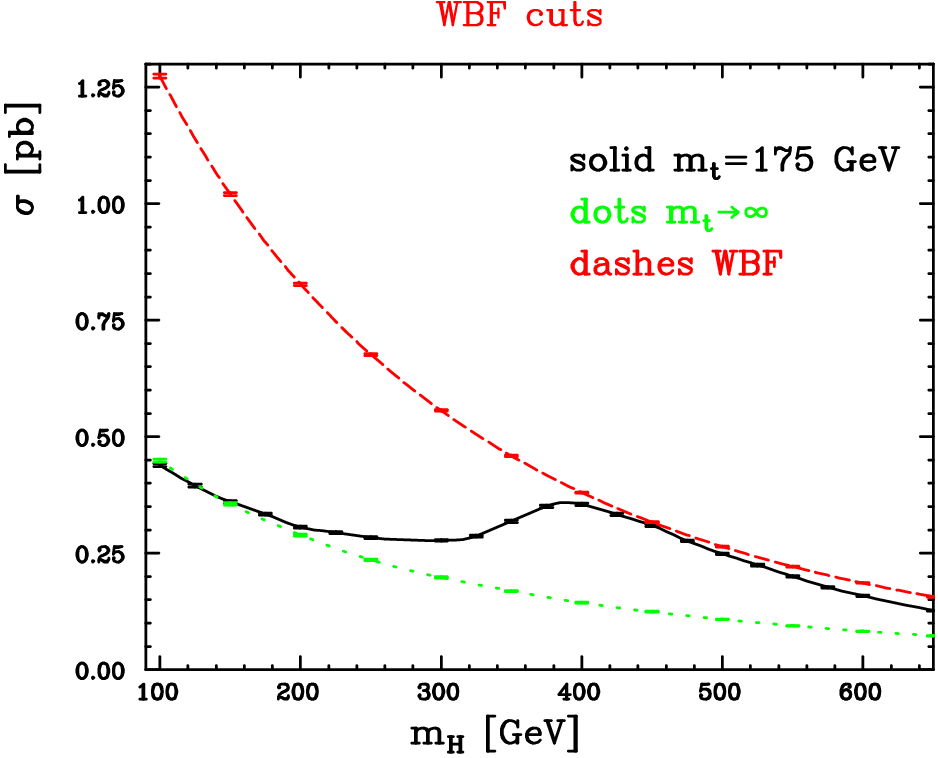}  }
{\it Figure 3.28: Higgs production with 2 jets at the LHC as a function of
$M_H$, in gluon fusion with $m_t\!=\!175$ GeV (solid line) and $m_t\!\to\! 
\infty$ (dotted line) and in vector boson fusion (dashed). The left (right) 
part shows the cross section with the inclusive (WBF) cuts \cite{Vittorio}. }
\vspace*{-0.2cm}
\end{figure}

Recently, associated Higgs production with 3 jets has been calculated in the
$m_t \to \infty$ limit \cite{pp-ggHqqq}. This is part of the NLO corrections 
to the Higgs+2 jet process,  which exhibits a strong dependence on the
renormalization scale $\mu_R$ since it is a LO process but of ${\cal
O}(\alpha_s^4)$. The very complicated virtual corrections need to be derived
which, when combined with the Higgs+3 jet real corrections, will hopefully
stabilize the scale dependence of the $H+2j$ rate.

\subsection{Associated Higgs production with heavy quarks} 

\subsubsection{The cross sections at the tree level}

The process where the Higgs boson is produced in association with heavy quark
pairs, $pp \to Q \bar Q H$ \cite{pp-Htt-LO,pp-Htt-LO1}, with the final
state quarks being either the top quark or the bottom quark [in this case,
see also Refs.~\cite{pp-Hbb-LO1,pp-Hbb-LO} for instance], is the most involved 
of all SM Higgs production mechanisms.  At tree--level, it
originates from $q\bar q$ annihilation into heavy quarks with the Higgs boson
emitted from the quarks lines; this is the main source at the Tevatron. At
higher energies, when the gluon luminosity becomes important, the process
proceeds mainly through gluon fusion, with the Higgs boson emitted from both
the external and internal quark lines. A generic set of Feynman diagrams is
shown in Fig.~3.29; those which are not shown differ only in the way the Higgs
line is attached to the final state quark line and the gluon symmetrization in
the last diagram.  

\begin{center}
\vspace*{-.6cm}
\hspace*{-4cm}
\SetWidth{1.}
\begin{picture}(300,100)(0,0)
\ArrowLine(0,25)(35,50)
\ArrowLine(0,75)(35,50)
\Gluon(35,50)(80,50){3.2}{5.5}
\Line(80,50)(115,25)
\Line(80,50)(115,75)
\DashLine(105,65)(130,47){4}
\Text(-2,35)[]{$\bar{q}$}
\Text(-2,65)[]{$q$}
\Text(55,65)[]{$g$}
\Text(122,20)[]{$Q$}
\Text(122,80)[]{$\bar{Q}$}
\Text(120,45)[]{\bH}
\Text(105,65)[]{\bb}
\hspace*{9mm}
\Gluon(150,25)(185,50){3.2}{5.5}
\Gluon(150,75)(185,50){3.2}{5.5}
\Gluon(185,50)(230,50){3.2}{5.5}
\ArrowLine(230,50)(265,25)
\ArrowLine(230,50)(265,75)
\DashLine(250,40)(270,50){4}
\Text(203,65)[]{$g$}
\Text(148,65)[]{$g$}
\Text(148,35)[]{$g$}
\Text(249,39)[]{\bb}
\hspace*{-5mm}
\Gluon(290,25)(330,25){3}{4.5}
\Gluon(290,75)(330,75){3}{4.5}
\Line(330,25)(330,75)
\DashLine(330,50)(365,50){4}
\ArrowLine(330,25)(375,25)
\ArrowLine(330,75)(375,75)
\Text(330,50)[]{\bb}
\end{picture}
\end{center}
\vspace*{-6mm}
{\it Figure 3.29: Generic Feynman diagrams for associated Higgs production 
with heavy quarks in hadronic collisions, $pp \to q\bar q, gg \to Q\bar QH$, 
at LO.}
\vspace*{2mm}

Added to the complication that one has to calculate the amplitude of 10 Feynman
diagrams and square them, one has to deal with the rather involved phase space
with three massive particles in the final state. This is the reason why the
complete analytical expression of the cross section is not available in the
literature. If only $q \bar q$ annihilation is considered, which is a good
approximation in the case of the Tevatron, the matrix element squared in terms
of the momenta $q (\bar q), p (\bar p)$ and $k$ of respectively, the initial
light and final heavy quark (antiquark) and the Higgs particle, with $Q^2=(q +
\bar q)^2 = (p + \bar p + k)^2 = \hat s$, is given by  \cite{pp-Htt-LO1}
\beq
\left| {\cal M}_{q \bar q} \right|^2 &=& \frac{32}{(2k.\bar p +M_H^2)
(2k.p +M_H^2)} \left\{ Q^2 (Q.k)^2 \left[ 1+ \frac{ Q^2 (4m_Q^2 -M_H^2)} 
{(2k.\bar p +M_H^2)(2k.p +M_H^2)} \right] \right.  \non \\
&&+ \left( \frac{1}{2} Q^2 m_Q^2 - 2 (p.q)(p.\bar q) \right) \left[ Q^2 -4m_Q^2 
+M_H^2 + \frac{ 2k. Q (4m_Q^2- M_H^2)}{2k. \bar p +M_H^2} \right]  \non \\
&&+ \left( \frac{1}{2} Q^2 m_Q^2 - 2 (\bar p. q)( \bar p.\bar q) \right) \left[ 
Q^2 -4m_Q^2 +M_H^2 + \frac{2k. Q (4m_Q^2- M_H^2)}{2k. p +M_H^2} \right]\non \\
&&- \left. ( Q^2 +M_H^2 - 4m_Q^2) \bigg[ 2(p. q) (\bar p . \bar q) +
2 (p.  \bar q) ( \bar p . q) - Q^2 (p . \bar p) \bigg] \right\} 
\eeq
where the coupling factors, $g_s^4 (\sqrt{2} m_Q^2 G_\mu)$, with $g_s^2=4\pi
\alpha_s$, have been removed. \s

For the gluon fusion diagrams, denoting by $\epsilon_1$ and $\epsilon_2$ the
polarization four--vectors of the gluons and by $g_1$ and $g_2$ their 
four--momenta, the various amplitudes are given by [the generators $T^a$ and
the SU(3) structure constants $f^{abc}$ are discussed in \S1.1.1]  
\cite{pp-Htt-LO1}
\beq 
{\cal M}_{gg}^a &=& -T_{ik}^a T_{kj}^b \, \bar u^j(p)\, \frac{ \slash 
\hspace*{-2mm}k+ \slash \hspace*{-2mm} p+m_Q}{2 p. k+ M_H^2 } \slash 
\hspace*{-2mm} \epsilon_2 \frac {- \slash \hspace*{-2mm} \bar{p} +
\slash \hspace*{-2mm} g_1 +m_Q }{- 2g_1 . \bar p } \slash \hspace*{-2mm}
\epsilon_1 \, v^i (\bar p) + \left[ \begin{array}{c} 
g_1 \leftrightarrow g_2 , \epsilon_1  \leftrightarrow \epsilon_2 \\
g_1 \leftrightarrow g_2 , \epsilon_1  \leftrightarrow \epsilon_2 , 
p \leftrightarrow \bar p \\
p \leftrightarrow \bar p \end{array} \right] \non \\
{\cal M}_{gg}^b &=& -T_{ik}^a T_{kj}^b \, \bar u^j(p)\, \slash \hspace*{-2mm}
\epsilon_2 \frac{ \slash \hspace*{-2mm} p - \slash \hspace*{-2mm} g_2 +m_Q}
{-p . g_2 } \frac { -\bar \slash \hspace*{-2mm} p + \slash \hspace*{-2mm} g_1 
+m_Q }{-g_1. \bar p} \slash \hspace*{-2mm} \epsilon_1 \, v^i (\bar p) + 
\left[ g_1\leftrightarrow g_2 ,\epsilon_1 \leftrightarrow \epsilon_2 \right] \\
{\cal M}_{gg}^c &=& if^{abc} T_{ij}^c \, \bar u^j(p)\, \frac{\slash 
\hspace*{-2mm} \epsilon_1 \slash \hspace*{-2mm} \epsilon_2  Q^\lambda}{\hat s} 
\bigg[  2 g_1^\nu g_2^{\lambda \mu}\! +\!  (g_2\! - \! g_1)^\lambda g^{\mu
\nu}\! -\!  2g_2^\mu g^{\nu \lambda} \bigg] \frac 
{\slash \hspace*{-2mm} \bar p +
\slash \hspace*{-2mm} k -m_Q }{2k. \bar p
+M_H^2} \, v^i (\bar p) + [p \leftrightarrow \bar p] \hspace*{5mm} \non 
\eeq 
where, again, the product of couplings $g_s^2 (\sqrt 2 m_Q^2 G_\mu)^{1/2}$ has 
been factorized out. The gluon polarization vectors obey the transversality 
condition $\epsilon_i . g_i=0$ and SU(3) gauge invariance can be checked by 
making the substitutions $\epsilon_i \to g_i$; one can also use the gauge 
condition $\epsilon_1 . g_2= \epsilon_2 . g_1$ to simplify the calculation. 
In the amplitude squared, summed over the final quarks color and spin and 
averaged over the gluon color and polarizations,
\beq 
|{\cal M}_{gg}|^2 = \frac{1}{256} \sum_{\rm spin,\, color} | 
{\cal M}_{gg}^a + {\cal M}_{gg}^b + {\cal M}_{gg}^c |^2
\eeq
one has to perform the trace over the $\gamma$ matrices and the sum over the
indices of the QCD Gell--Mann matrices and the $f^{abc}$ structures functions 
\beq
(T_{ik}^a T_{kj}^b)^2 = 24 \ , \ (f^{abc} T_{ij}^c)^2 = 12 \ , \
(T_{ik}^a T_{kj}^b) (f^{abc} T_{ij}^c)^2 = 0
\eeq 
The average over the gluon polarizations, since the gluons are massless, should
be performed in an axial gauge and, for instance, one can 
use\footnote{Alternatively, one can use the simpler Feynman gauge for the 
summation over polarization, $\sum \epsilon_i^\mu \epsilon_i^\nu=-g^{\mu \nu}$,
but two additional diagrams with ghosts replacing the gluons in the initial 
state have to be added.}
\beq
\sum_{\lambda_i =1}^2 \epsilon_i^\mu(g_i, \lambda_i) \epsilon_i^\nu 
(g_i, \lambda_i)= -g^{\mu \nu} +\frac{2}{\hat{s}} ( g_1^\mu g_2^\nu +g_1^\nu 
g_2^\mu) 
\eeq
The obtained expression for the amplitude squared is too long to be reproduced.
One has then to integrate over the phase space to obtain the cross section at
the partonic level for each subprocess $ij \to Q\bar Q H$ [with ${\small ij= q 
\bar q , gg}$]
\beq 
\hat \sigma_{\rm LO}^{ij} = \frac{1}{2 \hat s} \frac{\alpha_s^2 G_\mu m_Q^2}
{\sqrt{2} \pi^3 } \int \frac{ {\rm d} ^3 p } { 2 p_0}
\frac{ {\rm d} ^3 \bar p}{2 \bar p_0} \frac{ {\rm d} ^3 k } {2 k_0}
\delta^4 ( Q- p -\bar p -k) \left[ \sum |{\cal M}_{ij}|^2 \right]
\eeq
These partonic cross sections have then to be folded with the quark and gluon 
luminosities to obtain the full cross section at the hadronic level
\beq
 \sigma_{\rm LO}=  \int \sum_{i,j} \frac{1}{1+\delta_{ij} } \bigg(
{\cal F}_i^p (x_1, \mu_F) {\cal F}_j^p (x_2, \mu_F) \hat \sigma_{\rm 
LO}^{ij} (x_1, x_2, \mu_F) + [ 1 \leftrightarrow 2 ] \bigg) {\rm d}x_1 
{\rm d}x_2
\eeq
where ${\cal F}_i^p$ are the distributions of the parton $i$ in the proton 
defined at the factorization scale $\mu_F$. The factor $1/(1+ \delta_{ij})$ 
accounts for the identical gluons in the initial state. \s

The cross section for associated $t\bar tH$ and $b\bar bH$ production are 
shown for the Tevatron and LHC energies in Fig.~3.30. 
The MRST set of parton densities has been used and the renormalization and
factorization scales have been identified with $\mu_R=\mu_F=m_Q+\frac{1}{2}M_H$
and $\frac{1}{4}M_H$ for the $t\bar t H$ and $b \bar b H$ cases, respectively.
The pole masses of the top and bottom quarks have been fixed to $m_t=178$ 
GeV and $m_b=4.88$ GeV. As can be seen,  the $p \bar p \to t \bar t H$ cross 
section at the Tevatron is of the order of 5 fb for small Higgs masses,
$M_H \sim 120$ GeV, dropping to a level below 1 fb for $M_H \sim 180$ GeV, 
when the phase space becomes too narrow. At the LHC, the cross section is two 
orders of magnitude larger as a consequence of the higher energy, higher gluon 
luminosity and larger phase space. It varies however strongly with $M_H$,   
dropping by more than one order of magnitude when $M_H$ varies from 120 to 250 
GeV. The detection aspects at LHC \cite{pp-gHtt-pheno,pp-Htt-gamma,pp-Htt-gamma1,pp-Htt-bb,pp-Htt-VV,pp-Htt-WW,pp-Htt-tau} and Tevatron \cite{pp-Htt-phenoT} 
will be discussed in \S3.7. \s

Surprisingly, the $p p \to b \bar b H$ cross sections are slightly
larger than the corresponding cross sections for $p p \to t \bar t H$
at both the Tevatron and the LHC, for small enough Higgs masses. 
This is mere consequence of the larger available phase space at the Tevatron 
and the participation of the low $x$ gluons in the case of the LHC. The cross 
sections become comparable at moderate Higgs masses and, eventually, the $p 
p \to t \bar t H$ process dominates at higher Higgs masses, but the
production rates become then  too small.  

\begin{figure}[!h]
\begin{center}
\vspace*{-2.8cm}
\hspace*{-2.5cm}
\psfig{figure=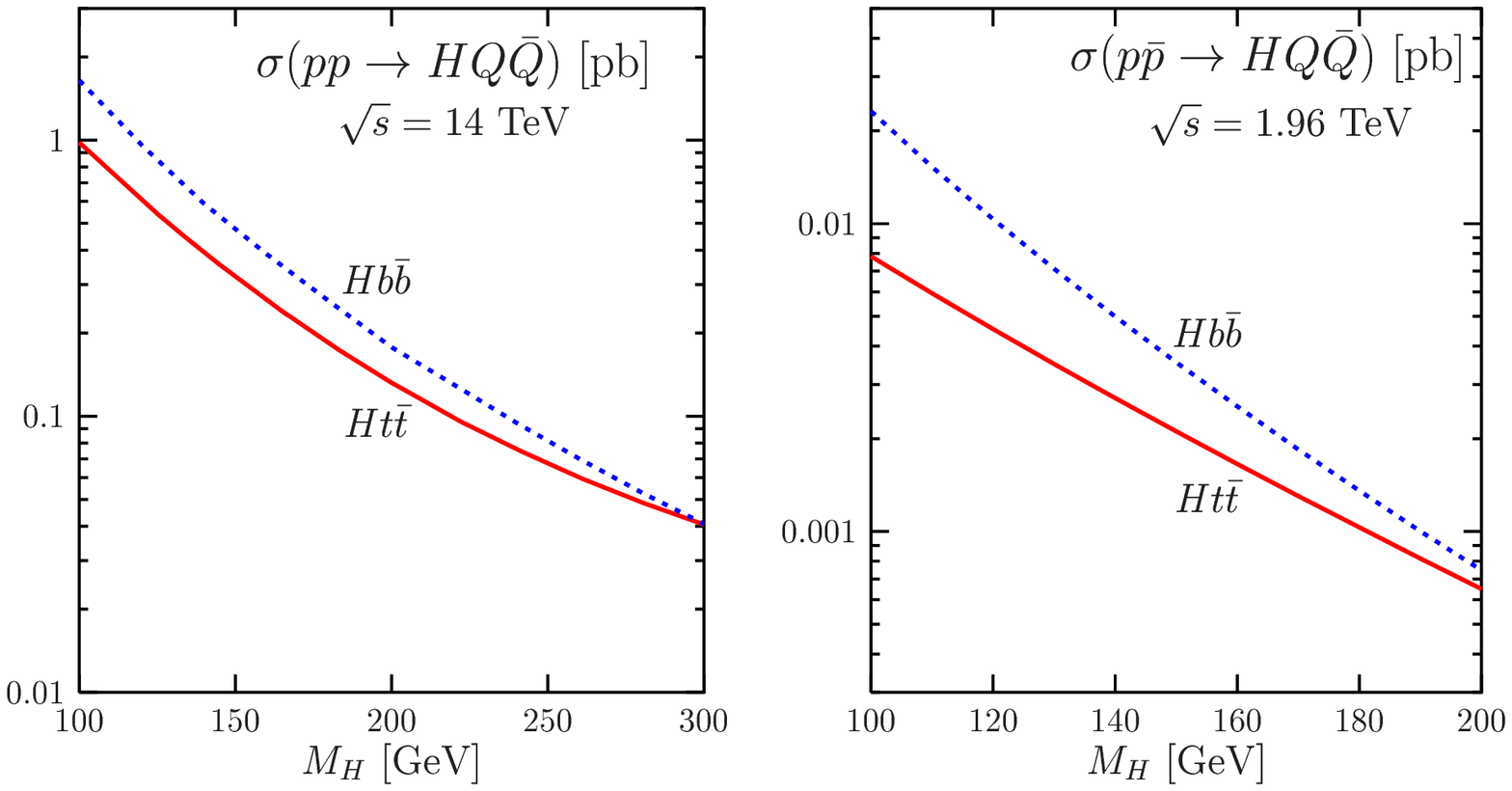,width=18cm}
\end{center}
\vspace*{-14.7cm}
{\it Figure 3.30: The $t\bar t H$ and $b\bar b H$ production cross sections 
at the LHC (left) and the Tevatron (right). The pole quark masses in the Yukawa
couplings are set to $m_t=178$ GeV and $m_b=4.88$ GeV and the MRST PDFs are 
used. The renormalization and factorization scales have been set to $\mu_{R,F}=
 m_t+\frac{1}{2} M_H$ for $pp \to t\bar tH$ and $\mu_{R,F}=  \frac{1}{2} m_b+  
\frac{1}{4}M_H$ for $pp \to b\bar b H$.} 
\vspace*{-9mm} 
\end{figure}

\subsubsection{The ttH cross section at NLO}

\subsubsection*{\underline{The calculation at NLO}}

As we have seen just before, it was already rather difficult to calculate the
$pp \to Q \bar Q H$ cross section at LO. At NLO, the task becomes formidable. 
The computation of these NLO corrections, which  was another {\it tour de
force}, has been been completed only very recently, by two independent groups
\cite{Htt-NLO-DESY,Htt-NLO-US}. The Feynman diagrams which
contribute to the NLO QCD corrections can be divided into four gauge invariant
subsets, some representative examples of which are presented in Fig.~3.31: $a)$
virtual gluonic corrections to the $q \bar q g$ and $ggg$ vertices as well as
to the initial and final quark and gluon self--energies, $b)$ vertex and box
virtual corrections where the Higgs boson is emitted from the internal lines
and where a final state gluon is emitted and splits into $Q\bar Q$ pairs, $c)$
pentagonal $q\bar q$ and $gg$ initiated diagrams where the Higgs boson is
emitted from the heavy quark internal lines, and finally $d)$ real corrections
where an additional gluon is emitted in the final state in all possible ways,
and which involve additional $qg$ and $\bar q g$ scattering diagrams which do
not occur at the tree--level.\s

\begin{figure}[!h]
\SetScale{0.5}
\SetWidth{1.1}
{\unitlength 0.5pt 
\hspace*{2cm}
\begin{picture}(220,120)(-80,-10)
\ArrowLine(20, 80)(50,50)
\ArrowLine(50, 50)(20,20)
\ArrowArc(100, 50)(20,0,180)
\ArrowArc(100, 50)(20,180,0)
\Gluon(50,50)(80,50){4}{3}
\Gluon(120,50)(150,50){4}{3}
\Vertex(50,50){3}
\Vertex(80,50){3}
\Vertex(120,50){3}
\Vertex(150,50){3}
\Vertex(175,25){3}
\DashLine(175, 25)(200,25){5}
\ArrowLine(150,50)(200,100)
\ArrowLine(175,25)(150, 50)
\ArrowLine(200, 0)(175, 25)
\put( 3,80){$q$}
\put( 3,20){$\bar q$}
\put(208, 95){$Q$}
\put(208,20){\bH}
\put(208,-3){$\bar{Q}$}
\put(-70, 85){\red{\bf a)}}
\end{picture} 
\hspace*{2em}
\begin{picture}(220,120)(-80,-10)
\Gluon(20,  0)(50,20){4}{4}
\Gluon(20,100)(50,80){4}{4}
\ArrowLine(50,80)(50,20)
\ArrowLine(50,20)(100,50)
\ArrowLine(100,50)(50,80)
\Gluon(100,50)(140,50){4}{4}
\Vertex(50,20){3}
\Vertex(50,80){3}
\Vertex(100,50){3}
\Vertex(140,50){3}
\Vertex(165,25){3}
\DashLine(165, 25)(190,25){5}
\ArrowLine(140,50)(190,100)
\ArrowLine(165,25)(140, 50)
\ArrowLine(190, 0)(165, 25)
\put( 3,100){$g$}
\put( 3,  0){$g$}
\put(198, 95){$Q$}
\put(198,20){\bH}
\put(198,-3){$\bar{Q}$}
\end{picture} 
}
\\[.5em]
\noindent
{\unitlength 0.5pt 
\hspace*{2cm}
\begin{picture}(220,120)(-80,-10)
\ArrowLine(20, 80)(50,50)
\ArrowLine(50, 50)(20,20)
\Gluon(50,50)(80,50){4}{3}
\ArrowLine(80,50)(130,20)
\ArrowLine(130,20)(130,80)
\ArrowLine(130,80)(80,50)
\Gluon(130,80)(160,80){4}{3}
\Vertex(50,50){3}
\Vertex(80,50){3}
\Vertex(130,20){3}
\Vertex(130,80){3}
\Vertex(160,80){3}
\DashLine(130, 20)(200,10){5}
\ArrowLine(160,80)(200,100)
\ArrowLine(200,60)(160, 80)
\put( 3,80){$q$}
\put( 3,20){$\bar q$}
\put(208, 95){$Q$}
\put(208,57){$\bar{Q}$}
\put(208, 0){\bH}
\put(-70, 85){\red{\bf b)}}
\end{picture} 
\hspace*{2em}
\begin{picture}(220,120)(-80,-10)
\Gluon(20,  0)(50,20){4}{4}
\Gluon(20,100)(50,80){4}{4}
\ArrowLine(50,80)(50,20)
\ArrowLine(50,20)(100,20)
\ArrowLine(100,20)(100,80)
\ArrowLine(100,80)(50,80)
\Gluon(100,80)(140,80){4}{4}
\Vertex(50,20){3}
\Vertex(50,80){3}
\Vertex(100,20){3}
\Vertex(100,80){3}
\Vertex(140,80){3}
\DashLine(100, 20)(190,10){5}
\ArrowLine(140,80)(190,100)
\ArrowLine(190,60)(140, 80)
\put( 3,100){$g$}
\put( 3,  0){$g$}
\put(198, 95){$Q$}
\put(198,57){$\bar{Q}$}
\put(198, 0){\bH}
\end{picture} 
}
\\[.1em]
\noindent
{\unitlength 0.5pt 
\hspace*{.9cm}
\begin{picture}(170,120)(-50,0)
\ArrowLine(50, 95)( 90, 95)
\ArrowLine(90, 95)( 90,  5)
\ArrowLine( 90,  5)(50,  5)
\Gluon(90, 95)(140, 95){4}{5}
\Gluon(90,  5)(140,  5){4}{5}
\Vertex( 90, 95){3}
\Vertex( 90,  5){3}
\Vertex(140,  5){3}
\Vertex(140, 50){3}
\Vertex(140, 95){3}
\DashLine(140,50)(170,50){5}
\ArrowLine(140, 95)(170, 95)
\ArrowLine(140, 50)(140, 95)
\ArrowLine(140,  5)(140, 50)
\ArrowLine(170,  5)(140,  5)
\put(178, 90){$Q$}
\put(178,44){\bH}
\put(178, 2){$\bar{Q}$}
\put(32, 95){$q$}
\put(32,  2){$\bar q$}
\put(-40, 85){\red{\bf c)}}
\end{picture}
\hspace*{.5em}
\begin{picture}(170,120)(-50,0)
\Gluon(50, 95)( 90, 95){4}{4}
\Gluon(50,  5)( 90,  5){4}{4}
\ArrowLine(170,  5)(140,  5)
\ArrowLine(140,  5)(90,  5)
\ArrowLine( 90,  5)(90, 50)
\ArrowLine( 90, 50)(90, 95)
\ArrowLine(90, 95)(140, 95)
\ArrowLine(140, 95)(170, 95)
\Vertex( 90, 95){3}
\Vertex( 90,  5){3}
\Vertex(140,  5){3}
\Vertex( 90, 50){3}
\Vertex(140, 95){3}
\DashLine(90,50)(170,50){5}
\Gluon(140,  5)(140, 95){4}{9}
\put(178, 90){$Q$}
\put(178,44){\bH}
\put(178, 2){$\bar{Q}$}
\put(32, 95){$g$}
\put(32,  2){$g$}
\end{picture}
\hspace*{.5em}
\begin{picture}(170,120)(-50,0)
\Gluon(50, 95)( 90, 95){4}{4}
\Gluon(50,  5)( 90,  5){4}{4}
\Gluon(90, 95)(140, 95){4}{5}
\Gluon(90,  5)(140,  5){4}{5}
\Gluon(90, 95)( 90,  5){4}{9}
\Vertex( 90, 95){3}
\Vertex( 90,  5){3}
\Vertex(140,  5){3}
\Vertex(140, 50){3}
\Vertex(140, 95){3}
\DashLine(140,50)(170,50){5}
\ArrowLine(140, 95)(170, 95)
\ArrowLine(140, 50)(140, 95)
\ArrowLine(140,  5)(140, 50)
\ArrowLine(170,  5)(140,  5)
\put(178, 90){$Q$}
\put(178,44){\bH}
\put(178, 2){$\bar{Q}$}
\put(32, 95){$g$}
\put(32,  2){$g$}
\end{picture}
\\[-0.2em]
\hspace*{.3cm}
\begin{picture}(170,120)(-50,30)
\vspace{1.5cm}
\ArrowLine(20, 80)(50,50)
\ArrowLine(50, 50)(20,20)
\Gluon(50,50)(100,50){4}{5}
\Vertex(50,50){3}
\Vertex(100,50){3}
\Vertex(125,25){3}
\Vertex(125,75){3}
\DashLine(125, 25)(150,25){5}
\ArrowLine(125,75)(150, 75)
\ArrowLine(100,50)(125, 75)
\ArrowLine(125,25)(100, 50)
\ArrowLine(150, 0)(125, 25)
\Gluon(125,75)(150,100){4}{3}
\put(158,100){$g$}
\put( 8,80){$q$}
\put( 8,20){$\bar q$}
\put(158, 70){$Q$}
\put(158,20){\bH}
\put(158,-10){$\bar{Q}$}
\end{picture}
\put(-160, 50){\red{\bf d)}}
\hspace*{1em}
\begin{picture}(170,120)(-50,30)
\Gluon(50,100)(100,100){4}{5}
\Gluon(100,100)(150,100){4}{5}
\Gluon(100,100)(100,70){4}{3}
\Gluon(50,  0)(100,  0){4}{5}
\Vertex(100,100){3}
\Vertex(100, 70){3}
\Vertex(100,  0){3}
\Vertex(100, 35){3}
\DashLine(100, 35)(150,35){5}
\ArrowLine(100, 70)(150, 70)
\ArrowLine(100, 35)(100, 70)
\ArrowLine(100,  0)(100, 35)
\ArrowLine(150,  0)(100,  0)
\put(158,100){$g$}
\put(158, 65){$Q$}
\put(158,30){\bH}
\put(158,-5){$\bar{Q}$}
\put(35,100){$g$}
\put(35,  0){$g$}
\end{picture}
\hspace*{1em}
\begin{picture}(170,120)(-50,30)
\ArrowLine(50,100)(100,100)
\ArrowLine(100,100)(150,100)
\Gluon(100,100)(100,70){4}{3}
\Gluon(50, 25)(100, 25){4}{5}
\Vertex(100,100){3}
\Vertex(100, 70){3}
\Vertex(100, 25){3}
\Vertex(125, 25){3}
\DashLine(125, 25)(150, 25){5}
\ArrowLine(100, 70)(150, 70)
\ArrowLine(100, 25)(100, 70)
\ArrowLine(125, 25)(100, 25)
\ArrowLine(150, 0)(125, 25)
\put(158, 95){$q$}
\put(158, 65){$Q$}
\put(158,20){\bH}
\put(158,-10){$\bar{Q}$}
\put(35, 95){$q$}
\put(35, 25){$g$}
\end{picture}
\hspace*{1em}
\begin{picture}(170,120)(-50,30)
\ArrowLine(100,100)(50,100)
\ArrowLine(150,100)(100,100)
\Gluon(100,100)(100,70){4}{3}
\Gluon(50,  0)(100,  0){4}{5}
\Vertex(100,100){3}
\Vertex(100, 70){3}
\Vertex(100,  0){3}
\Vertex(100, 35){3}
\DashLine(100, 35)(150,35){5}
\ArrowLine(100, 70)(150, 70)
\ArrowLine(100, 35)(100, 70)
\ArrowLine(100,  0)(100, 35)
\ArrowLine(150,  0)(100,  0)
\put(158, 65){$Q$}
\put(158,30){\bH}
\put(158,-5){$\bar{Q}$}
\put(35,  0){$g$}
\put(158, 95){$\bar q$}
\put(35, 95){$\bar q$}
\end{picture}
}
\vspace*{1cm}

{\it Figure 3.31: Representative Feynman diagrams for the NLO QCD corrections
to $pp \to Q\bar Q H$.}
\label{fig:NLOdiags}
\vspace*{-5mm}
\end{figure}

Technically speaking, there are two main difficulties which arise when
attempting to perform such a calculation. The first one is that the pentagonal
one--loop 5--point functions \cite{5point-function} are rather difficult to
evaluate in $n$ dimensions since, not to mention the complicated tensorial
structure which has to be reduced to known scalar integrals, they involve
massive particles and have ultraviolet, soft and collinear singularities which
have to be calculated in the dimensional regularization scheme. The second
major difficulty is the extraction of the soft and collinear singularities in
the real part of the corrections, which involve 4 particles in the final state,
with three of them being massive. This leads to severe numerical instabilities
which have to be handled with great care \cite{Diople-Formalism}. Added to
this, a large number of Feynman diagrams and the associated counterterms have
to be evaluated. \s

Several methods have been devised to overcome these problems and a detailed
account can be found in the original papers of
Refs.~\cite{Htt-NLO-DESY,Htt-NLO-US}.  In both calculations, the
renormalization has been performed in a mixed  scheme, where the heavy quark
mass is defined on--shell and the running of the $\overline{\rm MS}$ strong
coupling constant includes only the light quarks and gluons with the heavy
quarks decoupled. The factorization has been performed in the $\overline{\rm
MS}$ scheme where the heavy quark is decoupled from the evolution of the parton
densities.  The two calculations have been compared against each other and,
although the methods which have been used are different, a perfect agreement
has been found.  In the following, we simply summarize the numerical results
which have been obtained in the case of $pp \to t\bar t H$.

\subsubsection*{\underline{Numerical results for $pp \to t\bar t H$}}

In Fig.~3.32, the LO and NLO cross sections for associated Higgs production
with top quarks are presented in the case of the Tevatron with a c.m. 
energy $\sqrt{s}=2$ TeV. The CTEQ4M\,(L) parton distribution functions at 
NLO (LO) have been used and the top quark mass and the strong
coupling constants have been fixed to $m_t=174$ GeV and $\alpha_s^{\rm
NLO}(M_Z) =0.116$. The left--hand side of the figure displays the LO and NLO
cross sections as a function of the Higgs mass with the renormalization and
factorization scales chosen to be $\mu_R=\mu_F=m_t$ and $2m_t$, while the 
right--hand side displays the variation of the cross sections with the 
renormalization and factorization scales for a Higgs boson with a mass 
$M_H=120$ GeV.\s 

One can see that over the entire Higgs mass range accessible at the Tevatron,
the NLO corrections decrease the production rate.  For $\mu_F=\mu_R=m_t$, the
$K$--factor defined as the ratio of the NLO and LO cross sections consistently
evaluated at their respective orders, is $K \sim 0.8$ and is nearly independent
of the Higgs mass\footnote{This is one of the few examples of $K$--factors
below unity. As discussed in Refs.\cite{Htt-NLO-DESY,Htt-NLO-Tev}, the reason 
lies in the fact
that at $\sqrt{s}=2$ TeV, the $t\bar tH$ final state is produced in the
threshold region, where gluon exchange between the top quarks gives rise to
Coulomb singularities $\propto \alpha_s /\beta_t$. Since the $t\bar tH$ final
state is in a color octet, these corrections are negative and decrease the Born
cross section.}.  However, for a scale choice $\mu_R=\mu_F=2m_t$, the NLO
corrections are very small.  This suggests a very strong dependence of the
cross section on the scale choice and, indeed, as is illustrated in the
right--hand side of Fig.~3.32 where the scales are varied from $m_t$ to $\sim
3m_t$, while the NLO cross section is rather stable, the LO cross section
changes by 60\% in this range. Thus, the NLO corrections are 
mandatory to stabilize the prediction for the production cross section.  

\begin{figure}[!h]
\begin{center}
\hspace*{-.7cm}
\mbox{
\epsfxsize=8.cm \epsfbox{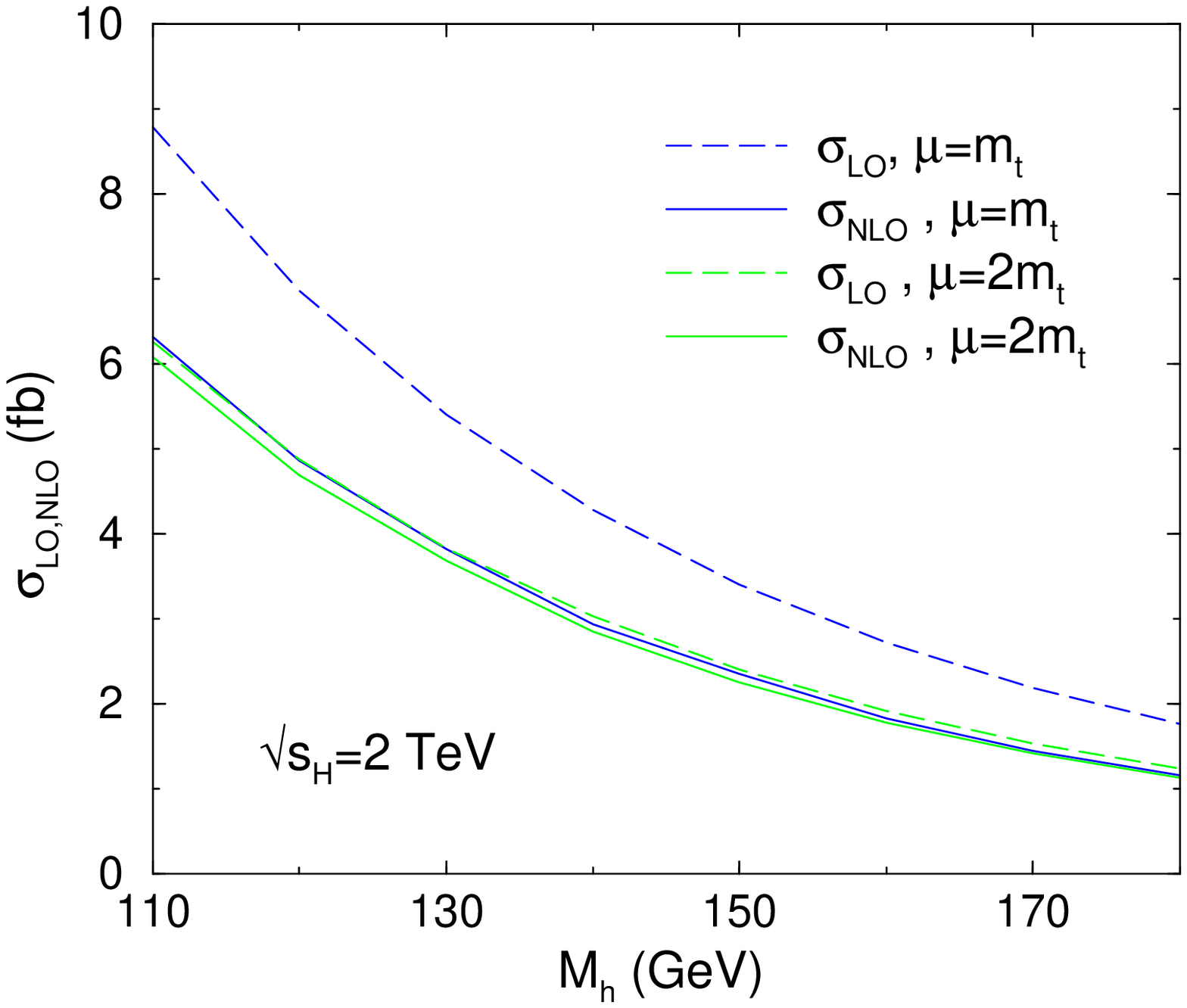} \hspace*{2mm} 
\epsfxsize=8.cm \epsfbox{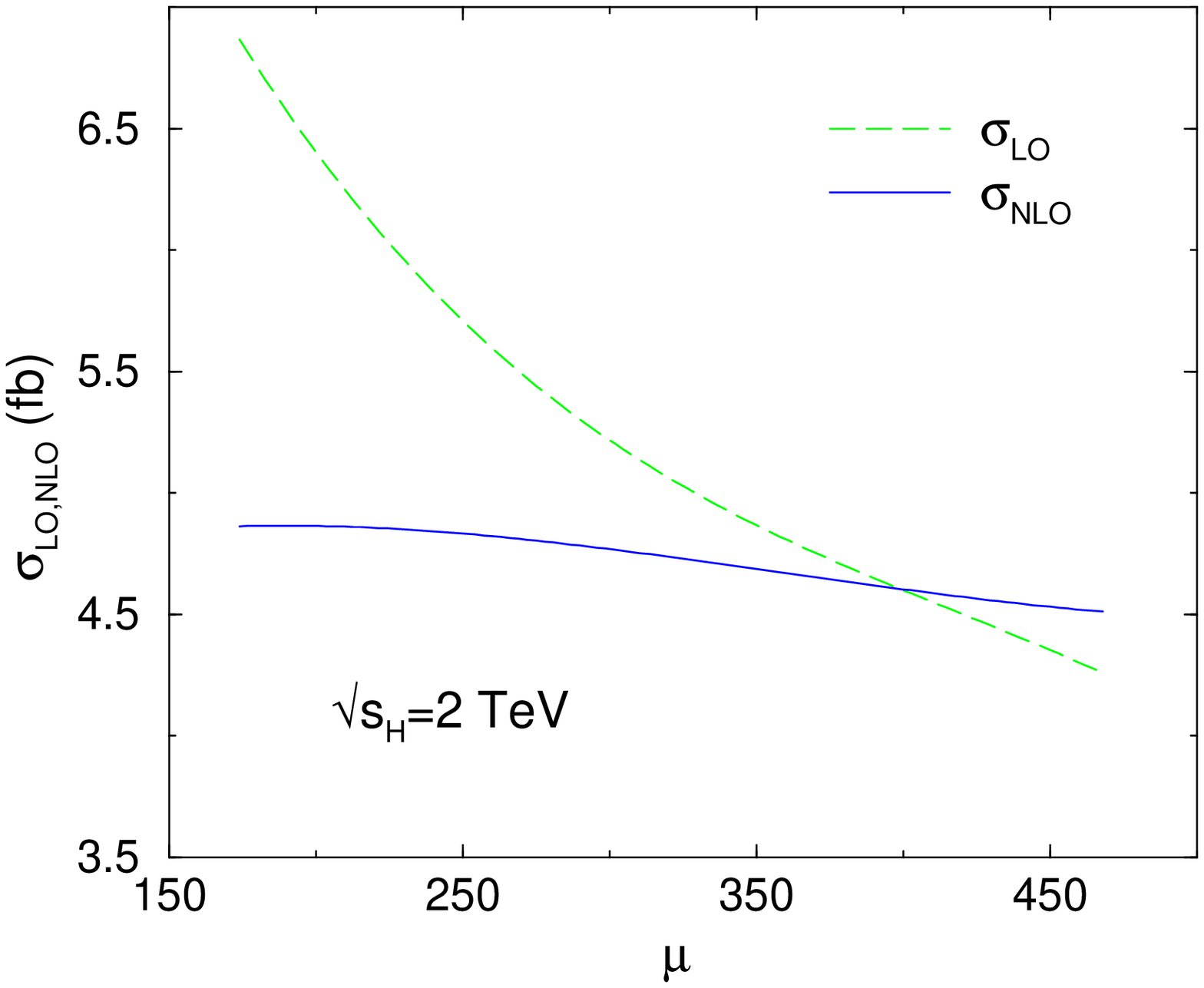} }
\end{center}
\vspace*{-.5cm}
{\it Figure 3.32: The $p\bar p \to t \bar tH+X$ production cross section at the 
Tevatron at both LO and NLO, as a function of the Higgs mass with two scale 
choices (left) and as a function of the scales $\mu= \mu_R=\mu_F$ for 
$M_H=120$ GeV (right); from Ref.~\cite{Htt-NLO-Tev}.}
\vspace*{-.4cm}
\end{figure}

In the subsequent figures, Figs.~3.33--3.34, we present the LO and NLO results
for the associated Higgs production with top quarks at the LHC, $pp\to t\bar t
H+X$, as derived in Ref.~\cite{Htt-NLO-DESY}. Besides the total hadronic cross
sections, the differential distributions in transverse momentum and rapidity of
the Higgs boson have been discussed in this case. Here, the MRST sets of parton
densities at LO and NLO have been adopted and the renormalization and
factorization scales are set to $\mu_R=\mu_F= \mu_0= \frac{1}{2} (2m_t+M_H)$
when they are not varied; the top--quark mass is also set to the old central
value $m_t=174$ GeV.  

\begin{figure}[!h]
\begin{center}
\hspace*{-5mm}
\mbox{
\epsfig{file=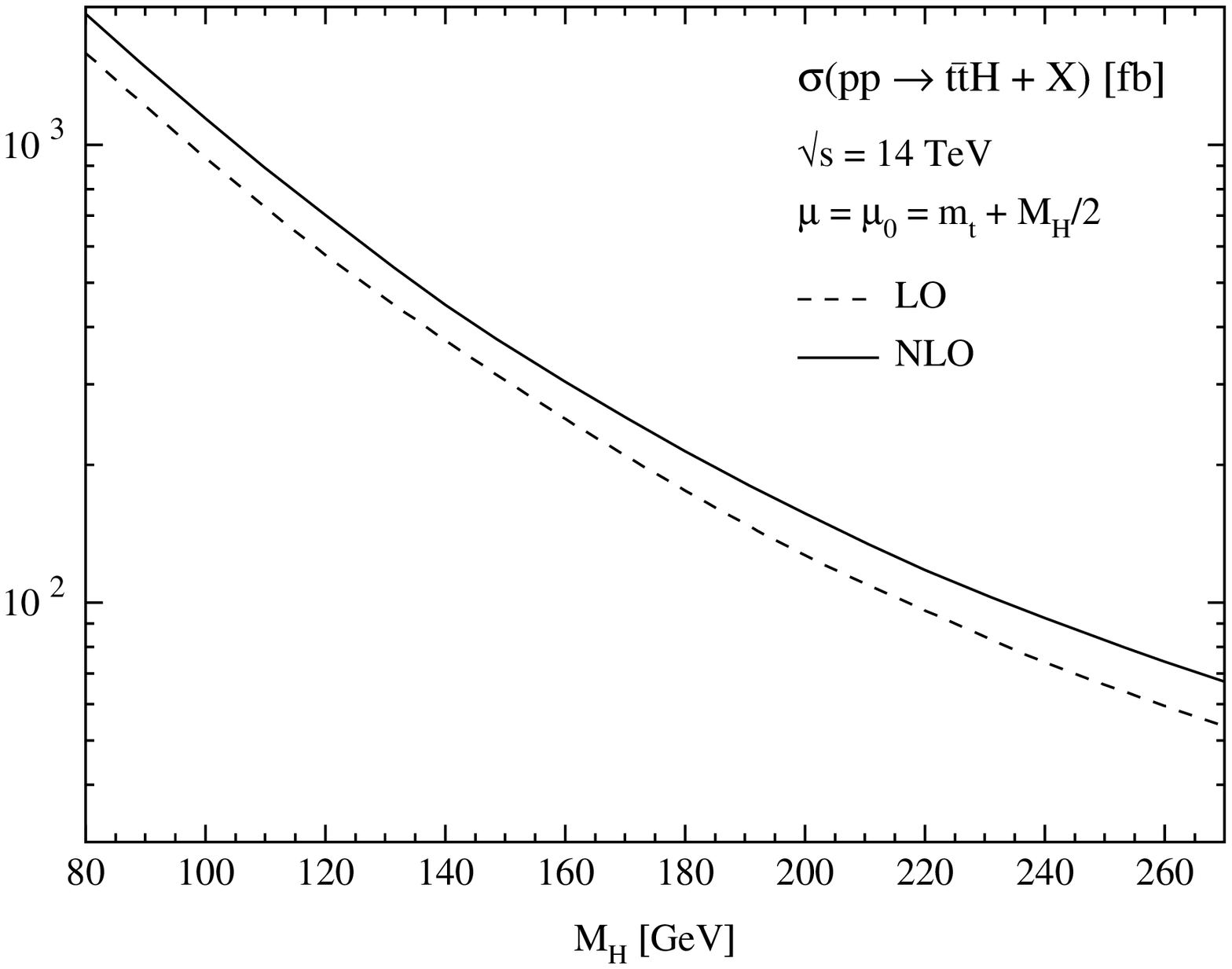,%
        bbllx=30pt,bblly=230pt,bburx=580pt,bbury=625pt,scale=0.44}\hspace*{-2mm}
\epsfig{file=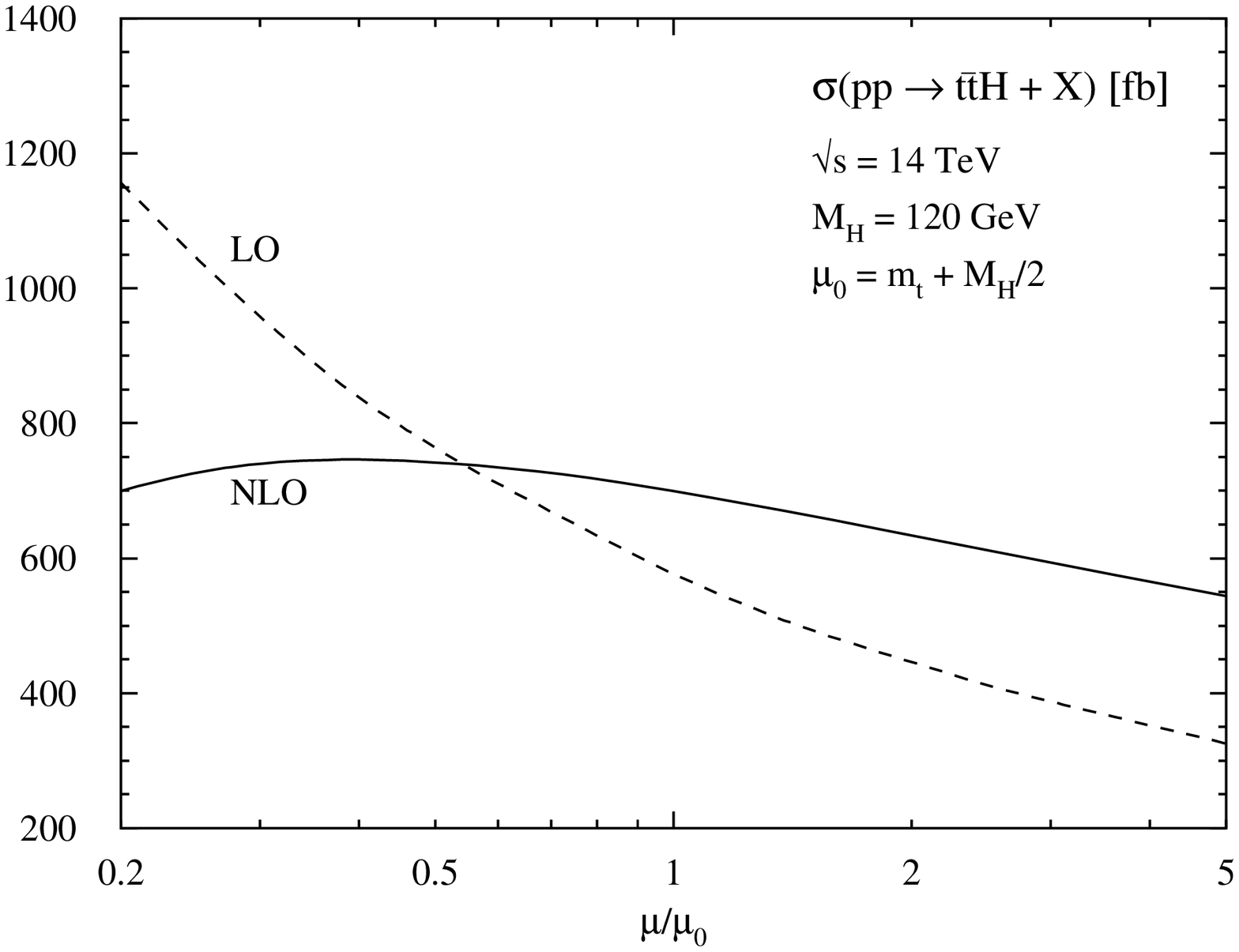,%
        bbllx=30pt,bblly=230pt,bburx=580pt,bbury=625pt,scale=0.44} }
\end{center}
\vspace*{-.2cm}
{\it Figure 3.33: Total cross section for $p p\to t\bar{t}H + X$
at the LHC in LO and NLO as a function of $M_H$ (left) and the variation 
with the scales for $M_H=120$ GeV (right); from Ref.~\cite{Htt-NLO-DESY}.}
\vspace*{-.3cm}
\end{figure}

Because the cross section at the LHC is dominated by the $gg$ fusion process,
it receives positive NLO corrections for the central renormalization and
factorization scale, $\mu_0=m_t+ \frac{1}{2}M_H$, as shown in the left--hand
side of Fig.~3.33. For this scale value, a factor $K \sim 1.2$ is obtained,
increasing to $K \sim 1.4$ when the choice $\mu_F=\mu_R=2\mu_0$ is made. As in
the case of the Tevatron, these values are nearly independent of the Higgs
boson mass in the displayed range. Again, and as is shown in the right--hand
side of the figure, the NLO corrections significantly reduce the
renormalization and factorization scale dependence and stabilize the
theoretical prediction for the cross section at the LHC. \s

The scale dependence of the rapidity and transverse momentum distributions of 
the Higgs boson is also significantly reduced at NLO and the shape of the 
distributions is practically constant when the scales are varied in a 
reasonable range. The ratio of the normalized NLO and LO distributions in
transverse momentum and rapidity, are shown in the inserts of, respectively, 
the left--hand and right--hand parts of Fig.~3.34 for $M_H=120$ GeV. In the 
former case, the default scale was set to the transverse mass, $\mu^2=p_{T,H}^2 
+ M_H^2$, which is a more natural choice for large transverse momenta. 
In this case, the NLO corrections are small for low values of the  Higgs 
transverse momentum, $p_{T,H} \lsim m_t$, but increase with increasing
$p_{T,H}$ values, reaching $\sim 30\%$ at the boundary of phase space where
the cross section is small. In the case of the normalized rapidity 
distribution, the NLO corrections are also very small in the central region
but they become negative and of the order of 10\% at the edge of phase space. 
A conclusion that one can draw from these figures, is that one cannot simply 
use a constant $K$--factor to describe these distributions. \s

Note that the transverse momentum and rapidity distributions of the
top and antitop quarks have been also studied; they are barely affected by the
NLO corrections once the scales have been properly chosen; see  
Ref.~\cite{Htt-NLO-DESY}. \s

\begin{figure}[!h] 
\begin{center}
\hspace*{-5mm}
\mbox{
\epsfig{file=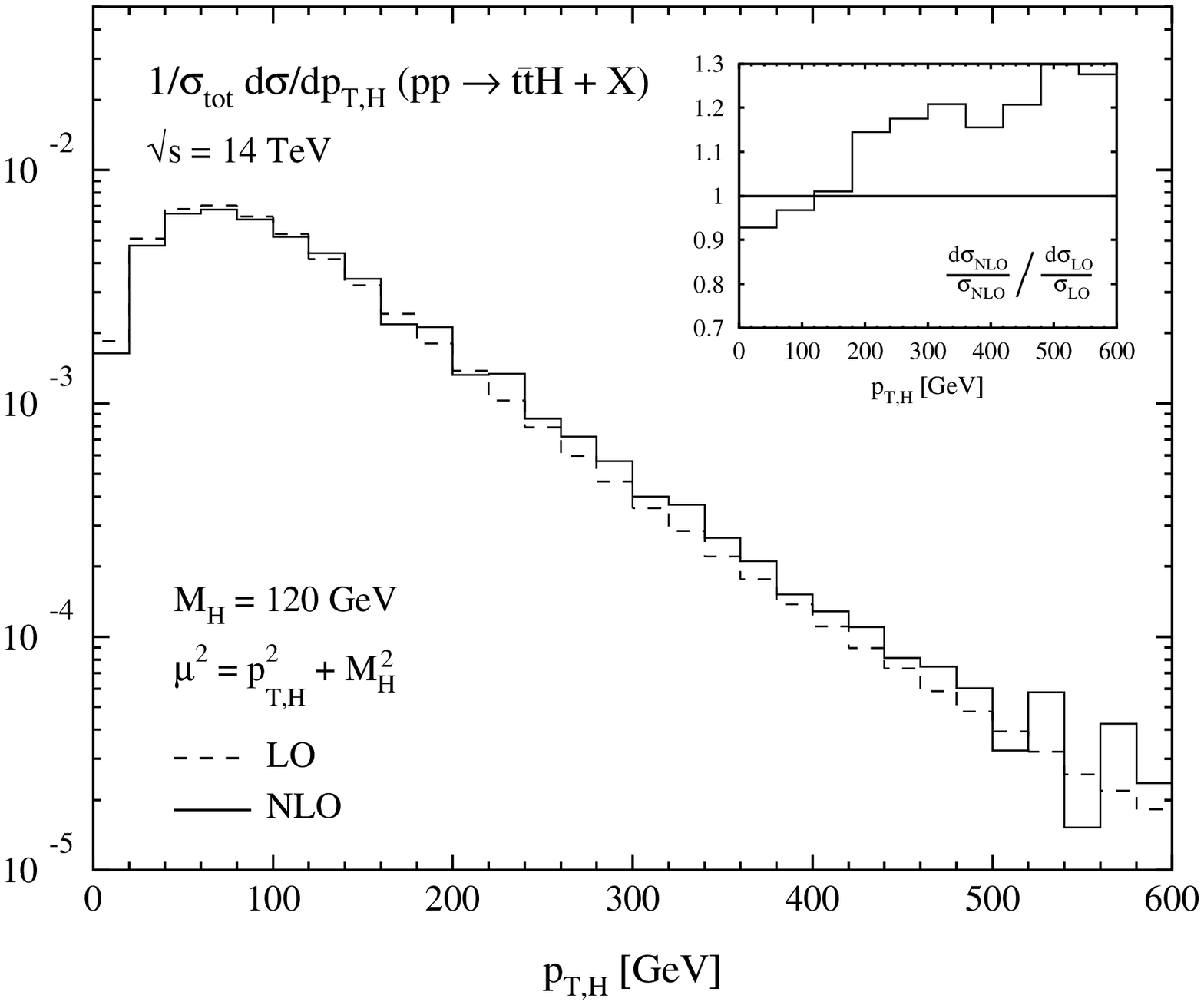,%
        bbllx=30pt,bblly=200pt,bburx=580pt,bbury=635pt,scale=0.44}\hspace*{-2mm}
\epsfig{file=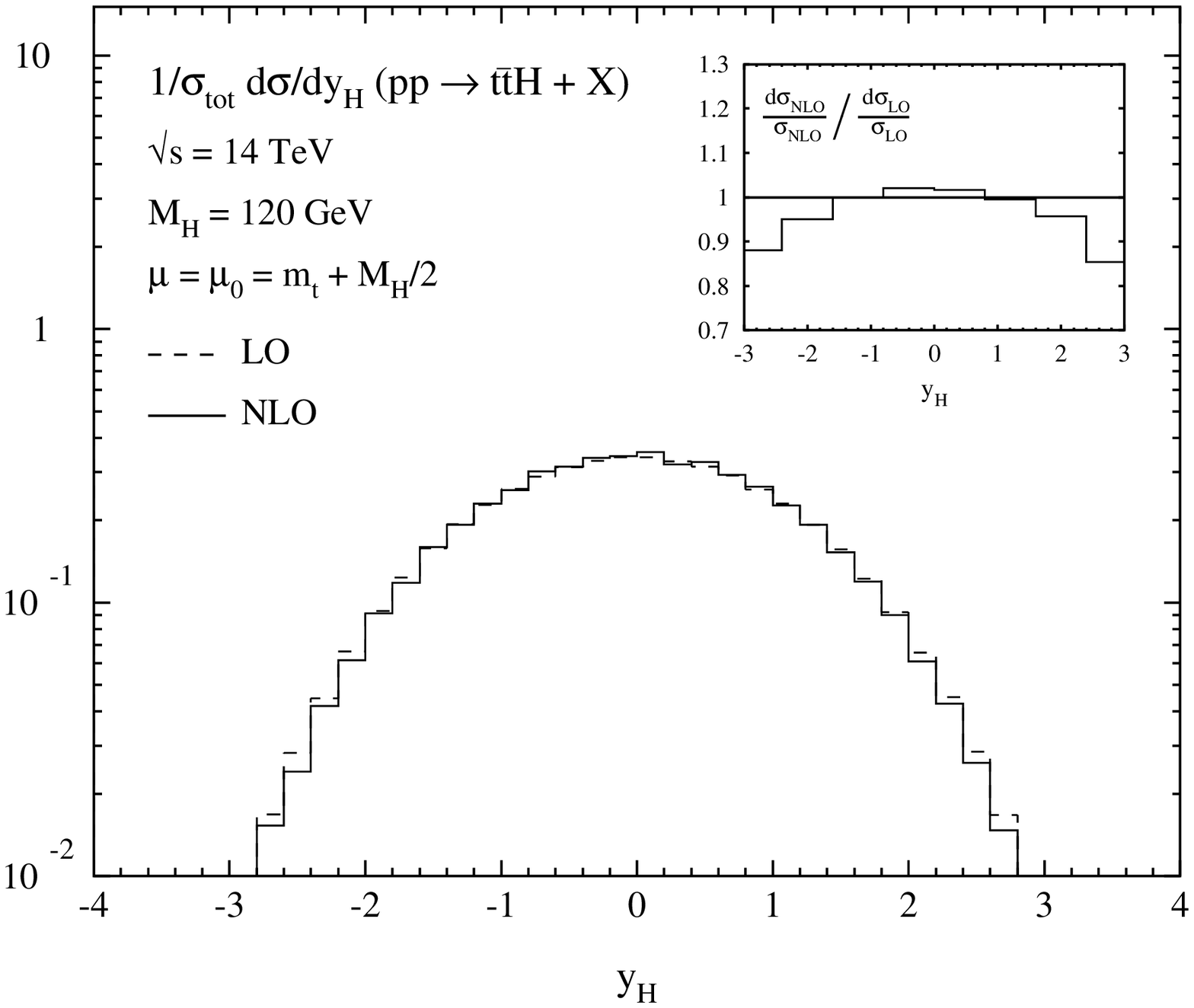,%
        bbllx=30pt,bblly=200pt,bburx=580pt,bbury=635pt,scale=0.44} }
\end{center}
\vspace*{-.5cm}
{\it Figure 3.34: Normalized transverse momentum (left) and rapidity (right)
distribution of the Higgs boson in the process $p p\to t\bar{t}H + X$ at the 
LHC in LO and NLO with $M_H\! =\! 120$ GeV. The inserts to the figures show the
ratio of the NLO to LO distributions; from Ref.~\cite{Htt-NLO-DESY}.}
\label{fig:ptH-lhc}
\end{figure}

\subsubsection*{\underline{The PDF uncertainties}}

Finally, let us discuss the PDF uncertainties in the prediction of the $pp \to 
H t\bar t$ cross section, restricting ourselves to the case of the LHC.
The  central values and the uncertainty band limits are shown for the CTEQ, 
MRST and Alekhin parameterizations in Fig.~3.35 as a function of $M_H$, using 
the procedure outlined in \S3.1.5. We also show in the insert, the spread 
uncertainties in the predictions when the cross sections are normalized to the 
values obtained using the CTEQ6M set. Since the NLO corrections have been 
calculated only recently and the presumably very complicated and slow programs 
are not yet publicly available, we simply use the program {\tt HQQ} of
Ref.~\cite{Michael-Web} for the LO cross section with scales $\mu_F=\mu_R= 
2m_t+M_H$
that we fold with the NLO PDFs. Although the overall normalization is different 
when including the NLO correction [one simply has to multiply by the $K$--factor
which is approximately 1.4 in this case], this procedure should describe the 
relative effects of the different PDFs at NLO with a rather good approximation.  

\begin{figure}[h]
\begin{center}
\vspace*{-2.6cm}
\hspace*{-2cm}
\psfig{figure=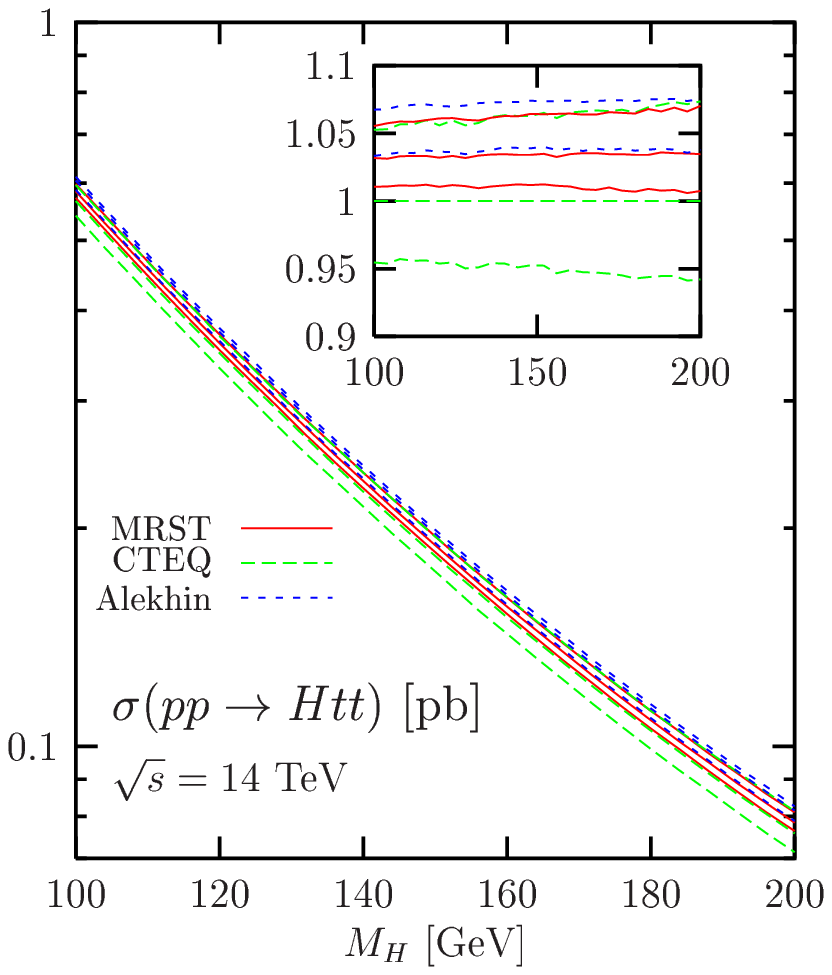,width=18cm}
\end{center}
\vspace*{-15.2cm}
{\it Figure 3.35: The CTEQ, MRST and Alekhin PDF uncertainty bands for the NLO
cross section $pp\to t \bar t H$ at the LHC; the scales have been fixed to
$\mu_F=\mu_R= 2m_t+M_H$. The insert shows the spread in the predictions 
compared to CTE6M; from Ref.~\cite{Samir}.}
\vspace*{-4mm} 
\end{figure}

As discussed above, the process is dominantly generated by gluon--gluon 
fusion at the LHC and, compared with the  process $gg  \ra H$ discussed in 
\S3.4 for a fixed  Higgs mass, a larger $Q^2$ is needed for this final 
state and the initial gluons should therefore have higher $x$ values. In
addition, the quarks that are involved in the subprocess $q\bar{q}\ra
t\bar{t}H$, which is also contributing,  are still in  the intermediate regime
because of the higher value $[x \sim 0.7$] at which the quark high--$x$ regime
starts.  This explains why the uncertainty band increases smoothly from 5\% to
7\% when the $M_H$ value increases from 100 to 200 GeV.

\subsubsection{The case of the bbH process} 

As seen in \S3.5.1, the production cross sections for the associated Higgs
production with bottom quarks are not that small, despite of the tiny $Hb\bar
b$ Yukawa coupling.  However, the dominant $b\bar b b\bar b$ signal final state
for a low mass Higgs boson decaying into $b\bar b$ pairs is rather complicated
to be isolated experimentally and suffers from a huge QCD jet background.  This
channel is therefore not considered as a discovery channel for the SM Higgs
boson at the Tevatron and the LHC.  Nevertheless, in extensions of the SM, such
as in minimal supersymmetric theories, the Higgs Yukawa coupling to bottom
quarks can be strongly enhanced, leading to large $b \bar b H$ production rates
which can exceed by far the cross sections in the $pp\to t\bar t H$ case, even
for high mass Higgs bosons. This channel will be discussed in some detail in
the second part of this review \cite{Tome2}. Here, we simply summarize the
impact of the NLO corrections, restricting ourselves to the inclusive total
rate generated via light quark annihilation and $gg$ fusion, $q\bar q, gg \to
b\bar b H$. \s

The calculation of the NLO correction to $b\bar b H$ production follows the same
lines as what has been discussed previously for $pp \to t\bar t H$ and the
results have been given in  Refs.~\cite{Hbb-NLO1,Hbb-NLO2}. There is, however, 
a major difference between the two cases \cite{pp-Hbb-pheno1}: because of the 
small $b$--quark mass, the cross section $\sigma (gg \to b\bar b H)$ develops 
large logarithms\footnote{The issue of resumming these large logarithms and
stabilizing the scale dependence of the cross section using heavy quark
distribution functions has been discussed in Ref.~\cite{pp-Hbb-NLO3}.},
$\log(Q^2/m_b^2)$, with the scale $Q$ being typically of the order of the
factorization scale $Q \sim M_H \gg m_b$. This leads  to large corrections,
part of which can be absorbed by choosing a low value for the factorization and
renormalization scales, $\mu_R = \mu_F \sim \frac{1}{4} (M_H+ 2m_b)$
\cite{pp-Hbb-pheno1,pp-Hbb-NLO3}. \s 

\begin{figure}[h]
\begin{center}
\mbox{
\includegraphics[bb=50 250 580 600,scale=0.44]{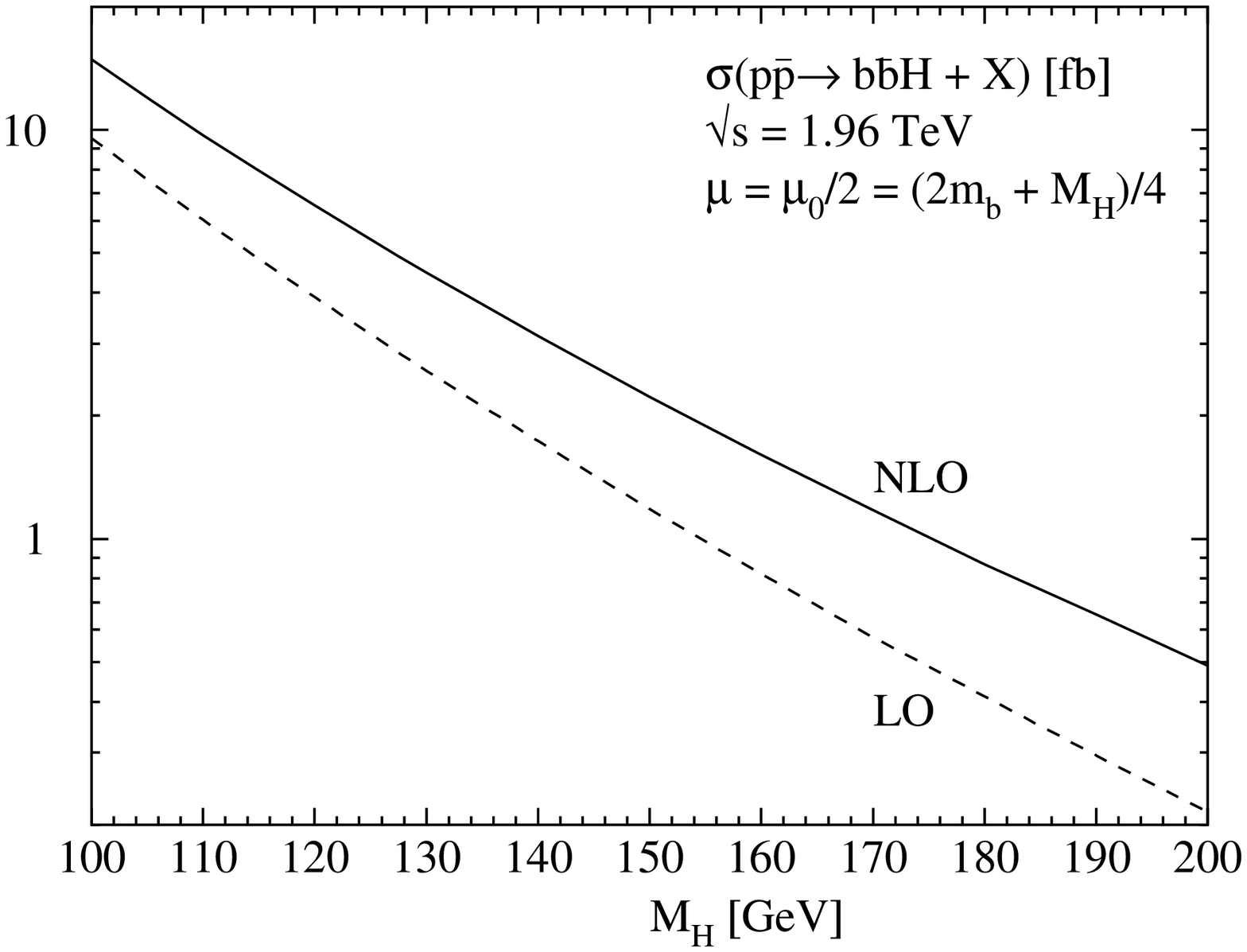}\hspace*{-2mm}
\includegraphics[bb=50 250 580 600,scale=0.44]{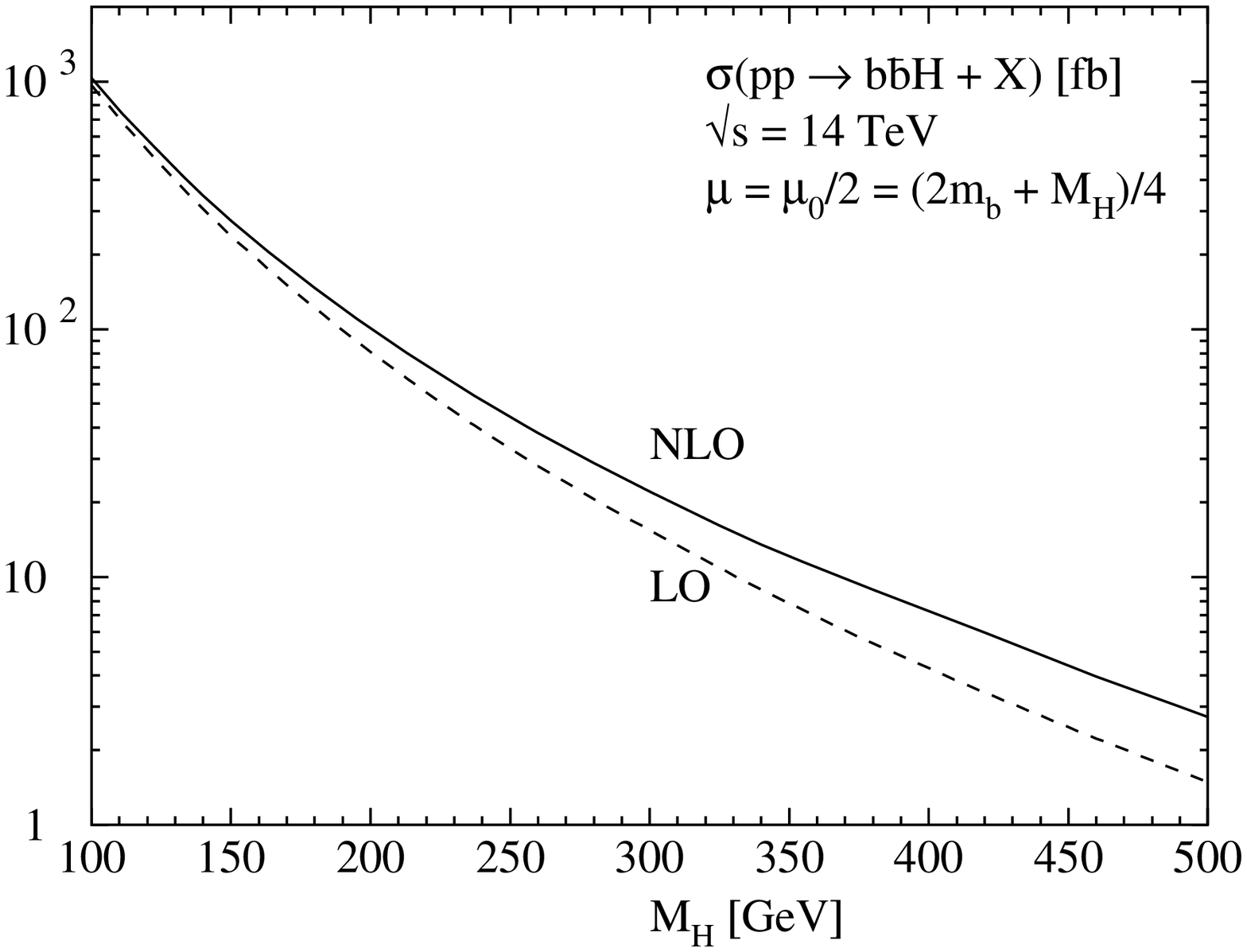} }
\end{center}
{\it Figure 3.36: Total inclusive cross sections for $pp \rightarrow b \bar b 
H+X$ at the Tevatron (left) and the LHC (right) as a function of $M_H$ with the
factorization and renormalization scales set to $\mu_R = \mu_F \sim \frac{1}
{4} (M_H+ 2m_b)$. The running $b$--quark mass, with a starting pole value
$m_b=4.88$ GeV,  has been used in  the Higgs coupling and the CTEQ6 PDFs
are adopted; from \cite{Hbb-NLO1}.}
\vspace*{2mm}
\end{figure}

The NLO cross sections are shown at Tevatron and LHC energies in Fig.~3.36 as a
function of the Higgs mass for this scale choice and compared to the LO cross
sections. In both cases, the running $b$--quark mass at the scale of the Higgs
mass, with the starting pole mass being $m_b =4.9$ GeV, has been used for 
the Yukawa coupling.  As can be seen, even with this scale choice, the
NLO corrections are large, with $K$--factors ranging from 1.6 to 2.6 at the
Tevatron and 1.1 to 1.8 at the LHC. The scale variation is still strong
even at NLO and further work is needed to improve the theoretical prediction of
the $b\bar b H$ production rate.

\subsubsection{Associated Higgs production with a single top quark} 

Since the phase space for $t\bar t H$ production is too penalizing, in
particular at the Tevatron, it has been suggested to consider the process where
the Higgs boson is produced in association with a single top or antitop quark
\cite{ppHt-3oldpapers,ppHt-Scott}
\beq
pp / p\bar p \to t H +X
\eeq
The expectation is that the  cross section can be comparable to that of the
$t\bar tH$ process, similarly to what occurs for top quark production 
in hadronic collisions where the rate for single top quark is not 
much smaller than that for top quark pair production, the ratio of the two 
being of the order of 1/3 \cite{ppSingleTop}. There are three types of 
contributions to this production channel, as shown in Fig.~3.37 where a few 
generic Feynman diagrams are presented: 

\begin{itemize}
\vspace*{-3mm}

\item[$a)$] $q \bar q'$ annihilation with $s$--channel $W$ boson exchange, 
which leads to the three--body final state involving a Higgs boson and a
$b t$ pair;  
\vspace*{-2mm}

\item[$b)$] $t$--channel fusion of a light quark and a bottom parton from the 
proton sea which, through $W$ exchange, leads to the $q tH$ final state; 
\vspace*{-2mm}

\item[$c)$] the scattering of gluons with again bottom 
partons from the proton sea and which lead to $tW H$ final states. 
\vspace*{-2mm}
\end{itemize}

In the language of gluon initiated production, the two last processes are in 
fact the higher--order mechanisms $gg \to bH+X$ with four final state particles
but with one $b$--quark integrated out. Note that in all three channels, the 
Higgs boson can be radiated not only from the top quark lines but also from 
the $W$ boson [as well as from the $b$--quark] lines.\s

\begin{figure}[h!]
\begin{center}
\vspace*{-.8cm}
\hspace*{-4cm}
\SetWidth{1.}
\begin{picture}(300,100)(0,0)
\ArrowLine(0,25)(35,50)
\ArrowLine(0,75)(35,50)
\Photon(35,50)(80,50){3.2}{5.5}
\Line(80,50)(115,25)
\Line(80,50)(115,75)
\DashLine(105,65)(130,47){4}
\Text(-2,35)[]{$\bar{q}'$}
\Text(-2,65)[]{$q$}
\Text(55,65)[]{$W$}
\Text(122,20)[]{$t$}
\Text(122,80)[]{$\bar{b}$}
\Text(120,45)[]{\bH}
\Text(105,65)[]{\bb}
\hspace*{9mm}
\ArrowLine(150,25)(190,25)
\ArrowLine(150,75)(190,75)
\Photon(190,25)(190,75){3.2}{5.5}
\Line(190,75)(240,75)
\ArrowLine(190,25)(240,25)
\DashLine(220,75)(240,50){4}
\Text(203,50)[]{$W$}
\Text(148,65)[]{$b$}
\Text(148,35)[]{$q$}
\Text(240,35)[]{$q'$}
\Text(245,69)[]{$t$}
\Text(223,55)[]{$H$}
\Text(219,74)[]{\bb}
\hspace*{-5mm}
\ArrowLine(290,25)(330,25)
\Gluon(290,75)(330,75){3}{4.5}
\Line(330,25)(330,75)
\DashLine(355,75)(380,50){4}
\Photon(330,25)(380,25){3.2}{5.5}
\Line(330,75)(380,75)
\Text(355,75)[]{\bb}
\hspace*{4.8cm}
\Text(148,65)[]{$g$}
\Text(148,35)[]{$b$}
\Text(240,35)[]{$W$}
\Text(245,69)[]{$t$}
\Text(221,55)[]{$H$}
\end{picture}
\end{center}
\vspace*{-1cm}
{\it Figure 3.37: Generic Feynman diagrams for associated Higgs production 
with a single top quark in hadronic collisions: $a)$ $q \bar{q}' \to \bar{b} t 
H$, $b)$ $q b \to q' t H$ and $c)$  $g b \to W^- t H$.}
\vspace*{-4mm}
\end{figure}

These processes have been revisited in Ref.~\cite{ppHt-Scott} and the
production cross sections are shown in Fig.~3.38 for the Tevatron (left) and
LHC (right) as a function of the Higgs mass. The rates for the three channels
are shown separately and compared with the $pp \to t\bar t H$ cross section. 
The renormalization and factorization scales are set to the Higgs mass and the
CTEQ5 set of PDFs has been used. Unfortunately, and contrary to the $t\bar t$
case, the rates for associated Higgs production with a single top quark are in
general much smaller than those of $t\bar t H$ production. At the Tevatron, all
channels lead to cross sections that are two orders of magnitude smaller.  At
the LHC, this is also the case for the $s$--channel $q\bar q' \to \bar b tH$
and $Wt H$ associated production.  However, for low Higgs masses, the
$t$--channel $qb \to q' tH$ cross section is suppressed only by a factor of 10
compared to $t\bar t H$ production and for larger masses, $M_H \sim 300$ GeV,
the two process have comparable but rather low rates.  \s

Focusing on the latter channel, where for $M_H \lsim 150$ GeV approximately
$10^4$ events can be collected at the LHC for ${\cal L}= 100$ fb$^{-1}$ before
cuts and efficiency losses are applied, the signals and the various backgrounds
have been studied in Ref.~\cite{ppHt-3oldpapers} for a Higgs boson decaying
into two photons and in Ref.~\cite{ppHt-Scott} where the more copious $H\to
b\bar b$ decays have been considered. The observation of a Higgs
boson in the first channel is certainly not possible since the $\gamma \gamma$
branching ratio is of ${\cal O}(10^{-3})$. In the configuration where the Higgs
boson decays into $b\bar b$ and the top quark into $Wb \to \ell \nu b$, the
yield depends on the number of $b$--quarks that are to be tagged: for
three $b$--tags, the background from $t\bar t j$ is overwhelming, while for
four $b$--tags, several backgrounds with rates that are comparable to the
signal are present.  The conclusion of Ref.~\cite{ppHt-Scott} is that a Higgs
signal is unlikely to be observed in this channel, except in extensions of the
SM where the production cross section can be enhanced.

\begin{figure}[t]
\begin{center}
\vspace*{-0.3cm}
\hspace*{-.02cm}
\mbox{
\epsfxsize=8.cm \epsfbox{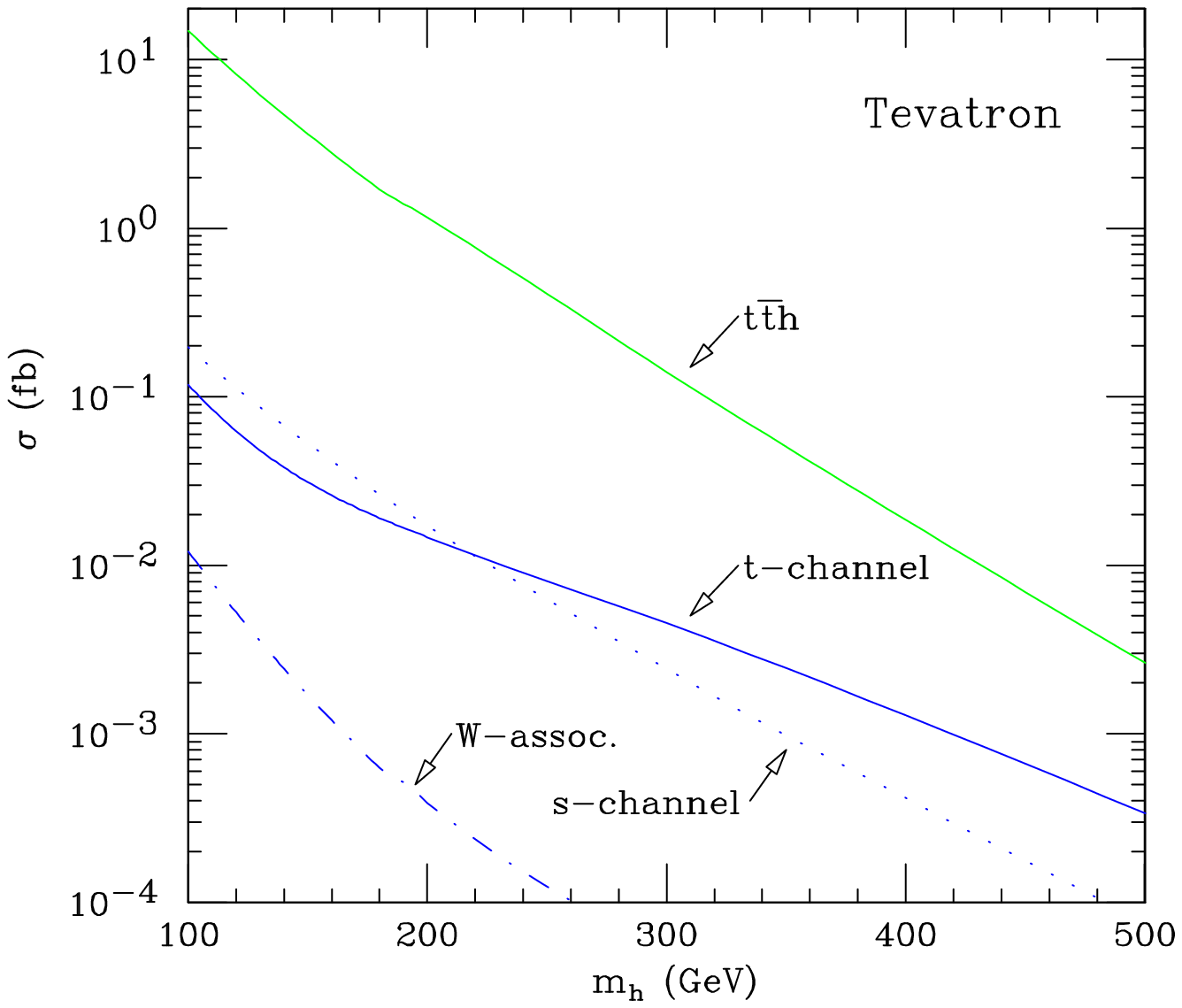}
\epsfxsize=8.cm \epsfbox{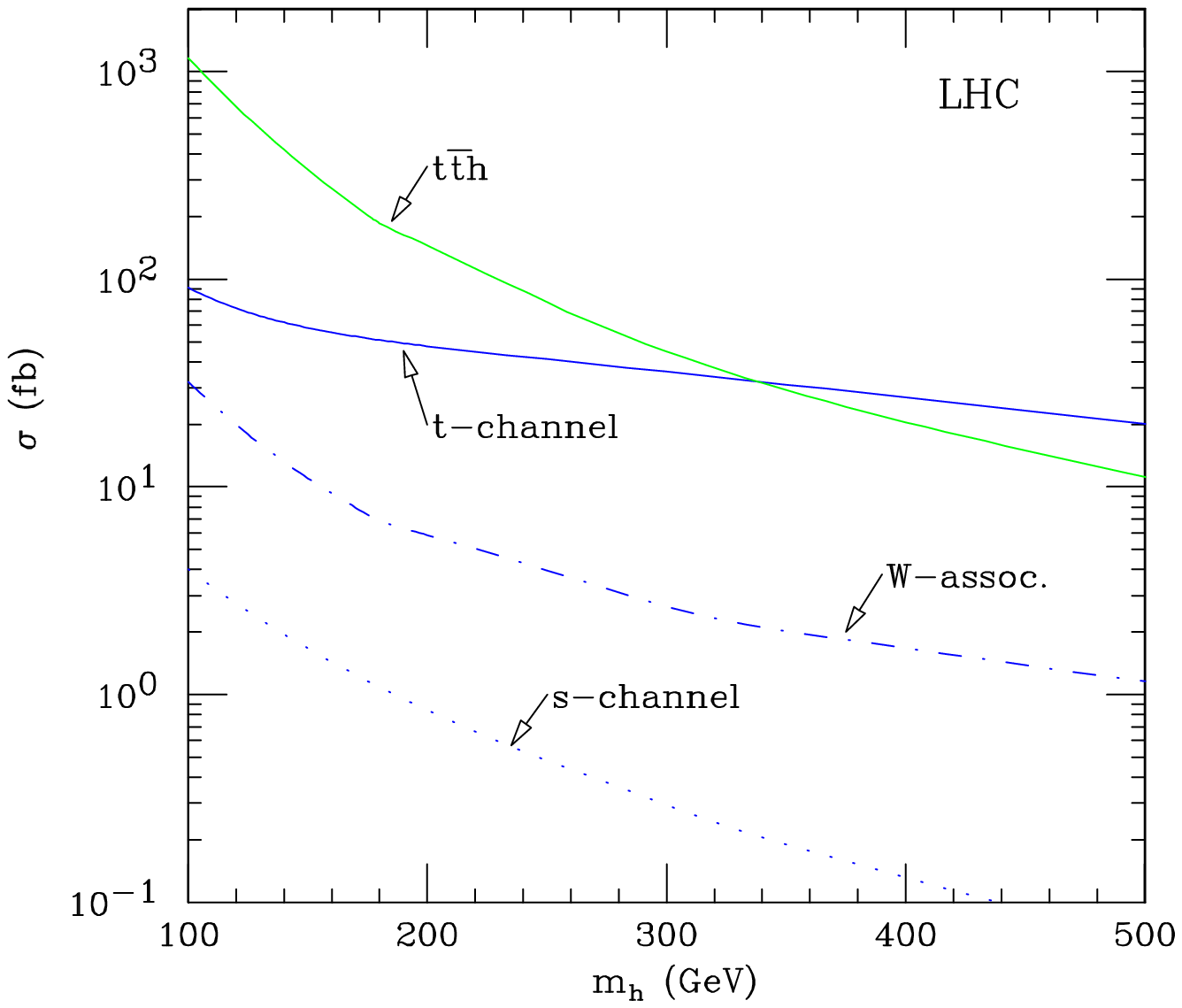} }
\vspace*{0cm}
\end{center}
\vspace*{-4mm}
\nn {\it Figure 3.38: The cross sections for Higgs plus single top production
at the Tevatron (left) and at the LHC (right), in the $t$--channel, $s$--channel
and $W$--associated  processes; for comparison the cross section for $t\bar{t}H$
is also shown. The CTEQ5L set of PDFs is used and the renormalization and 
factorization scales are set to $M_H$; from Ref.~\cite{ppHt-Scott}.} 
\vspace*{-5mm}
\end{figure}

\newpage

\subsection{The higher--order processes} 

\subsubsection{Higgs boson pair production}

In hadronic collisions, Higgs particles can be pair produced  in three main
processes\footnote{Triple Higgs production, which probes the quadrilinear
Higgs coupling, has a too small cross section \cite{gg-HHH}.}:\s

\nn $a)$ the gluon--gluon fusion mechanism which is mediated by loops of
third generation heavy quarks that couple strongly to the Higgs boson 
\cite{pp-ggHH-LO,pp-ggHH-LO1}
\beq
gg \to HH
\eeq
$b)$ double Higgs--strahlung from either a $W$ or a $Z$ boson  
\cite{pp-HHV,pp-DKMZ}
\beq
q\bar{q} & \to  V^* & \to VHH 
\eeq
$c)$ the $WW/ZZ$ fusion processes which lead to two Higgs particles and two jets
\cite{pp-VVHH,pp-VVH-Abas,pp-DKMZ}
\beq
qq & \to V^* V^* qq & \to HHqq
\eeq
The Feynman diagrams for these  processes are shown in Fig.~3.39 and, as can be
seen, one of them involves the trilinear Higgs boson coupling, $\lambda_{HHH}=3
M_H^2/v$, which can be thus probed in principle. The other diagrams involve
the couplings of the Higgs boson to fermions and gauge bosons and are probed in
the processes discussed in the previous sections.

\begin{picture}(100,90)(-30,-5)
\hspace*{5mm}
\SetWidth{1.1}
\put(-50, 60){\red{\bf (a)}}
\put(77, 27){\bb}
\put(117,27){\rb}
\Gluon(0,60)(40,60){3}{6}
\Gluon(0,0)(40,0){3}{6}
\ArrowLine(40,60)(80,30)
\ArrowLine(80,30)(40,0)
\ArrowLine(40,0)(40,60)
\DashLine(80,30)(120,30){5}
\DashLine(120,30)(150,60){5}
\DashLine(120,30)(150,0){5}
\put(155,60){\bH}
\put(155,-5){\bH}
\put(96,35){\bH}
\put(-15,-3){$g$}
\put(-15,60){$g$}
\put(50,30){$Q$}
\hspace*{8cm}
\Gluon(0,0)(40,0){3}{6}
\Gluon(0,60)(40,60){3}{6}
\ArrowLine(40,60)(90,60)
\ArrowLine(90,0)(40,0)
\ArrowLine(40,0)(40,60)
\ArrowLine(90,60)(90,0)
\DashLine(90,60)(130,60){5}
\DashLine(90,0)(130,0){5}
\put(87, 57){\bb}
\put(87, -3){\bb}
\put(135,60){\bH}
\put(135,-3){\bH}
\put(-15,-3){$g$}
\put(-15,60){$g$}
\put(60,30){$Q$}
\end{picture}

\vspace*{-3mm}
\begin{center}
\hspace*{-14cm}
\vspace*{-1.8cm}
\SetWidth{1.1}
\begin{picture}(300,100)(0,0)
\ArrowLine(150,25)(185,50)
\ArrowLine(150,75)(185,50)
\Photon(185,50)(230,50){3.5}{5.5}
\Photon(230,50)(265,25){3.5}{5.5}
\DashLine(230,50)(250,60){4}
\DashLine(250,60)(265,75){4}
\DashLine(250,60)(265,45){4}
\put(120, 75){\red{\bf (b)}}
\put(227,47){\bb}
\put(247,57){\rb}
\Text(145,30)[]{$\bar q$}
\Text(145,70)[]{$q$}
\Text(210,65)[]{$V^*$}
\Text(275,30)[]{$V$}
\Text(275,75)[]{\bH}
\Text(275,50)[]{\bH}
\ArrowLine(295,25)(330,50)
\ArrowLine(295,75)(330,50)
\DashLine(375,50)(410,25){4}
\Photon(375,50)(410,75){3.5}{5.5}
\Photon(330,50)(375,50){3.5}{5.5}
\DashLine(390,55)(410,45){4}
\put(373,47){\bb}
\put(387,52){\bb}
\vspace*{3mm}
\ArrowLine(445,25)(480,50)
\ArrowLine(445,75)(480,50)
\Photon(480,50)(525,50){3.5}{5.5}
\DashLine(525,50)(560,40){4}
\Photon(525,50)(560,75){3.5}{4.5}
\DashLine(525,50)(560,25){4}
\put(522,47){\bb}
\end{picture}
\vspace*{9.mm}
\end{center}
\begin{center}
\hspace*{-14cm}
\SetWidth{1.}
\begin{picture}(300,100)(0,0)
\hspace*{1cm}
\ArrowLine(150,25)(195,25)
\ArrowLine(150,75)(195,75)
\ArrowLine(195,25)(240,15)
\ArrowLine(195,75)(240,85)
\Photon(195,25)(195,75){3.5}{5.5}
\DashLine(195,50)(225,50){4}
\DashLine(225,50)(255,60){4}
\DashLine(225,50)(255,40){4}
\put(95, 75){\red{\bf (c)}}
\put(193,47){\bb}
\put(223,47){\rb}
\Text(145,30)[]{$q$}
\Text(145,70)[]{$q$}
\Text(245,20)[]{$q$}
\Text(245,80)[]{$q$}
\Text(210,65)[]{$V^*$}
\Text(210,35)[]{$V^*$}
\Text(270,60)[]{\bH}
\Text(270,40)[]{\bH}
\ArrowLine(295,25)(340,25)
\ArrowLine(295,75)(340,75)
\ArrowLine(340,25)(385,15)
\ArrowLine(340,75)(385,85)
\DashLine(340,60)(385,65){4}
\Photon(340,25)(340,75){3.5}{6.5}
\DashLine(340,40)(385,35){4}
\put(337,57){\bb}
\put(337,37){\bb}
\ArrowLine(425,25)(470,25)
\ArrowLine(425,75)(470,75)
\ArrowLine(470,25)(515,15)
\ArrowLine(470,75)(515,85)
\DashLine(470,50)(515,65){4}
\Photon(470,25)(470,75){3.5}{6.5}
\DashLine(470,50)(515,35){4}
\put(467,47){\bb}
\Text(330,-5)[]{\it Figure 3.39: Feynman diagrams for Higgs pair production in 
hadronic collisions.} 
\end{picture}
\vspace*{3.mm}
\end{center}
We briefly discuss these processes in this subsection, restricting
ourselves to the case of the LHC where the phase space is not too penalizing. 

\subsubsection*{\underline{The gluon--gluon fusion mechanism}}

The large number of gluons in high--energy proton beams implies that the
gluon--gluon fusion mechanism is the dominant process for Higgs boson pair
production. As for single Higgs production in this mechanism, the coupling
between gluons and Higgs bosons is mediated by heavy quark loops. In the SM, 
the top quark loop is dominating while the bottom quark loop gives a small
but non--negligible contribution. \s

In terms of the  trilinear Higgs coupling, $\lambda_{HHH}'=3M_H^2/M_Z^2$ [note 
the change in the normalization], the partonic cross section at leading order 
is given by \cite{pp-ggHH-LO1}
\begin{equation}
\hat \sigma_{\rm LO}(gg \to HH) = \int_{\hat t_-}^{\hat t_+} d\hat t \,
\frac{G_\mu^2 \alpha_s^2(\mu_R)}{256 (2\pi)^3} \left\{ \left| \frac{M_Z^2  
\lambda_{HHH}'}{\hat s - M_H^2}F_T + F_B \right|^2 + \left|G_B \right|^2 
\right\}
\end{equation}
with the Mandelstam variables for the parton process given by
\begin{equation}
\hat s = Q^2 \ \ , \ 
\hat t /\hat u  = - {\footnotesize 1 \over 2} \left[Q^2-2M_H^2 \mp Q^2 \beta_H 
\cos\theta \right]
\end{equation}
where $\theta$ is the scattering angle in the partonic c.m. system with 
invariant mass $Q$ and, as usual, $\beta= \sqrt{1-4M_H^2/Q^2}$. $\mu_R$ is the 
renormalization scale which, together with the factorization scale, will be 
identified to $\hat{s}$ and the integration limits correspond to 
$\cos\theta=\pm 1$ and $\hat  t_\pm = -\frac{1}{2} \left[ Q^2 - 2 M_H^2 \mp 
Q^2 \beta_H \right]$.
The proton cross section is derived by folding the parton cross section 
$\hat{\sigma}(gg\to HH)$ with the gluon luminosity
\beq
\sigma (pp \to HH) = \int_{4M_H^2/s}^1 d\tau 
\frac{d{\cal L}^{gg}}{d\tau} \hat{\sigma} (gg \to HH; \hat{s} = \tau s)
\eeq

The dependence on the quark masses is contained in the triangle and box 
functions $F_T, F_B$ and $G_B$.  The expressions of these form factors 
with the exact dependence on the quark masses can be found in 
Refs.~\cite{pp-ggHH-LO,pp-ggHH-LO1}. In the limit where the Higgs boson is much 
lighter or much heavier than the internal quark $Q$, the coefficients take a 
very simple form \cite{pp-ggHH-LO1}
\beq
M_H \ll 4 m_Q && F_T \simeq \frac{2}{3}\ , \ \  F_B \simeq -\frac{2}{3} \ , 
\ \ G_B \simeq 0 \non \\
M_H \gg 4 m_Q && F_T \simeq 
-\frac{m_Q^2}{\hat{s}} \bigg[ \log \frac{m_Q^2}{\hat{s}} +i\pi \bigg] \ , \ 
F_B \sim G_B \simeq 0
\eeq

As one might have expected from single Higgs production, the QCD radiative
corrections are particularly important for this production channel and must  be
included.  They have been determined in the heavy quark limit $M_H^2 \ll 4
m_Q^2$, where one can use the low energy theorem to determine the effective
$Hgg$ and $HHgg$ couplings in the triangle and box contributions, when the top
quark is integrated out. One can then use these effective couplings to calculate
the interaction of the light gluon and quark fields, as discussed previously. 
The $K$--factor was found to be $K\approx 1.9$ in the Higgs mass range between
100 and 200 GeV \cite{pp-ggHH-NLO}. A $K$--factor of  similar size is generally
expected for larger Higgs masses and even beyond the top--quark threshold, as
it was the case for the $gg \to H$ process.

\subsubsection*{\underline{The vector boson fusion and strahlung mechanisms}}

At high energies, on expects double Higgs production in the vector
boson fusion channel to have a substantial cross section since the longitudinal
vector bosons have couplings which grow with energy.  The calculation of the
full $2 \to 4$ process, $qq \to qq HH$, is rather complicated. However, one can
use the equivalent longitudinal vector boson approximation in which one 
calculates the cross section for  the $2 \to 2$ process 
\beq
V_L V_L \to HH
\eeq
Taking into account only the dominant longitudinal vector boson contribution,  
denoting by $\beta_{V,H}$ the $V,H$ velocities in the c.m.\ frame, the
production amplitude is given by
\beq
{\cal M}_{LL} &=& \frac{G_\mu \hat{s}}{\sqrt{2}} 
\, \left\{ (1 \!+ \beta_V^2) \left[ 1 \! + \, 
\frac{M_Z^2 \lambda_{HHH}'}{(\hat{s}-M_H^2)} \right] \right.  \\ 
&&  \left. + \frac{1}{\beta_V \beta_H}  \left[ 
\frac{(1-\beta_V^4)+ 
(\beta_V - \beta_H \cos\theta)^2}{\cos\theta - x_V} -
\frac{(1-\beta_V^4) + 
(\beta_V + \beta_H \cos\theta)^2}{\cos\theta + x_V} \non 
\right] \right\}
\label{WW--HHamp}
\eeq
with the variable $x_V$ defined as $x_V = (1- 2 M_H^2/\hat{s})/(\beta_V
\beta_H)$, $\theta$ the scattering angle in the $VV$ c.m. frame and 
$\hat{s}^{1/2}$ the invariant energy of the $VV$ pair. \s

Squaring the amplitude and integrating out the angular dependence, one obtains 
the cross section for the $V_L V_L \to HH$ subprocess, 
\beq
\hat \sigma (V_L V_L \to HH) &=& \frac{G_\mu^2 M_V^4}{8\pi \hat{s}} 
\frac{\beta_H} {\beta_V (1-\beta_V^2)^2} \int_{-1}^1 {\rm d}\cos \theta
\left| {\cal M}_{LL} \right|^2 
\eeq
which has then to be folded with the longitudinal vector boson luminosity 
spectra eq.~(\ref{WW-effective}) to obtain the $qq \to HHqq$ cross section, 
which again has to be convoluted with the parton densities to obtain the full 
hadronic cross section
\beq
\sigma (pp \to HHqq) = \int_{4M_H^2/s}^1 d\tau 
\frac{d{\cal L}^{qq}}{d\tau} \; 
\hat{\sigma} (qq \to HH qq; \hat{s} = \tau s)
\eeq
The result obtained in this way is expected to approximate the exact result 
within about a factor of two for low Higgs masses and very high energies
\cite{pp-VVHH,pp-VVH-Abas}. \s

In the case of the double Higgs--strahlung mechanisms, $q\bar{q} \to HHV$, the
production cross sections are expected to be rather small.  This can be guessed
by looking at the cross section for single  Higgs--strahlung: for $M_H \sim
200$ GeV [which in terms of phase space would correspond to the production of 
two Higgs bosons with a mass of 100 GeV], it is of the order of 30 fb, and 
there will be still an additional suppression by the electroweak coupling factor
in the case of double Higgs--strahlung. The analytical expressions will be 
given in the next section when this process will be discussed at $\ee$ 
colliders, where it is more relevant.

\subsubsection*{\underline{The cross sections at the LHC}}

The total cross sections for the pair production of Higgs bosons in the three
processes are shown in Fig.~3.40 as a function of the Higgs mass in the 
range $M_Z \lsim M_H \lsim 2M_Z$. In the $gg$ case, the full dependence
on the quark mass has been taken into account and the $K \sim 1.9$ factor  has
been included. Note that the NLO QCD corrections to the double Higgs production
in association with a vector boson and in the vector boson fusion channels, are
the same as the respective processes for single Higgs production and will
increase the cross sections by, respectively, $\sim 30\%$ and $\sim 10\%$; 
they have not been included. \s

As expected, gluon--gluon fusion dominates over the other mechanisms and has a
cross section larger than 10 fb for this Higgs mass range. The $WW/ZZ$ fusion
mechanisms are the next important channels, but with cross sections which are
one order of magnitude smaller; $WW$ fusion dominates over $ZZ$ fusion at a
ratio $WW/ZZ\approx 2.3$. The cross sections for double Higgs--strahlung are
relatively small as it follows from the scaling behavior of the cross sections
which drop $\sim 1/\hat{s}$. The cross sections for Higgs--strahlung off $W$ and
$Z$ bosons are combined in the figure and their their relative size is close to
$W/Z\approx 1.6$.  \s

The vertical arrows indicate the sensitivity of the production cross sections
to the size of the trilinear Higgs coupling; they correspond to a modification
of the coupling $\lambda'_{HHH}$ by the rescaling coefficient
$\kappa=\frac{1}{2} \to \frac{3}{2}$. 

\begin{figure}[htbp]
\begin{center}
\epsfig{figure=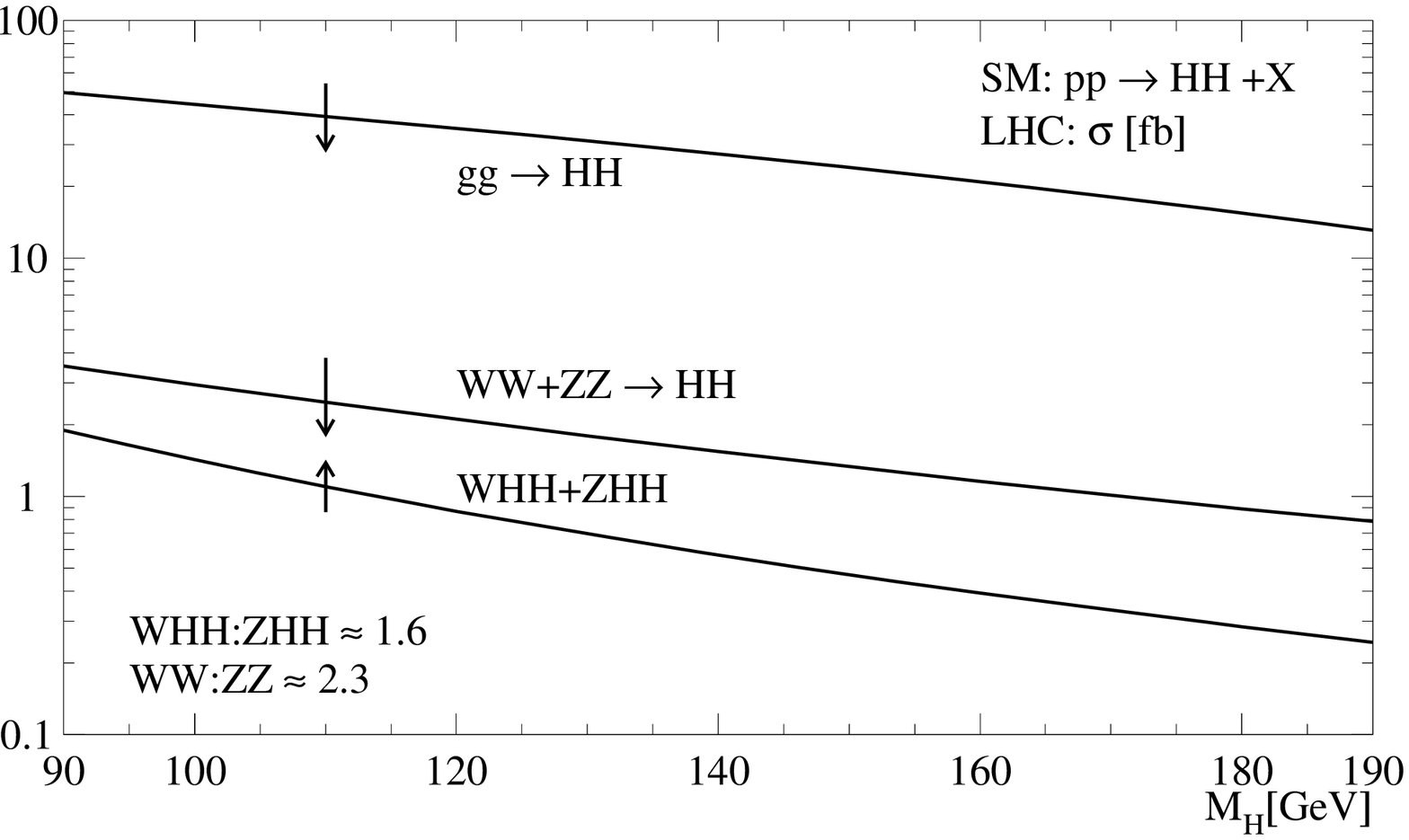,width=12cm,height=8cm}\\[3mm]
\end{center}
\vspace*{-3mm}
{\it Figure 3.40: The cross sections for gluon fusion, $gg \to HH$, the $WW/ZZ$
fusion $qq \to qqWW/ZZ \to HH$ and the double Higgs--strahlung $q\bar{q} \to 
WHH+ ZHH$ in the SM as a function of $M_H$. The vertical arrows  correspond to a
variation of the trilinear Higgs coupling from $\frac{1}{2}$  to $\frac{3}{2}$ 
of the SM value, $\lambda'_{HHH}=3 M_H^2/M_Z^2$; from Ref.~\cite{pp-DKMZ}.}
\end{figure}

\subsubsection{Higgs production in association with gauge bosons}

\subsubsection*{\underline{Higgs production in association with gauge boson
pairs}}

In high--energy collisions, the pair production of massive vector bosons $pp
\to VV'$, with $V,V'=W,Z (\gamma)$, has a very large cross section. In view of 
these rates, it is tempting to consider the possibility of emitting an 
additional Higgs particle from one of the gauge boson  lines \cite{pp-HVV,DWP}
\beq 
qq \, / \, q\bar{q} &\to & W^+W^- H \,  , \, ZZH \, , \, W^\pm ZH
\ \ {\rm and} \ \ qq \, / q\bar{q} \to  \gamma Z H \ , \gamma W^\pm H  
\label{qqHVV}
\eeq
The hope is that the suppression by the additional electroweak coupling factor 
might be compensated by the initially large production rate for gauge bosons. 
Formally, these processes are of the same order, ${\cal O}(G_\mu^3)$, as Higgs 
production in the $WW/ZZ$ fusion mechanisms and the suppression by the phase
space should not be too drastic at high enough energies. 

\vspace*{-.7cm}
\begin{center}
\SetWidth{1.}
\begin{picture}(300,100)(0,0)
\hspace*{-2cm}
\ArrowLine(0,25)(40,25)
\ArrowLine(0,75)(40,75)
\Line(40,25)(40,75)
\Photon(40,25)(85,25){3.5}{5.5}
\Photon(40,75)(85,75){3.5}{5.5}
\DashLine(60,75)(90,50){4}
\put(57,72){\bb}
\Text(0,35)[]{$q$}
\Text(0,65)[]{$q$}
\Text(30,50)[]{$q$}
\Text(97,50)[]{\bH}
\Text(95,30)[]{$V$}
\Text(95,70)[]{$V$}
\ArrowLine(130,25)(165,50)
\ArrowLine(130,75)(165,50)
\Photon(165,50)(210,50){3.5}{5.5}
\Photon(210,50)(245,25){3.5}{5.5}
\Photon(210,50)(245,75){3.5}{5.5}
\DashLine(235,65)(260,47){4}
\put(232,62){\bb}
\Text(185,65)[]{$V$}
\Text(258,20)[]{$V$}
\Text(258,80)[]{$V$}
\Text(270,50)[]{\bH}
\Line(295,25)(330,50)
\Line(295,75)(330,50)
\Photon(330,50)(375,50){3.5}{5.5}
\Photon(312,62)(350,75){3.5}{5.5}
\Photon(375,50)(415,25){3}{5}
\DashLine(375,50)(415,75){4}
\put(372,47){\bb}
\Text(355,65)[]{$V$}
\Text(420,37)[]{$V$}
\Text(420,67)[]{\bH}
\Text(320,75)[]{$\gamma$}
\Text(210,-1)[]{\it Figure 3.41: Diagrams for associated  Higgs boson 
production with two gauge bosons.} 
\vspace*{1.mm}
\end{picture}
\end{center}

As shown in Fig.~3.41, where some generic Feynman diagrams are displayed, the
processes proceed through $s$--channel gauge boson and/or $t$--channel quark
exchanges. Strictly speaking, the processes with additional final state
photons which have enough large $p_T$ to be observed, should be viewed as the
ISR part of the electroweak corrections to the $q\bar{q} \to HV$
processes as discussed in  \S3.2. However, they are interesting since, besides
the fact that the final  state contains an additional photon which can be
tagged, they can have larger rates compared to the parent processes which drop
like $1/\hat{s}$ at high energies. The processes not involving photons 
are genuine higher--order processes, though at high energies they can also be 
viewed as a kind of ``$V$ bremsstrahlung" correction to the main mechanisms 
$q \bar{q} \to HV$.\s

\begin{figure}[htbp]
\begin{center}
\epsfig{figure=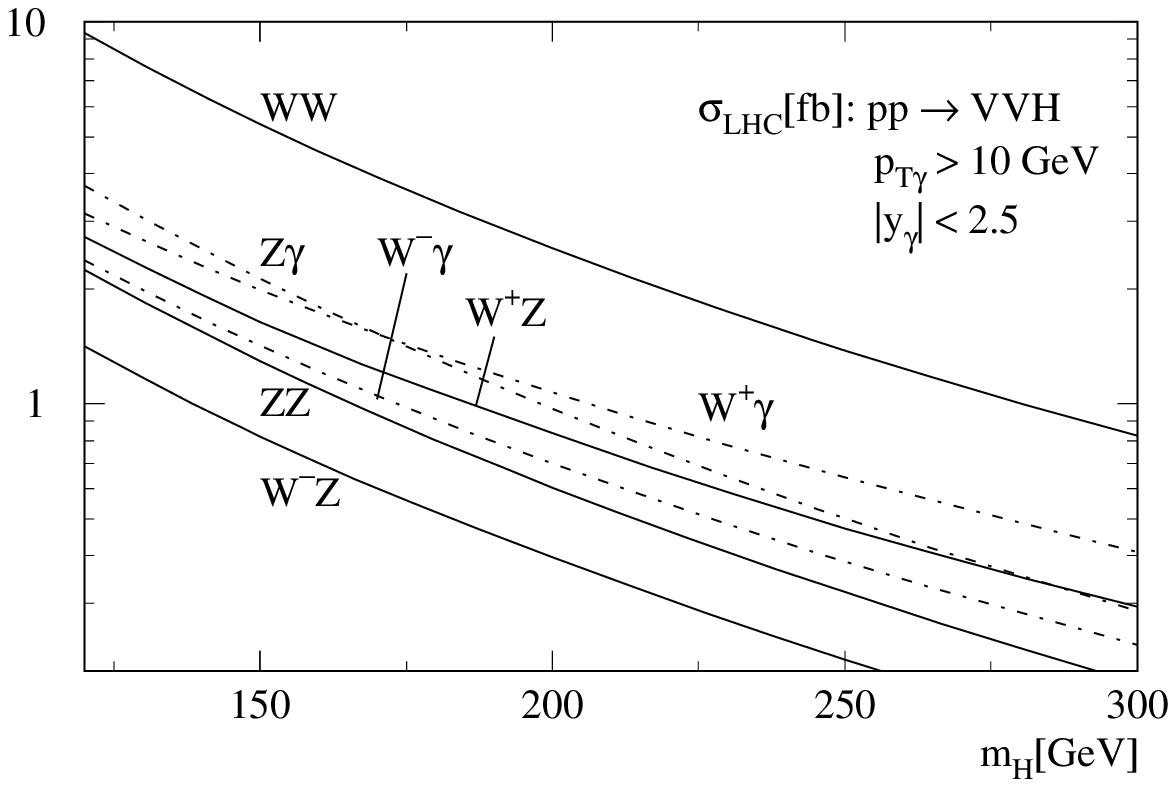,width=10cm}\\[3mm]
\end{center}
\vspace*{-3mm}
{\it Figure 3.42: The total cross sections for the associated production of the
Higgs boson with a pair of gauge bosons at the LHC, $pp \to HVV$, as a 
function of $M_H$; from Ref.~\cite{DWP}.}
\vspace*{-3mm}
\end{figure}

The cross sections for these processes have been evaluated in 
Ref.~\cite{pp-HVV} and updated recently \cite{DWP} using {\tt MadGraph}
\cite{MadGraph}. They are shown 
in Fig.~3.42 for the energy relevant at the  LHC as a function of $M_H$. 
For the final states involving photons, the cuts $p_T^\gamma \geq 10$ GeV 
and $|\eta^\gamma| \leq 2.5$ have been applied. The CTEQ6 PDFs have been used 
and the scales were set at $\mu_R^2 = \mu_F^2 =\hat s$. 
The largest cross section is obtained for the $pp \to HWW$ process,
as anticipated from the fact that the $WW$ cross section is dominant
at the LHC, being at the level of $\sigma (HWW) \sim 10$ fb for low mass 
Higgs values and decreases slowly to reach $\sim 1$ fb for $M_H \simeq
300$ GeV. Thus, it is larger than for double Higgs production in the 
strahlung and fusion processes. The cross sections for the other processes
are a factor of 3 to 5 smaller but, except for $\sigma (HWZ)$, they are
above the femtobarn level for $M_H \lsim 160$ GeV. \s

In view of the smallness of the signal cross sections, these processes cannot
of course be considered as Higgs discovery channels [in particular since the 
backgrounds from triple gauge boson production might be large]. However, once 
Higgs particles have been detected in the dominant detection channels, they 
could allow for additional tests and measurements, such as the determination of
the $HWW$ coupling from $pp \to HWW \to WWWW$ for instance. 
 
\subsubsection*{\underline{Higgs production in association with a gauge boson
and two jets}}

Associated Higgs production with a gauge boson and two quarks in
hadronic collisions 
\beq
qq \to HW qq \ , \  HZ qq \ , \  H\gamma qq
\label{eq:HVqq}
\eeq
originates from several sources, as shown in Fig.~3.43 where some Feynman 
diagrams are displayed, with the starting point being the fusion of the vector 
bosons producing a gauge or a Higgs boson. The  production of $H\gamma qq$ final
states occurs only through the $qq \to WWqq \to Hqq$ process, with the
photon emitted from the quark or the internal $W$ lines, which is part
of the photonic corrections to the initial mechanism. Note that an additional 
source might come from the $pp \to HV$ process, with the emission of two jets 
in the final state: this also is part of the NNLO QCD corrections to 
Higgs--strahlung that we have already discussed.  

\begin{figure}[h]
\begin{center}
\begin{picture}(100,90)(-30,-5)
\hspace*{-11cm}
\SetWidth{1.1}
\ArrowLine(150,25)(195,25)
\ArrowLine(150,75)(195,75)
\ArrowLine(195,25)(240,15)
\ArrowLine(195,75)(240,85)
\Photon(195,25)(195,75){3.5}{6}
\Photon(195,50)(230,50){3.5}{4} 
\DashLine(230,50)(265,65){4}
\Photon(230,50)(265,35){3.5}{4}
\put(227,47){\bb}
\Text(145,30)[]{$q$}
\Text(145,70)[]{$q$}
\Text(245,20)[]{$q$}
\Text(245,80)[]{$q$}
\Text(210,65)[]{$V^*$}
\Text(210,35)[]{$V^*$}
\Text(275,65)[]{\bH}
\Text(275,35)[]{$V$}
\hspace*{5mm}
\ArrowLine(295,25)(340,25)
\ArrowLine(295,75)(340,75)
\ArrowLine(340,25)(385,15)
\ArrowLine(340,75)(385,85)
\DashLine(340,60)(385,65){4}
\put(337,57){\bb}
\Photon(340,25)(340,75){3.5}{6.5}
\Photon(342,40)(385,35){3.5}{5.5}
\ArrowLine(425,25)(470,25)
\ArrowLine(425,75)(470,75)
\Line(470,25)(515,15)
\Line(470,75)(515,85)
\DashLine(470,50)(515,50){4}
\put(467,47){\bb}
\Photon(470,25)(470,75){3.5}{6.5}
\Photon(485,77)(515,65){3}{4}
\end{picture}
\vspace*{-10.mm}
\end{center}
{\it Figure 3.43: Feynman diagrams for associated $HVqq$ production 
in hadronic collisions.} 
\vspace*{-3mm}
\end{figure}

The cross sections for these processes have been calculated sometime ago
\cite{pp-HVqq} both exactly and in the longitudinal $W$ approximation [for the
energy which was relevant for the late SSC] and the output was that the latter
approximation gives results which are only about a factor of two different from
the exact result. This is similar to  Higgs pair production in the vector boson
fusion channels $qq \to V^* V^* \to qqHH$, discussed previously. In fact, for
$pp \to HZqq$, the analogy between the two processes is complete since the $Z$
boson can be treated as a neutral Goldstone boson $w_0$, which has exactly the
same coupling as the Higgs boson as can be seen from the effective potential
eq.~(\ref{Vequivalence}). The cross are not that small for such higher--order
mechanisms: in the case of $pp \to HWqq$, they almost reach the level of 100 fb
for Higgs masses in the low range [which is only one order of magnitude smaller
than the Higgs--strahlung $pp \to HW$ cross section] and decrease only slowly
with $M_H$.\s

\begin{figure}[htbp]
\begin{center}
\epsfig{figure=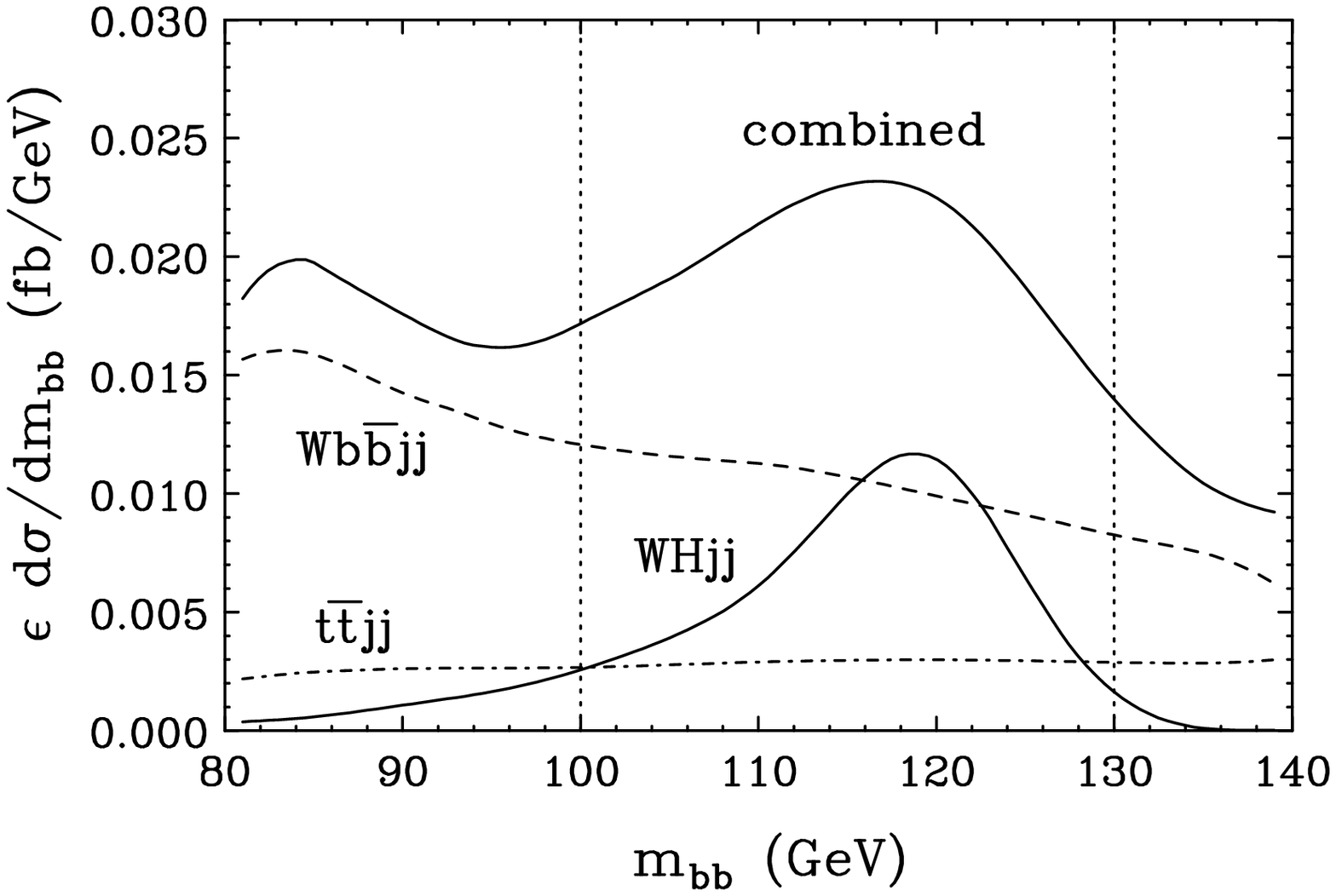,width=8cm,angle=90}\\[3mm]
\end{center}
\vspace*{-3mm}
{\it Figure 3.44: The $b\bar{b}$ invariant mass distribution of the $WHjj$ 
signal for $M_H = 120$~GeV and the $W b\bar{b}jj, t\bar{t}jj$ backgrounds 
after cuts; the combined signal and backgrounds are also shown. The vertical 
dotted lines denote the mass bin used for calculating the statistical 
significance of the signal; from Ref.~\cite{pp-HVqq-Rain}.}
\vspace*{-3mm}
\end{figure}

More recently, a detailed study of the signal for the $pp \to HWqq$ process has
been performed at the LHC \cite{pp-HVqq-Rain}, in the channel where the Higgs
decays into bottom quarks and the $W$ boson leptonically. Applying cuts that
are similar to those of the vector boson fusion process discussed in \S3.3.3
and assuming reasonable efficiency for tagging the $b$ quarks and resolution
for the reconstruction of the $b\bar{b}$ invariant mass, the various
backgrounds [in particular the $pp \to tt \to bbWW$ and the QCD $Wbbjj$ final
states] can be reduced at a level comparable to the signal as shown in
Fig.~3.44. This would allow for the extraction of the $Hb\bar b$ Yukawa
coupling with a reasonable accuracy if a high luminosity is available.  

\subsubsection{More on higher--order processes}

There are several other higher--order processes for Higgs production at hadron
colliders, but they lead to extremely small cross sections at the LHC and, {\it
a fortiori}, at the Tevatron. We briefly discuss some of them for completeness. 

\subsubsection*{\underline{Higgs production in association with a
photon}}

Higgs boson and photon final states in hadronic collisions \cite{pp-Hgamma} may
originate from two main sources. An obvious possibility is the  direct
production from light quarks, $ qq \to H \gamma$, where the Higgs boson is
emitted from the quark lines. Because the Yukawa couplings are very tiny, the
cross section are negligible. An exception might be the case of bottom quarks;
however, besides the fact that the $Hb\bar b$ Yukawa coupling is  still small,
there is a suppression from the $b$ density in the hadron. In fact, the cross
section is comparable to the one for charm quark, the  suppression of the
Yukawa coupling $m_c/m_b$ being partly compensated by the larger $c$--parton 
density and by the electric charge. For low Higgs masses, $M_H\sim 100$ GeV, 
the cross sections are at the femtobarn level at the LHC and one to two orders 
of magnitude smaller at the Tevatron. Since the dominant contribution is coming
from $b \bar{b}$ initial state, this process is anyway equivalent to the
processes $b\bar b \to H$ and $gg \to b\bar b H$ with two  undetected $b$
quarks, with the radiation of an ISR photon. \s

Another possibility to generate the Higgs plus photon final state is via loop
diagrams in quark--antiquark annihilation [the corresponding process with
initial state  gluons, $gg \to H \gamma$, is forbidden by Furry's theorem
similarly to the $H \to \gamma \gamma g$ decay discussed in \S2.3].  There
are triangular diagrams, when the $q\bar{q}$ state annihilates into a virtual
photon or $Z$ boson in the $s$--channel and which involve virtual top quarks
and $W$ bosons and box diagrams with $W$ bosons and light quarks running in the
loop. Since the process is of ${\cal O}(G_\mu^4)$ and because of the 
suppression by the loop factor, the cross sections are extremely small: at 
the LHC they are at the level of 0.1 fb and they are much lower at the 
Tevatron \cite{pp-Hgamma}. 

\subsubsection*{\underline{Loop induced Higgs pair production in $q\bar{q}$
annihilation}}

Similarly to the $gg$ fusion process, $gg \to HH$, which provides the largest
cross section for Higgs boson pairs at the LHC, one can produce pairs of Higgs
particles in $q\bar{q}$ annihilation. Because the Higgs couplings to the light
quarks are small and since CP conservation forbids a $ZHH$ coupling at the
tree--level, the entire contribution to this process originates from loop
diagrams. In fact, as a result of chiral symmetry, only box diagrams involving
quarks and massive gauge bosons, from which the Higgs particles are emitted, 
are present. The process is thus not sensitive to the trilinear Higgs coupling.
\s

This calculation has been performed in Ref.~\cite{pp-qqHH} and, as one might 
have expected, because of the lower luminosity for quarks than for gluons 
at high energies, the cross sections are much smaller than those of the $gg \to
HH$ production process. At LHC energies  the difference is at least one order 
of magnitude. At the Tevatron, the cross sections will be anyway very small
because of the reduced phase space.  The annihilation of $q\bar{q}$ states is, 
thus, not an important process for double Higgs boson production. 

\subsubsection*{\underline{Higgs pair production with heavy quarks}}

Similarly to the double Higgs boson production in the $WW/ZZ$ fusion processes,
one might take advantage of the large Yukawa coupling to top quarks to produce
two Higgs bosons emitted from the top quark lines in the process $gg/q\bar q
\to t\bar t$. An interesting feature is that there is a contributing diagram
where a Higgs boson is emitted from the top quark lines and then splits into
two Higgs bosons. This process is therefore sensitive to the trilinear Higgs
coupling and, despite of the suppression by the electroweak couplings,  one
might hope for a compensating  enhancement of the cross section due to the
presence of the Higgs boson exchange in the $s$--channel.
The complete calculation for this four massive particle final state is rather
complicated\footnote{One can estimate  the order of magnitude of the
cross section, by naively treating the heavy top quarks as partons inside the
hadron and considering at the partonic level the process  of heavy top quark
annihilation into two Higgs bosons, $t \bar{t} \to HH$. This calculation has
been performed in Ref.~\cite{pp-qqttHH0} some time ago [at the time where the
top quark was believed to have a mass of the order of 50 GeV and where the SSC
was still expected to operate] and the output was that, even for hadronic c.m.
energies of $\sqrt{s}= 40$ TeV, the ``partonic" cross sections folded with
luminosities which may be overestimated by a factor of ten, lead to a total
rate which is at the level of the cross section for the longitudinal $W$ boson
fusion  into two Higgs bosons.} and has been performed numerically in 
Ref.~\cite{pp-qqttHH}. At the LHC,
the cross section is at the level of 1 fb for $M_H \sim 120$ GeV and, thus, 
of the same order as $VHH$ production and much smaller than $gg \to HH$ 
production. The large backgrounds make it impossible to extract any signal
even with extremely high luminosities \cite{pp-qqttHH}.

\subsubsection*{\underline{Rare decays of the top quark}}

The huge cross section of the process $gg/q\bar{q} \to t\bar{t}$
allows  to produce $10^{7}$ to $10^{8}$ top quark pairs per year at the LHC. 
This large number of events could be used to look for very rare decays of this
particle. If the Higgs boson is not too heavy, $M_H < m_t$, the decay $t \to
cH$ can occur through loop diagrams. Starting from the flavor changing
transition  $t \to c$, which is mediated by loops involving $W$ bosons and
down--type [mainly bottom] quarks, one can attach a Higgs boson either to the
external  top quark line or to the internal $W$ boson or $b$ quark lines. 
However, because the decay is suppressed by three powers of the Fermi
constant $G_\mu$ and by the GIM mechanism, the branching ratio is extremely
small BR$(t \to cH) \lsim 10^{-13}$ for $M_H \gsim 100$ GeV \cite{pp-t-H}. 
In view of the experimental bound $M_H \gsim 115$ GeV, the parent decay process
$t \to bWH$ \cite{Three-Body2,pp-t-H2} is now kinematically closed.

\subsubsection{Diffractive Higgs boson production}

Diffractive processes in (anti)proton collisions are those in which color 
singlet objects are exchanged between the high energy initial protons, which 
allow them to be diffracted \cite{BL-diffr,diffr1,diffr2}. This can occur, for 
instance, when two gluons are exchanged in the $t$--channel by the initial 
protons: this neutralizes the color
and allow the two protons to remain intact and continue their way. Higgs 
production occurs in the emission from the $t$--channel exchanged particles and,
in the case of $t$--channel gluons, this occurs through the usual $ggH$ vertex 
mediated by heavy quark loops. The signature is then two protons which are 
produced at very large rapidities and a centrally produced Higgs particle 
\beq 
p + p \to p + H + p
\eeq
where the $+$ sign means that  there is a large rapidity gap between the 
particles. In addition, if one tags the initial hadrons [using the so--called 
roman pot detectors], these diffractive processes can be selected and result 
in a very clean signal [the backgrounds will be discussed later]: a Higgs boson
in the central region, and nothing else.\s

There are many models for diffractive Higgs production in the literature,
starting from the Bialas--Landshoff exclusive model \cite{BL-diffr}.  They all
involve a mixture of perturbative and non perturbative QCD physics which is not
very well understood yet.  In the context of Higgs physics, hard diffractive
production in the central region are the most interesting ones. We briefly
summarize the main features of some processes for a light Higgs boson decaying
mainly into $b\bar b$ pairs, following Refs.~\cite{Valery-myths,diffr-Houches}
where a detailed account is given and to which we refer for earlier work. 
Figure 3.45, taken from Ref.~\cite{Valery-myths}, illustrates three processes
for double diffractive Higgs production in hadronic collisions that one can
partly discuss in the familiar terms of perturbative QCD. We use the
terminology of this reference.\s

\begin{figure}[h]
\vspace*{-2mm}
\begin{center}
\epsfig{figure=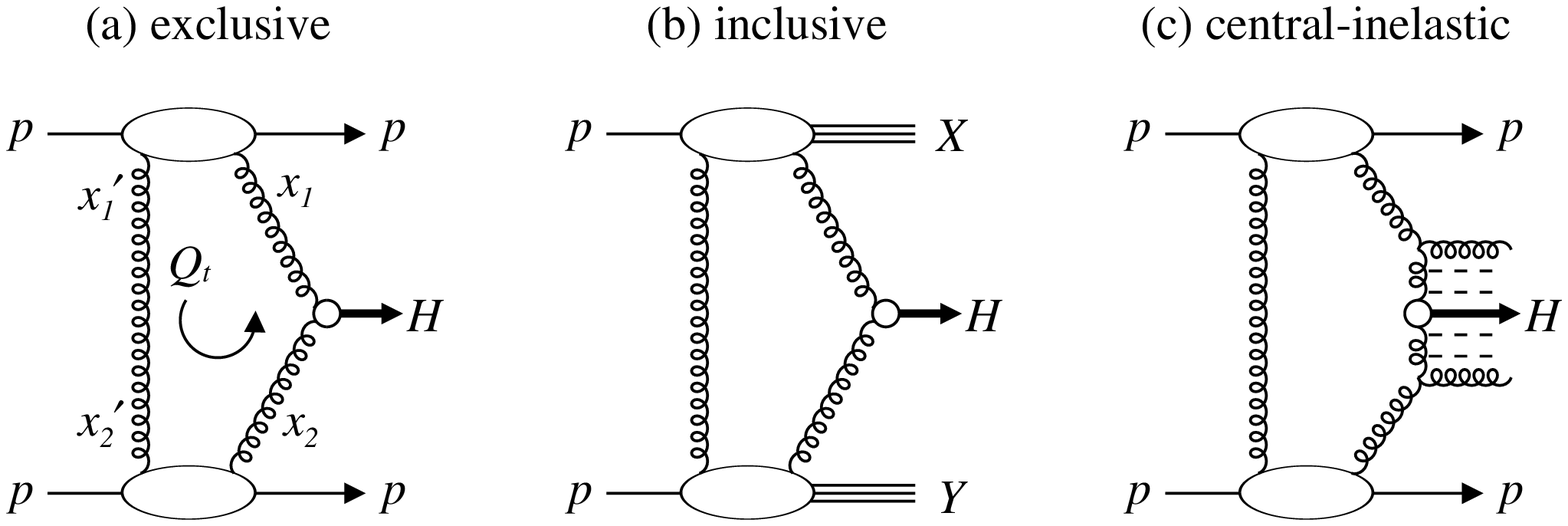,height=4.5cm,width=14.cm}
\end{center}
\vspace{-.2cm}
{\it Figure 3.45: Examples of processes for double--diffractive Higgs 
production in $pp$ collisions.}
\vspace{-.3cm}
\end{figure}

In the central exclusive double diffractive processes shown in Fig.~3.45a, the
Higgs boson is produced alone and is separated from the outgoing hadrons by
large rapidity gaps. If the latter are tagged, the Higgs mass can be determined
either by measuring the missing mass $M_{\rm inv}$ of the system or by
reconstructing the $H\to b\bar b$ decays mass peak $M_{b\bar b}$; one can then
match the two measurements, $M_{\rm inv}\!=\!M_{b \bar b}\!=\!M_H$ which
provides a strong kinematical constraint.  Moreover, an interesting feature is
that in the production vertex, the polarization vectors of the gluons are
correlated in a such way that with the resulting effective luminosity, only
spin--zero particles with positive parity, i.e. with $J^{\rm PC}=0^{++}$
quantum numbers, can be primarily produced [the cross section for CP--odd
states is strongly suppressed]. On the other hand, the background
from $gg \to b\bar b$ for instance is strongly suppressed, $\hat \sigma (gg \to
b\bar b) \propto \alpha_s^2 m_b^2/\hat s$.  The Higgs spin and parity quantum 
numbers can therefore be checked in this process \cite{diff-spin}
with only an ambiguity with $2^{++}$ states remaining.  Unfortunately, the model
predicts rather low Higgs production cross sections: for a Higgs boson with a
mass $M_H \sim 120$ GeV, they are of the order of 0.2 fb at the Tevatron and 3
fb at the LHC; see Table 3.1. The uncertainty in the prediction is also large,
a factor of 2 at the LHC for instance.\s

\begin{table}[h]
\renewcommand{\arraystretch}{1.3}
\begin{center}
\begin{tabular}{|l||c|c|c|} \hline
Model  & Exclusive (a) & Inclusive (b) & Central inel. (c) \\ \hline
Tevatron &  0.2 & 1  & 0.03 \\
LHC      &  3   & 40 & 50 \\ \hline
\end{tabular}
\end{center}
{\it Table 3.1: Higgs boson production cross sections in fb at the Tevatron and the LHC for $M_H \sim 120$ GeV in the various diffractive models of Fig.~3.45;
from Ref.~\cite{Valery-myths}.}
\vspace*{-3mm}
\end{table}

In central inelastic production, Fig.~3.45c, there is an additional radiation
accompanying the Higgs boson in the central region, but the latter is still
separated from the final hadrons by large rapidity gaps\footnote{In fact, in
the terminology of Ref.~\cite{Robi-Diff} which is becoming the standard one, it
is the process of Fig.~3.45c which is called the inclusive diffractive
process.}.  This leads in general to a much larger Higgs production cross
section at the LHC; see Tab.~3.1 [at the Tevatron all processes have too small
cross sections to be useful]. However, the background from $gg \to b\bar b$ is
also very large since there is no more the selection rule for spin--zero
particle production and the signal to background ratio is then very low. In
addition, one cannot use the missing mass technique to measure the Higgs 
mass [it has been suggested recently \cite{diff-Peshanski} to trigger on the
remnants to improve the $S/B$ ratio and to reconstruct the mass]. 
Nevertheless, besides the fact that these processes are actually the ones which
have been experimentally observed, since the CDF dijet data indicate the
presence of an additional soft hadronic radiation \cite{CDF-diffr}, they need
to be considered, first because they are potential backgrounds to the exclusive
process, and second because pseudoscalar Higgs bosons can be only produced in
these mechanisms. \s

In inclusive production [according to the terminology of
Ref.~\cite{Valery-myths}], Fig.~3.45b, the previous discussion on the process
of Fig.~3.45c also applies with the important exception that both incoming
protons are allowed to dissociate. This process has not received much attention
in the literature  as it has not the advantages of central exclusive
diffraction.  At the LHC, the production rates \cite{Valery-myths} are of the 
same order as in the previous case; Tab.~3.1.\s 

As mentioned previously, the treatment of diffractive processes involves a
mixture of perturbative and non--perturbative aspects of QCD. The rapidity gaps
for instance may be associated with the exchange of an effective Pomeron which
can be either a QCD Pomeron [which, at lowest order, is a $gg$ state] or a
phenomenological Pomeron fitted from e.g.~the HERA data. The non perturbative
aspect in exclusive diffraction arises when one attempts to calculate the
survival probabilities $S^2$ of the rapidity gaps, when secondary particles are
produced in the soft rescattering of the spectator partons and populate these
gaps. This probability is not universal and depends on the initial energy and
the considered final state. A recent estimate gives $S^2\!\sim\! 0.02$
while diffractive deep--inelastic processes at HERA and diffractive dijet
production at the Tevatron suggest, respectively, $S^2 \sim 1$ and $\sim 0.1$. 
Note that another probability for the gaps to be occupied arises from hard
gluon radiation in $gg \to H$ for instance; the latter can be, however,
calculated in perturbative QCD.\s

These non--perturbative aspects generate rather uncertain predictions of the
various models and, until recently, the spread in the predictions ranging over
several orders of magnitude.  A critical comparison of the various predictions
has been performed in Ref.~\cite{Valery-myths}, where it has been attempted to
explain the origin of the large differences. The conclusion was that either
different diffractive processes have been considered or important effects, such
as higher--order QCD corrections, have been neglected.  Many of the models, in
particular those which predict large Higgs production rates, are already ruled
out by Tevatron data on diffractive dijet production. Besides these theoretical
issues, experimental problems such as the possibility of triggering on the
events and the integration of the roman pot detectors within the machine,
remain still to be solved; see Ref.~\cite{diffr-Houches} for instance. \s

Note that the expectation for the clean exclusive central Higgs
production process can be checked at the LHC itself, since the main ingredients
which are needed for the calculation of the Higgs signal cross section are
involved in the calculation of dijet production with large rapidity gaps, $pp
\to p+ {\rm dijet}+p$. Since the latter can be measured from the side bands, one
can improve the prediction of the Higgs cross section. Other checks
can be performed \cite{diffr2}.\s 

In summary, diffractive processes in hadronic collisions provide an additional
means to produce the Higgs boson at the LHC. The double exclusive production
process allows a good measurement of the Higgs mass and a check of the SM Higgs
spin and parity quantum numbers [besides the selection which favors $0^{++}$ 
states, one can also use, for instance, the azimuthal asymmetry of the scattered
protons], which are notoriously difficult to verify at hadron colliders, as
will be discussed shortly. The production rates are, however, still uncertain
and the experimental conditions not yet established.  Many studies are being
performed and the situation might become more clear in a near future.  

\subsection{Detecting and studying the Higgs boson} 

\subsubsection{Summary of the production cross sections} 

The cross sections for Higgs boson production in the main channels,
eqs.~(3.1--3.4), are summarized in Fig.~3.46 at the Tevatron Run II with a c.m.
energy $\sqrt{s}=1.96$ TeV and in Fig.~3.47 for the LHC with $\sqrt{s}=14$ TeV
as functions of the Higgs boson mass, an update of
Refs.~\cite{Review-Michael,xs-old-update,xs-talks}.  They include the full NLO
QCD corrections which have been discussed earlier. The MRST sets of parton
densities \cite{MRST2001E} has been used for the cross sections. As inputs, we
use the central values for the fermion and gauge boson masses given in
eq.~(\ref{allmasses}), in particular we use $m_t=178$ GeV, while the strong
coupling constant is chosen to be $\alpha_s (M_Z)=0.119$ to match the value
that is incorporated in the PDFs at NLO. Here also, we will denote sometimes by
$pp$ both $pp$ and $p\bar p$ reactions and by ${\cal L}$ both ${\cal L}$ and
$\int\!{\cal L}{\rm d}\!t$.\s

The cross sections eqs.~(3.1) to (3.4) have been calculated using, respectively,
the NLO {\sc Fortran} codes {\tt V2HV, VV2H, HIGLU} and the LO code {\tt HQQ} 
of Ref.~\cite{Michael-Web} which are publicly available. A few remarks to 
explain how these cross sections have been obtained are in order:\s

-- In the $gg \to H$ mechanism, we display the NLO cross sections which have
been calculated for arbitrary quark masses, since the NNLO calculation is valid
only in the heavy top quark limit [although it is expected to be a good
approximation for the entire range if the Born amplitude contains the full
$m_t$ dependence].  However, we have set the renormalization and factorization
scales at $\mu_R\!=\!\mu_F=\! \frac{1}{2} M_H$. As discussed previously, in
this way the NLO (NNLO) correction increases (decreases) and the full result
approaches the total cross section at NNLO.  We have verified that the values
that we obtain are in a very good agreement with the NNLO values given for $M_H
\leq 300$ GeV in Ref.~\cite{ggH-NNLO-resum}.\s

-- In the case of the Higgs--strahlung processes, $pp \to HW$ and $pp \to HZ$, 
we incorporate the NLO and NNLO QCD corrections, including the $gg \to HZ$ 
component in the latter production process, as well as the electroweak 
radiative corrections [where we removed the kinks near the $2M_V$ thresholds]. 
For the PDFs, we will use the approximate densities which are included in the 
MRST2002 set and which approach very closely the exact result.\s

-- For the $pp \to t\bar{t} H$  production process, the NLO corrections have
become available only recently and the {\sc Fortran} codes for calculating 
the cross sections at this order are not yet publicly available. We therefore 
use the LO program {\tt HQQ} but we choose a scale for which the LO  
cross sections approach the NLO ones, i.e. $\mu_R=\mu_F\sim \frac{1}{2}M_H 
+ m_t$. We then multiply the cross sections by constant $K$--factors of 1.2 and 
0.8 for the LHC and the Tevatron, respectively, to approach the exact result. 
We have verified that, for $m_t=174$ GeV, the obtained rates are in a very good
agreement with the  NLO ones given in Ref.~\cite{Htt-NLO-DESY}.\s

-- The cross section for the vector boson fusion process $pp \to Hqq$ has
been calculated at NLO with the scales fixed to $\mu_R=\mu_F = Q_V$. No
kinematical cut has been applied. 

\begin{figure}[!h]
\begin{center}
\vspace*{-.7cm}
\hspace*{-1.cm}
\psfig{figure=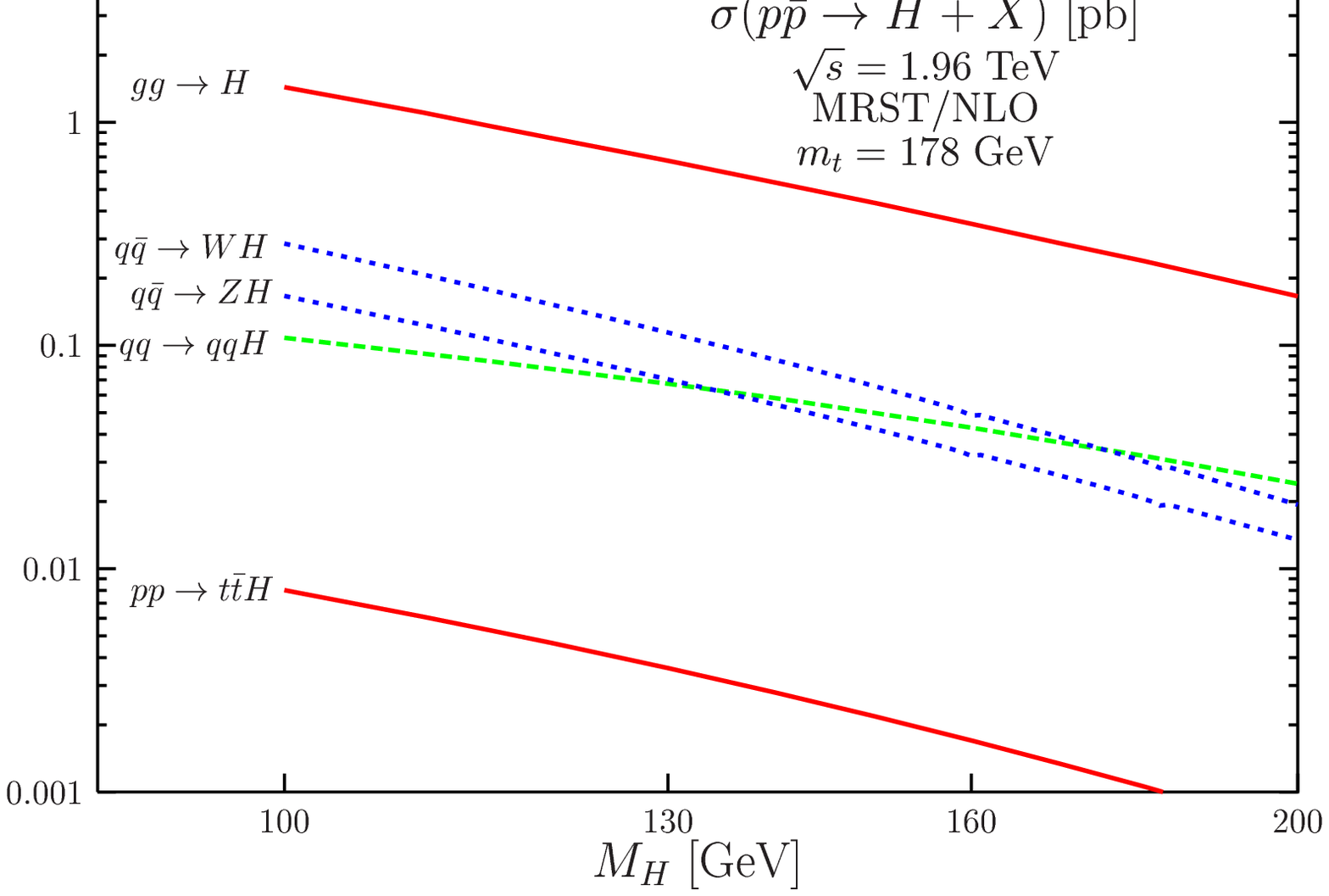,width=17cm}
\end{center}
\vspace*{-14.2cm}
{\it Figure 3.46: The Higgs boson production cross sections at the Tevatron  
in the dominant mechanisms as a function of $M_H$. They are (almost) at NLO
with $m_t=178$ GeV and the MRST set of PDFs has been used. The scales are
as described in the text.}
\vspace*{-.8cm}
\end{figure}

\begin{figure}[!h]
\begin{center}
\hspace*{-1.cm}
\psfig{figure=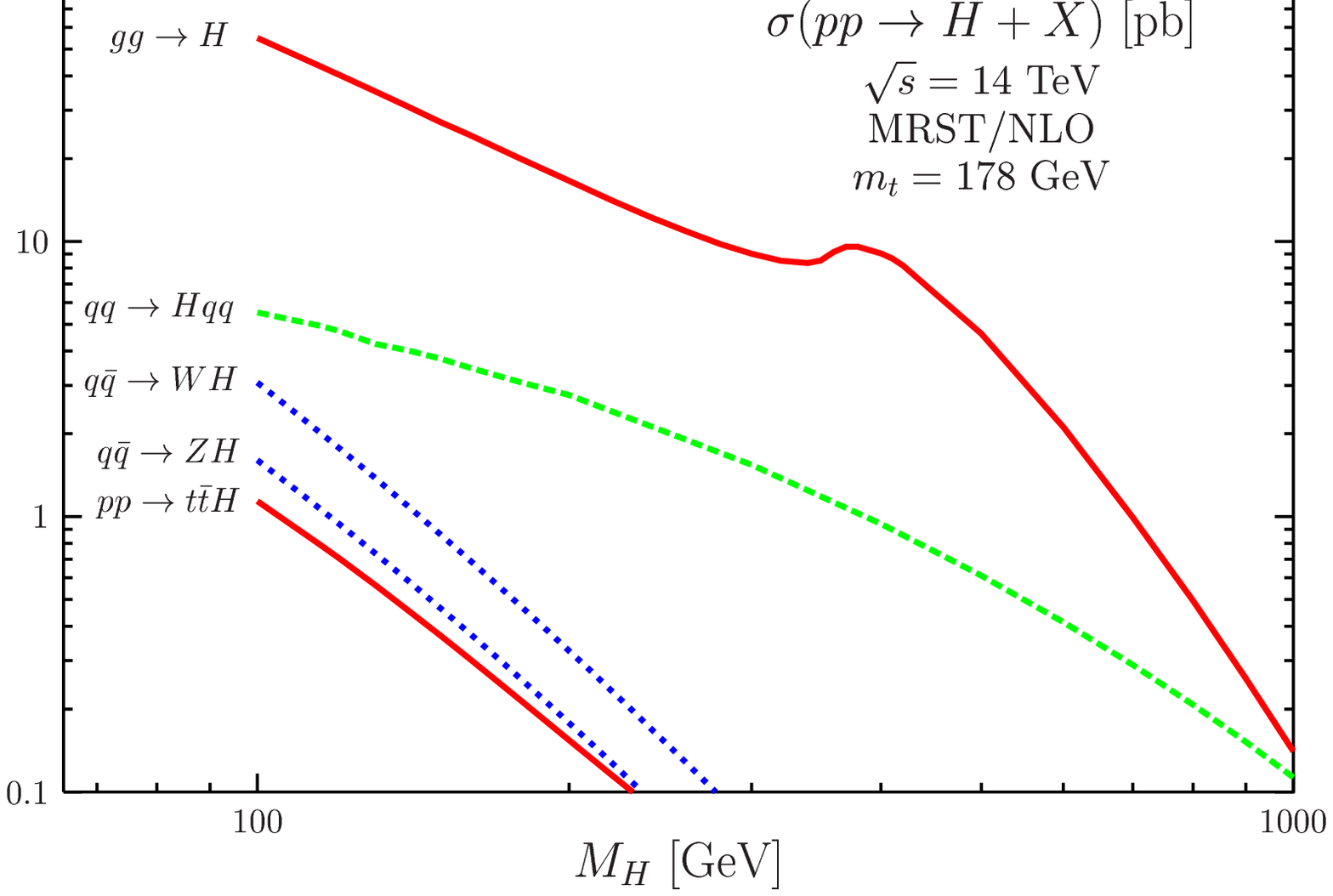,width=17cm}
\end{center}
\vspace*{-14.2cm}
\centerline{\it Figure 3.47: The same as  Fig.~3.46 but for the LHC.}
\vspace*{-7mm}
\end{figure}

\newpage

In Table 3.2, we display the numerical values of the cross sections for 
selected values of the Higgs mass that are relevant for the Tevatron and 
the LHC, as they might serve as useful inputs in other studies. The 
various SM input parameters are as discussed above.

\begin{table}[!h]
\begin{center}
\renewcommand{\arraystretch}{1.22}
\vbox{\columnwidth=26pc
\begin{tabular}{|c||c|c|c|c|c|c|}\hline
$M_H$ [GeV] & \ $\sigma(HW)$ \ & \ $\sigma(HZ)$ \ &   \ $\sigma (Hqq)$ \ & 
$\sigma(gg\to H)$ & \  $\sigma (H t \bar t)$ \\ \hline \hline
115 & 0.178  & 0.107 & 0.085 & 0.96 & 0.0053  \\
120 & 0.153 & 0.093 & 0.078 & 0.85 & 0.0047  \\
130 & 0.114 & 0.070 & 0.067 & 0.67 & 0.036  \\
140 & 0.086 & 0.054 & 0.058 & 0.54 & 0.0028  \\
150 & 0.065 & 0.042 & 0.050 & 0.43 & 0.0022  \\
160 & 0.048 & 0.032 & 0.043 & 0.35 & 0.0017  \\
170 & 0.039 & 0.026 & 0.037 & 0.29 & 0.0013  \\
180 & 0.030 & 0.020 & 0.032 & 0.24 & 0.0010  \\
200 & 0.019 & 0.013 & 0.024 & 0.17 &  --  \\
\hline
\end{tabular}
}
\end{center}
\vspace*{-5mm}
\end{table}

\begin{table}[!h]
\begin{center}
\renewcommand{\arraystretch}{1.22}
\vbox{\columnwidth=26pc
\begin{tabular}{|c||c|c|c|c|c|c|}\hline
$M_H$ [GeV] & \ $\sigma(HW)$ \ & \ $\sigma(HZ)$ \ & \ $\sigma (Hqq)$ & 
$\sigma(gg\to H)$ & \  $\sigma (H t \bar t)$ \ \\ \hline \hline
115 & 1.89 & 1.01 & 4.93 &  43.32  &    0.79     \\
120 & 1.65 & 0.89 & 4.72 &  40.25  &    0.70    \\
130 & 1.28 & 0.70 & 4.24 &  35.04  &    0.56    \\
140 & 1.00 & 0.55 & 4.01 &  30.81  &    0.45   \\
150 & 0.79 & 0.44 & 3.76 &  27.22  &    0.37   \\
160 & 0.62 & 0.35 & 3.49 &  24.44  &    0.31   \\
170 & 0.52 & 0.29 & 3.26 &  21.97  &    0.25   \\
180 & 0.42 & 0.24 & 3.07 &  19.87  &    0.21   \\
200 & 0.30 & 0.17 & 2.76 &  16.61  &    0.15   \\
300 & 0.04 & 0.07 & 1.54 &  9.02   &     --    \\
400 & -- & -- & 0.94    &  9.05   &     --    \\
500 & -- & -- & 0.61    &  4.62   &     --    \\
600 & -- & -- & 0.41    &  2.12   &     --    \\
700 & -- & -- & 0.29    &  0.99   &     --    \\
800 & -- & -- & 0.21    &  0.49   &     --    \\
900 & -- & -- & 0.15    &  0.26   &     --    \\
1000 & -- & -- & 0.11   & 0.14    ,&     --   \\
\hline
\end{tabular}
}
\end{center}
\vspace*{0mm}
{\it Table 3.2: Numerical values for SM Higgs production cross sections at 
the Tevatron (upper part) and the LHC (lower part) in picobarns for selected 
values of the Higgs mass.  These values have been obtained as in 
Figs.~3.46--3.47 and as explained in the text. }
\vspace*{-0.2cm}
\end{table}

As can be seen, in the interesting mass range favored by the electroweak
precision data, 100 $\lsim M_H \lsim 250$ GeV, the dominant production process
of the SM Higgs boson  at the LHC is by far the gluon--gluon fusion mechanism, 
the cross section being of the order a few tens of pb. In fact, this process
dominates all the way up to Higgs masses of the order of 1 TeV, where the cross
section is still sizable, $\sigma ( gg \to H) \sim 0.1$ pb.  It is followed by
the $WW/ZZ$ fusion process which has a  cross section of a few pb in the
interesting Higgs mass range above and which reaches the level of the $gg$
fusion cross section for very large $M_H$ values.  The cross sections for the
associated production with $W/Z$ bosons or $t\bar{t}$ pairs are one to three
orders of magnitude smaller than the $gg$ cross section and these processes
are only relevant in the mass range $M_H \lsim 250$ GeV.  For the luminosities
expected at the LHC, a very large sample of Higgs particles can be thus
collected before selection cuts are applied.\s

At the Tevatron, the most relevant production mechanism is the associated
production with $W/Z$ bosons [the $WH:ZH$ cross section ratio is approximately
$1.5$ for $M_H \lsim 200$ GeV], where the cross section is slightly less than
250 fb for $M_H \sim 120$ GeV when both processes are summed, leading to
$2.500$ Higgs events for the maximal luminosity expected at the Tevatron, $\int
{\cal L}=10$ fb$^{-1}$; the cross section drops to the level of less than 30 fb
for Higgs masses larger than 200 GeV. The $WW/ZZ$ fusion cross sections are of
the same order in the mass range $M_H \lsim 100$--200 GeV, while the cross
sections for associated production with $t\bar{t}$ pairs are rather low, being
less than 10 fb already for $M_H \sim 115$ GeV. The $gg$ fusion mechanism has
the largest production cross section, reaching the picobarn level for low Higgs
masses,  but suffers from a very large QCD two--jet background as will be
discussed later. \s

A huge effort, which already started in the late seventies, has been devoted to
the search of suitable signals to detect the Higgs boson at hadron colliders
and to suppress the various corresponding backgrounds, which are in general
very large. It is an impossible task to present here a detailed account of the
large number of theoretical and experimental studies which have been performed
in this context. In the next subsection, we simply summarize the Higgs
detection channels which are established since already some time, mostly
relying on the report of the Higgs working group in the case of the Tevatron
\cite{Higgs-TeV} [see \cite{pp-HV-expT} for earlier work] and in the case of
the LHC\footnote{The analyses at the LHC that will be discussed here will be
mostly based on Monte--Carlo \cite{PYTHIA,HERWIG} simulations which take into
account the parametrized \cite{ATLASFAST,CMSJET} or full detector response
\cite{ATLAS-TDR,CMS-TDR-True}.},
on the ATLAS Technical Design Report \cite{ATLAS-TDR} and CMS Technical
Proposal \cite{CMS-TDR} with some updates made in
Refs.~\cite{ATLAS-review,CMS-review,ATLAS+CMS} as well as on the joint
theoretical and experimental studies which have been performed at the three Les
Houches \cite{Houches1999,Houches2001,Houches2003} and at the 2001 Snowmass
\cite{Snowmass2001} Workshops [where some of the references to the original
work can be found]; see also Ref.~\cite{Karl-new}. Some of the important
backgrounds will be briefly mentioned and a detailed account can be found in
various reviews \cite{Top-LHC,EW-LHC,QCD-LHC,Houches-QCD}; see also
Ref.~\cite{Zepp-HC01}. For earlier work, we refer the reader to  {\it The Higgs
Hunter's Guide} where the pioneering analyses and a complete set of earlier
references can be found.  

\vspace*{-3mm}
\subsubsection{Higgs signals and backgrounds at the Tevatron and the LHC}

\vspace*{-1mm}
\subsubsection*{\underline{The $pp \to HW/HZ$ channel}}

It has been realized a long time ago \cite{pp-Galison,pp-EHLQ} that the
associated Higgs production with $W/Z$ bosons, with the latter decaying into
leptons $\ell = e^\pm, \mu^\pm$, is a potential channel for detecting the Higgs
particle at high--energy hadron colliders [the LHC and the late SSC]. This has 
been confirmed later for the LHC  in parton--level analyses in the case of the
photonic Higgs boson decays \cite{pp-HW-laa0,pp-HW-laa1}. Also more recently,
it has been shown that this production channel  is the most promising detection
mode at the Tevatron Run II for a relatively light Higgs boson which decays
dominantly into $b\bar b$ pairs \cite{pp-HW-bb-TeV,Gunion-Han}.\s 

In principle, the hadronic decays of the companion vector bosons cannot be used
[unless the Higgs boson itself does not decay into hadrons] as they are
overwhelmed by the huge irreducible QCD backgrounds. Since the branching
fraction BR$(W \to \ell \nu) \sim 20\%$ is larger than BR$(Z \to \ell \ell)
\sim 6\%$, and the cross section for $q \bar q' \to WH$ is a factor $\sim 1.5$
larger than for $q \bar q \to ZH$, the process $p\bar p\to HW \to H\ell \nu$
leads to five times more interesting events than the corresponding $p\bar p \to
HZ \to H\ell \ell $ process. Both channels have to be summed, however, to
increase the statistics.  In addition, the neutrino decays of the $Z$ boson
which have a substantial rate, BR($Z \to \nu \bar{\nu}) \sim 18\%$, can also be
considered.  The final signals depend on the decay modes of the Higgs boson
and, thus, on its mass and are summarized below.  \s

$\bullet$ \underline{$H \to b\bar{b}$}: the dominant Higgs decay mode for
$M_H \lsim 135$ GeV, leads to the final states $\ell \nu  b\bar{b}$, $\ell \ell
b\bar{b}$ and $\nu \bar{\nu} b\bar{b}$ that exhibit distinctive signatures
[isolated leptons and/or missing energy] which can be used at the Tevatron
where the backgrounds are not too large. The latter mainly originate from the
production of vector bosons plus two--jets, $p\bar p \to Vjj$ and in particular
$Wb\bar b$ \cite{pp-bkg-Vjj,pp-bkg-vecbos}, vector boson pairs, $p\bar p \to
VV$ \cite{pp-bkg-vecbos,pp-bkg-WW,pp-bkg-ZZ}, top quark pairs, $p\bar p\to
t\bar t$ \cite{pp-bkg-tt} and single top quarks $p\bar p \to t+X$
\cite{ppSingleTop}.  These  processes are known at least to NLO in QCD. 
$b$--tagging as well as the reconstruction of the $b\bar b$ invariant mass peak
are crucial to reject them.  Results based on the the SHW simulation
\cite{SHW-Tevatron} which gives the average response of the upgraded CDF and
D\O\ detectors in a simple way, show that these processes and, in particular, 
$p\bar p \to WH \to \ell \nu b\bar b$, are viable at the Tevatron
\cite{Higgs-TeV}. In the case of the $\nu \bar \nu b\bar b$ channel, a
significant $b\bar b$ background remains [which in Ref.~ \cite{Higgs-TeV} has
been assumed to be equal to the sum of the remaining backgrounds]. The
separation of the signal and backgrounds was optimized using neutral network
techniques which lead to an appreciable increase of the signal significance
\cite{NN-Tevatron,NN-Tevatron-U}. The $WH \to \ell \nu b \bar b$ channel has
been also discussed for the LHC \cite{pp-Galison,pp-HW-bb-LHC} but due to the
much larger QCD background, it is not considered alone as a clear discovery
channel
\cite{ATLAS-TDR,CMS-review,pp-HW+Htt-Froidevaux,pp-HW-bb-E,pp-HWHbb+Httbb-sim}.
The significance of the signal is at the level of $\sim 3\sigma$ for $M_H=120$
GeV with ${\cal L}=30$ fb$^{-1}$ \cite{pp-HW-bb-E} when the $Wb\bar b$ and
$t\bar t$  backgrounds have been sufficiently suppressed. This significance can
be, however, added to the one from $pp \to t\bar t H$ when it leads
to the same final state as will be discussed later.\s
 
$\bullet$ \underline{$H \to WW^{(*)}$}: which becomes the dominant Higgs
decay mode for $M_H \gsim 135$ GeV has also been considered both at the
Tevatron and the LHC \cite{pp-HW-lWW-LHC+Tev}. It receives irreducible
backgrounds from triple vector boson production, $pp \to WW+W/Z$
\cite{pp-bkg-VVV}, in addition to those from vector boson and $t\bar t$ pair
production. Distinctive signatures are trilepton $\ell \ell \ell$ events, like
sign dileptons and two jets $\ell^\pm \ell^\pm jj$ as well as a high--$p_T$
lepton pair plus missing $E_T$. At the Tevatron, the small production cross
section, the low luminosity as well as the small branching ratio $WWW \to 3\ell
\sim 10^{-2}$, make that only a few trilepton events can be observed even for 
30 fb$^{-1}$ data; the $\ell^\pm \nu \ell^\pm \nu jj$ signal
is only a factor of three larger.  The rates are thus too small for this
channel to be useful. The channel is more promising at the LHC, in particular
for a Higgs boson in the mass range $M_H \sim 160$--180 GeV, where it decays
almost 100\% of the time into $WW$ final states, and where the production cross
section is still large. Detailed simulations have shown that a significance
$S/\sqrt{B} \gsim 5$ can be obtained in the channel $pp \to HW \to \ell^\pm \nu
\ell^\pm \nu jj$  or $3\ell$ with a high luminosity ${\cal L}= 100$ fb$^{-1}$
\cite{pp-WHWW-sim}. \s

$\bullet$ \underline{$H \to ZZ^{(*)}$}: has a too small branching ratio for
$M_H \lsim 180$ GeV, when one of the $Z$ bosons is virtual. Above the $ZZ$
threshold, the $HV$ production cross section is very small at the Tevatron.
At the LHC, the cross section is still sizable and, once the leptonic branching
fractions of the $Z$ and $W$ boson in $HW$ production have been taken into 
account, a rate of $\sim 2$ fb can be obtained for $M_H\sim 200$ GeV before 
applying cuts. The few hundred $\ell \ell \ell \nu jj$ events which could be 
collected in the high luminosity regime might allow to detect the signal. To 
our knowledge, no simulation has been performed for this channel alone.\s 

$\bullet$ \underline{$H \to \gamma \gamma$}: is a decay mode that is too rare
to be useful at the Tevatron but it is the main detection channel at the LHC in
the low Higgs mass range for this production process. In fact, it was the first
channel in $HV$ production which has been shown in parton--level analyses to be
viable \cite{pp-HW-laa0,pp-HW-laa1}. The backgrounds are similar to those which
affect the process $gg \to H \to \gamma \gamma (j)$ to be discussed later: the
reducible ones are small \cite{pp-aal-Dubinin} and the irreducible $\gamma
\gamma \ell+\eslash$ and $\gamma \gamma \ell \ell$ backgrounds can be
suppressed by requiring high--$p_T$ and well isolated photons and lepton(s). 
Early analyses have shown that this signal is indeed viable at the LHC for a
Higgs boson in the low mass range \cite{ATLAS-TDR,CMS-TDR,pp-HWaa-sim} but the
signal has a small significance and should be combined with the one from $pp
\to t\bar t H \to \ell \gamma \gamma+ X$ as will be seen shortly.  A recent CMS
simulation \cite{pp-HWaa-simNew} has shown that in a one year of LHC at
high--luminosity,  a $5\, (4) \, \sigma$ significance can be obtained for the
signal if $M_H \lsim 135$ (150) GeV.\s

$\bullet$ \underline{$H \to \tau \tau$}: has a branching ratio of only a few 
percent for $M_H \lsim 135$ GeV and one cannot afford to let the associated 
gauge bosons to decay leptonically. The $pp \to ZH/WH \to jj \tau \tau$ channel
has been considered in a parton level analysis \cite{pp-HW-Mrenna} for the
Tevatron with the result that a significant improvement of $\tau$ identification
and a large luminosity might allow to detect the signal for low mass Higgs 
bosons if one can trigger on these events. No discussion of this channel has
been made in the Tevatron study of Ref.~\cite{Higgs-TeV} nor at the
LHC, though.   

\vspace*{-3mm}
\subsubsection*{\underline{The gluon--gluon fusion channel}}

This process, having the largest production cross section, has been considered
for a long time as being the most efficient one in the search for the Higgs
boson at the LHC. However, it appeared quite early that these searches cannot
be made in the dominant hadronic $H \to b\bar{b}$ and $H \to WW/ZZ \to 4j$ decay
channels because of the extremely large QCD jet backgrounds. The $H \to \tau^+
\tau^-$ signature in the low Higgs mass range is also very difficult to extract
at the LHC [and also at the Tevatron] because of these backgrounds. One has
then to rely on rare Higgs decays which provide clean signatures involving 
photons and/or leptons for which the backgrounds are smaller but far from being 
negligible.\s

$\bullet$ \underline{$H \to \gamma \gamma$}: has been proposed rather early
\cite{pp-Galison,pp-Wudka} and is the ``silver" detection channel for a Higgs
boson with a mass below 150 GeV \cite{pp-Haa}.  The reducible QCD background
from jets faking photons is huge and a rejection factor of ${\cal O}(10^6)$ is
needed to bring it down to the level of the irreducible one from direct $q\bar
q \to \gamma \gamma +X$ production and the loop induced channel $gg\to\gamma
\gamma+X$ which provides a 50\% contribution.  These have been studied in great
detail \cite{pp-Haa-bckg1} and the state--of--the--art higher--order results
are contained in the program {\tt DIPHOX} \cite{Diphox} which also includes the
fragmentation effects. These backgrounds can, in principle, be determined by
measuring the two--photon invariant mass distribution ${\rm d}\sigma/ {\rm d}
M_{\gamma \gamma}$ on both sides of the resonance peak. However, they need to
be precisely calculated for the evaluation of the detection significance and
when it comes to measurements of the Higgs properties \cite{Zepp-HC01}.  A
reconstruction efficiency of about 75\% can be achieved for a single photon and
for $M_H \sim 130$ GeV, the final signal to background ratio is of the order of
$1/30$ in a window of $M_{\gamma \gamma}\sim 2$ GeV. However, since the decay
is rare, a large amount of luminosity needs to be collected. One could then
use, in addition, the $pp \to Hj$ signal \cite{ggHp-dubinin} as the $gg \to
\gamma \gamma g$ background with a hard jet has been found to be much smaller.
In fact, at low luminosities, the combination of all $H \to \gamma \gamma$
channels is required: not only the $pp \to \gamma \gamma$ and $pp \to \gamma
\gamma j$ channels but also the channel $pp \to \gamma\gamma +\ell$ where the
additional lepton comes from the decay of a $W$ boson in the associated $HW$
production process discussed previously or in $t\bar t H$ production with $t
\to bW \to b\ell \nu$ as will be seen later.\s

$\bullet$ \underline{$H \to ZZ^{(*)}$}: in the high mass region, $M_H >2M_Z$,
the decay $H \to ZZ \to 4\ell$ is the ``gold--plated" mode
\cite{pp-Galison,pp-Wudka,pp-EHLQ,pp-HWW-Theory}, allowing for Higgs detection
up to masses of ${\cal O}(1\,$TeV) \cite{ATLAS-TDR,pp-HZZ-4l,pp-HZZ-4l-new}.
The main background is due to continuum $ZZ$ production which is known rather
precisely \cite{pp-bkg-WW,pp-bkg-ZZ} but which can be also directly measured
from the side bands of the resonance peak and interpolated to the signal
region. For $ M_H \gsim 600$ GeV, high enough luminosities are required  since
BR$(H\to ZZ \to 4\ell) \sim 0.1\%$ is small and the total Higgs width becomes
large.  To increase the statistics,  one can use in addition the $H \to ZZ \to
\ell \ell \nu \nu$ decays \cite{pp-HZZ-llnnTheory} in which the signal appears
as a Jacobian peak in the missing transverse energy spectrum. Additional
backgrounds from $Zj$ events \cite{pp-bkg-Vjj}, where the ~$\eslash_T$ is due
to neutrinos from semi--leptonic $b$ decays for instance, need to be considered
\cite{pp-HZZ-llnn}. Allowing one of the $Z$ bosons to be virtual, the discovery
reach can be extended down to masses $M_H \sim 120$ GeV using $H \to ZZ^* \to
4\ell$ decays \cite{pp-HZZ-4low}, except in the range $M_H \sim 2M_W$--$2M_Z$
where BR($H\to ZZ^*$) is too small. In this case, a very sharp peak can be
observed in the $4\ell$ invariant mass distribution. Here, additional
backgrounds from $t\bar t$ \cite{pp-bkg-tt} and $Zb\bar b$ \cite{pp-bkg-Vjj}
production contribute besides $ZZ^*, Z\gamma^*$ events.\s 

$\bullet$ \underline{$H \to WW^{(*)}$}: leading to $ \ell \ell \nu \nu$ final
states turned out to be one of the most promising detection modes of a light
Higgs boson at the LHC, i.e. from $M_H \sim 2M_Z$ down to $M_H \sim 120$ GeV
\cite{pp-HWW-lnln,pp-HWW-lnln2}, and it is even a potential discovery mode at 
the Tevatron \cite{pp-WW-TeV}. Indeed, BR($H \to WW)$ is appreciable if not 
largely dominating in this mass range and the clean leptonic decays represent 
4\% of the initial $WW$ sample. Since the Higgs mass cannot be reconstructed in
this process, the signal should be observed from a clear excess of events above
backgrounds which need, thus, to be known rather precisely. The most important
source is due to $W$ boson \cite{pp-bkg-WW} and top quark \cite{pp-bkg-tt} pair
production. The latter can be removed with suitable cuts, while for the former
one needs, in addition, to take advantage of the characteristic spin
correlations in the $H \to WW^* \to \ell \nu \ell \nu$ decays
\cite{CPHVVpol,pp-HWW-lnln}: the azimuthal separation of the charged
leptons, for instance, peaks at smaller values for the signal than for the $WW$
background.  A clear signal has been established at the LHC for Higgs masses 
down to $M_H \sim 120$ GeV if enough luminosity is collected
\cite{pp-HWW-lnln,pp-HWW-lnln2}.  At higher Higgs masses, the additional channel
$H\to WW\to \ell\nu jj$, eventually combined with $H\to ZZ\to \ell\ell jj$ and 
with the $H\to ZZ\to \ell \ell\nu \bar \nu$ channel discussed previously, would
extend the discovery reach to masses up to 1 TeV at high luminosities, after
reducing the enormous $t\bar t$ and $W+$jets backgrounds \cite{H-WW-lnjj-high}.
At the Tevatron, high $p_T$ lepton pairs plus missing energy $\ell \ell \nu
\bar \nu$ and like--sign dileptons plus jets $\ell^\pm \ell^\pm jj$ in $gg \to
H\to WW^*$, when combined with similar events in $p\bar p \to HW/HZ$ associated
production, would allow to detect the Higgs boson at the $3\sigma$ level for
masses up to $M_H \sim 180$ GeV with 30 fb$^{-1}$ data \cite{pp-WW-TeV}.\s 

$\bullet$ \underline{$H \to \tau^+ \tau^-$}: has been proposed long ago
\cite{pp-ggH-tau-old,pp-Hgg-PT} in associated $gg \to Hg$ production
where the additional jet provides a significant transverse momentum to the
$\tau^+\tau^-$ system.  To our knowledge, the process has not been considered
for the LHC in a realistic experimental simulation [at least not in positive
terms]; see Ref.~\cite{pp-ggH-tau}, however. The process has been discussed for
the Tevatron \cite{pp-tau-TeV} but, again, it needs a better identification of 
the $\tau$--leptons and resolution on the missing $E_T$ \cite{Higgs-TeV}.\s

$\bullet$ \underline{$H \to \mu^+ \mu^-$}: the signal in this very rare decay
channel, BR($H \to \mu^+\mu^-) \sim 10^{-4}$ for $M_H \sim 115$--140 GeV,  
is rather clean but it needs a very large amount of luminosity: ${\cal L}= 300$
fb$^{-1}$ is required for a 3$\sigma$ signal in the Higgs mass range $M_H \sim 
120$--140 GeV \cite{pp-ggH-mu} [for lower masses one is still sensitive to the
tail of the huge Drell--Yan $pp \to \gamma^*/Z \to \mu^+ \mu^-$ background]. 
This process is, thus, more appropriate for the SLHC or VLHC.\s 

$\bullet$ \underline{$H \to t\bar t$}: suffers from the huge $t\bar t$
continuum background which has to be evaluated in a large mass window as the
Higgs total width is large. It has been shown in
Refs.~\cite{pp-Htt-gg,pp-Htt-ggA} that the surplus from Higgs events produces a
dip--peak structure in the $gg \to t\bar t$ invariant mass spectrum which could
have been observable at the LHC if the Higgs total width were smaller [as in
extensions of the SM where the Higgs has reduced couplings to the vector
bosons]. This is unfortunate as this process would allow to probe directly the
$Ht\bar t$ couplings and to check, for instance, the presence of anomalous
interactions and/or CP violation \cite{CPHttpol,CPHttpolA}.

\subsubsection*{\underline{The $WW/ZZ$ fusion channel}}

This channel is not considered as being viable at the Tevatron. In the study of
Ref.~\cite{Higgs-TeV}, it has been shown that even with a good resolution on
the  $b\bar b$ invariant mass, $\Delta m_{b\bar b} = \pm 10$ GeV, the signal to
background ratio in the $p\bar p \to  qq b \bar b$ channel is of ${\cal
O}(10^{-3})$ within a 20 GeV $m_{b\bar b}$ bin. For cleaner decay modes of the
Higgs boson, the $p\bar p \to Hqq$ production cross section is too small to be
useful: for ${\cal L}\sim 10$ fb$^{-1}$ for instance, only two $H \to \gamma
\gamma$ events and four dileptons events from $H \to \tau^+ \tau^-$ are
expected before acceptance cuts are applied. \s

At the LHC, the cross section is two orders of magnitude larger and the double
forward jet tagging as well as the central soft--jet vetoing [the latter still
needs more studies to be more firmly established] discussed previously help to
drastically suppress the various large backgrounds.  Applying the specific
vector boson fusion cuts discussed in \S3.3.3, the signal cross section is
still large, a few picobarns for Higgs masses in the range $M_H=100$--200 GeV,
while the signal to background ratio is of order one. In addition, these
specific cuts allow to distinguish between this mechanism and the $gg \to H+2j$
process as discussed in \S3.4.4 [only $\sim 10\%$ of the latter is left after
cuts].  Adding the fact that it is theoretically rather clean, since the
$K$--factors, the renormalization and factorization scale dependence as well as
the PDF uncertainties are rather small, this process will thus play a key role
when it comes to extract the Higgs couplings to the SM particles at the LHC. 
For this purpose all possible decay channels of the Higgs boson must be
considered. \s

$\bullet$ \underline{$H \to \tau^+ \tau^-$:} is a promising channel for
$M_H\sim 120$--140 GeV if enough luminosity, ${\cal L}\sim 30$ fb$^{-1}$, is
available. This has been established first with parton level analyses
\cite{Zepp-tau} which were later confirmed by detector simulations
\cite{ATLAS-review,AzuelosHC01,Exp:Hqq-tau}. In particular, for $M_H \sim 125$
GeV, a statistical  significance of $2.3, 2.5$ and $\sim 4.5\sigma$ can be
achieved in the channels $qqH \to qq \tau \tau \to qq ee/\mu \mu + \eslash +X$,
$qq e\mu + \eslash +X$ and $qq \ell h + \eslash +X$, respectively, for the
luminosity quoted above \cite{AzuelosHC01}, leading to a combined significance
of $\sim 6\sigma$. The $\tau^+ \tau^-$ invariant mass can be determined at the
level of 10\% which would allow to measure the backgrounds [the major ones
being QCD and electroweak $Zjj$ production with $Z\to \tau^+ \tau^-$, in
addition to the usual $VV$ and $t\bar t$ processes] from the side bands. 
$\tau$--polarization effects \cite{tauola} are useful to discriminate the
decays $H \to \tau^+ \tau^-$ from the Drell--Yan $\gamma^*, Z^* \to \tau^+
\tau^-$ background.\s

$\bullet$ \underline{$H \to \gamma \gamma$}: as shown in a parton level
analysis \cite{Zepp-gamma}, this is a rather clean channel with backgrounds
which can be measured directly from the data. However, since the decay is rare,
the channel needs a  high luminosity and eventually has to be combined with
other production processes [such as the $gg \to H \to \gamma \gamma (j)$ and
$pp \to WH \to \ell\gamma \gamma$ channel discussed previously] to allow for  a
significance that is larger than $5\sigma$ for masses below $M_H\lsim 150$ GeV.
For instance, in the CMS simulation performed in Ref.~\cite{pp-Hqq-Dubinin} and
where only the irreducible $\gamma \gamma jj$ background has been included with
an assumed two--photon invariant mass window of $\pm 3$ GeV, it has been found
that a statistical significance of 3--5$\sigma$ can be obtained in the mass
range $M_H=115$--140 GeV with ${\cal L}=30$ fb$^{-1}$ data. \s

$\bullet$ \underline{$H \to WW^{(*)}$}: although feasible and competitive
\cite{Zepp-WW}, this channel might prove to be rather difficult since one
cannot reconstruct the Higgs mass peak and, thus, measure the background from
the side bands. The most important backgrounds, $pp \to t\bar t+$ jets and
$WWjj$ production, need therefore to be known precisely; QCD+EW $\tau\tau jj$
production \cite{pp-bkg-Vjj} can be removed with suitable cuts.  Recently, this
mode has also been studied experimentally and the prospects are rather good
\cite{ATLAS-review,Exp:Hqq-WW,High-mass-range}. In the ATLAS analysis of the
$qqH \to qq \ell \nu \ell \nu$ channel \cite{ATLAS-review}, the signal [with
the usual specific vector boson fusion cuts, the contributions of the $pp \to
WH,ZH$ and $t\bar tH$ processes to this topology are small] and the $t\bar t$
plus zero, one and two--jet backgrounds [the other important background,
$pp \to \gamma^*/Z+ X$ with $Z \to \tau^+ \tau^-$, can be rejected by requiring
a high $\ell \ell \nu$ transverse mass] have been studied. It has been shown
that a significance larger than $3\sigma$ can be obtained for a luminosity of 
${\cal L}= 10$ fb$^{-1}$ in both the $e\mu +X$ and $ee/\mu \mu+X$ channels in
the Higgs mass range above $M_H\sim 130$ GeV, i.e.\ when BR($H\to WW^*$) is
large enough.  Combining these channels with the $\ell\nu jj$ mode [and with
the standard $\gamma \gamma$ and $ZZ^*$ channels], one then obtains a $\sim
5\sigma$ significance for the $M_H$ and ${\cal L}$ values  above
\cite{Karl-new}.  In fact, the $pp \to qqH \to \ell \nu jj$ channel becomes
very powerful at higher Higgs masses
\cite{ATLAS-review,CMS-review,High-mass-range,Exp:Hqq-WW-qqlnHeavy}. With a
slightly different optimization of the cuts than at low Higgs masses, one can
arrive at a signal significance that is larger that $5\sigma$ in the entire
mass range $M_H \sim 200$--800 GeV with a luminosity of ${\cal L}= 30$
fb$^{-1}$ \cite{Exp:Hqq-WW-qqlnHeavy}. \s

$\bullet$ \underline{$H \to ZZ$}: this channel has also been considered in
experimental simulations \cite{Exp:Hqq-WW-qqlnHeavy,Exp:Hqq-ZZ}, but it cannot
be used below the $M_H =2M_Z$ threshold as the $H\to ZZ^*$ branching ratio is
very small in view of the not so large production rate. In addition, since the
rates in the $H \to ZZ \to 4\ell$ channel are also very tiny, on has to
consider the final states $\ell \ell \nu \nu$ and $\ell \ell jj$. These
processes receive large backgrounds, in particular from the process $Z+4j$ in
the second case where one has $S/B \sim 1/3$ at $M_H \sim 300$ GeV, after all
cuts have been imposed. In the high Higgs mass range, these channels can be
useful, but they need again very high luminosities. For instance, it has been 
shown in Ref.~\cite{Exp:Hqq-WW-qqlnHeavy} that a significance of more
than $5\sigma$ can be achieved for the $qqH \to \ell^+ \ell^- \nu \nu jj$
signal with a luminosity of ${\cal L}= 30$ fb$^{-1}$ in the Higgs mass window
$M_H \sim 500$--800 GeV.\s

$\bullet$ \underline{$H \to \mu^+\mu^-$}: again, the signal in this channel is 
very clean but the branching ratio for the decay is too small. A large amount 
of luminosity, ${\cal L}= 0.5$--1 ab$^{-1}$, is required for a  3$\sigma$ 
signal in the Higgs mass range below 140 GeV \cite{VVH-mu}. This signal should 
be combined with the $\mu^+ \mu^-$ sample obtained from gluon--gluon fusion
discussed previously.\s 

$\bullet$ \underline{$H \to b\bar{b}$}: this channel suffers from a huge $4j$
QCD background which can possibly be measured using the side bands of the
$b\bar{b}$ invariant mass; in addition, it has a major problem with overlapping
events. A preliminary parton level analysis \cite{VVH-bb} shows that with
reasonable assumptions but with a very large luminosity,  ${\cal L}= 600$
fb$^{-1}$, one can obtain a signal to square--root background ratio of
$S/\sqrt{B} \sim 3$ for a mass $M_H \sim 120$ GeV. However, it is not yet very
clear if it possible to trigger efficiently on this channel
\cite{VVH-bb-trigger}.  

\subsubsection*{\underline{The $pp \to t\bar t H$ channel}}
 
Finally, Higgs production in association with top quarks has a strongly
decreasing cross section with increasing $M_H$ which makes the process useful
only in the low mass range, $M_H \lsim 135$ GeV, when the
$\gamma \gamma$ and $b\bar b $ Higgs decays are relatively important [see
below, however]. In addition, one needs to have at least one of the $W$ bosons
from $t\to bW$ which decays leptonically, to trigger efficiently on the events 
and suppress the QCD backgrounds.
Since the rates are rather low, a large luminosity is required, in
particular at the Tevatron.\s

$\bullet$ \underline{$H \to b\bar{b}$}: associated Higgs production with $t
\bar t$ pairs \cite{pp-Htt-bb} is the only channel in which it has been firmly
established that the Higgs decays into $b\bar b $ pair can be extracted from
the backgrounds at the LHC \cite{pp-HW+Htt-Froidevaux}. A clear evidence of the
$4b$ tagged jet and lepton signal above the $W+\,$ jets and $t\bar t+jj$
backgrounds [$b$--tagging is of course crucial here] can be obtained for $M_H
\lsim 130$ GeV if enough luminosity, ${\cal L}\gsim 100$ fb$^{-1}$, is
collected \cite{pp-ttHbb-sim}. A clear reconstruction of the $H \to b \bar b$
mass peak is difficult because of the combinatorial background from the signal
itself and the reconstruction of the top quark decays might be needed.  The
fully hadronic final state $pp \to t\bar t H \to q\bar q q\bar q b\bar b b\bar
b$ would double the number of $pp \to t\bar tH$ signal events
\cite{pp-Htt-jets} but one still needs a proper evaluation of the eight jet QCD
background. At the Tevatron, the channel is more challenging as the production 
rate is very small but a signal might be visible if a very high enough 
luminosity is collected \cite{pp-Htt-phenoT,Higgs-TeV}.\s 

$\bullet$ \underline{$H \to \gamma \gamma$}: the decay is too rare for the
Tevatron, but it can be detected at the LHC when an additional lepton from the
$t \to bW$ decay is present \cite{pp-Htt-gamma}. The process, again, gives a
narrow mass  peak which is visible for $M_H \lsim 140$ GeV when the $pp \to t
\bar tH$ production rate and the $H\to \gamma \gamma$ branching ratio are
sizable enough \cite{pp-Htt-gamma}. With the additional charged lepton, the
backgrounds are manageable \cite{pp-Htt-gamma1} after suitable cuts [it can
also be measured from the side bands], but the statistics have to be added to
those obtained in the search of the $\gamma \gamma$ peak in the three other
Higgs production channels $pp \to H\to\gamma \gamma, \gamma \gamma+j$, $pp\to
qqH \to qq \gamma \gamma$ and  $pp\to HW \to \gamma \gamma \ell \nu$ to obtain
a significant signal at moderate luminosities.\s 

$\bullet$ \underline{$H \to WW^{(*)}$}: with $\ell \ell \nu \nu$ final states,
has been suggested recently to extend the reach of the $t \bar tH$ process to
Higgs masses above $\sim 140$ GeV \cite{pp-Htt-WW}. This mode receives very
large $t \bar t Wjj$ and  $t \bar t  \ell \ell$ + jet backgrounds [and smaller
ones from $t\bar tWW$ and $t\bar t t\bar t$ final states] which need to be
accurately determined as the invariant Higgs mass peak cannot be reconstructed
and one would have to rely on a counting of the signal versus the background
events. It has been recently shown that a signal can be observed
\cite{pp-ttHWW-sim} but further investigations are needed to confirm that the
backgrounds can be indeed reduced to a low level.\s

$\underline{H \to ZZ^*}$ has been discussed in Ref.~\cite{pp-Htt-VV} but it 
has too small rates for $M_H \lsim 2M_Z$ if leptonic $Z$ decays are selected; 
to our knowledge, no simulation for this channel is available.\s

$\bullet$ \underline{$H \to \tau \tau$}: this channel has also been discussed 
\cite{pp-Htt-tau} in a parton--level analysis. It seems extremely challenging 
and, again, no detailed experimental simulation has been performed.  

\subsubsection{Discovery expectations at the Tevatron and the LHC}

At the Tevatron, the required luminosity to discover or exclude a SM Higgs
boson, combining all channels in the processes $p\bar p \to HV$ and $gg \to H$
discussed previously, and the results of both CDF and D\O\ experiments, is
shown in Fig.~3.48 as a function of $M_H$ \cite{Higgs-TeV}. With 10 fb$^{-1}$
luminosity per experiment, a 3$\sigma$ evidence for a Higgs boson can be
achieved for $M_H \lsim 125$ GeV and, in the absence of any signal, a 95\% CL
exclusion limit can be set up to Higgs masses of order 180 GeV. However, for
discovery, only 30 fb$^{-1}$ data per experiment will allow to observe a
$5\sigma$ signal for $M_H \lsim 130$ GeV, slightly above the LEP2 Higgs mass
bound. Unfortunately, these large luminosities are not expected to be reached
in Run II. \s

\begin{figure}[!h]
\vspace*{-0.5cm}
\begin{center}
\includegraphics[width=13.cm]{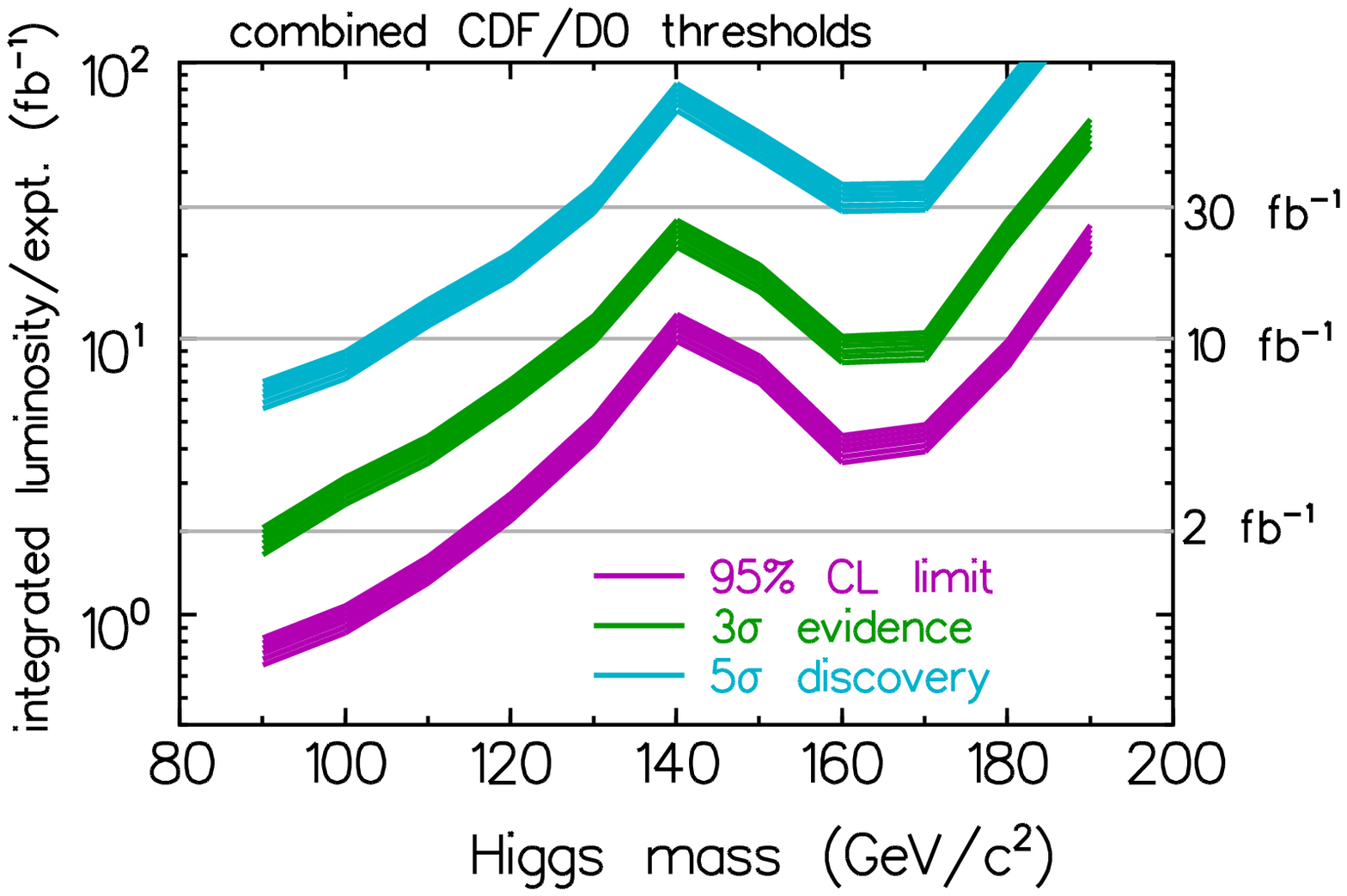}
\vspace*{-0.6cm}
\end{center}
{\it Figure 3.48: The integrated luminosity required per experiment at the 
Tevatron, to either exclude a SM Higgs boson at 95\% CL or observe it at the 
$3\sigma$ or $5\sigma$ level; from 
Ref.~\cite{Higgs-TeV}.} 
\vspace*{-0.5cm}
\end{figure}

At the LHC, the significance of the signals in the various Higgs production and
decay channels are shown as a function of $M_H$ in Figs.~3.49 and 3.50. 
The ATLAS plot in the left--hand
side of Fig.~3.49 shows the significance for an integrated luminosity of ${\cal
L}= 100$ fb$^{-1}$ in the ``standard" search channels where the vector boson
fusion processes are not yet included. The detection in this case relies mostly
on the $gg \to H$ production mechanism with the decays $H\to \gamma \gamma,
WW^{(*)}$ and $ZZ^{(*)}$ [where one of the vector boson is allowed to decay
hadronically in the high Higgs mass range], supplemented by the processes $pp
\to t\bar  tH$ with $H \to \gamma \gamma, b\bar{b}$ and $pp \to WH$ with $H \to
\gamma \gamma$. As can be seen, the significance is above 10 in the entire
Higgs mass range when the various channels are combined. The significance is
the smallest in the low mass range, $M_H \lsim 130$ GeV, when the $H \to
VV^{*}$ decays are not yet dominant. This is exemplified in the right--hand
side of the figure where the significance is shown in the mass range below
$M_H=200$ GeV but with the luminosity ${\cal L}= 30$ fb$^{-1}$ which is
expected at an earlier stage. The updated analysis now includes the vector boson
fusion channels with the decays $H\to \tau \tau$ and $H\to WW^*$ which  lead to
a substantial increase of the total significance. Note that the $K$--factors,
which would have significantly increased the signal for the $gg \to H$
process that is mostly used at high $M_H$, have unfortunately not been
included [see the discussion below].

\begin{figure}[!h]
\vspace*{-1.cm}
\begin{center}
\epsfig{file=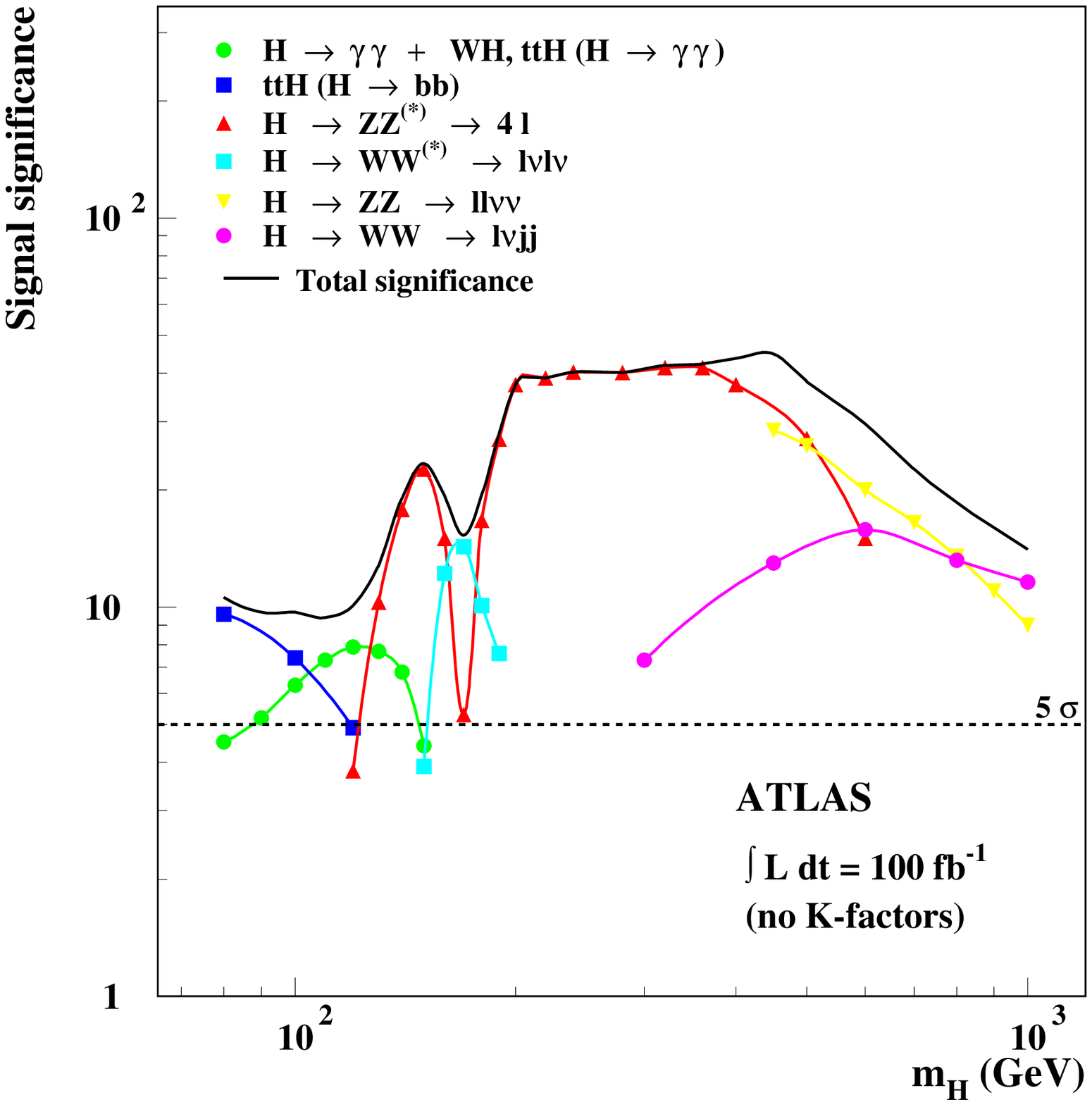,width=8.cm,height=8cm}\hspace*{2mm}
\epsfig{file=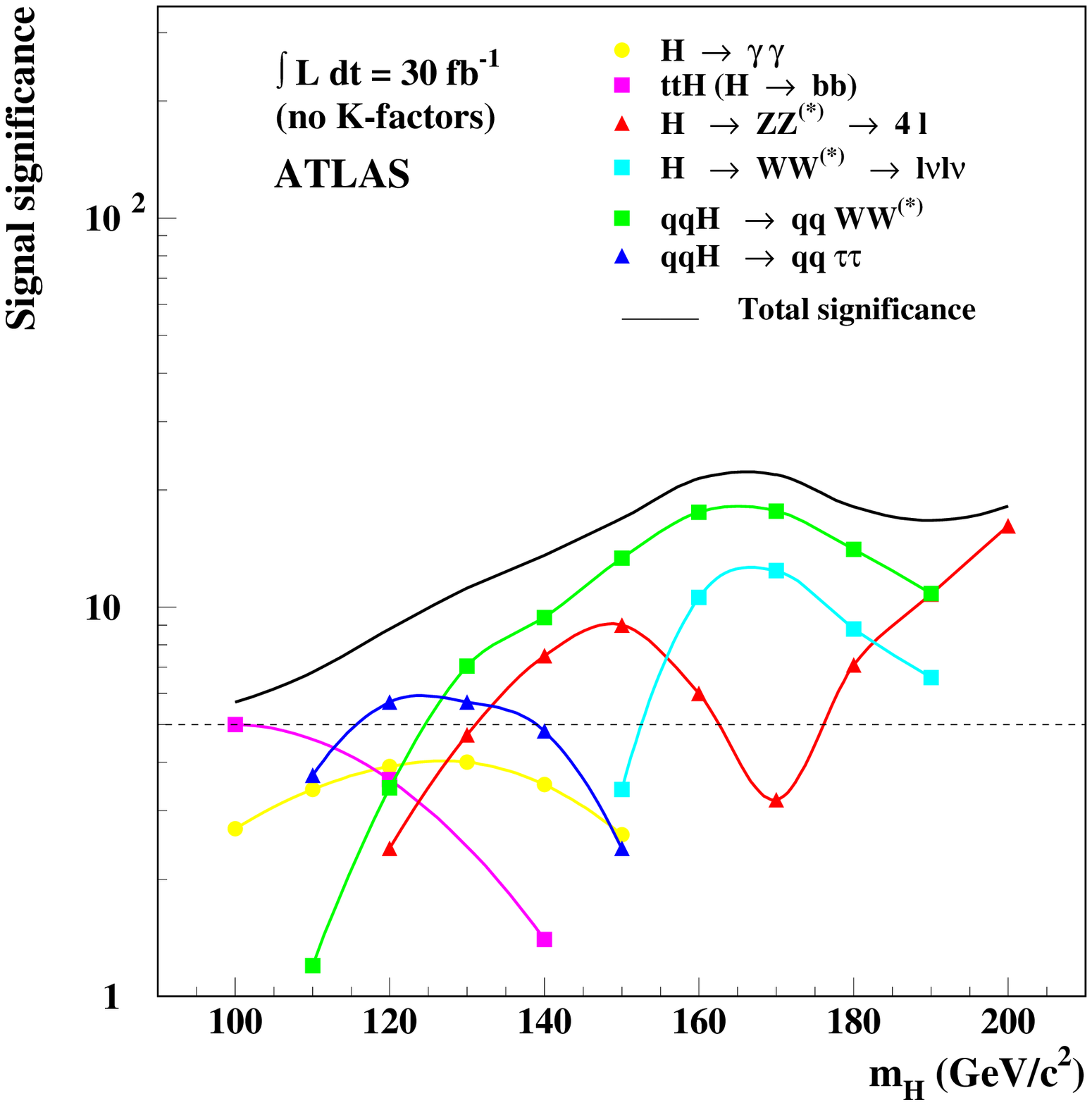,width=9.cm,height=9.cm}
\end{center}
\vspace*{-.2cm}
{\it Figure 3.49: The significance for the SM Higgs boson discovery in various 
channels in ATLAS as a function of $M_H$. Left:  the significance for 100 
fb$^{-1}$ data and with no vector boson fusion channel included and right: 
for 30 fb$^{-1}$ data in the $M_H \leq 200$ GeV range with the $qq \to qqH$ 
channels included \cite{ATLAS-review}.}
\vspace*{-.4cm}
\end{figure}

The CMS plot in Fig.~3.50 shows the integrated luminosity that is needed to
achieve a $5\sigma$ discovery signal in the various detection channels. Here,
the vector boson fusion process with all relevant Higgs decays, $H \to \gamma
\gamma, \tau \tau, WW^{(*)},ZZ^{(*)}$, has been included [together with the
$K$--factors for the $gg\to H$ process]. As can be seen, a minimal luminosity
of 10 fb$^{-1}$ is necessary to cover the low Higgs mass range down to $M_H
\sim 115$ GeV and the high mass range up to $M_H \sim 800$ GeV when all
channels are combined. One can see also
that the vector boson fusion channels add value in the entire Higgs mass range.
In particular, the $qq \to Hqq$ processes with $H \to  WW,ZZ$ are also very 
useful in the high Higgs mass range. \s

\begin{figure}[!h]
\vspace*{-.5cm}
\begin{center}
\epsfig{file=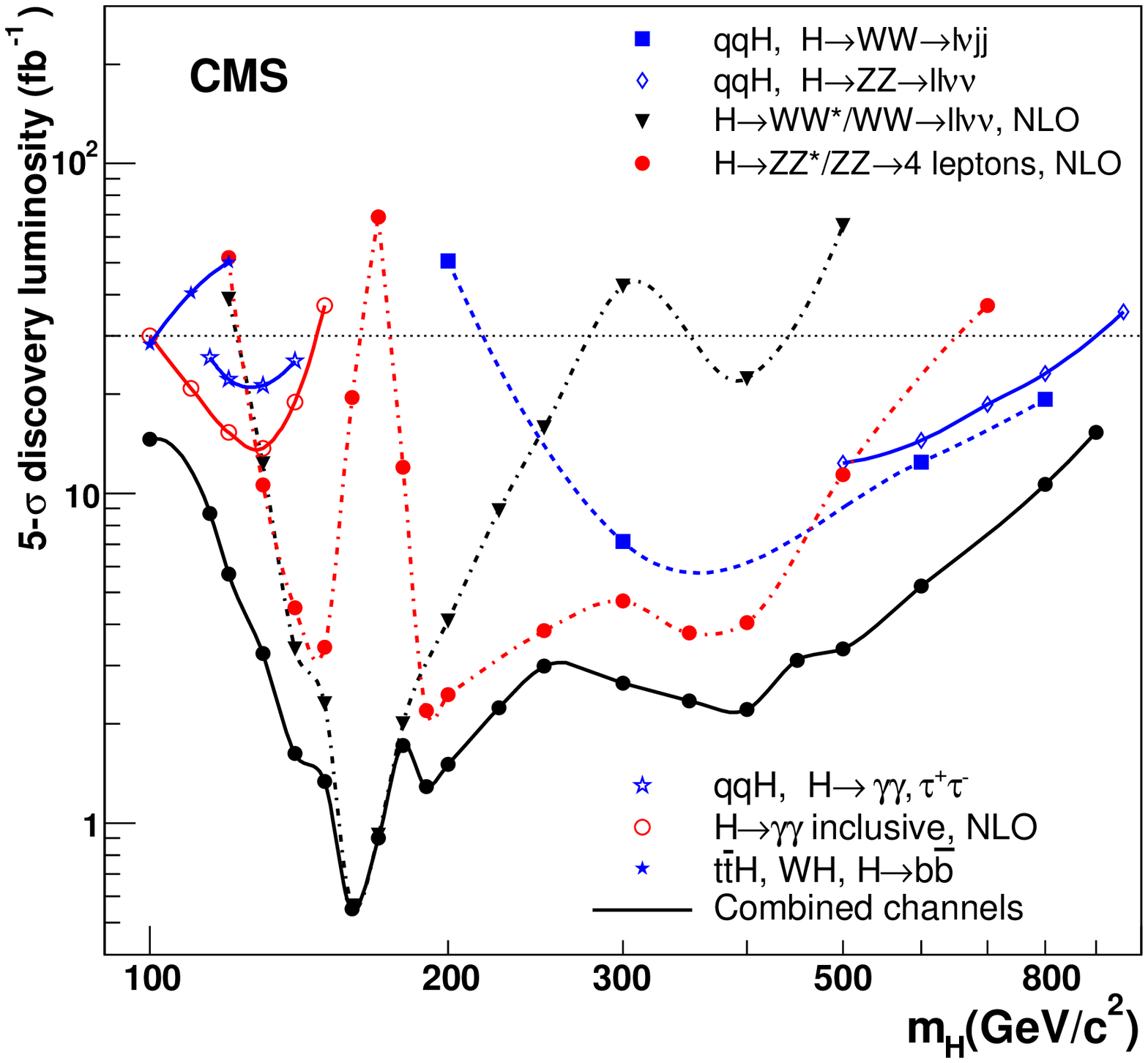,width=12.cm,height=9cm}
\end{center}
\vspace*{-.2cm}
{\it Figure 3.50: The required integrated luminosity that is needed to achieve 
a $5\sigma$ discovery signal in CMS using various detection channels as a
function of $M_H$ \cite{CMS-review}.}
\vspace*{-.2cm}
\end{figure}

Thus, the SM Higgs boson in its entire mass range will be found at the LHC
provided that a luminosity larger that $\int {\cal L}= 30$ fb$^{-1}$ is
collected and the performances of the detectors are as expected. For higher
luminosities, this can be done in various and sometimes redundant channels,
therefore strengthening the signal and providing great confidence that it is
indeed a scalar Higgs boson which has been observed. However, at low
luminosities, and in particular in the low Higgs mass range $M_H \lsim 135$
GeV, several channels must be combined in order to establish a clear evidence
for the Higgs particle. The interesting question which can be asked is thus: at
which stage this integrated luminosity will be collected.\s

Before closing this section, let us make a digression about the $K$--factors. 
The inclusion of the higher--order radiative corrections to the Higgs production
cross sections and distributions, which is theoretically indispensable to
stabilize the scale dependence and to allow for precise predictions as it has
been discussed at length in the previous sections, can be also very important
in the experimental analyses. Indeed, not only they increase [in general] the
size of the discovery signals and, thus, their significance, but they also can
change the kinematical properties of the processes under study, leading to
different selection efficiencies and, thus, to a different number of collected
events. This is particularly the case in the $gg\to H$ process where large 
$K$--factors appear and where the Higgs transverse momentum is generated at 
higher orders, when additional jets which balance this $p_T$ are produced.\s

Of course, the $K$--factors can be included for the signal only if they are also
available for the backgrounds and there are at least two situations in which 
this holds: 

\begin{itemize}
\vspace*{-2mm}
\item[$i)$] The signal appears as a narrow peak in an invariant mass 
distribution and, thus, the corresponding backgrounds can be precisely measured
from the side bands and safely extrapolated to the signal region. This is the 
case of the important $H \to \gamma \gamma$ and $H \to ZZ^{(*)} \to 4\ell$ 
detection channels for instance.  
\vspace*{-2mm}

\item[$ii)$] When estimates of signal significances are made before having the
data or in the case where the invariant Higgs mass peak cannot be 
reconstructed and one would have to rely on a counting of the number of signal 
versus background events, the $K$ factors can be included if the backgrounds 
are also known at the same level of accuracy as the signal. This is clearly 
the case for many background processes such as $\gamma \gamma, WW, ZZ$ and 
$tt$ production which are known at least to NLO accuracy.\s 
\vspace*{-2mm}
\end{itemize}

Furthermore, the $K$--factors should not only be implemented in the total
normalization of the signal and backgrounds, but also in the various
kinematical distributions when they are strongly affected by the higher--order
corrections\footnote{This is not always the case. In Ref.~\cite{K-Mellado}, the
search sensitivity in the process $gg \to H \to ZZ \to 4\ell$ has been shown to
depend mainly on the signal and background cross sections as well as on the
detector performance and the selection cuts and not, for instance, on
additional jet activity. A simple scaling of the signal and background rates
with their respective $K$--factors leads,  therefore, to reasonable results. 
It has been shown that in this particular case, one needs 30--35\% less
integrated luminosity to achieve a given signal significance when the
$K$--factors are included.}. Ideally, this has to be performed at the level of
Monte--Carlo event generators which are required in practice to obtain a
realistic final state with fragmented particles and underlying events. This is
not a trivial task and there are many ongoing discussions on this topic; see
Ref.~\cite{Houches-QCD} as an example. Fortunately, besides the fact that
NLO parton level Monte--Carlo programs start to appear 
\cite{pp-ggH-eta2,pp-MCFM,MC-WWNLO}, this can be
performed in an effective way even in MC event generators
\cite{K-Mellado,K-Michael}: differential effective $K$--factors can be defined
for relevant kinematical variables and used to reweight individual events with
reconstructed jets coming from a LO Monte--Carlo event generator\footnote{For
instance, in Ref.~\cite{K-Michael}, the channel $gg \ra H \ra W W \ra \ell 
\nu \ell \nu$ has been considered and the higher--order QCD corrections
have been taken into account by using this reweighting procedure, allowing to
combine event rates obtained with the {\tt PYTHIA} event generator with the most
up--to--date theoretical predictions for the $p_T$ spectra of the Higgs signal
and the corresponding $WW$ background. An experimental effective $K$--factor of
$\sim 2$ has been obtained in the range $M_H=140$--180 GeV, which is only about
15\% smaller than the theoretical inclusive $K$--factor. This led to a
considerable increase of the statistical significance of the Higgs discovery
in this specific channel.}.\s

Thus, all $K$--factors [which have been determined after a very hard
theoretical work]  should ultimately be included in the experimental analyses
as they allow a more accurate prediction of the discovery potential and often
lead to a better cut optimization.  Apparently, we are finally heading to this
direction.  

\subsubsection{Determination of the Higgs properties at the LHC}

Once a convincing signal for a Higgs boson has been established, the next step
would be to determine its properties in all possible details and to establish 
that the particle is indeed the relic of the electroweak symmetry breaking 
mechanism and that it has the features that are predicted in the SM, that is:
it is a spin--zero particle with $J^{\rm PC}= 0^{++}$ quantum numbers and that
it couples to fermions and gauge boson proportionally to their masses. 
Ultimately, the scalar Higgs potential responsible for the symmetry breaking
should be reconstructed by precisely measuring the trilinear and quartic Higgs 
self--couplings. At the LHC, several important measurements can be performed 
as is briefly summarized below. 

\subsubsection*{\underline{The Higgs mass and total decay width}}

The Higgs mass can be measured with a very good accuracy 
\cite{pp-meas-mass+width}.
In the range below $M_H \lsim 400$ GeV where the total width is not too large,
a relative precision of $\Delta M_H/M_H \sim 0.1$\% can be achieved in the
channel $H \to ZZ^{(*)} \to 4\ell^\pm$ if 300 fb$^{-1}$ luminosity is
collected by ATLAS and CMS. This is shown in Fig.~3.51 where
the relative precision is displayed as a function of $M_H$ and where the
statistical and some systematical errors are included \cite{Mass-Fabiola}.\s 

\begin{figure}[!h]
\vspace{-.9cm}
\begin{center}
\mbox{\epsfig{file=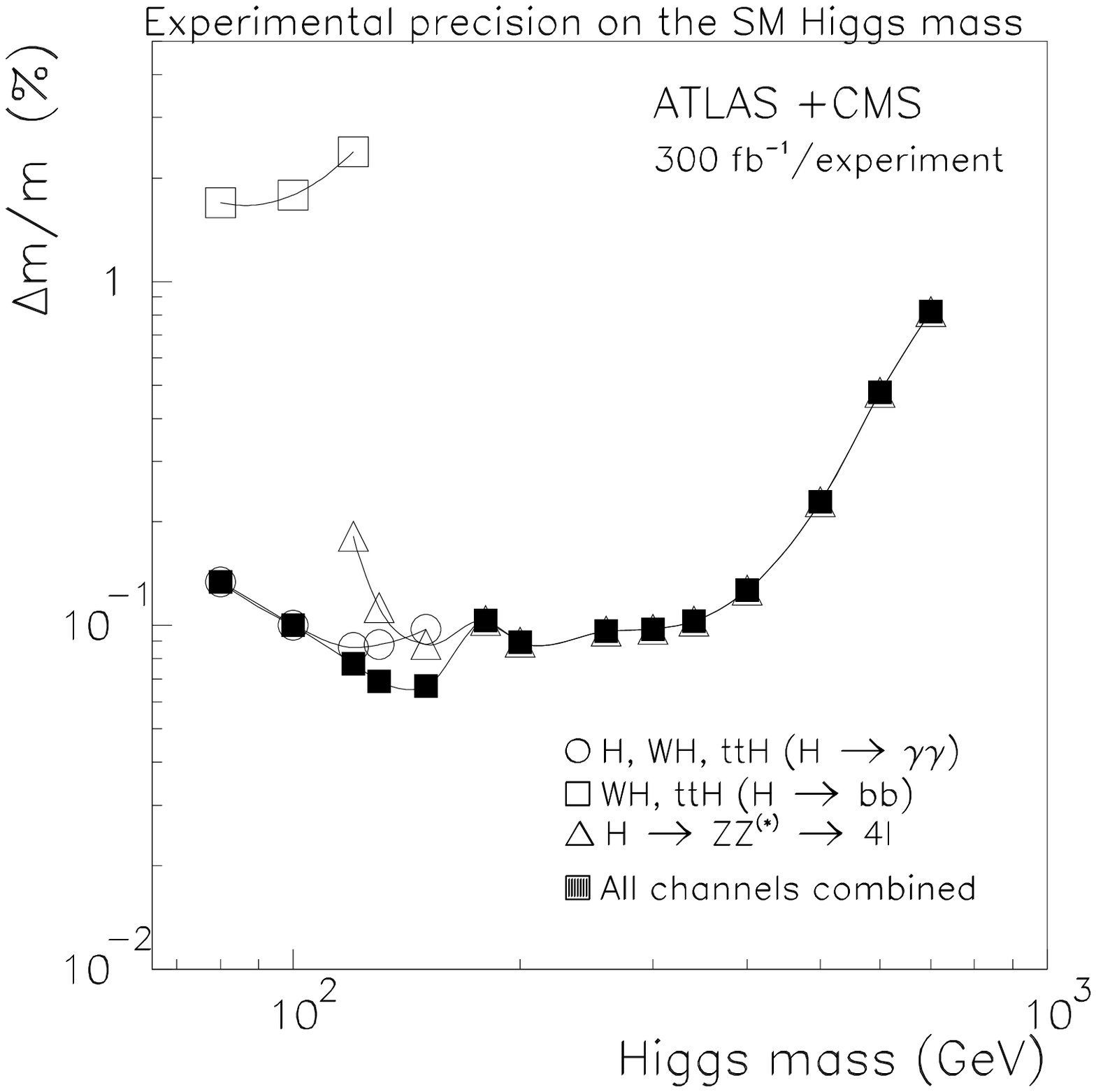,width=8.3cm} \hspace*{-2mm}
\epsfig{file=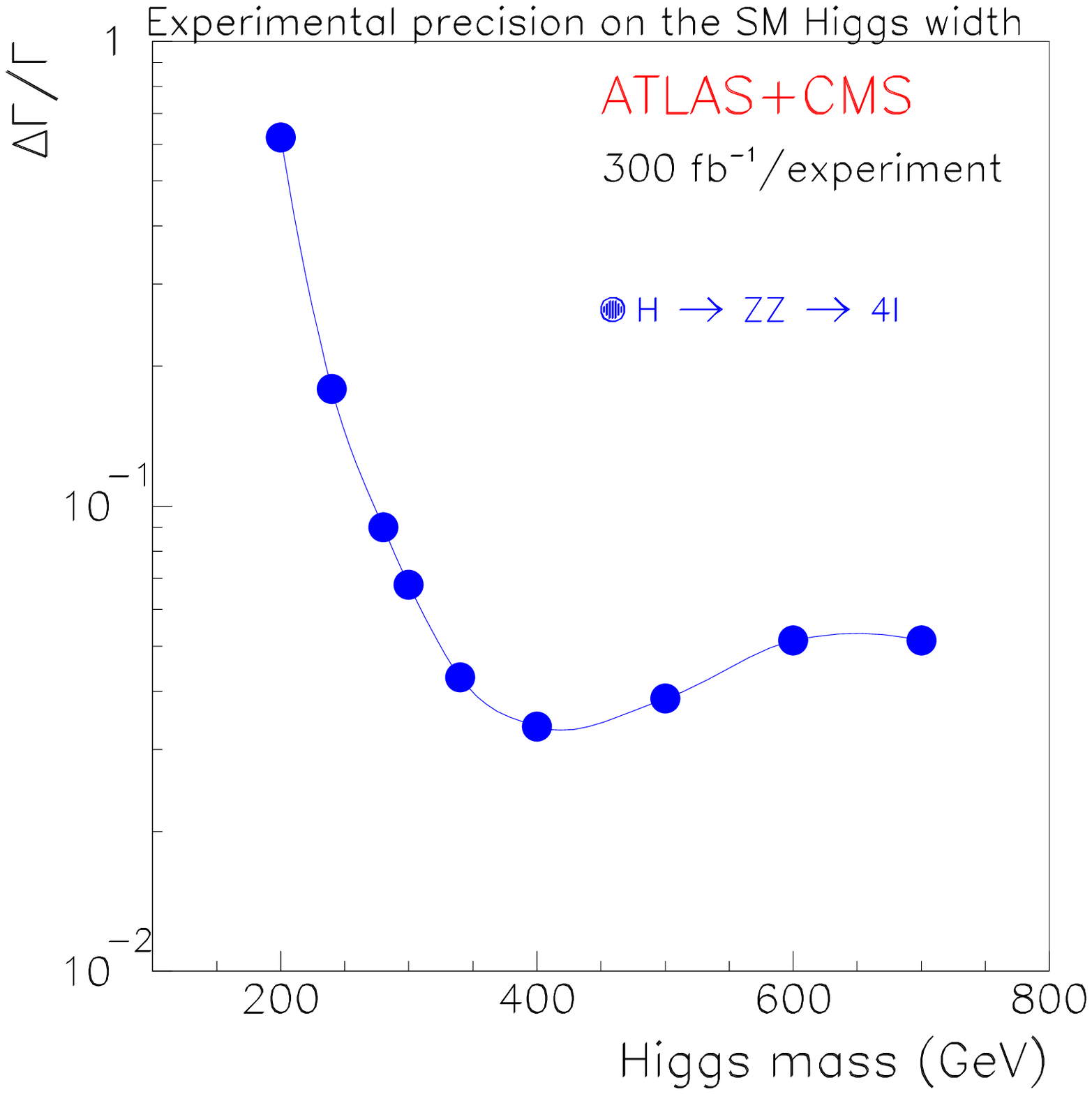,width=8.3cm}}
\end{center}
\vspace{-.4cm}
{\it Figure 3.51: Expected errors on the measurement of the Higgs boson mass 
(left) and total decay width (right) at the LHC as a function of $M_H$, 
combining both ATLAS and CMS with a luminosity of 300 fb$^{-1}$ per experiment;
from Ref.~\cite{Mass-Fabiola}.}
\label{fig:Hmass}
\vspace{-.5cm}
\end{figure}

In the low Higgs mass range, a slight improvement can be obtained by
reconstructing the sharp $H \to \gamma \gamma$ peak. In the range $M_H \gsim
400$ GeV, the precision starts to deteriorate because of the smaller production
rates which increase the statistical error.  However a precision of the order of
$1$\% can still be achieved for $M_H\sim 700$ GeV if theoretical errors, such
as width effects, are not taken into account.  \s

Using the same process, $H \to ZZ \to 4\ell^\pm$, the total decay width of the
Higgs boson can be measured for masses above $M_H \gsim 200$ GeV when it is
large enough to be resolved experimentally. While the precision is rather poor
near this mass value, approximately $60\%$,  it improves to reach the
level of $\sim 5$\% around $M_H \sim 400$ GeV and the precision stays almost
constant up to a value $M_H\sim 700$ GeV \cite{pp-meas-mass+width}. This is
shown in the right--hand side of Fig.~3.51 where the relative precision on
$\Gamma_H$ is displayed as a function of $M_H$ with 300 fb$^{-1}$ luminosity
for the combined ATLAS and CMS experiments \cite{Mass-Fabiola}.  

\subsubsection*{\underline{The Higgs spin and parity quantum numbers}}

As seen previously, if a high enough luminosity is collected at the LHC, a
Higgs boson in the low mass range, $M_H \lsim 135$ GeV, will be detected
through its $H \to \gamma \gamma$ decay mode. This observation will immediately
rule out the spin possibility $J=1$ by Yang--Landau's theorem, and will fix the charge
conjugation to be positive C$=+$ \cite{Yang-theorem}. This argument cannot be
generalized to Higgs production in the $gg$ fusion mechanism or to Higgs decays
into gluons, $gg \leftrightarrow H$, since gluons cannot be reasonably
distinguished from light quark jets. \s

For higher Higgs masses when the $\gamma \gamma$ decay becomes too rare, the
observation of the Higgs boson in the decays $H \to WW^*, ZZ^*$ provides some
information. Indeed, as discussed in \S2.2, these decays are sensitive to the
spin--zero nature of the Higgs boson, if one of the gauge bosons is virtual. 
The invariant mass ($M_*$) spectrum of the off--shell gauge boson in $H \to
VV^*$, see eq.~(\ref{dGHVV*}), is proportional to the velocity $d\Gamma/dM_*
\sim \beta \sim \sqrt{(M_H -M_V)^2 -M_*^2}$, and therefore decreases steeply
with $M_*$ just below the kinematical threshold; see Fig.~2.12. This is
characteristic of a spin--zero particle decaying into two vector bosons, and
rules out all spin assignments except for two cases, $J^P=1^+$ and $2^-$.  This
is shown in the left--hand side of Fig.~3.52 where the threshold behavior of
$d\Gamma/dM_*$ is displayed for the $\sim 200$ signal events which are expected
for $M_H=150$ GeV and $ {\cal L} =300$ fb$^{-1}$ [histogram] and compared with
the prediction for the SM Higgs and for two examples of spin $1$ and $2$ cases
\cite{CPHVVchoi}.\s 

The spin--correlations, which are useful to discriminate between the signal $gg
\to H \to WW^*$ and $pp \to WW$ background \cite{pp-HWW-lnln} for instance, can
be used to determine the Higgs boson spin at the LHC. In practice, however, the
complete final state
must be reconstructed and one has to rely on the decays $H\to ZZ^* \to 4\ell$
which have rather low rates. The two remaining configurations $J=1^+$ and $2^-$
which are not probed, as well as the CP--odd $0^-$ case, can be discriminated
against the Higgs spin by looking at the angular distribution in the decays $H
\to VV^{(*)} \to 4f$ given in eq.~(\ref{HVV*distr}), and experimentally
observing a $\sin^2 \theta_1 \sin^2 \theta_3$ correlation and not observing the
$(1+\cos^2\theta_{1,3}) \sin^2 \theta_{3,1}$ correlation 
\cite{Bargeretal,CPHVVchoi}.\s  

\begin{figure}[htb!]
\vspace*{-.5cm}
\begin{center}
\mbox{
\includegraphics[clip=true,trim=5 5 5 5,width=8.cm]{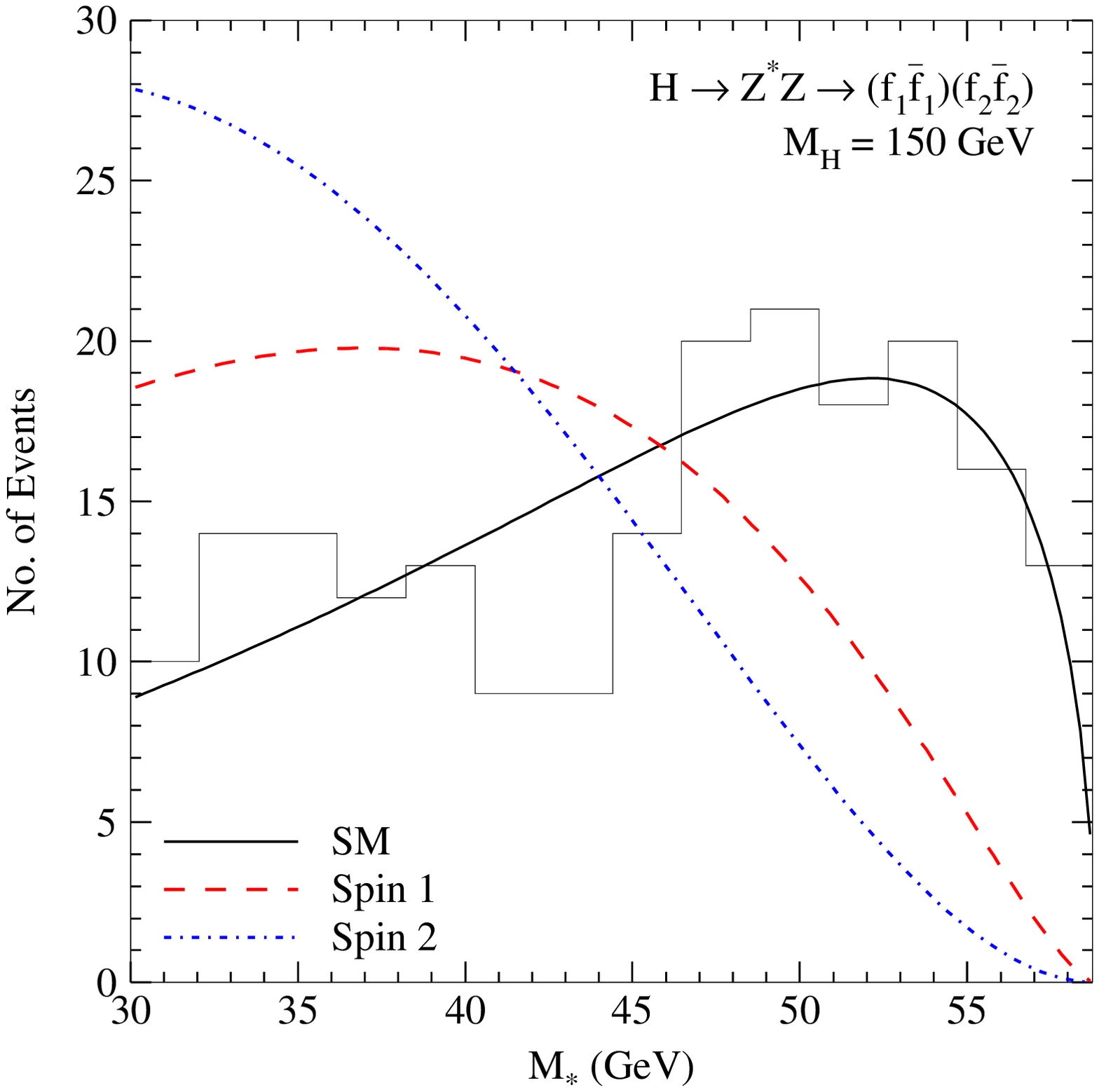} \hspace*{-3mm}
\includegraphics[clip=true,trim=5 5 5 5,width=8.cm]{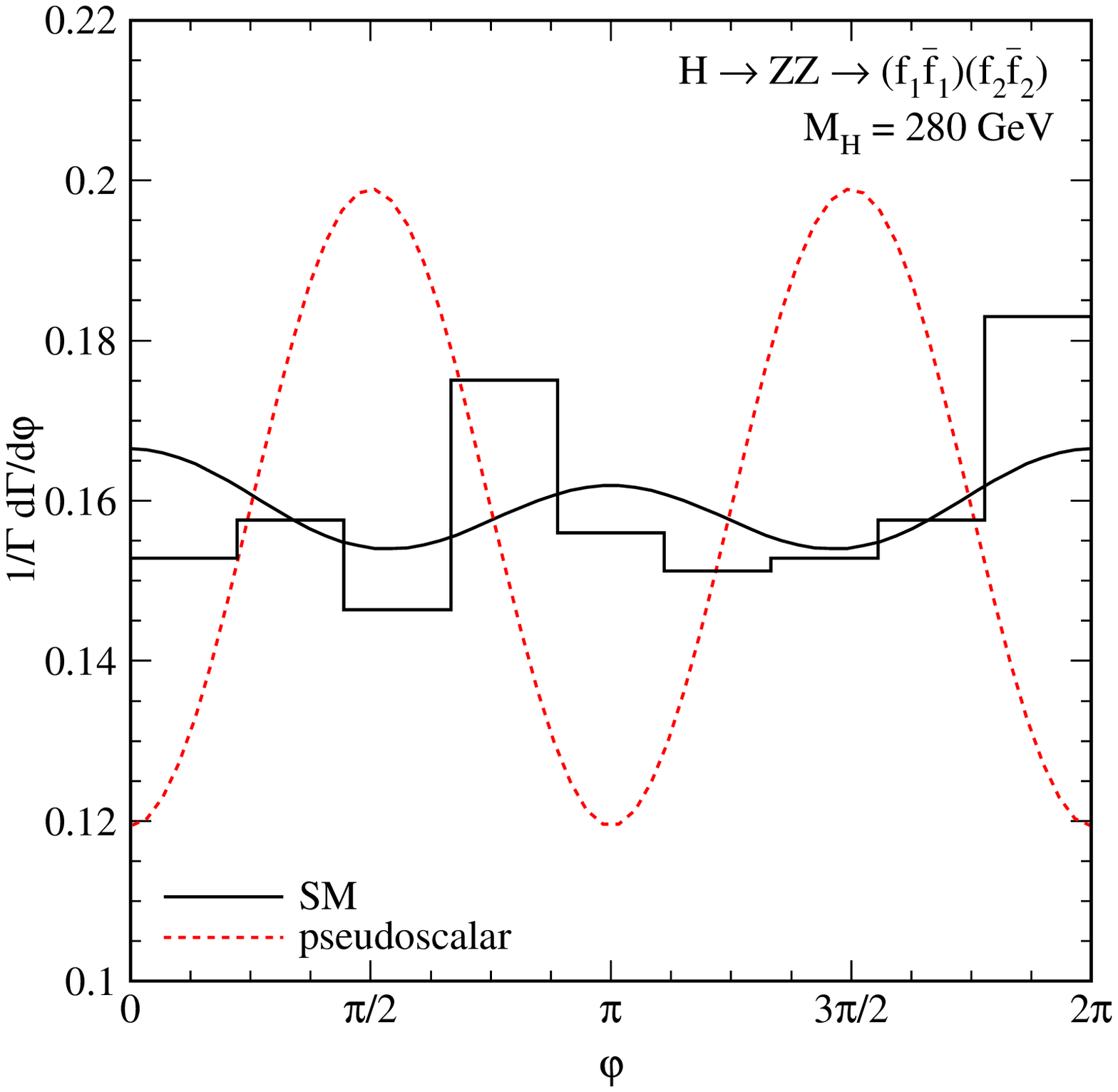} }
\end{center}
\vspace*{-.5cm}
{\it Figure 3.52: The threshold behavior of the differential distribution
$d\Gamma/dM_*$ for the SM Higgs and two spin examples of $J=1$ and $2$ for
$M_H=150$ GeV (left) and the azimuthal distributions $d\Gamma/d\phi$ for the SM
and a pseudoscalar Higgs bosons for $M_H=280$ GeV (right). The histograms have
been obtained with $\int {\cal L} dt = 300\, {\rm fb}^{-1}$ at the LHC, with
efficiencies and cuts included according to an ATLAS simulation of 
Ref.~\cite{CPHVV-LHC2}; from 
Ref.~\cite{CPHVVchoi}. }
\vspace*{-.2cm}
\end{figure}
\noindent

In fact, the angular correlations are also sensitive to the parity of the Higgs
boson as seen in \S2.2.4 and can discriminate between the CP--even SM Higgs case
and the pseudoscalar Higgs case. In particular, the dependence on the azimuthal
angle is very different, as it can be seen from eq.~(\ref{HVV*azimuth}) and in
Fig.~2.13. The same simulations as previously \cite{CPHVVchoi,CPHVV-LHC2} have
been performed for $M_H=280$ GeV and the distribution $d\Gamma/d\phi$ is shown 
in Fig.~3.52 as a function of the azimuthal angle for the 900 expected events 
at the LHC for this Higgs mass. A clear discrimination between the CP--even and
CP--odd cases can be made in this case.  \s

The Higgs CP properties and the structure of the $HVV$ coupling can also be
determined in the vector boson fusion process, $qq \to qqH$,  by
looking at the azimuthal dependence of the two outgoing forward tagging jets
\cite{WVB-spin}. The analysis is independent of the Higgs mass 
and decay modes but might be difficult because of background problems 
\cite{pp-ggHqq,WVB-spin-back}. \s

However, there is a theoretical caveat in this type of analyses
\cite{JFG-Snowmass96}: if a Higgs boson is observed with substantial rates in
channels where it couples to vector bosons, it is very likely that it is
CP--even since the $VV$ couplings of a pure CP--odd state are generated only
through loop corrections. The decisive test of the CP properties should be
therefore to verify that the SM Higgs boson is pure CP--even and rule out the
small loop--induced CP--odd component. This becomes then a very high precision 
test which is very challenging at the LHC. \s

The couplings of the Higgs boson to fermions provide a more democratic probe of
its CP nature since in this case, the CP--even and CP--odd components can have
the same magnitude. One thus has to look at channels where the Higgs boson is
produced and decays through these couplings. Discarding the possibility of
$H\to b\bar{b}$ and $\tau^+ \tau^-$ decays in the $gg\to H$ production channel,
which have very large backgrounds, one has to rely on Higgs production in $pp
\to t\bar tH$ with $H\to \gamma \gamma$ and eventually $b\bar b$. Techniques
based on the different final states distributions in the production of a scalar
or a pseudoscalar Higgs boson have been suggested in
Refs.~\cite{Spin-pp-G1,Spin-pp-Field} to discriminate between the two scenarios
or a mixture.  However these channels are rather difficult as we have seen
previously.  With very large luminosities ${\cal L}= 600$ fb$^{-1}$ and for a
rather light Higgs boson, $M_H \sim 100$ GeV, an equal mixture of CP--even and
CP--odd couplings [with a total coupling squared equal to the SM one] can be
probed at a few $\sigma$ level \cite{Spin-pp-Field}. But again, this method does
not allow to check precisely the CP--even purity of the SM Higgs boson, at 
least in this particular channel. Central exclusive diffractive Higgs production
\cite{diff-spin,John-new} might provide the solution; \S3.6.4.  

\subsubsection*{\underline{The measurement of the Higgs couplings at the LHC}}

The determination of the Higgs couplings to gauge bosons and fermions is
possible at the LHC through the measurement of the cross sections times
branching ratios, $\sigma \times  {\rm BR}$, given by the event rate in the 
various search channels \cite{Zepp-meas,Duhrssen,Duhrssen2,Snow-meas,SLHC+VLHC};
for earlier analyses see Ref.~\cite{ATLAS-TP,CMS-TDR,ATLAS-TDR}. However,
the accuracy in this determination is rather limited because of the small
statistics that one obtains after applying the cuts that suppress the large
backgrounds which are often plagued with uncertainties, and the various
systematical errors such as the common uncertainty in the absolute luminosity.
In addition, when one attempts to interpret the measurements, theoretical
uncertainties from the limited precision on the parton densities and from the
higher--order radiative corrections or scale dependence should be taken into
account. Furthermore, the couplings which can be measured will critically
depend on the Higgs boson mass. For instance, in the mass range above $M_H \sim
2M_W$, only the couplings to gauge bosons can be accessed directly and 
the $Ht\bar t$ coupling can be probed indirectly.\s

The cross sections times branching ratios which can be measured in various
channels at the LHC are shown in Fig.~3.53 for Higgs masses below 200 GeV
\cite{Zepp-meas}. The $gg$ fusion (solid lines), the expectations for weak
boson fusion with a parton level analysis (dashed lines) and the associated $pp
\to t\bar{t}H , H\to b\bar{b}$ (dotted lines) channels are for a luminosity of
200 fb$^{-1}$. The channels $pp \to t\bar{t}H \to t\bar t WW^*$ (red--dotted
lines) assume a luminosity of 300 fb$^{-1}$. In this figure, as well as in the
subsequent discussion, only the statistical errors are taken into account. A
precision of the order of 10 to 20\% can be achieved in some channels, while
the vector boson fusion process, $pp \to H qq \to W W qq$, leads to accuracies
of the order of a few percent. \s 

These $\sigma \times  {\rm BR}$ can be translated into Higgs partial widths in
the various decay channels $\Gamma_X \equiv \Gamma( H\to XX)$ \cite{Duhrssen},
which are proportional to the square of the Higgs couplings, $g_{HXX}^2$.
However, in the case of the vector boson fusion mechanism,  which has
contributions from $ZZ \to H$ and $WW \to H$, the $HZZ$ and $HWW$ couplings
cannot be disentangled.  One then has to assume that they are related by SU(2)
symmetry as is the case in the SM [an assumption which can be tested with a
20\% accuracy in $gg\to H \to ZZ^*$ {\it versus} $gg \to H\to WW^*$ but for
large enough $M_H$]. With this assumption, one can perform ratios of partial
widths $\Gamma_{X_i} /\Gamma_{X_j}$, in which some common theoretical and
experimental errors will cancel. This is shown in Fig.~3.54 (left) for a
luminosity of 200  fb$^{-1}$, where the relative accuracy on the ratios of
$\sigma \times  {\rm BR}$ of the production and decay channels discussed above
can be formed. Again, measurements at the level of 10--20\% can be made in
some cases. \s

One can indirectly measure the total Higgs width $\Gamma_H$ and thus derive the
absolute values of the partial widths $\Gamma_X$ by making additional
assumptions besides $g_{HWW}/g_{HZZ}$ universality: $i)$ $\Gamma_b/\Gamma_
\tau$ is SM--like [with an error of $\sim 10\%$ corresponding to the
uncertainty on the $b$--quark mass] since both fermions have the same isospin
and $ii)$ the branching ratio for Higgs decays into unexpected channels is
small [in the SM, this error is less than about $3\%$ and corresponds to the
missing BR$(H \to c\bar c)$] so that $1-\Gamma_{X_i}/\Gamma = \epsilon \ll 1$. 
The Higgs boson total width $\Gamma_H$ can be then determined and the partial
widths $\Gamma_X$ as well. \s

\begin{figure}[!h]
\vspace*{1mm}
\begin{center}
\includegraphics[width=9.cm, angle=90]{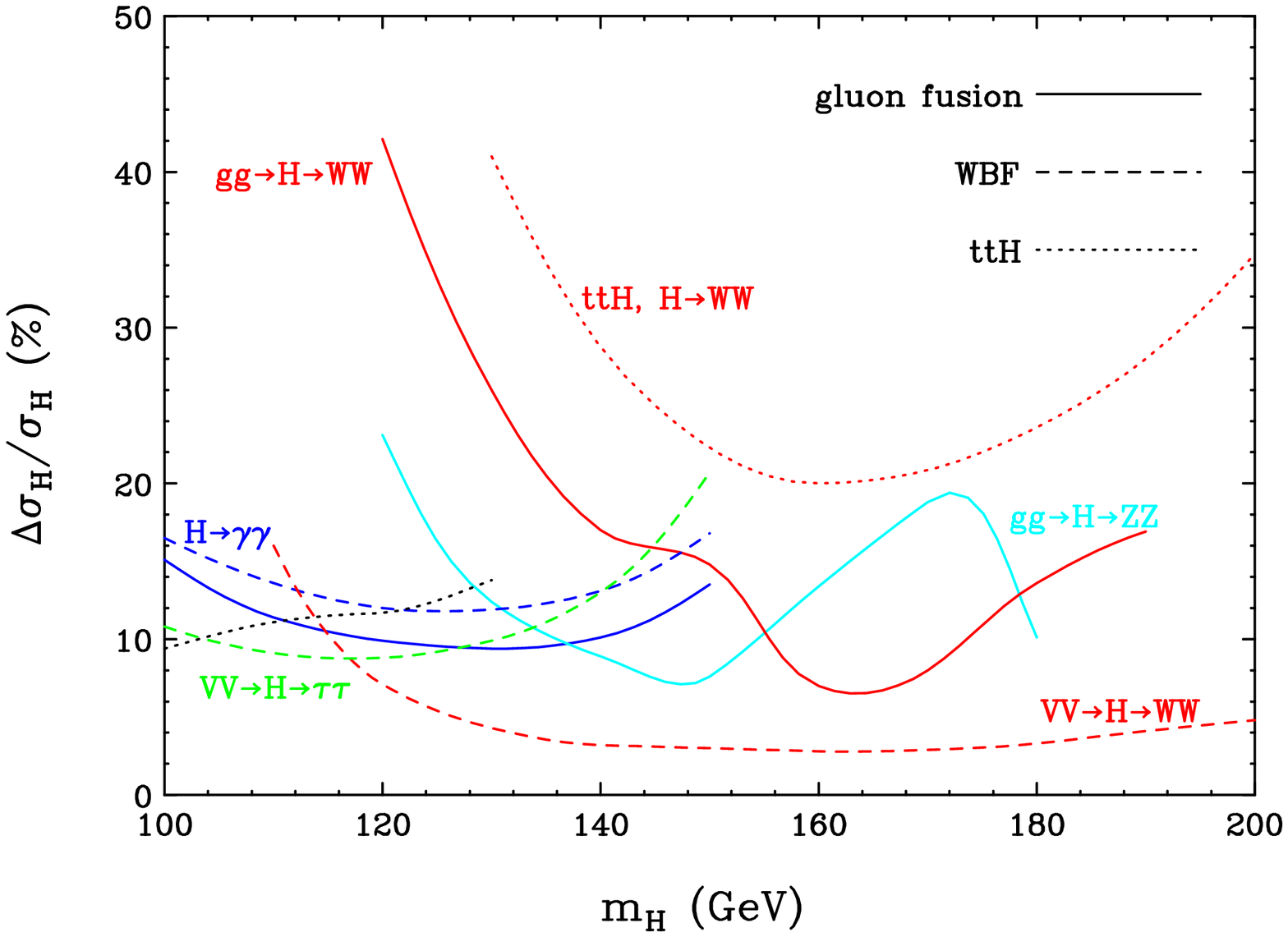}
\end{center}
\vspace*{-1mm}
{\it Figure 3.53: Expected relative errors on the determination of 
$\sigma \times {\rm BR}$ for various Higgs boson search channels at the LHC 
with 200--300 fb$^{-1}$ data; from Ref.~\cite{Zepp-meas}.}
 \label{fig:delsigh}
\vspace*{-.0cm}
 \end{figure}

\begin{figure}[!h]
\vspace*{2mm}
\begin{center}
\includegraphics[width=9.0cm,angle=90]{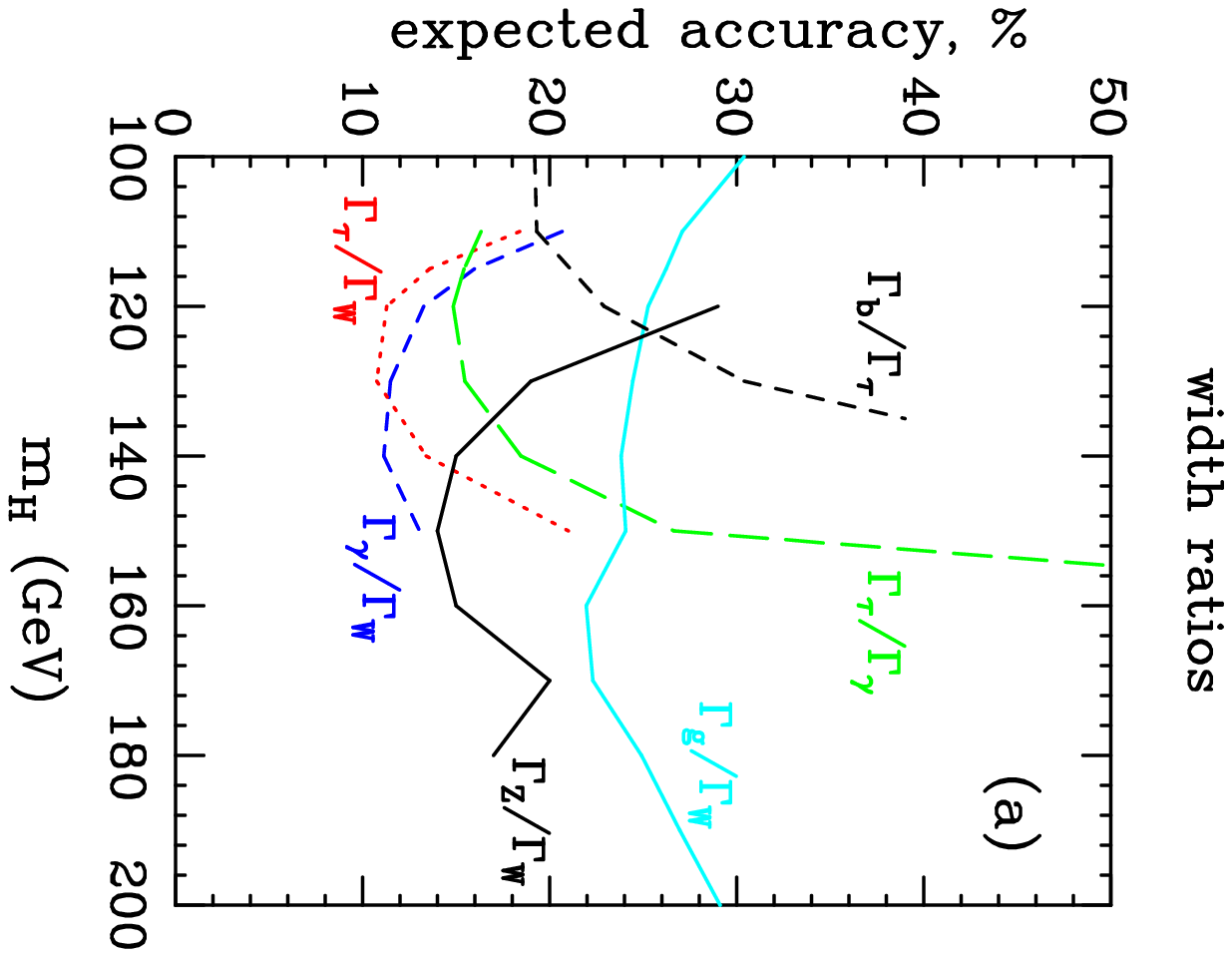} \hspace*{1cm}
\includegraphics[width=9.0cm,angle=90]{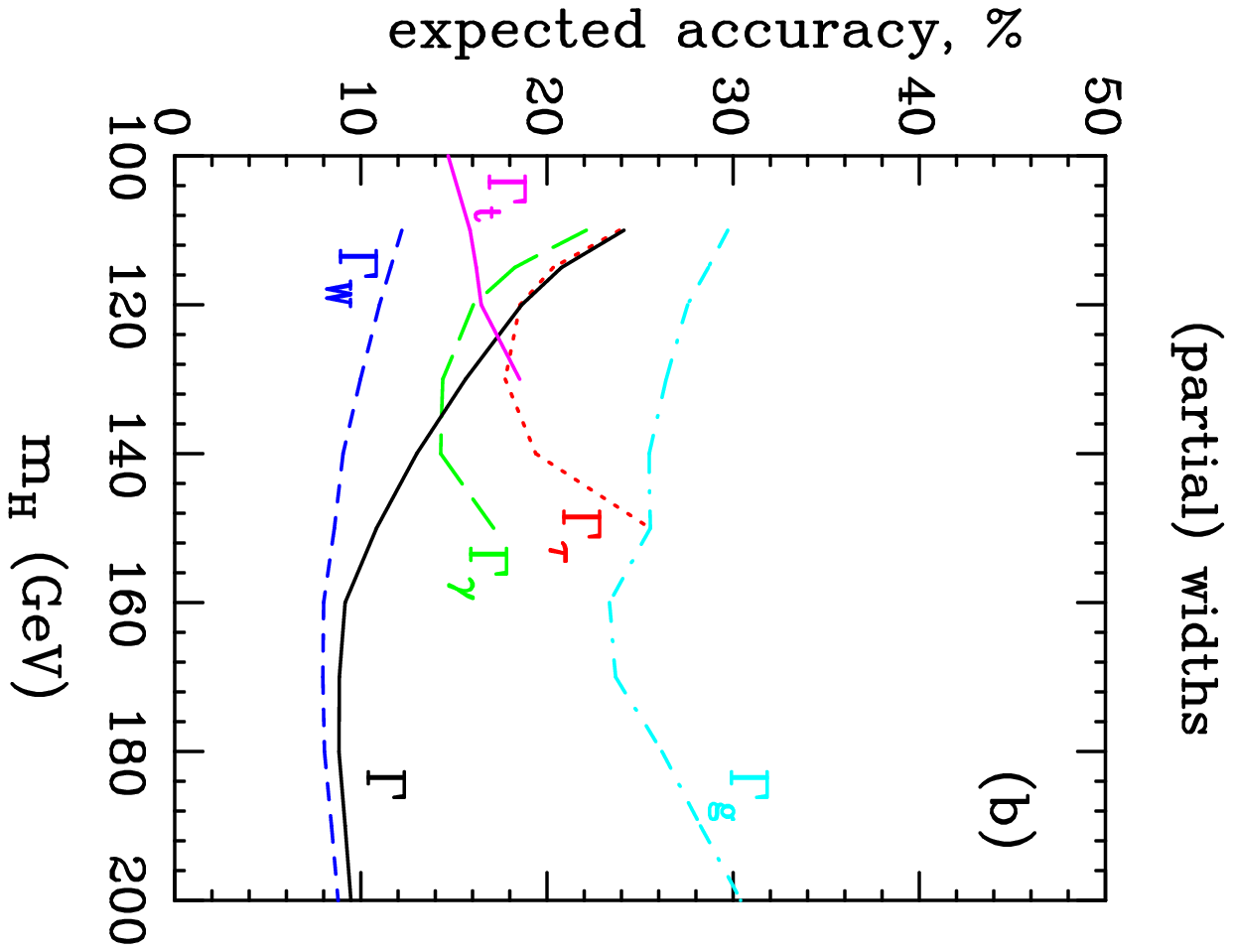}
\end{center}
\vspace*{-.2cm}
{\it Figure 3.54: Relative accuracy expected at the LHC with a luminosity of 
200 fb$^{-1}$ for various ratios of Higgs boson partial widths (left) and 
the indirect determination of partial and total widths $\tilde\Gamma_i$ and 
$\Gamma$ with the assumptions discussed in the text (right); from 
Ref.~\cite{Duhrssen}.} 
\vspace*{-.2cm}
\end{figure}

The expected accuracies are shown in the right--hand side of Fig.~3.54.  They
are at the level of 10 to 30\% depending on the final states and on $M_H$, and
translate to an accuracy on the couplings of the order of 5 to 15\%
\cite{Duhrssen}.  Detailed experimental analyses accounting for the backgrounds
and for the detector efficiencies, as well as further theoretical studies for
the signal and backgrounds, have to be performed to confirm these values.  

\subsubsection*{\underline{The Higgs self--coupling}}

The trilinear Higgs boson self--coupling $\lambda_{HHH}$ is too difficult to be
measured at the LHC because of the smallness of the $gg\to HH$ [and, {\it a 
fortiori}, the $VV \to HH$ and $qq \to HHV$] cross sections and the very large 
backgrounds \cite{HHH-LHC,HHH-WW,HHH-bb}; see also Refs.~\cite{HHH-LHC-E} and
\cite{HHH-LHC-R} for an earlier and more recent analysis, respectively.  A
parton level analysis \cite{HHH-WW} has been recently performed in the channel
$gg\to HH \to (W^+W^-)(W^+W^-) \to (jj \ell \nu) (jj \ell \nu)$ and $(jj \ell
\nu) (\ell \ell \nu \nu)$ with same sign dileptons, including all the relevant
backgrounds which, as one might have expected, are significantly large. At the
LHC, the statistical significance of the signals, once most of the  backgrounds
are removed, is very small, even with an extremely high luminosity. However, it
was found that the distribution of the invariant mass of the visible final
state particles peaks at much higher values for the backgrounds than for the
signal, independently of the value of the trilinear coupling; see the left--hand
side of Fig.~3.55.\s
 
\begin{figure}[!h] 
\vspace*{2mm}
\begin{center}
\includegraphics[width=8cm,height=8cm]{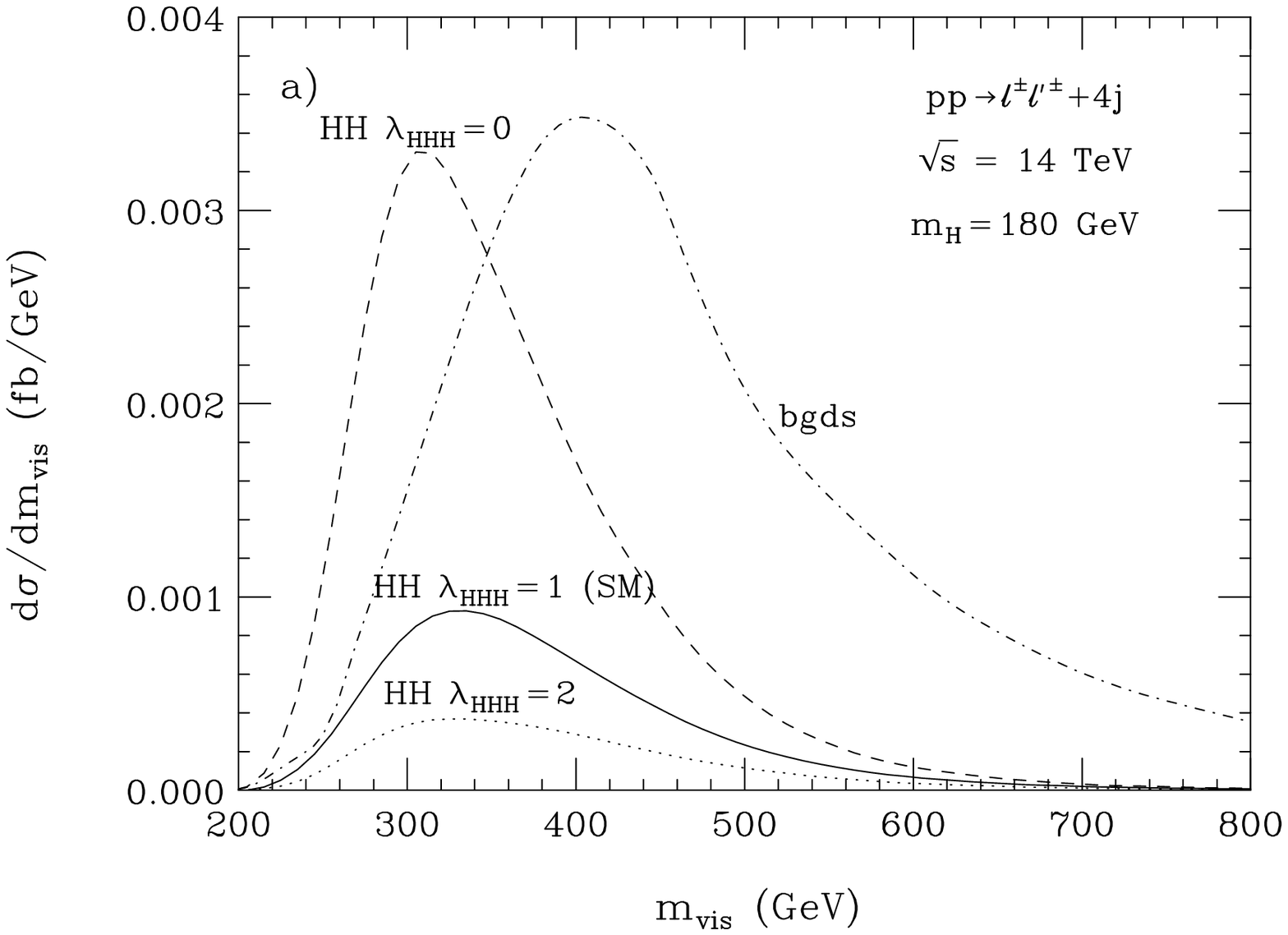} 
\includegraphics[width=8cm,height=8cm]{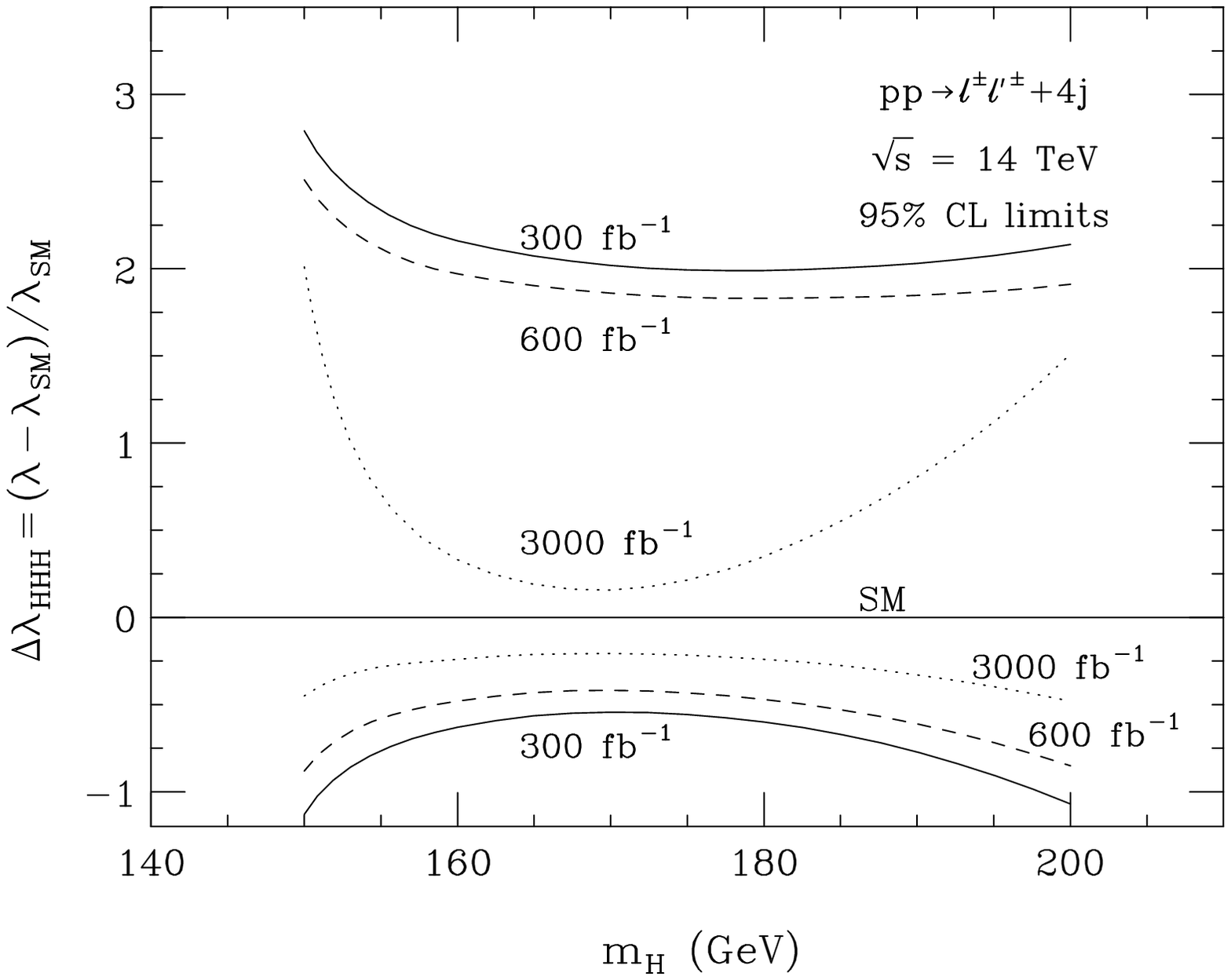} 
\vspace*{-2mm}
\end{center}
{\it Figure 3.55: 
The visible mass distribution of the signal for $pp\to \ell^\pm{\ell}^\pm+4j$ 
for $M_H=180$~GeV at the LHC for various $\lambda_{HHH}$ values 
and for the combined backgrounds (left). Limits achievable at $95\%$ CL for 
$\Delta\lambda_{HHH}=(\lambda-\lambda_{SM})/\lambda_{SM}$ in $pp\to\ell^\pm{
\ell'}^\pm+4j$ at the LHC for various integrated luminosities (right); 
from Ref.~\cite{HHH-WW}.}
\vspace{-2mm}
\end{figure}

This observation can be used to set limits on the Higgs self--coupling. For a
luminosity of 300 fb$^{-1}$ one can check a non--vanishing value of
$\lambda_{HHH}$ at the 95\% CL if the Higgs boson mass happens to lie in the
range 150--200 GeV.  Much more luminosity would be needed to perform a decent
measurement; see the right--hand side of Fig.~3.55.  For lower Higgs masses,
$M_H \lsim 140$ GeV, one would have to rely on the dominant decays $ HH\to 4b$
not to lose too much statistics, but in view of the formidable backgrounds,
this process seems to be hopeless at the LHC. The channel $H \to b\bar b
\tau \tau$ is only slightly easier \cite{HHH-bb}.  

\subsubsection{Higher luminosities and higher energies}

Some of the detection signals as well the measurements discussed previously
would greatly benefit from an increase of the LHC luminosity. As mentioned in
the beginning of this chapter, there are plans to achieve an instantaneous
luminosity of ${\cal L}=10^{35}\, {\rm cm^{-2} s^{-1}}$ at $\sqrt{s} \simeq 14$
TeV, while keeping the present dipole and magnets. This would allow to collect
6 ab$^{-1}$ for both the ATLAS and CMS experiments after three years of data
tacking. This SLHC option will allow to probe rare production and decay
processes of the Higgs particle.  A brief summary of the interesting physics
which can be performed at such a machine in the context of the SM Higgs boson
is as follows \cite{SLHC,SLHC+VLHC}: \s

-- $H \to \mu^+ \mu^-$: we have seen that with the present LHC design
luminosity, this rare decay can be observed only at the $3\sigma$ level, even
with 600 fb$^{-1}$ of data. With 6 ab$^{-1}$ data, the process can be observed
at the 5\,$\sigma$ level for $M_H$ in the range 120 to 140 GeV and would allow
the first measurement of the Higgs coupling to second generation fermions.\s

-- $H\to Z\gamma$: this decay has not been mentioned in the previous 
discussion because it is too rare: if the $Z$ boson decays leptonically, 
the branching fraction for this mode is about $2 \times 10^{-4}$. 
With 6 ab$^{-1}$ data, the $gg \to H \to Z\gamma \to \ell \ell\gamma$ process 
can be observed at the $\sim 10 \sigma$ level for a Higgs boson in the mass 
range $M_H=120$--150 GeV and would provide complementary information to the
$H \to \gamma \gamma$ decay channel.\s

-- The measurement of the ratios of Higgs couplings discussed before is mostly
statistics limited. Provided that detector performances are not significantly
reduced in the high luminosity environment, these ratios of couplings can be
probed at the level of 10\% accuracy, and even below in some cases, if the
the sample of 6 ab$^{-1}$ data is collected. This is shown in Fig.~3.56 where
the combined ATLAS+CMS accuracies in the direct [with tree--level processes]
and indirect measurements [that is, involving the loop induced processes $gg 
\to H$ and $H \to \gamma \gamma$ which are indirectly sensitive to the Higgs
couplings to the top quark, and in the case of $H \to \gamma \gamma$ also to 
the $HWW$ coupling] are shown for a luminosity of 3 ab$^{-1}$ per 
experiment and compared to what can be achieved with only 300 fb$^{-1}$ 
data per experiment.\s 

\begin{figure}
\vspace*{3mm}
\begin{center}
\mbox{
\includegraphics[width=8cm,height=8cm,clip]{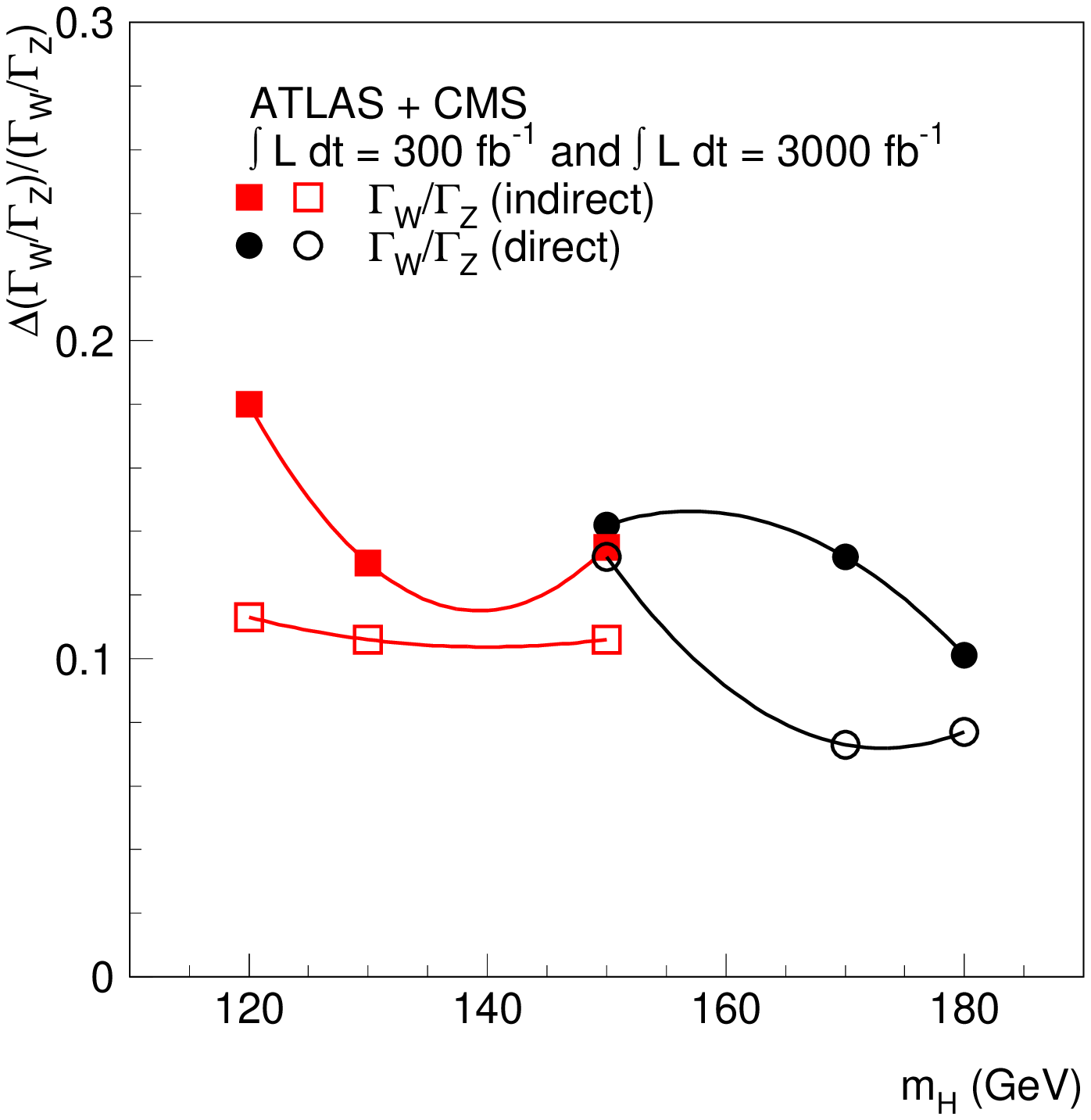}\ \
\includegraphics[width=8cm,height=8cm,clip]{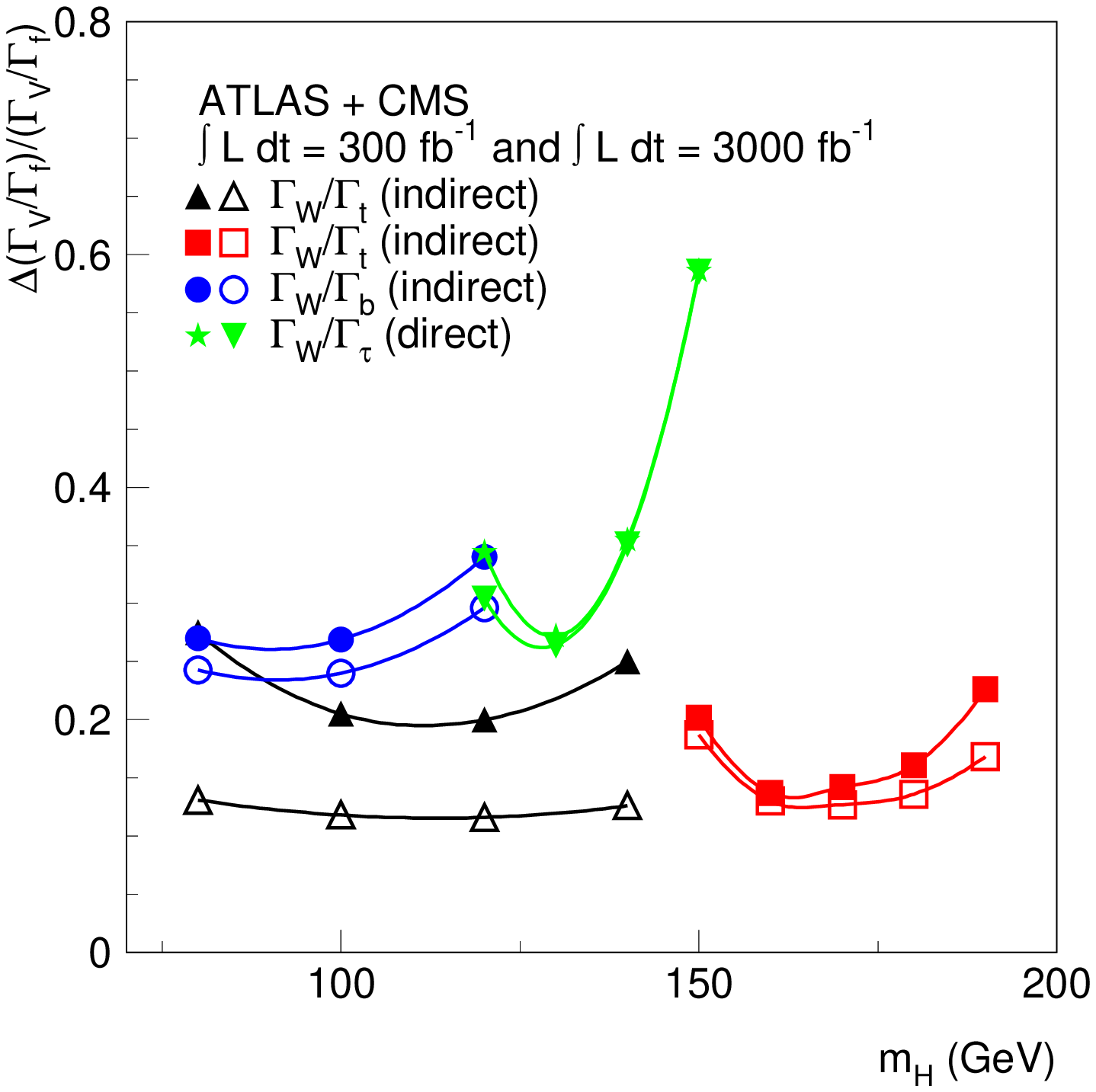} }
\end{center}
\vskip -.3cm 
\nn {\it Figure 3.56: Expected uncertainties on the measured ratios of the 
Higgs boson widths to final states involving  gauge bosons (left) and gauge
bosons and fermions (right) as a function of the Higgs mass. The closed (open) 
symbols are for the two two experiments and 3000\, (300)~fb$^{-1}$ data per 
experiment. Indirect measurements use the loop induced processes $gg 
\to H$ and $H \to \gamma \gamma$; from Ref.~\cite{SLHC}.}
\vspace*{-3mm}
\end{figure}

-- The most important window that a  sample of 6 ab$^{-1}$ data could open
would be the measurement of the Higgs self--coupling $\lambda_{HHH}$. As we
have seen previously, this important coupling cannot be probed with the
presently planed luminosity. The same parton--level simulation mentioned 
previously \cite{HHH-WW} has shown that a signal for the process $gg \to HH \to WWWW \to
\ell^\pm \ell^\pm \nu \nu jjjj$ can be observed with a 5.5 (3.8) significance 
for $M_H=170 \, (200)$ GeV with ${\cal L}=6$ ab$^{-1}$, allowing to probe
$\lambda_{HHH}$.  As can be seen in Fig.~3.55, the trilinear coupling  could be
measured with a statistical error of about 25\% in the Higgs mass window
between 160 and 180 GeV in the channel $pp\to \ell^\pm \ell^\pm jj$ with 3
ab$^{-1}$ data. \s

The precision on the various measurement discussed above can be improved by
increasing the luminosity of the collider but, also, by raising the c.m. energy
which leads to an increase of the Higgs boson production rates in most
processes.  This is explicitly shown in Fig.~3.57, where the cross sections for
the various production processes for a single Higgs boson (upper curves) and
for Higgs pairs (lower curves) are displayed as a function of $\sqrt{s}$ for a
Higgs  mass of 120 GeV. As can be seen, the $gg\to H$ cross section for
instance increases by almost two orders of magnitude compared to the LHC when
the energy of the collider is raised to $\sqrt{s}=200$ TeV. The cross
sections for Higgs pair production also tremendously increase and for the
vector boson fusion processes, $pp \to HHqq$, they reach the level of 
$0.1$ pb at c.m. energies of the order of $\sqrt{s}=200$ TeV.

\begin{figure}[!h]
\begin{center}
\vspace*{3mm}
\includegraphics[width=4.in]{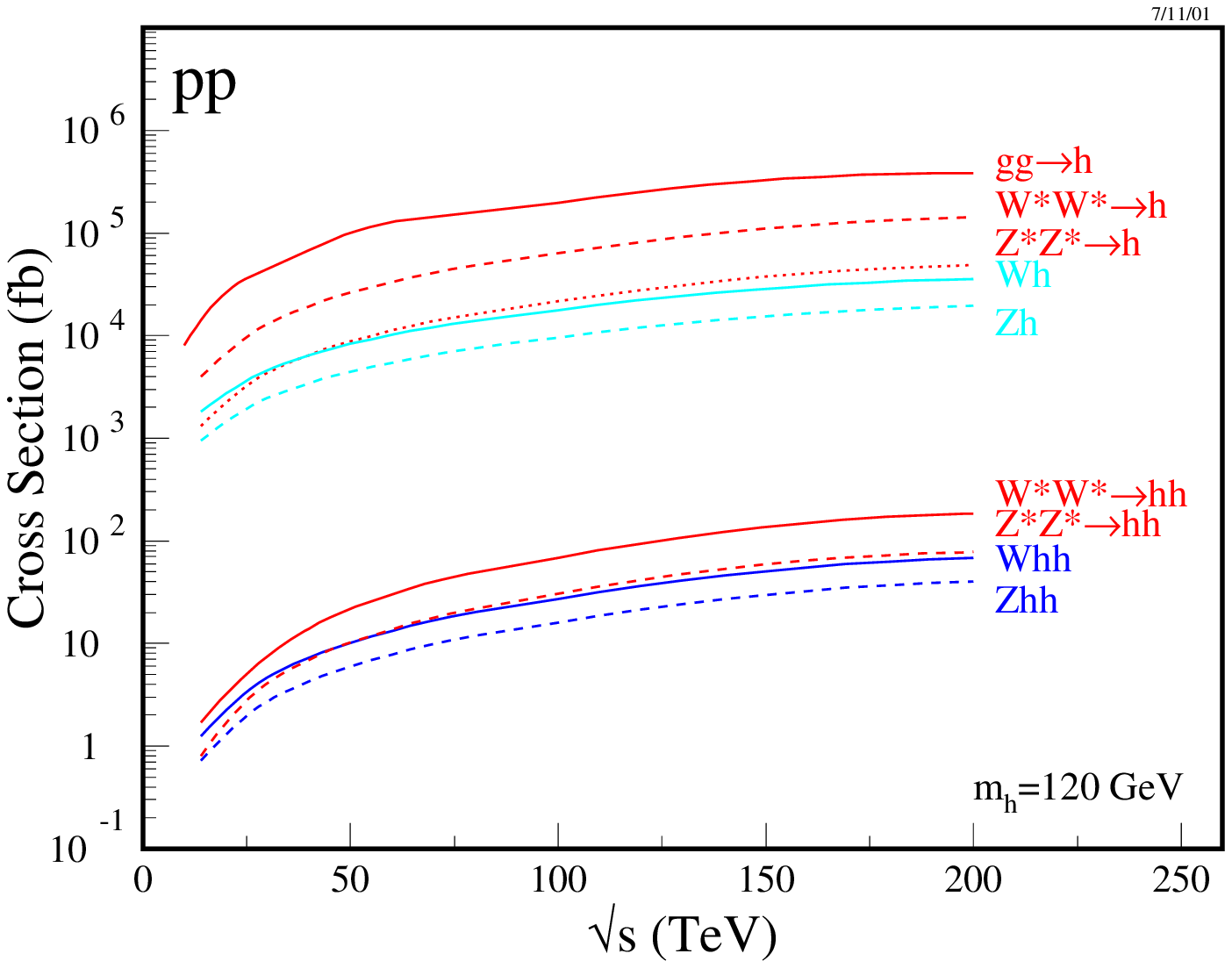}
\end{center}
\vspace*{-2mm}
{\it Figure 3.57: Total cross sections for single and double Higgs boson 
production in various processes as a function of $\sqrt{s}$ for $pp$ 
collisions and $M_H=120$~GeV; from Ref.~\cite{SLHC+VLHC}.}
\vspace*{-1mm}
\end{figure}

One can then probe the rare decays of the Higgs boson and measure more precisely
its couplings to fermions and gauge bosons and its self--coupling, in much the
same way as it has been discussed for the SLHC. The accuracies in the
determination of some couplings of the SM Higgs boson will for instance start
to approach the few percent level.\s 

In fact, with a luminosity of 100 fb$^{-1}$,
a VLHC running at $\sqrt{s}=50$ TeV will be comparable and in some cases
superior to the SLHC.  The potential of the two options has been
discussed and compared in specific examples in Ref.~\cite{SLHC+VLHC} to
which we refer for details. Note, however, that these accuracies cannot
compete with those that can be achieved at high--energy $e^+ e^-$ linear
colliders [which are expected to operate either before or at the same time]
and to which we turn our attention now. 

\newpage


\section{Higgs production at lepton colliders}
\setcounter{equation}{0}
\renewcommand{\theequation}{4.\arabic{equation}}

\subsection{Lepton colliders and the physics of the Higgs boson}

\subsubsection{Generalities about $\ee$ colliders}

The $\ee$ collision \cite{ee-basics-review} is a very simple reaction, with a
well defined initial state and rather simple topologies in the final state. It
has a favorable signal to background ratio, leading to a very clean
experimental environment which allows to easily search for new phenomena and to
perform very high--precision studies as has been shown at PEP/PETRA/TRISTAN and
more recently at SLC and LEP.  In particular, the  high--precision studies of 
the properties of the $Z$ boson at LEP1 and SLC, and the determination of the
properties of the $W$ boson at LEP2,  have laid a very solid base for the
Standard Model as was discussed in \S1.\s

The physical processes in $\ee$ collisions are in general mediated by
$s$--channel photon [for charged particles] and $Z$ boson exchanges with cross
sections which scale as the inverse of the center of mass energy squared,
$\sigma \propto 1/s$, and $t$--channel gauge boson or electron/neutrino
exchange, with cross sections which may rise like $\log(s)$. In these
$t$--channel processes, only particles which couple directly to the electron
are involved at lowest order.  The $s$--channel exchange is the most
interesting process when it takes place: it is democratic, in the sense that it
gives approximately the same rates for weakly and strongly interacting matter
particles, and for the production of known and new particles, when the energy
is high enough.\s

However, in this channel, the rates are low at high energies and one needs to
increase the luminosity to compensate for the $1/s$ drop of the interesting
cross sections. At $\sqrt{s} \sim 1$ TeV, a luminosity ${\cal L}\sim 10^{34}$ 
cm$^{-2}\,$s$^{-1}$ is required, which for a run time of $10^7$s a year leads
to an integrated luminosity  of $\int {\cal L} \sim 100$ fb$^{-1}$ per year, to
produce $10^4$ muon pairs as at PEP and PETRA. Such a luminosity is necessary
to allow for thorough data analyses, including cuts on the event samples and
allowing for acceptance losses in the detectors. At higher energies, the
luminosity should be scaled as $s$ to generate the same number of events.\s

Because of synchrotron radiation which rises as the fourth power of the c.m. 
energy in circular machines, $\ee$ colliders beyond LEP2 must be linear
machines \cite{Technology-choice}, a type of accelerator which has been
pioneered by the SLC.  Various reference designs of future high--energy $\ee$
colliders in the TeV range are being studied in Europe (TESLA \cite{TESLAtdr}
at DESY and CLIC \cite{CLICtdr} at CERN), the United States (NLC \cite{NLCtdr})
and in Japan (JLC \cite{JLCtdr}). Two technologies have been proposed for the
next linear collider with center of mass energies up to $\sqrt{s}=1$ TeV: one
based on superconducting acceleration modules at moderate frequency, and
another based on  warm acceleration structures operating at high radio
frequencies.  A third and rather new approach is based on a two--beam scheme
where high current and low energy beams create the acceleration field for the
high--energy electron--positron beams. The technology for this collider, which
could reach c.m. energies in the multi--TeV range after presumably some halts
at intermediate energies, is still to be proved. \s

To achieve the  large luminosities which are targeted at  these machines, many
technological challenges need to be overcome. For instance, the beams have to
be focused to extremely small dimensions near the interaction point and very
high  acceleration gradients are needed to reach center of mass energies in the
TeV range. In the case of the TESLA machine, for which a Technical Design
Report  is already available, the main parameters of the beams  are summarized
in Table 4.1. In the case of the JLC and NLC machines, the designs have been
worked out in detail and are documented in a zeroth order  Technical Design
Report. The two--beam acceleration scheme is being followed at CERN and it is
hoped that before the end of this decade, the technical concept can be proved; 
this multi--TeV collider is thus expected to be a next generation machine. \s

\begin{table}[htbp]
\begin{center}
\renewcommand{\arraystretch}{1.5}
\hskip3pc
\vbox{\columnwidth=26pc
\begin{tabular}{|ccc|c|c|c|}\hline
Parameter & Label & Units & 500 GeV &800 GeV & $\gamma \gamma/500$ GeV\\ \hline 
Luminosity & ${\cal L}$ & ${\rm 10^{34}cm^{-2}s^{-1}}$ & 3.4 & 5.8 & 0.6 \\
Number of bunches   & $n_b$ & & 2820 & 4886 & 2820 \\ 
Pulse train length  & $T_P$ & $\mu {\rm s}$ & 950 & 860 & 950  \\ 
Repetition rate & $f_{\rm rep}$ & ${\rm Hz}$ & 5 & 5 & \\ 
Acceleration gradient & $E_{\rm acc}$ & ${\rm MV/m}$ & 23.4 & 35 & 23.4 \\ 
Beamstrahlung & $\delta_E$ & \% & 3.2 & 4.3 & $-$  \\ \hline
\end{tabular}
}
\end{center}
\vspace*{1mm}
{\it Table 4.1: Main parameters of the TESLA Linear Collider for the energies
$\sqrt{s} =500$ and 800 GeV, with $\sqrt{s_{ee}}=500$ GeV for the $\gamma 
\gamma$ option of the machine.}\bigskip
\vspace*{-.2cm}
\end{table}

For details on the future machines and on the issues related to the foreseen
and planned detectors, we refer the reader to Refs.~\cite{eeReviews,Rolf-para}. 
In the following, we briefly list a few important physics points about 
these future linear $\ee$ colliders  \cite{TESLA,LC-Notes,NLC,JLC,CLIC}: \s 

$\bullet$ One should have the possibility to adjust the c.m. energy of the 
colliders in order to make detailed studies and, for instance, to maximize the 
cross section for Higgs  boson production in some particular channels or scan 
the threshold for $W$ boson and top quark pair production, or for some newly 
produced particles. \s

$\bullet$ The requirement of a high--luminosity is achieved by squeezing the
electrons and positrons into bunches of very small transverse size, leading to
beamstrahlung which results into beam energy loss and the smearing of the
initially sharp $\ee$ c.m. energy. Since the precise knowledge of the initial
state energy is very important for precision studies [in particular in some
channels where one would need missing mass techniques], beamstrahlung should be
reduced to a very low level, as is already the case in narrow  beam designs.
\s

$\bullet$ The longitudinal polarization of the electron [and, to a lesser
extent, positron] beam should be easy to obtain as has already been shown at
the SLC. Degrees of polarization of the order of 80--90\% and 40--50\% for,
respectively, the electron and positron beams are expected. The longitudinal
polarization might be important when it comes to make very precise measurements
of the properties of Higgs bosons and to suppress some large backgrounds [in
particular from $W$ bosons] to its production signals \cite{LC-polarization}. \s

$\bullet$ By building a bypass for the transport of the electron and positron
bunches, for instance, very high luminosities can also be obtained at energies
in the range of 100 GeV. Operating on the $Z$ boson resonance, $10^{9}$ $Z$
bosons can be produced, a sample which is two orders of magnitude larger than
the one obtained at LEP1. This GigaZ machine, in particular since longitudinal
polarization will be available, could significantly improve the precision tests
of the SM which have been performed in the previous decade \cite{LC-GigaZ}. \s

$\bullet$ Last but not least, the linear collider can run in three additional
modes. First, one just needs to replace the positron bunches by electron
bunches to have an $e^- e^-$ collider. Then, by illuminating the initial lepton
bunches by laser photons, one can convert the original collider into an $e
\gamma$ or $\gamma \gamma$ collider, with a  comparable total center of mass
energy and luminosity as the initial lepton collider
\cite{gamma-machine1,gamma-machine2}. Higgs particles can be produced as
$s$--channel resonance  at $\gamma \gamma$ colliders
\cite{gamma-Rev-old,gamma-Rev-TESLA,gamma-Rev-NLC,gamma-Jose} and this mode
will be very useful to address problems such as the Higgs boson couplings to
photons and its CP properties. These options, will be also considered in this
chapter. \s

Very recently, the International Technology Recommendation Panel has
recommended \cite{Technology-choice} that the next linear $e^+e^-$ machine,
which should and hopefully will be a joint project, the International Linear
Collider (ILC), is based on superconducting radio--frequency cavities. The
machine should, in a first step, run at energies between $\sqrt{s}=200$ and 500
GeV with an integrated luminosity of 500 fb$^{-1}$ in the four first years,
have the possibility of 80\% polarized electron beams and two interaction
regions with easy switching.  In a second phase, the machine should run at an
energy of $\sqrt{s}=1$ TeV with an integrated luminosity of 1 ab$^{-1}$ in four
years. As options, the panel recommended that the machine should possibly run
in the $e^- e^-$ mode, have 50\% positron polarization, the possiblity to
operate near the $M_Z$ and $2M_W$ thresholds, and  the possibility to run in
the $e\gamma$ and $\gamma \gamma$ modes. \s

In our study of the physics of the Higgs boson at $\ee$ linear colliders, we
will assume for the numerical analyses three values for the c.m. energy,
$\sqrt{s}=0.5, 1$ and 3 TeV which will correspond to the two phases discussed
above and to the subsequent CLIC phase, and an integrated luminosity ${\cal L}
\sim 500$ fb$^{-1}$.  We will also consider briefly the GigaZ option and in
some detail the option of turning the machine into to a $\gam$ collider, the
particularities of which are summarized in the following subsection. Finally,
future muon colliders will be discussed in the last section of this chapter. 

\subsubsection{The photon colliders}

The Compton scattering of laser photons with energies $\omega_0$ in the eV
range with high--energy electrons, $E_e \sim {\cal O}(100~{\rm GeV})$, leads to
a tight bunch of back--scattered high--energy photons
\cite{gamma-machine1,gamma-machine2}. The kinematics of the process is governed
by the dimensionless parameter $x =4 \omega_0 E_e/ m_e^2$.  The fraction of
energy carried by the back--scattered photon, $y=\omega/E_e$, is maximal for
$y_{\rm max}= x/(1+x)$. The highest energy, compared to the $e^-$ beam energy
is therefore obtained for very large values of the parameter $x$. However, to
prevent the creation of $\ee$ pairs in the annihilation of the laser and
scattered photons, one demands that  $x \lsim x_0=4.83$. For this value, the
photon collider can have as much as $\sim 80$\% of the  energy of the original
$\ee$ collider. The scattering angle of the obtained photons is given by
$\theta(y) \simeq m_e( 1+x)/E_e \times \sqrt{y_{\rm max}/y-1}$ and is of the
order of a few micro--radians. \s

The energy spectrum of the back--scattered photon
\beq
f_c (y)= \sigma_c^{-1} \, {\rm d} \sigma_c (y)/ {\rm d}y
\eeq
depends on the product of the  the mean helicity of the initial electron 
$\lambda_e$ and on the degree of circular polarization of the laser photon
$P_\gamma$ with $ -1 \leq 2 \lambda_e P_\gamma \leq +1$. It is defined by 
the differential Compton cross section \cite{gamma-machine2}
 \beq
\frac{d\sigma_c}{dy}  = \frac{\pi\alpha^2}{x m_e^2} [f_0 + 2\lambda_e P_\gamma 
f_1 + 2\lambda_e P_{\gamma '} f_2 + P_\gamma P_{\gamma '} f_3]
\eeq
where the dependence on the polarization of the back--scattered photon 
$P_{\gamma'}$ has been retained. In terms of the variable $r=y/[x(1-y)]$, the 
functions  $f_i \ (i=0,..,3)$ read
\beq
f_0 = \frac{1}{1-y} + 1-y - 4r(1-r) \;, \quad f_1 = xr (1-2r)(2-y) \non\\
f_2 = xr \bigg[ 1+(1-y)(1-2r)^2\bigg] \;, \quad f_3 = (1-2r) \bigg[ \frac{1}
{1-y}+1-y \bigg]
\eeq
When the polarization of the scattered photon is discarded, the integrated 
Compton cross section can be cast into the form 
\beq
&& \hspace*{2cm} \sigma_c= \sigma_c^{\rm np} + 2\lambda_e P_\gamma
\sigma_c^{\rm p} \non \\ 
\sigma_c^{\rm np} &=& \frac{\pi\alpha^2}{x m_e^2} \Bigg[ \frac{1}{2} + 
\frac{8}{x} -\frac{1}{2(1+x)^2} + \bigg( 1-\frac{4}{x}-\frac{8}{x^2} 
\bigg) \ln (x+1) \Bigg] \non \\
\sigma_c^{\rm p} &=& \frac{\pi\alpha^2}{x m_e^2} \Bigg[ -\frac{5}{2} + 
\frac{1}{x+1} - \frac{1}{2(x+1)^2} + \bigg(1+\frac{2}{x}\bigg) 
\ln (x+1) \Bigg]
\eeq
By selecting a given polarization of the initial $e^-$ and laser beams, one 
can have different shapes for the energy distribution of the scattered photons:
a flat distribution if the electron and the laser have the same polarization, 
or an almost  monochromatic  distribution peaked at $y_{\rm max}$ if they have 
opposite polarization. The latter scenario if of course very interesting. \s

Because of the small scattering angle $\theta$ of the photons, the luminosity
of the spectrum depends on the the conversion distance, i.e. the distance
between the intersecting point of the laser and the electron beam and the
interaction point, as well as on the size and shape of the electron beam.  A
geometrical factor $\rho= b \theta/a$ takes into account the non--zero
conversion distance $b$ and the radius $a$ of the assumed round electron beam
[typically, the sizes are $a \sim {\cal O}(10^2~{\rm nm})$ and $b \sim {\cal
O}(1~{\rm cm})]$. If $\rho$ is much larger than unity, only the photons with
high energy can meet at the interaction point, while for $\rho \ll 1$,  photons
with various energies collide and give a rather broad spectrum. \s

When the Compton backscattered photons, that we will denote by $\gamma_1$ and 
$\gamma_2$, are taken as initial states, the cross section for the process
$\gamma \gamma \to X$  with polarized photons reads
\beq
{\rm d}\hat{\sigma}_{\gamma \gamma} = \sum_{i,j=0}^4 \, \xi_{1i} \xi_{2j} \, 
{\rm d}\hat{\sigma}_{ij} 
\eeq
in the $\xi_{1i}$ and $\xi_{2j}$ photon Stokes parameter basis, with 
zeroth components such that $\xi_{10}=\xi_{20}=1$. The event rate d$N$ can be
then written as
\beq
{\rm d} N \, = \, {\rm d}{\cal L} \, \langle {\rm d} \hat{\sigma}_{\gamma 
\gamma}  \rangle \, = \, {\rm d}{\cal L} \, \sum_{i,j=0}^3  \, \langle \xi_{1i}
\xi_{2j} \rangle \, {\rm d} \hat{\sigma}_{ij}
\eeq 
where d${\cal L}$ is the differential $\gamma \gamma$ luminosity and  the
average $\langle \xi_{1i} \xi_{2j} \rangle$ is along the interaction region 
[only the diagonal terms in the product are relevant in general]. \s

For circularly polarized laser beams, one has for the average Stokes parameters 
\beq
\langle \xi_{12} \xi_{22} \rangle = \xi_{12} \xi_{22}  \ , \ \langle \xi_{13} 
\xi_{23} \rangle = - \langle \xi_{11} \xi_{21} \rangle \ll 1
\eeq
so that the 
event rate can be written in terms of the luminosities corresponding to the 
$J_Z=0$ and $J_Z=\pm 2$ scattering channels, as
\beq
{\rm d}N \, = \, {\rm d} {\cal L}^{J_Z=0} \, {\rm d} \hat{\sigma}_{J_Z=0} \, 
+ \, {\rm d} {\cal L}^{J_Z=\pm 2 } \, {\rm d} \hat{\sigma}_{J_Z=\pm 2} 
\hspace*{2cm} \non \\
{\rm d} {\cal L}^{J_Z=0}  =  \frac{1}{2}  {\rm d}{\cal L} \, 
(1 + \langle \xi_{12} \xi_{22} \rangle ) \ , \ 
{\rm d} {\cal L}^{J_Z= \pm 2} = \frac{1}{2}  {\rm d}{\cal L} \, 
(1 - \langle \xi_{12} \xi_{22} \rangle ) 
\eeq

With this polarization, a  broad luminosity spectrum can be achieved by using
electrons and laser photons with like--handed helicities and a small value,
$\rho \sim 0.6$, which leads to low energetic backscattered photons in the
interaction region. In contrast, a sharp spectrum peaking near $y_{\rm max}$
can be obtained using opposite--handed electrons and laser photons in a more
restrictive interaction region $\rho \sim 3$; see Fig.~4.1 (left). The events
in the $J_Z=0\,(J_{Z}= \pm 2)$ channels can be enhanced (suppressed) by
choosing the laser and electron beams so that $x_0=4.83$, which in addition,
maximizes the collider energy.\s

\begin{figure}[htbp]
\begin{center}
\hspace*{-5mm}
\epsfig{file=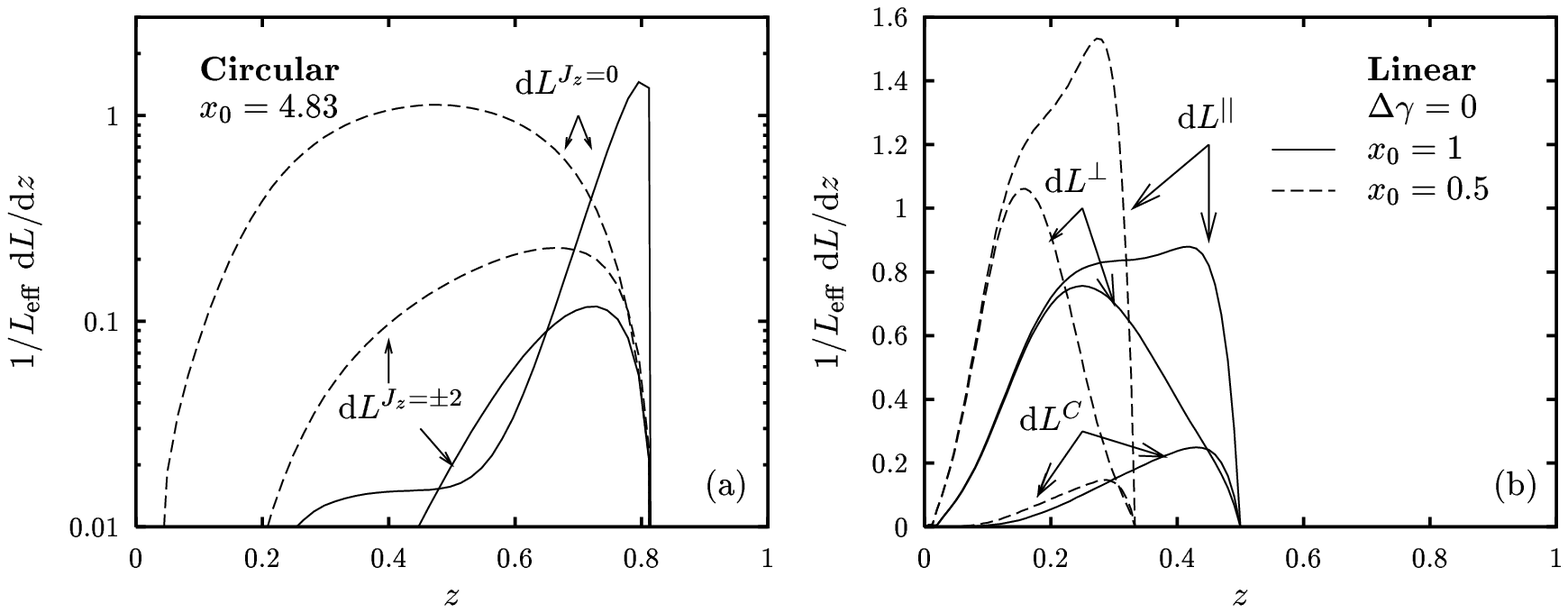,angle=0,width=1.\linewidth}
\vspace*{-3mm}
\end{center}
{\it Figure 4.1: Normalized $\gamma$ luminosities as functions of
$z= \sqrt{s_{\gamma \gamma} }/ \sqrt{s_{\ee } }$ for left:  circularly
polarized lasers with $x_0=4.83$ and the solid (dashed) lines are for
opposite-handed (like-handed) photons and electrons with $\rho=3\, (0.6)$, 
and  right:  linearly polarized lasers with $\Delta\gamma=0$, $\rho=0.6$
and $x_0=1\,  (0.5)$ for the solid (dashed) lines. The lasers are assumed
to be completely polarized and the electrons 85\% longitudinally polarized,
and the configurations for both collider arms are the same; from 
Ref.~\cite{gamma-Jose}.}
\end{figure}

For linearly polarized laser beams, neglecting $\rho \ne 0$ effects for 
simplicity, the average Stokes parameters are
\beq 
\langle \xi_{12} \xi_{22} \rangle &\simeq& \langle \xi_{12} \rangle
\langle \xi_{22} \rangle \, = \, 4 \lambda_{e^-} \lambda_{e^+}  \, c_1 
c_2 \non \\
\langle \xi_{13} \xi_{23} -\xi_{11} \xi_{21} \rangle & \simeq & \langle
\xi_{13} \rangle\langle \xi_{23} \rangle -
\langle \xi_{11} \rangle\langle \xi_{21} \rangle=
P_{1t} P_{2t} \, {\ell}_1 {\ell}_2 \, \cos 2 (\Delta\gamma) 
\eeq 
where  $P_{ti}$ are the mean linear laser polarizations  while $c_i$ and 
$\ell_i$ are the induced circular and linear polarizations of the backscattered
photons; $\Delta\gamma$ is the angle between the planes of maximal linear 
polarization of the two lasers. The circular polarizations $c_i$ and linear 
polarizations $\ell_i$ are large for, respectively,  high and low values of the
parameter $x$, and both increase with $y$. The event rate in this case is 
given by
\beq
 {\rm d}N = {\rm d} {\cal L}^{||}\, {\rm d} \hat{\sigma}_{||}
+ {\rm d} {\cal  L}^{\perp}\, {\rm d}\hat{\sigma}_\perp + \frac{1}{2} {\rm d} 
{\cal L}^C \, \left( {\rm d}\hat{\sigma}_{J_Z=0} - {\rm d} \hat{\sigma}_{J_Z=2}
\right) \hspace*{3cm}  \\
{\rm  d} {\cal L}^{||} = \frac{1}{2} {\rm d} {\cal L}\,  (1+\langle
\xi_{13} \xi_{23}-\xi_{11} \xi_{21} \rangle) \, , \ 
{\rm d}{\cal  L}^{\perp} = \frac{1}{2} {\rm d} {\cal  L} \,  (1-\langle
\xi_{13} \xi_{23} -\xi_{11} \xi_{21} \rangle) \, , \
{\rm d} {\cal L}^C = {\rm d} {\cal L}\,  \langle \xi_{12} \xi_{22} \rangle \non
\eeq
For this polarization, one has to make a compromise between having a 
good separation of the $||$ and $\perp$ components, which needs a small value
of $x$, and having a high energy which needs a larger value since the available
energy is proportional to $x/(x+1)$. \s

For more details on the main features of the $\gamma \gamma$ machines, such
as energy, luminosity distributions, polarization, etc...,  see the  
reviews given in Refs.~\cite{gamma-Rev-TESLA,gamma-Rev-NLC,gamma-Jose}.

\subsubsection{Future muon colliders}

The concept of $\mu^+\mu^-$ colliders, although introduced already in the late
sixties \cite{mu-Budker}, has been taken very seriously only in the last
decade. A Muon Collider Collaboration (MCC) \cite{mu-machine,mu-machine1} has 
been
formed in the US in the mid--nineties to complete the R\&D that is required to
determine whether a muon collider is technically feasible and, in the case of a
positive answer, to propose a design for a First Muon Collider. In the
late nineties, the European community joined the project and a study report on
the feasibility of a muon collider at CERN has been produced
\cite{mu-machine2}. A three--step scenario for a muon collider is presently
foreseen \cite{mu-machine1,mu-machine2}:\s 

-- $i)$ A first step would be an intense proton source for producing muons 
which will be then captured, cooled, accelerated and stored. In the 
storage ring, they then decay and they allow to produce high--intensity 
and high--quality neutrino beams which could be used to perform detailed studies
of neutrino oscillations and neutrino--nucleon scattering, as well as some
physics with stopped muons such as the measurement of the muon magnetic and 
electric dipole moments and the search for some rare $\mu$ decays.\s  

-- $ii)$ The second step would be a $\mu^+ \mu^-$ collider with a center of 
mass energy in the range $\sqrt{s} \sim 100$--200 GeV. This collider could do 
the same physics as an $e^+e^-$ collider and it will be a Higgs factory 
that would possibly allow to study in more detail the properties of the Higgs 
particles that have been produced at the LHC and at the ILC.\s

-- $iii)$ A final step would be then to operate the muon collider at the
maximum possible c.m. energy and to probe the physics of the multi--TeV scale. 
For instance, energies up to $\sqrt s \sim 7$ TeV could be reached with the 
facilities that are available at CERN. However, with the present designs [and 
not to mention the very high luminosities which need to be achieved in this 
case], the radiation induced by the neutrinos is extremely high for c.m. 
energies in excess of a few TeV and poses a very serious problem. Major 
technological developments are therefore required to reach this high--energy 
step.\s 

In this report, we will be interested only in the second phase of the muon
collider, that is, the Higgs factories with c.m. energies $\sqrt s \lsim 200$
GeV. Compared to an $\ee$ machine, the main advantages of a muon collider as
far as Higgs physics is concerned \cite{mu-Rev1,mu-Rev2,mu-Rev3,mu-Rev4}, are
principally due to the fact that the muon is much heavier than the electron,
$m_\mu/m_e \sim 200$: the Higgs boson coupling to muons is much larger than the
coupling to electrons, yielding significantly larger rates for $s$--channel
Higgs production at muon colliders, $\mu^+ \mu^- \to H$ [the production rate in
this channel is of course negligible in $\ee$ collisions].\s

Another advantage of $\mu^+\mu^-$ colliders, compared to $\ee$ colliders, is
the very precise knowledge of the beam energy spectrum which would allow for
very high precision analyses of the mass, total width and peak cross section of
the produced Higgs resonance. According to the analyses performed in
Ref.~\cite{mu-machine1,mu-machine2}, the energy can be tuned with a precision
$\Delta E_{\rm b}/E_{\rm b} \sim \times 10^{-6}$ [i.e. 100 keV for $\sqrt 
s\!=\! 100$ GeV] and values ten times smaller seem possible. 
The small amount of beamstrahlung [which, in $\ee$ collisions, induces an
energy loss of a few percent that is difficult to measure very precisely] and
bremsstrahlung [again due to the larger mass of the muon compared to the
electron] could lead to a relative beam energy spread or resolution of the
order of $R= \sigma_{ E_{\rm b} }/E_{\rm b} \sim 10^{-3}$ down to $R= 3 \times
10^{-5}$ and which could be known with an accuracy of $ \Delta \sigma_{E_{\rm
b} }/E_{\rm b} \sim 0.5\%$. Such a small energy spread is very important when
performing a scan around the very narrow Higgs resonance, $\Gamma_H \sim 
2$ MeV for $M_H \sim 100$ GeV. In addition, since synchrotron radiation is also
very small, one can still use the available circular machines. 
The energy calibration can be made by spin precession as the muons that are
produced in the weak decays of pions are 100\% polarized, leading to a natural
longitudinal polarization of approximately 30\% which, however, drops to the
level of $\sim 20\%$ due to the handling before injection into the collider. The
drawback, compared to $\ee$ machines, is that it is difficult to maximize this 
polarization without an important loss in luminosity and that a muon collider 
cannot be turned into a $\gamma \gamma$ or $\mu \gamma$ collider.\s

Nevertheless, the design of the machine is still at an early stage and  many
problems remain to be solved \cite{mu-machine1,mu-machine2}.  In addition, the
delivered luminosity which can be achieved is still uncertain, and it depends
strongly on the baseline parameters of the collider; Tab.~4.2. There is, for 
instance, a particularly strong dependence on the beam energy resolution. As 
can be seen from the table, at $\sqrt s=100$ GeV, the estimates indicate that
only ${\cal 
L} \sim 10^{31}\, (10^{32})\, {\rm cm}^{-2}{\rm s}^{-1}$ can be obtained for a
resolution of $R= 0.003\%\, (0.1\%)$, leading to an integrated luminosity 
$\int {\cal L}=0.1\, (1)~{\rm fb}^{-1}$ per year. The luminosity, however can 
substantially be increased with energy reaching, for $R \sim 0.1$\%, values of 
the order of ${\cal L}\sim 10^{33}\, (10^{35})\, {\rm cm}^{-2}{\rm s}^{-1}$ for 
$\sqrt{s} \sim 0.4\, (3)$ TeV; see Table 4.2 and the details given in 
Refs.~\cite{mu-machine1,mu-machine2}. \s

\begin{table}[ht]
\begin{center}
\begin{tabular}{|l|c|c|ccc|} 
  \hline
c.m.\ energy      & 3 TeV            & 400 GeV & \multicolumn{3}{c|}{100 GeV} \\
  \hline
p power (MW)    & 4                & 4   &  \multicolumn{3}{c|}{4} \\
$1/\tau_\mu$ (Hz)& 32               & 240 &  \multicolumn{3}{c|}{960} \\
$\mu$/bunch      & $2\times10^{12}$ & $2\times10^{12}$ & \multicolumn{3}{c|}{$2\times10^{12}$} \\
circumference (m)& 6000             & 1000& \multicolumn{3}{c|}{350} \\
$\langle B\rangle$ (T)     & 5.2              & 4.7 & \multicolumn{3}{c|}{3} \\
$n^{\rm effective}_{\rm turns}$&785         & 700 & \multicolumn{3}{c|}{450} \\
6-D $\epsilon_{6,N} \times10^{-10}$ ($\pi$m$^3$) &$1.7$& $1.7$ &  
\multicolumn{3}{c|}{1.7}\\
$R=\delta p/p$ (\%)& 0.16             & 0.14& 0.12 & 0.001 & 0.003 \\
RMS $\epsilon_T$ ($\pi$ mm-rad)
                 & 0.05            & 0.05 & 0.085& $0.0195$ & 0.29 \\
$\beta^*$ and $\sigma_z$    (cm)   & 0.3             & 2.6  & 4.1& 9.4 & 14.1 \\
$\sigma_r$ ($\mu$m)& 3.2           & 26   & 86   & 196 & 294 \\ \hline
Luminosity (${\rm cm^2s^{-1} }$)& $7\times10^{34}$ & $10^{33}$ 
& $1.2\times10^{32}$ & $2.2 \times10^{31}$ & $10^{31}$  \\ \hline
\end{tabular}
\end{center}
\nn {\it Table 4.2: Possible parameter sets for a $\mu^+ \mu^-$ Higgs factory
at $\sqrt s\!=\! M_H\!=\!100$ GeV as expected by the MCC \cite{mu-machine1}; 
higher energy machines are also shown for comparison.}
\vspace*{-3mm}
\end{table}

\subsubsection{Higgs production processes in lepton collisions}

In $\ee$ collisions with center of mass energies beyond LEP2, the main
production mechanisms for Higgs particles are the Higgs--strahlung process 
\cite{EGN,LQT,Bjorken-process,Petcov,Higgs-strahlung} and the $WW$ fusion 
mechanism \cite{Petcov,VVH-Cahn,VVH-DW,VVH-Altarelli,VVH-Kilian,WWH-ee-Dawson,VVH-Hikasa}, 
depicted in Fig.~4.2,
\begin{eqnarray}
{\rm Higgs-strahlung \ process} : & & \ee \longrightarrow (Z^*) 
\longrightarrow Z \, H \\
{\rm WW \ fusion \ process} : & & \ee \longrightarrow \bar{\nu}\nu \, (W^*W^*)
\longrightarrow \bar{\nu}\ \nu \, H 
\end{eqnarray}

\begin{center}
\begin{picture}(300,100)(0,0)
\SetWidth{1.}
\ArrowLine(0,25)(40,50)
\ArrowLine(0,75)(40,50)
\Photon(40,50)(90,50){4}{5.5}
\DashLine(90,50)(130,25){4}
\Text(90,50)[]{\bb}
\Photon(90,50)(130,75){4}{5.5}
\Text(-10,20)[]{$e^-$}
\Text(-10,80)[]{$e^+$}
\Text(70,65)[]{$Z^*$}
\Text(139,20)[]{\bH}
\Text(139,80)[]{$Z$}
\ArrowLine(200,25)(240,25)
\ArrowLine(200,75)(240,75)
\ArrowLine(240,25)(290,0)
\ArrowLine(240,75)(290,100)
\Photon(240,25)(280,50){4}{5.5}
\Photon(240,75)(280,50){4}{5.5}
\DashLine(280,50)(320,50){4}
\Text(280,50)[]{\bb}
\Text(190,20)[]{$e^-$}
\Text(190,80)[]{$e^+$}
\Text(275,72)[]{$W^*$}
\Text(275,27)[]{$W^*$}
\Text(300,60)[]{\bH}
\Text(300,10)[]{$\nu_e$}
\Text(300,90)[]{$\bar{\nu}_e$}
\Text(150,-15)[]{\it Figure 4.2: The dominant Higgs production mechanisms 
in high--energy $\ee$ collisions.} 
\end{picture}
\vspace*{7.mm}
\end{center}
There are several other mechanisms in which Higgs bosons can be produced in 
$\ee$ collisions: the $ZZ$ fusion process 
\cite{VVH-Cahn,VVH-DW,VVH-Hikasa,VVH-Altarelli,ZZH-Kilian}, the 
radiation off heavy top quarks \cite{ee-ttH0,ee-ttH} and the double Higgs boson
production process either in Higgs--strahlung or $WW/ZZ$ fusion 
\cite{pp-HVV,pp-VVHH,ee-HHZ,HH-Barger,ee-DKMZ}
\begin{eqnarray}
{\rm ZZ \ fusion \ process} : & & \ee \lra e^+ e^- (Z^*Z^*) \lra e^+ e^- \, 
H \\
{\rm radiation \ off \ heavy\ fermions}:& & \ee \lra (\gamma^*,Z^*) \lra f 
\bar{f} \, H \\
{\rm double\ Higgs\ production}: & & \ee \lra Z H H \, , \, \ell \ell HH 
\end{eqnarray}
These are, in principle, higher--order processes in the electroweak coupling 
with production cross sections much smaller than those of the Higgs--strahlung 
process and the $WW$ fusion channel [for $ZZ$ fusion, only at low energies]. 
However, with the high luminosity planned 
for future linear colliders, they can be detected and studied. These processes 
are extremely interesting since they allow for the determination of some of the 
fundamental properties of the Higgs particle, such as its self--coupling and 
its Yukawa coupling to top quarks. \s

There also other higher--order processes in which Higgs particles can be
produced in $\ee$ collisions, but with even smaller production cross sections
than those mentioned previously: associated production with a photon, $\ee \ra
H+ \gamma$ \cite{ee-Hgamma}, loop induced Higgs pair production, $\ee \ra
HH$ \cite{ee-HHloop}, associated production with vector bosons, $\ee \ra
VV + H$ \cite{ee-HVff,ee-HVV}, and associated production with a gauge boson 
and two fermions, $\ee \ra VH + f\bar{f}$ \cite{ee-HVff}. Except possibly for 
the two latter processes, the cross sections are in general below the femtobarn 
level and, thus, too small for the processes to be detected at future machines, 
unless extremely high--luminosities are made available. \s

Higgs particles can be produced as $s$--channel resonances 
\cite{gamma-Rev-old,BBC}
[among other possibilities  which will be also discussed] in the 
$\gam$ option of future $\ee$ linear colliders 
\beq
\gamma \gamma \lra H 
\eeq
allowing the measurement of the important $H \gamma \gamma$ coupling. In the 
$e \gamma$ option, one can also produce the Higgs boson in the channel
$e \gamma \to \nu_e W^- H$ \cite{egamma-H}. \s

Finally,  on can also produce the Higgs boson as an $s$--channel resonance at 
future muon colliders \cite{mu-Rev1,mu-schannel}
\beq
\mu^+ \mu^- \lra H
\eeq

In the following sections, we discuss the dominant production processes in some
detail and summarize the main features of the subleading processes.  We first
focus on $\ee$ linear colliders in the  $\ee$ option, and discuss in more
details the physics potential at the first phase with center of mass energies
around $\sqrt{s} \sim 500$ GeV \cite{ee-Review-old,ee-Review-DESY};
occasionally, we will comment on the benefits of raising the energy of the
machine. The case of the $\gamma \gamma$ option of the machine, as well as the
physics at future muon colliders will be postponed to, respectively, the
previous--to--last and last sections.\s 

Since $\ee$ colliders are known to be high--precision machines as
demonstrated at LEP and SLC, the theoretical predictions have to be rather
accurate and thus the radiative corrections to the Higgs production  
processes have to be taken into account. The one--loop electroweak and QCD 
radiative corrections to the most important production mechanisms have been 
completed only recently 
\cite{RCHZ,RCWW1,RCWW2,RCWW3,RCZZ,RCTTqcd1,RCTTqcd2,RCTTew1,RCTTew2,RCTTew0,RCZHH1,RCZHH2,RCWWHH} and their main effects will be summarized.\s

In addition, the main motivation of future $\ee$ in the sub--TeV 
energy range is the detailed exploration of the electroweak symmetry breaking 
mechanism and the thorough study of the fundamental properties of the Higgs 
particle, in particular the spin and parity quantum numbers. At least in the 
main processes, we study the energy and the angular dependence of the 
cross sections as well as the angular correlations of the final decay products,
and confront, whenever possible, the predictions for the $J^{\rm PC}=0^{++}$ 
case of the SM Higgs particle to what would be expected if the Higgs were a 
pseudoscalar boson with $J^{\rm PC}=0^{+-}$ spin--parity assignments. We also 
discuss the measurements of the Higgs mass and total decay width, the Higgs 
couplings to fermions and gauge bosons, and the Higgs self--coupling which 
allows for the reconstruction of part of the scalar potential that is 
responsible of the spontaneous breaking of the electroweak symmetry. \s

Some particular points relevant to this section have been already discussed in 
the context of hadron colliders or in the section on the decays of the Higgs 
particle. However, some important features will be rediscussed in the context 
of lepton colliders, to make the section more complete and self--contained. 

\newpage

\subsection{The dominant production processes in $\ee$ collisions}

\subsubsection{The Higgs--strahlung mechanism}

\subsubsection*{\underline{The production cross section}}

The production cross section for the Higgs strahlung process is given by
\beq 
\sigma(\ee \ra ZH) = \frac{G_\mu^2 M_Z^4}{96 \pi s} (\hat{v}_e^2+\hat{a}_e^2) \ 
\lambda^{1/2} \frac{ \lambda+ 12M_Z^2/s}{(1-M_Z^2/s)^2} 
\eeq
where, as usual, $\hat{a}_e=-1$ and $\hat{v}_e=-1+4s_W^2$ are the $Z$ charges 
of the electron and $\lambda^{1/2}$ the usual two--particle phase--space 
function
\beq
\lambda=(1-M_H^2/s-M_Z^2/s)^2-4M_H^2M_Z^2/s^2
\eeq
The production cross section is shown in  Fig.~4.3 as a function of the Higgs
mass for the values of the c.m energy $\sqrt{s}=0.5,1$ and 3 TeV.  At
$\sqrt{s}=500$ GeV, $\sigma(\ee \to HZ)  \sim $ 50 fb for $M_H \sim 150$ GeV,
leading to a total of $\sim $ 25.000 Higgs particles that are created at an
integrated luminosity of $\int {\cal L} =500\ {\rm fb}^{-1}$, as expected for 
future machines. The cross section  scales as the inverse of the c.m. energy,
$\sigma \sim 1/s$ and, for moderate Higgs masses, it is larger   for
smaller c.m.  energies. The maximum value of the cross section for a given 
$M_H$ value is at $\sqrt{s} \sim M_Z +\sqrt{2}M_H$. An energy 
of the order of $\sqrt{s} \sim 800$ GeV is needed to cover the entire Higgs
boson mass range allowed  in the SM, $M_H \lsim 700$ GeV.  

\begin{figure}[!h]
\begin{center}
\vspace*{-1.2cm}
\hspace*{-1.cm}
\psfig{file=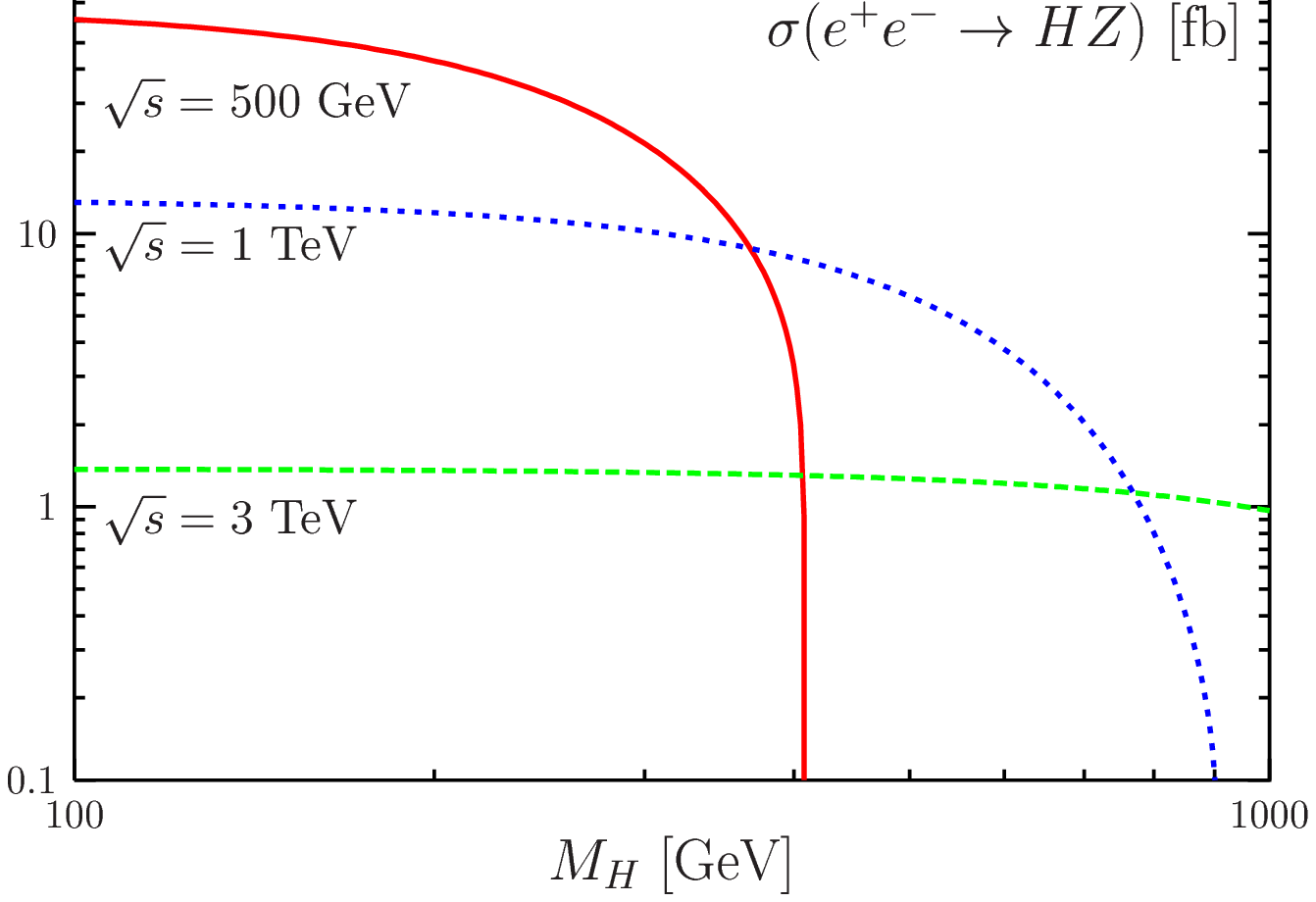,width=18.cm} 
\end{center}
\vspace*{-16.3cm}
\nn {\it Figure 4.3: Higgs boson production cross sections in the 
Higgs--strahlung mechanism in $\ee$ collisions with c.m. energies 
$\sqrt{s}=0.5,1$ and 3 TeV as a function of $M_H$.} 
\vspace*{-.5cm}
\end{figure}

\subsubsection*{\underline{The energy dependence}}

The recoiling $Z$ boson in the two--body reaction $\ee \ra ZH$ is
mono--energetic, $E_Z = (s-M_H^2+M_Z^2)/(2\sqrt{s})$, and the mass of the Higgs
boson can be derived from the energy of the $Z$ boson, $M_H^2 =s -2\sqrt{s} E_Z
+M_Z^2$, if the initial $e^+$ and $e^-$ beam energies  are sharp. \s

The excitation curve rises linearly with the phase--space factor 
$\lambda^{1/2}$, which is characteristic to the production of a scalar 
particle in association with a $Z$ boson  
\beq
\sigma (\ee \to HZ) \sim \lambda^{1/2} \sim \sqrt{s -(M_H+M_Z)^2} 
\eeq
This behavior for the $J^{\rm PC}=0^{++}$ SM Higgs boson can be compared with
the case of a CP--odd Higgs boson $A$ with $J^{\rm PC}=0^{+-}$ quantum 
numbers and with couplings given in \S2. The total production cross section
for the process $\ee \to ZA$ \cite{Bargeretal,ee-HZ-stong}
\begin{eqnarray}
\sigma (\ee \to ZA) = \eta^2\frac{G_\mu^2M_Z^6}{48\pi M_A^4} \, 
(\hat{a}_e^2+\hat{v}_e^2) \, \frac{\lambda^{3/2}}{(1-M_Z^2/s)^2}
\end{eqnarray}
has a momentum dependence $\sim \lambda^{3/2}$ that is characteristically
different  from the $ZH$ cross section near threshold. This is illustrated in
Fig.~4.4, where the behavior near the production threshold for the assignments
$J^{\rm PC}= 0^{++}$ and $0^{+-}$ is shown for a Higgs mass $M_H=120$ GeV.\s

\begin{figure}[!h]
\begin{center}
\vspace*{-1.cm}
\hspace*{-1.cm}
\psfig{file=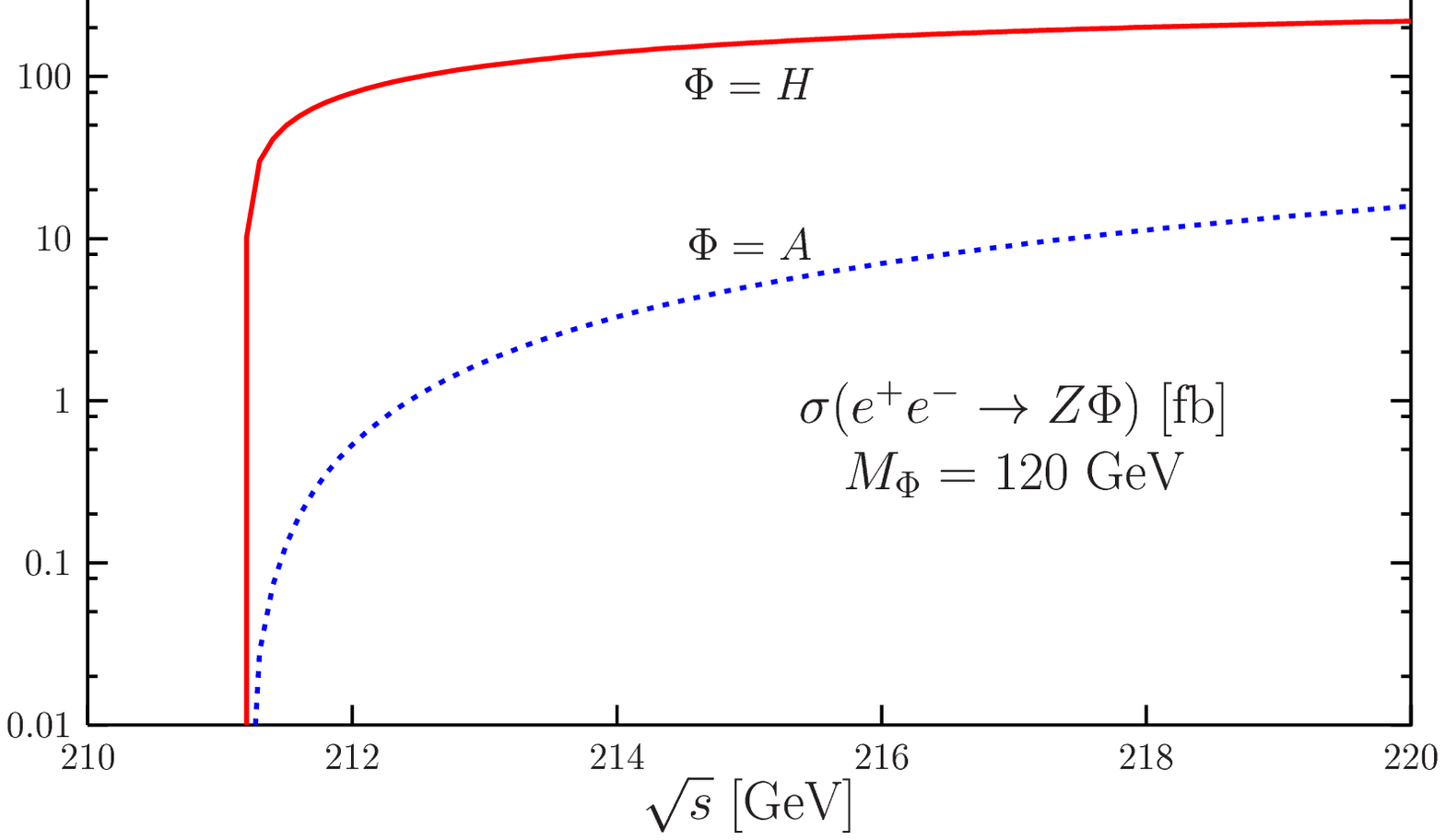,width=13.cm,height=23cm} 
\end{center}
\vspace*{-14.9cm}
\nn {\it Figure 4.4: The $\ee \to Z\Phi$ cross section energy dependence near
the threshold for the two parity cases $\Phi=H$ and $\Phi=A$ [with $\eta=1]$ 
with $M_\Phi=120$ GeV.}
\vspace*{-.2cm}
\end{figure}

In fact, as discussed in Ref.~\cite{ee-Hspin}, the linear threshold behavior of
the SM Higgs boson rules out not only the quantum number $J^{\rm P}=0^{-}$ but
also $J^{\rm P}=1^{-}, 2^{+}$ and higher spin $3^\pm,  \cdots$, which rise
with higher powers of  $\lambda$ too.  The production of states with the two
remaining spin--parity assignments $J^{\rm P}=1^+, 2^+$ can be ruled out using 
the angular correlations as is discussed hereafter.  

\subsubsection*{\underline{ The angular distribution}}

The angular distribution of the $Z/H$ bosons in the bremsstrahlung process is 
also sensitive to the spin of the Higgs particle \cite{ee-HZ-ang}. The explicit
form of the angular distribution, with $\theta$ being the scattering angle, is 
given by
\begin{eqnarray}
\frac{ {\rm d}\sigma (\ee \to ZH)} { {\rm d} \cos \theta} \sim \lambda^2
\sin^2\theta +8M_Z^2/s  \ \stackrel{ s \gg M_Z^2} \longrightarrow  \ 
\frac{3}{4} \sin^2\theta
\end{eqnarray}
and approaches the spin--zero distribution asymptotically, $\propto \sin^2
\theta$, in accordance with the equivalence theorem which requires that the 
production amplitude becomes equal to the amplitude where the $Z$ boson is 
replaced by the neutral Goldstone boson $w_0$. Thus, for high energies, the $Z$ 
boson is produced in a state of longitudinal
polarization
\begin{eqnarray}
\frac{\sigma_L} {\sigma_L +\sigma_T} = 1- \frac{8M_Z^2}{12M_Z^2+\lambda s}
\end{eqnarray}
Let us again confront the characteristics of a $J^{\rm PC}=0^{++}$ state with
those of a pseudoscalar Higgs boson $A$. In the  process $\ee \to ZA$, the 
angular distribution is given by 
\begin{eqnarray}
\frac{ {\rm d}\sigma (\ee \to ZA)} { {\rm d} \cos \theta} \sim 1+ \cos^2\theta
\end{eqnarray}
independent of the energy. The $Z$ boson in the final state is purely
transversally polarized, so that the cross section need not be $\sim
\sin^2\theta$ in this case.\s

If the Higgs particle were a mixture $\Phi$ of scalar and pseudoscalar bosons, 
with a coupling to the virtual and real $Z$ bosons given by 
\beq
g_{ZZ\Phi}=g_{ZZH} \bigg( g_{\mu \nu} + i \eta M_Z^{-2}\epsilon_{\mu \nu 
\rho \sigma} p^\sigma_{Z} p^\rho_Z \bigg)
\eeq
the angular distribution of $\ee \to \Phi Z$ would read [$A_f= 2a_f 
v_f/(a_f^2+a_f^2)$ as usual]
\begin{eqnarray}
\frac{ {\rm d}\sigma (\ee \to Z\Phi)} { {\rm d} \cos \theta} \sim 1+ 
\frac{s\lambda^2}{8M_Z^2} \sin^2\theta + \eta A_e \frac{s \lambda}{M_Z^2}
\cos \theta + \eta^2 \frac{s^2\lambda^2}{M_Z^4} (1+\cos^2\theta)
\label{HZ:angular}
\end{eqnarray}
The presence of the interference term proportional to $\eta$ is a clear 
indication of CP--violation in the Higgs sector. One can thus define an
observable \cite{ee-HZ-cos}, conveniently written as, 
\beq
\langle O \rangle = 2 {\rm Re} \bigg( \frac{ {\cal M} (\ee \to ZH) 
{\cal M}^* (\ee \to ZA) } { | {\cal M} (\ee \to ZH)| ^2 } \bigg) \propto  
\eta A_e \frac{s \lambda}{M_Z^2}
\label{Oobservable}
\eeq
which quantifies the amount of this CP--violation. \s

\begin{figure}[!h]
\begin{center}
\vspace*{-2.2cm}
\hspace*{-1.cm}
\psfig{file=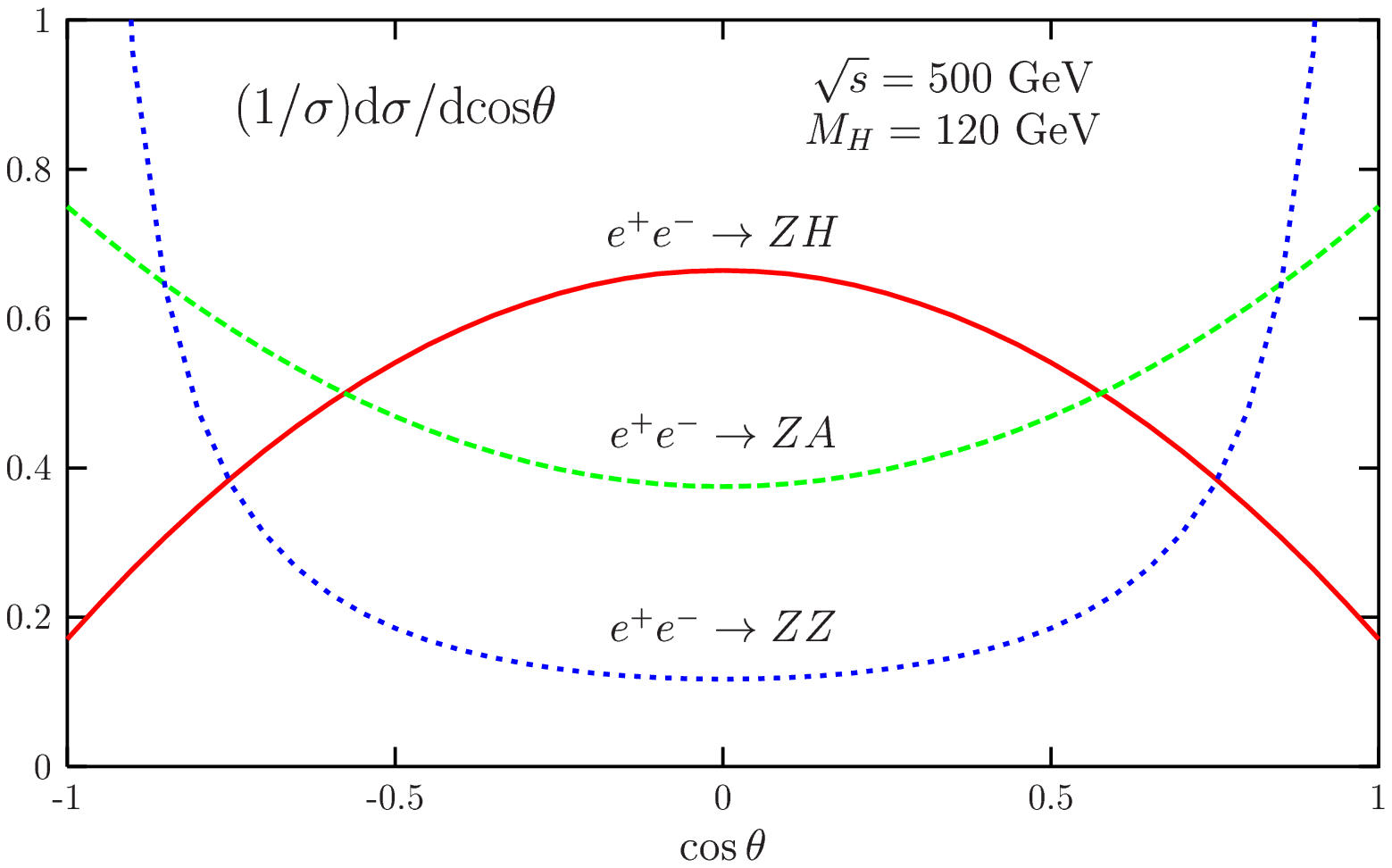,width=16.cm} 
\end{center}
\vspace*{-13.cm}
\nn {\it Figure 4.5: Angular distribution in the process  $\ee \to HZ$ for
$\sqrt{s}=500$ GeV and $M_H=120$ GeV. The distributions for  the CP--odd Higgs
and $\ee \to ZZ$ cases are also shown.}
\end{figure}

The angular momentum structure specific to Higgs production can also directly
be  confronted experimentally with the one of the process $e^+e^- \rightarrow  
ZZ$ that is distinctly different. Mediated by electron exchange in the
$t$--channel, the amplitude for this process is built--up by many partial waves,
peaking in the forward/backward directions. The two angular distributions, 
together with the angular distribution for the CP--odd Higgs case, $\ee \to AZ$,
are compared with each other in Fig.~4.5 which demonstrates the specific 
character of the SM Higgs production process. 

\subsubsection*{\underline{The angular correlations}}

The pattern for the $Z$ boson polarization in the $\ee \to HZ,HA$ and $ZZ$ 
processes can be checked \cite{Bargeretal,ee-HZ-stong}: while the distribution 
of the fermions in the $Z 
\to  f\bar{f}$ rest frame with respect to the $Z$ flight direction is given by 
$\sin^2\theta_*$ for longitudinally polarized $Z$ bosons, it behaves as 
$(1\pm \cos\theta_*)^2$ for transversally polarized states, after averaging 
over the azimuthal angles. The definition of the polar angles  $\theta$ and 
$\theta_*$ is shown in Fig.~4.6; the azimuthal angle $\phi_*$ is the angle 
between the plane of the $f\bar{f}$ from $Z$ decays and the Higgs decay 
products. \s

Including the azimuthal angles, the final angular correlations may be written
for the process $e^+e^- \rightarrow  ZH$ with $Z\rightarrow  f \bar{f}$ as
\cite{Bargeretal}
\begin{eqnarray}
\frac{ {\rm d}\sigma (\ee \to ZH)} { {\rm d} c_\theta {\rm d} c_{\theta_*}
{\rm d} \phi_*} &\sim & s^2_\theta s^2_{\theta_*} -\frac{1}{2 \gamma} s_{2
\theta} s_{2\theta_*} c_{\phi_*} + \frac{1}{2\gamma^2} [(1+c^2_\theta)(1
+c^2_{\theta_*})+s^2_\theta s^2_{\theta_*} c_{2\phi_*} ] \nonumber \\
&&- 2A_e A_f\frac{1}{\gamma} \left[ s_\theta s_{\theta_*} c_{\phi_*} -
\frac{1}{\gamma} c_\theta c_{\theta_*} \right]
\end{eqnarray}
where $s_\theta= \sin \theta$ $etc$, $A_f=2v_fa_f/(v_f^2+a_f^2)$ and $\gamma^2
=E^2/M_Z^2= 1+ \lambda s/4M_Z^2$. As before, $\theta$ is the polar $Z$ angle 
in the laboratory frame, $\theta_*$ the polar fermion angle in the $Z$ rest 
frame and $\phi_*$ the corresponding azimuthal angle with respect to the 
$\ee \to ZH$ production plane. After integrating out the polar angles 
$\theta$ and $\theta_*$, one finds the familiar $\cos \phi_*$ and $\cos 2 
\phi_*$ dependence discussed in \S2.2.4 associated with P--odd and even 
amplitudes, respectively
\begin{eqnarray}
\frac{ {\rm d}\sigma (\ee \to ZH)} { {\rm d} \phi_*} \sim 1+a_1 \cos \phi_*
+ a_2 \cos 2\phi_* \non \\
a_1= -\frac{9\pi^2}{32}\, A_e A_f\, \frac{\gamma}{\gamma^2+2}\, \, 
,\qquad a_2= \frac{1}{2}\,\frac{1}{\gamma^2+2}
\end{eqnarray}
The azimuthal angular dependence disappears for high energies $\sim 1/\gamma$
as a result of the dominating longitudinal polarization of the $Z$ boson.\s

Note again the characteristic difference to the $0^{+-}$ case, 
$e^+e^- \rightarrow  ZA \to f\bar{f} A$ \cite{Bargeretal,CP-full}
\begin{eqnarray}
\frac{ {\rm d} \sigma (\ee \to ZA)} { {\rm d} c_\theta {\rm d} c_{\theta_*}
{\rm d} \phi_*} &\sim & 1+ c^2_\theta c^2_{\theta_*} -\frac{1}{2}
s^2_\theta s^2_{\theta_*} -\frac{1}{2} s^2_\theta s^2_{\theta_* }
c_{2\phi_*} + 2 A_e A_f c_\theta c_{\theta_*}
 \end{eqnarray}

\begin{center}
\vspace*{.5cm}
\hspace*{-4cm}
\begin{picture}(550,200)(0,0)
\SetWidth{1.5}
\LongArrow(220,100)(317,100)
\LongArrow(420,100)(323,100)
\LongArrow(320,100)(360,150)
\LongArrow(320,100)(280,50)
\SetWidth{1.}
\DashLine(360,150)(400,200){5}
\LongArrow(360,150)(390,110)
\LongArrow(360,150)(330,190)
\Line(200,210)(200,100)
\Line(200,50)(200,40)
\Line(200,100)(200,50)
\Line(200,210)(440,210)
\Line(200,40)(440,40)
\Line(440,210)(440,40)
\SetWidth{0.8}
\LongArrowArc(320,100)(23,0,51.34)
\LongArrowArc(345,155)(23,20,105.44)
\SetWidth{1.1}
\Text(240,88)[]{\large\color{red} $e^-$}
\Text(400,88)[]{\large\color{red} $e^+$}
\Text(330,130)[c]{\large\color{red} $Z$}
\Text(310,70)[c]{\large\color{blue} $H$}
\Text(353,112)[c]{\large\color{black} $\theta$}
\Text(360,185)[c]{\large\color{black} $\theta_*$}
\end{picture}
\end{center}
\vspace*{-1.2cm}
\nn {\it Figure 4.6: The definition of the polar angles $\theta, \theta_*$ 
in  the process $\ee \to ZH \to H f\bar{f}$.}
\vspace*{3mm}

This time, the azimuthal dependence is P--even and independent of the
energy in contrast to the $0^{++}$ case; after integrating out the polar
$\theta, \theta_*$ angles, one obtains
\begin{eqnarray}
\frac{ {\rm d}\sigma (\ee \to ZA) }{ {\rm d} \phi_* }\sim 1 - \frac{1}{4}
\cos 2\phi_*
\end{eqnarray}

The production of the two states with $J^{\rm
P}=1^+, 2^-$ quantum numbers, which also lead to a $\beta$ behavior near the
kinematical threshold as in the $0^+$ case, can be ruled out using the angular
correlations as they lead to $(1+c_\theta^2)s_{\theta_*}^2$ and
$(1+c_{\theta_*}^2)s_{\theta}^2$ distributions which are absent in the SM Higgs
case \cite{CPHVVchoi}.\s 

We can thus conclude that the angular analysis of the Higgs production in
$e^+e^- \rightarrow  Z^* \rightarrow  ZH$ with $Z \rightarrow  f \bar{f}$,
together with the threshold behavior of the cross section, allows stringent
tests of the $J^{\rm PC}=0^{++}$ quantum numbers of the Higgs boson in the low
and intermediate mass range. In the high mass range, $M_H \gsim 2M_W$, when
the Higgs boson decays almost exclusively into two vector bosons, the Higgs
spin--zero and parity can be checked not only in the production process $\ee\to
HZ$, but also in the decay processes $H\to VV \to 4f$ as discussed in \S2.2.4. 
The full correlations between the final decay products $\ee \to ZH \to ZVV \to 
6f$ has not been yet worked out explicitly because of the rather complicated 
six fermion final state. 
  
\subsubsection{The WW fusion process} 

\subsubsection*{\underline{The production cross section}}

The $WW$ fusion process 
\cite{VVH-Cahn,VVH-DW,VVH-Hikasa,VVH-Altarelli,VVH-Kilian,Petcov}
is most important for  small values of the ratio $M_H/\sqrt{s}$, {\it i.e.} 
high energies where the cross  section grows $\sim M_W^{-2}$log$(s/M_H^2)$.  
The production cross section, discussed in \S3.3  at hadron colliders, can be 
more conveniently written as 
\beq 
\sigma= \frac{G_\mu^3 M_V^4}{64 \sqrt{2} \pi^3} \int_{\kappa_H}^1 {\rm d}x
\int_x^1 \frac{ {\rm d}y}{[1+(y-x)/ \kappa_V]^2} \left[ (\hat{v}_e^2+\hat{a}_e
^2)^2 f(x,y) + 4 \hat{v}_e^2 \hat{a}_e^2 g(x,y) \right] \label{WWxsection} 
\eeq 
\vspace*{-6mm}
\beq 
f(x,y) &=& \left(\frac{2x}{y^3} -\frac{1+2x}{y^2} +\frac{2+x}{2y}
-\frac{1}{2} \right)\left[ \frac{z}{1+z} -\log (1+z) \right]
+\frac{x}{y^3}\frac{z^2(1-y)} {1+z} \non \\ 
g(x,y) &=& \left(-\frac{x}{y^2} +\frac{2+x}{2y} -\frac{1}{2} \right) 
\left[ \frac{z}{1+z} -\log (1+z) \right]
\non 
\eeq 
with $\kappa_H =M_H^2/s, \kappa_V=M_V^2/s , z=y(x-\kappa_H)/(\kappa_Vx)$ and 
$\hat{v}, \hat{a}$ the electron couplings to the massive gauge bosons, $\hat{v
}_e=\hat{a}_e=\sqrt{2}$ for  the $W$ boson. [Note that in the effective 
longitudinal $W$ approximation, and as discussed in \S3.3.5, one obtains a 
simple result for the cross section of this process, but  which is
twice larger than the exact result for small Higgs boson masses.].\s

The production cross section is shown in Fig.~4.7 as a function of $M_H$ at c.m
energies $\sqrt{s}=0.5, 1$ and 3 TeV. For Higgs masses in the intermediate
range, the cross section is comparable to the one of the Higgs--strahlung
process at $\sqrt{s}=500$ GeV, leading to $\sim 25.000$ events for the expected
luminosity ${\cal L} =500 \ {\rm fb}^{-1}$, and is larger at higher energies.  

\begin{figure}[!h]
\begin{center}
\vspace*{-1.cm}
\hspace*{-1.cm}
\psfig{file=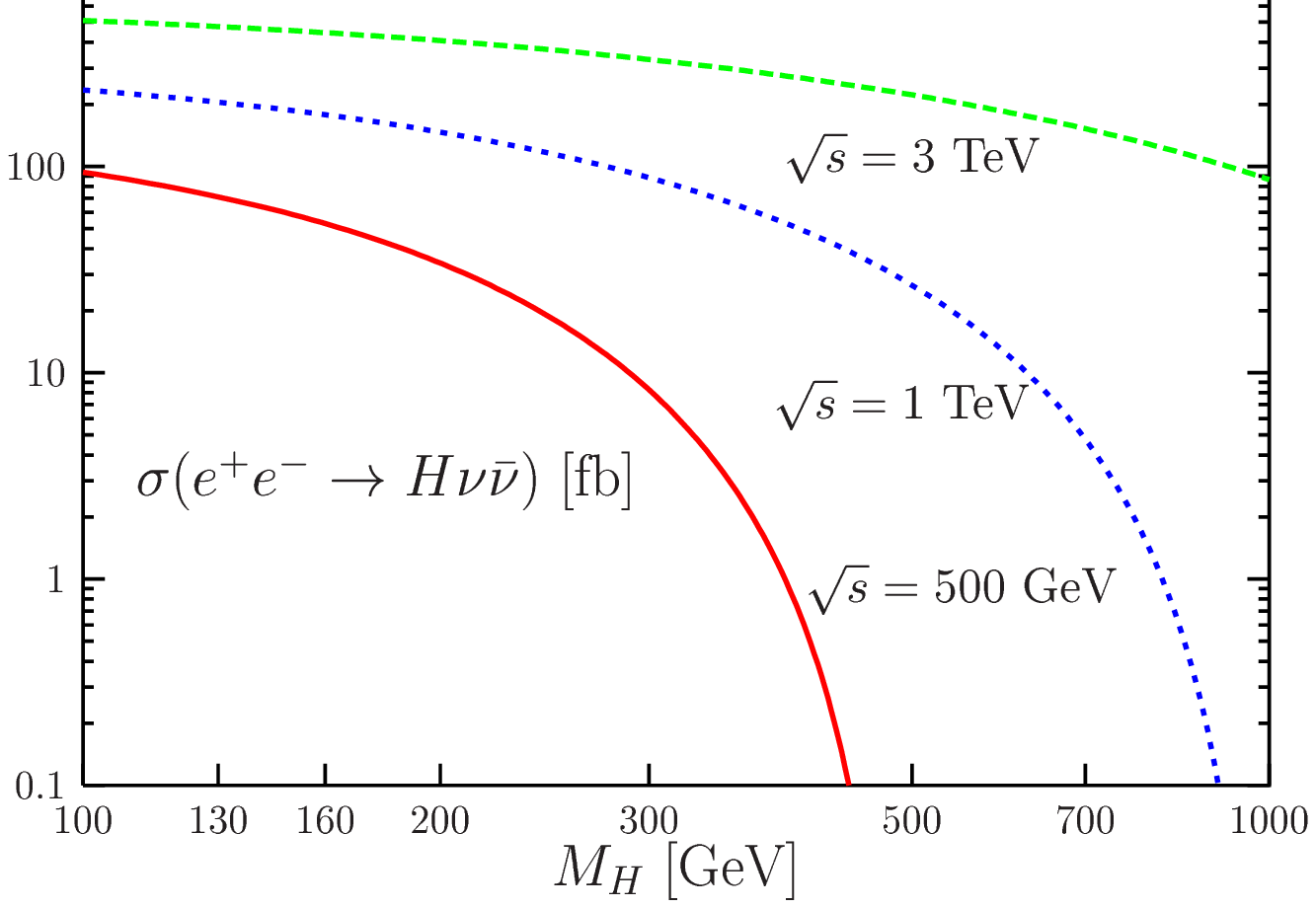,width=18.cm} 
\end{center}
\vspace*{-16.4cm}
\nn {\it Figure 4.7: The Higgs production cross section in the $WW$ fusion  
mechanism in $\ee$ collisions with  c.m. energies $\sqrt{s}=0.5,1$ and 3 TeV 
as a function of $M_H$.} 
\vspace*{-.5cm}
\end{figure}

\subsubsection*{\underline{The full cross section with the interference with 
Higgs--strahlung}}

The overall cross section that will be observed experimentally for  the process
$e^+e^- \to H\!+\!\bar\nu\nu$ will not be due to the $WW$ fusion process only,
but  part of it will come from the Higgs--strahlung  process,  $\ee \to HZ$,
with the $Z$ boson decaying into the three types of neutrinos.  A compact
expression for the full cross section of the Higgs--strahlung and $WW$ fusion 
mechanisms, including the interference terms,  has been derived in the general 
case by choosing the energy $E_H$ and the polar angle $\theta$ of the Higgs 
particle as the basic variables in the $e^+e^-$ c.m.\ frame. Decomposing  the
total contribution into three parts,  the contributions $3\times{g}_S$  from
Higgs--strahlung with $Z$ decays into three types of neutrinos, ${g}_W$
from $WW$ fusion, and ${g}_I$ from the interference term between  fusion
and Higgs--strahlung for $\bar\nu_e\nu_e$ final states, one has  for energies
$\sqrt{s}$ above the $Z$ resonance \cite{VVH-Altarelli,VVH-Kilian}

\begin{eqnarray}
&&  \frac{d\sigma(\ee \to H\bar\nu\nu)}{dE_H\,d\cos\theta}
  = \frac{G_\mu^3 M_Z^8p_H}{\sqrt2\,\pi^3s}
  \left(3\,g_S + g_I + g_W \right) \\
g_S &=& \frac{\hat{v}_e^2+\hat{a}_e^2}{96}\; \frac{ss_\nu + s_1s_2}
{\left(s-M_Z^2\right)^2 \left[(s_\nu-M_Z^2)^2 + M_Z^2\Gamma_Z^2\right]} \ , \ \
g_W= \frac{c^8_W}{s_1 s_2 r} \, {\cal H}_+ \non   \\
g_I &=& \frac{(\hat{v}_e+\hat{a}_e)c_W^4}{8}\; \frac{s_\nu-M_Z^2}
{\left(s-M_Z^2\right) \left[(s_\nu-M_Z^2)^2 + M_Z^2\Gamma_Z^2\right]}
\, {\cal H}_I
\end{eqnarray}
where all the abbreviated quantities have been defined in 
eq.~(\ref{pp:VVHabrev}), the factor ${\cal H}_+$ in eq.~(\ref{pp:VVH+H-}), 
while the factor ${\cal H}_I$ for the interference term is given by
\beq
{\cal H}_I= 2 - (h_1+1)\log\frac{h_1+1}{h_1-1}
             - (h_2+1)\log\frac{h_2+1}{h_2-1}
           +\, (h_1+1)(h_2+1)\frac{\ell}{\sqrt{r}}
\eeq
To derive the total cross section $\sigma(e^+e^-\to H\bar\nu\nu)$, the
differential cross section must be integrated over $\theta$ and $E_H$, with the
boundary conditions given in eq.~(\ref{pp:VVHbound}). The two main components 
and the total cross section for $e^+e^-\to H\bar\nu\nu$ are displayed in 
Fig.~4.8 as a function of the c.m. energy  for $M_H=115$ and 150 GeV. 
One can see that Higgs-strahlung is dominant a lower energies, $WW$ fusion at
higher energies, and the  interference term is small except in the cross over 
regions. \s

\begin{figure}[hbtp]
\vspace*{2mm}
\centerline{\includegraphics[width=.51\textwidth]{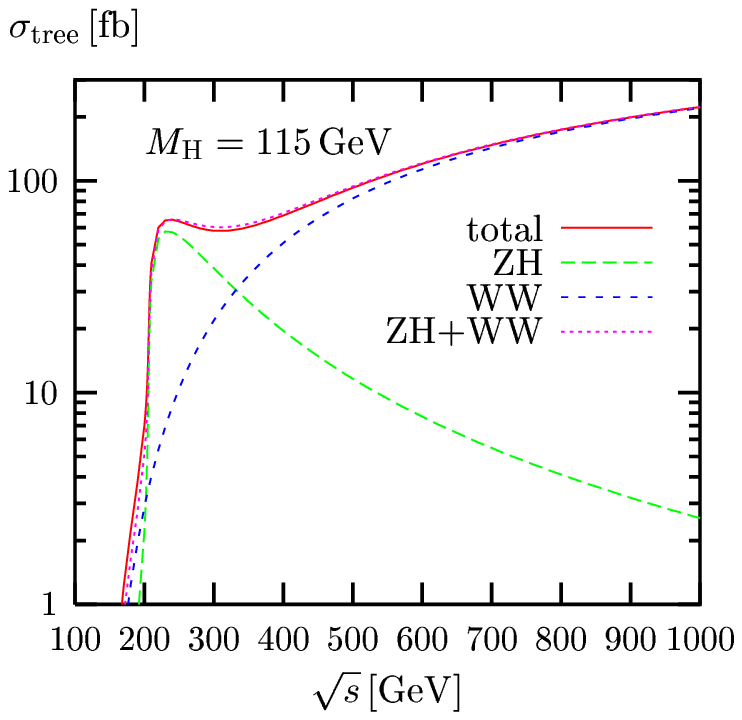}
\includegraphics[width=.51\textwidth]{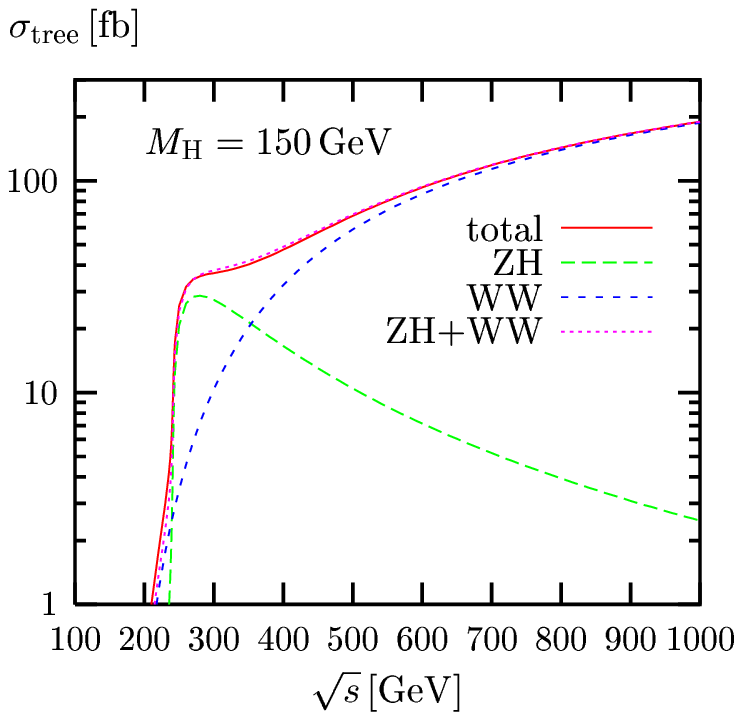}}  
\vspace*{1mm}
{\it Figure 4.8: The production cross section for the process $e^+e^-\to
H\bar\nu\nu$ as a function of $\sqrt{s}$ for $M_H=115$ and 150 GeV. 
The three components, i.e. Higgs--strahlung, $WW$ fusion, their sum, and
the total cross section including the interference term, are shown;
from Ref.~\cite{RCWW1}.}
\end{figure}%

At $\ee$ colliders, the initial $e^\pm$  beams can be polarized longitudinally.
The Higgs--strahlung and $WW$ fusion require opposite helicities of the $e^-$ 
and $e^+$ beams.~Denoting~$\sigma_{U,L,R}$ the cross sections for 
unpolarized $e^-/e^+$, $e^-_L/e^+_R$ and $e^-_R/e^+_L$, respectively, one 
obtains \cite{VVH-Kilian}
\begin{eqnarray}
  \sigma_U  \propto 3\, g_S + g_I + g_W \ , \ \
  \sigma_L  \propto 6\, g_S + 4\,g_I + 4\,g_W\ , \ \   
  \sigma_R \propto 6\, g_S
\end{eqnarray}
The cross section for $WW$ fusion of Higgs particles increases by a
factor four, compared with unpolarized beams, if left--handed electrons
and right--handed positrons are used.  By using right--handed electrons,
the $WW$ fusion mechanism is switched off. The interference term
cannot be separated from the $WW$ fusion cross section.

\subsubsection{The electroweak radiative corrections}

To have a full control on the production cross sections of the Higgs--strahlung 
and $WW$ fusion processes, in view of the high--precision tests which can be 
performed using them, the electroweak radiative corrections must be taken 
into account. These corrections, consisting of virtual and real corrections 
with the emission of an additional photon in the final or initial state (ISR), 
have been completed recently. Note, however, that at high--energy linear 
colliders, in addition to ISR, one has also to take into account the beam 
energy spread and beamstrahlung. The latter is machine dependent and will 
smear out the c.m. energy and the system moves along the beam axes; it must
be thus suppressed  as strongly as possible in order to allow for high--quality
analyses which are often based on kinematical constraints derived 
from the precise knowledge of the initial beam energies. 

\subsubsection*{\underline{The Higgs--strahlung process}}

At one--loop order, the radiative corrections to the Higgs--strahlung 
process consist of  self--energy, vertex and box corrections to the tree--level
amplitude and the emission of an additional photon in the initial state;
Fig.~4.9.  The corrections have been calculated some time ago \cite{RCHZ} and
reanalyzed recently in the context of the full $\ee \to H\bar{\nu} \nu$ 
process \cite{RCWW1,RCWW2,RCWW3}. Let us summarize the salient features of these
corrections.

\begin{center}
\vspace*{-.2cm}
\hspace*{-12.cm}
\SetWidth{1.1}
\begin{picture}(300,80)(0,0)
\ArrowLine(100,25)(140,50)
\ArrowLine(100,75)(140,50)
\Photon(140,50)(165,50){3.2}{4.5}
\ArrowLine(165,50)(200,75)
\ArrowLine(165,50)(200,25)
\Line(200,75)(200,25)
\DashLine(200,75)(240,75){4}
\Photon(200,25)(240,25){3.2}{4.5}
\Text(202,75)[]{\bb}
\Text(100,60)[]{$e^+$}
\Text(100,40)[]{$e^-$}
\Text(150,63)[]{$\gamma,Z$}
\Text(190,50)[]{$f$}
\Text(250,30)[]{$Z$}
\Text(250,70)[]{\bH}
\ArrowLine(270,25)(310,25)
\ArrowLine(270,75)(310,75)
\ArrowLine(310,25)(310,75)
\Photon(310,25)(355,25){3.2}{5.5}
\Photon(310,75)(355,75){3.2}{5.5}
\Photon(355,75)(355,25){3.2}{5.5}
\DashLine(355,75)(390,75){4}
\Text(357,75)[]{\bb}
\Photon(355,25)(390,25){3.2}{5}
\ArrowLine(420,25)(460,50)
\ArrowLine(420,75)(460,50)
\Photon(460,50)(500,50){3.2}{5.5}
\Photon(500,50)(540,25){3.2}{5.5}
\DashLine(500,50)(540,75){4}
\Text(502,50)[]{\bb}
\Photon(440,65)(470,75){3.2}{5.}
\Text(477,70)[]{$\gamma$}
\Text(330,-3)[]{\it Figure 4.9: Generic diagrams for the ${\cal O}(\alpha)$
corrections to the process $\ee \to HZ$.}
\vspace*{2.mm}
\end{picture}
\end{center}
\vspace*{-0.mm}

The photonic corrections to the initial state, that is vertex and self--energy
corrections with photon exchange as well as photon radiation (ISR) can be
implemented using the structure function approach discussed in \S1.2.1; see,
eq.~(\ref{QEDradiator}).  The fermionic corrections which  are contained in
the running of the QED constant $\alpha$ for the light fermions,
eq.~(\ref{Deltaalpha}), and the correction to the $\rho$ parameter for the
heavy top quark, eq.~(\ref{deltarho}), can be incorporated by using the
improved Born approximation (IBA): starting with the Born cross section defined
in terms of the bare electromagnetic coupling $\alpha(0)$, one performs the
substitution $\pi \alpha(0) \to \sqrt{2}G_F M_W^2 (1-M_W^2/M_Z^2)$ which
absorbs the correction $\Delta r \simeq  \Delta \alpha - 3 \Delta  \rho$.  One
has also to include the additional corrections to the $HZZ$ vertex and in
particular the heavy top contributions, $\delta^t_{HZZ}$ in
eq.~(\ref{WWHvertex}).  The largest part of the weak correction is then
absorbed into the couplings and the remaining  corrections should be in
principle rather small \cite{RCWW1}. \s

The overall correction to the tree--level $\ee \to HZ$ amplitude,  including
an additional term that is logarithmic in the top quark mass, is given by
\cite{RCreviewEW}
\beq
K_{\ee \to HZ}^t \simeq 1 + \frac{\alpha}{4\pi s_W^2}\frac{1}{g_i} 
\left[ \frac{1}{8} \left( 6 \frac{c_W}{s_W} + g_i  \right)
\frac{m_t^2}{M_W^2} + \frac{3 -2s_W^2}{3c_Ws_W} \log \frac{m_t}{M_W} \right]
\eeq
These factors correct in fact the amplitudes with left-- and right--handed
electrons with couplings $g_{L}=(2s_W^2- 1)/(2s_W c_W)$ and $g_R= s_W/c_W$. 
At low and moderate energies, this approximation is rather good. However, at
high energies, it turns out that this expression in the heavy--top quark limit 
does not reproduce exactly the full $m_t$ dependent result, as a consequence 
of the presence of the box contributions  which depend both on $s$ and $m_t$.

\subsubsection*{\underline{The WW fusion process}}

Since already at the tree--level the $WW$--fusion mechanism is a three--body
final state production process [which was thus not trivial to handle], the
calculation of the one--loop radiative corrections is a real challenge. 
Indeed, not only one has to deal with the numerous diagrams involving
self--energy, vertex and box corrections [due to the additional final state,
the number of such diagrams is much larger than for a $2\to 2$ process like
Higgs--strahlung], one has to consider in addition one--loop corrections
involving pentagonal diagrams which are extremely difficult to handle, and
corrections with real photon emission, leading to four particles in the final
state which are rather involved; see Fig.~4.10. To these complications, one has
to add the fact that to derive the full corrections to the $\ee \to H\nu
\bar{\nu}$ final state, both the $WW$ fusion mechanism and the Higgs--strahlung
process with $Z \to \nu \bar{\nu}$ have to be considered and added coherently.

\begin{center}
\vspace*{-5mm}
\hspace*{-12.8cm}
\SetWidth{1.1}
\begin{picture}(300,100)(0,0)
\ArrowLine(110,25)(140,25)
\ArrowLine(110,75)(140,75)
\ArrowLine(140,25)(200,20)
\ArrowLine(140,75)(200,80)
\Photon(140,25)(165,35){3.2}{3.5}
\Photon(140,75)(165,65){3.2}{3.5}
\ArrowLine(165,35)(190,50)
\ArrowLine(165,65)(190,50)
\Line(165,35)(165,65)
\DashLine(190,50)(230,50){4}
\Text(192,50)[]{\bb}
\Text(120,65)[]{$e^+$}
\Text(120,35)[]{$e^-$}
\Text(190,70)[]{$\bar \nu$}
\Text(190,30)[]{$\nu$}
\Text(175,50)[]{$f$}
\Text(220,60)[]{\bH}
\Text(150,40)[]{$W$}
\Text(150,60)[]{$W$}
\ArrowLine(270,25)(310,25)
\ArrowLine(270,75)(310,75)
\ArrowLine(310,25)(310,75)
\Photon(310,25)(355,25){3.2}{5.5}
\Photon(310,75)(355,75){3.2}{5.5}
\Photon(355,75)(355,25){3.2}{5.5}
\ArrowLine(355,25)(395,20)
\ArrowLine(355,75)(395,80)
\DashLine(355,50)(390,50){4}
\Text(357,50)[]{\bb}
\ArrowLine(420,25)(460,25)
\ArrowLine(420,75)(460,75)
\ArrowLine(460,25)(510,20)
\ArrowLine(460,75)(510,80)
\Photon(460,25)(500,50){3.2}{5.5}
\Photon(460,75)(500,50){3.2}{5.5}
\Photon(440,75)(480,95){3.2}{5.5}
\DashLine(500,50)(540,50){4}
\Text(502,50)[]{\bb}
\Text(330,1)[]{\it Figure 4.10: Generic diagrams for the ${\cal O}(\alpha)$
corrections to the $WW$ fusion process.}
\vspace*{0.mm}
\end{picture}
\end{center}
\vspace*{-1mm}

The challenge of deriving these corrections has been met by three groups. In
Ref.\cite{RCWW2}, the calculation was performed using {\tt GRACE-LOOP}
\cite{GRACE}, an automatic calculation system. In Ref.~\cite{RCWW3}, the
results have been obtained as a {\tt MAPLE} output using the program  {\tt
DIANA} \cite{Diana} without an explicit evaluation. In Ref.~\cite{RCWW1}, the
calculation has been performed in two independent ways, using the program {\tt
FeynArts} \cite{Feynarts} to generate the Feynman  graphs, and using {\tt
Mathematica} to express the amplitudes in terms of standard matrix elements or
using the package {\tt FormCalc} \cite{Formcalc} based on {\tt Form} 
\cite{Form}. We briefly summarize the main results of these calculations, 
mostly relying on Ref.\cite{RCWW1}. \s

The ISR corrections stemming from the radiation of a photon from the initial
$\ee$ states and from the intermediate $W$ bosons, can again be obtained in the
structure function approach either at ${\cal O}(\alpha)$  or including 
higher--order corrections. The running of the electromagnetic constant due to 
the light fermion contributions [because the cross section  is proportional to
$\alpha^3$, this leads to a $\sim 18$\% change of the cross section]  can be
included using the IBA discussed previously. Finally,  since the $WW$--fusion
cross section gets its main contribution from  small momenta $W$ bosons, the
loop corrections are mainly determined  by the $\nu_e eW$ and $HWW$ vertices at
zero--momentum transfer.  The  correction to the  $e\nu_eW$ vertex is well
described by $\Delta r$ and  the $HWW$ vertex correction is given by
$\delta^t_{HWW}$ in eq.~({\ref{WWHvertex}). It turns out that these corrections
largely cancel the corresponding ones when $G_\mu$ is used in the tree--level
expression of the amplitude and one obtains a small remaining piece \cite{RCWW1}
\beq
K_{\ee \to H \nu \bar{\nu} }^{t} = 1 - \frac{5 \alpha}{16
\pi s_W^2} \frac{m_t^2}{M_W^2} 
\eeq
which approximates the fermionic contribution to the amplitude quite 
well. To this correction, one has to add the bosonic contribution for which
no simple approximation is possible.

\vspace*{-2mm}
\subsubsection*{\underline{Numerical results}}

The final output of the calculation is shown in Fig.~4.11, where the radiative
corrections to the Higgs strahlung process [left figure] and the the $WW$ 
fusion mechanism [right figure], without the small interference terms, are shown
as a function of $\sqrt{s}$ for $M_H=150$ GeV. The various components, the 
fermionic contribution, the bosonic contribution, the  initial state radiation
at ${\cal O}(\alpha)$ and beyond, are displayed. \s

In the case of $WW$ fusion, the ISR corrections, the bulk of which comes from
${\cal O}(\alpha)$ contributions, are negative for all energies as a
consequence of the decrease of the effective c.m. energy which leads to a
smaller cross section. The fermionic corrections are negative and small, being
at the level of $-2$\%, while the bosonic corrections range from $+1$\%   near
the production threshold to $-3$\% at high energy.  For the Higgs strahlung
process, at high enough energies $\sqrt{s}\gsim 500$ GeV, the fermionic
contribution is positive and almost constant, $+10\%$, while the bosonic 
contribution is negative and large, increasing in absolute value with 
$\sqrt{s}$. The largest correction is due to the ${\cal O}(\alpha)$ ISR [
the contribution of higher--orders is again very small], which increases the 
cross section by $20$\% for $\sqrt{s}=1$ TeV.\s

\begin{figure}%
\centerline{\includegraphics[width=.5\textwidth]{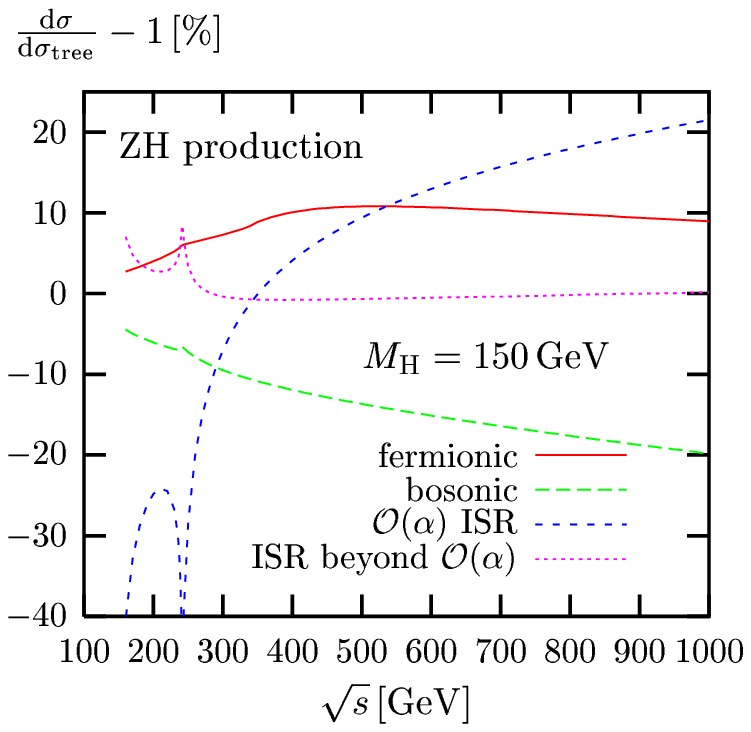}
\includegraphics[width=.5\textwidth]{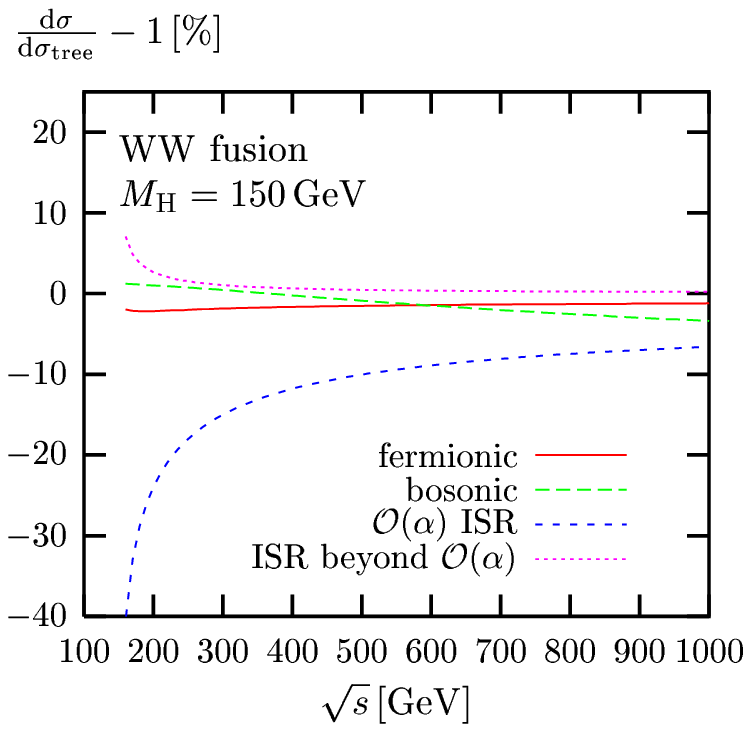}}
\vspace*{3mm}
{\it Figure 4.11: Relative electroweak corrections to the Higgs--strahlung 
$\ee \to HZ$ and to $WW$ fusion $\ee \to H \nu\bar{\nu}$ processes resulting 
from ISR at ${\cal O}(\alpha)$  and beyond, fermion loops, and non--ISR 
bosonic corrections as a function of  $\sqrt{s}$ for $M_H = 150$ GeV; from 
Ref.~\cite{RCWW1}.}
\label{fi:contrchannels}
\vspace*{-3mm}
\end{figure}%
 
Adding the channel where the neutrinos are coming from the Higgs--strahlung
process and the small  interference term, one obtains the total production
cross section for the full $\ee \to H \nu \bar{\nu}$ process. The relative
corrections to the lowest  order cross section for the various components are
shown in Fig.~4.12 for $M_H=115$ and 150 GeV as a function of $\sqrt{s}$. Below
the threshold, the correction to the $ZH$ channel are large and negative,
reaching $\sim -20$\%, rise fastly near threshold, and at $\sqrt{s}=1$ TeV 
reach the level of $\sim 20\, (10)$\% for $M_H=115\, (150)$ GeV. The
corrections to the $WW$ fusion channel rise also sharply at the threshold but
reach quickly a plateau at a level of $-10$\% beyond $\sqrt{s}=500$ GeV. The
corrections to the complete process follow those of the $WW$ component at high
energy and those of the $HZ$ process at low energies, a consequence of the
relative magnitude of the two processes at tree--level. They are always
negative, being of order $-10$\% at $\sqrt{s} \gsim 350$ GeV.  

\begin{figure}[!h]
\vspace*{-1mm}
\centerline{\includegraphics[width=.5\textwidth]{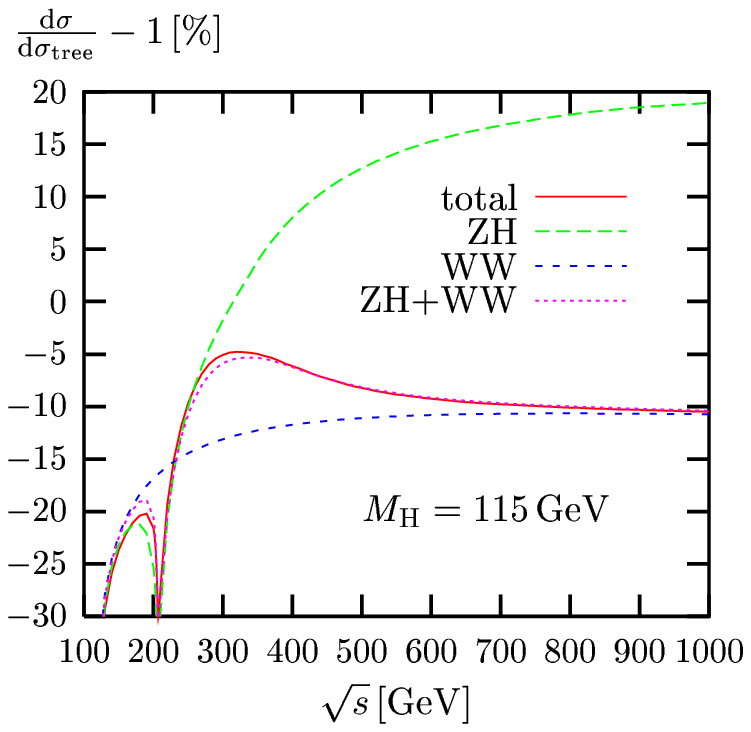}
\includegraphics[width=.5\textwidth]{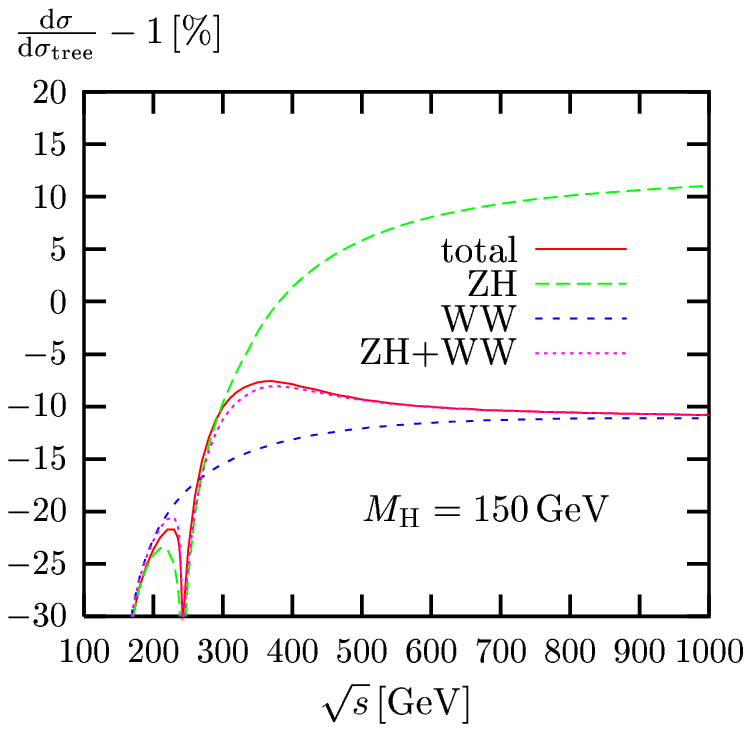}}
\vspace*{1mm}
{\it Figure 4.12: Relative  corrections to the complete process $\ee \to H 
\nu\bar{\nu}$ and the contributions of the various components as a function
of $\sqrt{s}$ and for $M_H=115$ and $150$ GeV; from~\cite{RCWW1}.}
\label{fi:corr}
\vspace*{-1.1cm}
\end{figure}

\subsection{The subleading production processes in $\ee$ collisions}

\subsubsection{The ZZ fusion mechanism}

The cross section for the $ZZ$ fusion mechanism, $\ee \to e^+e^- (Z^*Z^*) \to
\ee H$, Fig.~4.13, is given by the same expression in eq.~(\ref{WWxsection})
for the $WW$ fusion mechanism with the vector boson  $V=Z$ having the usual
couplings to the electron $\hat{v}_e=-1+4s_W^2 , \hat{a}_e=-1$. The total
production  cross section is about an order of magnitude smaller than the cross
section for $WW$ fusion, $\sigma(WW \to H)/\sigma (ZZ \to H)  \sim 16 c_W^2
\sim 9$,  a mere consequence of the fact that the neutral current couplings are
smaller than the charged current couplings. The lower rate, however, could be
at least partly compensated by the clean signature of the $\ee$ final state.
The cross section is shown in Fig.~4.14 as a function of $M_H$ for the c.m.
energies $\sqrt{s}=0.5$, 1 and 3 TeV. It follows the same trend as the $WW$
fusion cross section.  

\vspace*{-5mm}
\begin{center}
\begin{picture}(300,100)(0,0)
\SetWidth{1}
\SetScale{1.}
\vspace*{-1.6cm}
\hspace*{-5cm}
\ArrowLine(200,25)(240,25)
\ArrowLine(200,75)(240,75)
\ArrowLine(240,25)(290,15)
\ArrowLine(240,75)(290,85)
\Photon(240,25)(280,50){3.2}{5.5}
\Photon(240,75)(280,50){3.2}{5.5}
\DashLine(280,50)(320,50){4}
\Text(282,50)[]{\bb}
\Text(190,30)[]{$e^-$}
\Text(190,70)[]{$e^+$}
\Text(240,60)[]{$Z^*$}
\Text(240,40)[]{$Z^*$}
\Text(330,50)[]{\bH}
\Text(300,20)[]{$e^-$}
\Text(300,80)[]{$e^+$}
\Text(290,-5)[]{\it Figure 4.13: Higgs boson production in the $ZZ$ fusion 
mechanism in $\ee$ collisions.} 
\vspace*{0.mm}
\end{picture}
\end{center}

\begin{figure}[!h]
\begin{center}
\vspace*{-1.1cm}
\hspace*{-1.cm}
\psfig{file=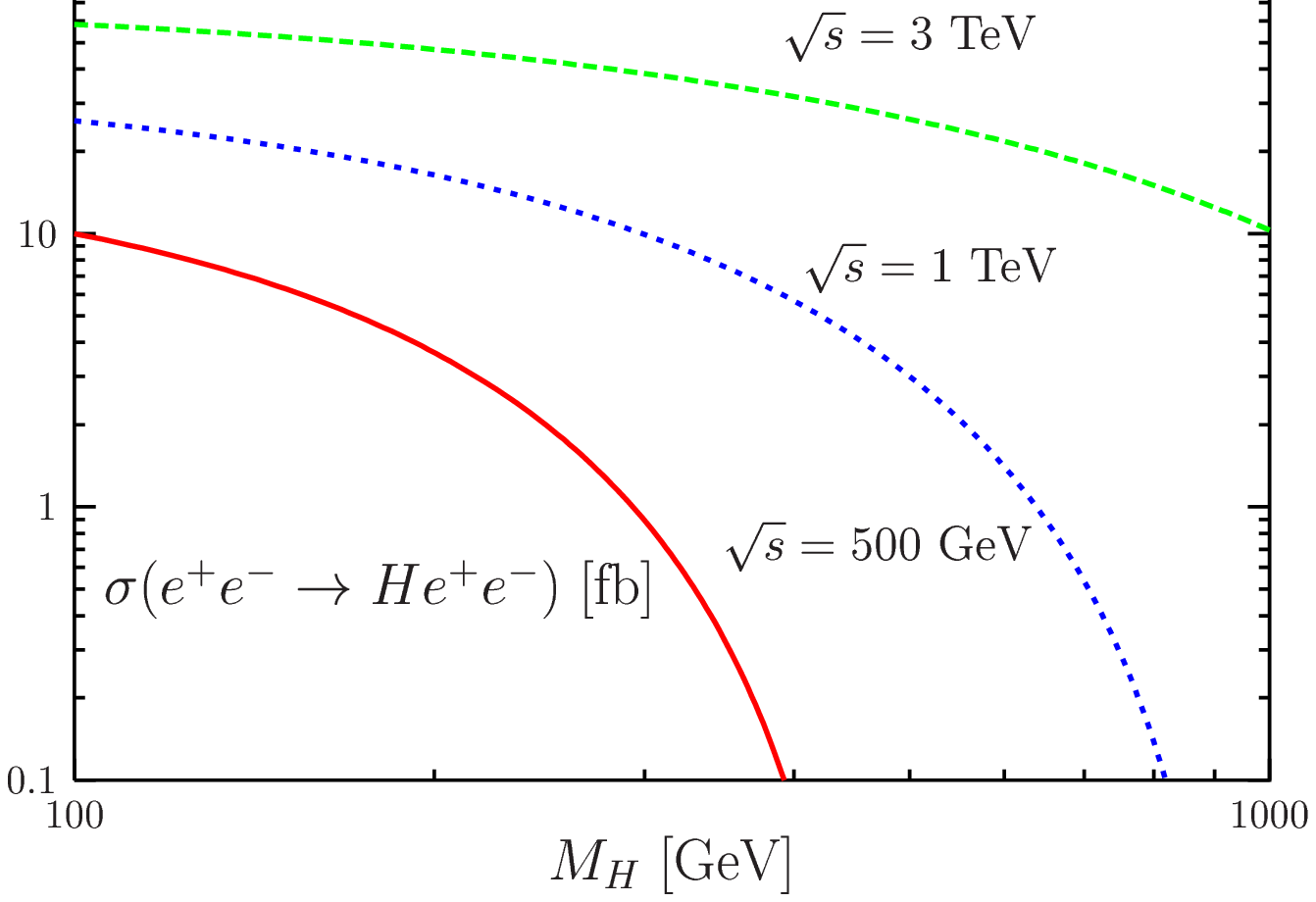,width=16.cm} 
\end{center}
\vspace*{-14.6cm}
\nn {\it Figure 4.14: Higgs production cross sections in the $ZZ$ fusion  
mechanism in $\ee$ collisions with  c.m. energies $\sqrt{s}=0.5,1$ and 3 TeV 
as a function of $M_H$.} 
\vspace*{-.3cm}
\end{figure}

Similarly to the $WW$ fusion case,  the overall cross section for the process
$e^+e^- \to H + e^+e^-$ receives contributions $g_S$ from Higgs--strahlung with
$Z \to \ee$, $g_{Z_\pm}$ from $ZZ$ fusion, and $g_I$ from the interference term 
between fusion and Higgs--strahlung \cite{ZZH-Kilian}
\begin{equation}\label{totalZZ}
  \frac{d\sigma(\ee \to He^+e^-)}{dE_H\,d\cos\theta}
  = \frac{G_\mu^3 M_Z^8p_H}{\sqrt2\,\pi^3s}
  \left(g_S + g_I + g_{Z+} + g_{Z-} \right)
\end{equation}
with
\begin{eqnarray}
  g_S &=& \frac{\left(\hat{v}_e^2+\hat{a}_e^2\right)^2}{192}\;
    \frac{ss_e + s_1s_2}{\left(s-M_Z^2\right)^2
               \left[(s_e-M_Z^2)^2 + M_Z^2\Gamma_Z^2\right]}\non \\
  g_I &=& \frac{\left(\hat{v}_e^2+\hat{a}_e^2\right)^2+4\hat{v}_e^2
  \hat{a}_e^2}{64}\;
    \frac{s_e-M_Z^2}{\left(s-M_Z^2\right) 
                \left[(s_e-M_Z^2)^2 + M_Z^2\Gamma_Z^2\right]} {\cal H}_I
    \nonumber\\
  g_{Z+} &=& \frac{\left(\hat{v}_e^2+\hat{a}_e^2\right)^2+4\hat{v}_e^2
 \hat{a}_e^2} {32\, s_1 s_2 r}\, {\cal H}_+    \ , \
  g_{Z-} = \frac{\left(\hat{v}_e^2-\hat{a}_e^2\right)^2}{32\, s_1 s_2 r}\,
        (1 - c_\chi){\cal H}_-
\end{eqnarray}
where the same abbreviations as in the formulas for the $W$ fusion case, with
the appropriate replacements, $\nu\to e$ and $W\to Z$, have been used. Again, 
the three components and the total cross sections follow the same trend as in 
the case of the $WW$ fusion process. \s

The calculation of the one--loop  radiative corrections to this process follows
the same lines as the one for the companion process $\ee \to H \nu \bar{\nu}$,
the only difference being that there are additional diagrams where photons are
exchanged between the initial and final state electrons and positrons, and also
between the final state $\ee$ pair. The corrections have been calculated using
the {\tt GRACE-LOOP} \cite{GRACE} system, and the result has recently appeared 
in Ref.~\cite{RCZZ}. They are summarized in Fig.~4.15 as a function of $\sqrt 
s$  for three Higgs mass values.\s

After subtracting the photonic corrections which decrease the cross section by
about 5\% for $\sqrt{s} \gsim 350$ GeV, as shown in the left--hand side of 
the figure, one obtains a rather small electroweak correction: when the 
tree--level cross section is expressed in terms of $G_\mu$, the correction is 
${\cal O}(-5\%)$ at $\sqrt{s}=350$--500 GeV and varies very little with energy 
to reach $-4\%$ at 1 TeV, as can be seen in the right--hand side of Fig.~4.15. 
The correction is also almost independent of the Higgs mass in the chosen 
range, $M_H \sim 100$--200 GeV. The correction factor when $\alpha$ 
is used as input at the tree--level is also shown.\s

\begin{figure}[!h]
\vspace*{-2mm}
\begin{center}
\mbox{
\includegraphics[width=0.5\textwidth,height=8.cm]{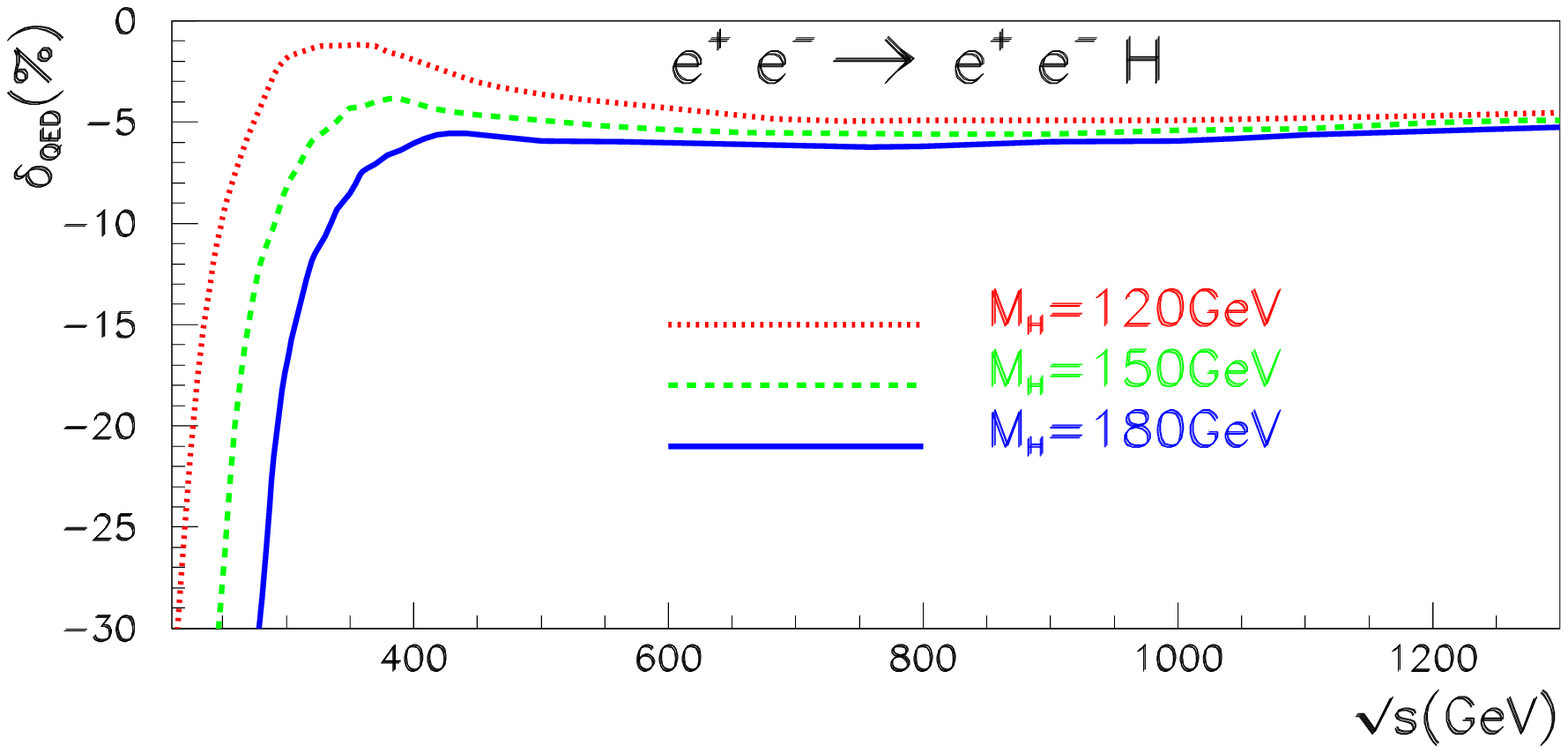}
\includegraphics[width=0.5\textwidth,height=8.cm]{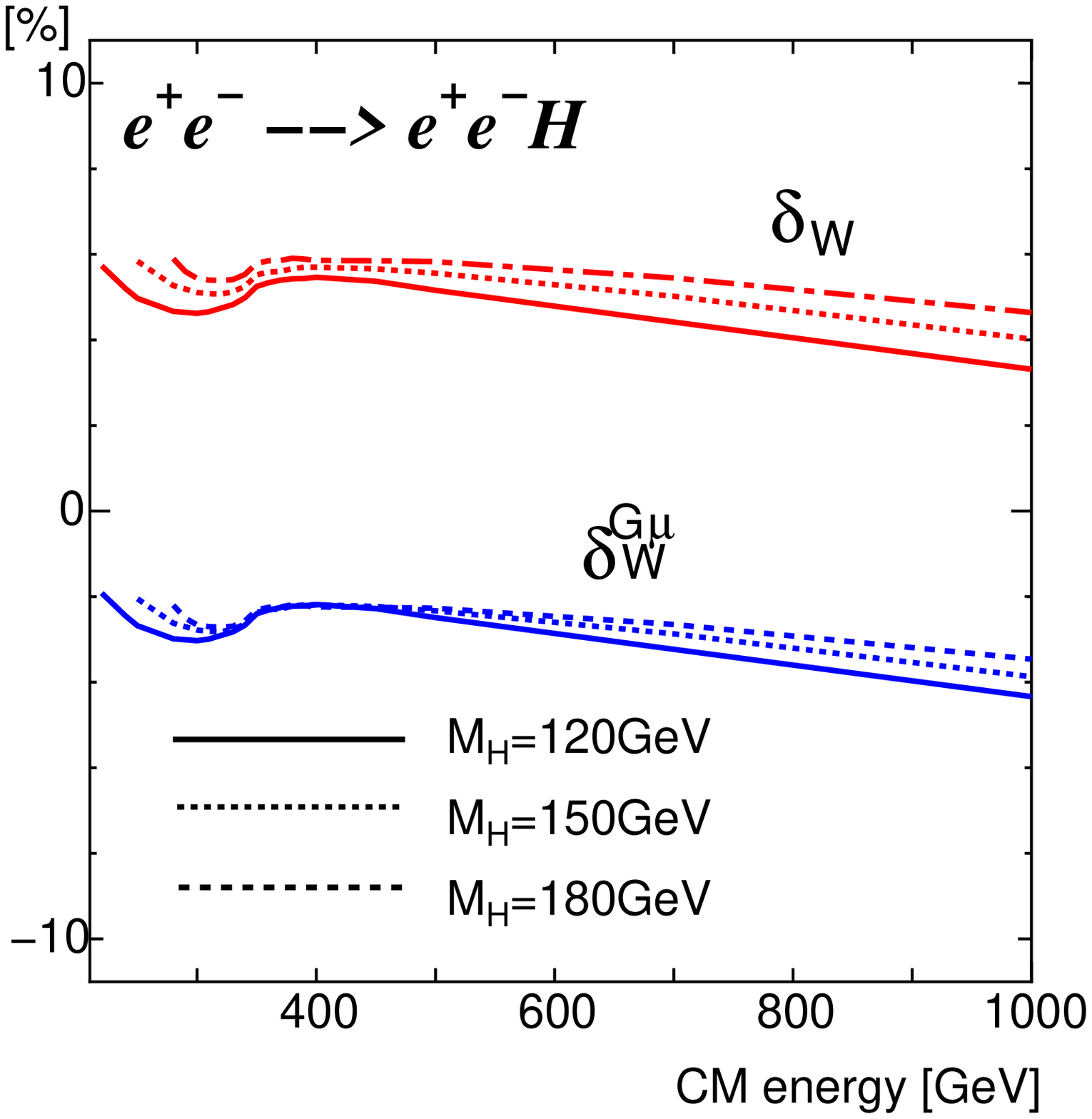} }
\end{center}
\vspace*{-4mm}
\nn {\it Figure 4.15: The photonic corrections (left) and the genuine 
electroweak radiative corrections in the $G_\mu$ and $\alpha$ schemes (right)
for the process $\ee \to H\ee$ as a function of the c.m. energy for $M_H =120$,
150 and 180 GeV; from Ref.~\cite{RCZZ}.}
\vspace*{-2mm}
\end{figure}

For the process $e^+e^-\to He^+e^-$, the pattern for the polarized and 
unpolarized cross sections is slightly more complicated than for the $WW$ 
fusion process \cite{ZZH-Kilian}
\begin{eqnarray}
  \sigma_U &\propto&
	g_S + g_I + g_{Z+} + g_{Z-}\ , \  \sigma_{LL} = \sigma_{RR} \propto
	2\,g_{Z-} \non \\
  \sigma_{LR/RL} &\propto&
	2\frac{(\hat v_e\pm \hat a_e)^2}{(\hat v_e^2+\hat a_e^2)}g_S
	+ 2\frac{(\hat v_e\pm\hat a_e)^4}{(\hat v_e^2+\hat a_e^2)^2+4\hat 
        v_e^2 \hat a_e^2}
	\left(g_I + g_{Z+}\right) \non 
\end{eqnarray}
Since $\hat v_e \sim -1+4s_W^2 \ll \hat a_e$, the difference between 
$\sigma_{RL}$ and $\sigma_{LR}$ is, however, strongly suppressed and one 
obtains $\sigma_{LR} \simeq \sigma_{LR}= 2 (g_S+ g_I+ g_{Z_+})$.\s
 
Finally, let us note that in the $e^-e^-$ option of future high--energy linear 
colliders, one can produce Higgs bosons in a similar channel \cite{VVH-Hikasa}
\beq
e^- e^- \lra e^- e^- (Z^*Z^*) \lra e^- e^- H
\eeq
The production cross section [up to some statistical factors due to the 
identical initial and final states] and the main features of the process are 
the same as those discussed above for the $\ee$ option of the machine.

\subsubsection{Associated production with heavy fermion pairs}

\subsubsection*{\underline{The process at the tree--level}}

In the SM, the associated production of Higgs bosons with a pair of heavy 
fermions, $\ee \to H f \bar{f}$ \cite{ee-ttH0,ee-ttH}, proceeds through two 
set of diagrams: those
where the Higgs boson is radiated off the $f$ and $\bar{f}$ lines, and a 
diagram where the Higgs boson is produced in association with a $Z$ boson 
which then splits into an $f\bar{f}$ pair; Fig.~4.16.\\[-1.2cm]

\begin{center}
\hspace*{-4cm}
\vspace*{-1.cm}
\SetWidth{1.}
\begin{picture}(300,100)(0,0)
\ArrowLine(0,25)(35,50)
\ArrowLine(0,75)(35,50)
\Photon(35,50)(80,50){3.2}{5.5}
\ArrowLine(80,50)(115,25)
\ArrowLine(80,50)(115,75)
\DashLine(105,65)(130,47){4}
\Text(-5,30)[]{$e^+$}
\Text(-5,70)[]{$e^-$}
\Text(55,65)[]{$\gamma,Z$}
\Text(123,30)[]{$f$}
\Text(123,70)[]{$\bar{f}$}
\Text(137,55)[]{\bH}
\Text(105,65)[]{\bb}
\ArrowLine(150,25)(185,50)
\ArrowLine(150,75)(185,50)
\Photon(185,50)(230,50){3.2}{5.5}
\ArrowLine(230,50)(265,25)
\ArrowLine(230,50)(265,75)
\DashLine(250,40)(270,50){4}
\Text(203,65)[]{$\gamma,Z$}
\Text(249,39)[]{\bb}
\ArrowLine(295,25)(330,50)
\ArrowLine(295,75)(330,50)
\Photon(330,50)(375,50){4}{5.5}
\Photon(375,50)(400,60){3}{4.5}
\DashLine(375,50)(410,25){4}
\ArrowLine(400,60)(420,75)
\ArrowLine(400,60)(420,50)
\Text(355,65)[]{$Z$}
\Text(377,50)[]{\bb}
\Text(210,3)[]{\it Figure 4.16: Diagrams for the associated production of 
Higgs bosons with a fermion pair.}
\end{picture}
\vspace*{0.mm}
\end{center}
\vspace*{.99cm}

Since  the fermion and Higgs boson masses must be kept non--zero, the total 
cross section for these processes is quite involved. However, the Dalitz  
density, once the angular dependence is integrated out, can be written in a 
simple and compact form \cite{ee-ttH}
\beq
\frac{{\rm d}\sigma (\ee \to f\bar{f}H)}{{\rm d}x_1 {\rm d}x_2} =
\frac{\bar{\alpha}^2 N_c}{12 \pi s} \hspace*{-4mm} & & \left\{ \left[
Q_e^2 Q_f^2 + \frac{2 Q_e Q_f {v}_e {v}_f}{1-z}+\frac{( {v}_e^2
 + {a}_e^2) ({v}_f^2 + {a}_f^2)}{(1- z)^2} \right]G_1 \right.
\\
&&+  \left. \frac{{v}_e^2+ {a}_e^2}{(1- z)^2} \left[ {a}_f^2
\sum_{i=2}^{6} G_i + {v}_f^2 (G_4+G_6) \right] + \frac{Q_e Q_f {v_e}
{v}_f}{1-z} G_6 \right\} \non
\label{ttHxsection}
\eeq
with $\bar{\alpha} \equiv \alpha(s) \sim 1/128$, $N_c$ the color factor and 
${v}_e, {a}_e$ the usual couplings of fermions to the $Z$ boson,
eq.~(\ref{Zffcouplings}).  $z$ is the scaled mass of the $Z$ boson, $z=M_Z^2/s$,
and we will use later on the scaled masses $f =m_f^2/s$ and $h=M_H^2/s$.  $x_1=2
E_f/\sqrt{s}$ and $x_2=2E_{\bar{f}}/ \sqrt{s}$ are the reduced energies  of the
$f$ and $\bar{f}$ states; we will also use the Higgs scaled energy,
$x_H=2E_H/\sqrt{s}=2-x_1-x_2$, as well as  the variables $x_Z$ and $x_{12}$ 
defined by $x_Z=x_H-1-h+z$ and $x_{12}=(1-x_1) (1-x_2)$. In terms of these
variables and the $g_{Hff}=m_t/v$ and  $g_{HZZ}=2M_Z/v$ Higgs couplings,
the coefficients $G_i$,  with $i=1$--6, are given by 
\beq
G_1 &=& \frac{g_{Hff}^2}{x_{12}} \bigg[ x_H^2 - h \bigg( \frac{x_H^2}{x_{12}}
+ 2 (x_H -1-h) \bigg) + 2f \bigg(  4(x_H-h) + \frac{x_H^2}{x_{12}} (4f -h+2)
\bigg) \bigg] \non \\
G_2 &=&  - 2 \frac{g_{Hff}^2}{x_{12}} \bigg[ x_{12}(1+x_H)  - h (x_{12}
+ 2x_H + 8f -2h)  + 3f x_H \bigg( \frac{x_H}{3} +4 + \frac{x_H}{x_{12}} 
(4f -h) \bigg) \bigg] \non \\
G_3 &=& 2 \frac{g_{HZZ}^2}{x_Z^2} \bigg[ f(4h-x_H^2-12z)+ \frac{f}{z} 
( 4h -x_H^2) (x_H-1-h+z) \bigg] \non \\
G_4 &=& 2 \frac{g_{HZZ}^2}{x_Z^2} z \bigg[ h +x_{12} + 2(1-x_H)+4f) \bigg]
\non \\
G_5 &=& - \frac{g_{Hff} g_{HZZ}}{x_{12} x_Z} \, 4 x_H \frac{m_f}{M_Z}\, 
\bigg[ (x_{12} -h) (x_H-1-h) + f(12z-4h+x_H^2)  - 3z \bigg( h- 2\frac{x_{12}}
{x_H} \bigg) \bigg] \non \\
G_6 &=& - \frac{g_{Hff} g_{HZZ}}{x_{12} x_Z} \, 4 z \frac{m_f}{M_Z}\,
\bigg[ x_H(h-4f-2) -2x_{12} + x_H^2  \bigg]
\eeq
Integrating over the fermion energies, with the boundary conditions similar to 
that given in eq.~(\ref{eq:dalitzbound}), one obtains the total production 
cross section. In the case of $\ee \to t\bar{t}H$, it is shown in Fig.~4.17 as 
a function of $M_H$ for three c.m. energy values $\sqrt{s}=0.5, 1$ and 3 TeV.

\begin{figure}[!h]
\begin{center}
\vspace*{-2.2cm}
\hspace*{-1.cm}
\psfig{file=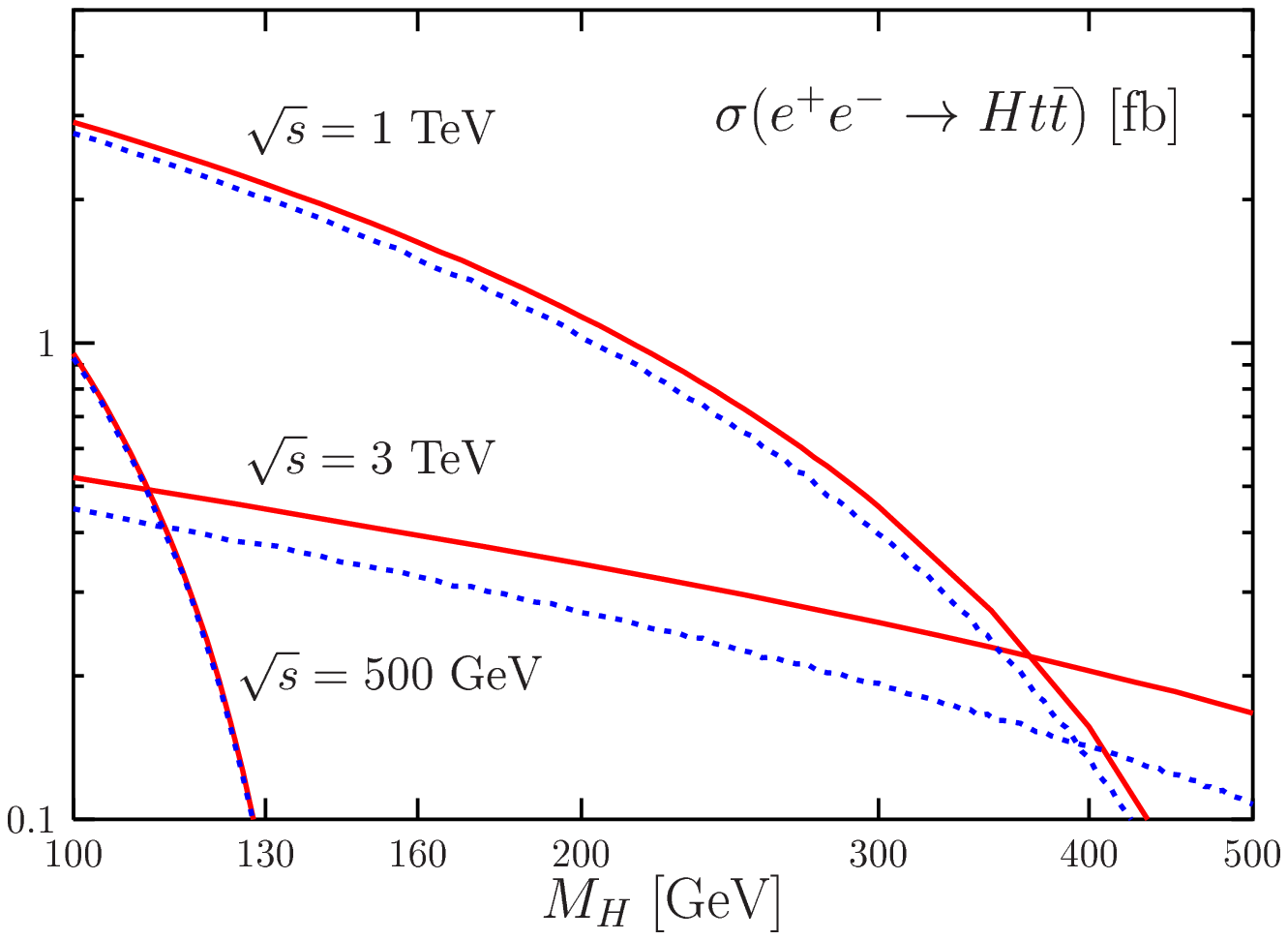,width=16.cm} 
\end{center}
\vspace*{-13.5cm}
\nn {\it Figure 4.17: The cross section for the associated production of the 
Higgs boson with $t\bar{t}$ pairs in $\ee$ collisions with  c.m. energies 
$\sqrt{s}=0.5,1$ and 3.TeV. The dotted lines are when only the contributions 
with the Higgs radiated off the top quark lines is taken into account.} 
\vspace*{-.3cm}
\end{figure}

While the cross section is in general small for the lowest  c.m. energy
$\sqrt{s}=500$ GeV, it is more important at $\sqrt{s}=1$ TeV as a result of the
larger available phase--space. For $\sqrt{s}=3$ TeV, it becomes  again smaller
as it scales like $1/s$. The cross section is at the  level of a few to a
fraction of a femtobarn, depending on the Higgs mass and the  c.m.  energy and
therefore, this process requires high--luminosities.  The $t\bar{t}H$ final
state in this associated production mechanism is generated almost exclusively
through Higgs--strahlung off top quarks. As shown in Fig.~4.17, the additional
contributions from Higgs bosons emitted by the $Z$ line are very small,
amounting, for $\sqrt{s} \leq 1$ TeV,  to only a few percent. In addition,
since top quark pair production in $\ee$ collisions at high energy is known to
be dominated by photon exchange, the bulk of the cross section is generated by
the $\ee \to \gamma^* \to t\bar{t}H$ subprocess.  This process thus allows the
determination of the important Yukawa coupling of the Higgs boson to top quarks
in an almost unambiguous way.

\vspace*{-2mm}
\subsubsection*{\underline{The radiative corrections}}

The QCD corrections to the process  $\ee \to t\bar{t}H$, consist of the top
vertex and self--energy corrections and the emission of an additional gluon in
the final state, $\ee \to t\bar{t}H+g$. The rather involved analytical
expression of the cross section at NLO can be found in
Refs.~\cite{RCTTqcd1,RCTTqcd2}; see also Refs.~\cite{RCTTew1,RCTTew2}. The 
corrections can be interpreted in an easy way and be given analytically in two 
kinematical regimes \cite{RCTTqcd1}.  

$(i)$ In the case where the invariant $t\bar{t}$ mass is close to the threshold,
the rescattering diagrams generated by the gluon exchange between the two
quarks gives rise to a correction that is proportional to $\alpha_s/\beta_t$, 
where $\beta_t$ is the top quark velocity which vanishes at the threshold in 
the $t\bar{t}$ rest frame. The $K$--factor in this case is given by
\cite{RCTTqcd1}
\beq
K^{\rm thresh}_{\ee \to t\bar{t}H}= 1 + 64 \alpha_s/(9\pi) \, \pi m_t \, 
\left[ (\sqrt{s}-M_H^2)^2 -4m_t^2 \right]^{-1/2}  
\eeq
This pole is regularized by the vanishing phase--space at threshold in the 
leading order cross section, once it is integrated over the 3--body phase 
space. \s

$(ii)$ At high energies, these rescattering corrections become less important.
For the dominant component of the $\ee \to t\bar{t}H$ process, i.e. Higgs
radiation off top quarks, the correction can be crudely estimated in the
limit $s \gg m_t^2 \gg M_H^2$: the radiation of a low mass Higgs boson can be 
separated from the top quark production process. The cross section can then 
be approximated by the product of the probability of producing top quark pairs 
[which at high energies, is given by the well--known factor $1+ \alpha_s/\pi$]  
and the probability for the splitting processes $t \to t+H$ and $\bar{t} 
\to \bar{t}H$ [which at this order, gives a factor $-2\alpha_s$ for each 
state]. The net result will be then an NLO coefficient factor \cite{RCTTqcd1}
\beq
K^{\rm high-en.}_{\ee \to t\bar{t}H}= 1 - 3\alpha_s/\pi
\eeq
leading to a correction factor, $K \sim 0.9$ at high energies. The QCD
correction factor is shown in Fig.~4.18 as a function of the c.m. 
energy for $M_H=150$ GeV. \s

The electroweak corrections have been calculated only recently by two of the
groups that evaluated the correction to the $WW$ fusion process
\cite{RCTTew1,RCTTew2}.  The calculation's techniques are the same as those
discussed previously.  [There is a third calculation performed in
Ref.~\cite{RCTTew0} but the results differ from those of the two other 
calculations at large c.m. energies and at the threshold.] The results are 
also shown in Fig.~4.18  together with the QCD corrections, as a function of 
the c.m. energy and for $M_H=150$ GeV.\s

As can be seen, the weak bosonic corrections are at the level of
$+10\%$  close to the $2m_t+M_H$ threshold and drop rapidly with increasing
energy to reach $-20\%$ at $\sqrt{s}=1.5$ TeV. The fermionic corrections are
approximately $+10\%$ over the entire energy range.  The QED corrections, which
include the full photonic and the higher--order ISR corrections are large and
negative near threshold and rise with the energy to reach a few percent at
$\sqrt{s}=1.5$ TeV. At energies above $\sqrt{s} \sim 600$ GeV, the fermionic, 
weak bosonic and QED contributions partly cancel each other, leading to a total
electroweak correction that is almost constant and of the order of $-10\%$.
This is of the same order as the QCD correction far enough from the production
threshold.  The total cross section at NLO, in which both the QCD and 
electroweak corrections are included, is thus 10 to 15\% smaller than at 
tree--level for $\sqrt{s}\gsim 750$ GeV; see the right--hand side of 
Fig.~4.18.  

\begin{figure}
\begin{center}
\mbox{
\includegraphics[width=.5\textwidth]{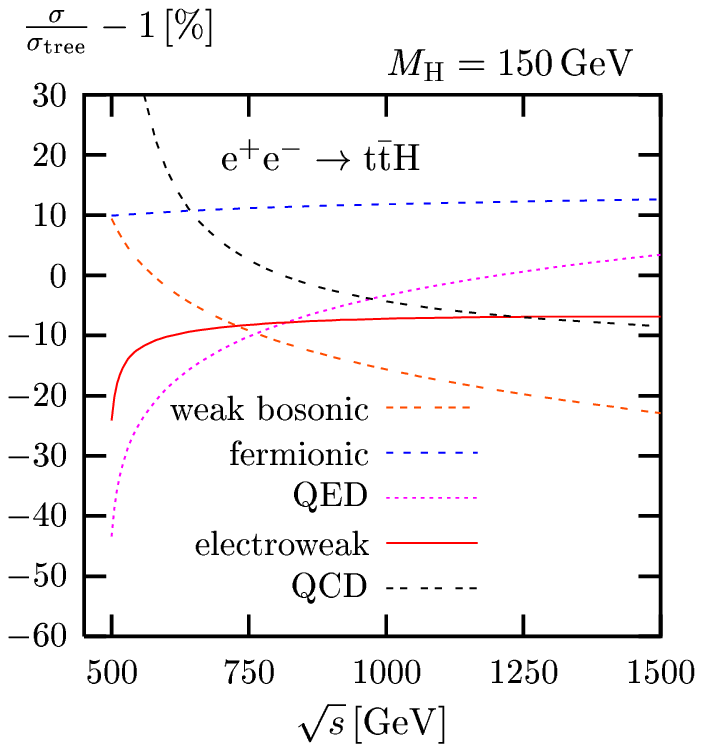}
\includegraphics[width=.5\textwidth]{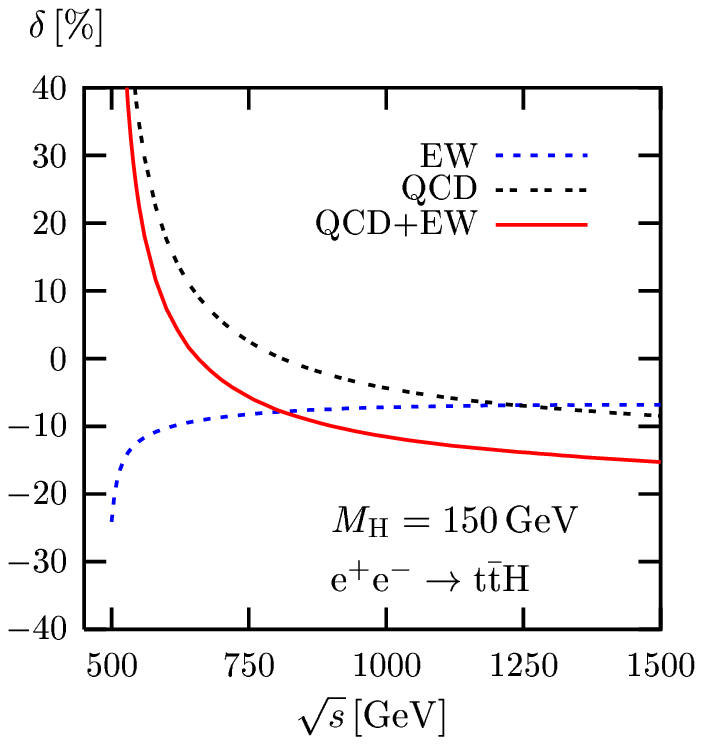} }
\end{center}
\vspace*{1mm}
\nn {\it Figure 4.18: The QCD and the various components and the electroweak 
radiative corrections (left) and the total QCD and electroweak corrections 
(right) for the process $\ee \to t\bar{t} H+X$ as a function of the c.m. 
energy for $M_H = 150$ GeV; from Ref.~\cite{RCTTew1}.}
\end{figure}

\subsubsection*{\underline{The pseudoscalar case and the Higgs CP properties}}

If the Higgs boson were of pseudoscalar nature, with couplings to fermions as 
given in eq.~(\ref{Affcp}), the dominant contribution to the cross section of 
the process $\ee\to f\bar{f}A$ would be also due to the Higgs radiation off the 
heavy fermion that are produced mainly through photon exchange. The expression 
of the Dalitz density ${\rm d} \sigma(\ee \to f\bar{f}A)/{\rm d}x_1 {\rm d} x_2$
will be still as in eq.~(\ref{ttHxsection}), with the coefficients $G_1$ and 
$G_2$ given by [here $a=M_A^2/s$] \cite{ee-ttH,ee-ttA}
\beq
G_1 &=& \frac{g_{Aff}^2}{x_{12}} \bigg[ x_A^2 - a \bigg( \frac{x_A^2}{x_{12}}
(1+2f) +2(x_A -1 -a) \bigg) \bigg] \non \\
G_2 &=&  -2 \frac{g_{Aff}^2}{x_{12}} \bigg[ x_{12}(1+ x_A)-a(x_{12}-4f+2x_A-2a)
 + f \frac{x_A^2}{x_{12}} (x_{12}-3a ) \bigg] 
\eeq
while the contributions of $G_{3}$--$G_6$ can be neglected [note that, in
two--Higgs doublet models, additional contributions to this process might come 
from other channels]. As can be seen, because the top quark is massive, 
the Dalitz density is different from the CP--even Higgs case by terms of
${\cal O}(m_t^2/s)$ which, for moderate c.m. energies, are not that small. 
This feature provides an additional means to discriminate between a scalar
and a pseudoscalar Higgs boson and even, to probe CP violation in the 
$t\bar{t}$--Higgs couplings when both components are present; for a detailed 
discussion, see Ref.~\cite{ee-ttHspin}.\s

If one assumes general Higgs couplings to top quarks compared to the SM, ${\cal
L} (Htt) = (a+ib \gamma_5) g_{Htt}$ [and also to the $Z$ boson, ${\cal L} (HZZ)
= c g_{HZZ} g_{\mu \nu}$, when the diagram $\ee \to HZ^*$ with $Z^* \to t
\bar{t}$ is included, since its contribution needs not to be small relative to
the dominant ones in extensions of the SM], one would have a rather involved
dependence of the $\ee \to t\bar{t}H$ cross section on the phase space.  The
differential cross section can be written in a general form as d$\sigma/$d$\Phi
= \sum_i d_i f_i(\Phi)$, where $\Phi$ is the final state phase--space
configuration and $d_i$ are combinations of the Higgs coupling parameters
$a,b,c$ [in the SM, only the combinations $d_i\!=\! a^2,ac$ and $c^2$ will be
present with $a\!=\!c\!=\!1$]. An optimal technique has been proposed in
Ref.~\cite{ee-ttHspin} for determining the coefficients $d_i$ of the cross
section by using appropriate weighting functions $w_i(\Phi)$ such that $\int
\omega_i (d\sigma/d\Phi)=d_i$, with the additional requirement that the
statistical error in the extraction of the coefficients is minimized.  

\subsubsection{Higgs boson pair production}

To establish the Higgs mechanism experimentally, once the Higgs particle is
discovered, the characteristic self--energy potential of the SM  must be 
reconstructed. This task requires the measurement of the trilinear and
quartic self--couplings of the Higgs boson, $\lambda_{HHH}=3 M_H^2/v$ 
and  $\lambda_{HHHH}=3 M_H^2/v^2$. The trilinear Higgs coupling can be 
measured directly in pair production of Higgs particles in $\ee$ collisions 
and several mechanisms can be exploited. Higgs pairs can be produced 
through double Higgs--strahlung off $Z$ bosons 
\cite{ee-HHZ,HH-Barger,ee-DKMZ,ee-H3-pheno}
\beq 
\ee \to Z^* \lra ZHH
\eeq
and vector boson [mostly $W$ boson] fusion into two Higgs bosons 
\cite{pp-VVHH,HH-Barger,ee-DKMZ}
\beq
\ee \to V^* V^* \lra \ell \ell HH
\eeq 
The Feynman diagrams for the two processes are shown in Fig.~4.19 and, as can 
be seen, one of them involves the triple Higgs interaction. The other
diagrams  are generated by the gauge interactions familiar
from single Higgs production in the dominant processes.\s

\vspace*{-5mm}
\begin{center}
\hspace*{-14cm}
\vspace*{-1.8cm}
\SetWidth{1.}
\begin{picture}(300,100)(0,0)
\ArrowLine(150,25)(185,50)
\ArrowLine(150,75)(185,50)
\Photon(185,50)(230,50){3.5}{5.5}
\Photon(230,50)(265,25){3.5}{5.5}
\DashLine(230,50)(250,60){4}
\DashLine(250,60)(265,75){4}
\DashLine(250,60)(265,45){4}
\put(120, 75){\red{\bf (a)}}
\put(227,47){\bb}
\put(247,57){\rb}
\Text(145,30)[]{$e^+$}
\Text(145,70)[]{$e^-$}
\Text(210,65)[]{$Z^*$}
\Text(275,30)[]{$Z$}
\Text(275,75)[]{\bH}
\Text(275,50)[]{\bH}
\ArrowLine(295,25)(330,50)
\ArrowLine(295,75)(330,50)
\DashLine(375,50)(410,25){4}
\Photon(375,50)(410,75){3.5}{5.5}
\Photon(330,50)(375,50){3.5}{5.5}
\DashLine(390,55)(410,45){4}
\put(373,47){\bb}
\put(387,52){\bb}
\vspace*{3mm}
\ArrowLine(445,25)(480,50)
\ArrowLine(445,75)(480,50)
\Photon(480,50)(525,50){3.5}{5.5}
\DashLine(525,50)(560,40){4}
\Photon(525,50)(560,75){3.5}{4.5}
\DashLine(525,50)(560,25){4}
\put(522,47){\bb}
\end{picture}
\vspace*{9.mm}
\end{center}
\begin{center}
\hspace*{-14cm}
\SetWidth{1.}
\begin{picture}(300,100)(0,0)
\hspace*{1cm}
\ArrowLine(150,25)(195,25)
\ArrowLine(150,75)(195,75)
\ArrowLine(195,25)(240,15)
\ArrowLine(195,75)(240,85)
\Photon(195,25)(195,75){3.5}{5.5}
\DashLine(195,50)(225,50){4}
\DashLine(225,50)(255,60){4}
\DashLine(225,50)(255,40){4}
\put(90, 75){\red{\bf (b)}}
\put(193,47){\bb}
\put(223,47){\rb}
\Text(145,30)[]{$e^+$}
\Text(145,70)[]{$e^-$}
\Text(245,20)[]{$e^+$}
\Text(245,80)[]{$e^-$}
\Text(210,65)[]{$W^*$}
\Text(210,35)[]{$W^*$}
\Text(270,60)[]{\bH}
\Text(270,40)[]{\bH}
\ArrowLine(295,25)(340,25)
\ArrowLine(295,75)(340,75)
\ArrowLine(340,25)(385,15)
\ArrowLine(340,75)(385,85)
\DashLine(340,60)(385,65){4}
\Photon(340,25)(340,75){3.5}{6.5}
\DashLine(340,40)(385,35){4}
\put(337,57){\bb}
\put(337,37){\bb}
\ArrowLine(425,25)(470,25)
\ArrowLine(425,75)(470,75)
\ArrowLine(470,25)(515,15)
\ArrowLine(470,75)(515,85)
\DashLine(470,50)(515,65){4}
\Photon(470,25)(470,75){3.5}{6.5}
\DashLine(470,50)(515,35){4}
\put(467,47){\bb}
\Text(330,-5)[]{\it Figure 4.19: Higgs pair production in the bremsstrahlung 
and $WW$ fusion processes.}
\end{picture}
\vspace*{3.mm}
\end{center}

The complete reconstruction of the SM Higgs potential requires the measurement 
of the quadrilinear coupling $\lambda_{HHHH}$ which can be accessed directly 
only through the production of three Higgs bosons, $\ee \ra ZHHH$ and $\ee \ra
\bar{\nu}_e \nu_e HHH$.  However, these cross  sections are reduced by two to
three orders of magnitude compared to the  corresponding double Higgs production
channels, and are therefore too  small to be observed at future $\ee$ 
colliders even with the large luminosities which are planned [see \S4.3.4].  

\vspace*{-3.mm}
\subsubsection*{\underline{The double Higgs--strahlung}}

The differential cross section for the process of double Higgs-strahlung, $\ee 
\to ZHH$, after the angular dependence is integrated out, can be cast into
the form \cite{ee-DKMZ}
\beq 
\frac{{\rm d} \sigma (e^+ e^- \to ZHH)}{{\rm d} x_1 {\rm d} x_2} = 
\frac{G_\mu^3 M_Z^6}{384 \sqrt{2} \pi^3 s}
\frac{(\hat a_e^2 + \hat v_e^2)}{(1- \mu_Z)^2}\, {\cal Z} 
\eeq 
where the electron--$Z$ couplings are defined as usual, 
eq.~(\ref{Zffcouplings}). 
$x_{1,2} =2 E_{1,2}/\sqrt{s}$ are the scaled energies of the two Higgs
particles, $x_3 = 2 - x_1 -x_2$ is the scaled energy of the $Z$ boson, and we
define $y_i = 1 - x_i$; the scaled masses are denoted by $\mu_i = M_i^2/s$. In
terms of these variables, the coefficient ${\cal Z}$ may be written as
\beq 
{\cal Z}\!&\!=\!&\!\frac{1}{8} a^2 f_0 +
\frac{1}{4 \mu_Z (y_1+\mu_H -\mu_Z)} \left[ 
\frac{f_1}{y_1+\mu_H -\mu_Z} + \frac{f_2}{y_2+\mu_H -\mu_Z} 
+ 2\mu_Z \, a \,  f_3 \right] + \Bigg\{ y_1 \leftrightarrow y_2 \Bigg\} 
\non \\
&& {\rm with} \ \ 
a = \frac{\lambda_{HHH}'}{y_3+\mu_Z-\mu_H} + \frac{2}{y_1+\mu_H -\mu_Z} + 
\frac{2}{y_2+\mu_H -\mu_Z} + \frac{1}{\mu_Z} 
\eeq
The coefficients $f_i$ are given by
\beq
f_0 &=& \mu_Z[(y_1+y_2)^2 + 8\mu_Z] \non\\
f_1 &=& (y_1-1)^2(\mu_Z-y_1)^2-4\mu_H y_1(y_1+y_1\mu_Z-4\mu_Z) + \mu_Z(\mu_Z-4\mu_H)(1-4\mu_H)-\mu_Z^2  
\non\\
f_2 &=& [\mu_Z(1+\mu_Z - y_1 -y_2 - 8\mu_H)-(1+\mu_Z)y_1 y_2](2+2\mu_Z 
-y_1-y_2) \non\\
& & {}+ y_1 y_2[y_1 y_2 + \mu_Z^2+1+4\mu_H (1+\mu_Z)]
+ 4\mu_H \mu_Z(1+\mu_Z+4\mu_H)+ \mu_Z^2 
\non\\
f_3 &=& y_1(y_1-1)(\mu_Z-y_1)-y_2(y_1+1)(y_1+\mu_Z)+2\mu_Z
(\mu_Z+1-4\mu_H) 
\eeq
The first term in the coefficient $a$ includes the scaled trilinear coupling 
$\lambda_{HHH}'=3 M_H^2/M_Z^2$. The other terms are related to 
sequential Higgs--strahlung and the 4 gauge--Higgs boson coupling; the
individual terms can easily be identified by examining the propagators.\s

The production cross section, which is a binomial in the self--coupling
$\lambda_{HHH}$,  is shown in Fig.~4.20 as a function of the Higgs mass for
three c.m. energies $\sqrt{s} = 0.5,1$ and 3 TeV. It is of the order of  a
fraction of a femtobarn when it is not too much suppressed by phase--space and,
because it is mediated by $s$ channel gauge boson exchange and scales like
$1/s$, it is higher at lower energies for moderate Higgs masses.  In addition,
since the process is mediated by $Z$--boson exchange, the cross section is
doubled if oppositely polarized electron and positron beams are used. The cross
section 
for the $ZHH$ final state is rather sensitive to the $\lambda_{HHH}$ coupling: 
for $\sqrt{s}\!=\!500$ GeV and $M_H\!=\!120$ GeV for instance, it varies by 
about 20\% for a 50\% variation of the trilinear coupling as shown in the 
figure.

\begin{figure}[!h]
\begin{center}
\vspace*{-2.3cm}
\hspace*{-1.cm}
\psfig{file=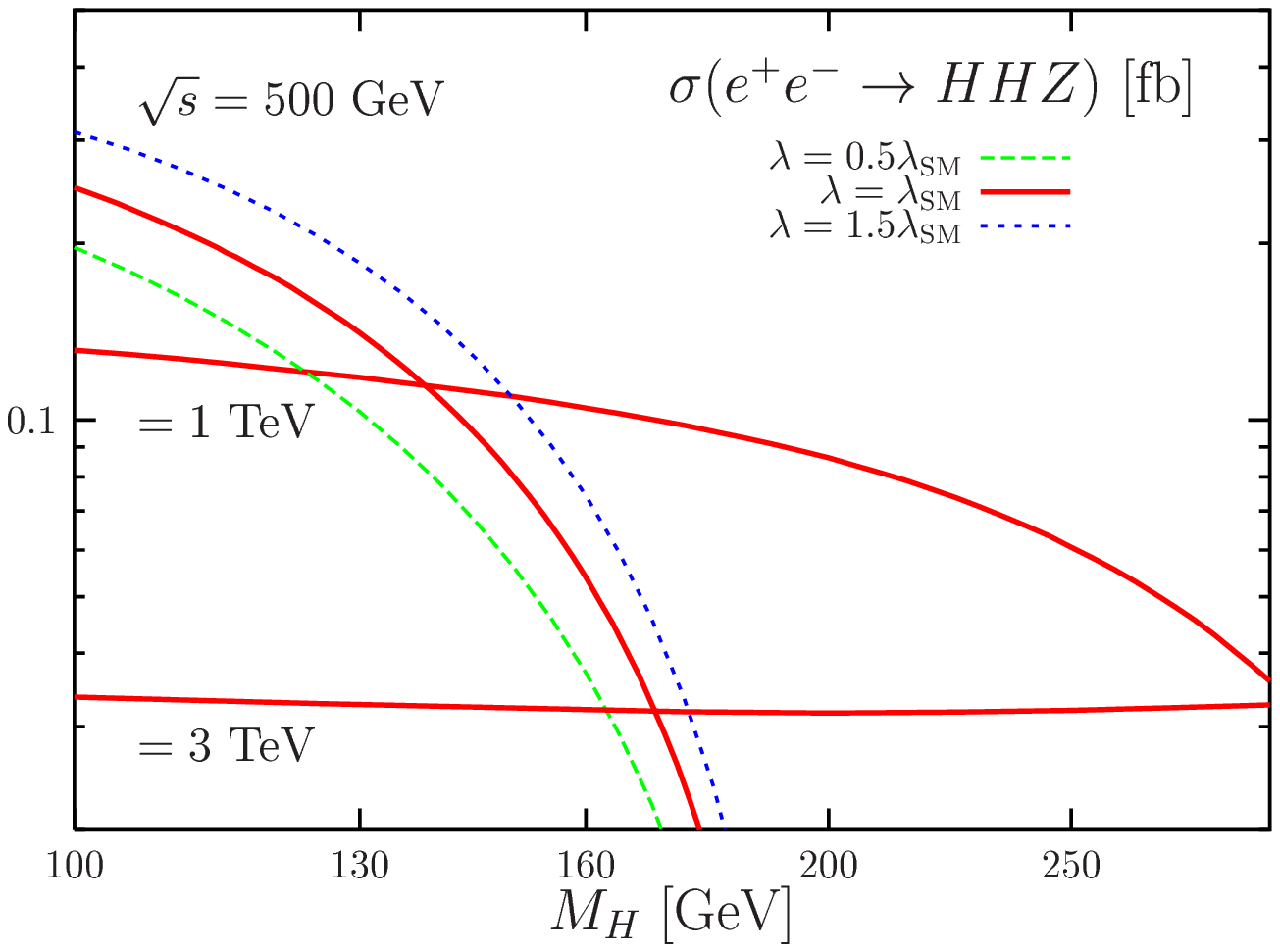,width=17.cm} 
\end{center}
\vspace*{-14.3cm}
\nn {\it Figure 4.20: The cross section for double Higgs--strahlung in 
$\ee$ collisions, $\ee \to HHZ$,  at c.m. energies $\sqrt{s}=0.5,1$ and 3 TeV 
as a function of $M_H$. Shown for $\sqrt s=500$ GeV are the effects of a
variation of the trilinear coupling by  50\% from its SM value.}
\vspace*{-.3cm}
\end{figure}

The one--loop radiative corrections to the double Higgs--strahlung process are
also very involved to calculate since, already at the tree--level, one has to
deal with three massive particle in the final state and, thus, one has to
consider pentagonal diagrams and four--body finals states at NLO. They have
again been calculated recently by two independent groups \cite{RCZHH1,RCZHH2},
with results that agree reasonably, in particular at low energies. The QED
corrections follow the same trend as what has been observed in the case of the
$\ee \to t\bar{t}H$ process for $M_H=150$ GeV: they are very large and negative
for c.m.  energies near the production threshold, $\sim -40\%$ at $\sqrt{s}
\sim 400$ GeV, and decrease in absolute value to reach the level of a few
percent above $\sqrt{s} \sim 600$ GeV, $\sim +5\%$ at 1.5 TeV; see the left
panel of Fig.~4.21.  For the pure weak corrections, when calculated using
$\alpha$ in the Born term, they are rather small not exceeding $\sim +5\%$ near
the threshold and at moderate c.m. energies when the cross section is maximal;
see the right panel of Fig.~4.21. At higher energies, the weak corrections turn
negative and increase in size to reach $\sim -10\%$ at $\sqrt{s}=1.5$ TeV. The
weak corrections calculated in the IBA are also shown (dotted lines). As in the
case of the $\ee \to HZ$ parent process, this approximation fails to reproduce
the magnitude of the weak corrections, especially at high energies. The
approximate top quark mass correction to the Higgs self--coupling does also not
reproduce the bulk of the weak correction.  

\begin{figure}[!h]
\begin{center}
\mbox{
\includegraphics[width=8cm,height=8cm]{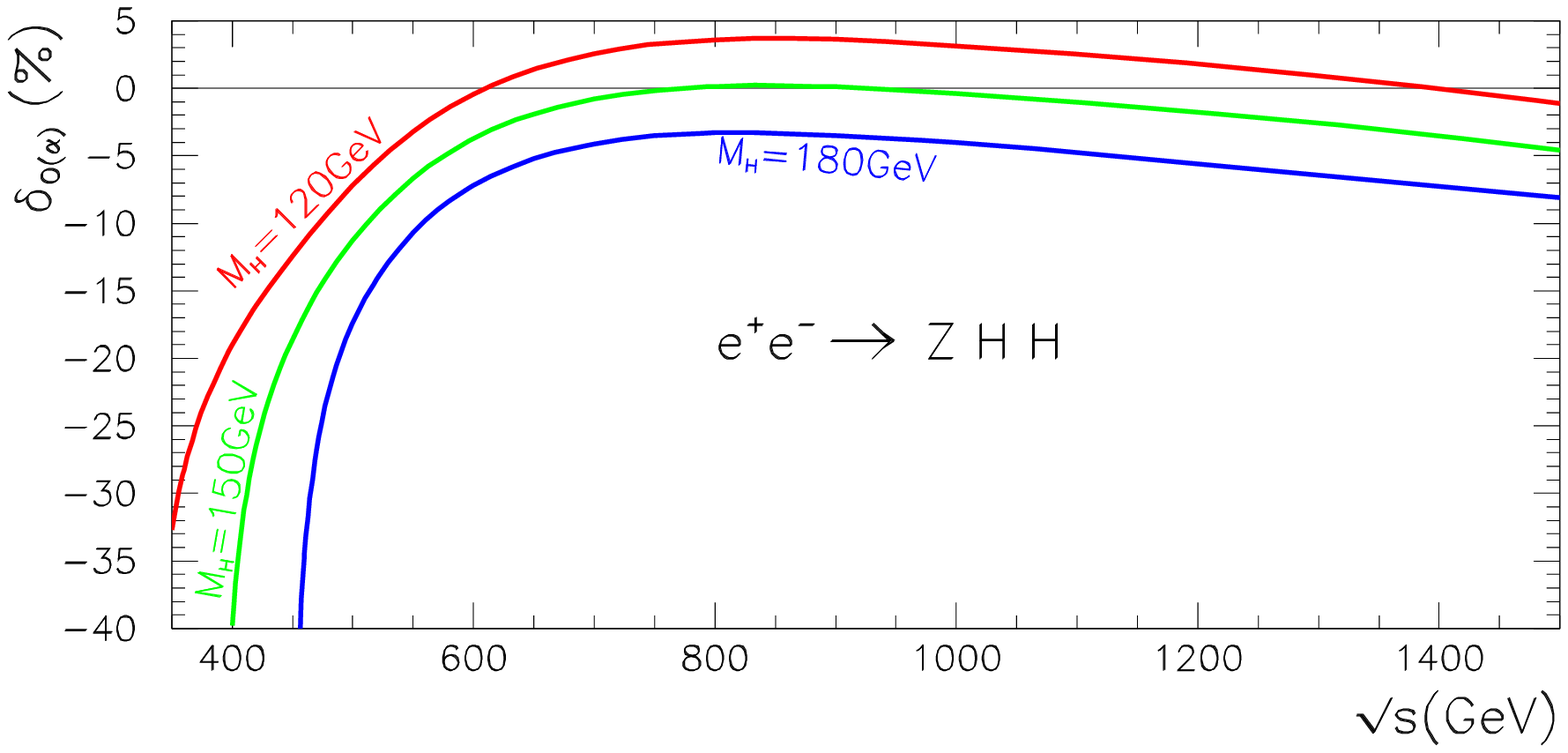}
\includegraphics[width=8cm,height=8cm]{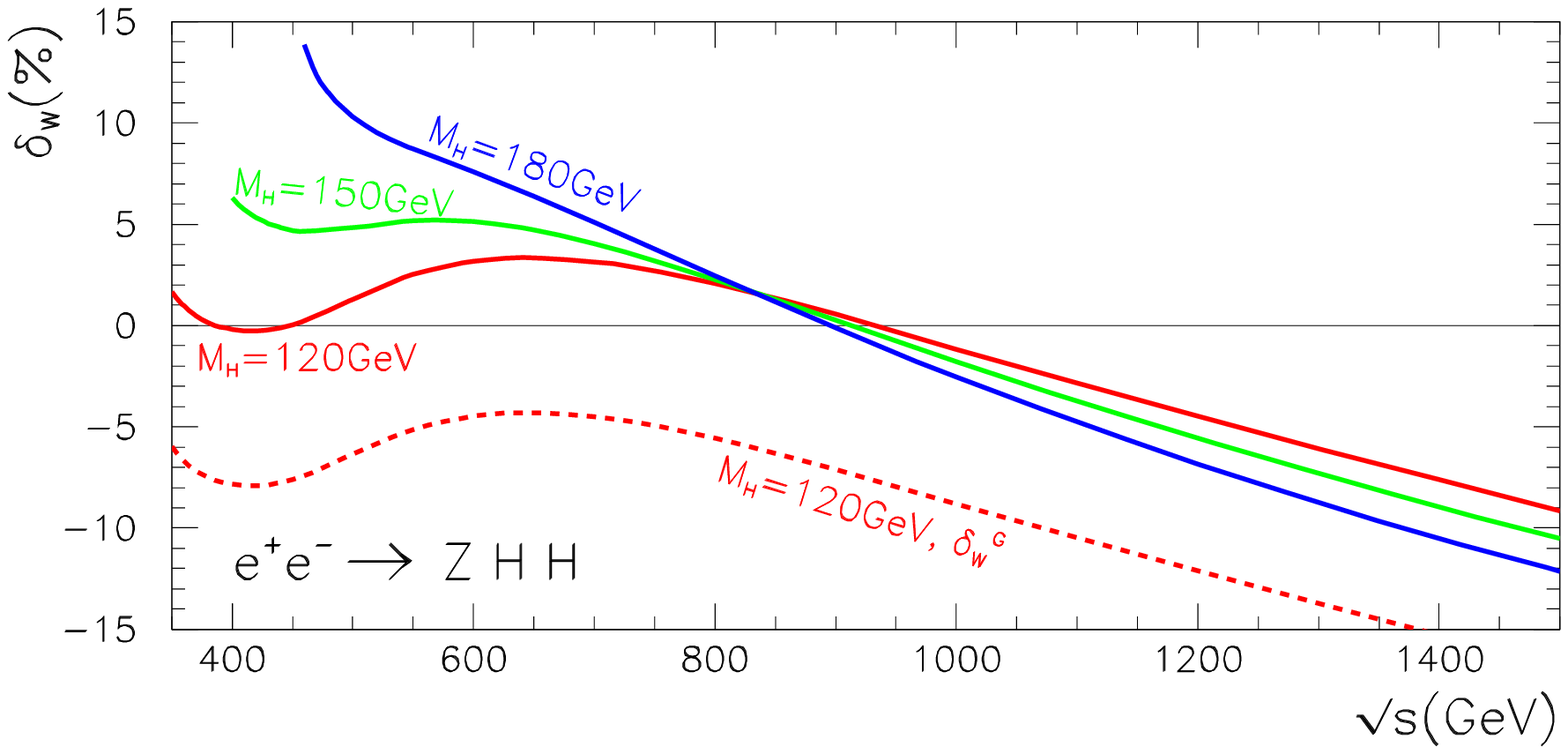}}
\end{center}
\vspace*{-6mm}
\nn {\it Figure 4.21:  The full ${\cal O}(\alpha)$ relative correction (left
panel) and the relative electroweak correction $\delta_W$ (right panel) as a 
function of the c.m. energy for $M_H=120,150,180$ GeV; the genuine weak 
correction in the IBA is presented for $M_H=120$ GeV (dotted line)
\cite{RCZHH1}.}
\vspace*{-5mm}
\end{figure}

Note that the correction to the invariant mass distribution of the Higgs pair,
which can be a means to isolate the $HHH$ vertex since the two Higgs bosons 
originate from the decay of an off--shell scalar particle \cite{gam-WWHH},
has also been calculated and found to be small.

\vspace*{-2mm}
\subsubsection*{\underline{The $WW$ fusion process}}

At high energies, double Higgs boson production in the $WW$ fusion channel,
$\ee \to \nu \bar{\nu}HH$ \cite{pp-VVHH,HH-Barger}, provides the largest cross 
section for  Higgs bosons in the intermediate mass range, in particular when the
initial beams are polarized. [Again, the $ZZ$ fusion channel has a cross section
that is one order of magnitude smaller compared to $WW$ fusion as a 
result of the smaller $Z$ couplings to electrons]. The cross
section for this four--particle final state is very involved but it can be
roughly estimated in the equivalent $W$ boson approximation, $WW \to HH$. 
Taking into account only the dominant longitudinal $W$ contribution, denoting
by $\beta_{W,H}$ the $W,H$ boson velocities in the c.m.\ frame, we define the
variable $x_W = (1- 2 M_H^2/\hat{s})/(\beta_W \beta_H)$ with  $\hat{s}^{1/2}$
is the invariant energy of the $WW$ pair.  The amplitude ${\cal M}_{LL}$ has
been given in eq.~(\ref{WW--HHamp}) when this process was discussed at hadron
colliders and, integrating out the angular dependence, the corresponding total
cross section reads \cite{ee-DKMZ,gam-WWHH} 
\beq
\hat{\sigma}_{LL} &=& \frac{G_F^2 M_W^4}{4\pi \hat{s}} \frac{\beta_H}
{\beta_W (1-\beta_W^2)^2} \Bigg\{ (1+\beta_W^2)^2 \left[1 + 
\frac{\lambda_{HHH}'}
{(\hat{s} -M_H^2)/M_Z^2} \right]^2 \non\\
& + &\frac{16}{(1+\beta_H^2)^2-4\beta_H^2 \beta_W^2} 
\left[ \beta_H^2(
-\beta_H^2 x_W^2+4\beta_W \beta_H x_W -4\beta_W^2)
 + (1+\beta_W^2 -\beta_W^4)^2 \right] \non \\
& + &\frac{1}{\beta_W^2 \beta_H^2} \left( \ell_W +
 \frac{2x_W} {x_W^2-1} \right) 
 \left[ \beta_H ( \beta_H x_W-4\beta_W) (1+ \beta_W^2-
\beta_W^4 +3 x_W^2 \beta_H^2) \right. \non\\ 
&+& \left.  \beta_H^2 x_W 
(1-\beta_W^4 +13 \beta_W^2) 
- \frac{1}{x_W}(1+\beta_W^2-\beta_W^4)^2 \right] 
+  \frac{2(1+\beta_W^2)}{ \beta_W \beta_H} 
\left[1 + \frac{\lambda_{HHH}}{(\hat{s} -M_H^2)/M_Z^2} \right] \non \\
&\times & \left[ \ell_W ( 1+\beta_W^2-\beta_W^4 -2\beta_W \beta_H x_W +
\beta_H^2 x_W^2)
 +2\beta_H (x_W \beta_H  -2\beta_W ) \right] 
\Bigg\} 
\eeq
with $ \ell_W = \log [(x_W-1)/(x_W+1)]$. After folding the cross section of the
subprocess with the longitudinal $W_L$ spectra given in eq.~(\ref{WW-spectra}),
one obtains the total $e^+ e^-$ cross section in the effective $W_LW_L$ 
approximation, which exceeds the exact value of the $\ee \to \nu \bar{\nu}HH$ 
cross section by  about a factor 2 to 5 depending on the collider energy and 
the Higgs mass. \s

\begin{figure}[!h]
\begin{center}
\vspace*{-1.cm}
\hspace*{-1.cm}
\psfig{file=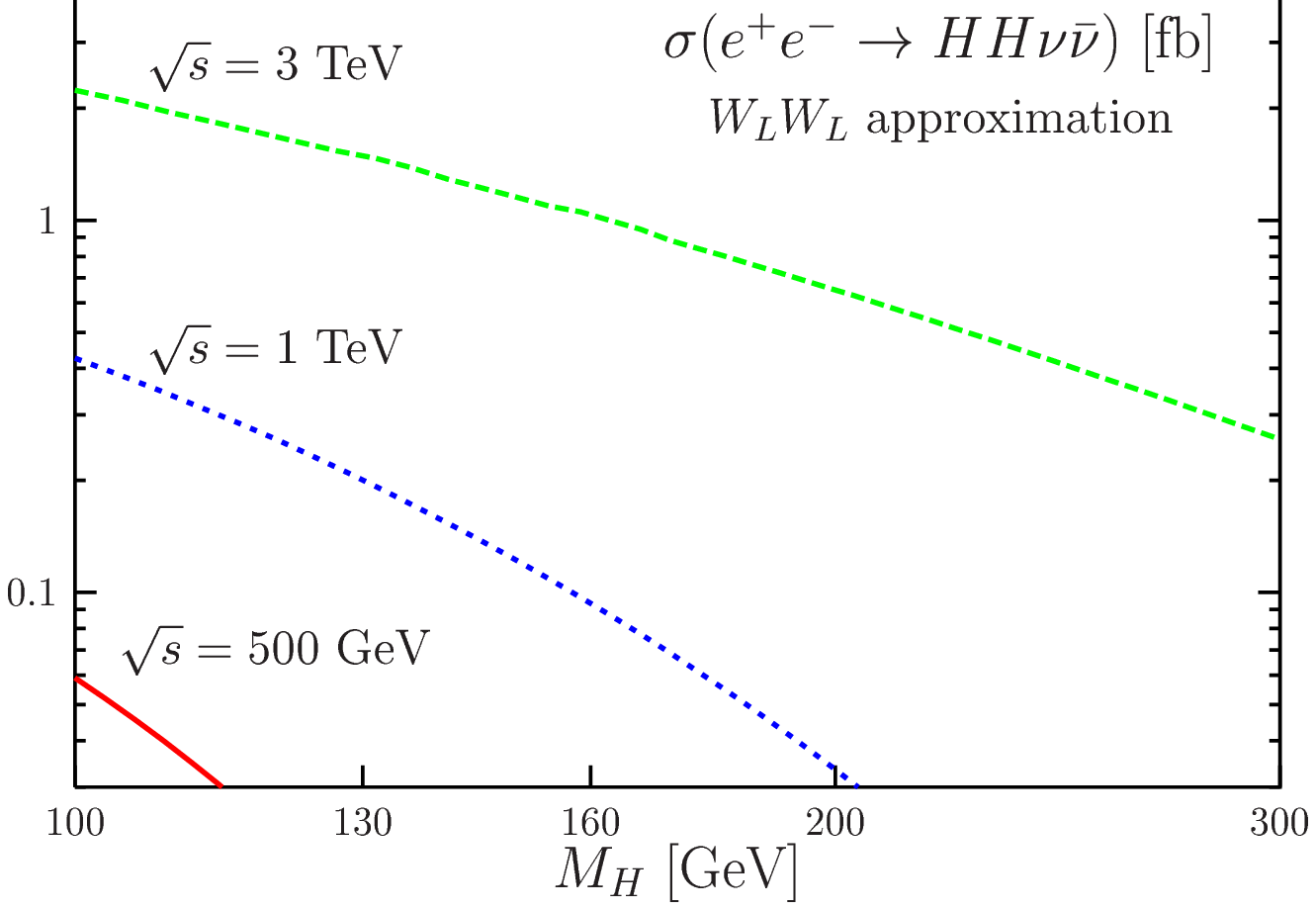,width=17.cm} 
\end{center}
\vspace*{-15.5cm}
\nn {\it Figure 4.22: The cross section for the $W_LW_L \to HH$ process in $\ee$
collisions with at c.m. energies $\sqrt{s}=0.5,1$ and 3 TeV as a function of 
$M_H$.}
\vspace*{-.1cm}
\end{figure}

The cross section is shown in Fig.~4.22 as a function of $M_H$ for $\sqrt{s}
=0.5,1$ and 3 TeV. As expected, the fusion cross sections increase with rising
energy.  Again, there is a significant variation of the cross section with a
variation of $\lambda_{HHH}$. The transverse components of the $W$ bosons give
rather small contributions through $W_T W_T \to HH$ for large Higgs masses.
Note that the ${\cal O}(\alpha)$ corrections have been also calculated using
{\tt GRACE-LOOP} and a preliminary result has appeared in Ref.~\cite{RCZZ}; the
corrections are of ${\cal O}(10\%)$.

\subsubsection{Other subleading processes in $\ee$ collisions} 

Finally, there are other subdominant higher--order Higgs production processes:
the associated production with a photon, the loop induced as well as some 
tree--level higher--order double Higgs production, the associated Higgs 
production with gauge boson pairs and the associated production with two 
fermions and a gauge boson.  We briefly summarize the main features of 
these processes for completeness.

\vspace*{-2mm}
\subsubsection*{\underline{Higgs production in association with two gauge 
bosons}}

Similarly to what one observes at hadron colliders, in high--energy $\ee$ 
collisions, $W$ pair production, $\ee \to W^+W^-$, has a very large cross 
section. This
is also the case of $\ee \to ZZ$ and $Z \gamma$ production\footnote{As noted 
before, the process with the additional final state photon should be viewed 
as part of the radiative corrections to the Higgs--strahlung process [the same 
remark holds for the process $\ee \to \nu_e \bar{\nu}_e H\gamma$ to be 
discussed later, which is part of the QED correction to the $WW$ fusion 
mechanism]. However, this process can be discussed on its own since here 
the photon is required to be detected and the $\ee \to HZ\gamma$ process can 
have a comparable rate than the parent process which scales as $1/s$ at 
high  energies, as the ISR photon will decrease the effective c.m. energy.}, 
which are mediated by  $t$--channel electron exchange. It is thus 
tempting to take advantage of these large production rates and consider the 
emission of an additional  Higgs particle from one of the gauge boson  lines
\beq 
\ee \to W^+W^- H \  , \ ZZH \ , \ Z\gamma H 
\eeq
as shown in Fig.~4.23.  The hope is that the suppression by the additional 
electroweak factor might be compensated by the initially large production 
rates.

\vspace*{-.7cm}
\begin{center}
\SetWidth{1.}
\begin{picture}(300,100)(0,0)
\hspace*{-2cm}
\ArrowLine(0,25)(40,25)
\ArrowLine(0,75)(40,75)
\Line(40,25)(40,75)
\Photon(40,25)(85,25){3.5}{5.5}
\Photon(40,75)(85,75){3.5}{5.5}
\DashLine(60,75)(90,50){4}
\put(57,72){\bb}
\Text(0,35)[]{$e^+$}
\Text(0,65)[]{$e^-$}
\Text(30,50)[]{$\ell$}
\Text(97,50)[]{\bH}
\Text(95,30)[]{$V$}
\Text(95,70)[]{$V$}
\ArrowLine(130,25)(165,50)
\ArrowLine(130,75)(165,50)
\Photon(165,50)(210,50){3.5}{5.5}
\Photon(210,50)(245,25){3.5}{5.5}
\Photon(210,50)(245,75){3.5}{5.5}
\DashLine(235,65)(260,47){4}
\put(232,62){\bb}
\Text(185,65)[]{$V$}
\Text(258,20)[]{$V$}
\Text(258,80)[]{$V$}
\Text(270,50)[]{\bH}
\ArrowLine(295,25)(330,50)
\ArrowLine(295,75)(330,50)
\Photon(330,50)(375,50){3.5}{5.5}
\Photon(310,60)(350,75){3.5}{5.5}
\Photon(375,50)(415,25){3}{5}
\DashLine(375,50)(415,75){4}
\put(372,47){\bb}
\Text(355,65)[]{$V$}
\Text(420,37)[]{$V$}
\Text(420,67)[]{\bH}
\Text(320,75)[]{$\gamma$}
\Text(210,-1)[]{\it Figure 4.23: Diagrams for associated  Higgs boson 
production with two gauge bosons.} 
\vspace*{1.mm}
\end{picture}
\end{center}

This turns out to be quite true \cite{ee-HVff,ee-HVV,DWP}: at least for the
process $\ee \to Z\gamma H$ [where one has to apply a cut on the transverse
momentum $p_T \gsim 5$ GeV of the photon] and for the $\ee \to W^+W^- H$
mechanism, the cross sections are quite sizable. At $\sqrt{s}= 800$ GeV and
for $M_H\sim 100$--200 GeV, they are at the level of a few fb as shown in
Fig.~4.24. With the expected luminosity ${\cal L}=$ 500 fb$^{-1}$, they could 
lead to more than 1000 events which are rather clean. For masses $M_H \sim 300$
GeV, they are still at the level of 1 fb, which is only one order of magnitude
smaller than the Higgs--strahlung process at these values of $M_H$ and
$\sqrt{s}$.  Again, as one might have expected, the production rate for the
$\ee \to ZZH$ process is an order of magnitude smaller than that of the  $\ee
\to WWH$ process. Note that the cross sections for these processes do not
become larger at higher energies. \s

Once the Higgs particle has been detected in the main channels, these processes
could be useful: in conjunction with the dominant Higgs--strahlung and $WW$
fusion processes, they would allow to test the quartic couplings involving
Higgs and gauge bosons and, for instance, to probe directly the $HZW^+W^-$ and
$H\gamma W^+W^-$ couplings and even, potentially, C--violating $HZZZ$ and
$H\gamma ZZ$ couplings which are absent in the SM.  

\begin{figure}[htbp]
\vspace*{1mm}
\begin{center}
\epsfig{figure=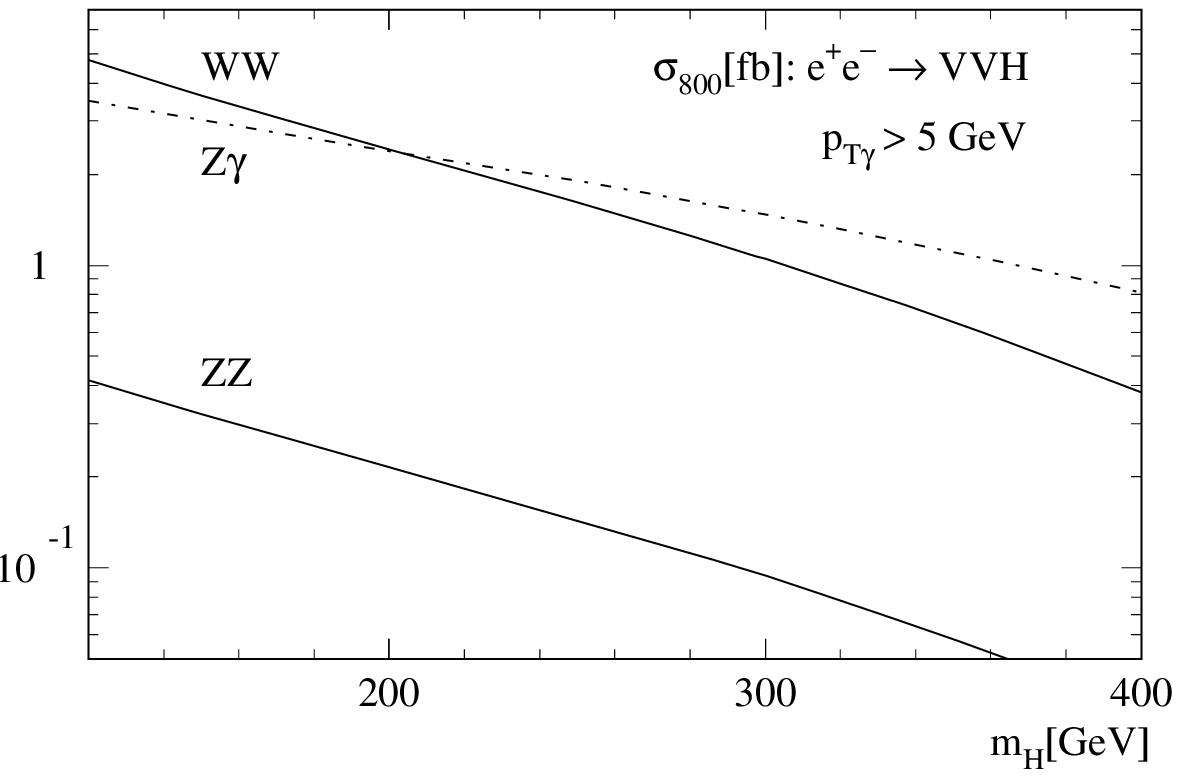,width=12cm}\\[3mm]
\end{center}
\vspace*{-2mm}
{\it Figure 4.24: The cross sections for the associated production of the
Higgs boson with a pair of gauge bosons, $\ee \to HVV$, as a function of 
$M_H$ at $\sqrt{s}=800$ GeV; from \cite{DWP}.}
\end{figure}

\subsubsection*{\underline{Higgs production in association with a gauge boson
and two leptons}}

Also as in the case of the LHC, Higgs bosons can be produced in association
with a gauge boson and two leptons in the fusion processes \cite{ee-HVff,DWP} 
\beq 
\ee \to \nu_e e^\pm W^\mp H \ , \ \nu_e \bar{\nu}_e \gamma H \ , \
\nu_e \bar{\nu}_e Z H 
\eeq
with some generic Feynman diagrams shown in Fig.~4.25. \s

\begin{figure}[h]
\vspace*{2mm}
\begin{center}
\begin{picture}(100,90)(-30,-5)
\hspace*{-11cm}
\SetWidth{1.1}
\ArrowLine(150,25)(195,25)
\ArrowLine(150,75)(195,75)
\ArrowLine(195,25)(240,15)
\ArrowLine(195,75)(240,85)
\Photon(195,25)(195,75){3.5}{6}
\Photon(195,50)(230,50){3.5}{4} 
\DashLine(230,50)(265,65){4}
\Photon(230,50)(265,35){3.5}{4}
\put(227,47){\bb}
\Text(145,30)[]{$e^+$}
\Text(145,70)[]{$e^-$}
\Text(245,20)[]{$\ell$}
\Text(245,80)[]{$\ell$}
\Text(210,65)[]{$V^*$}
\Text(210,35)[]{$V^*$}
\Text(275,65)[]{\bH}
\Text(275,35)[]{$V$}
\hspace*{5mm}
\ArrowLine(295,25)(340,25)
\ArrowLine(295,75)(340,75)
\ArrowLine(340,25)(385,15)
\ArrowLine(340,75)(385,85)
\DashLine(340,60)(385,65){4}
\put(337,57){\bb}
\Photon(340,25)(340,75){3.5}{6.5}
\Photon(342,40)(385,35){3.5}{5.5}
\ArrowLine(425,25)(470,25)
\ArrowLine(425,75)(470,75)
\ArrowLine(470,25)(515,15)
\ArrowLine(470,75)(515,85)
\DashLine(470,50)(515,50){4}
\put(467,47){\bb}
\Photon(470,25)(470,75){3.5}{6.5}
\Photon(485,77)(515,65){3}{4}
\end{picture}
\vspace*{-5.mm}
\end{center}
{\it Figure 4.25: Feynman diagrams for the associated production of a  Higgs 
boson with a gauge boson and two leptons in $\ee$ collisions.} 
\vspace*{-.2cm}
\end{figure}

Since, as previously discussed, the parent fusion processes $\ee \to H \ell
\ell$ have rather large production cross sections at high energies, one might
hope again that the emission of an additional gauge boson will still lead to a
reasonable event rate, similarly to the case of double Higgs boson production
in the vector boson fusion channels $\ee \to HH \ell \ell$ discussed in the
preceding section. These processes have been considered in
Ref.~\cite{ee-HVff} and are being updated \cite{DWP}. The cross sections for
$\ee \to \nu \bar \nu ZH$  and $\ee \to \nu e WH$ are shown in Fig.~4.26 as a
function of the c.m. energy for $M_H=160$ GeV.  As can be seen, they follow the
general trend of vector boson fusions mechanisms and increase with energy
and/or lower  Higgs masses.  They are quite sizable since, for $\ee \to \nu_e
e^\pm W^\mp H$, the cross section reaches almost the level of 10 fb at $\sqrt
s \sim 1$ TeV for $M_H \sim 120$ GeV. The cross section is a factor of $\sim 
5$ smaller in the case of the $\ee \to \nu_e \bar{\nu}_e Z H$ mechanism and is 
even smaller in the case of the $\ee \to \ee Z H$ process for which is not 
shown. 

\begin{figure}[!h]
\begin{center}
\vspace*{-2.6cm}
\hspace*{-1.cm}
\psfig{file=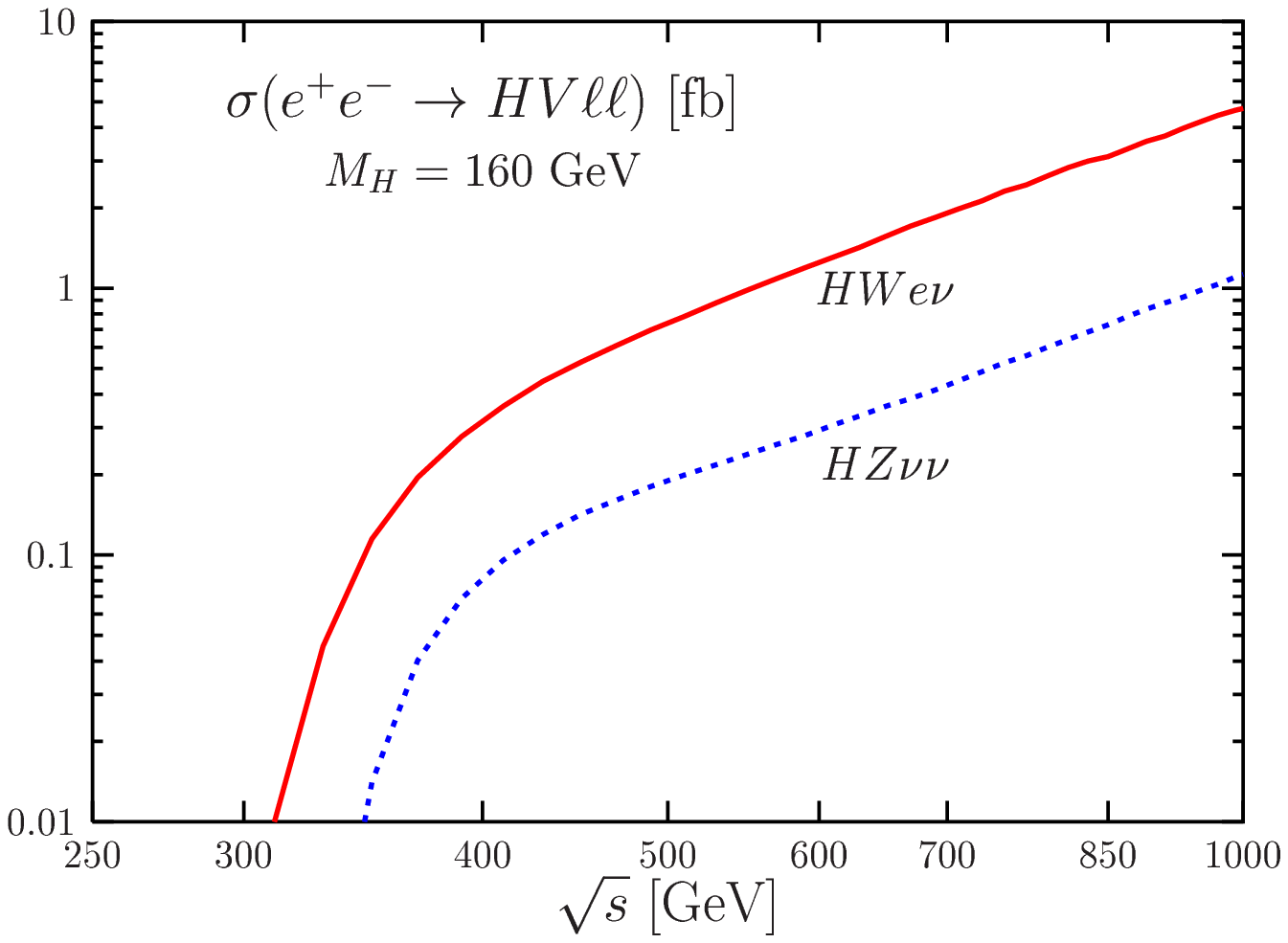,width=17.cm} 
\end{center}
\vspace*{-13.9cm}
{\it Figure 4.26: The cross sections for associated production of the
Higgs boson with a gauge boson and two leptons, $\ee \to HV \ell \ell$, as 
a function of $\sqrt{s}$ for $M_H =160$ GeV. They have been obtained using
the program {\tt WHIZARD} \cite{Whizard}.}
\vspace*{-7mm}
\end{figure}

\subsubsection*{\underline{Higgs production in association with a photon}}

In the SM, the process where a Higgs boson is produced in association with a
photon, $\ee \to H\gamma$ \cite{ee-Hgamma}, proceeds through $s$--channel 
$\gamma^* \gamma H$
and $Z^* \gamma H$ vertex diagrams, but additional $t$--channel vertex and box
diagrams involving $W$/neutrino and $Z$/electron exchange also occur;
Fig.~4.27. The $s$--channel contributions involve the same form factors as the
effective couplings for the $H \to Z\gamma, \gamma \gamma$ decays discussed in
S2.3, but with one of the two photons  and the $Z$ boson being virtual, with an
effective mass $M_{Z^*}=\sqrt{s}$.\\[-1.cm]

\begin{center}
\hspace*{-7.4cm}
\SetWidth{1.}
\begin{picture}(300,100)(0,0)
\ArrowLine(95,25)(130,50)
\ArrowLine(95,75)(130,50)
\ArrowLine(175,50)(200,25)
\ArrowLine(175,50)(200,75)
\ArrowLine(200,25)(200,75)
\Photon(130,50)(175,50){3.2}{5.5}
\Photon(200,25)(240,25){3.2}{5.5}
\DashLine(200,75)(240,75){4}
\put(198,73){\bb}
\Text(100,60)[]{$e^+$}
\Text(100,40)[]{$e^-$}
\Text(150,65)[]{$\gamma,Z$}
\Text(245,30)[]{$\gamma$}
\Text(245,65)[]{\bH}
\ArrowLine(270,25)(310,25)
\ArrowLine(270,75)(310,75)
\ArrowLine(310,25)(310,75)
\Photon(310,25)(355,25){3.2}{5.5}
\Photon(310,75)(355,75){3.2}{5.5}
\Photon(355,75)(355,25){3.2}{5.5}
\Photon(355,25)(395,25){3.2}{5.5}
\DashLine(355,75)(395,75){4}
\Text(300,50)[]{$\nu_e$}
\Text(340,50)[]{$W$}
\Text(400,30)[]{$\gamma$}
\Text(400,65)[]{\bH}
\put(353,73){\bb}
\Text(260,-2)[]{\it Figure 4.27: Diagrams for associated Higgs production 
with a photon in $\ee$ collisions.} 
\vspace*{0.mm}
\end{picture}
\end{center}
\vspace*{0cm}

Since it is a higher--order process in the electroweak coupling, the cross
section is rather small, $\sigma(\ee \to H\gamma) \sim 0.05$
fb for $M_H \sim100$--200 GeV at $\sqrt{s}=500$ GeV. However, since the photon
is mono--chromatic, the signal is very clean  allowing for a reasonable hope to
isolate these events if enough luminosity is collected a future high--energy
colliders.  Note that the longitudinal polarization of both electron and
positron beams will increase the cross sections by about a factor of 4 compared
to the unpolarized case. This process would then allow  for an alternative way
to probe the induced $H\gamma \gamma$  and $HZ\gamma$ couplings and,
potentially, to probe the heavy particles involved in the loops.  

\vspace*{-2mm}
\subsubsection*{\underline{Loop induced double Higgs production}}

Due to CP invariance, the $ZHH$ coupling is absent in the SM and the  process
$\ee \to Z \to HH$ does not occur  at tree--level but only through  loop
contributions \cite{ee-HHloop}. Because of orbital  momentum conservation, the
amplitudes for the vertex diagrams with $s$--channel $\gamma$ and $Z$ bosons
giving rise to two $H$ bosons vanish [only the  contribution of the
longitudinal component of the $Z$ boson survives but it is proportional to the
electron mass and is thus negligible].  In addition, because of chiral symmetry
for $m_e=0$,  the diagrams  involving the $He^+e^-$ vertices give zero
contributions. The contribution of vertices involving the $HHVV$ interaction
give also contributions that are proportional to $m_e$ or $m_{\nu_e}$.
Therefore, in the SM, the process $\ee \to HH$ can be generated only through
box diagrams involving $W$/neutrino and $Z$/electron virtual states, Fig.~4.28.

\vspace*{-.5cm}
\begin{center}
\hspace*{-18.4cm}
\SetWidth{1}
\begin{picture}(300,100)(0,0)
\ArrowLine(270,25)(310,25)
\ArrowLine(270,75)(310,75)
\ArrowLine(310,25)(310,75)
\Photon(310,25)(355,25){3.2}{5.5}
\Photon(310,75)(355,75){3.2}{5.5}
\Photon(355,75)(355,25){3.2}{5.5}
\DashLine(355,25)(395,25){4}
\DashLine(355,75)(395,75){4}
\put(353,73){\bb}
\put(353,23){\bb}
\Text(280,67)[]{$e^+$}
\Text(280,33)[]{$e^-$}
\Text(300,50)[]{$\ell$}
\Text(340,50)[]{$V$}
\Text(400,37)[]{\bH}
\Text(400,67)[]{\bH}
\hspace*{6cm}
\ArrowLine(270,25)(310,25)
\ArrowLine(270,75)(310,75)
\ArrowLine(310,25)(310,75)
\Photon(310,25)(355,25){4}{5.5}
\Photon(310,75)(355,75){4}{5.5}
\Photon(355,75)(355,25){4}{5.5}
\DashLine(355,25)(395,75){4}
\DashLine(355,75)(395,25){4}
\put(353,73){\bb}
\put(353,23){\bb}
\Text(400,37)[]{\bH}
\Text(400,67)[]{\bH}
\Text(220,0)[]{\it Figure 4.28: Feynman diagrams for loop induced Higgs pair
production.} 
\vspace*{0.mm}
\end{picture}
\end{center}
\vspace*{-2mm}

Again, because of the additional electroweak factor, the production cross 
sections are rather small. Except when approaching the $M_H=2M_W$ 
threshold, where there is a small increase, the cross section is practically
constant and amounts to $\sim 0.2$ fb at $\sqrt{s}=500$ GeV  for $M_H \sim
100$--200 GeV. With left--handed polarization of the electron beam, the cross 
section is increased by a factor of two, while for left--handed electrons and
right--handed positrons, it increases by a factor of four; these simple
factors are due to the fact that, as usual the contribution of the box with $W$ 
exchange is much larger than that with the $Z$ exchange. With a very high
luminosity, one might hope that the final state can be isolated. A deviation 
from the SM  expectation would signal a breakdown of CP--invariance or the
existence of new particles contributing to the loop diagrams. 

\subsubsection*{\underline{Higher order tree--level multi Higgs production}}

Finally, there are also higher--order processes for double Higgs production
which occur at the tree--level. Besides the $ZZ$ fusion process $\ee \to
HH \ee$ which, as mentioned previously, has a cross section that is one order 
of magnitude smaller than that of the $WW$ fusion process [for $M_H \sim 100$ 
GeV, the $\ee \to \ee HH$ cross section barely reaches the level of $\sim 0.1$ 
fb even at very high energies, $\sqrt s \sim 2$ TeV], one has the following 
reactions with $V=W,Z$ and $\ell=e,\nu_e$:
\beq
{\rm associated\ double\ Higgs\ production\ with\ two\ gauge\ bosons}
&:& \ee \lra VVHH \non  \\
{\rm associated\ double\ Higgs\ production\ with\ t\bar t\ pairs}
&:& \ee \lra t \bar t HH 
\eeq 

\begin{figure}[h!]
\vspace*{-7mm}
\begin{center}
\SetWidth{1.}
\begin{picture}(300,100)(0,0)
%
\ArrowLine(0,25)(40,25)
\ArrowLine(0,75)(40,75)
\Line(40,25)(40,75)
\Photon(40,25)(85,25){3.5}{5.5}
\Photon(40,75)(85,75){3.5}{5.5}
\DashLine(60,75)(90,50){4}
\DashLine(90,50)(120,65){4}
\DashLine(90,50)(120,35){4}
\put(57,72){\bb}
\Text(0,35)[]{$e^+$}
\Text(0,65)[]{$e^-$}
\Text(30,50)[]{$\ell$}
\Text(75,50)[]{\bH}
\Text(125,60)[]{\bH}
\Text(125,42)[]{\bH}
\Text(95,25)[]{$V$}
\Text(95,75)[]{$V$}
%
%
%
\hspace*{8mm}
\ArrowLine(130,25)(165,50)
\ArrowLine(130,75)(165,50)
\Photon(165,50)(210,50){3.5}{5.5}
\ArrowLine(210,50)(245,25)
\ArrowLine(210,50)(245,75)
\DashLine(235,65)(260,50){4}
\DashLine(260,50)(285,65){4}
\DashLine(260,50)(285,35){4}
\put(232,62){\bb}
\Text(258,70)[]{$t$}
\Text(258,30)[]{$\bar t$}
\end{picture}

\vspace*{-3mm}
\nn {\it Figure 4.29: Higher order double Higgs production processes at 
the tree--level.}
\end{center}
\vspace*{-6mm}
\end{figure}

Some Feynman diagrams for these reactions [those which involve the trilinear
Higgs interaction] are displayed in Fig.~4.29. The production cross sections
for these processes have been calculated in
Refs.~\cite{ee-H3-Ilyn} using the package {\tt CompHEP}
\cite{Comphep} for the automatic evaluation of the full set of amplitudes and,
as expected, they are very small. The $\ee \to WWHH$ cross section is at the
level of 0.03 fb at $\sqrt s \sim 700$ GeV even for a Higgs mass as
low as $M_H \sim 65$ GeV, while the rate for $\ee \to ZZHH$ is again one
order of magnitude smaller. In the case of the 
$\ee \ra t \bar t HH$ process \cite{ee-H3-Ilyn,ee-H3-Sampayo}, the cross
section is at the level of $6\, (15)$ ab at a c.m. energy $\sqrt s =0.8\,
(1.6)$ TeV for $M_H\sim 130$ GeV and $m_t=175$ GeV. Thus, about 10 of such
events could be produced if very high luminosities, ${\cal L}\sim 1$ ab$^{-1}$,
can be collected at these energies. \s

In the case of triple Higgs production processes, which would allow for the
determination of the quartic Higgs coupling, the cross section are
unfortunately too small as mentioned earlier. In the $\ee \to ZHHH$ process
\cite{ee-DKMZ,ee-H3-Ilyn}, for instance, the signal amplitude squared involving
the four--Higgs coupling [as well as the irreducible Higgs--strahlung 
amplitudes]
is suppressed by a factor $[\lambda_{HHHH}^2/16\pi^2]/[\lambda_{HHH}^2/M_Z^2]
\sim 10^{-3}$ relative to $\ee \to ZHH$, not to mention the phase--space
suppression due to the additional final--state heavy particle. The cross
sections are below the atobarn level: $\sigma (HHHZ) \sim 0.44$ ab for $M_H
\sim 110$ GeV and $\sqrt s \sim 1$ TeV and are not very sensitive to a
variation of the self--coupling: $\sigma (HHHZ) \sim 0.41\, (0.46)$ ab when
$\lambda_{HHHH}$ is altered by a factor $\frac12 \, (\frac32)$ 
\cite{ee-DKMZ}.  The fusion
process $\ee \to HHH \nu \bar \nu$ has also a very small cross section, $\sigma
(HHH \nu \bar \nu) \sim 0.4$ ab at $\sqrt s =3$ TeV \cite{ee-H3-Battaglia}.  

\subsection{Higgs studies in $\ee$ collisions}

In this section, we  summarize the precision tests of the SM Higgs sector
which can be performed at an $\ee$ machine operating in the 350--1000 GeV
energy range. We also briefly discuss the additional precision studies
which can be made by moving to higher energies at CLIC and by revisiting the
physics at the $Z$ resonance in the GigaZ option. We will almost exclusively
rely on the detailed studies which have been performed for the TESLA Technical
Design Report \cite{TESLA,LC-GigaZ} and on the very recent analyses of the CLIC
Physics working group \cite{CLIC}, since they involve realistic simulations of
the experimental environments\footnote{For the TESLA analyses in particular,
the backgrounds, the beamstrahlung and detector response have been taken into
account, generally using  programs such as {\tt CompHEP} \cite{Comphep} or {\tt
WHiZard} \cite{Whizard} in addition to the usual Monte-Carlo generators
\cite{PYTHIA,HERWIGee,HZHA}, {\tt Circe} \cite{Circe} and {\tt SIMDET}
\cite{Simdet} or {\tt BRAHMS} \cite{Brahms}; see Ref.~\cite{Ronan}.}.  We refer
to these two reports for more details and for more references on the original
work.  We will also mention some updated analyses which appeared
during the Linear Collider Workshops held in Amsterdam \cite{Desch} and Paris
\cite{LCWS}. Complementary material can be found in the reports of the American
Linear Collider working group \cite{NLC} and of the JLC working group
\cite{JLC}, as well as in the detailed reviews given in
Refs.~\cite{ee-Review-old,LHC-LC}.

\subsubsection{Higgs boson signals}

As discussed in the previous sections, the main production mechanisms for SM
Higgs particles are the Higgs--strahlung process $\ee \to ZH$ and the $WW$
fusion process $\ee \to \bar{\nu}_e \nu_e H$. Subleading production channels
are the $ZZ$ fusion mechanism, $\ee \to \ee H$,  the associated production with
top quarks $\ee \to t\bar{t}H$ and double Higgs production in the strahlung
$\ee \to HHZ$ and fusion $\ee \to \bar{\nu} \nu HH$ processes which, despite
the small production rates, are very useful when it comes to the study of the 
Higgs properties. The other production processes, although some of them have
substantial cross sections such as $\ee \to HW^+W^-$ and  $\nu_e e^\pm W^\mp
H$, will not [at least in the context of the SM] provide any additional
information and we will ignore them in the following discussion.\s 

The cross sections have been given previously, but we summarize them again in
Fig.~4.30 for four c.m. energies $\sqrt{s}=350$ GeV, 500 GeV, 1 TeV and 3 TeV,
as functions of the Higgs mass. They have been obtained with the {\sc fortran}
code {\tt HPROD} \cite{HPROD}. We should mention that these cross sections do
not include the radiative corrections which have been discussed in this chapter
[except that we work in the IBA which absorbs some of the electroweak
corrections], and no photon ISR nor beamstrahlung effects have been taken into
account. However, since these corrections and effects are rather small, except
in peculiar regions of the phase space [such as for $\ee \to t\bar t H$ near
threshold and $\ee \to HZ$ at $\sqrt s \gg M_H$], these numbers approach the
exact results to better than 5 to 10\% depending on the process, and this
approximation is sufficient for most of the purposes that one can have before
the experiments actually start. In Table 4.3, we display the numerical values
of the cross sections for selected values of the Higgs mass at the two
different energies $\sqrt{s}=500$ GeV and 1 TeV.\s 

\begin{figure}[!h]
\begin{center}
\vspace*{-2.2cm}
\hspace*{-3.5cm}
\psfig{file=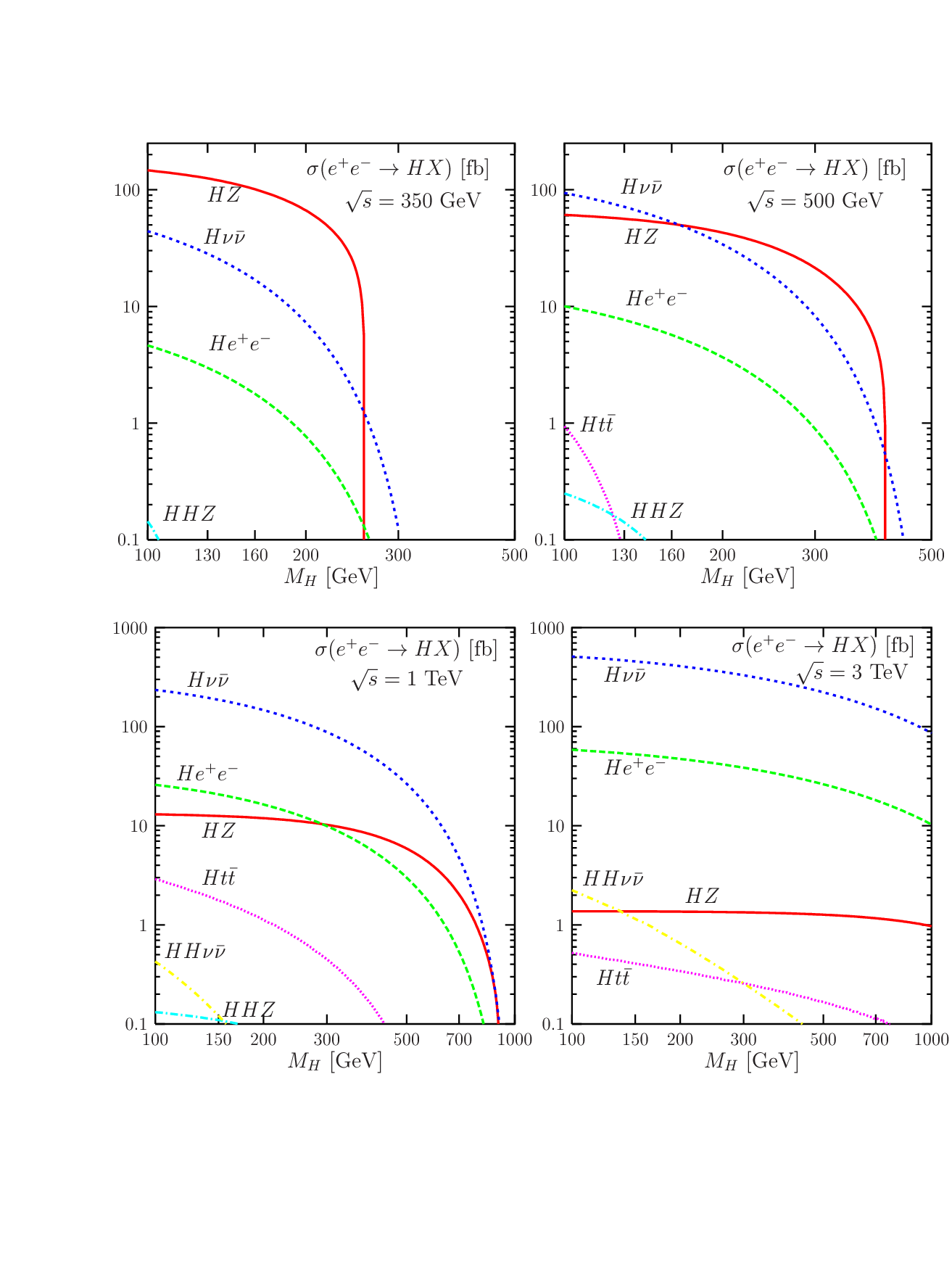,width=18cm} 
\end{center}
\vspace*{-4.2cm}
\nn {\it Figure 4.30: Production cross sections of the SM Higgs boson in $\ee$ 
collisions in the dominant and subdominant processes as a function of the Higgs
mass for four center of mass energies, $\sqrt{s}=350$ GeV, 500 GeV, 1 TeV and 
3 TeV. Radiative corrections, initial state radiation and beamstrahlung effects
are not included. The cross sections have been obtained with the program {\tt 
HPROD} \cite{HPROD}. }
\vspace*{-.6cm}
\end{figure}

\begin{table}[htbp]
\begin{center}
\renewcommand{\arraystretch}{1.2}
\vbox{\columnwidth=26pc
\begin{tabular}{|c||c|c|c|c|c|c|}\hline
$M_H$ (GeV) & \ $\sigma(HZ)$ \ & \ $\sigma(H\nu_e \bar{\nu}_e)$ \ &   \ 
$\sigma (H\ee)$ \ & $\sigma(H t \bar t)$ & \  $\sigma (HHZ)$ \ & $\sigma(
HH\nu\bar \nu)$ \\ \hline \hline
115 &  58.67 & 81.98 &  8.77 & 0.36 & 0.19  & 0.03  \\
120 &  57.91 & 78.30 &  8.38 & 0.23 & 0.18  & 0.02  \\
130 &  56.31 & 71.28 &  7.64 & 0.07 & 0.14  & 0.01  \\
140 &  54.61 & 64.71 &  6.95 &  --  & 0.11  & --    \\
150 &  52.83 & 58.58 &  6.30 &  --  & 0.08  & --     \\
160 &  50.96 & 52.88 &  5.69 &  --  & 0.05  & --     \\
170 &  49.03 & 47.60 &  5.13 &  --  & 0.03  & --     \\
180 &  47.03 & 42.71 &  4.60 &  --  & 0.02  & --     \\
200 &  42.88 & 34.03 &  3.67 &  --  &  --   & --     \\
300 &  21.38 & 8.26  &  0.89 &  --  &  --   & --     \\
400 &  3.24  & 0.73  &  0.07 &  --  &  --   & --     \\
\hline
\end{tabular}
}
\end{center}
\vspace*{-5mm}
\end{table}

\begin{table}[!h]
\begin{center}
\renewcommand{\arraystretch}{1.2}
\vbox{\columnwidth=26pc
\begin{tabular}{|c||c|c|c|c|c|c|}\hline
$M_H$ (GeV) & \ $\sigma(HZ)$ \ & \ $\sigma(H\nu_e \bar{\nu}_e)$ \ &   \ 
$\sigma (H\ee)$ \ & $\sigma(H t \bar t)$ & \  $\sigma (HHZ)$ \ & $\sigma(
HH\nu\bar \nu)$ \\ \hline \hline
115 & 12.90 & 219.54 & 24.26 & 2.50 & 0.12 &  0.30 \\
120 & 12.86 & 214.58 & 23.73 & 2.38 & 0.12 &  0.27  \\
130 & 12.76 & 204.92 & 22.70 & 2.16 & 0.12 &  0.21 \\
140 & 12.66 & 195.60 & 21.70 & 1.96 & 0.11 &  0.16 \\
150 & 12.55 & 186.63 & 20.73 & 1.79 & 0.11 &  0.12 \\
160 & 12.44 & 178.01 & 19.80 & 1.63 & 0.10 &  0.10 \\
170 & 12.32 & 169.72 & 18.90 & 1.49 & 0.10 &  0.07 \\
180 & 12.19 & 161.76 & 18.03 & 1.36 & 0.10 &  0.06 \\
200 & 11.92 & 146.78 & 16.40 & 1.14 & 0.09 &  0.03 \\
300 & 10.22 &  88.19 &  9.93 & 0.45 & 0.03 &  -- \\
400 &  8.13 &  50.32 &  5.68 & 0.16 & --   &  -- \\
500 &  5.89 &  26.55 &  3.00 & 0.04 &      &  -- \\
600 &  3.78 &  12.41 &  1.40 & --   & --   &  -- \\
700 &  2.03 &  4.75  &  0.53 & --   & --   &  -- \\ 
800 &  0.81 &  1.24  &  0.14 & --   & --   &  -- \\ 
\hline
\end{tabular}
}
\end{center}
\vspace*{0mm}
{\it Table 4.3: Numerical values for SM Higgs production cross sections [in fb]
in $\ee$ collisions at two center of mass energies $\sqrt{s}=500$ GeV (top) and
$\sqrt{s}=1$ TeV (bottom) for selected values of the Higgs boson mass. These 
numbers have been obtained with the program {\tt HPROD} \cite{HPROD} and no
radiative correction nor beamstrahlung is included. }
\vspace*{-3mm}
\end{table}

As previously mentioned, the Higgs--strahlung cross section scales as $1/s$ and
therefore dominates at low energies, while the one of $WW$ fusion mechanism
rises like $\log(s/M_H^2)$ and becomes more important at high energies. At
$\sqrt{s} \sim 500$ GeV, the two processes have approximately the same cross
sections, ${\cal O} (50~{\rm fb})$ for the interesting Higgs mass range 115 GeV
$\lsim M_H \lsim$ 200 GeV.  With an integrated luminosity ${\cal L} \sim 500$
fb$^{-1}$, as expected at the TESLA machine for instance, approximately 30.000
and 40.000 events can be collected in, respectively, the $HZ$ and $\nu \bar \nu
H$ channels for $M_H \sim 120$ GeV.  This sample is more than enough to observe
the Higgs particle and to study its properties in great detail.  \s

In the Higgs--strahlung process, the recoiling $Z$ boson, which can be tagged
through its clean $\ell^+ \ell^-$ decays, with $\ell=e$ or $\mu$, but also
through decays into quarks which have a much larger statistics, is
mono--energetic and the Higgs mass can be derived from the energy of the $Z$
boson since the initial $e^\pm$ beam energies are sharp when the effect of
beamstrahlung is strongly suppressed.  Therefore, it will be easy to separate
the signal from the backgrounds \cite{ee-HZ-backg,ee-HZ-backg1}. In the low
mass range, $M_H \lsim 140$ GeV, the process leads to $b\bar{b}q\bar{q}$ and
$b\bar{b}\ell \ell$ final states, with the $b$--quarks being efficiently tagged
by means of micro--vertex detectors. In the mass range where the decay $H \ra
WW^*$ is dominant, the Higgs boson can be reconstructed by looking at the $\ell
\ell + \,$4--jet or 6--jet final states, and using the kinematical constraints
on the fermion invariant masses which peak at $M_W$ and $M_H$, the backgrounds
are efficiently suppressed. Also the $\ell \ell q\bar q  \ell \nu$ and 
$q\bar q q\bar q  \ell \nu$ channels are easily accessible. \s

It has been shown in detailed simulations \cite{TESLA} that only a few
fb$^{-1}$ data are needed to obtain a 5$\sigma$ signal for a Higgs boson with a
mass $M_H \lsim 150$ GeV at a 500 GeV collider, even if it decays invisibly [as
could happen in some extensions of the SM].  In fact, for such small masses, it
is better to move to lower energies where the Higgs--strahlung cross section is
larger. Fig.~4.31 shows the reconstructed Higgs mass peaks in the strahlung
process at $\sqrt{s}=350$ GeV with a luminosity ${\cal L}=500$ fb$^{-1}$ for
$M_H=120$ GeV in the decay $H \to q\bar{q}$ and for $M_H=150$ GeV in the decay
$H \to WW^*$. At this energy and integrated luminosity, Higgs masses up to $M_H
\sim 260$ GeV can be probed in this channel.\s

\begin{table}[h!]
\vspace*{-2mm}
\begin{center}
\renewcommand{\arraystretch}{1.3}
\begin{tabular}{|c||c|c||c|}
\hline
$M_H$ (GeV) & 350~GeV & 500~GeV & 1000~GeV\\
\hline
\hline
120    & 4670 & 2020 & ~377 \\
180    & 2960 & 1650 & ~365 \\
250    & 230  & 1110 & ~333 \\ \hline
Max $M_H$   & 258 &  407 & 730 \\
\hline
\end{tabular}
\end{center}
{\it Table 4.4: Expected number of signal events for 500 fb$^{-1}$ for the 
Higgs-strahlung channel with dilepton final states $e^+e^- \rightarrow
Z H \rightarrow \ell^+ \ell^- X$, at different $\sqrt{s}$
and $M_H$ values. The last line is for the maximum $M_H$ value yielding more 
than 50 signal events in this final state. The numbers for $\sqrt{s}\!=\!1$ TeV
do not include the selection cuts and ISR corrections of \cite{TESLA}.} 
\label{tab:discovery}
\vspace*{-2mm}
\end{table}

\begin{figure}[ht!]
\begin{center}
\begin{tabular}{c c}
{{\epsfig{file=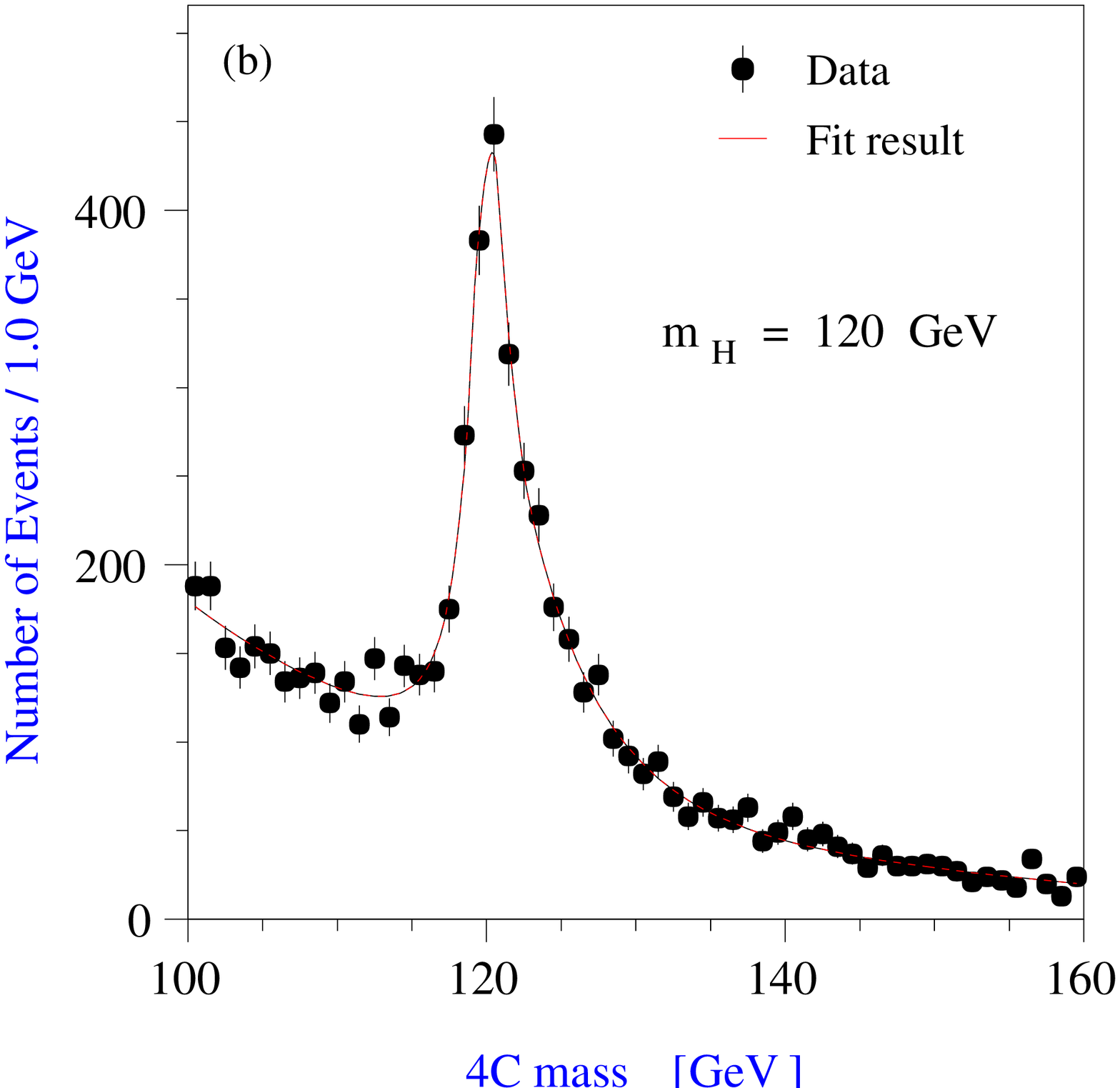,width=0.45\linewidth}}} &
{{\epsfig{file=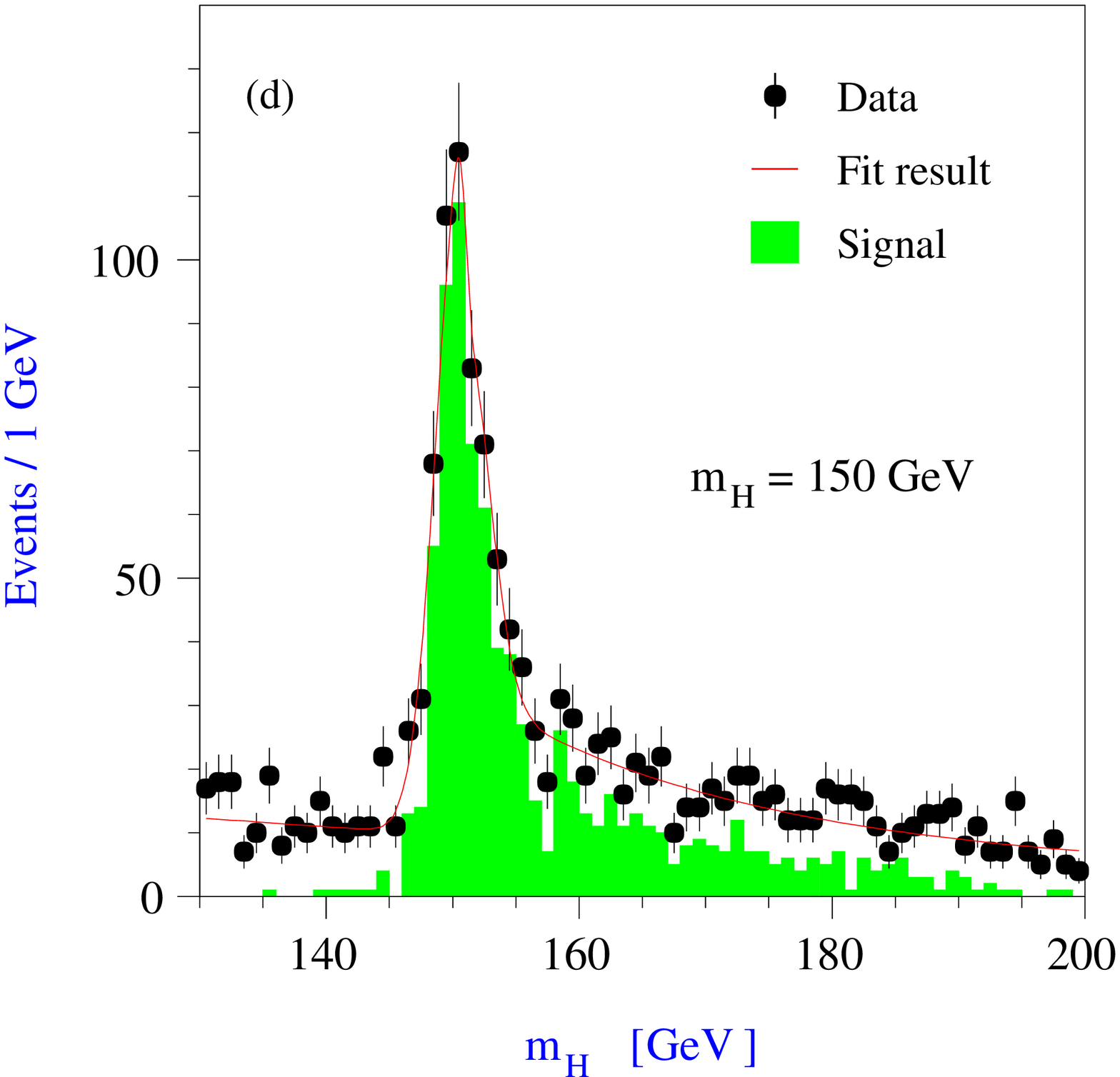,width=0.45\linewidth}}} \\
\end{tabular}
\end{center}
\vspace*{-2mm}
{\it Figure 4.31: The Higgs mass peak reconstructed in different channels with 
constrained fits for two values of $M_H$, an integrated luminosity of 500\,fb$^
{-1}$ and $\sqrt{s} =350$ GeV in $HZ \rightarrow q \bar q \ell^+ \ell^-$ at 
$M_H = 120~{GeV}$ (left) and $HZ \rightarrow W^+W^- \ell^+ \ell^-$ at $M_H = 
150~{GeV}$ (right); from Ref.~\cite{TESLA}.  }
\vspace*{-3mm}
\end{figure}

Moving to higher energies, Higgs bosons with masses up to $M_H \sim 400$ GeV
can be discovered in the strahlung  process at an energy of 500 GeV and with
a luminosity of 500 fb$^{-1}$. For even higher masses, one needs to increase
the c.m. energy of the collider and, as a rule of thumb, Higgs masses up to
$\sim 80$\% $\sqrt{s}$ can be probed. This means that a 1 TeV collider can 
probe the entire SM Higgs mass range, $M_H \lsim 700$ GeV. Table 4.4 shows the 
maximal Higgs mass values which can be reached at various c.m. energies by 
requiring at least 50 signal events in the process $\ee\to HZ \to H\ell \ell$.\s

The $WW$ fusion mechanism offers a complementary production channel.  In the
low Higgs mass range where the decay $H\to b\bar{b}$ is dominant, flavor tagging
plays an important role to suppress the 2--jet plus missing energy background. 
The $\ee \to H\bar{\nu}\nu \to b\bar{b}\bar{\nu}\nu$ final state can be
separated from the corresponding one in the Higgs--strahlung process $\ee \to
HZ \to b\bar{b}\bar{\nu}\nu$ \cite{WWH-sep} by exploiting their different
characteristics in the $\nu \bar{\nu}$ invariant mass which are measurable
through the missing mass distribution; see Fig.~4.32. The polarization of the
electron and positron beams, which allow to switch on and off the $WW$ fusion
contribution, can be very useful to control the systematic uncertainties. \s
 
For larger Higgs boson masses, when the decays $H \to WW^{(*)},ZZ^{(*)}$ are
dominant, the main backgrounds are $WW(Z)$ and $ZZ(Z)$ production which have
large cross sections at high energies and eventually $t\bar t$, but again, they
can be suppressed using kinematical constraints from the reconstruction of the
Higgs mass peak. \s For even higher masses, when the Higgs boson decays into
$t\bar{t}$ final states, the $\ee \to t\bar{t}$ and $ t\bar{t} \ee$ backgrounds
can be reduced to a manageable level by exploiting the characteristics of the
$\nu \bar \nu b\bar b  WW$ signature.

\begin{figure}[ht!]
\begin{center}
{{\epsfig{file=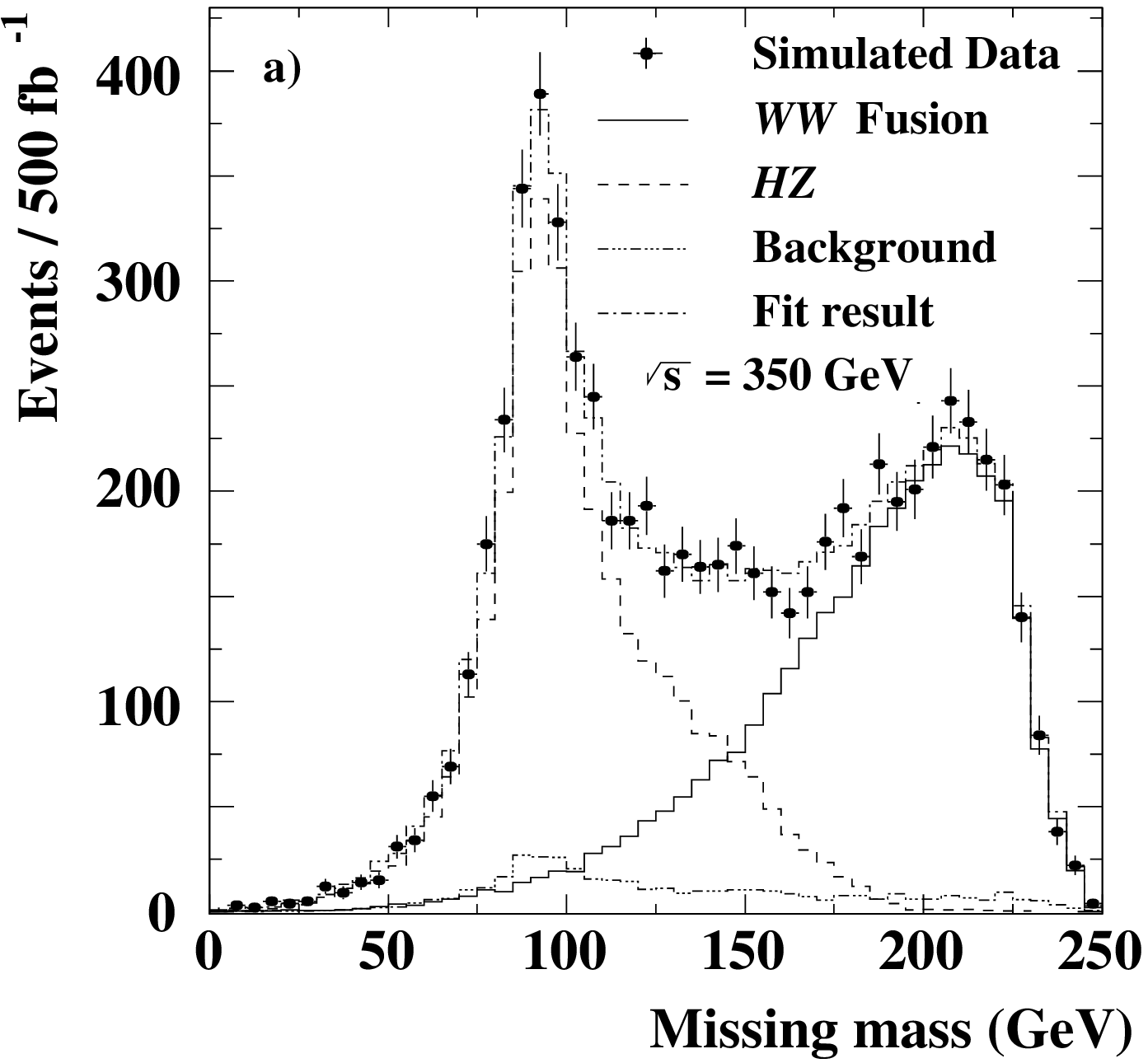,width=0.45\linewidth}}} 
{{\epsfig{file=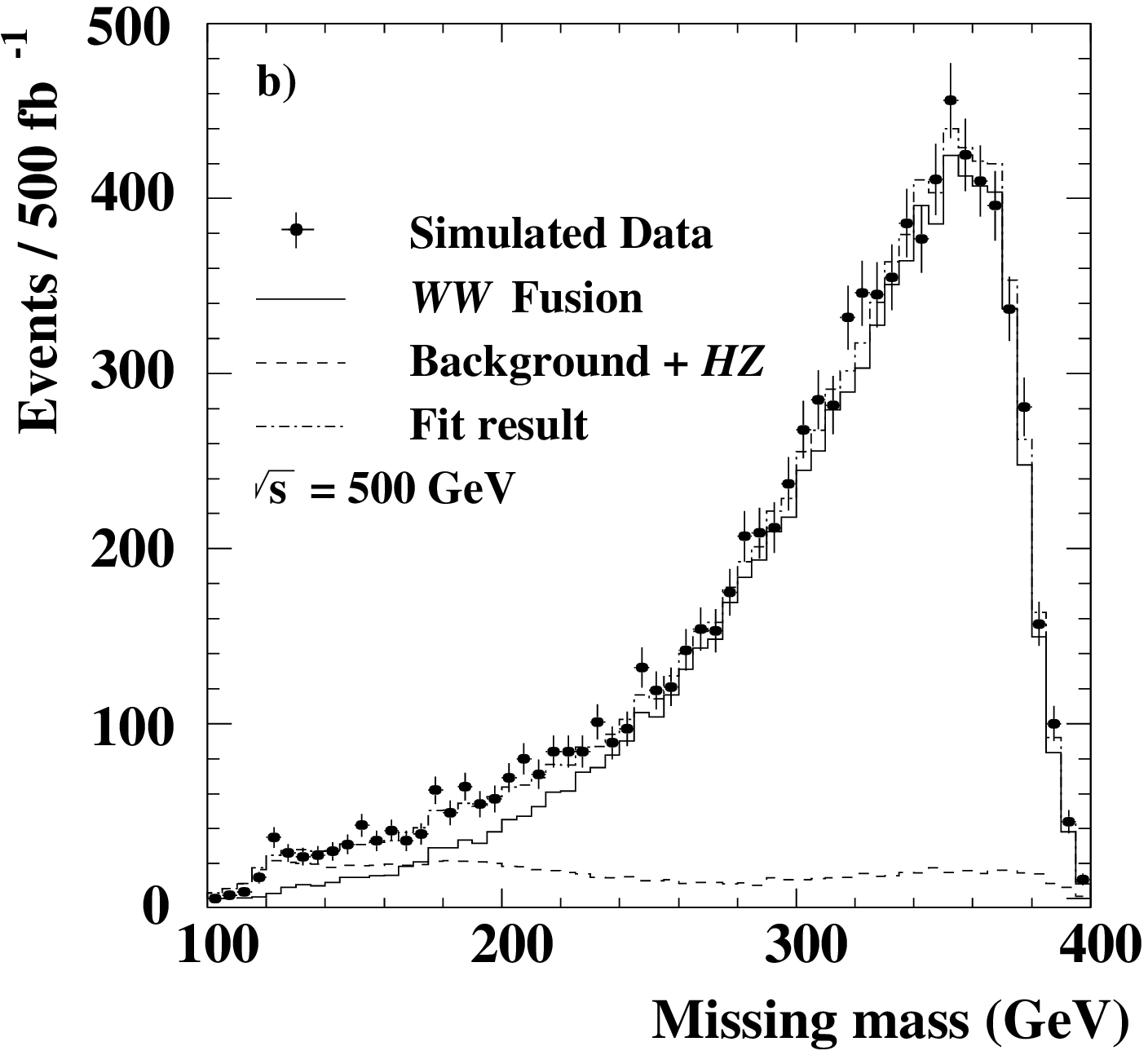,width=0.45\linewidth}}} 
\end{center}
\vspace*{-2mm}
{\it Figure.~4.32: 
The missing mass distribution in the $\nu \bar{\nu} b \bar b$ final state at 
$\sqrt{s} =350~{GeV}$ (left) and 500 GeV (right) for $M_H = 120~{GeV}$ 
in $WW$ fusion, Higgs--strahlung and the interference, as well as for the 
background. The $WW$ fusion contribution is measured from a fit to the shape 
of this distribution; from Ref.~\cite{TESLA}.  }
\vspace*{-5mm}
\end{figure}

Turning to the subleading processes,  we have seen that the $ZZ$ fusion
mechanism has a cross section that is one order of magnitude smaller than $WW$
fusion, a result of the smaller neutral couplings compared to the charged
current couplings. However, the full final state can be reconstructed in this
case.  At c.m. energies above 1 TeV, the cross section exceeds the one of the
Higgs strahlung process so that $\ee \to H \ee$ can be used instead for model
independent searches by tagging the $\ee$ pair and reconstructing the missing
mass \cite{LCWS}. \s

The associated production with top quarks has a very small cross
section at $\sqrt{s}=500$ GeV due to the phase space suppression but at
$\sqrt{s}=800$ GeV it can reach the level of a few femtobarn.  For $M_H \lsim
140$ GeV, the spectacular final state signal, $W^+W^- b\bar{b} b \bar{b}$, has
large backgrounds which can be suppressed by tagging the $b$--quarks and 
reconstructing the Higgs mass. The statistics are nevertheless very small 
and one has to resort to a neural network analysis to isolate the signal from
the remaining backgrounds. For higher Higgs masses, the final state 
$H t \bar t \to 4W b\bar b$ has also large backgrounds, which are nevertheless
manageable using again a neutral network. \s

The cross section for the double Higgs production in the strahlung process is
at the level of $\sim \frac12$ fb for a light Higgs at $\sqrt{s} =500$ GeV and 
is smaller at higher energies. The large backgrounds from four and six fermion
events can be suppressed for $M_H \lsim 140$ GeV by using the characteristic
signal of four $b$--quarks and a $Z$ boson, reconstructed in both leptonic an
hadronic final to increase the statistics, and using $b$--tagging. For higher
Higgs masses, the dominant final state is $Z+4W$.  In contrast, the cross
section for the $\ee \to \nu_e \bar{\nu}_e HH$ is extremely small at $\sqrt{s}
=500$ GeV but reaches the fb level at $\sqrt{s} =3$ TeV.  

\subsubsection{Precision measurements for a light Higgs boson}

Once the Higgs boson is found, it will be of great importance to explore all
its fundamental properties. This can be done at great details in the clean
environment of $\ee$ linear colliders: the Higgs boson mass, its spin  and
parity quantum numbers and its couplings to fermions, massive and massless
gauge bosons as well as its trilinear self--couplings can be measured with very
good accuracies. The measurements would allow to probe in all its facets  
the electroweak symmetry breaking mechanism. 

\vspace*{-2mm}
\subsubsection*{\underline{The Higgs boson mass}}
\vspace*{-1mm}

Many of the properties of the SM Higgs boson can be determined in a model
independent way by exploiting the recoil mass technique in the strahlung
process, $\ee \to HZ$.  The measurement of the recoil $\ee$ or $\mu^+ \mu^-$
mass in $\ee \ra ZH\ra H\ell \ell$, allows a very good determination of the
Higgs  mass \cite{meas-mass1,meas-mass2}. At $\sqrt{s}=350$ GeV and with a
luminosity of ${\cal L}= 500$ fb$^{-1}$, a precision of $ \Delta M_H \sim 70$
MeV can be reached for a Higgs mass of $M_H \sim 120$ GeV.  The precision can
be increased to $\Delta M_H \sim 40$ MeV by using in addition the hadronic
decays of the $Z$ boson which have more statistics \cite{meas-mass2}. 
Accuracies of the order of $\Delta M_H \sim 80$ MeV can also be reached for
$M_H$ values between 150 and 180 GeV when the Higgs boson decays mostly into
gauge bosons [see Ref.~\cite{LCWS1}, however].  The reconstructed Higgs mass
peak is shown in Fig.~4.33 at a 350 GeV collider in the two channels $HZ
\rightarrow b \bar b q \bar q$ for $M_H = 120$ GeV} and $HZ \rightarrow W^+W^-
q \bar q$ for $M_H=150$ GeV.  The obtained accuracy on $M_H$ is
a factor of two better than the one which could be obtained at the LHC.

\begin{figure}[ht!]
\begin{center}
\begin{tabular}{c c}
{{\epsfig{file=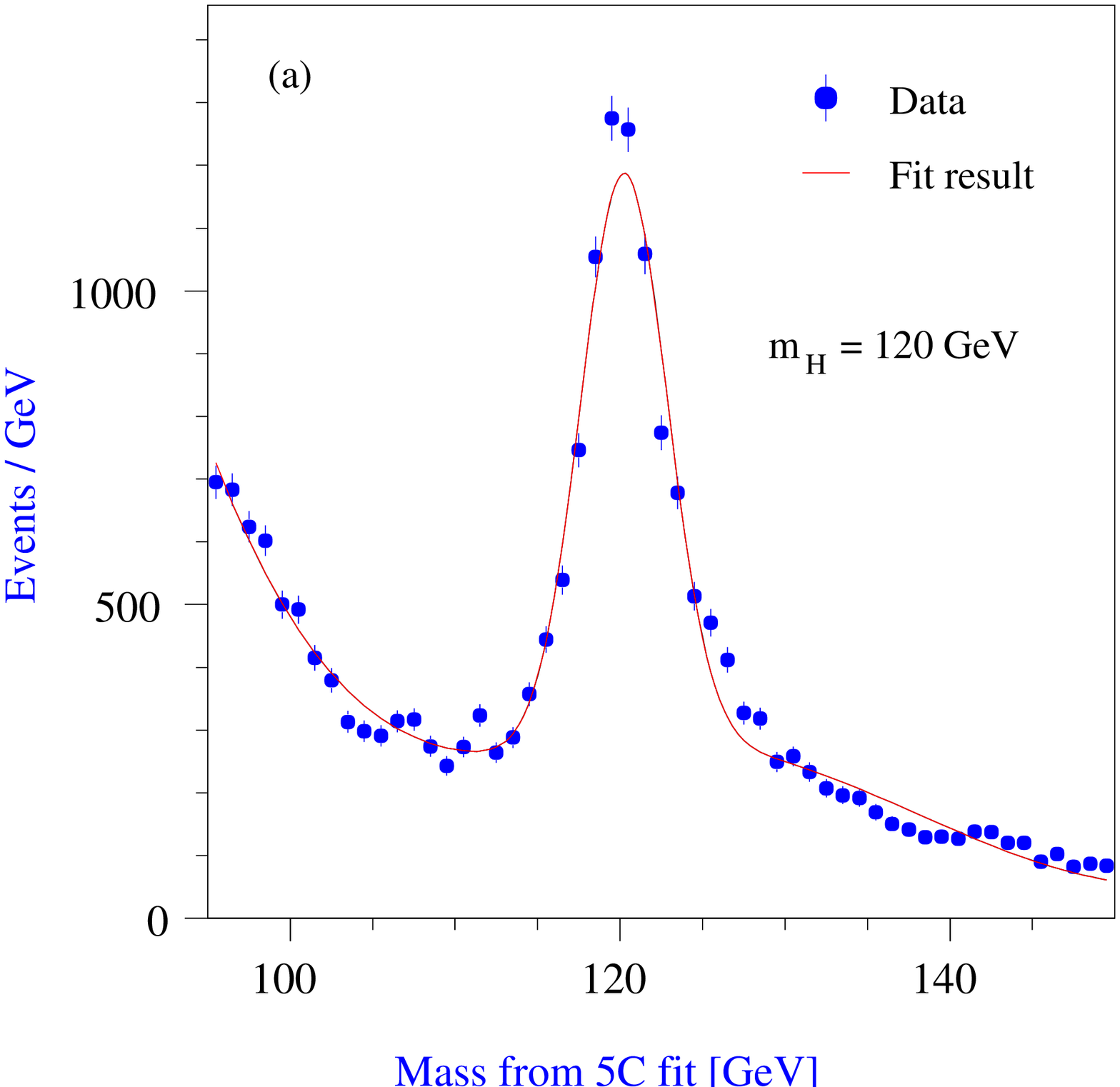,width=0.45\linewidth}}} &
{{\epsfig{file=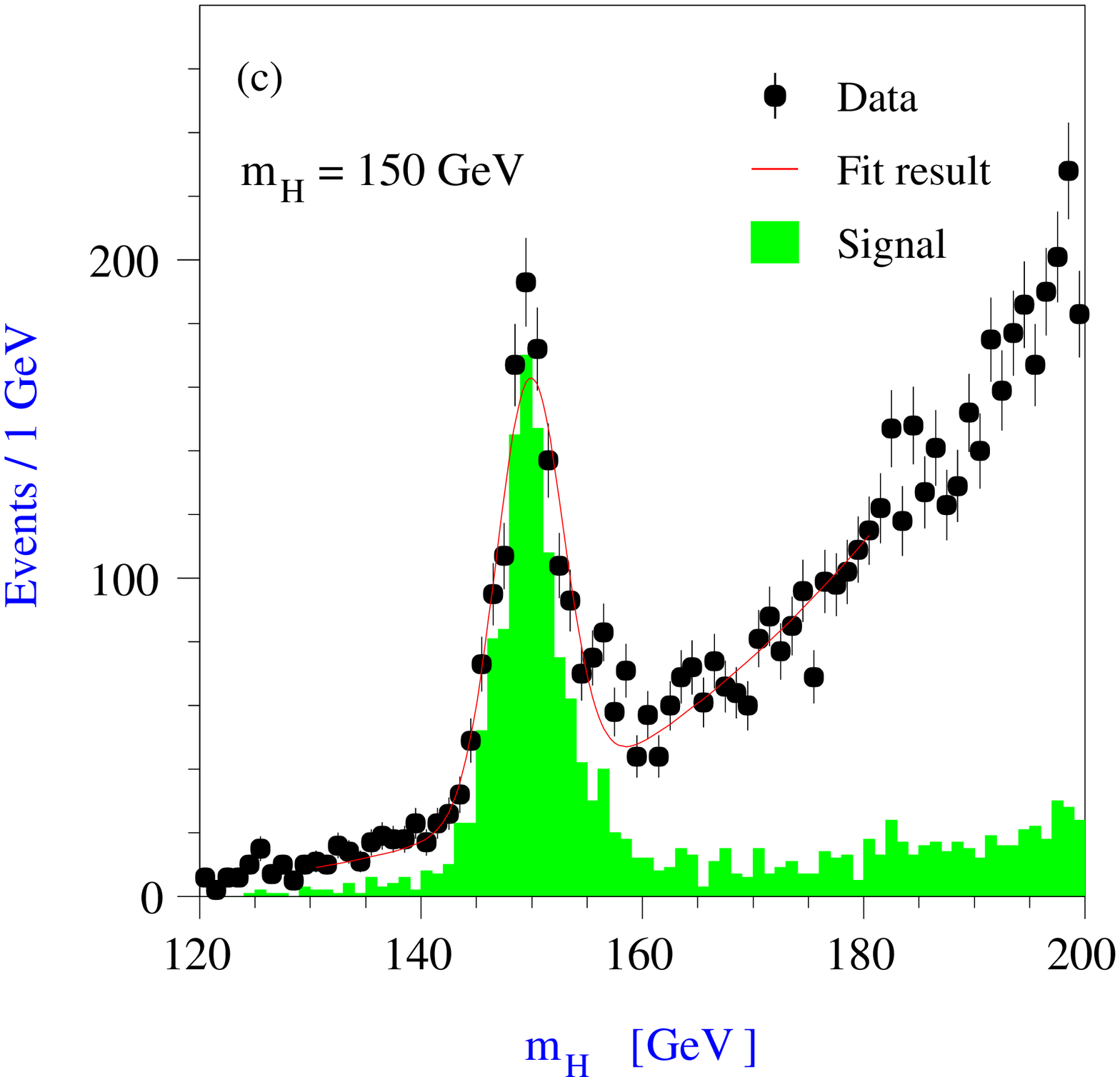,width=0.45\linewidth}}} \\
\end{tabular}
\end{center}
\vspace*{-2mm}
{\it Figure 4.33: The Higgs mass peak reconstructed in 
different channels with constrained fits for two values of $M_H$, an integrated
luminosity of 500\,fb$^{-1}$ and $\sqrt{s} =350~{GeV}$ in 
$HZ \rightarrow b \bar b q \bar q$ at $M_H = 120~{GeV}$ (left) and
$HZ \rightarrow W^+W^- q \bar q$ at $M_H = 150~{GeV}$ (right);
from Ref.~\cite{TESLA}.  }
\vspace*{-3mm}
\end{figure}

\subsubsection*{\underline{The Higgs spin and parity}}
\vspace*{-1mm}

The determination of the $J^{\rm P}=0^{+}$ quantum number of the SM Higgs boson
can also be performed in the strahlung process. As discussed in \S4.2.1, the
measurement of the rise of the cross section near threshold, $\sigma (\ee \to
HZ) \propto \lambda^{1/2}$, rules out $J^{\rm P}=0^{-}, 1^{-}, 2^{+}$ and higher
spin $3^\pm,  \cdots$, which rise with higher powers of the velocity 
$\lambda^{1/2}$. A threshold scan with a luminosity of 20 fb$^{-1}$ at three
center of mass energies is sufficient to distinguish the various behaviors;
Fig.~4.34.  The production of states with the two remaining $J^{\rm P}=1^+,
2^-$ quantum numbers can be ruled out using the angular correlations of the
final state $\ee \to HZ \to 4f$.\s

The angular distribution of the $Z/H$ bosons in the Higgs--strahlung process is
also sensitive to the spin--zero of the Higgs particle: at high--energies, the
$Z$ is longitudinally polarized and the distribution follows the $\sim \sin^2
\theta$ law which unambiguously characterizes the production of a $J^P=0^+$
particle, since in the case of a pseudoscalar Higgs boson, the angular
distribution would behave as $1 +\cos^2\theta$. Assuming that the Higgs
particle is a mixed CP--even and CP--odd state with $\eta$ parameterizing the
mixture, the angular distribution given by eq.~(\ref{HZ:angular}) can be
checked experimentally. This is shown in the right--hand side of Fig.~4.34,
where one can see that the parameter $\eta$ can be measured to a precision of
3--4 percent, which is the typical size of electroweak radiative corrections
which, in CP--conserving models, could generate the CP--odd component of the
$ZZ\Phi$ coupling. Note that the Higgs $J^{\rm PC}$ quantum numbers can also be
checked by looking at correlations in the production $\ee \ra HZ \ra 4f$ or in
the decay $H \ra WW^*, ZZ^* \ra 4f$ processes, just as in the LHC case but with
more accuracy at the LC since one can use the larger hadronic modes of the $W$
and $Z$ bosons.\s

\begin{figure}[h!]
\begin{center}  
\begin{minipage}{7cm}
\psfig{figure=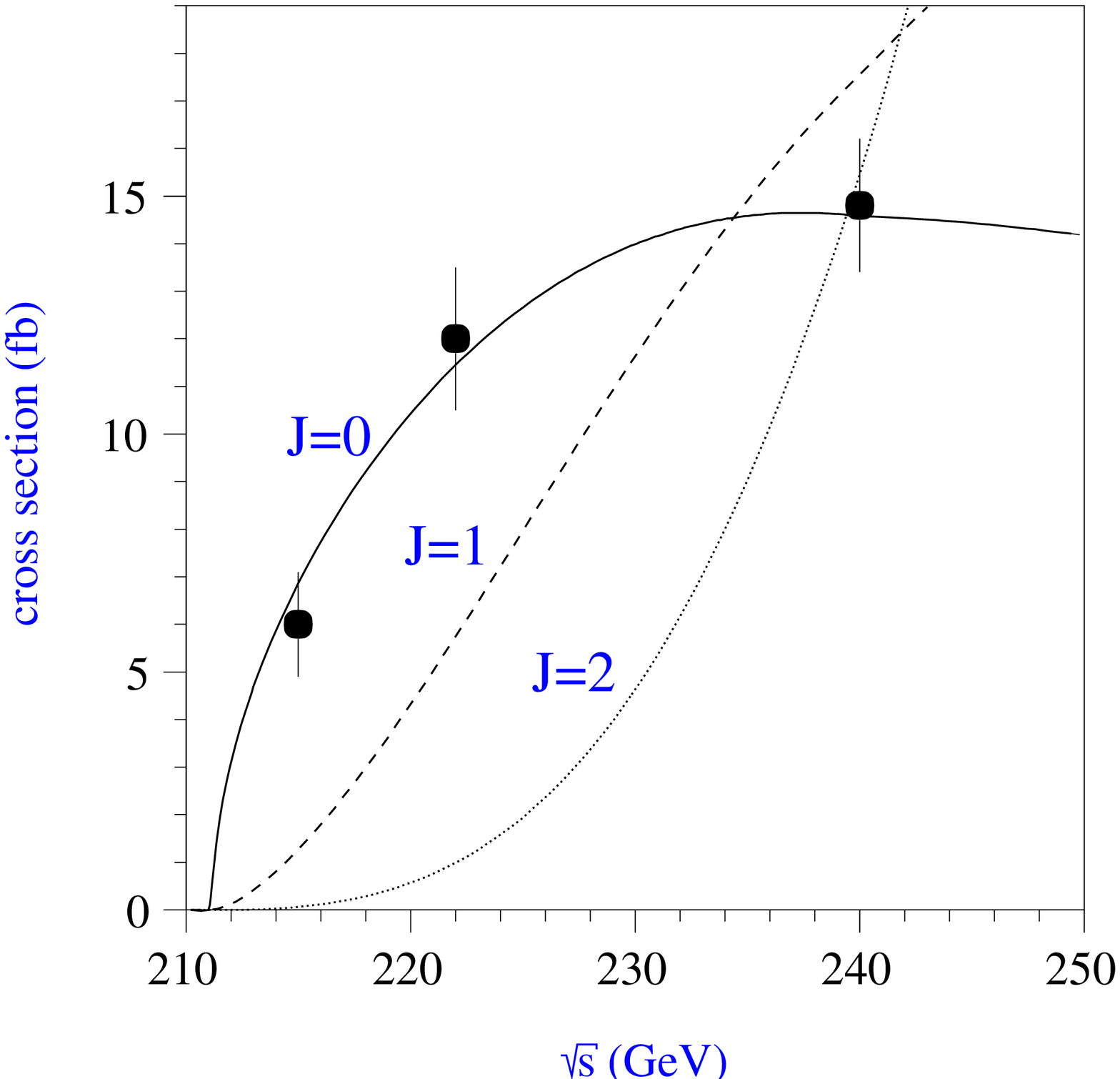,width=7.cm} 
\end{minipage}
\hspace*{10mm}
\begin{minipage}{7cm}
\vspace*{-13.mm}
\psfig{figure=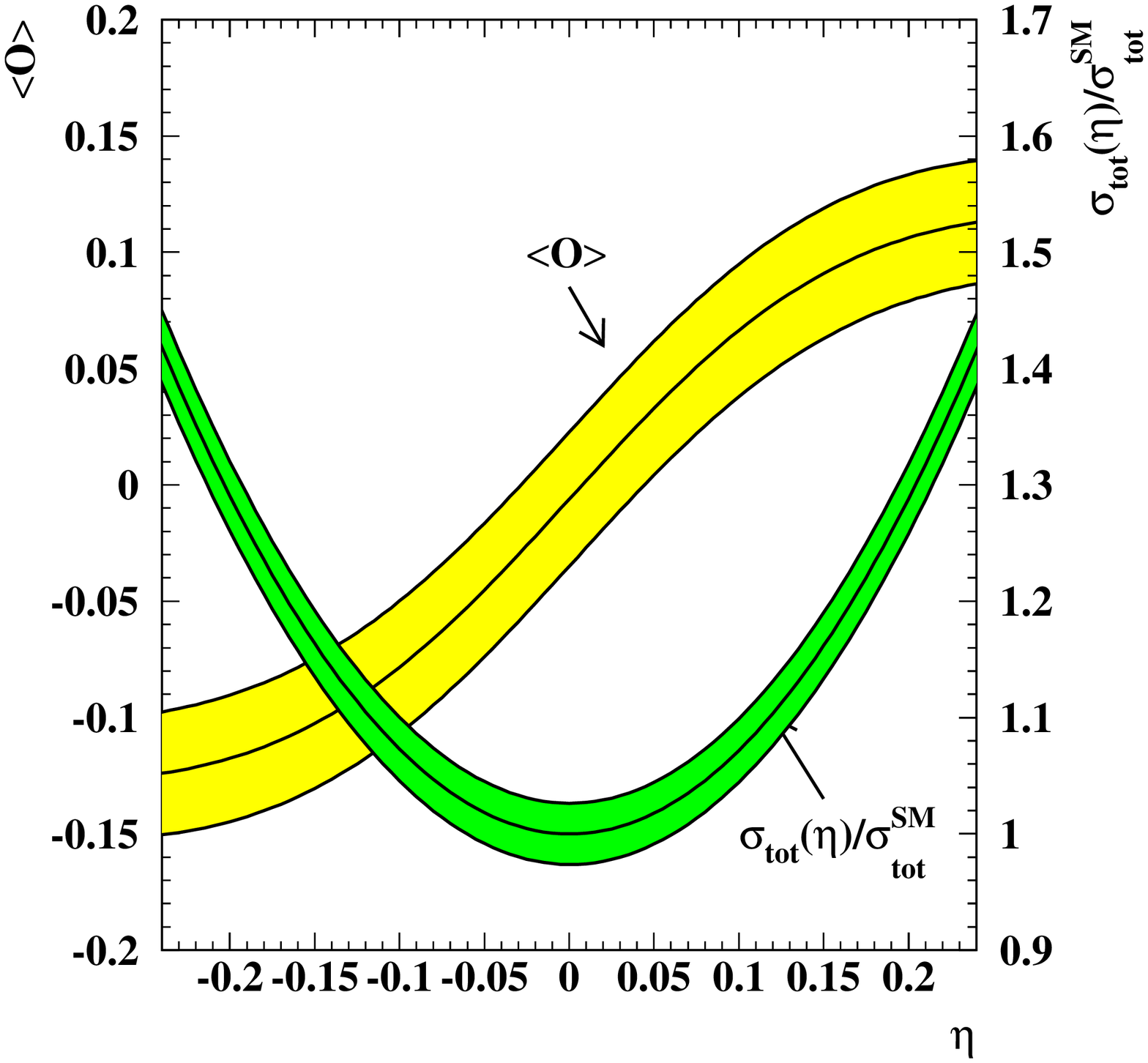,width=7.2cm}
\end{minipage}
\end{center}
\vspace*{-3.mm}
\nn {\it Figure 4.34: The $\ee \to ZH$ cross section energy dependence near
threshold for $M_H=120$ GeV for spin $0^+, 1^-$ and $2^+$ bosons 
\cite{meas-spin1} (left). The 
dependence of $\sigma(\ee\to HZ)$ and the observable $\langle O \rangle $ 
defined in eq.~(\ref{Oobservable}) on the parameter $\eta$ with the 
shaded bands showing the $1\sigma$ uncertainties at $\sqrt{s}=$ 350 GeV and 
500 fb$^{-1}$ \cite{meas-spin2} (right).}
\vspace*{-2.mm}
\end{figure}

The CP nature of the Higgs boson would be best tested in the couplings to 
fermions, where the scalar and pseudoscalar components might have comparable 
size. Such tests can be performed in the decay channel $H \ra \tau^+ \tau^-$ 
for $M_H \lsim 140$ GeV by studying the spin correlations between the final 
decay products of the two $\tau$ leptons \cite{CPHff1,CPHff2}. The 
acoplanarity angle between 
the decay planes of the two $\rho$ mesons produced from $\tau^+$ and $\tau^-$,
which can be reconstructed in the Higgs rest frame using the $\tau$ lifetime
information, is a very sensitive probe, allowing a discrimination between a 
CP--even and CP--odd state at the 95\% CL for $M_H=120$ GeV at the usual energy
and luminosity \cite{CPHff3}; using the additional information from the $\tau$ 
impact parameter significantly improves this determination. \s   

If the observed Higgs boson is a mixture of CP--even and CP--odd states, with
a coupling $g_{\Phi \tau \tau} =g_{H\tau \tau} (\cos\phi+ i\sin\phi \gamma_5)$
with $\phi=0$ in the SM Higgs case, the angular distributions in the $\tau^\pm 
\to \rho^\pm \nu$ decays allow to measure the mixing angle with an accuracy of $
\Delta \phi\sim 6^\circ$. This is shown in Fig.~4.35, which displays the 
distribution
of the acoplanarity angle $\varphi^*$ between the decay planes of the $\rho^+$
and $\rho^-$ in the rest frame of the pair, for several values of the mixing 
angle $\phi$,  as a result of a simulation for $\sqrt{s}=350$ GeV and  
${\cal L}=1$ ab$^{-1}$. \s 

\begin{figure}[!h]
\begin{center} 
{\epsfig{file=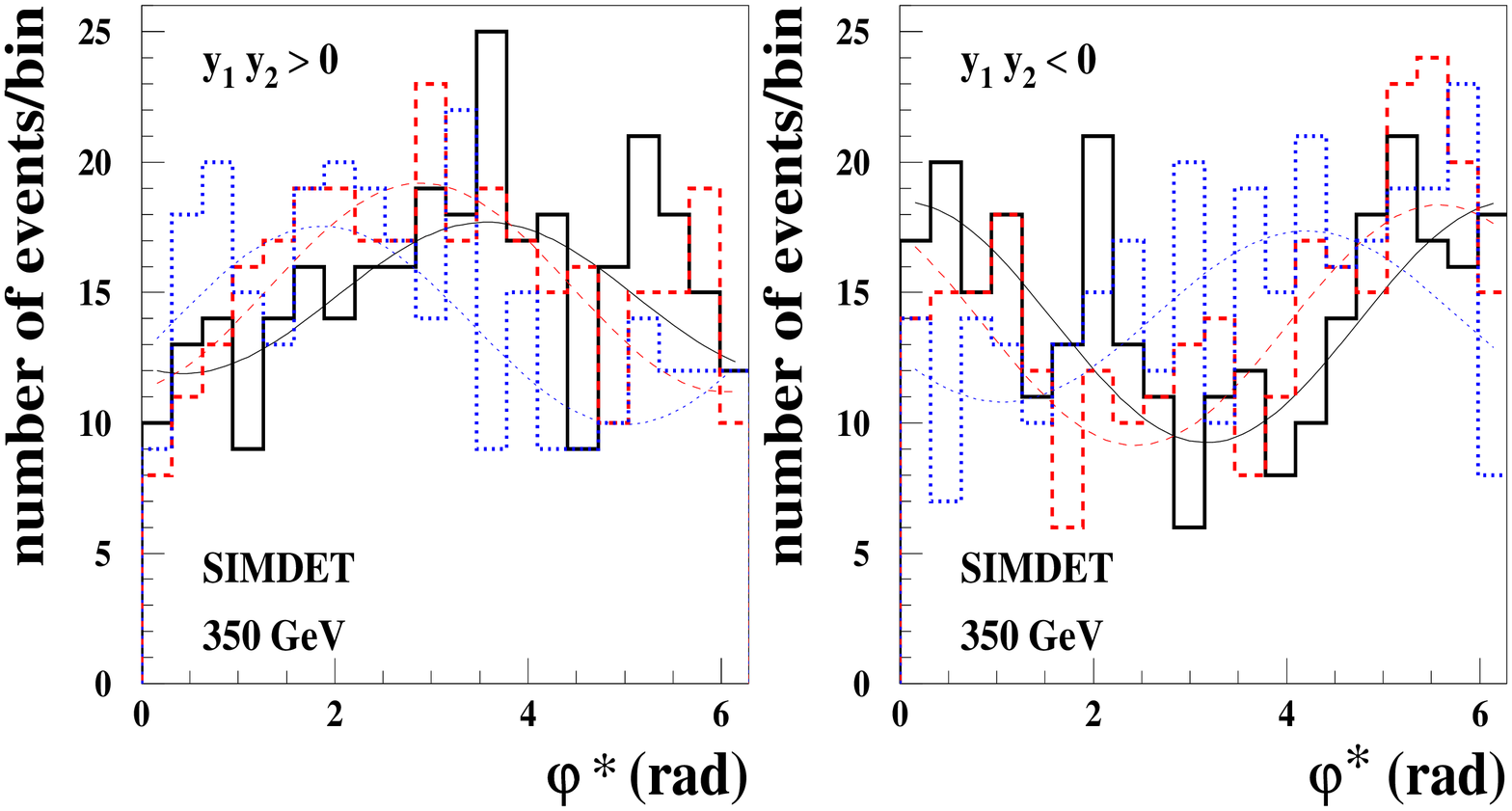,width=160mm,height=73mm}}
\end{center} 
\vspace*{-2.mm}
{\it Figure 4.35: Distribution of the reconstructed acoplanarity angle 
$\varphi^*$ for $\phi = 0$ (full histogram), $\phi = \pi/8$ (dashed) and 
$\phi = \pi/4$ (dotted) for $y_1 y_2 > 0$ (left) and $y_1 y_2 < 0$ (right) with 
$y_{1,2}=(E_{\pi^{\pm}}-E_{\pi^{0}}) / (E_{\pi^{\pm}}+E_{\pi^{0}})$; the lines 
indicate the results of the fits; from \cite{Desch}.}
\label{aco-shape-2ab-rec}
\end{figure}  

For heavier Higgs bosons, when the $H \to \tau^+ \tau^-$ becomes too small,
these studies cannot be performed anymore. A promising channel would be 
the decay $H \to t\bar{t}$ for $M_H >2m_t$, but no realistic simulation 
of the potential of this channel has been performed. 
Finally, and as discussed in \S4.3.2, the differential cross section in 
associated production with top quarks, $\ee \to t\bar t H$, is sensitive
to the CP nature of the Higgs boson, though no analysis has been performed 
to verify at which extent this information can be experimentally extracted. 

\newpage 
\subsubsection*{\underline{The Higgs couplings to gauge bosons}}
\vspace*{-1mm}

The fundamental prediction that the Higgs couplings to $ZZ/WW$ bosons are
proportional to the masses of these particles can be easily verified 
experimentally since  these couplings can be directly determined by measuring 
the production cross sections in the bremsstrahlung and the fusion processes.  
$\sigma(\ee \ra HZ \to H \ell^+ \ell^-)$ can be measured by analyzing the recoil
mass against the $Z$ boson and provides a determination of the $g_{HZZ}$ 
couplings independently of the decay modes of the Higgs boson. Adding the
two lepton channels, one obtains a statistical accuracy of less than 3\%
at $\sqrt{s}\sim 350$ GeV and with $\int {\cal L}=500$ fb$^{-1}$ 
\cite{meas-mass1}. \s
       
The coupling $g_{HWW}$ for $M_H\lsim 2M_W$ can determined from the measurement
of the total cross section of the process $\ee \to W^* W^* \nu \bar \nu \to
H\nu \bar{\nu}$ which, as discussed previously, can be efficiently separated
from the $\ee \ra HZ \to H \nu \bar \nu$ channel and from the backgrounds, see
Fig.~4.32.  A precision of also less than 3\% can be achieved for $M_H=120$
GeV, but at a slightly higher energy, $\sqrt{s}\sim 500$ GeV, where the
production rate is larger \cite{meas-HWW}. The precision becomes worse for
increasing Higgs mass as a result of the falling cross section.\s

The accuracies which can be achieved are shown in Tab.~4.5 for three Higgs 
masses
and the precision on the Higgs couplings is half of these errors, since the
cross sections scale as $g_{HVV}^2$. Thus, a measurement of the Higgs couplings
to gauge bosons can be performed at the statistical level of 1 to 2\% and would
allow to probe the quantum corrections.  

\begin{table}[hbt]
\renewcommand{\arraystretch}{1.2}
\begin{center}
\begin{tabular}{|c|c|c|c|}
\hline
Channel & $M_H=120$ GeV & $M_H=140$ GeV & $M_H=160$ GeV \\ \hline
$\sigma (\ee \to HZ)$       & 2.5\% & 2.7\% & 3.0 \%  \\ 
$\sigma(\ee \to H \nu \bar \nu)$ & 2.8\% & 3.7\% & 13 \%  \\ \hline 
\end{tabular}
\end{center}
\vspace{.1cm}
{\it Table 4.5: Relative precision in the determination of the SM Higgs cross 
sections for 120 GeV $\leq M_H \leq 160$ GeV with ${\cal L}=500$ fb$^{-1}$ 
at $\sqrt{s} = 350$ and 500 GeV; from Ref.~\cite{TESLA}.}
\vspace{-.5cm}
\end{table}

\subsubsection*{\underline{The Higgs decay branching ratios}}
\vspace*{-1mm}

The measurement of the branching ratios of the Higgs boson
\cite{BRs-early-studies,BRs-NLC,LCWS2,Brient,Barklow,ee-pp-ex,ee-pp-ex-G,ee-pZ-ex,ee-mu-ex
, ee-bb-ex,ee-Hinv-ex} is of utmost importance. For Higgs masses below $M_H
\lsim 150$ GeV a large variety of branching ratios can be measured at the
linear collider, since the $b\bar b, c\bar c$ and $gg$ final states can be very
efficiently disentangled by means of vertex detectors \cite{ZVTOP}. The
$b\bar{b}, c\bar{c}$  and $\tau^+ \tau^-$ fractions allow to measure the
relative couplings of the Higgs boson to these fermions and to check the
prediction of the Higgs mechanism that they are indeed proportional to fermion
masses. In particular, BR$(H \ra \tau^+ \tau^-) \sim m_{\tau}^2/3\bar{m}_b^2$
allows such a test in a rather clean way.  The gluonic branching ratio is
indirectly sensitive to the $t\bar{t}H$ Yukawa coupling and would probe the
existence of new strongly interacting particles that couple to the Higgs and
which are too heavy to be produced directly.  The branching ratio of the loop
induced $\gamma \gamma$ and $Z\gamma$ Higgs decays are also very sensitive to
new heavy particles and their measurement is thus very important. The branching
ratio of the Higgs decays into $W$ bosons starts to be significant for $M_H
\gsim 120$ GeV and allows to measure again the $HWW$ coupling in an independent
way. In the mass range 120 GeV $\lsim M_H \lsim 180$  GeV, the $H \to ZZ^*$
fraction is too small to be precisely measured, but for higher masses it is
accessible and allows an additional determination of the $HZZ$ coupling. \s

There are two methods to measure the Higgs branching ratios: first by measuring
the event rate in the Higgs--strahlung process for a given final state 
configuration and then dividing by the total cross section which is measured
from the recoil mass, and second, by selecting a sample of unbiased events in 
the $\ee \to HZ$ recoil mass peak and determining the fraction of events 
corresponding to a given final state decay. The first case, which is called the 
indirect method, has been used to study the Higgs branching ratios for the
TESLA TDR \cite{TESLA,LCWS2} while the second one, called the direct method,
appeared only recently \cite{Brient}. Both methods give rather similar results
but, since they are almost independent, these results may be combined to 
provide a significant improvement of the expected accuracies. \s

\centerline{\psfig{figure=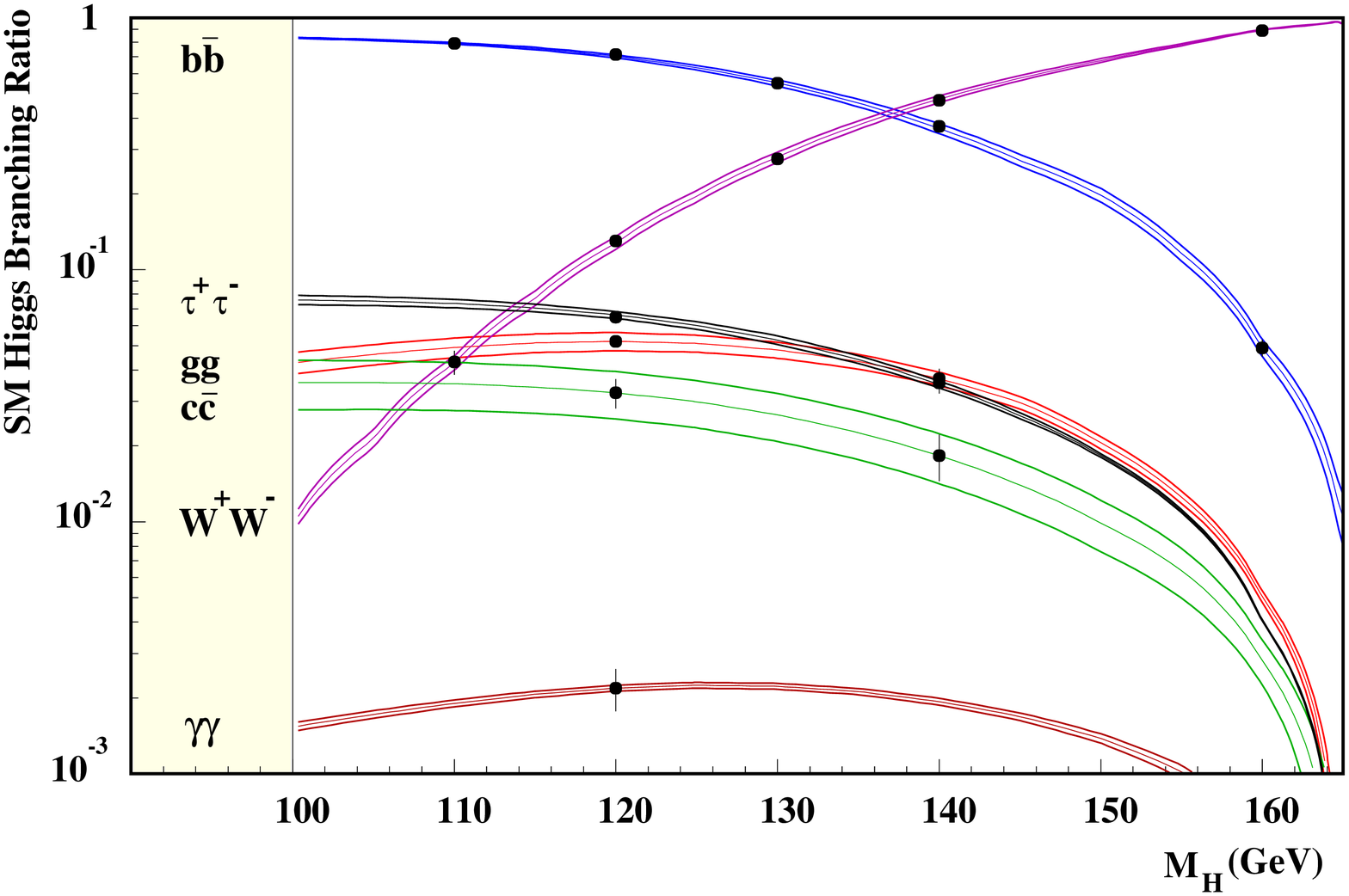,width=14.cm}}
\vspace*{1.mm}
\nn {\it Figure 4.36: The theoretical predictions [with the bands due to the 
uncertainties in the measurement of the quark masses and $\alpha_s$] 
and the experimental accuracy [the points with error bars] for the SM Higgs 
branching ratios at $\sqrt{s}=350$ GeV with 500 fb$^{-1}$; from 
Ref.~\cite{TESLA}.}
\vspace*{2.mm}

The expected accuracies on the Higgs branching fractions are shown in Fig.~4.36
and in Table 4.6 [the low--energy (LE) numbers at the left] mostly at 
$\sqrt{s}=350$ GeV and with 500 fb$^{-1}$ integrated luminosity for $M_H \leq 
160$ GeV. The $b\bar{b}, c\bar{c}, \tau^+ \tau^-,gg$ and $WW$ branching ratios 
of the Higgs boson can be measured with a very good accuracy. For the mass 
value $M_H=120$ GeV and using the indirect method, one obtains an accuracy of,
respectively, 2.4\%, 8.3\%, 5\%, 5.5\% and 5.1\%. When combined with the direct 
method measurements labeled LE(D), the errors decrease quite significantly. The
uncertainties in the measurements become larger when approaching the $WW$ 
threshold: at $M_H \sim 160$ GeV, only the $b\bar b, WW$ and $ZZ$ fractions are 
accessible, with still a poor accuracy in the latter case.
For $M_H \sim 200$ GeV, a higher energy $\sqrt{s}=500$ GeV is needed to
compensate for the falling cross section, and the precision is good only
for the $WW$ and $ZZ$ channels. For the $H\to b\bar b$ decays, an energy of 
800 GeV and 1 ab$^{-1}$ data are required to reach the quoted precision of
17\%. \s

\begin{table*}[hbt]
\renewcommand{\arraystretch}{1.2}
\begin{center}
\begin{tabular}{|c|ccc|cc|cc|cc|}
\hline
$M_H$ [GeV] &  \multicolumn{3}{c|}{120}&  \multicolumn{2}{c|}{140}& 
 \multicolumn{2}{c|}{160}     &  \multicolumn{2}{c|}{200} \\ \hline
Decay mode  & \multicolumn{9}{c|}{Relative Precision (\%)} \\ 
            &LE &LE(D)&  HE & LE  & HE  & LE  & HE   &  LE  & HE\\ \hline
$b\bar b$   &2.4 &1.5 & 1.6 & 2.6 & 1.8 & 6.5 & 2.0  & 17.  & 9.0 \\
$c\bar c$   &8.3& 5.8 & --  & 19. & --  &     &      &      &    \\
$\tau\tau$  &5.0&4.1 & --  & 8.0 & --  &     &      &      &     \\ \hline
$gg$        &5.5&3.6 & 2.3 & 14.0& 3.5 &  -- & 14.6 &      &      \\
$WW$        &5.1& 2.7 & 2.0 & 2.5 & 1.8 & 2.1 & 1.0  &  3.5 &2.5    \\
$ZZ$        &  &   &     &     &     & 16.9& --   &  9.9 & --    \\ \hline 
$\gamma\gamma$&23& 21.& 5.4 & --  & 6.2 &  -- & 24   &      &       \\
$Z\gamma$   &  &   &     &27.  & --  &     &      &      &       \\
$\mu\mu$    &  30 & & -- &     &     &     &       &      & \\ \hline
\end{tabular}
\end{center}
\vspace*{-1mm}
{\it Table 4.6: Summary of expected precisions on Higgs boson branching ratios 
from existing studies within the ECFA/DESY workshops (LE) \cite{Desch} 
obtained for 500 fb$^{-1}$ at $\sqrt{s}=350$ GeV, except for $M_H=200$ GeV 
where $BR(WW)$ and BR$(ZZ)$ are measured at $\sqrt{s}=500$ GeV and BR$(bb)$ 
which uses 1 ab$^{-1}$ at 800 GeV, as in the case of BR($\mu \mu$).
LE stands for the measurement with the indirect method, while LE(D) is for the 
combined measurements of the direct and indirect methods \cite{Brient}. HE is 
the combination of the measurements from the direct method with the NLC 
results obtained for 1 ab$^{-1}$ at $\sqrt{s}=1$ TeV \cite{Barklow}.}
\vspace*{-2mm}
\end{table*}

In the low Higgs mass range, even the rare decays into $\gamma \gamma$ and $Z
\gamma$ final states can be measured with an accuracy of approximately 5 to 20\%
\cite{ee-pp-ex,ee-pZ-ex,Barklow}. The very rare decay into muon pairs is also
measurable, though with a rather poor accuracy, by going to high energies and
taking advantage of the enhanced production rates in $\ee \to H\nu \bar \nu$
\cite{ee-mu-ex}. A luminosity of 1 ab$^{-1}$ is necessary to probe all these
rare decay modes of the Higgs boson.\s

Finally, invisible Higgs decays can also be probed with a 
very good accuracy, thanks to the missing mass technique. One can also look 
directly for the characteristic signature of missing energy and momentum.
Recent studies show that in the range 120 GeV $\lsim M_H \lsim$ 160 GeV, 
an accuracy of $\sim 10\%$ can be obtained on a invisible decay with a branching
ratio of 5\% and a $5\sigma$ signal can be seen for a branching ratio as low 
as 2\% \cite{ee-Hinv-ex}.\s

Moving to higher energies, $\sqrt{s}=1$ TeV, the larger rate for 
the $WW$ fusion process helps improving the accuracy on the main decay 
branching ratios and even search for rare decays 
[as it was the case for $H\to \mu^+ \mu^-$]. In the right--hand side of 
Table 4.6, the HE numbers stand for measurements performed at this energy 
and with 1 ab$^{-1}$ data, when combined with the respective 
measurements at low energies \cite{Barklow}. As can be seen the accuracy on 
some decay branching ratios, in particular BR$(H\to b\bar{b}, \gamma \gamma)$, 
can be significantly improved. 

\subsubsection*{\underline{The Higgs total decay width}}
\vspace*{-1mm}

The total decay width of the Higgs boson, for $M_H \gsim 200$ GeV, is large 
enough to be accessible directly from the reconstruction of the Higgs  boson 
lineshape. For smaller Higgs masses, the total decay is less than 1 GeV and it
cannot be resolved experimentally.  However, it can be determined indirectly by
exploiting the relation between the total and partial decay widths for some
given final states. For instance, in the decay $H\to WW^*$, the total decay
width is given by $\Gamma_H = \Gamma(H \to WW^*)/{\rm BR}(H \to WW^*)$.
One can then combine the direct measurement of the $H \to WW^*$ branching ratio
discussed above and use the information on the $HWW$ coupling from the $WW$
fusion cross section to determine the partial decay width $\Gamma (H\to WW^*)$. 
Alternatively, on can exploit the measurement of the $HZZ$ coupling from the 
production cross section of the Higgs--strahlung process, since the mass reach
is higher than in $WW$ fusion,  and assume SU(2) invariance to relate the two 
couplings, $g_{HWW}/g_{HZZ} = 1/\cos\theta_W$. The accuracy on the total decay
width measurement follows then from that of the $WW$ branching ratio and the
$g_{HWW}$ coupling. 

\begin{table}[hbt]
\renewcommand{\arraystretch}{1.2}
\begin{center}
\begin{tabular}{|c|c|c|c|}
\hline
Channel & $M_H=120$ GeV & $M_H=140$ GeV & $M_H=160$ GeV \\ \hline
$g_{HWW}$ from $\sigma(\ee \to H \nu \nu)$& 6.1\% & 4.5\% & 13.4 \%  \\ 
$g_{HWW}$ from $\sigma(\ee \to H Z)      $ & 5.6\% & 3.7\% & 3.6 \%  \\ 
\hline \hline
BR$(WW)$ at $\sqrt{s}=1$ TeV & 3.4\% & 3.6\% & 2.0 \%  \\ \hline
\end{tabular}
\end{center}
{\it Table 4.7: Relative precision in the determination of the SM Higgs decay 
width with $\int {\cal L}=500$ fb$^{-1}$ at $\sqrt{s} = 350$ GeV using
the two methods described in the text \cite{TESLA}. The last line shows the 
improvement which can be obtained when combining these results with those
which can be extracted from measurements at $\sqrt{s}\sim 1$ TeV with $\int 
{\cal L}=1$ ab$^{-1}$ \cite{Barklow}.}
\vspace{-.3cm}
\end{table}

As shown in Tab.~4.7, in the range 120 GeV $\lsim M_H \lsim$ 160 GeV, an 
accuracy  ranging from 4\% to 13\% can be achieved on $\Gamma_H$ if the
$HWW$ coupling is measured in the fusion process. This accuracy greatly
improves for higher $M_H$ values by assuming SU(2) universality which
allows to use the $HWW$ coupling as derived from the strahlung process.
If in addition a measurement of BR($H\to WW)$ is performed at higher energies
and combined with the previous values, the accuracy on the total Higgs width 
will greatly improve for high masses.\s

Note that the same technique would allow to extract the total Higgs decay 
width using the $\gamma \gamma$ decays of the Higgs boson together with 
the cross section from $\gamma \gamma \to H \to b\bar b$ as measured at a 
photon collider. This is particularly true since the measurement of BR($\gamma 
\gamma)$ at $\sqrt s \sim 1$ TeV is rather precise, allowing the total width 
to be determined with an accuracy of $\sim 5\%$ with this method for
$M_H=120$--140 GeV independently of the $WW$ measurement.

\vspace*{-.2cm}
\subsubsection*{\underline{The Higgs Yukawa coupling to top quarks}}

The Higgs Yukawa coupling to top quarks, which is the largest coupling in the
electroweak SM, is directly accessible in the process where the Higgs is
radiated off the top quarks, $\ee \ra t\bar{t}H$, since the contribution from
the diagram where the Higgs boson is radiated from the $Z$ line,  $\ee \to HZ
\to Ht\bar{t}$, is very small; Fig.~4.17. Because of the limited phase space,
this measurement can only be performed at high energies $\sqrt{s} \gsim 500$
GeV.  For $M_H \lsim 140$ GeV, the Yukawa coupling  can be measured in the
channel $W Wb\bar{b}b\bar{b}$ with the $W$ bosons decaying both leptonically
and hadronically to increase the statistics; $b$--tagging is essential in this
mass range \cite{Strasbourg,ee-ttH-exp}. For higher Higgs masses,  $M_H
\gsim 140$ GeV, the channels with $b\bar b+4W$ have to be considered, with
again, at least two $W$ bosons decaying hadronically, leading to 2 leptons plus
6 jets and one lepton plus 8 jets, respectively. The complexity of the final
states and the small statistics requires a neural network analysis
\cite{Strasbourg}.

\centerline{\psfig{figure=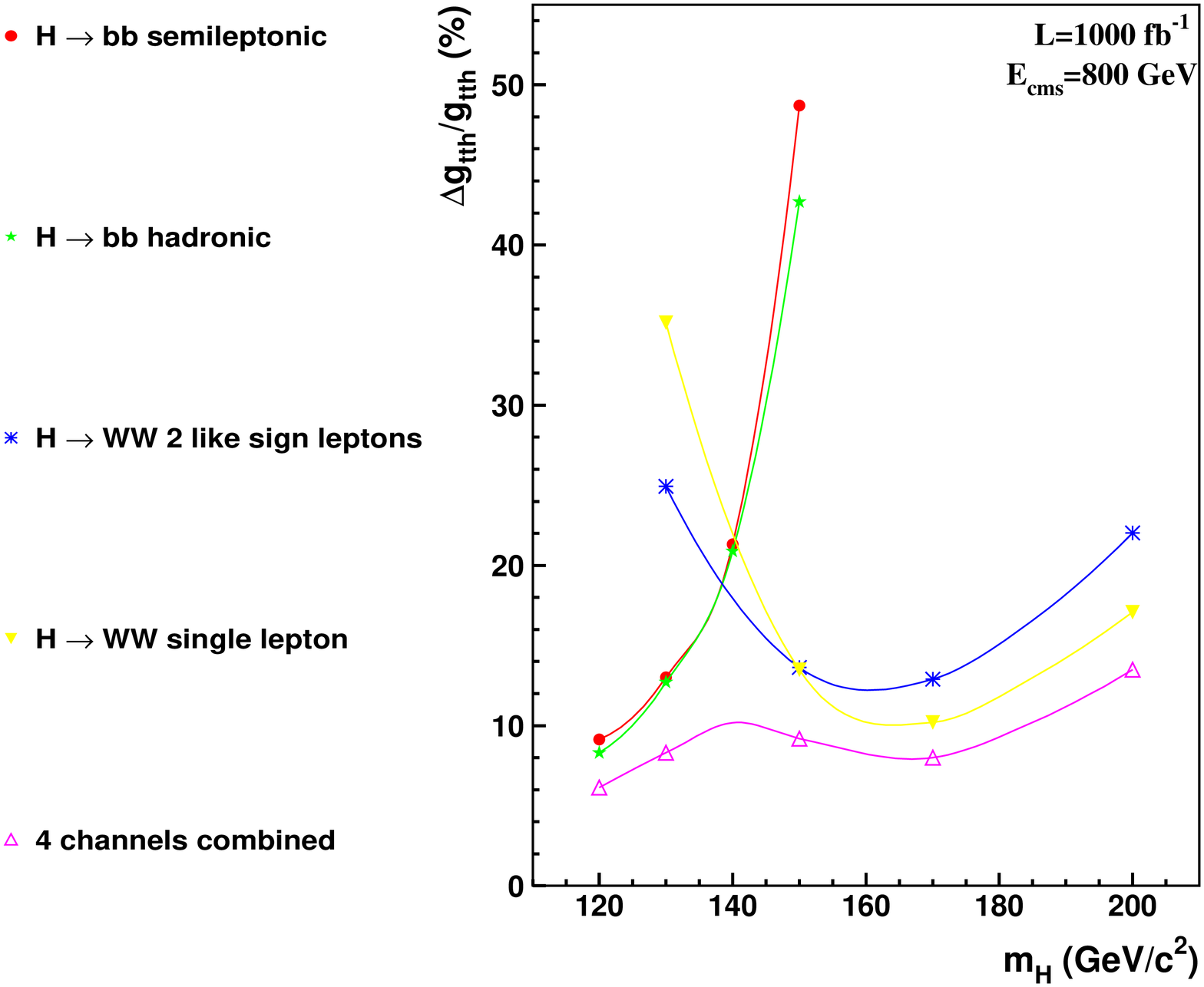,width=15cm,height=7.4cm}}
\vspace*{-.cm}
\nn {\it Figure 4.37: Expected accuracies for the measurement of the $Ht\bar t$
coupling as a function of $M_H$ in the process $\ee \to t \bar t H$ for 
$\sqrt{s}= 800$ GeV and 1 ab$^{-1}$ in various decay channels. A 5\% 
systematical error is assumed on the normalization of the background; 
from Ref.~\cite{Strasbourg}.} \s

The expected accuracies on the $H t \bar t$ Yukawa coupling are shown in
Fig.~4.37 from Ref.~\cite{Strasbourg} as a function of the Higgs mass, for
$\sqrt s = 800$ GeV and a luminosity of 1 ab$^{-1}$. Assuming a 5\%
systematical uncertainty on the normalization of the background, accuracies on
the $Ht\bar t$ Yukawa coupling of the order of 5\% can be achieved for Higgs
masses in the low range. A 10\% measurement is possible up to Higgs masses of
the order of 200 GeV. \s 

For large masses, $M_H \gsim 350$ GeV, the $Ht \bar{t}$ coupling can be derived
by measuring the $H \ra t\bar{t}$ branching ratio with the Higgs boson produced
in the strahlung and $WW$ fusion processes \cite{Hagiwara,Hagiwara0}. A 
detailed 
simulation, performed for the TESLA TDR in the latter channel, shows that
once the $t\bar t$ and $\ee t\bar t$ backgrounds are removed by requiring four 
light jets and two $b$ quarks in the final state in addition to the missing 
energy, an accuracy of the order of 5\% (12\%) for a Higgs mass of 400 (500) 
GeV can be achieved on the top quark Yukawa coupling, again at $\sqrt{s}= 
800$ GeV and with ${\cal L} \sim 1$  ab$^{-1}$ data \cite{Httexp}. 

\subsubsection*{\underline{The trilinear Higgs coupling}}

The measurement of the trilinear Higgs self--coupling, which  is the
first non--trivial probe of the Higgs potential and, probably, the most decisive
test of the electroweak symmetry breaking mechanism, is possible in the double
Higgs--strahlung process. For Higgs masses in the range 120 GeV $\lsim M_H 
\lsim 140$ GeV, one has to rely on the $b\bar b$ decays and the cross section 
in the $\ee \to HHZ \to \bar{b}b \bar{b}b+\ell^+ \ell^-$ or $q\bar{q}$ 
channels is rather small, see Fig.~4.20, while the four and six fermion 
background are comparatively very large. \s

The excellent $b$--tagging efficiencies and the energy flow which can be
achieved at future linear colliders makes it possible to overcome the
formidable challenge of suppressing the backgrounds, while retaining a
significant portion of the signal.  Accuracies of about 20\% can be obtained
on the measurement of the $\ee \to HHZ$ cross section in the mass range below 
140 GeV; see the left--hand side of Fig.~4.38. Neural network analyses allow 
to improve the accuracy of the measurement from 17\% to 13\% at a Higgs mass 
$M_H=120$ GeV and to obtain a 6$\sigma$ significance for the signal
\cite{Clermont-Ferrand}; see also Ref.~\cite{HHH-baur}. \s

Since the sensitivity of the process $\ee \to HHZ$ to the trilinear Higgs 
coupling is diluted by the additional contributions originating from diagrams 
where the Higgs boson is emitted from the $Z$ boson lines, only an accuracy of 
$\Delta \lambda_{HHH} \sim 22\%$  can be obtained for $M_H=120$ GeV at an 
energy of  $\sqrt{s}\sim 500$ GeV with an integrated luminosity of
${\cal L} \sim 1$  ab$^{-1}$. The accuracy becomes worse for higher Higgs 
masses. In particular, for $M_H \gsim 140$ GeV, the $H\to WW^*$ decays must 
be used, leading to the even more complicated $4W$+$2f$ final state topologies.
No experimental analysis of this topology has been attempted yet.\s

Also in this case, one can proceed to higher energy and take advantage of the
$WW$ fusion process $\ee\to HH \nu \bar \nu$ \cite{Yamashita,Yamashita0} which
has a larger cross section, in particular with longitudinally polarized $e^\pm$
beams. The estimated sensitivity of the trilinear Higgs couplings to $\sqrt{s}$
is shown in the right--hand side of Fig.~4.38 for $M_H=120$ and 150 GeV with
polarized electron beams and no efficiency loss \cite{Yamashita}. It is
dominated by Higgs--strahlung at low energy and $WW$ fusion for $\sqrt{s} \gsim
700$ GeV. A recent simulation at $\sqrt{s}=1$ TeV which combines both the $\ee
\to HHZ$ and $\ee \to HH\nu \bar \nu$ processes with  $HH\to 4b$ final states,
assuming a 80\% $e^-_L$ polarization and a luminosity of 1 ab$^{-1}$, shows
that an accuracy of $\Delta \lambda_{HHH}/\lambda_{HHH} \sim 12\%$ may be
achieved if the trilinear coupling is SM--like \cite{Yamashita}. The relative
phase of the coupling and its sign, may be also measured from the interference
terms \cite{LCWS,Yamashita}. \\ 

\centerline{\psfig{figure=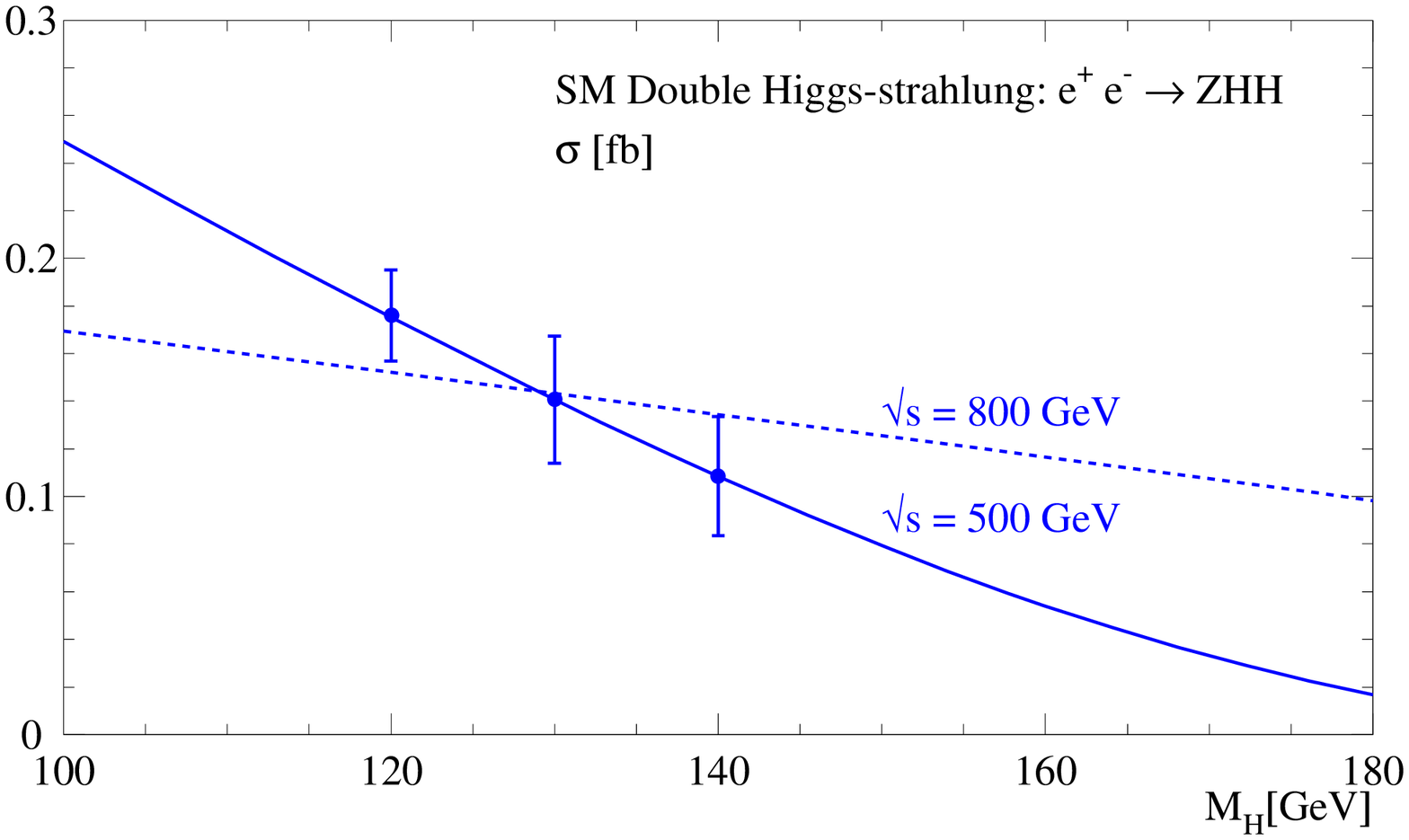,width=7.5cm,height=7cm} 
\psfig{figure=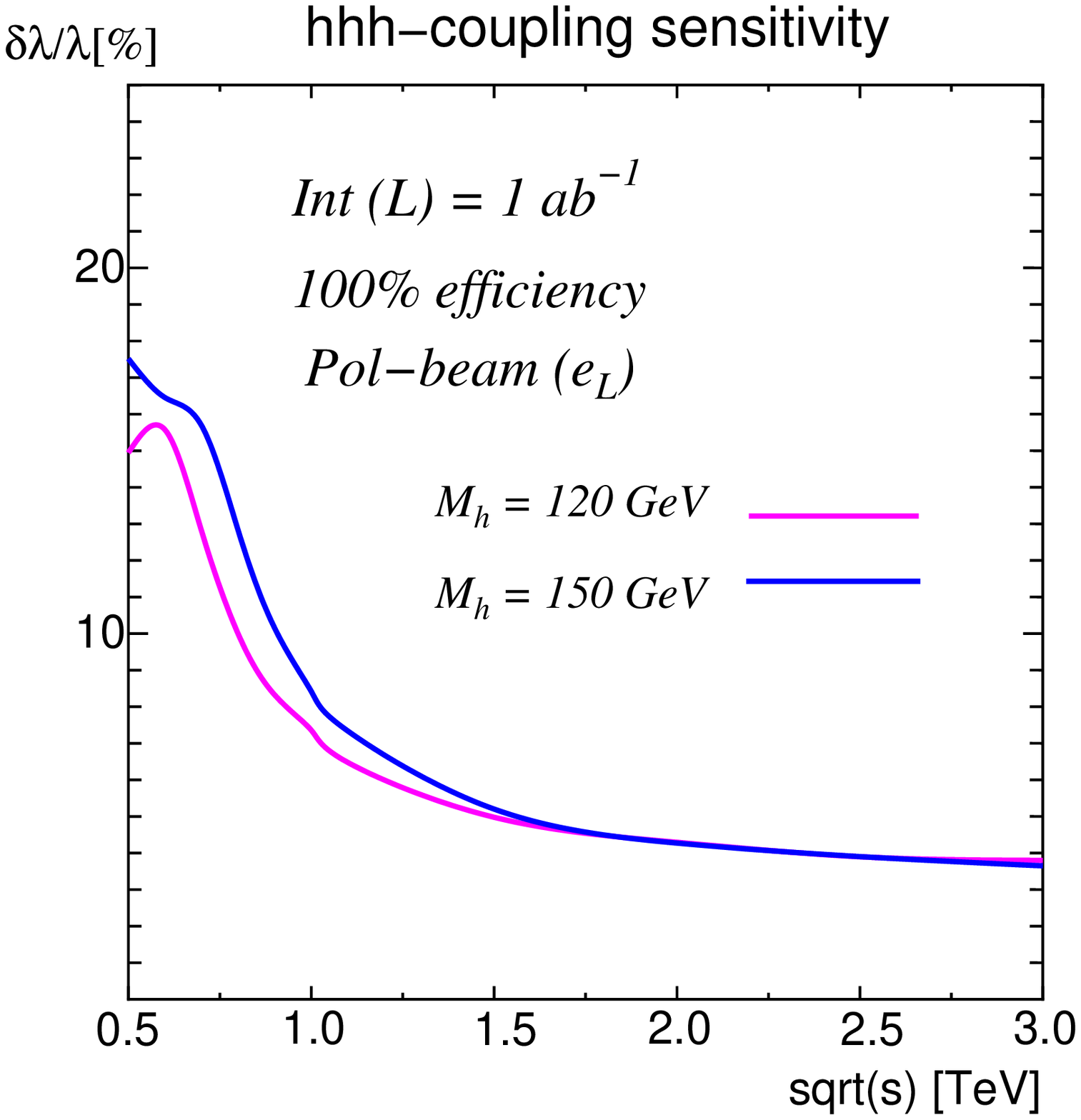,width=7.5cm,height=7cm}}
\vspace*{.mm}
\nn {\it Figure 4.38: The accuracy in the determination of $\sigma(\ee \to 
HHZ)$ for several Higgs masses at $\sqrt{s} = 500$ GeV with ${\cal L} =1$  
ab$^{-1}$ (left) \cite{Clermont-Ferrand} and the sensitivity of $\lambda_{HHH}$
to the c.m. energy for  ${\cal L} =1$  ab$^{-1}$, $P_L(e^-)=100 \%$ and without
efficiency corrections (right) \cite{Yamashita}.}

\vspace*{-2mm}
\subsubsection*{\underline{Expectations for a heavier Higgs boson}}

Finally, let us make a few remarks about a Higgs boson that is heavier than
$2M_Z$, which has been recently discussed in Ref.~\cite{ee-H-heavy}. In this
case, all decay channels other than $H \to WW,ZZ$ are not accessible
experimentally. The only exceptions are the $b\bar b$ decays for masses $M_H
\lsim 200$ GeV and the $t\bar t$ decays for $M_H \gsim 350$ GeV. However, the
Higgs boson mass and its total decay width, as well as the production cross
sections which provide the couplings to gauge bosons, can be obtained from the
lineshape. Typical accuracies on these parameters are shown in Table  4.8 at a
c.m. energy of 500 GeV with 500 fb$^{-1}$. The accuracies of the $WW$ and $ZZ$
branching are also shown for the same energy and luminosity
[other decay channels have not been discussed yet]. Thus, relatively
precise measurements can also be performed for heavier Higgs particles.

\begin{table}[hbt]
\renewcommand{\arraystretch}{1.2}
\begin{center}
\begin{tabular}{|c|c|c|c||c|c|}
\hline
$M_H$(GeV) & $\Delta \sigma$(\%)&$\Delta M_H$(\%)&$\Delta\Gamma_H$ (\%) &
$\Delta$BR$(WW)$ (\%)  & $\Delta$BR$(ZZ)$ (\%) \\ \hline
200 & 3.6 & 0.11& 34 & 3.5 & 9.9   \\ 
240 & 3.8 & 0.17& 27 & 5.0 & 10.8   \\
280 & 4.4 & 0.24& 23 & 7.7 & 16.2   \\
320 & 6.3 & 0.36& 26 & 8.6 & 17.3   \\ \hline
\end{tabular}
\end{center}
\vspace*{-0mm}
{\it Table 4.8: Expected precision on heavier Higgs lineshape parameters with 
500 b$^{-1}$ at $\sqrt{s}=500$ GeV \cite{Desch} and on the $WW/ZZ$ branching 
ratios with 1 ab$^{-1}$ at $\sqrt{s}=1$ TeV \cite{Barklow}. }
\vspace{-.5cm}
\end{table}

\newpage

\subsubsection{Combined measurements and the determination of the 
couplings}

Once the Higgs production cross sections and the various decay branching ratios
have been measured, one can derive the Higgs boson couplings to
fermions and gauge bosons. This is a crucial test for the experimental 
verification that the Higgs mechanism is responsible for the generation of the
masses of the particles. Since some of the couplings can be determined in
different ways, while other determinations are partially correlated, a global
fit to the various observables is highly desirable to extract the Higgs 
couplings in a model independent way. Such a fit would optimize the collected
information and takes properly into account all the experimental correlations
between the various measurements. \s

A dedicated program called {\sc hfitter} \cite{Hfitter}, based on {\sc hdecay}
\cite{HDECAY} for the calculation of the Higgs decay branching ratios, has been
developed for this purpose. It uses as inputs the production cross sections
$\sigma(\ee \to HZ)$, $\sigma(\ee \to H\nu \bar{\nu})$ and $\sigma(\ee \to
t\bar{t}H)$, and the branching ratios into $WW, \gamma \gamma, b\bar{b},
c\bar{c}, \tau^+ \tau^-$ and $gg$. It uses the full covariance matrix for the
correlated measurements, and the non--correlated measurement of the Higgs
self--coupling from $\sigma( \ee \to HHZ)$ can be added.  The results for the
accuracies on the Higgs couplings to fermions, gauge bosons and the
self--coupling are displayed in Table 4.9 for $M_H=120$ GeV and 140 GeV at a
c.m. energy of 500 GeV with a luminosity of 500 fb$^{-1}$  [except again for
the measurement of $g_{Htt}$ which has been performed at  $\sqrt{s}=800$ GeV
with a luminosity of 1 ab$^{-1}$; the same luminosity is also used for the
measurement of $\lambda_{HHH}$].  For completeness, we also display the errors
on the Higgs boson mass, its total decay width and its CP--even component
[$\Delta {\rm CP}$ represents the relative deviation from the 0$^{++}$ case],
which have been measured at $\sqrt{s}=350$ GeV  with the same luminosity ${\cal
L} =500$ fb$^{-1}$. \s 

\begin{table}[!h]
\vspace*{5mm}
\renewcommand{\arraystretch}{1.6}
\begin{center}
\begin{tabular}{|l|c|c|}
\hline
Quantity & $M_H$ = 120 GeV & $M_H$ = 140 GeV \\
\hline \hline
$\Delta M_H$ & $\pm$ 0.00033 & $\pm$ 0.0005 \\ 
$\Gamma_H$ & $\pm$ 0.061 & $\pm$ 0.045 \\ 
$\Delta {\rm CP}$ & $\pm$ 0.038 & -- \\ \hline
$\lambda_{HHH}$ & $\pm$ 0.22       &  $\pm$ 0.30 \\ 
$g_{HWW}$ & $\pm$ 0.012       &  $\pm$ 0.020           \\
$g_{HZZ}$ & $\pm$ 0.012       & $\pm$ 0.013            \\ 
$g_{Htt}$ & $\pm$ 0.030       & $\pm$ 0.061            \\
$g_{Hbb}$ & $\pm$ 0.022       & $\pm$ 0.022             \\
$g_{Hcc}$ & $\pm$ 0.037       & $\pm$ 0.102            \\ 
$g_{H\tau\tau}$ & $\pm$ 0.033       & $\pm$ 0.048            \\ \hline
\end{tabular}
\end{center}
\vspace*{2mm}
{\it  Table 4.9: Relative accuracy on Higgs couplings 
obtained from a global fit. An integrated luminosity of
500\,fb$^{-1}$ at $\sqrt{s} = 500$ GeV is assumed except for the measurement of
$g_{Htt} (\lambda_{HHH})$, which assumes 1000\,fb$^{-1}$ at $\sqrt{s} =$ 800 
(500) GeV in addition. On top of the table we display the accuracies on the
Higgs mass, the total width and its CP--component as obtained at $\sqrt{s}=350$
GeV with 500 fb$^{-1}$.}
\end{table}

As can be seen, an $\ee$ linear collider in the energy range $\sqrt{s}= 
350$--800 GeV and a high integrated luminosity, ${\cal L} \sim 500$ fb$^{-1}$,
is a very high precision machine in the context of Higgs physics.  This
precision would allow the determination of the complete profile of the SM Higgs
boson, in particular  if its mass is smaller than $\sim 140$ GeV. It would also
allow to distinguish the SM Higgs particle from a scalar particle occurring in
some of its extensions, with a very high level of confidence. \s

Thus, very precise measurements can be performed at the next linear collider
allowing the detailed exploration of the electroweak symmetry breaking
mechanism and the determination of the fundamental properties of the Higgs boson
in the SM.  We have seen in the previous section on hadron colliders that 
while the SM Higgs boson will undoubtedly be produced at the LHC, the detailed
study of its properties will be a difficult task in the rather hostile hadronic
environment. Due to the limited signal statistics for some channels, the large
backgrounds and various systematic uncertainties, the LHC can provide only some
ratios of Higgs couplings  [as well as the Higgs mass and the total decay width
for $M_H \gsim 200$ GeV, which can be measured rather well]. The measurement of
the various absolute couplings can be performed only at an $\ee$ collider. 
There is therefore a clear complementarity between the LHC and the linear 
collider Higgs physics programs. \s

From the previous discussions, one can single out two physics points for which
$\ee$ colliders have some weakness: the determination of the total width is
rather poor [without the $\gamma \gamma$ option] for low mass Higgs bosons and
the CP--quantum numbers cannot be determined in a very convincing way for $M_H
\gsim 140$ GeV when the $H \to \tau^+ \tau^-$ decay mode becomes too rare.
Unambiguous tests of the CP properties of the Higgs boson can be performed at 
photon colliders in the loop induced process $\gamma \gamma \ra H$ or at muon
colliders in the process $\mu^+ \mu^- \to H$, if suitable polarization of the
initial beams is available.  The measurement of $\Gamma_H$ can benefit from the
precise determination of the Higgs photonic width at $\gamma \gamma$ colliders.
However, it is at the muon collider that extremely good accuracies on
$\Gamma_H$ can be obtained by simply performing a threshold scan around the
Higgs resonance produced in $\mu^+ \mu^- \to H$.  These topics will be
addressed in detail in the next section. Before that, we first briefly 
summarize the benefits of raising and lowering the energy of the $\ee$ collider.

\subsubsection{Measurements at higher and lower energies}

\subsubsection*{\underline{Measurements at CLIC}}

Some of the previously discussed measurements can significantly benefit from an
increase of
statistics. This can be obtained not only by increasing the luminosity, but 
also by raising the energy. Indeed, at the c.m. energies relevant for CLIC, 
$\sqrt{s} \sim 3$ TeV, the cross section for the $WW$ fusion process becomes 
extremely large. If the luminosity is also scaled with $s$, a sample of more 
than one million Higgs particles can be collected for ${\cal L}=3$ ab$^{-1}$. 
Some of the previous measurements could thus be performed with more accuracy  
and new ones could be made possible. Examples of such measurements at CLIC are 
as follows \cite{CLIC}: \s

$i)$ With $\cal L =$ 3 ab$^{-1}$ at a c.m. energy of 3 TeV, 400 $H \to \mu^+
\mu^-$ events can be collected for $M_H=120$ GeV.  This sample would allow the
measurement of the Higgs couplings to muons to better than 5\%
[the precision drops to 10\% for $M_H=150$ GeV due to the smaller branching
ratio]. The dimuon signal can be isolated from the important $WW, WW\nu \bar
\nu, ZZ \nu\bar \nu$ backgrounds with a statistical significance which is
rather large; see the left--hand side of Fig.~4.39.  This would be the first
precise measurement of the Higgs couplings to second generation fermions since,
as seen previously, although the $Hc\bar{c}$ coupling can be determined with
the same accuracy, the associated theoretical uncertainties are rather large. \s

$ii)$ The $H\to b \bar{b}$ branching ratio becomes very small in the 
intermediate and high Higgs mass ranges, and at $\sqrt{s}=500$ GeV, it cannot 
be determined to better than 10\% for $M_H \sim 200$ GeV. At $\sqrt{s}=3$ TeV,
the signal to background ratio is very favorable at these masses, as shown in 
the right--hand side of Fig.~4.39, and the rather large number 
of events to be collected at CLIC would allow a measurement of the $Hb\bar{b}$ 
coupling with an accuracy of 5\% for Higgs masses up to about $M_H=250$ GeV. 
\s

\begin{figure}[!h]
\vspace*{-0.8cm}
\begin{center}
\epsfig{file=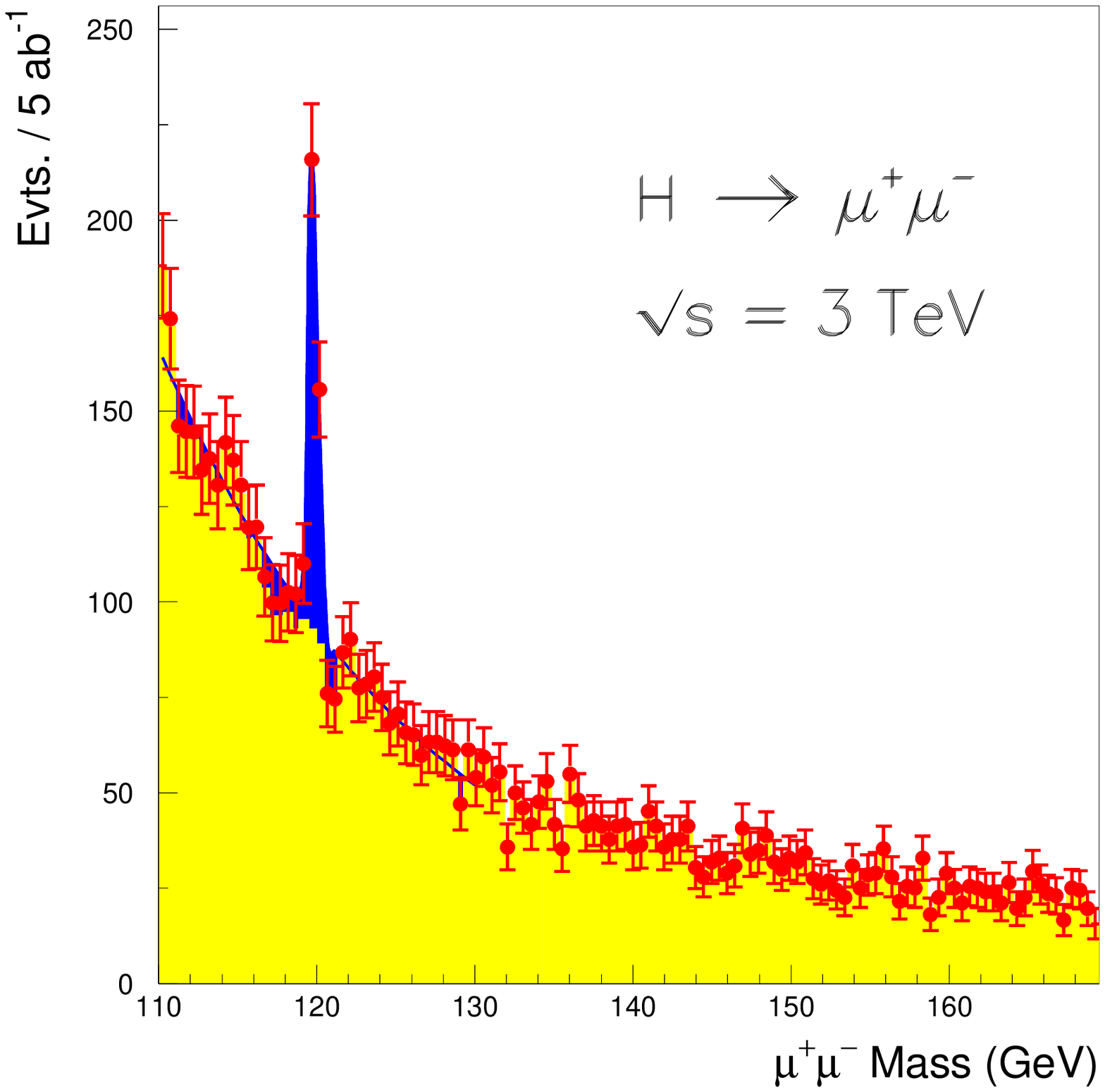,width=8cm,height=7.4cm,clip} 
\epsfig{file=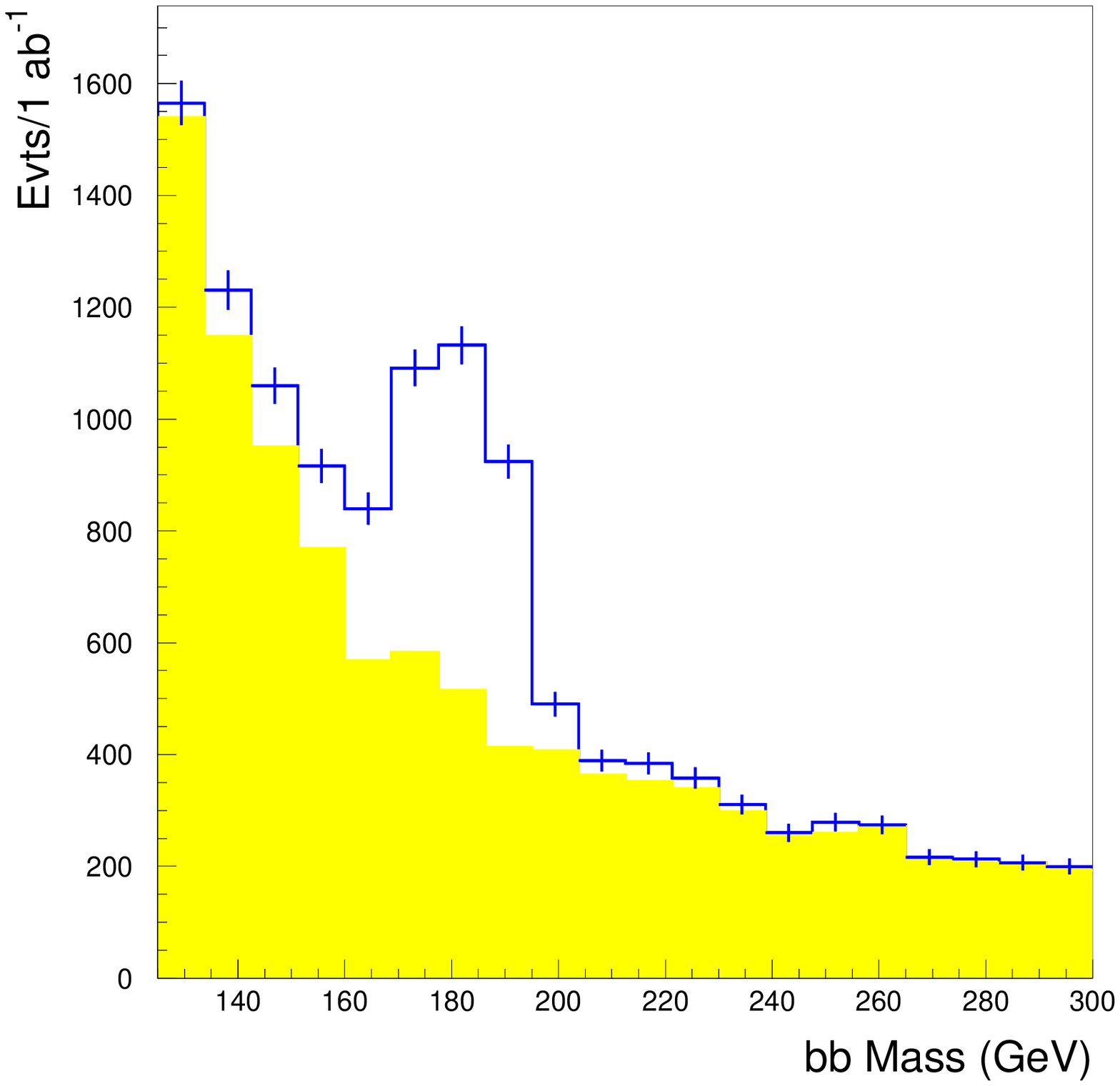,width=8cm,height=7.4cm} 
\end{center}
\vspace*{-0.5cm}
{\sl Figure 4.39: The reconstructed signals for $e^+e^- \to \nu \bar \nu H \to
\nu \bar \nu  \mu^+\mu^-$ for $M_H=120$ GeV (left) and $e^+e^- \to \nu \bar 
\nu  H \to  \nu \bar \nu b\bar{b}$ for $M_H=200$ GeV (right) at CLIC with 
$\sqrt{s}$=3~TeV \cite{ee-H3-Battaglia}.}
\vspace*{-0.5cm}
\end{figure}

$iii)$ The trilinear Higgs coupling can be measured in the $WW$ fusion process,
$\ee \to \nu \bar{\nu}HH$, for which the cross section reaches the level of a
few fb at energies around 3 TeV. A relative accuracy of $\sim 10\%$ can be
obtained on this coupling for Higgs masses up to 250 GeV. Contrary to what 
occurs in the process $\ee \to
HHZ$, the interference between the diagram involving the self--Higgs coupling
and the others, is negative.  The sensitivity to $\lambda_{HHH}$ can be
enhanced by studying the angle $\theta^*$ of the $H^* \to HH$ system
in its rest frame: because of the scalar nature of the Higgs boson, the $\cos 
\theta^*$ distribution is flat for $H^* \rightarrow HH$ while it is peaked in 
the forward direction for the other diagrams \cite{gam-WWHH}; see the 
left--hand side of Fig.~4.40. From a fit of the distribution one can perform
a very nice determination of the $\lambda_{HHH}$
coupling as shown in the right--hand side of Fig.~4.40.  Note that the
quadrilinear Higgs couplings remains elusive, even at c.m.  energies of 5 TeV.
\s

\begin{figure}[!h]
\vspace*{-0.8cm}
\begin{center}
\includegraphics[scale=0.35]{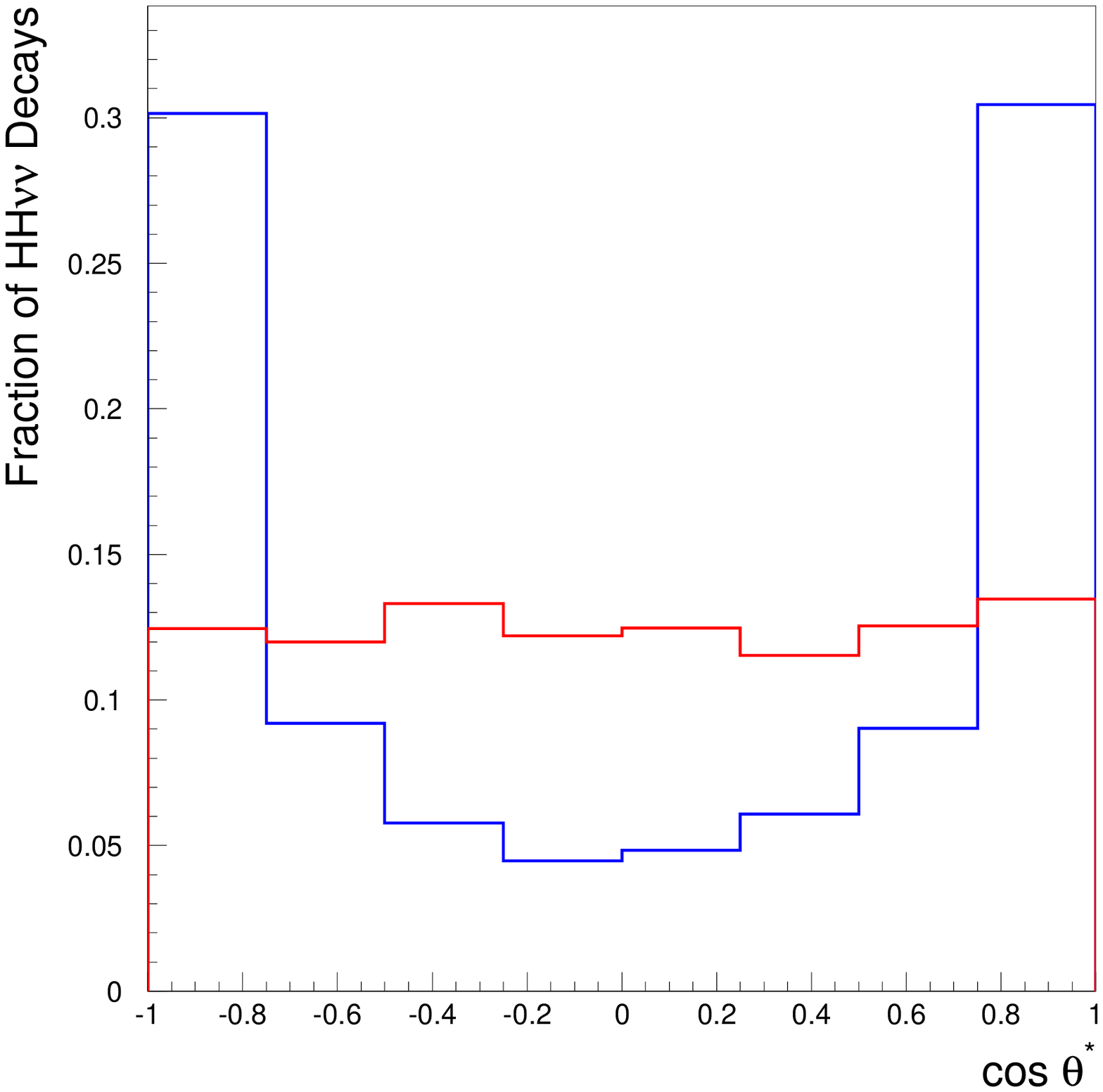} 
\includegraphics[scale=0.35]{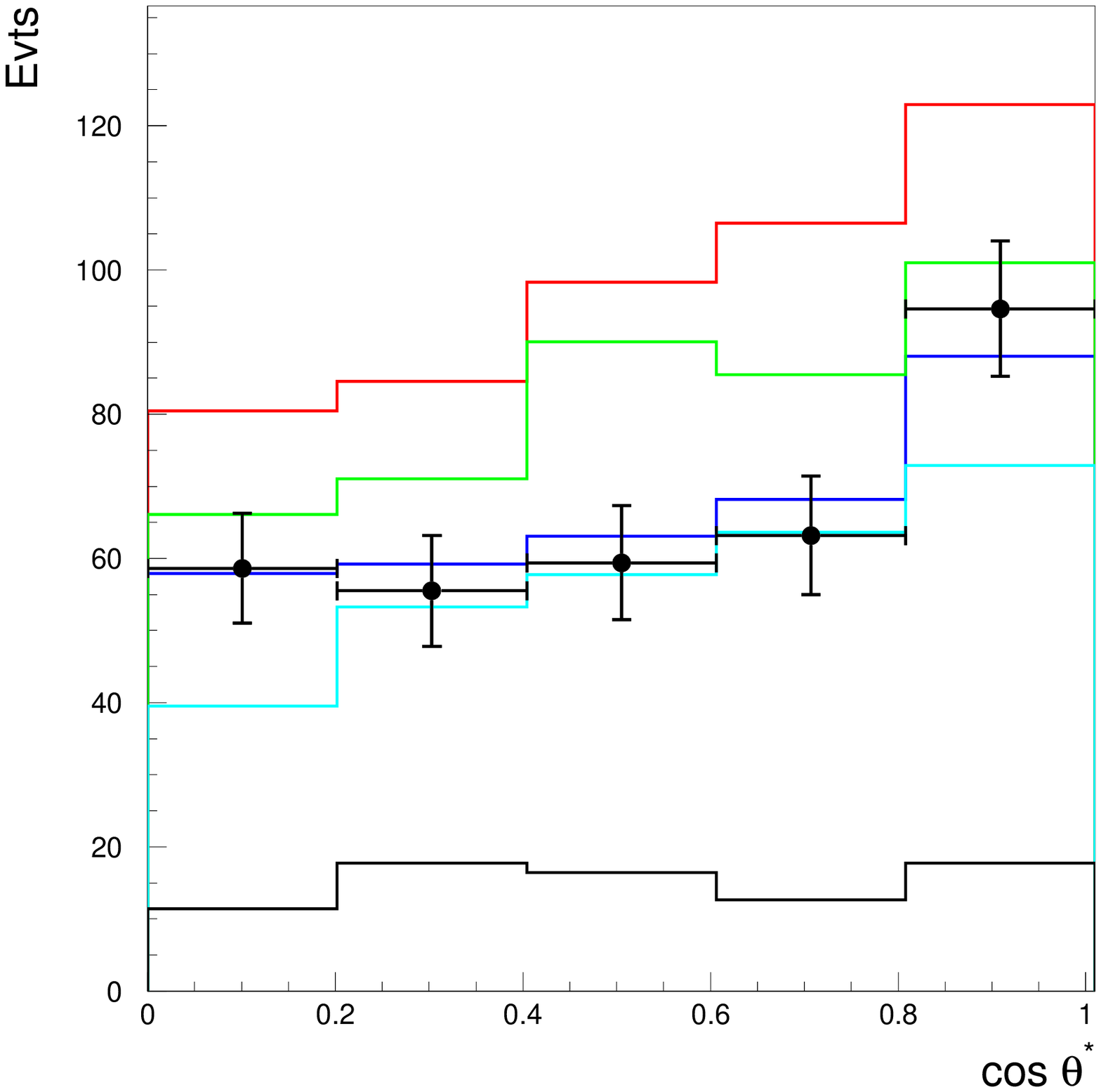}
\end{center}
\vspace*{-0.8cm}
{\sl Figure 4.40: The $\cos \theta^*$ distribution in the process $\ee 
\to  HH\nu\bar{\nu}$ due to the diagram containing the triple Higgs vertex 
(red/light grey) and other diagrams (blue/dark grey) for $M_H=120$
GeV at $\sqrt{s}$=3~TeV (left) and the reconstructed $|\cos \theta^*|$ 
distribution for $\lambda_{HHH}/\lambda_{HHH}^{\rm SM}=$1.25,1.0,0.75 and 0.5 
from bottom to top, with the points with error bars showing the expectation 
for 5~ab$^{-1}$ of data (right); from Ref.~\cite{ee-H3-Battaglia}.} 
\vspace*{-0.3cm}
\end{figure}

The higher energy of the collider can also be very useful in the case where
the Higgs boson is very heavy. For $M_H \sim 700$ GeV and beyond, the cross
sections in the Higgs--strahlung and $WW$ fusion processes are small at
$\sqrt{s} \sim 1$ TeV [see Fig.~4.30] and do not allow to perform detailed
studies. At CLIC energies, $\sqrt{s}=3$ TeV, one has $\sigma (\ee \to H\nu \bar
\nu) \sim 150$ fb which allows for a reasonable sample of Higgs particles to
be studied. In addition, the cross section for the $ZZ$ fusion process is large
enough, $\sigma (\ee \to H\ee) \sim 20$ fb for $M_H \sim 700$ GeV, to allow for
model independent Higgs searches in much the same way as in the
Higgs--strahlung process at low energies, since the forward electron and
positron can be tagged, and the mass recoiling against them can be
reconstructed. The high energy available at CLIC will also be important to
investigate in detail a possible strongly interacting Higgs sector scenario, as
will be discussed in another part of this review.  

\subsubsection*{\underline{The GigaZ and MegaW options}}

The high luminosities available at the next generation of $\ee$ colliders
would allow to collect more than $10^{9}$ $Z$ bosons in one year by running 
at energies close to the resonance. The same luminosity would allow to collect
more than $10^6$ $W$ boson pairs near the $WW$ threshold. These samples are
two orders of magnitude larger than those obtained at LEP1 and LEP2 and can 
be used to significantly improve the high--precision tests of the SM which
have been performed in the last decade \cite{LC-GigaZ}. \s

At GigaZ, using the possibility of polarizing the electron/positron beams, one
can measure the longitudinal left--right asymmetry
$A_{LR}=2a_ev_e/(a_e^2+v_e^2) \sim 2(1-4\sin^2\theta_{\rm eff}^{\rm lep})$ with
a very high statistical accuracy in hadronic and leptonic $Z$ decays. Using the
Blondel scheme \cite{Blondel}, the asymmetry can be obtained from the cross
sections when the polarization of both the electron and positron beams
$P_{e^\pm}$ are used in the various combinations,  $\sigma = \sigma_{\rm unpol}
[1 - P_{e^+} P_{e^-} + A_{LR} ( P_{e^+} - P_{e^-} )]$, leading to a
systematical error of about $10^{-4}$. This corresponds to a measurement of the
electroweak mixing angle with a precision 
\beq
\Delta \sin^2\theta_{\rm eff}^{\rm lep} \simeq 1.3 \times 10^{-5}
\eeq
which is one order of magnitude more accurate than the presently measured 
value, $\sin^2\theta_{\rm eff}^{\rm lep} = 0.2324 \pm 0.00012$. The measurement
of the total and partial $Z$ decay widths and the various polarization and/or 
forward--backward asymmetries can be significantly improved. In particular, the
measurement of the ratio of leptonic to hadronic $Z$ decay widths with an 
expected accuracy of $\Delta R_\ell/R_\ell \sim 0.05\%$, would allow a clean 
measurement of the strong coupling constant to better than $\Delta \alpha_s
\simeq 0.001$. \s

On the other hand, one can perform a scan around the $WW$ threshold, where
the cross section for $W$ pair production rises quickly, $\sigma(\ee \to W^+W^-)
\sim \beta$, allowing an accurate measurement of the $W$ boson mass. With an
integrated luminosity of only ${\cal L} \simeq 100$ fb$^{-1}$ at $\sqrt{s}\sim 
2M_W$ and a 6 point scan, the mass can be measured with an accuracy
\beq
\Delta M_W \simeq 6~{\rm MeV}
\eeq
which is six times better than the present measurement, $M_W = 80.449 \pm 
0.034$ GeV, and almost three times the precision which can be reached at 
the LHC and at the LC.\s 

Since the top quark mass, which leads to the major part of the theoretical
uncertainties in the present high--precision observables, will be measured with
an accuracy of $\Delta m_t \simeq 200$ MeV at the LC and that $\alpha_s$ will be
known more precisely at this time, $\Delta \alpha_s \simeq 0.001$, the only
dangerous source of errors from SM inputs will be the hadronic uncertainty in
$\Delta \alpha$. One might hope that with the low energy $\ee$ experiments
which will be performed in the future, the error will reduce to $\Delta 
\alpha^{\rm had} \simeq 5 \times 10^{-5}$. Taking into account also the error 
 $\Delta M_Z \simeq 2$ MeV on the $Z$ boson mass, which at this level of 
precision induces an error on $\sin^2\theta_{\rm eff}^{\rm lep}$ which is of 
the same size as the experimental error, the future total theoretical 
uncertainties on the two observables from the various sources are estimated 
to be \cite{GigaZ-P}
\beq 
\Delta \sin^2\theta_{\rm eff}^{\rm lep} \simeq \pm 3 \times 10^{-5} \ , \
\Delta M_W \simeq \pm 3~{\rm MeV}
\eeq
The very small experimental and theoretical errors on these two parameters
will allow to test the SM on much more solid grounds than in the past and to 
isolate the effects of the Higgs boson in the electroweak radiative corrections
with an incredible accuracy. This is exemplified in the left--hand side of 
Fig.~4.41 where the expected accuracy in the determination of the Higgs  
mass at GigaZ/MegaW in the plane $M_H$--$m_t$, together with the allowed bands 
for $\sin^2\theta_{\rm eff}^{\rm lep}$ and $M_W$, are shown. The central 
values of the various input parameters and the Higgs mass, as well as the area 
labeled ``now", are for the measurements which were available in the year 2000. 
One can simply notice the vast improvement which can be made at the 
GigaZ/MegaW option, where one can indirectly measure the Higgs boson mass 
with a precision of $\Delta M_H/M_H \sim 7\%$ \cite{GigaZ-P}. One can also use 
the direct measurement of the Higgs boson mass at the LC (and LHC) with
$\Delta M_H \simeq 50$ MeV, to predict the value of $\sin^2\theta_{\rm eff}^{\rm
lep}$ and $M_W$ and to check the consistency of the SM, as shown in the
right--hand side of Fig.~4.41. Because of the high--precision which can be 
reached at GigaZ/MegaW, the improvement compared to the present situation and 
even after LHC/LC is again spectacular.

\begin{figure}[ht!]
\begin{center}
\begin{tabular}{c c}
{{\epsfig{file=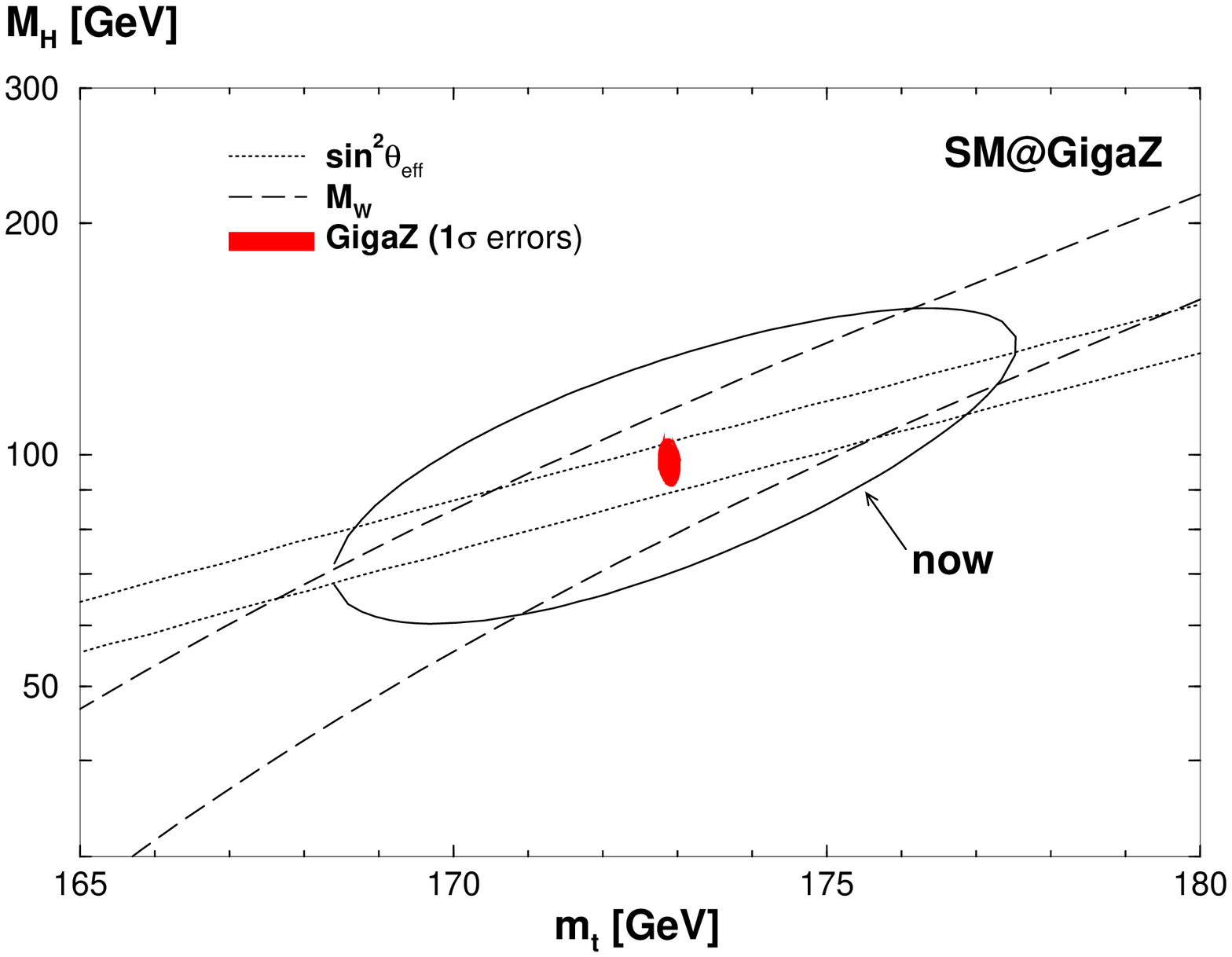,width=0.5\linewidth}}} &
{{\epsfig{file=./sm4/GigaZ2.eps,width=0.45\linewidth}}} \\
\end{tabular}
\end{center}
{\it Figure 4.41: The allowed region in the $M_H$--$m_t$ plane 
after the precision measurements at GigaZ/MegaW compared to the situation 
in the year 2000 (left) and the theoretical predictions for $\sin^2\theta_{\rm 
eff}^{\rm lep}$ and $M_W$ for three $M_H$ values compared to the experimental
measurements at LEP2/Tevatron, LHC/LC and after GigaZ/MegaW. The various 
theoretical and experimental uncertainties are as discussed in the text; 
from Ref.~\cite{GigaZ-P}. }
\end{figure}


\subsection{Higgs production in  $\gamma \gamma$ collisions}

As discussed in \S4.1.2, future high--energy $\ee$ linear colliders can be made
to run in the $e \gamma$ or $\gamma \gamma$ modes by using Compton back
scattering of laser light off the high--energy electron beams
\cite{gamma-machine1,gamma-machine2}.  These colliders will have practically
the same energy, up to $\sim 80$\%, as the original $\ee$  collider and a
luminosity that is somewhat smaller.  One of the best motivations for turning
to the $\gamma\gamma$ mode of the linear collider is undoubtedly the study of
the properties of the Higgs boson, which can be produced as a resonance in the
$s$--channel \cite{gamma-Rev-old,BBC,gamma-Rev-TESLA,gamma-Rev-NLC}. In this 
context, two main
features which are difficult to study in the $\ee$ mode can be investigated at
such colliders: first, the  precise measurement of the $H\gamma\gamma$ coupling
\cite{gam-Hff,gam-Hff-Warsaw,gam-HWW,gam-HZZ,gam-HVV-Warsaw} and second, the 
determination of the CP--properties of the Higgs boson 
\cite{CPHff1,gam-HVV-Warsaw,gam-Gun,gam-Htt-Roh,gam-Htt-Asa1,gam-Htt-Asa2,gam-Htt-rev}. 
Several other studies can also be made, such as the measurement of the Higgs 
boson
self--coupling and its Yukawa coupling to top quarks, although these studies
can be already performed in $\ee$ collisions [and, in general, in a much
cleaner way]. 

\subsubsection{Higgs boson production as an s--channel resonance}

The production cross section for the process $\gamma \gamma \to X$ with
initial state polarized photons, can be written in the helicity basis as
\beq
{\rm d} \hat{\sigma}_{\gamma \gamma}  = \sum_{i,j,k,l=\pm} \, \rho^1_{ik} \,  
\rho^2_{jl} M_{ij} M^*_{kl} \, {\rm d}\Gamma
\eeq
where $M_{ij}$ are the invariant scattering amplitudes with photon helicities
$i,j=\pm 1$ and ${\rm d}\Gamma$ the phase space element divided by the
incoming  flux. Comparing to the cross section written in the Stokes parameter
basis, the elements of the photon polarization density matrix are such that
$\rho^i_{ \pm \pm}= \frac{1}{2} (1 \pm \xi_{i2}), \rho^{i}_{+-} = \rho^{i*}_{-+}
= - \frac{1}{2} (\xi_{i3}-i \xi_{i1})$. 
The unpolarized cross section is 
\beq
{\rm d} \hat{\sigma}  &=& {\rm d} \hat{\sigma}_{00} \, =  \, \frac{1}{4} {\rm d}
\Gamma \, \left( \left| M_{++}  \right|^2 + \left| M_{--} \right|^2 
+ \left| M_{+-}  \right|^2 + \left| M_{-+} \right|^2 \right) \non \\
&=& \frac{1}{2} ({\rm d} \hat{\sigma}_{J_Z=0} + {\rm d} \hat{\sigma}_{J_Z=\pm2})
\, = \, \frac{1}{2} ({\rm d} \hat{\sigma}_{||} + {\rm d} \hat{\sigma}_{\perp}) 
\eeq
where ${\rm d} \hat{\sigma}_{J_Z=0} \, ({\rm d} \hat{\sigma}_{J_Z=\pm 2})$ 
are the cross sections for photons with total helicity $0\, (\pm 2)$ and 
${\rm d} \hat{\sigma}_{||}\, ({\rm d} \hat{\sigma}_{\perp})$ is for
parallel (orthogonal) linear photon polarizations. \s

In the case of a spin--zero particle, the production occurs through the $J_Z=0$
channel. In terms of the Higgs total decay width $\Gamma_H$, the width into two
photons $\Gamma(H \to \gamma \gamma)$ and into a given final state,
$\Gamma( H \to X)$, the cross section for the subprocess $\gamma \gamma \to
H$ is given by
\beq
\hat{\sigma} (W)= 8\pi \frac{ \Gamma (H \to  \gamma  \gamma)\Gamma(H \to X)}
{(W^2-M_H^2)^2+ M_H^2 \Gamma_H^2 } \, (1+ \lambda_1 \lambda_2)
\eeq
where $W$ is the invariant mass of the $\gamma \gamma$ system.   Using the same
photon helicities $\lambda_1 \lambda_2=1$ projects out the $J_Z=0$ component 
and therefore maximizes the Higgs cross section. \s

For masses  below $M_H \sim 2M_Z$, the Higgs boson is very narrow with a total
decay width $\Gamma_H \lsim 1\,$GeV and, therefore, the detector resolution
should be taken into account. When the Higgs  width can be neglected,  a
rather simple way to obtain the effective signal cross section is to introduce
a Gaussian smearing of the  $\gamma\gamma$ invariant mass $W$ \cite{BBC}
\beq
N_{\rm eff}= {\cal L}_{\rm eff}\ \frac{{\rm d} \sigma^{\rm eff}}{{\rm d} W} 
(W) = \int^{y_m\sqrt{s_{e^+e^-}}}_{M_X} {\rm d} W' 
\frac{1}{\sqrt{2\pi}\delta}\exp \left(-\frac{(W'-W)^2}{2\delta^2}\right)
\frac{{\rm d} {\cal L}}{{\rm d} W'} \ \langle \hat{\sigma}(W') \rangle
\label{gamma-eff}
\eeq
and select events within bins of invariant masses $M_H \pm \Delta$, where
the Higgs mass is assumed to be precisely known already.  In the previous
expression, ${\cal L}_{\rm eff}$ and $y_m \sqrt{s_{\ee}}$ are the effective
luminosity and the maximum energy of the $\gamma \gamma$ collider and $\delta$
is one sigma of the detector resolution for $W$.  The cross section for the
signal process $\gamma\gamma \to H \to X$  can be written as [for $\Gamma_H \ll
M_H$, $\Gamma_H M_H [(W^2-M_H^2)^2+M_H^2\Gamma_H^2]^{-1} \simeq \frac{\pi}
{2M_H} \delta (W- M_H)]$
\beq
\hat{\sigma}_{\rm signal}(W) = 4\pi^2\frac{\Gamma(H \to\gamma\gamma)
{\rm BR}(H \to X)}{M^2_H} (1+\lambda_1\lambda_2)\delta(W-M_H)
\eeq
Inserting this expression in eq.~(\ref{gamma-eff}), and selecting the events 
in the bin $M_H \pm \Delta$, one has
\beq
{\cal L}_{\rm eff}\ \sigma^{\rm eff}_{\rm signal} (M_H) = R(\Delta/\delta) \
\left. \frac{\rm d \cal L}{\rm dW }^{J_Z=0}\right|_{W=M_H} \
4\pi^2\frac{\Gamma(H \to \gamma\gamma){\rm BR}(H \to X)}{M^2_H}
\label{gamma-signal}
\eeq
with $R(\Delta/\delta)$ being the Gaussian error function giving the fraction 
of signal events contained in the bin $M_H\pm\Delta$ [for instance, for $\Delta
=2\delta$ one has $R \simeq 0.95$]. \s

The effective background, $\gamma \gamma \to X$, for an effective invariant 
mass of the two--photon system $W= M_H$ can be approximated by
\beq
N^{\rm eff}_{\rm bckg}(W) \simeq 2 \Delta \frac{{\rm d} {\cal L}}{{\rm d}W}
\langle {\rm d} \hat{\sigma}_{\rm bckg} (W) \rangle
\eeq
if one assumes a smooth enough distribution of two--photon invariant masses 
weighted with luminosity distributions. \s

To have a large effective cross section for the Higgs boson signal, the
$\gamma \gamma$ energy must be tuned at the peak, $0.8 \sqrt{s}_{\ee} \sim M_H$
for a perfect spectrum, while the luminosity with circularly polarized laser 
photon and electron beams are chosen so that  they have opposite handedness 
with $x=4.83$. The $J_Z=0$ events containing the signal are then enhanced, 
while the $J_Z= \pm2$ events are suppressed \cite{gamma-machine2,gam-Kuhn}. \s

The measurement of the $\Gamma(H \to\gamma\gamma) \times {\rm BR}(H \to X)$ 
rate and, thus, the $H\gamma \gamma$ coupling squared if the branching ratio 
is known, will follow from eq.~(\ref{gamma-signal}) if the effective luminosity
and the Higgs mass are specified, and from the signal and background rates. 
The statistical error in the decay width times branching ratio determination is
\beq
\Delta (\Gamma \times {\rm BR}) /(\Gamma \times {\rm BR}) =({\cal L}_{\rm eff}
)^{-1} S^{-1}\sqrt{S+B}
\eeq

\subsubsection*{\underline{Low mass Higgs boson}}

In the low mass range,  $M_H \lsim 130$ GeV, the Higgs boson will mainly decay
into $b\bar{b}$ final states, $H \to b\bar{b}$, and the main source of
background is the continuum production of $b$-- and $c$--quark pairs
\cite{gam-ff}, including gluon radiation which leads to fake two--jet  events
\cite{gam-ff-NLO1}. The total cross section for heavy quark production, $\gamma
\gamma \to q\bar{q}$, with a polar cut in the center of mass of the 
two--photon system $|\cos \theta| <c$ is given, at the tree--level, by
\beq
\hat \sigma_{J_Z=0} (W) &=& \frac{12\pi\alpha^2 Q^4_q}{W^2}\frac{8m_q^2}{W^2}
\left( 1- \frac{2m_q^2}{W^2} \right) \left[ \frac{1}{2}\log \frac{1+c\beta}
{1-c\beta}+\frac{c\beta}{1-c^2\beta^2} \right] \non \\
\hat\sigma_{J_Z=2}(W) &=& \frac{12\pi\alpha^2 Q^4_q}{W^2}
\left[ \frac{1}{2}(5-\beta^4) \log \frac{1+c\beta}{1-c\beta}
-c\beta\left( 2+\frac{(1-\beta^2)(3-\beta^2)}{1-c^2\beta^2}
\right) \right] 
\label{ggqq:bckg}
\eeq
with the quark velocity $\beta=\sqrt{1-4m^2_q/W^2}$ and electric charge $Q_q$.
One can choose $c=0.7$ which helps to eliminate many background events which
are peaked in the forward and backward directions, with only a moderate loss of
the signal events. In addition, as can be seen, the contribution of the $J_Z=0$
channel is proportional to $m_q^2/W^2$ and is therefore strongly suppressed
\cite{gam-ff}. Choosing the configuration where $\lambda_1 \lambda_2=1$ helps 
to suppress the two--jet background, while it maximizes the signal cross 
section; see e.g. Refs.~\cite{gamma-machine2,gam-Kuhn}.  \s

The background cross sections receive important QCD radiative corrections
\cite{gam-ff-NLO1,gam-ff-NLO2}, which are particularly large for the $J_Z=0$ 
component to which additional continuum $q\bar{q}g$ final states contribute 
[one can select slim two--jet final state configurations to suppress this gluon
radiation  contribution to the $J_Z=0$ amplitude], and also non--negligible
electroweak corrections \cite{gam-ff-ew}. The radiative corrections to the 
signal
cross section discussed in \S2.3.1, and the corrections to the interference
between the signal and background cross sections \cite{gam-ff-Maggie} have to
be taken into account. In addition, one has to consider low energy $\gamma
\gamma \to$ hadrons processes which contribute to the overlaying events  
\cite{Manuel+Rohini}. The overlaying events are peaked in the forward and
backward directions and can be suppressed  by the angular cut. $b$--tagging is
of course mandatory and one can take advantage of the fact that the Higgs boson
is produced almost at rest so that the total longitudinal momentum of the 
visible particles is smaller than the total visible energy.\s

A full simulation, which uses a realistic spectrum for the photon collider and
includes the overlaying $\gamma \gamma \to$ hadrons events, as well as a
realistic $b$--tagging, has been recently performed \cite{gam-Hff-Warsaw}. The
signal  and backgrounds  events have been generated with all the relevant
higher--order corrections and including the fragmentation into hadrons, and the
expected response of the detector has been taken into account. Cuts such as 
those
discussed above have been applied and the output is shown in Fig.~4.42 where
the energy of the original collider, $\sqrt{s_{ee}}=210$ GeV leading to a
yearly luminosity of ${\cal L}_{\gamma \gamma}=410$ fb$^{-1}$, has been
optimized for the production of a Higgs boson with a mass $M_H=120$ GeV.\s

The left--hand side of the figure displays the reconstructed invariant mass
distribution of the selected $b\bar b$ events, when it is corrected to take 
into account the effect of the undetected neutrinos. The Higgs boson signal
as well as the $b\bar b (g)$ and $c \bar c (g)$ backgrounds including the 
overlaying events are displayed for $M_H=120$ GeV at the luminosity of 410 
fb$^{-1}$. The most precise measurement of the $H\to \gamma \gamma$ width is
obtained in the mass window 110--150 GeV which is indicated. With
the assumed luminosity, about 7000 signal events are reconstructed with 
about 9000 background events surviving the cuts, leading to a signal over 
background ratio of order one. Therefore, a statistical accuracy of 1.8\% can 
be achieved on the measurement of $\Gamma (H \to \gamma \gamma) \times {\rm BR}
(H\to b \bar{b})$. The right--hand side of the figure shows the accuracy 
of the measurement of $\Gamma (H \to \gamma \gamma) \times {\rm BR}
(H\to b \bar{b})$ for various Higgs mass values, with and without the inclusion 
of the overlaying events (OE). Again, this is the result of a full simulation 
where the energy of the initial collider has been optimized to produce a Higgs 
boson with a mass $M_H=130,140,150$ and 160 GeV. A precision of 2--7\% can be 
obtained in the entire Higgs mass range  $M_H=120$--160 GeV. \s

\begin{figure}[thb]
\vspace*{-.5cm}
{\centering \resizebox*{!}{0.38\textheight}
{\includegraphics{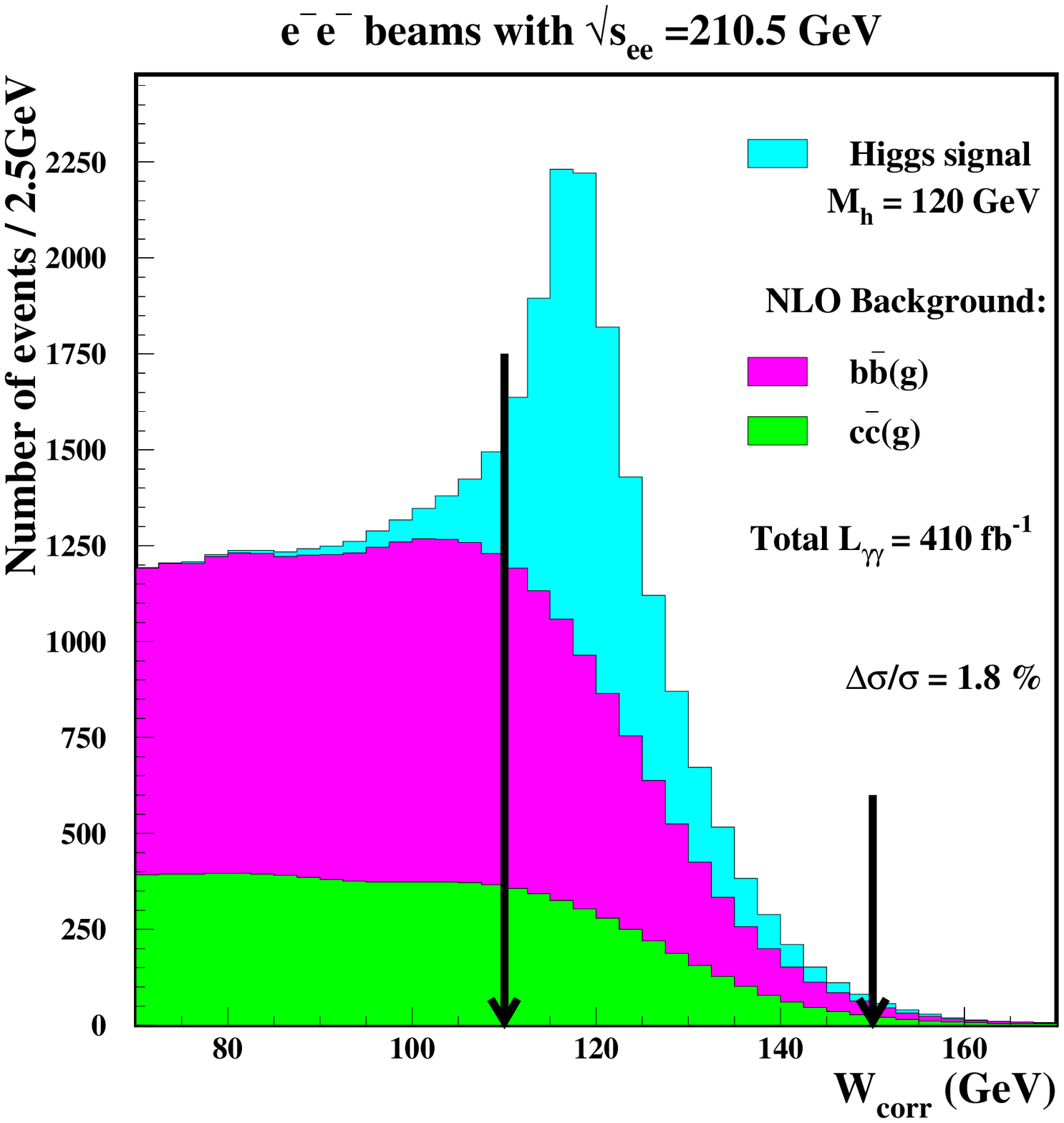}}
\hspace*{-5mm} \resizebox*{!}{0.38\textheight}
{\includegraphics{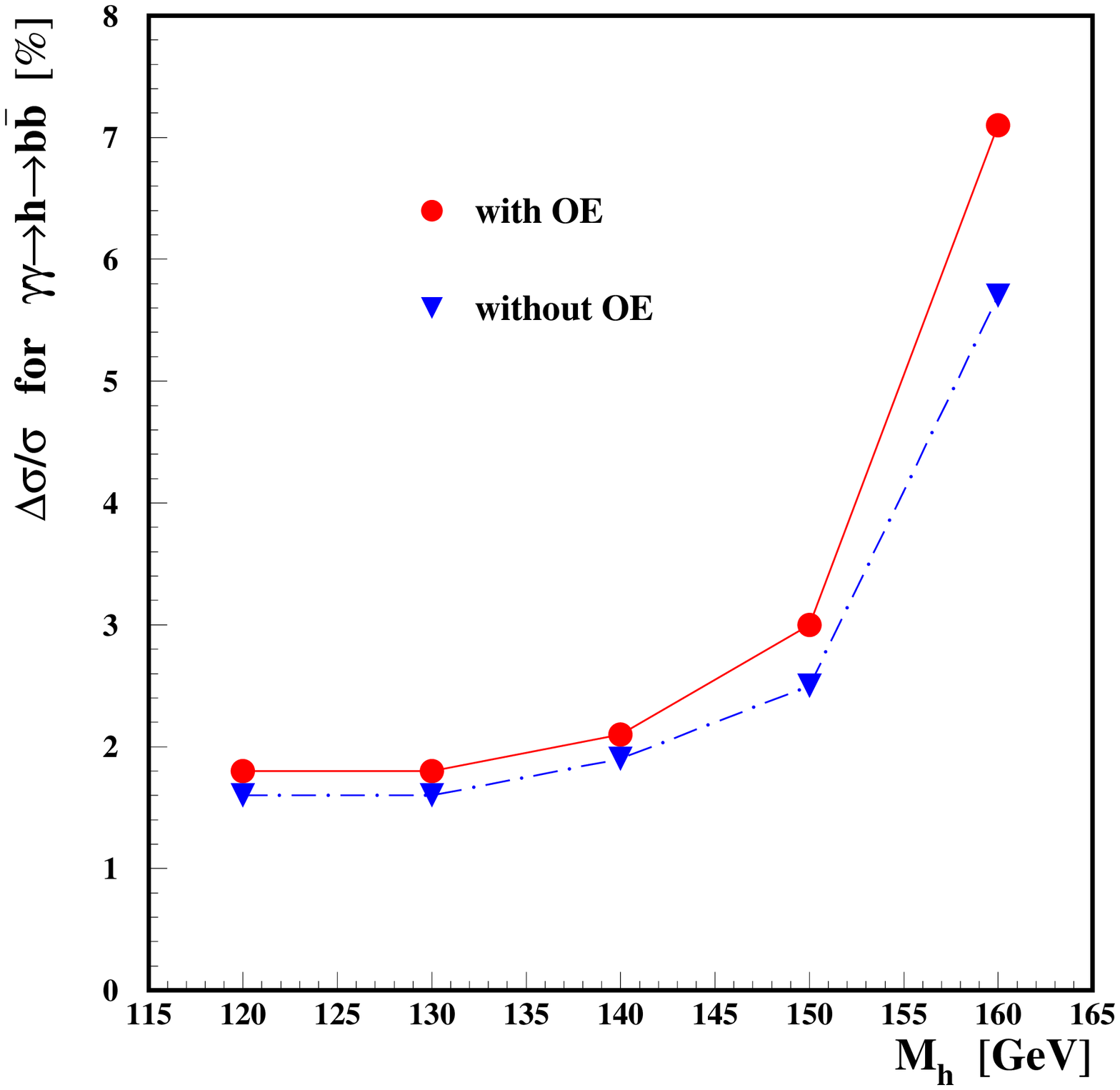}} }
\nn {\it Figure 4.42: 
The reconstructed invariant mass distribution of the $b\bar b$ signal and 
the $b\bar b (g)$ and $c \bar c (g)$ backgrounds for $M_H=120$ GeV at the 
luminosity of 410 fb$^{-1}$ (left) and the accuracy of the measurement of 
the cross section $\sigma (\gamma \gamma \to H\to b \bar{b})$ for various Higgs 
mass values, with and without the inclusion of the overlaying events (right);
from Ref.~\cite{gam-Hff-Warsaw}.}
\vspace*{-.2cm}
\end{figure}

From the measurement of the branching ratio of the Higgs decays into bottom
quarks which, as seen previously, can be made with an accuracy of 1.5\%
for $M_H=120$ GeV [see Table 4.6], the partial decay width $\Gamma (H \to
\gamma \gamma)$ can be extracted with a precision of 2.3\%. With a precise
measurement of the $H\to \gamma \gamma$ branching ratio in the $\ee$ mode
of the collider, one can determine the Higgs total width with an accuracy of 
the order of 10\%. 
 
\subsubsection*{\underline{Heavier Higgs bosons}}

For masses larger than $M_H \sim 140$ GeV, the Higgs boson decays 
predominantly into massive gauge bosons, $H \to WW$ and $H\to ZZ$,  
the branching ratios being $\sim 2/3$ and $1/3$, respectively, for the
charged and  neutral decays for $M_H$ above the $ZZ$ threshold. The total
Higgs decay width becomes significant, being of the order of $\Gamma_H \sim 
1.5\,(8)$ GeV for $M_H=200\, (300)$ GeV, and cannot be neglected anymore. 
However, the total production cross section of such heavy Higgs particles 
is of the same order as the one discussed previously, once the energy of 
the $\gamma \gamma$ collider is tuned to the Higgs boson mass. \s

The backgrounds for the production of such a Higgs boson at $\gamma \gamma$
colliders are vector boson production, $\gamma \gamma \to W^+ W^-$ and $\gamma
\gamma \to ZZ$. The former process occurs at the tree--level and has an
extremely large cross section, $\sigma(\gamma \gamma \to W^+W^-) \sim {\cal
O}(10^2~{\rm pb})$ in both the $J_Z=0$ and $J_Z=\pm 2$ channels
\cite{gam-WW,Othergam-WW}. This background cannot therefore be very efficiently
suppressed by  selecting only the $J_Z=0$ channel in which the Higgs boson is
produced. The only region where the signal and backgrounds have similar rates
is for $M_H \sim 170$ GeV, where the Higgs boson decays almost 100\% of the
time into $WW$ bosons, while the background cross sections are not yet too
large since they increase with higher photon c.m. energy \cite{gam-HWW}. \s

In the case of $ZZ$ boson final states, the background is generated only at the
one--loop level \cite{gam-HZZ} since the $Z$ boson is neutrally charged and 
does not couple directly to photons. It is therefore much less dangerous than 
the $WW$
background: for c.m.  energies of the order of $\sqrt{s}_{\gamma \gamma} \sim
200-300$ GeV, the cross section is at the level $\sigma(\gamma \gamma \to ZZ)
\sim {\cal O}(10^2~{\rm fb})$ in the $J_Z=0$ channel. Therefore, for $M_H \gsim
180$ GeV where the $ZZ$ Higgs branching ratio becomes significant, the cross
section is dominated by the Higgs boson contribution. \s

For photons colliding with a total angular momentum $J_Z=0$, the interference
between the signal  $\gamma \gamma \to H \to VV$ and the background $\gamma
\gamma \to VV$ must be taken into account. For $WW$ final states, the
interference is very large: for $M_H \gsim 200$ GeV, this term is negative and
is larger than the resonant contribution from the Higgs boson, leading to a
decrease of the total $WW$ cross section. For $ZZ$ production, the interference
term is rather small, although it has visible effects, resulting for instance
in an asymmetric Higgs resonance.  Thus,  in addition to the extraction of the
$H\gamma \gamma$ couplings as in the $H \to b\bar{b}$ case discussed before,
these processes could in principle allow for the determination of the phase of 
the $H \gamma \gamma$ amplitude via a measurement of the interference term 
which is sensitive to it.   \s

A detailed simulation has been also performed in these two channels 
\cite{gam-HVV-Warsaw} and the analysis follows the same lines as what has 
been previously discussed in the case of $\gamma \gamma \to H \to b\bar b$. 
The cuts have been optimized to select the final states $H \to WW \to q\bar q 
q \bar q$ and $H \to ZZ \to q\bar q \ell \ell$. The center of mass energy of 
the original electron collider has been tuned to optimize Higgs production:
for $\sqrt{s_{ee}}=305\, (500)$ GeV, which is the optimal choice for $M_H=200 
\, (350)$ GeV, a luminosity of about 600 (1000) fb$^{-1}$ can be collected
in a photon collider such as the one discussed for TESLA. Once the 
distributions of the reconstructed invariant masses for $\gamma \gamma \to
WW$ and $ZZ$ are obtained experimentally, one can fit the simulated mass 
distributions with the width $\Gamma_{\gamma \gamma}$ and the phase 
$\phi_{\gamma \gamma}$ as being the only free parameters. \s

The output is shown in Fig.~4.43 where the statistical accuracies expected for 
the $\Gamma_{\gamma \gamma}$ width and the $\phi_{\gamma \gamma}$ phase are 
displayed for four examples of Higgs masses $M_H=200, 250, 300$ and 350 GeV. 
The solid thick light (yellow) line shows for comparison the prediction in a 
specific two--Higgs doublet model (2HDM). As can be seen for low Higgs masses, 
$M_H \sim 200$ GeV, the width can be measured with a precision $\Delta \Gamma_{
\gamma \gamma} \simeq 3\%$ which is similar to the accuracy obtained in the case
of $H\to b \bar b$. For this Higgs mass value, the phase can be measured 
with an accuracy of $\Delta \phi_{\gamma \gamma} \sim 35$ mrad. For higher
Higgs masses, the uncertainties increase and for $M_H=350$ GeV, they are
a factor of three larger. Note that the phase is mainly constrained by the $WW$ 
process as expected, while the width is more accurately measured in the channel 
$ZZ \to q\bar q \ell \ell$ as the background is smaller. Thus, it is
only the combination of the two processes which allow to determine both 
parameters.

\begin{figure}[h]
  \begin{center}
  \epsfig{figure=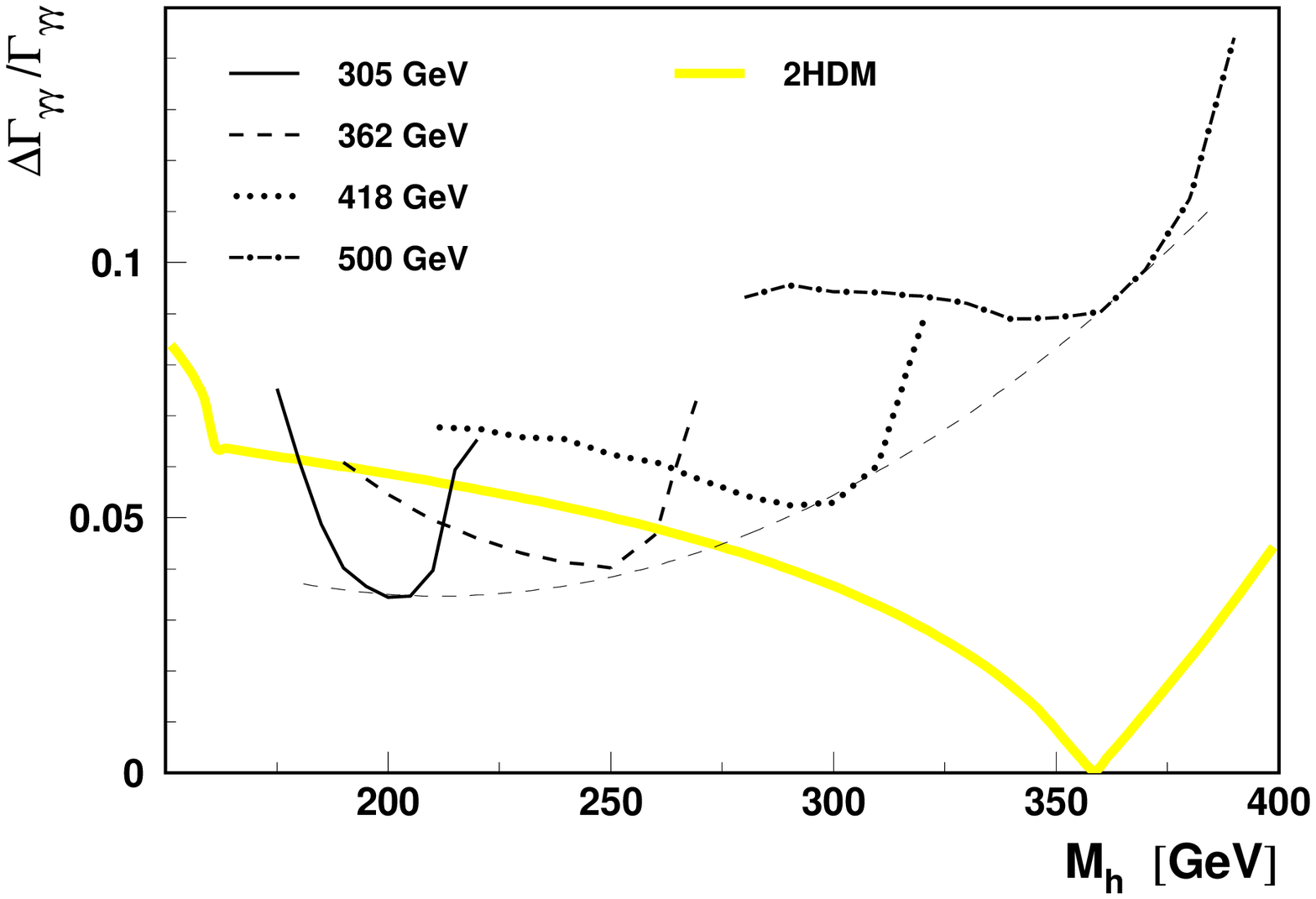,width=8cm,clip=}
  \epsfig{figure=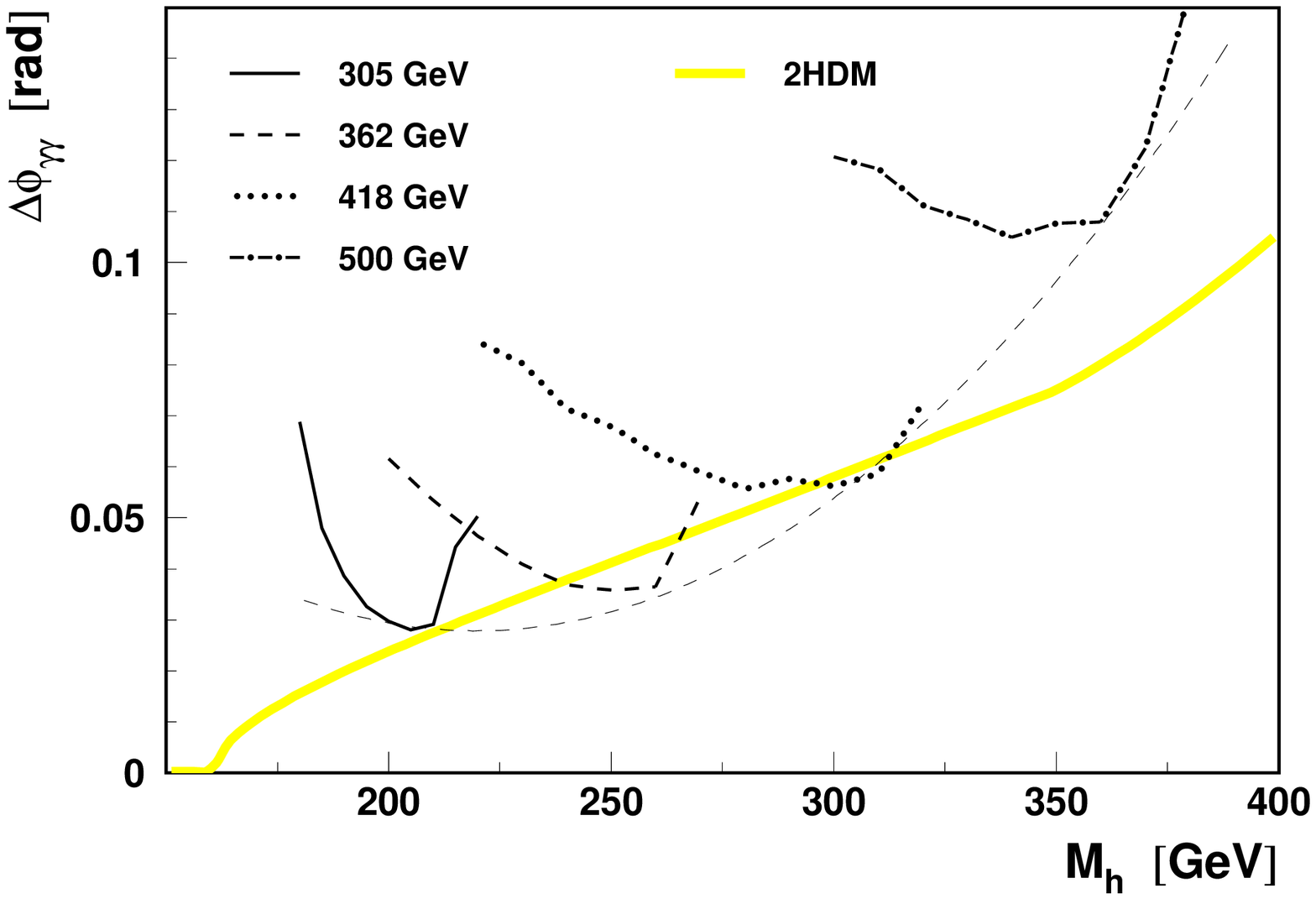,width=8cm,clip=}
  \end{center}
  \vspace{-0.2cm}
{\it Figure 4.43: Expected statistical errors in the determination of the Higgs
width $\Gamma_{\gamma \gamma}$ (left) and phase $\phi_{\gamma \gamma}$ 
(right) at a photon collider, from the simultaneous fit to the observed $W^+ 
W^-$ and $ZZ$ mass spectra. The yellow (thick light) band shows the prediction 
in a 2HDM \cite{gam-HVV-Warsaw}. }
  \vspace{-0.2cm}
\end{figure}

For even heavier Higgs bosons, $M_H \gsim 350$ GeV, the $H \to t\bar{t}$ decays
can be in principle exploited. However, the branching fraction is not very
large, BR($H \to t\bar{t}) \sim 15$\% for $M_H \simeq 400$ GeV, and becomes even
smaller for higher masses. The main background
process $\gamma \gamma \to t\bar{t}$ has a much larger cross section  [which is
still given by eq.~(\ref{ggqq:bckg})] compared to $b$--quark production, first
because of the larger charge $Q_t=+2/3$  with the cross section being
proportional to $Q_q^4$, and second, because the $J_{Z}=0$  contribution is not
suppressed since the top quark mass is of the same order as the effective
$\gamma \gamma$ energy. Furthermore, the total width of the Higgs boson 
becomes too large, $\Gamma_H \sim 30$ GeV  for the previous mass value, and the
particle is not a narrow resonance anymore; because of this large $\Gamma_H$ 
value, one has to integrate the continuum background over a rather large bin. \s

For all these  reasons, the process $\gamma \gamma \to H \to t\bar{t}$ is 
expected to be a rather difficult channel to exploit. However, it can provide 
some valuable information on the CP properties of the produced Higgs particle
\cite{gam-Htt-Roh,gam-Htt-Asa1,gam-Htt-Asa2,gam-Htt-rev}, a subject to which we turn our 
attention now.


\subsubsection{Measuring the CP--properties of the Higgs boson}

\subsubsection*{\underline{Measurements using photon polarization}} 

The general amplitude for the production of a spin--zero Higgs particle in
two--photon collisions, $\gamma \gamma \to H $,  cand be written in terms
of the CP--even and CP--odd components of the $H \gamma \gamma$ coupling which
are proportional to, respectively, $(\epsilon_1 \cdot \epsilon_2)$ and
$(\epsilon_1 \times \epsilon_2)$, as
\beq
M_{\lambda_1 \lambda_2} =  (\epsilon_1 \cdot \epsilon_2) {\cal C}_+ 
\, + \, (\epsilon_1 \times \epsilon_2) {\cal C}_-
\eeq
where ${\cal C}_+\, ({\cal C}_-)$ are the CP--even (odd) contributions to the
amplitude. Four independent functions describe the process out of
the 16 helicity amplitudes present in the general case
\beq
{\rm d} \hat{\sigma}_{00} + {\rm d}\hat{\sigma}_{22} &=& \frac{1}{2} {\rm d}
\Gamma \, \left( \left| M_{++} \right|^2 + \left| M_{--} \right|^2 \right) 
= |{\cal C}_+|^2 + |{\cal C} _-|^2 \non \\
{\rm d} \hat{\sigma}_{20} + {\rm d}\hat{\sigma}_{02} &=& \frac{1}{2} {\rm d}
\Gamma \, \left( \left| M_{++} \right|^2 - \left| M_{--} \right|^2 \right) 
= -2 {\rm Im} \left( {\cal C}_+ {\cal C}^*_- \right) \non \\
{\rm d} \hat{\sigma}_{31} + {\rm d}\hat{\sigma}_{13} &=&  {\rm d}
\Gamma \, {\rm Im} \left( M_{++} M_{--}^* \right) 
= -2 {\rm Re} \left( {\cal C}_+ {\cal C}_-^* \right) \non \\
{\rm d} \hat{\sigma}_{33} + {\rm d}\hat{\sigma}_{11} &=&  {\rm d}
\Gamma \, \left( M_{++} M^*_{--} \right) 
= |{\cal C}_+|^2 - |{\cal C}_-|^2 
\eeq
One can then define the asymmetries \cite{gam-Gun}
\beq
{\cal A}_1 &= & \frac{ \left| M_{++} \right|^2 - \left| M_{--} \right|^2 } 
{\left| M_{++} 
\right|^2 + \left| M_{--} \right|^2} = - \frac{2{\rm Im} ({\cal C}_+ {\cal 
C}_-^*)}{ |{\cal C}_+|^2 + |{\cal C}_-|^2}  \non \\
{\cal A}_2 &=& \frac{ 2 {\rm Im} (M_{++} M^*_{--})} {\left|M_{++} 
\right|^2 + \left| M_{--} \right|^2} = - \frac{2{\rm Re} ({\cal C}_+ {\cal 
C}_-^*)}{ |{\cal C}_+|^2 + |{\cal C}_-|^2}  \non \\
{\cal A}_3 &=& \frac{ 2{\rm Re} (M_{++} M^*_{--})} {\left|M_{++} 
\right|^2 + \left| M_{--} \right|^2} =  \frac{|{\cal C}_+|^2 - |{\cal 
C}_-^*|^2}{ |{\cal C}_+|^2 + |{\cal C}_-|^2}  
\eeq
and write the event rate as ${\rm d}N = {\rm d} {\cal L}^{J_Z=0} {\rm d} 
\hat{\sigma}$ with
\beq
{\rm  d} {\cal L}^{J_Z=0} = {\rm d} {\cal L} \bigg[ 1+\langle
\xi_{12} \xi_{22} \rangle + \langle \xi_{12}+ \xi_{22} \rangle {\cal A}_1
+  \langle \xi_{13} \xi_{21} + \xi_{11} \xi_{23} \rangle {\cal A}_2 
+  \langle \xi_{13} \xi_{23} - \xi_{11} \xi_{21} \rangle {\cal A}_3 \bigg]
\hspace*{0.6cm} 
\eeq
with the unpolarized cross section given by
\beq
{\rm d} \hat{\sigma}_{0}= \frac{1}{4} {\rm d} \Gamma \, \left( \left| M_{++}
\right|^2 + \left| M_{--} \right|^2 \right) 
\eeq
If ${\cal A}_1$ and ${\cal A}_2$ are both non--zero, then, CP is violated since 
the Higgs boson is a mixture of CP--even and CP--odd states. One can thus, by 
analyzing the spins of the final photons, probe CP--violation. If the Higgs 
boson is a definite CP--eigenstate, that is, a pure scalar or pseudoscalar
particle, one has ${\cal A}_1={\cal A}_2=0$ and ${\cal A}_3= \eta_{\rm C}$ with 
$\eta_{\rm CP}=1 \, (-1)$ for a CP--even (CP--odd) Higgs particle. The 
luminosity written above simplifies then to
\beq
{\rm  d} {\cal L}^{J_Z=0} &=& {\rm d} {\cal L} \, \bigg[ 1+\langle
\xi_{12} \xi_{22} \rangle + \eta_{\rm CP} \langle \xi_{13} \xi_{23} - \xi_{11}
\xi_{21} \rangle \bigg]
\eeq
In fact, if CP is conserved, one has $M_{++} = \eta_{\rm CP} M_{--}$ leading to
the relation between cross sections with parallel and orthogonal linear
polarizations for the photons, ${\rm d} \hat{\sigma}_{||}-{\rm d} \hat{\sigma}
_\perp=\eta_{\rm CP} \cdot ({\rm d}\hat \sigma_{||}+{\rm d}\hat \sigma_\perp)$.
This means that  only photons with parallel (orthogonal) linear polarizations
couple to scalars  (pseudoscalars).  Note that only if the lasers are
linearly polarized that it is possible to distinguish between the two CP
quantum numbers  since the relevant average for the Stokes parameters, 
$\langle\xi_{13} \xi_{23}-\xi_{11} \xi_{21} \rangle$, is negligible for 
circularly polarized lasers. \s

In practice, the asymmetry ${\cal A}_3$ is determined by making two runs and  
measuring the difference of the event rates for lasers with parallel
polarization, $\Delta \gamma =0$, and lasers with perpendicular polarization,
$\Delta \gamma= \frac{\pi}{2}$ \cite{gam-Gun}
\beq
\langle {\cal A}_3 \rangle = \frac{ \sigma^{\rm eff} (\Delta \gamma=0) - 
\sigma^{\rm eff} (\Delta \gamma = \frac{\pi}{2})} {\sigma^{\rm eff} (\Delta 
\gamma=0) + \sigma^{\rm eff} (\Delta \gamma = \frac{\pi}{2})}
\eeq
where the contamination from the background is taken into account
$\sigma^{\rm eff}= \sigma^{\rm eff}_{\rm signal}+ \sigma^{\rm eff}_{\rm bckg}$.
In terms of the electron and laser beam polarization, the asymmetry is given by
\beq
\langle {\cal A}_3 \rangle \simeq \frac{ 
\eta_{\rm CP} \, \sigma_{\rm signal}  \, P_{1t} P_{2t} \langle \ell_1 \ell_2 
\rangle} {\frac{1}{2} (1+ 4\lambda_{e^-} \lambda_{e^+} \langle c_1 c_2 \rangle)
(2 \hat{\sigma}^{\rm signal}+ \hat{\sigma}_0^{\rm bckg}) +
\frac{1}{2} (1- 4\lambda_{e^-} \lambda_{e^+} \langle c_1 c_2 \rangle )
\hat{\sigma}_2^{\rm bckg} }
\eeq
where the effects for $\rho \neq 0$ have been ignored for simplicity [there is
also a generally small contribution to the background in the numerator from 
the component  $\hat{\sigma}^{\rm bckg}_{||} - \hat{\sigma}^{\rm bckg}_{\perp} 
\propto m_q^4/W^4$]. As can be  seen, a very important role is played by the
linear laser polarization  $P_{it}$, the average of the induced linear
polarizations of the photons  [the asymmetry is directly proportional to the
product] and by the  longitudinal polarizations of the electron beams and the
induced circular polarization of the photons. \s

The statistical significance of the signal is given by 
\beq
N_{\rm SD} ({\cal A}_3) = \frac{ |\sigma^{\rm eff} (\Delta \gamma=0) - 
\sigma^{\rm eff} (\Delta \gamma  = \frac{\pi}{2}) |} { \sqrt{\sigma^{\rm eff} 
(\Delta \gamma=0) + \sigma^{\rm eff} (\Delta \gamma = \frac{\pi}{2}) }}
\times \sqrt{ {\cal L}^{\rm eff}}  
\eeq
With the machine parameters, polarization and luminosity discussed above, a
measurement at the level of 10\% can be made, allowing the
distinction between the two CP possibilities for the Higgs particle; see 
Ref.~\cite{gam-Gun} for an analysis where a realistic luminosity spectra and 
photon polarizations are taken into account. 

\subsubsection*{\underline{Measurements using angular distributions}} 

Another way to test the CP nature of the produced Higgs particle is to study 
the angular distributions in its decays. For a relatively heavy Higgs boson, 
$M_H \gsim 2M_Z$, this can be done in the final state $H \to WW,ZZ\to 4f$ in 
which, as discussed in \S2.2.4, the angular correlations between the final 
state fermions are very different in the case of scalar and pseudoscalar Higgs
particles. For instance, the azimuthal dependence on the angle $\Delta \phi$ 
between the decay planes of the two vector bosons is characteristically 
different in the $0^{++}$ and $0^{+-}$ cases \cite{Bargeretal,CPHVVchoi}. 
Another different observable is the correlation 
\beq
\zeta_{VV} =  \frac{\sin^2\theta_1\; \sin^2\theta_3} {(1+\cos^2\theta_1) \;
(1+\cos^2\theta_3)} 
\eeq 
where $\theta_1$ and $\theta_3$ are the polar angles of the two fermions from
the $V \to f\bar f$ decays defined in Fig.~2.11 and which corresponds to
the ratio of the angular distributions expected for the decay of a scalar and a
pseudoscalar particle in the limit $M_H \gg M_V$. \s 

A detailed simulation in the decay channels $H \to ZZ \to \ell \ell jj$ and $H
\to WW \to 4j$ has been performed  \cite{gam-HVV-Warsaw} along the same lines 
as for
the measurement of the amplitudes of the $HVV$ couplings and their phases 
discussed earlier. The output of this analysis is shown in Fig.~4.44 for the 
example of a Higgs boson with a mass $M_H=200$ GeV, produced in the $\gamma 
\gamma$ mode of an $\ee$ collider with initial energy of $\sqrt{s_{\ee}}=
305$ GeV and decaying into $ZZ \to \ell \ell jj$ final states.  The figure 
shows the number of expected events for a scalar and pseudoscalar Higgs boson
and for the non--resonant SM background, for a variation of the reconstructed 
azimuthal angle $\Delta \phi_{ZZ}$ (left) and the correlation $\zeta_{ZZ}$ 
(right). The points with error bars indicate the statistical precision of the 
measurements after a one year running of the photon collider with a luminosity 
of 600 fb$^{-1}$. \s

\begin{figure}[!h]
\vspace*{-2mm}
\begin{center}
\epsfig{figure=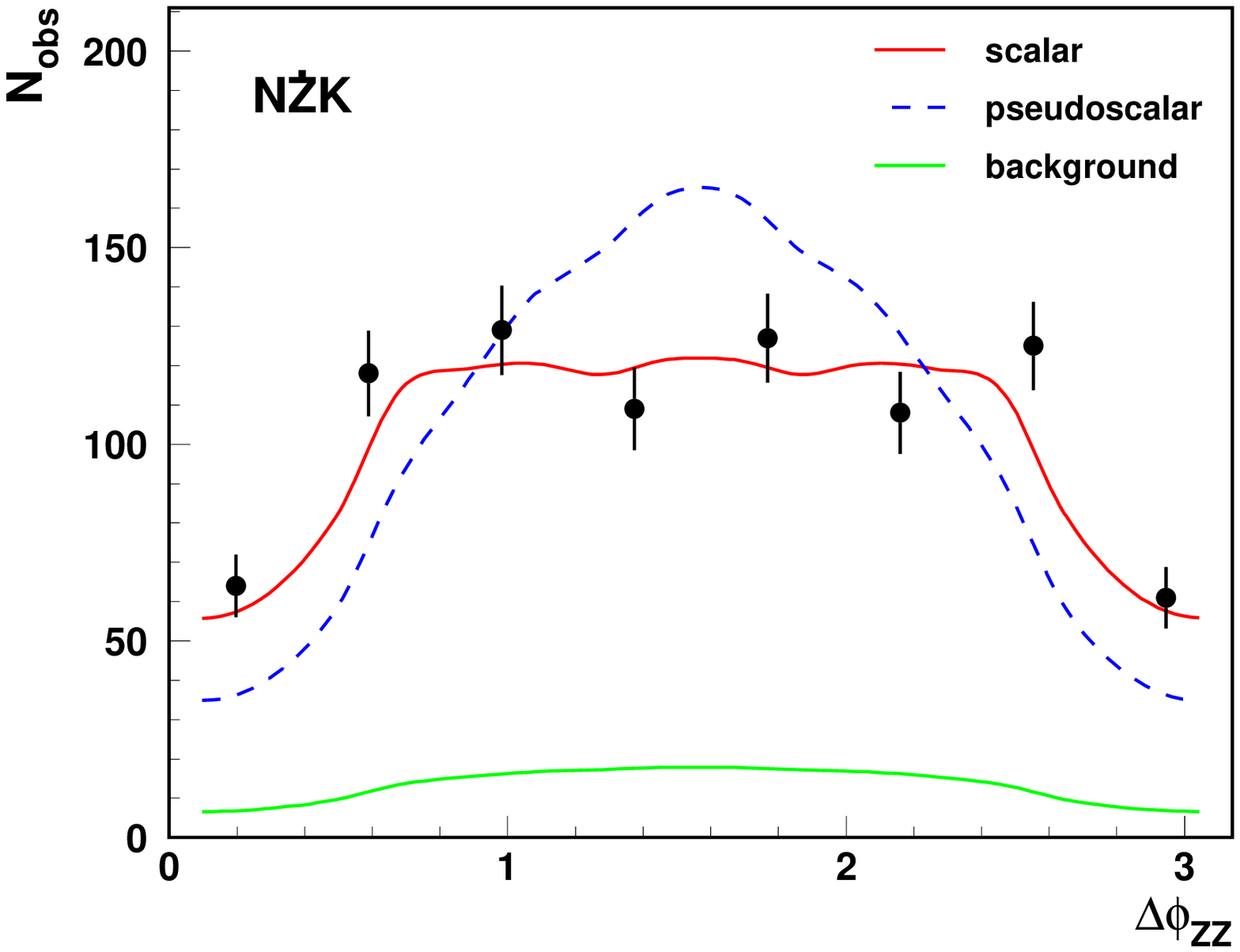,width=7.7cm,clip=}
\epsfig{figure=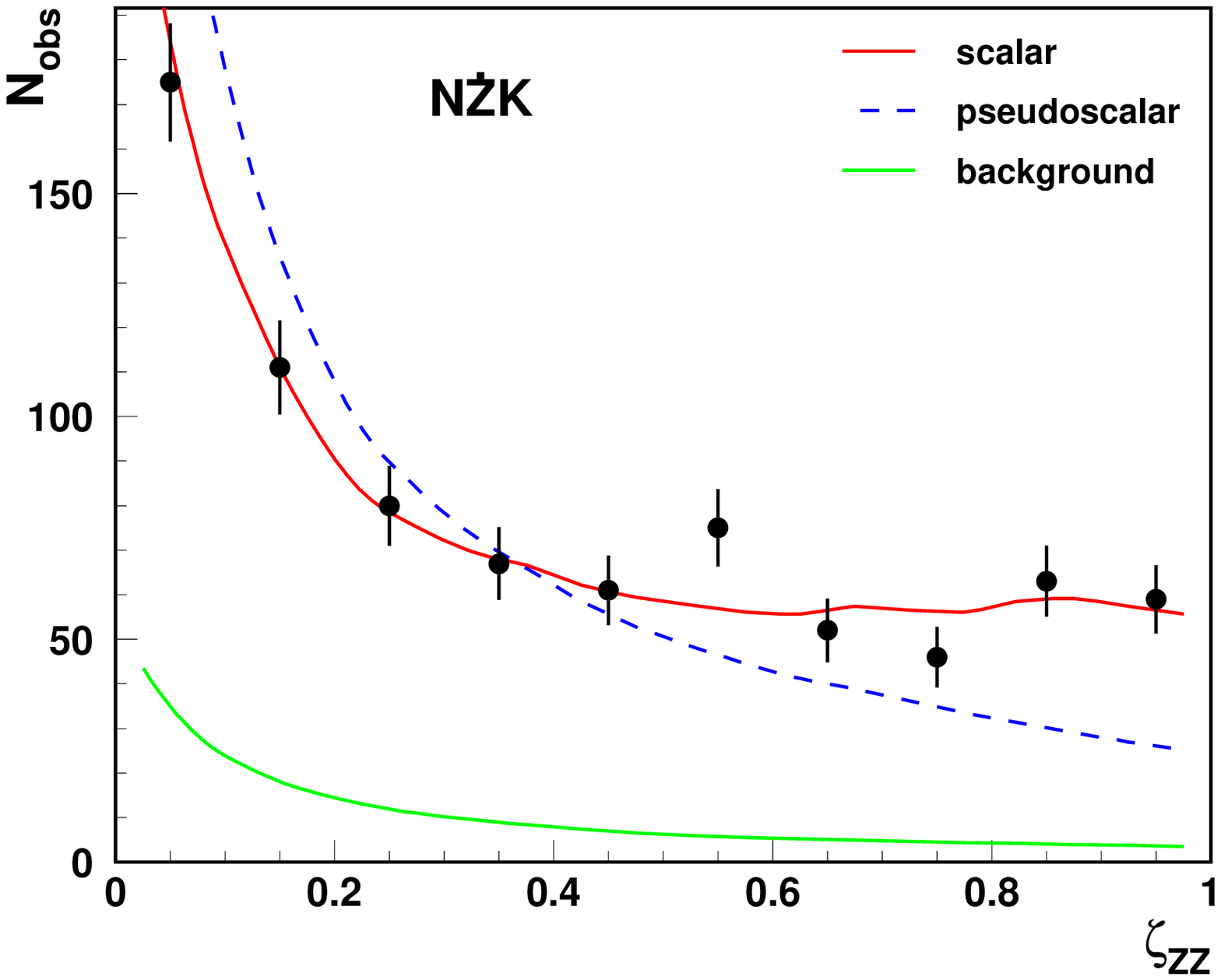,width=7.7cm,clip=}
\end{center}
\vspace*{-5mm}
 \nn {\it Figure 4.44: The measurement of the azimuthal angle  $\Delta 
\phi_{ZZ}$ and the correlation $\zeta_{ZZ}$ in the process $\gamma \gamma
\to H \to ZZ \to \ell \ell jj$ for $M_H=200$ GeV at a photon collider; from 
\cite{gam-HVV-Warsaw}. }
\vspace*{-5mm}
 \end{figure} 

If the $HVV$ coupling, including CP--violation, is parameterized as  $g_{HVV}= 
\lambda \cos \Phi$ with $\lambda=1$ and $\Phi=0$ in the SM, from a combined fit
of the $H \to WW/ZZ \to 4f$ events which includes a free variation of the two 
photon width and phase one can measure the absolute magnitude of the
coupling with a precision $\Delta \lambda/\lambda =2\%$ and the CP--violating
phase with a precision $\Delta \Phi= 50$ mrad, in the entire Higgs mass
range $M_H=200$--350 GeV \cite{gam-HVV-Warsaw}. \s 

Similar tests can be performed in the decay $H\to t\bar t$ for a heavier Higgs 
particle, $M_H >350$ GeV. In particular, the interference 
pattern of the resonant and the continuum amplitudes for the $\gamma\gamma 
\ra t \bar t$ process allows to check the parity of the Higgs boson and
the presence of CP--violation, by using circularly polarized colliding photons 
\cite{gam-Htt-Asa1}. Indeed, from the $t \bar t$ decay angular distribution 
one can built four convoluted observables $\Sigma_{1..4}$ 
\beq
\Sigma_i (\sqrt{s}_{\gamma\gamma})= \int {\rm d}\sqrt{s}_{\gamma\gamma} 
\sum_{\lambda_1,~\lambda_2} \left( \frac{1}{{\cal L} }\frac{ {\rm d}{\cal L}^{
\lambda_1\lambda_2}} {{\rm d} \sqrt{s}_{\gamma\gamma}} \right) \left( 
\frac{3\beta} {32\pi s_{\gamma\gamma}}
\int S^i_{\lambda_1 \lambda_2} (\theta, \sqrt{s}_{\gamma\gamma}){\rm d}
\cos\theta \right)
\end{eqnarray}
with $\theta$ being the polar angle of the $t$ momentum in the $\gamma\gamma$ 
c.m. frame and the first bracket corresponding to the normalized luminosity 
distribution for each of the photon $\lambda_1 \lambda_2$ helicity combinations.
The functions $S_{\lambda_1 \lambda_2}^i$ contain the information on the 
$\gamma\gamma \ra t\bar t$ helicity amplitudes
\beq
S^1_{\lambda_1 \lambda_2} = \left| M_{\lambda_1\lambda_2}^{RR} \right|^2\, , \
S^2_{\lambda_1 \lambda_2} = \left| M_{\lambda_1\lambda_2}^{LL} \right|^2\, , \
S^{3(4)}_{\lambda_1 \lambda_2} = 2{\rm Re} ({\rm Im}) \left[ 
M_{\lambda_1\lambda_2}^{RR} M _{\lambda_1\lambda_2}^{LL*} \right]
\eeq 
Writing the $\gamma \gamma \to t\bar t$ amplitudes as sums of the resonant 
and non--resonant contributions
\begin{eqnarray}
M_{\lambda \lambda}^{\sigma \sigma}= \left[ M_t \right]_{\lambda \lambda}^{
\sigma \sigma} + \left( \frac{\sqrt{s}_{\gamma\gamma}}{M_H} \right)^3
r_H \cdot i \left[ 1+{\rm exp} \left( 2i \tan^{-1} \frac{s_{\gamma\gamma}^2
-M_H^2}{M_H \Gamma_H} \right) \right]
\end{eqnarray}
the phase of the resonance amplitude is shifted by $r_H$ which is 
essentially the phase of the $\gamma\gamma H$ coupling when neglecting the 
phase in the $t \overline{t} H$ vertex. In the left--hand side of Fig.~4.45,
the four observables $\Sigma_{1..4 }$ for the production of scalar $H$ and 
pseudoscalar $A$ bosons with $M_{H,A}=400$ GeV, are shown for two values of the 
$\gamma\gamma H/A$ phase, arg$(r_{H,A})=0$ and ${\pi \over 4}$, and one can see
that the difference is significant enough to be measured experimentally. \s
\begin{figure}[ht]
\begin{center}
\begin{tabular}{ll}
\mbox {\epsfxsize=6.5cm \epsfysize=6.5cm \epsffile{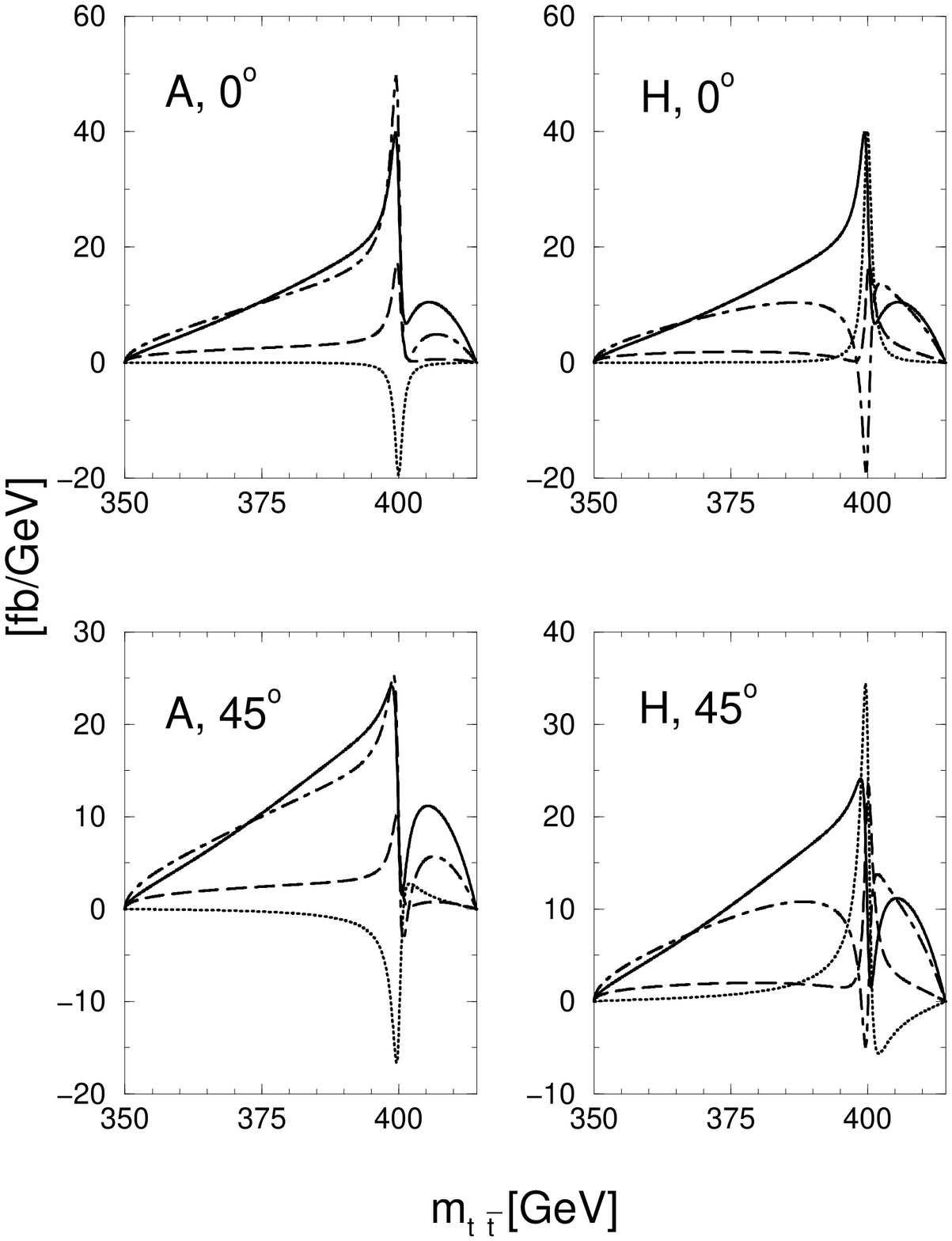} } 
\hspace*{-5mm} & \hspace*{-5mm} 
\includegraphics*[scale=0.45]{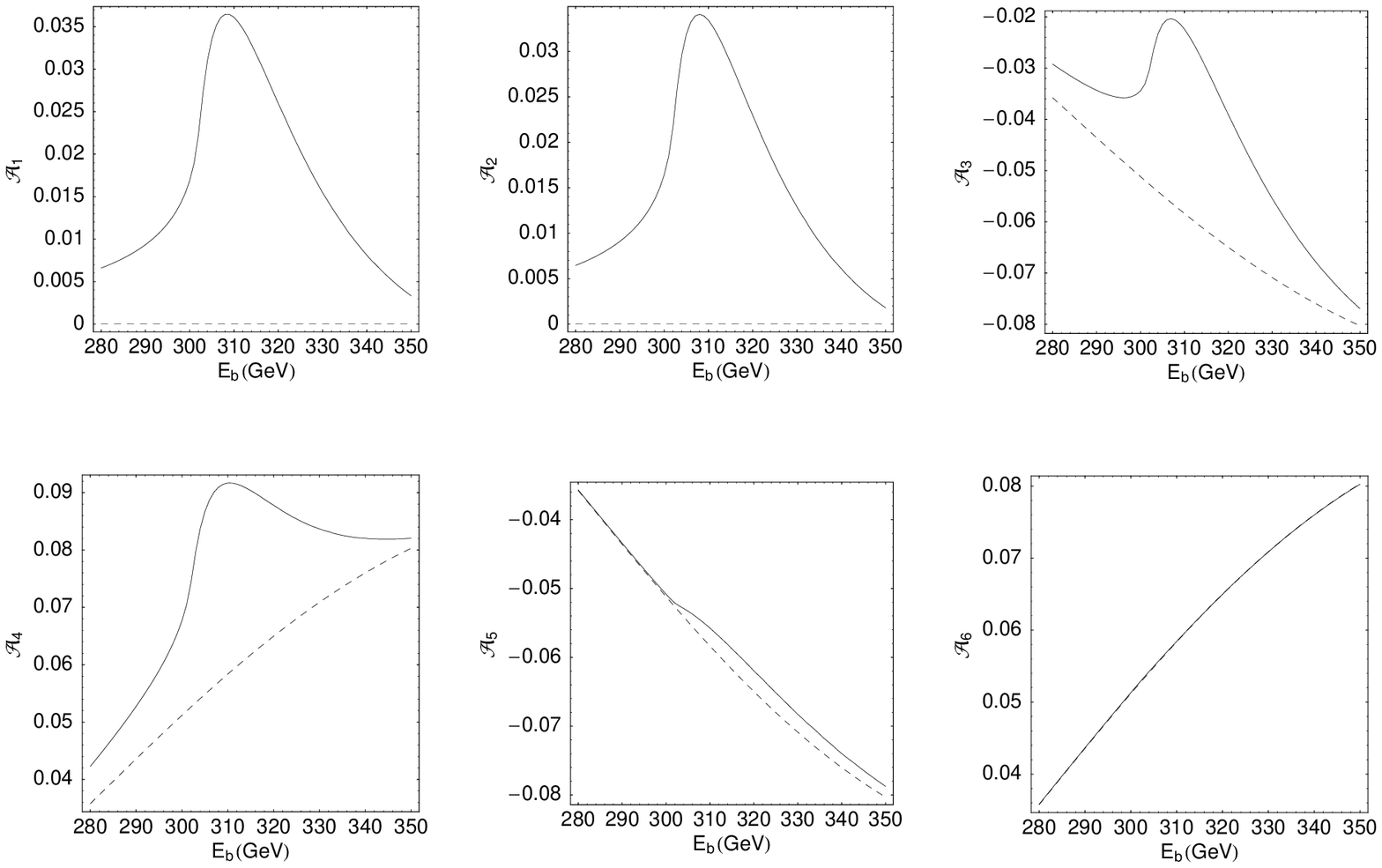}
\end{tabular}
\end{center}
\vspace*{-3mm}
\nn {\it Figure 4.45: Left: The observables $\Sigma_{1}$ (solid), $\Sigma_2$ 
(dashed), $\Sigma_3$ (dot--dashed) and $\Sigma_4$ (dotted) for the production 
of $H$ and $A$ bosons for  ${\rm arg}(r_{H,A})=0$ and ${\pi \over 4}$ with
$x=4.8$, $P_L=-1.0$ and $P_e=0.9$; from  Ref.~\cite{gam-Htt-Asa1}. Right:
The asymmetries ${\cal A}_{1..6}$ as a function of the $e^-$ beam energy
for continuum $\gamma \gamma \to t \bar t$ production (dotted line) and
when the resonant contribution with $M_{H,A}=500$ GeV is included (solid lines);
from Ref.~\cite{gam-Htt-Roh}.}
\vspace*{-3mm}
\end{figure}

Another possibility to probe the Higgs CP--quantum numbers in $\gam \to t\bar 
t$ production is to look at the net polarization of the $t/\bar t$ quarks
either with circularly polarized \cite{gam-Htt-Asa2} or linearly polarized 
photons \cite{gam-Htt-Roh}. In the latter case, the top polarization has been 
analyzed through the decay lepton energy and angular distributions in the decay
$t\to b \ell \nu$. The full differential distribution of the decay lepton
has been written and, in terms of the initial state $\ee$ polarizations
$\lambda_{\ee}=\pm1$ and final charge   of the decay lepton $e_{\ell^\pm}=
\pm 1$, one can obtain four cross sections $\sigma (\pm, \pm)$ from which
one can construct six asymmetries that are sensitive to the Higgs coupling
\cite{gam-Htt-Roh}
\beq
{\cal A}_{1/4}=  \frac{\sigma(+,\pm)-\sigma(-,-)}{\sigma(+,\pm)+\sigma(-,-)}
\, , \ 
{\cal A}_{2/3}=  \frac{\sigma(+,\mp)-\sigma(-,+)}{\sigma(+,\mp)+\sigma(-,+)}
\, , \ 
{\cal A}_{5/6}=\frac{\sigma(\pm,+)-\sigma(\pm,-)}{\sigma(\pm,+)+\sigma(\pm,-)}
\label{asymm}
\eeq
${\cal A}_{5/6}$ are charge asymmetries for a given polarization and 
vanish for zero--angle, which is not the case for the purely CP--violating
${\cal A}_{1/2}$ asymmetries;  ${\cal A}_{3/4}$ are the polarization
asymmetries for a given lepton charge. Note that the charge asymmetries do not
vanish in the case of the SM where only the non--resonant amplitude is taken
into account. The sensitivity of the six asymmetries to the $\gamma\gamma H/A$
coupling and to its possible CP--violating component is exhibited in the 
right--hand side of Fig.~4.45 for a specific point with $M_{H,A}=500$ GeV 
and a $\gamma \gamma H/A$ vertex which has both real and imaginary parts, 
as a function of the electron beam energy. As can be seen, the asymmetries can 
be large and in most cases different from  the asymmetries of the continuum
$\gamma \gamma \to t \bar t$ production. \s 

Finally, for  $M_H \lsim 140$ GeV, one can also study the CP--nature of the 
Higgs boson by looking at the polarization of the $\tau$ leptons produced in 
$\gamma \gamma \to H \to \tau^+ \tau^-$. One can again construct polarization 
asymmetries which probe  both the $H\gamma \gamma$ and $H\tau \tau$ couplings 
\cite{gam-Hll-Singh}

\subsubsection{Other Higgs production mechanisms}

Other processes than Higgs boson production as s--channel resonances have been
discussed in the context of $\gam$ colliders: Higgs pair production via
loop diagrams \cite{gam-HH,gam-HH1}, production  in association with vector
bosons \cite{gam-WWH, gam-WWHH,gam-HZ} and associated Higgs production
with top quarks \cite{gam-ttH}. At $e\gamma$ colliders, the Higgs boson can
also be produced in the reaction $e\gamma \to \nu_eW^+ H$
\cite{egam-H,egam-Hall}.  In this section, we briefly summarize the main
features of these processes, concentrating only on the magnitude of the cross
sections of the subprocesses [i.e. without folding with the photon luminosity
spectra].  

\subsubsection*{\underline{Higgs pair production: $\gamma \gamma \to HH$}}

The pair production of Higgs bosons in $\gamma \gamma$ collisions is induced by
loops of top quarks and $W$ bosons where two sets of diagrams are involved:
$(i)$ $s$--channel vertex diagrams where the intermediate Higgs particle splits
into two and which involves the trilinear Higgs coupling $\lambda_{HHH}$;
the contributions of these diagrams are essentially the same as those discussed
for $\gamma \gamma \to H$, except that here the Higgs particle is virtual, and 
($ii)$ box diagrams involving top quarks and $W$ bosons, as well as
vertex and self--energy diagrams which do not involve the trilinear Higgs
coupling but the quartic Higgs--gauge boson interaction. \s

The cross section has been calculated in Ref.~\cite{gam-HH} in the SM case; see
also Ref.~\cite{gam-HH1}. At small energies, $\sqrt{s}_{\gamma \gamma} \sim
500$ GeV, it is dominated by the top quark contribution. For photons having the
same helicities,  it is at the level of $\sim 0.5$ fb for $M_H \sim 100$ GeV,
decreases very slowly with $M_H$ and falls--off rapidly when approaching the
$\sqrt{s}_{\gamma \gamma} =2M_H$ threshold.  At higher energies,
$\sqrt{s}_{\gamma \gamma} \gsim 1$ TeV, the cross section is dominated by the
$W$ boson loop contribution which, contrary to the case of single Higgs boson,
interferes constructively with the top quark contribution for large enough
$M_H$.  While the cross section  is smaller than at 500 GeV for low $M_H$, it
increases with $M_H$ almost up to the kinematical boundary, where it reaches
values of the order $1\, (10)$ fb at $\sqrt{s}_{\gamma \gamma} \sim 1\, (2)$
TeV.  This is mainly due to the large triple and quartic Higgs couplings to the
Goldstone or $W_L$ bosons which grow as $M_H^2$.\s 

 For opposite photon helicities, the cross section has the same magnitude as 
in the same--helicity case for $M_H\sim 100$ GeV, but because in this case it  
is dominated by the contributions of transverse $W$ bosons it  falls off more
rapidly with increasing $M_H$ values even for high center of mass energies. At 
$\sqrt{s}=2$ TeV, there is bump for a very heavy Higgs boson. \s

The sensitivity of the production cross section to the trilinear Higgs coupling
$\lambda_{HHH}$ depends on the relative weight of the diagram with the exchange
of the Higgs boson in the $s$--channel and the other diagrams
\cite{gam-HH1}. For very heavy
Higgs bosons, $M_H \sim 500-800$ GeV, the cross section is very sensitive to the
coupling $\lambda_{HHH}$, in particular near the $\sqrt{s}_{\gamma \gamma} =2
M_H$ threshold where it is maximal: for $M_H \sim 700$ GeV, removing the
trilinear coupling leads to an increase of the cross section [which is
unfortunately rather small, being less than 1 fb] by about 60\%.  For smaller
$M_H$ values, the sensitivity is much weaker since the cross section at high
energies [where it is sizable] is dominated by the box contributions which do
not involve $\lambda_{HHH}$, while at low energies the rates are too small. 
Note finally, that a change of the trilinear Higgs coupling does not affect the
angular distribution of the Higgs pair production process. \s

Thus, at very high energies and for rather heavy Higgs bosons, on can  possibly
probe the trilinear Higgs coupling in the process $\gamma \gamma \to HH$. This
is complementary to the $\ee$ case where the coupling can be best probed for
low Higgs boson masses. However, to assess to which extent the coupling can be
measured, more detailed analyses are needed.  

\subsubsection*{\underline{Higgs production in association with top quarks: 
$\gamma \gamma \to t\bar{t} H$}}

The process $\gamma \gamma \to t\bar{t}H$ offers an additional opportunity to
probe the Yukawa coupling of the Higgs boson to top quarks 
\cite{gam-ttH,gam-ttH0}. Contrary to the
similar process in the $\ee$ case, where the Higgs boson can be radiated not 
only from the top quark lines but also from the $Z$ line in the Higgs strahlung
like process $\ee \to HZ^* \to Ht\bar{t}$ [although the  contribution of the
latter is very tiny as discussed earlier], in associated $Ht\bar{t}$ 
production in photon--photon collisions, the Higgs boson is only radiated 
from the top quark lines and the cross section is directly proportional to
the $Ht\bar{t}$ Yukawa coupling. \s

As in the case of $\ee$ collisions, the cross section for $t\bar{t}H$
production is rather small at $\sqrt{s}_{\gamma  \gamma} \sim 500$  GeV,
because of the limited phase space. It increases with energy and for $M_H \sim
100$ GeV it reaches the level of $\sigma (\gamma \gamma \to t\bar{t}H) = {\cal
O}(1~{\rm fb})$ at $\sqrt{s}_{\gamma \gamma} \sim 1$ TeV, where it begins to
flatten [this is opposite to the $\ee$ case, where $\sigma \propto 1/s$]. The 
cross section drops rapidly with increasing $M_H$ and
at a c.m. energy of 1 TeV it is one order of magnitude smaller for $M_H \sim
200$ GeV than for $M_H \sim 100$ GeV. \s

The $\gamma \gamma \to t\bar{t}H$ process can be used as a means to
determine the CP properties of the Higgs boson and to distinguish between
scalar and pseudoscalar particles and to probe CP--violation.  
In addition, associated Higgs production with lighter fermions, such as
$\tau$--leptons and $b$--quarks, which have larger cross sections in extensions
of the SM where the Higgs couplings to down--type fermions are enhanced, has
been discussed \cite{bb-H-gamma}.  

\subsubsection*{\underline{Higgs production in association with gauge 
bosons}}

As mentioned previously, the $\gam \to W^+ W^-$ production cross section is
enormous, being at the level of ${\cal O}(100~{\rm pb})$ for c.m. energies
around $\sqrt{s}_{\gamma \gamma} \sim 300-500$ GeV \cite{gam-WW}, and one could
attach one or even two additional Higgs bosons to the $W$ lines, while still
having sizable rates \cite{gam-WWH}. For a Higgs boson with a mass $M_H \sim
100$ GeV, the cross section for $\gamma \gamma \to W^+W^-H$ is about 20 fb for
$\sqrt{s}_{\gamma \gamma}= 500$ GeV and,  therefore, it is at level of the
cross section for the Higgs--strahlung process in $\ee$ collisions with the
same c.m.~energy. The cross section quickly rises with energy, to reach the
level of 400 fb for $\sqrt{s}_{\gamma \gamma}=2$ TeV, i.e. almost two orders of
magnitude larger than the Higgs--strahlung cross section which drops like
$1/s$, and of the same order as the dominant production mechanism, $\ee \to
H\nu \bar{\nu}$.  Compared to the processes for  associated Higgs production
with gauge bosons in $\ee$ collisions discussed previously, $\sigma( \gamma
\gamma \to W^+W^-H)$ is a factor of three larger than any of the $\ee \to HVV$
processes.  Note however, that this process does not provide any additional
information that could not be obtained in the $\ee$ option of the machine. \s

A channel that is, in principle, more interesting is when two Higgs particles
are produced in association with the $W$ boson pair. Indeed, similarly to the
$WW$ fusion mechanism $WW \to HH$, this process is sensitive to the trilinear
Higgs boson coupling since the Higgs particle produced in $\gamma \gamma \to
WWH$ can split into two Higgs bosons.  Unfortunately, the rates are too small
to be useful even with very high luminosities \cite{gam-WWHH}: for $\gam$
energies of the order of 1 TeV, $\sigma (\gamma \gamma \to W^+W^- HH) \sim
0.02$ fb for $M_H \sim 100$ GeV, and barely reaches 0.2 fb at $\sqrt{s}_{\gamma
\gamma}\sim 2$ TeV. Finally, note that the Higgs bosons can be also produced in
association with a $Z$ boson in the loop induced process $\gamma
\gamma \to HZ$ \cite{gam-HZ} where, in particular,  virtual top quarks and $W$ 
bosons contribute.  The cross section are, however, rather small: for
$\sqrt{s}_{\gamma \gamma}= 500$ GeV and $M_H \sim 100$ GeV, they are at the
level of 0.1 fb.  

\subsubsection*{\underline{Higgs production in $e\gamma$ collisions}}

Finally, let us close this discussion on Higgs physics at the photonic 
mode of future $\ee$ linear colliders by considering the other possible
option, the $e \gamma$ mode, that can be obtained by converting only one of
the electron beams into a very energetic back--scattered photon.  Higgs bosons
can be produced in $e\gamma$ collisions through bremsstrahlung off the $W$
lines,  $e^- \gamma \to \nu_e W^- H$ \cite{egam-H,egam-Hall}; the relevant 
diagrams are shown in Fig. 4.46. \s

\begin{center}
\hspace*{-1cm}
\vspace*{-1.cm}
\SetWidth{1.}
\begin{picture}(300,100)(0,0)
\ArrowLine(0,25)(35,50)
\Photon(0,75)(35,50){3.2}{5.5}
\ArrowLine(35,50)(80,50)
\Photon(80,50)(120,75){3.2}{5.5}
\ArrowLine(80,50)(120,25)
\DashLine(105,65)(130,47){4}
\Text(-10,30)[]{$e^-$}
\Text(-10,70)[]{$\gamma$}
\Text(55,65)[]{$e^-$}
\Text(125,20)[]{$\nu$}
\Text(130,80)[]{$W^-$}
\Text(137,55)[]{\bH}
\Text(105,65)[]{\bb}
\hspace*{1.5cm}
\ArrowLine(145,25)(185,25)
\ArrowLine(185,25)(230,25)
\Photon(145,75)(185,75){3.2}{5.5}
\Photon(185,75)(230,75){3.2}{5.5}
\Photon(185,75)(185,25){3.2}{7.5}
\DashLine(185,50)(230,50){4}
\Text(187,47)[]{\bb}
\Text(120,-2)[]{\it Figure 4.46: Diagrams for Higgs boson production in 
$e\gamma$ collisions.}
\vspace*{0.mm}
\end{picture}
\end{center}
\vspace*{1cm}

For a low mass Higgs boson, $M_H \sim 100$ GeV, the cross section for the
subprocess [again without folding with the photon spectrum] is at the level of
$\sim 40$ fb for $\sqrt{s}_{e\gamma}=500$ GeV and increases monotonically to
reach values of the order of 100 (300) fb for $\sqrt{s}_{e \gamma}= 1\, (2)$
TeV; i.e. the rates are comparable to those of the $WW$ fusion in $\ee$
collisions at high energies. While the variation of the cross section with the
Higgs mass is rather pronounced at low energies [$\sigma (e\gamma \to
\nu_e WH)$ drops by a factor of two when increasing $M_H$ from 100 to 150 GeV,
as a result of phase space reduction], it is very mild at higher energies. When
convoluting the cross sections  with the back--scattered photon flux, they are
reduced by about $50$\%  at $\sqrt{s_{e\gamma}}=500$ GeV and slightly less at
higher energies \cite{egam-Hall}.\s

The large Higgs production rates in this process could allow to perform
an independent determination of the $HWW$ couplings [which can be made already
in the  $\ee \to H \nu \bar{\nu}$ production and $H \to WW$ decay process if
the Higgs is not too heavy] and to probe anomalous contributions.
However, the environment of the collision should be well under control to match
the accuracy which can be achieved in the clean $\ee$  mode of the linear
collider. 

\subsection{Higgs production at muon colliders}

The ability of a future $\mu^+ \mu^-$ collider to investigate the Higgs sector
of the SM and its  extensions has been discussed in numerous papers; see for
instance the detailed reviews of
Refs.~\cite{mu-machine1,mu-machine2,mu-Rev1,mu-Rev2,mu-Rev3,mu-Rev4}.  In this
section, we simply summarize the main studies which have been performed
in this context, concentrating on the benefits of such a collider compared to
$\ee$ linear colliders for determining the properties of the Higgs particle.  

\subsubsection{Higgs production in the $s$--channel}

\subsubsection*{\underline{Resonant Higgs production at the tree--level}} 

In $\mu^+ \mu^-$ collisions, the resonance production cross section for a Higgs
boson decaying into a final state X is given, in terms of the partial decay 
widths, by
\beq
\sigma_H (\sqrt{s})= \frac{ 4\pi \Gamma(H \to \mu^+ \mu^-) \Gamma (H \to X)}
{(s-M_H^2)^2 + M_H^2 \Gamma_H^2}
\eeq
In practice, however, on has to include the Gaussian center of mass energy
spread $\sigma_{\sqrt{s}}$. Assuming a central c.m. energy value $\sqrt{s}$,
one obtains after convolution \cite{mu-schannel}
\beq
\overline{\sigma}_H (\sqrt{s}) = \frac{1}{2\pi \sigma_{\sqrt{s}}} \int \sigma_H (\sqrt{
\hat{s} }) \ {\rm exp} \left(  \frac{ - \left( \sqrt{\hat{s}} - \sqrt{s}
\right)^2 } {2 \sigma^2_{\sqrt{s}} } \right) \, {\rm d} \sqrt{\hat{s}}
\eeq
which, when the energy is tuned to the Higgs boson mass value, gives
\begin{equation}
\overline{\sigma}_H(\sqrt{s} \simeq M_H) = \frac{4\pi}{M^{2}_{H}}\frac{{\rm BR}  (H
\rightarrow \mu^+ \mu^-) {\rm BR} (H \rightarrow X)} {\Bigl[ 1+ \frac{8}  {\pi}
\Bigl(\sigma_{\sqrt{s}}/ \Gamma_H \Bigr)^2\Bigr]^{1/2}}					 \end{equation}
If the energy spread is much smaller than the Higgs boson total decay width,
the effective cross section is simply given
by 
\beq
\sigma_{\sqrt{ s}} \ll \Gamma_H \ : \quad 
\overline{\sigma}_H \simeq \frac{4\pi}{M_H^2} \, {\rm BR}(H \to \mu^+\mu^-)
{\rm BR}(H \to X)
\label{mu-gamma-small}
\eeq
while in the opposite case, the effective cross section reads
\beq
\sigma_{\sqrt{ s}} \gg \Gamma_H \ : \quad 
\overline{\sigma}_H \simeq \frac{4 \pi^2}{M_H^2}  \Gamma (H \to \mu^+ \mu^-)
{\rm BR} (H \to X) \times \frac{1} {2 \sqrt{2\pi} \sigma_{\sqrt{s}} } 
\label{mumuxsection2}
\eeq
One needs therefore a very small resolution to maximize the Higgs boson 
production rate. \s

Recalling that there is a trade between the luminosity delivered by the machine
and the energy resolution $R$ of the muon beams, \S4.1.3, the production rate  
can be maximized by choosing $R$ such that the energy spread $\sigma_{\sqrt{s}}$
is slightly  smaller than the Higgs boson total decay width, $\sigma_{\sqrt{s}}
\lsim \Gamma_H$, which in the SM corresponds to $R=0.003$\% for $M_H \lsim 120$
GeV. The energy spread can  be then more conveniently written as \cite{mu-Rev1}
\begin{equation}
\sigma_{\sqrt{s}} = 0.002 \, {\rm GeV} \, \Bigl( \, \frac{R}{0.003\%}\Bigr) \, 
\Bigl(\frac{\sqrt{s}}{100 \, \rm {GeV} }\Bigr)
\end{equation}
For Higgs bosons in the low mass range, $M_H \lsim 130$ GeV, a small resolution
$R=0.003$\% would be more advantageous.  In the intermediate Higgs
mass range, 130 GeV $\lsim M_H \lsim$ 160 GeV,  the Higgs boson is broad enough
and  one can use a resolution $R = 0.01$\% without too much loss of production
rates. In such a case, the cross sections are functions of the Higgs branching
fractions and Higgs masses and practically do not depend on $R$; this is even 
more true for Higgs bosons in the  high mass range, $M_H \gsim 180$ GeV. 
[See Table 2.1, for the Higgs total width and branching ratios for 
selected values of $M_H$.]

\subsubsection*{\underline{$\mu^+ \mu^- \to b\bar b$ and the radiative 
corrections}} 

For a light Higgs boson, $M_H \lsim 140$ GeV, the dominant decay  is $H \to b
\bar b$ and one has to consider the full process $\mu^+ \mu^- \to b\bar b$
which receives contributions from the resonant $\mu^+ \mu^- \to H \to b\bar b$
channel and continuum $\mu^+ \mu^- \to \gamma, Z \to b\bar b$ production;
Fig.~4.47a.  The latter is mediated by gauge boson $s$--channel exchange and
would act as a background.  

\begin{figure}[!h]
\setlength{\unitlength}{1pt}
\centerline{
\begin{picture}(120,85)(0,0)
\Text     ( -35,70)[l]{\red{\bf (a)}}
\ArrowLine(10,70)(40,40)
\ArrowLine(40,40)(10,10)
\Photon   (40,40)(80,40){2}{4}
\ArrowLine(80,40)(110,70)
\ArrowLine(110,10)(80,40)
\Vertex   (40,40){2.0}
\Vertex   (80,40){2.0}
\Text     ( -8,70)[l]{$\mu^-$}
\Text     ( -8,10)[l]{$\mu^+$}
\Text     (115,10)[l]{$\bar b$}
\Text     (115,70)[l]{$b$}
\Text     ( 50,28)[l]{$\gamma,Z$}
\end{picture}
\hspace*{2em}
\begin{picture}(120,85)(0,0)
\ArrowLine(10,70)(40,40)
\ArrowLine(40,40)(10,10)
\DashLine(40,40)(80,40){5}
\ArrowLine(80,40)(110,70)
\ArrowLine(110,10)(80,40)
\Vertex   (40,40){2.0}
\Vertex   (80,40){2.0}
\Text     ( -8,70)[l]{$\mu^-$}
\Text     ( -8,10)[l]{$\mu^+$}
\Text     (115,10)[l]{$\bar b$}
\Text     (115,70)[l]{$b$}
\Text     ( 50,28)[l]{$H$}
\end{picture} }
\vspace*{.5em}
\centerline{
\hspace*{2em}
\begin{picture}(120,85)(0,0)
\ArrowLine(10,70)(25,55)
\ArrowLine(25,55)(40,40)
\ArrowLine(40,40)(25,25)
\ArrowLine(25,25)(10,10)
\Photon   (40,40)(80,40){2}{4}
\DashLine(40,40)(80,40){4}
\ArrowLine(80,40)(110,70)
\ArrowLine(110,10)(80,40)
\Photon   (25,55)(25,25){2}{4}
\Vertex   (40,40){2.0}
\Vertex   (80,40){2.0}
\Vertex   (25,55){2.0}
\Vertex   (25,25){2.0}
\Text     ( -35,70)[l]{\red{\bf (b)}}
\Text     ( -8,70)[l]{$\mu^-$}
\Text     ( -8,10)[l]{$\mu^+$}
\Text     (115,10)[l]{$\bar b$}
\Text     (115,70)[l]{$b$}
\Text     ( 12,38)[l]{$\gamma$}
\Text     ( 40,28)[l]{$\gamma,Z,H$}
\end{picture}
\hspace*{2em}
\begin{picture}(120,85)(0,0)
\ArrowLine(10,70)(40,65)
\ArrowLine(40,65)(40,15)
\ArrowLine(40,15)(10,10)
\Photon   (40,65)(80,65){2}{4}
\Photon   (40,15)(80,15){2}{4}
\ArrowLine(80,65)(110,70)
\ArrowLine(80,15)(80,65)
\ArrowLine(110,10)(80,15)
\Vertex   (40,15){2.0}
\Vertex   (80,15){2.0}
\Vertex   (40,65){2.0}
\Vertex   (80,65){2.0}
\Text     ( -8,70)[l]{$\mu^-$}
\Text     ( -8,10)[l]{$\mu^+$}
\Text     ( 27,38)[l]{$\mu$}
\Text     (115,10)[l]{$\bar f$}
\Text     (115,70)[l]{$f$}
\Text     ( 87,38)[l]{$f$}
\Text     ( 55,77)[l]{$\gamma,Z$}
\Text     ( 55,03)[l]{$\gamma,Z$}
\end{picture}
\hspace*{2em}
\begin{picture}(120,85)(0,0)
\ArrowLine(10,70)(25,55)
\ArrowLine(25,55)(40,40)
\ArrowLine(40,40)(10,10)
\Photon   (40,40)(80,40){2}{4}
\DashLine   (40,40)(80,40){4}
\ArrowLine(80,40)(110,70)
\ArrowLine(110,10)(80,40)
\Photon   (25,55)(60,70){2}{4}
\Vertex   (40,40){2.0}
\Vertex   (80,40){2.0}
\Vertex   (25,55){2.0}
\Text     ( -8,70)[l]{$\mu^-$}
\Text     ( -8,10)[l]{$\mu^+$}
\Text     (115,10)[l]{$\bar b$}
\Text     (115,70)[l]{$b$}
\Text     ( 68,70)[l]{$\gamma$}
\Text     ( 40,28)[l]{$\gamma,Z,H$}
\end{picture}
}
\vspace*{.5em}
\centerline{
\begin{picture}(120,85)(0,0)
\ArrowLine(10,70)(40,40)
\ArrowLine(40,40)(10,10)
\Photon   (40,40)(80,40){2}{4}
\DashLine   (40,40)(80,40){4}
\ArrowLine(80,40)( 95,55)
\ArrowLine(95,55)(110,70)
\ArrowLine(110,10)(95,25)
\ArrowLine( 95,25)(80,40)
\Gluon   (95,55)(95,25){2}{4}
\Vertex   (40,40){2.0}
\Vertex   (80,40){2.0}
\Vertex   (95,55){2.0}
\Vertex   (95,25){2.0}
\Text     ( -35,70)[l]{\red{\bf (c)}}
\Text     ( -8,70)[l]{$\mu^-$}
\Text     ( -8,10)[l]{$\mu^+$}
\Text     (115,10)[l]{$\bar b$}
\Text     (115,70)[l]{$b$}
\Text     (102,38)[l]{$g$}
\Text     ( 40,28)[l]{$\gamma,Z,H$}
\end{picture}
\hspace*{2em}
\begin{picture}(120,85)(0,0)
\ArrowLine(10,70)(40,40)
\ArrowLine(40,40)(10,10)
\Photon   (40,40)(80,40){2}{4}
\DashLine(40,40)(80,40){5}
\ArrowLine(80,40)( 95,55)
\ArrowLine(95,55)(110,70)
\ArrowLine(110,10)(95,25)
\ArrowLine( 95,25)(80,40)
\Gluon   (95,55)(110,40){2}{3}
\Vertex   (40,40){2.0}
\Vertex   (80,40){2.0}
\Vertex   (95,55){2.0}
\Vertex   (95,25){2.0}
\Text     ( -8,70)[l]{$\mu^-$}
\Text     ( -8,10)[l]{$\mu^+$}
\Text     (115,10)[l]{$\bar b$}
\Text     (115,70)[l]{$b$}
\Text     (102,38)[l]{$g$}
\Text     ( 40,28)[l]{$\gamma,Z,H$}
\end{picture} }
\vspace*{2mm} 
\nn {\it Figure 4.47: Lowest-order diagrams for $\mu^-\mu^+\to b\bar b$ 
including the continuum and resonant channels (a) as well as the photonic
QED (b) and the final state QCD corrections (c).}
\vspace*{-1mm} 
\end{figure}

The photonic corrections which include the gauge invariant subset of initial 
and final state virtual corrections and box diagrams involving at least one 
photon as well a initial and final state photon radiation, Fig.~4.47b, and
the QCD corrections to the final state with virtual gluon exchange and 
gluon emission, Fig.~4.47c, have been calculated in Ref.~\cite{mu-RC} with
a careful treatment of both the $Z$ and Higgs boson resonances.  In the case
where no energy resolution is included, the results are shown in Fig.~4.48
for the production of SM Higgs bosons with masses $M_H=115$ GeV and 150 GeV. 
The tree--level couplings have been expressed in terms of $G_\mu$ to encapsulate
the leading electroweak correction and the running $b$--quark mass has been 
used in the $Hb\bar b$ coupling to absorb the bulk of the QCD corrections.\s

\begin{figure}[h]
\vspace*{-3mm} 
\setlength{\unitlength}{1cm}
\centerline{
\begin{picture}(15.5,8.0)
\put(-5.0,-15.1){\includegraphics{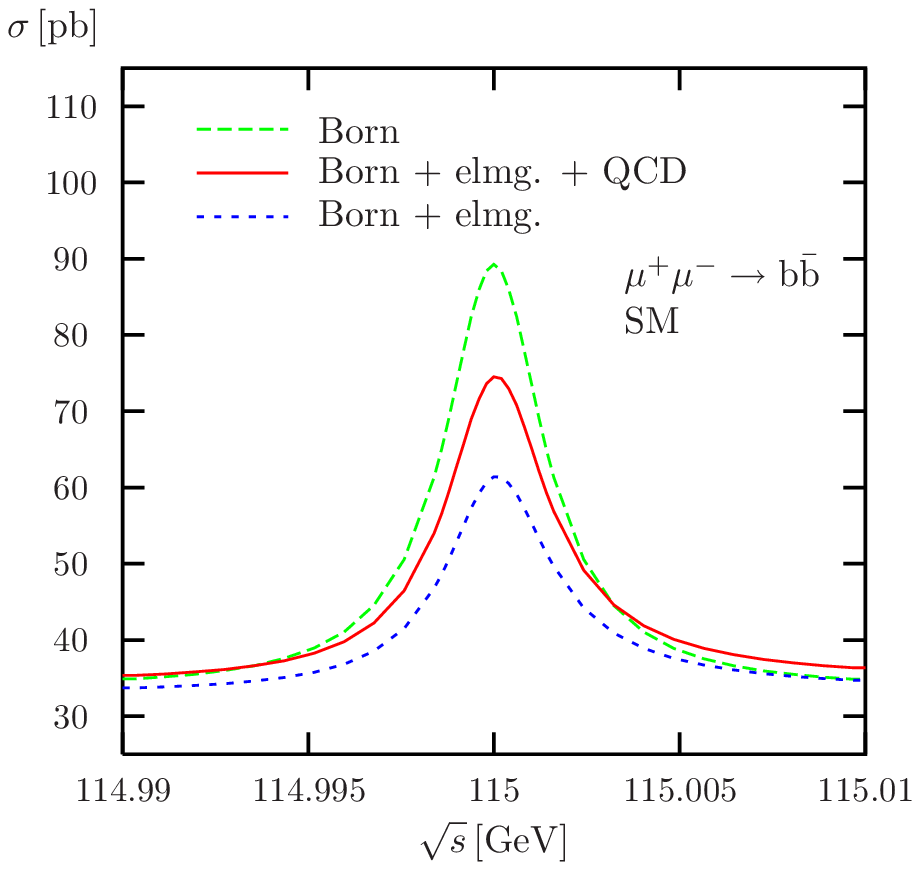}}
\put( 3.0,-15.1){\includegraphics{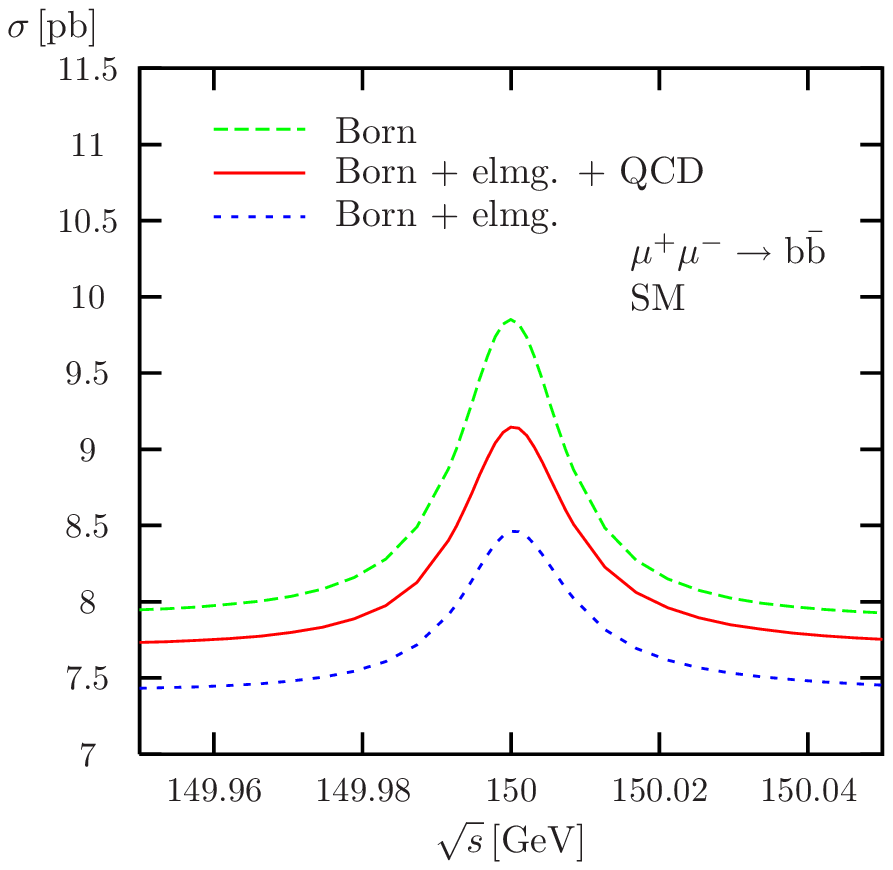}}
\end{picture} }
\vspace*{1mm}  
{\it Figure 4.48: The effective cross section for Higgs production in 
$\mu^+\mu^-\to b\bar b$ for $M_H=115$ and 150 GeV. Shown are the Born cross 
sections and the cross section with electromagnetic and QCD corrections. No 
energy resolution has been assumed; from Ref.~\cite{mu-RC}.} 
\vspace*{-3mm} 
\end{figure}

For the photonic corrections, the large ISR corrections from the radiative
return to the $Z$ resonance can be suppressed by requiring that the invariant
mass of the hadronic final state, thus including gluon radiation, should not
exceed 10 GeV  compared to the Higgs mass, $M_{\rm had} > \sqrt{s} -10$ GeV. 
[For  continuum production, the main difference between $\ee$ and $\mu^+
\mu^-$ collisions is due to the different leading logarithmic photonic
corrections, $\log (s/m_\ell^2)$, which lead to ISR effects that are roughly a
factor of two smaller in $\mu^+ \mu^-$ than in $\ee$ collisions.] With this cut,
the photonic corrections which are still dominated by ${\cal O}(\alpha)$ ISR 
turn negative and of order $- 5\, (10)\%$ for $M_H=115\, (150)$ GeV
for the continuum production and $\sim -50\%$ for the resonant production,
leading to a reduction of the resonance peak compared to the continuum
background.  The QCD corrections are positive and, as they are larger for the
Higgs mediated channel $[\sim 20\%$ as discussed in \S2.1.2] compared to $b\bar
b$ continuum production [$\sim \frac{\alpha_s}{\pi} \sim 4\%$], they tend to
enhance the resonance peak. \s

When including a beam energy resolution $R=0.003\%$, the relative impact of the 
radiative correction stays the same. However, the  signal peaks are suppressed,
in particular for small Higgs masses. For instance, the ratio is $\sigma_{
\sqrt{s}}/\Gamma_H \sim 0.7$ at $M_H=115$ GeV, compared to $\sigma_{\sqrt{s}}/
\Gamma_H \sim 0.2$ at $M_H=150$ GeV, as can be seen in Fig.~4.49.  

\begin{figure}[h]
\vspace*{-3mm} 
\setlength{\unitlength}{1cm}
\centerline{
\begin{picture}(15.5,8.0)
\put(-5.0,-15.1){\includegraphics{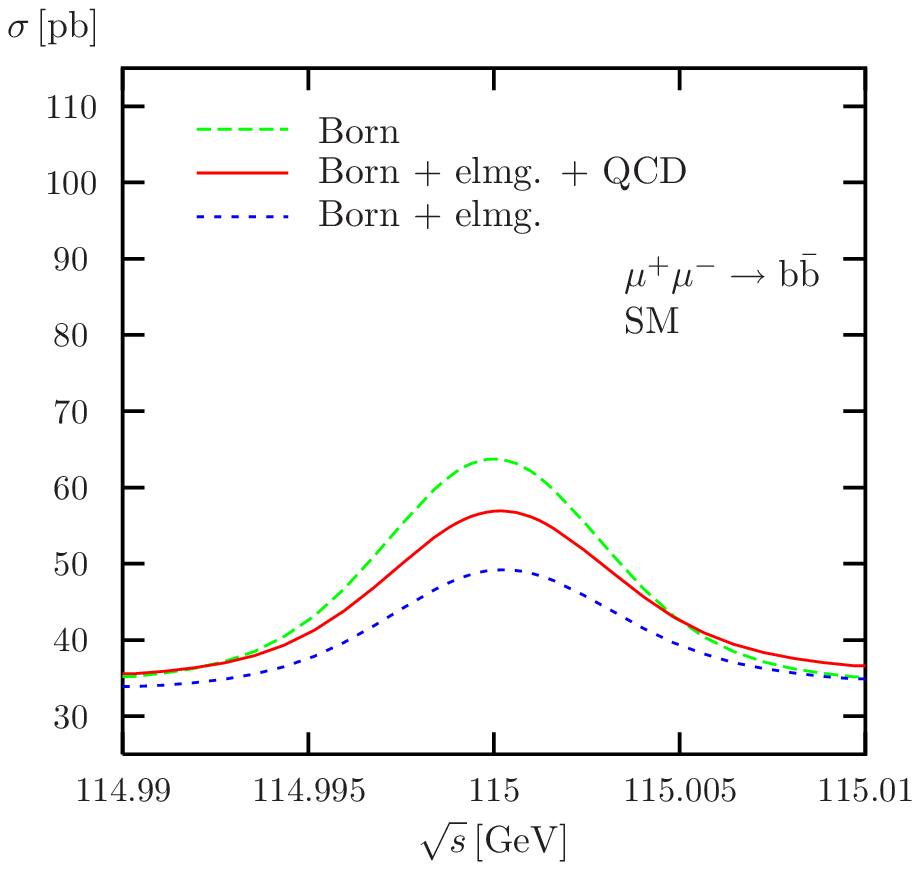}}
\put( 3.0,-15.1){\includegraphics{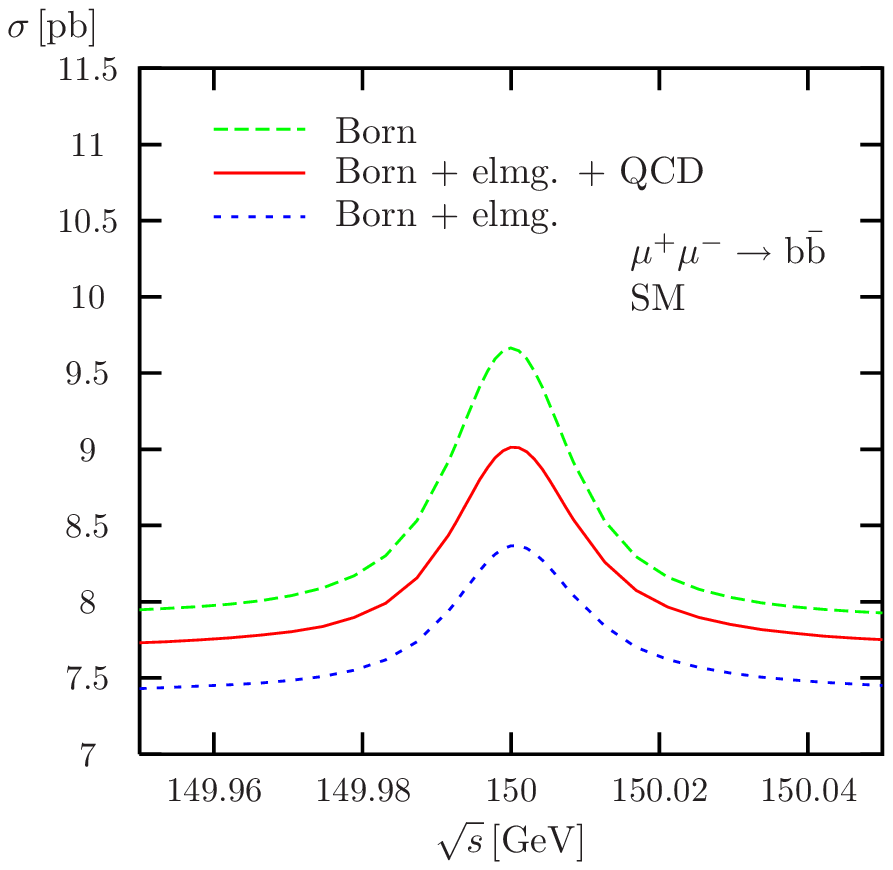}}
\end{picture} } 
\vspace*{1mm} 
{\it Figure 4.49: Same as in Fig.~4.48, but with an energy resolution $R=0.003
\%$; from \cite{mu-RC}.} 
\vspace*{-4mm} 
\end{figure}

\subsubsection*{\underline{Signals and backgrounds in $\mu^+ \mu^- \to
H \to b\bar b, WW, ZZ$}} 

In the main Higgs decay channels, $H \to b\bar b, WW, ZZ$, the cross sections 
for the signals and the corresponding SM backgrounds are shown in the left--hand
side of Fig.~4.50 as a function of the Higgs mass in the range $M_H=80$--160 
GeV for an energy resolution $R=0.003\%$. In the right--hand side of the 
figure, the luminosity that is required to observe the signal at the 5$\sigma$ 
level is displayed for the same energy resolution. Various cuts have been 
applied to reject part of the background [$b$--tagging, cuts to remove gauge
bosons in the forward and backward directions] and are discussed in 
Ref.~\cite{mu-Rev1} from which we borrowed the figure. \s

As can be seen, the $\mu^+ \mu^- \to H \to b\bar{b}$ signal rate is rather 
large for $M_H \lsim 140$ GeV, leading to ${\cal O}(10^4)$ events for a 
luminosity of ${\cal L}=1~{\rm fb}^{-1}$. The backgrounds from direct $\mu^+ 
\mu^- \to \gamma, Z^* \to b\bar{b}$ production\footnote{Since  the background is
practically constant in the window  $M_H \pm \sigma_{\sqrt{s}}$, one can
measure it below and above the resonance and eventually, subtract it if enough
luminosity is available} [the light quark--jet background can be removed with
$b$--tagging] is much larger than the signal for a Higgs boson in the mass
range $M_H \lsim 115$ GeV which is ruled out by the LEP2 negative searches [in
particular, for $M_H \sim 90$ GeV a huge background from the resonant
production $\mu^+ \mu^-  \to Z \to b\bar{b}$ is present], and is of comparable
size as the signal in the mass range 115 GeV $\lsim M_H \lsim$ 135 GeV. For 
larger masses, the signal  event drop dramatically because of the decrease of 
the $H \to b\bar{b}$  branching ratio. \s

In the case of gauge boson production, $\mu^+ \mu^- \to WW^*$ and $ZZ^*$, the 
event rates are much smaller than those of the $b\bar{b}$ final
states in the low Higgs mass range, as a result of the tiny branching ratios. 
For larger masses, $M_H \sim 140$ GeV, the $WW$ and $b\bar b$ cross sections
become comparable but the absolute rates are rather small;  for $M_H \gsim 160$
GeV, the cross sections are below the femtobarn level. The backgrounds from
continuum $\mu^+ \mu^- \to WW^*, ZZ^*$ production [once cuts have been applied
to remove for instance the forward and backward events which are rare in the
signal where the Higgs boson is centrally produced] do not exceed the signal
cross sections for $M_H \lsim 150$ GeV.  For higher Higgs masses, $M_H \gsim
160$ GeV, when the production of two real gauge bosons opens up kinematically,
the backgrounds become much larger than the resonant signal. \s

\begin{figure}[h]
\vspace*{-1mm}
\begin{center}
{\epsfig{file=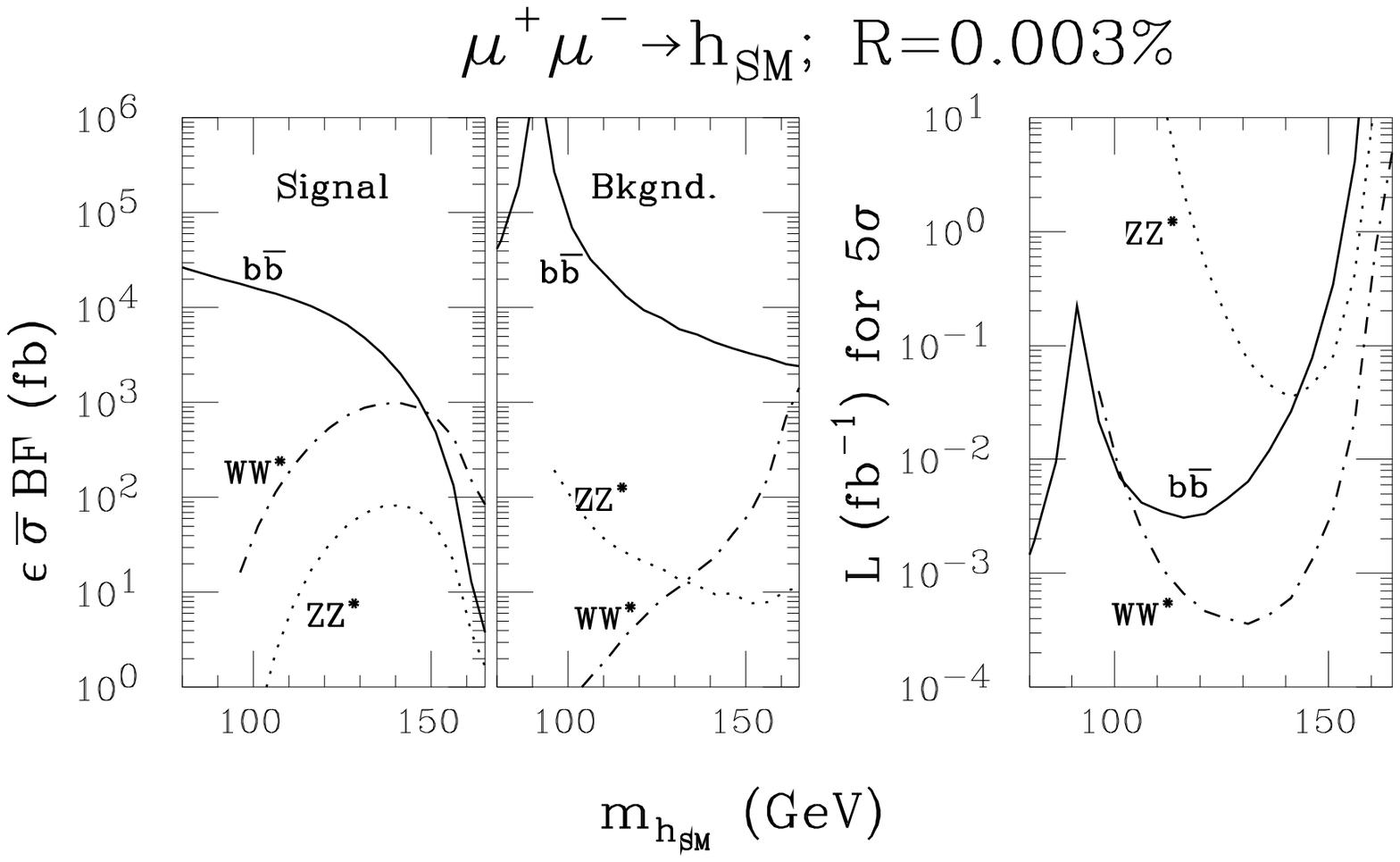,width=6in,height=4in}}
\end{center}
\vspace*{-6mm}
{\it Figure 4.50: The cross sections for the processes $\mu^+\mu^-\to b\bar{b}, 
WW, ZZ$ for signals and backgrounds as a function of $M_H$ for $R=0.003$\%
(left) and the luminosity required for a $5\sigma$ observation of the 
$\mu^+\mu^-\to H \to b\bar{b}, WW, ZZ$ signals (right); from 
Ref.~\cite{mu-Rev1}.} 
\vspace*{-2mm}
\end{figure}

For a SM Higgs boson with a mass $M_H\gsim 2M_W$, $s$--channel production
in $\mu^+ \mu^-$ colliders will, anyway, not be very useful since the total
width becomes large and the $H \to \mu^+ \mu^-$ decay branching fraction
drops drastically.  However, there are extensions of the SM in which Higgs
bosons can have relatively large masses but suppressed total  widths [this is 
the case of e.g. pseudoscalar Higgs bosons which do not couple to 
massive gauge bosons at tree--level]. In this case, the production rates 
are not very suppressed and muon colliders can be valuable tools in 
determining their properties as will be discussed in another part of this 
review.  

\subsubsection{Determination of the properties of a light Higgs boson}

In the SM, for Higgs bosons in the mass range $M_H \lsim 160$ GeV, important
measurements can be  performed at the muon collider in the  channels $\mu^+
\mu^- \to H \to  b\bar{b}, WW^*,ZZ^*$,  which  have sizable production rates as
shown previously, as well as in the channel $\mu^+ \mu^- \to H \to \tau^+
\tau^-$.   The Higgs mass, its total decay width and the cross section
for the various final states, which are sensitive to the branching fractions 
and thus the Higgs couplings, can be determined.\s

The Higgs mass can be measured by a straightforward scan in the vicinity
of $\sqrt{s}= M_H$.  The approximate values of $M_H$  would be already known
from measurements at $\ee$ and hadron colliders, or being measured at the muon
collider by producing  first the Higgs boson in the Higgs--strahlung channel,
$\mu^+ \mu^- \to HZ$.  The detection of the signal peak for a Higgs mass
$M_H=110$ GeV has been performed e.g. in Ref.~\cite{mu-Murray} and the output
is summarized in Fig.~4.51 which has been obtained with 10 pb$^{-1}$ data,
assuming that the beam energy spread is very small.  The Monte Carlo generator
{\tt PYTHIA} has been used to generate the $\mu^+\mu^- \to H \to b\bar b$ signal
and the $\mu^+ \mu^- \to q\bar q (\gamma)$ background events and a crude
estimate of detector effects [using a typical LEP detector] has been made. It
has been assumed that 80\% efficiency for $b$--quark tagging can be achieved
as expected at the LC for instance.  For such a Higgs mass, one is close the
$Z$ boson resonance and the backgrounds are rather large; they become smaller
when one moves to higher Higgs masses, but the Higgs branching ratio BR$(H\to 
b\bar b)$ becomes then smaller. In another analysis presented in
Ref.~\cite{mu-precision} but which takes into account the energy spread, it has
been shown that a precision of the order of $\Delta M_H \sim 0.1$ MeV for $M_H 
\simeq 115$ GeV can be achieved with $\sim 30$ data points with a luminosity 
${\cal L}=1.25~{\rm pb}^{-1}$ per point and a resolution $R=0.003$\%.\s

\begin{figure}[h]
\vspace*{-3mm}
\begin{center}
\epsfig{file=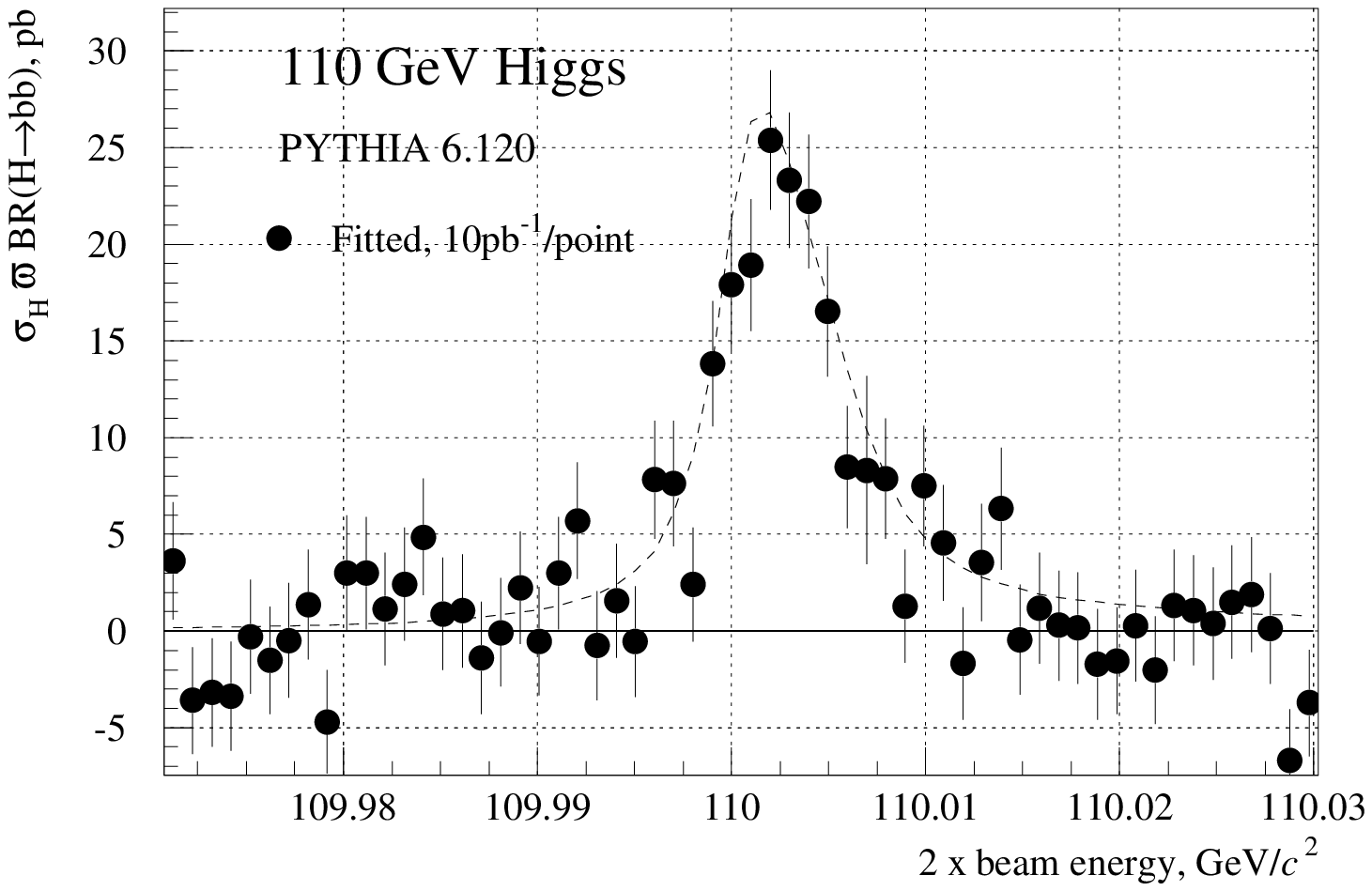,width=0.62\textwidth}
\end{center}
\vspace*{-2mm}
\nn {\it Figure 4.51: The $\mu^+ \mu^- \to b\bar b$ production cross section
as a function of the beam energy for $M_H=110$ GeV; the points corresponding 
to 10~pb$^{-1}$ data are superimposed and no beam energy spread is taken into
account; from Ref.~\cite{mu-Murray}.} 
\vspace*{-2mm}
\end{figure}

Since both the Higgs mass $M_H$ and its total width $\Gamma_H$ enter the cross
section value at the same time, and if one refrains from making any theoretical
assumption on the  partial decay widths, a model--independent determination of
$\Gamma_H$ is required. This determination can be made
\cite{mu-Rev1,mulineshape} by noting that if one adjusts the normalization of
the theoretical curve of $\Gamma_H$ at $\sqrt{s}=M_H$ in such a way that it
agrees with the experimental curve, then the wings of the theoretical curve are
increased (decreased) if $\Gamma_H$ is larger (smaller).  With precise
measurements at a central energy value $\sqrt{s}$ and at the wings, one can
measure $\Gamma_H$ through the ratio of the cross sections at the central peak
and on the wings, in which the partial decay widths cancel out. This method,
with a dedicated three point scan near the threshold, allows to measure $M_H$
at the same time with a precision that is expected to be better than the 
rough scan discussed above, with the same integrated luminosity.\s

However, for a light Higgs boson which might be very narrow, one could achieve
a very small beam energy resolution, $\sigma_{\sqrt{s}} \ll \Gamma_H$, only at
a cost of a low luminosity. In this case, it has been advocated to operate
the collider  at the Higgs peak with two different resolutions, $\sigma_{\sqrt{
s}}^{\rm min} \ll \Gamma_H$ and $\sigma^{\rm max}_{ \sqrt{s}} \gg \Gamma_H$ 
and determine the total width from the ratios of the peak cross sections
\cite{mulineshape}. Using eqs.~(\ref{mu-gamma-small})--(\ref{mumuxsection2}), 
one obtains
\beq
\overline{\sigma}_H (\sigma_{\sqrt{s}}^{\rm min}) /
\overline{\sigma}_H (\sigma_{\sqrt{s}}^{\rm max}) = [2\sqrt 2 \sigma_{\sqrt{
s}}^{\rm min}]/[\sqrt \pi \Gamma_H]
\eeq
Figure 4.52a, shows the results of a Monte-Carlo simulation \cite{mu-Rev4}
for the determination of the total
decay width of the Higgs boson as a function of its mass in the range between
100 and 160 GeV. Also shown in the figure are the spread in the c.m. energy for
a resolution of $R = 0.003\%$ (solid circles) and the spread that is obtained
if the resolution $R$ is varied in such a way that the beam energy spread is
always 40\% of the Higgs total decay width (open squares). Here, one assumes
that any value of the resolution $R$ can be obtained and that the luminosity
scales as $R^{2/3}$;  this procedure helps to optimize the Higgs production
rate and, hence, the statistical error on the production cross section.\s

Figure 4.52b displays the factor which reduces the peak cross section when the 
Gaussian distribution with a width $\sigma_{\sqrt s}$ is included. This 
signal reduction factor is given by
\beq
S_R = \eta \, A e^{A^2} \left( \sqrt \pi - 2 \int_0^A \, e^{-t^2} {\rm d}t
\right) \ , \quad A=   \frac{1}{2\sqrt{2}}\, \frac{\Gamma_H}{\sigma_{\sqrt s} }
\eeq
where $\eta$ is factor which takes into account the effects of ISR. With a 
fixed resolution, $R = 0.003\%$ (filled circles), the signal cross section is 
reduced by approximately a factor of two for low Higgs masses, $M_H \lsim 130$ 
GeV, while for masses close to $M_H \sim 160$ GeV, the Higgs total width 
becomes large enough and there is no reduction.  For an optimally varying $R$ 
value (open squares), the peak cross section is reduced by a constant factor 
$S_R \simeq 0.8$.\s 

\begin{figure}[h!]
\begin{center}
\epsfig{figure=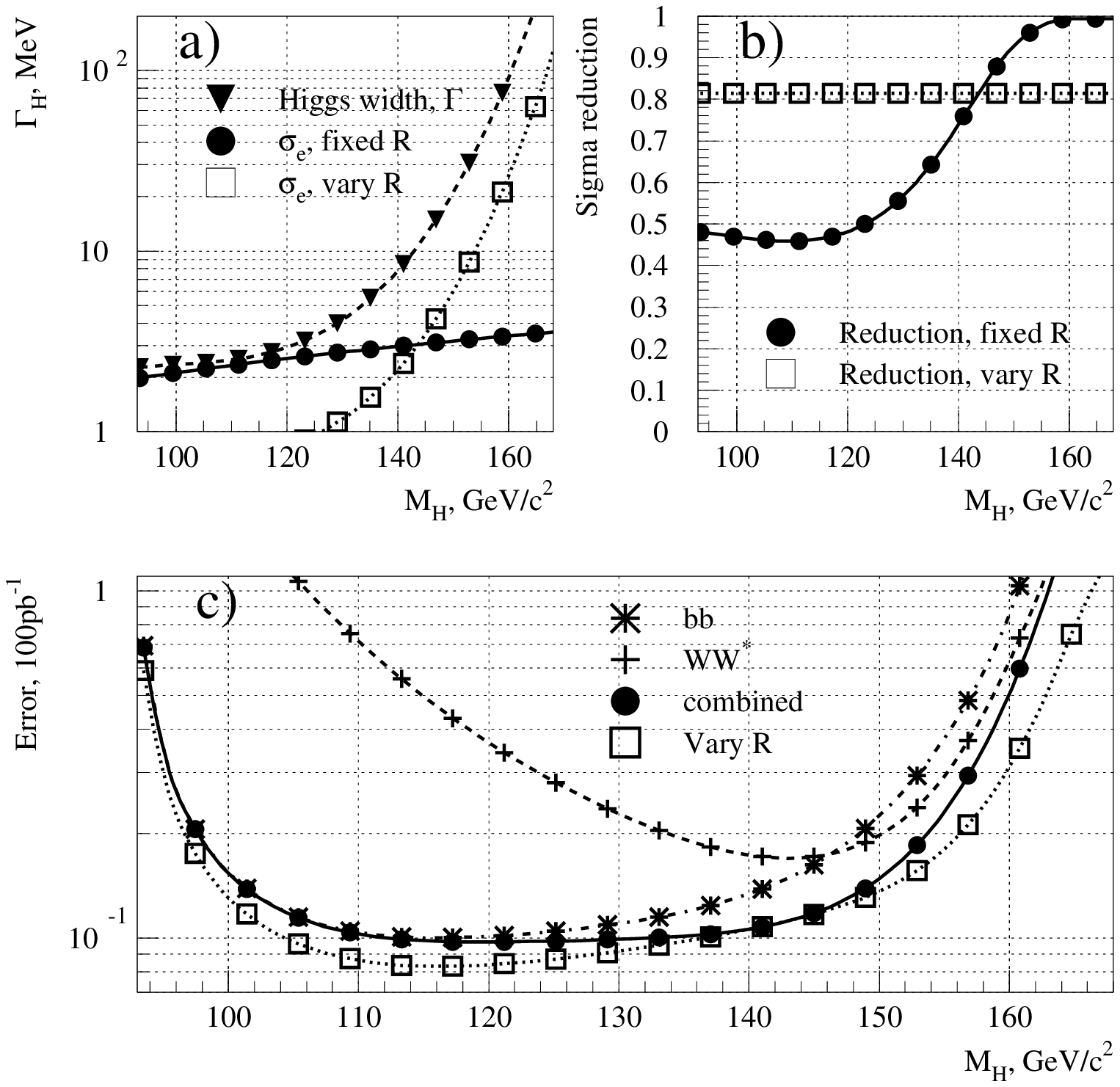,width=15cm,height=15cm}
\vspace*{-1.cm}
\end{center}  
\nn {\it Figure 4.52: a): The width of the SM Higgs boson as a function of its
mass (triangles), also shown are    the effects of a fixed c.m. energy spread
for $R=0.003\%$ (filled circles) and an optimal varying energy spread (open
squares). b) The cross section suppression factor due to the width of the
beams if $R = 0.003\%$ (filled circles) and for the optimal varying $R$ (open
squares). c): The fractional error with which the Higgs cross section can be
measured in the $b\overline{b}$ (stars) and $WW^*$ (crosses)  decay modes using
100~pb$^{-1}$ data with $R=0.003\%$; the solid circles show the accuracy with
which the peak cross sections can be extracted if the SM branching ratios are
assumed and the open squares show the errors obtained in the same running period
by optimizing $R$. From Ref.~\cite{mu-Rev4}.}
\vspace*{-.3cm}
\end{figure}

The peak cross sections for the processes $\mu^+ \mu^- \to b\bar b$ and $\mu^+
\mu^- \to WW^{(*)}$ are shown in Fig.~4.52c as a function of $M_H$ under the
same conditions  as above with an integrated luminosity of ${\cal L}=100$ pb$^{
-1}$. While the simulation of $b\bar b$ decays is as described previously, the
efficiency in the channel $H\to WW^*$ is based on a LEP2--type detector with the
conservative assumption that the spin information is not used to further reduce
the non--resonant $WW$ background.  As can be seen, the $b\bar b$ cross section
can be measured with a statistical accuracy of about 10\%, while a 20\%
accuracy can be achieved on the $WW^*$ cross section for $M_H=130$--150 GeV.
The accuracy will improve by a factor of two if the luminosity is increased to
${\cal L}=400$ pb$^{-1}$, which corresponds to four years of running, thereby
approaching those which can be achieved at the LC. The accuracy on the 
production cross section for the $ZZ$ final state is expected to be worse as 
BR$(H\to ZZ^*)$ is very small in this mass range. \s  

Once the Higgs mass, total decay width and peak cross sections have been
determined, one can measure the partial decay width  into muons $\Gamma
(H\to \mu^+ \mu^-)$ and the final state branching ratios BR$(H \to X)$.  From
the cross section eq.~(\ref{mumuxsection2}), they appear
as the product
\beq
B(X) = \Gamma(H \to \mu^+ \mu^-) \times {\rm BR}(H \to X)
\eeq
These measurements can be then combined with other precision measurements
performed at the LHC and/or at $\ee$ colliders to determine the couplings of
the Higgs boson in a model independent way. For instance, the Higgs partial
decay width into muons, if the measurements at the linear collider 
discussed in the previous section  are available, can be determined through
\cite{mu-Rev2}
\beq
\Gamma (H \to \mu^+ \mu^-) &=& \frac{B(X)} { {\rm BR}(H \to X)_{\rm LC} }
\eeq
with $b\bar{b}$ and $WW^*$ final states for instance, where the branching
ratios in the denominator can be measured precisely at the LC in the low Higgs
mass range,  or make use of the total decay width measured at muon colliders,
via
\beq
\Gamma (H \to \mu^+ \mu^-) &=& \frac{B(X) \times \Gamma_H }{\Gamma(H \to 
X)|_{\rm LC}}
\eeq
for the $WW^*$ and eventually $ZZ^*$ final states where the partial widths can 
be measured at the LC for large enough Higgs boson masses. The combination of 
all measurements allow a very precise test of the Higgs couplings to fermions 
and gauge bosons and, in particular, a precise determination of the Higgs 
couplings to second generation fermions. 

\subsubsection{Study of the CP properties of the Higgs boson}

\subsubsection*{\underline{Measurements in the decay $H \to \tau^+ \tau^-$}}

A very interesting process to study at muon colliders is $\mu^+ \mu^- \to
\tau^+ \tau^-$ \cite{mu-Htau,mu-Htau0}. The process proceeds
through $s$--channel photon and $Z$ boson exchanges as well as via $s$--channel
Higgs boson exchange. In the former channels, the production is similar to what
occurs in $\ee$ collisions and,  as discussed in \S1.2, the process has
characteristic forward--backward and left--right asymmetries. Assuming that the
$\mu^\pm$ beams have longitudinal  polarizations $P_\pm$ and  using
eqs.~(\ref{eexsection},\ref{eeafb},\ref{eealr}) for the cross
section and asymmetries,  one can write the differential cross section for
$\mu^+ \mu^- \to \gamma, Z^* \to \tau^+ \tau^-$ as
\beq
\frac{ {\rm d}\sigma_{\gamma,Z}}{ {\rm d} \cos \theta}=\frac{4\pi \alpha^2}{3s}
 \times \frac{3}{8} \sigma_U \left[ (1+ \cos^2 \theta) + \frac{8}{3} \cos 
\theta A_{FB}^{\rm eff}  \right]
\eeq
where $A_{FB}^{\rm eff}$ has the usual component $A_{FB}^\tau$ already 
discussed, but also a component which includes the information on the
longitudinal polarization and $A_{LR,FB}^\tau$ in eq.~(\ref{eealr})
\beq
A_{FB}^{\rm eff} = \frac{ A_{FB}^{\tau} + P_{\rm eff} A^\tau_{LR,FB} }
{1+ P_{\rm eff} A_{LR}^\tau } \ \ , \  P_{\rm eff}= \frac{P_+ - P_-}
{1- P_+ P_-}
\eeq
The angular distribution has a clear forward--backward asymmetry: it vanishes
for $\theta=\frac{\pi}{2}$ and is large and positive for $\theta=0$. For 
$\sqrt{s}=120$ GeV one has  $A_{FB} \sim 0.7$ and $A_{LR}=0.15$
\cite{mu-Htau}. \s

In the Higgs boson exchange channel, the differential cross section is
flat and is simply given, in terms of the effective cross section with
$\sqrt{s}=M_H$,  by
\beq
\frac{ {\rm d} \sigma}{ {\rm d} \cos \theta} = \frac{1}{2} \overline{\sigma}_H
\left (1 + P_+ P_- \right)
\eeq
Considering this channel as the signal and the $\gamma,Z$ exchange channel
as the background, the enhancement of the signal cross section compared to
the background is given by
\beq
\frac{S}{B} \sim \frac{1+ P_+ P_-}{ 1-P_+ P_- +(P_+ - P_-) A_{LR}^\tau}
\eeq
One can therefore use the polarization of the initial beams and the
forward--backward asymmetries to enhance the signal--to--background ratio. \s

One can also distinguish the signal from the background by using the final
state polarization of the $\tau$ leptons which are very different. We will
briefly discuss this point, following Ref.~\cite{mu-Htau} and recalling
the discussion of \S2.1.4 on $H \to \tau^+ \tau^-$ decays. In the two--body
decays of the $\tau$ lepton,  $\tau^- \to \pi^- \nu_\tau, \rho^- \nu_\tau,\ 
etc..$, defining  $\theta_i$ as the angle between the momenta of the $\tau$
lepton and the charged final  particle, $B$ as the branching ratio of the decay
and $P_\tau = \pm 1$ as the $\tau$ lepton helicity, the normalized differential 
decay rate in the $\tau$ rest frame is simply 
\beq
\frac{1}{\Gamma} \frac{ {\rm d}\Gamma_{i} }{ {\rm d}c_{\theta_i}} =
\frac{B_i}{2} (1+ r_i P_{\tau} \cos  \theta_i )
\eeq
with $r_i=1$ for decays into pions and $r_i= -(m_\tau^2 -2m_i^2)/
(m_\tau^2+ 2 m_i^2) \simeq 0.45$ for $i=\rho$. In the signal, $\mu^+ \mu^- \to 
H \to \tau^+ \tau^-$, the $\tau^-$ and $\tau^+$ helicities are correlated as 
$P_{\tau^-}\!=\!P_{\tau^+}\!=\!\pm 1$, and the spin correlated  differential 
cross section with polarized $P_\pm$ beams reads
\beq
\frac{ {\rm d}\sigma_H} { {\rm d}c_{\theta_i} {\rm d}c_{\theta_j} }
= (1+ P_- P_+) \, \sigma_H \, \frac{B_i B_j}{4} \, \left[ a_i a_j + b_i b_j
c_{\theta_i} c_{\theta_j} \right]
\eeq
reaching a maximum (minimum) for 
$c_{\theta_i}\!=\!c_{\theta_j}(c_{\theta_i}\!=-c_{\theta_j})\!=\pm \!1$, with 
the significance of the peaks depending on the  magnitude of $r_i$. In the case
of the decay  $\tau^-\to \rho^- \nu_\tau$, distinctive peaks in the distribution
can be seen  for $c_{\theta_{\rho^-}}$=$c_{\theta_{\rho^+}}=\pm 1$ and
$P_+=P_-=25$\%  \cite{mu-Htau}. \s

In contrast, in the standard channel $\mu^+ \mu^- \to \gamma, Z^* \to \tau^+
\tau^-$, the $\tau$ leptons are produced with helicities $P_{\tau^-} = -
P_{\tau^+} = \pm1$ and the number of left--handed and right--handed
$\tau$ leptons  are different because of the polarization left--right asymmetry.
The spin correlated  differential distribution in this case is 
 \beq
\frac{ {\rm d}\sigma_{\gamma,Z}} { {\rm d}c_{\theta_i} {\rm d}c_{\theta_j}}
&=& (1- P_- P_+) \sigma_{\gamma,Z} \, (1+ P_{\rm eff} A_{LR} ) \, \non \\
&\times & \frac{1}{4} B_i B_j \left[ (a_i a_j - b_i b_j c_{\theta_i} 
c_{\theta_j})+ (a_i b_j c_{\theta_j} - a_j b_i  c_{\theta_i}  )A_{LR}^{\rm eff} \right]
\label{mumucore-bckg}
\eeq
with
\beq
A_{LR}^{\rm eff}= \frac{A_{FB,LR}^\tau+ P_{\rm eff} A_{FB}^\tau }{1+ P_{\rm 
eff} A_{LR}^\tau } 
\eeq
Again, for the decay  $\tau^- \to \rho^- \nu_\tau$, the peaks in the
distribution  for $\cos  \theta_{\rho^-} = - \cos \theta_{\rho^+} = \pm 1$ can
be seen for  $P_+=P_-=25$\% \cite{mu-Htau}. The peaks occur in opposite
regions compared to the Higgs signal and the spin correlation in the signal is
symmetric, while it is not the case in the background as  a consequence
of the presence of the term $A_{LR}^{\rm eff}$ in eq.~(\ref{mumucore-bckg}). \s

Summing both $\rho\nu$ and $\tau \nu$ final states and using $R=0.0005$\%, 
$P_\pm=25$\% and ${\cal L}=1~{\rm fb}^{-1}$, one obtains the statistical
error $\epsilon\!=\!\sqrt{S+B}/S$ on the cross section measurement which
determines to which extent the $H \tau^+ \tau^-$ coupling can be measured. For
$M_H\!=\!110\, (130)$ GeV, one has $\epsilon \simeq 20$\% (30\%) showing that 
one can observe the resonant $\mu^+ \mu^- \to H \to \tau^+ \tau^-$ process 
above the continuum background and therefore possibly  measure the $H \tau^+ 
\tau^-$ coupling and check the Higgs boson spin. \s

Note that because of depolarization effects, this type  of analysis cannot be
performed in the decays $H\to b\bar b$, while for $H \to t\bar t$ the rates are
too small, the Higgs resonance being very broad as discussed earlier. On the
other hand, the CP quantum numbers of a relatively heavier Higgs boson, $M_H
\gsim 140$ GeV, can be studied in the decays $H\to VV^* \to 4f$ by looking at
angular distributions and correlations as discussed  in detail in \S2.2.4.  
 
\subsubsection*{\underline{CP Measurements with transverse polarization}}

It is expected that muon colliders will have a natural transverse polarization
of the order of 20\% for both the $\mu^+$ and $\mu^-$ beams. This polarization,
if maximized, could also provide an unambiguous test of the CP quantum numbers
of the Higgs boson \cite{CP-GunionGrz,CP-Gunion-GP,mu-pol}, similarly to the
case of $\gamma \gamma$  colliders previously discussed. Indeed, if one
considers a scalar particle with couplings to muons which have both CP--even
and CP--odd components, ${\cal L}(H \mu\mu) \propto H \bar{\mu} (a + i
b\gamma_5)\mu$, and assumes that the muon beams are 100\% transversally
polarized, with $\phi$ being the angle between the $\mu^+$ and $\mu^-$
transverse polarizations, the production cross section of the Higgs boson in
the $s$--channel reads 
\beq
\sigma(\mu^+ \mu^- \to H) \propto 1 - \frac{ a^2- b^2}{a^2+b^2} 
\cos \phi + \frac{2 a b}{ a^2 + b^2} \sin \phi
\eeq
If the Higgs boson is a mixture of CP--even and CP--odd states, both the
$a_+$ and $a_-$ components are non--zero and the asymmetry 
\beq
{\cal A}_1 = \frac{ \sigma( \phi= \frac{\pi}{2})-\sigma( \phi=- \frac{\pi}{2})}
{\sigma( \phi=\frac{\pi}{2})+\sigma( \phi=-\frac{\pi}{2})}=\frac{ 2 a b}
{a^2 + b^2} 
\eeq
is large if $a$ and $b$ have the same magnitude. A non--zero value of this
asymmetry would indicate a clear violation of CP symmetry. For a pure 
CP--eigenstate, one of the coefficients $a$ or $b$ is zero and the 
asymmetry
\beq
{\cal A}_2 = \frac{ \sigma( \phi= \pi) - \sigma( \phi=0)}
{ \sigma( \phi= \pi) + \sigma( \phi=0)} = \frac{a^2 - b^2}
{a^2 + b^2} 
\eeq
is either equal to $+1$ or $-1$, if the Higgs boson is, respectively a CP--even
or a CP--odd state. In the ideal world, this is an unambiguous  test of the CP
nature of the Higgs boson. However, the transverse polarization will most 
probably not be maximal
and background events will alter the signal and dilute the asymmetries. 
Thorough studies, including the machine and background aspects must be
performed to quantify the extent to which the Higgs boson CP--properties can be
measured; see  Ref.~\cite{CP-Gunion-GP} for such an attempt.



\begin{thebibliography}{999}
\bibitem{LCWS1} Recently, an analysis has shown that in the Higgs mass 
mesurement using the recoil mass technique, systematical errors from the beam 
energy scale and differential luminosity distributions lead to an additional 
combined uncertainty of 30 MeV for $M_H=120$ GeV; see A. Resperaza in 
Ref.~\cite{LCWS}.
\bibitem{LCWS2} Using an improved version of the TESLA detector simulation 
program {\tt SIMDET}, it has been shown that the charm tagging capabilities have
been overestimated in the analyses performed for the TESLA TDR and the  
precision on the the $b\bar b, c\bar c$ and $gg$ are thus worse than  
those presented in Ref.~\cite{TESLA}; see T. Kuhl in Ref.~\cite{LCWS}.
\end{thebibliography}
\end{document}